\documentclass{jpp-AAS}

\usepackage{graphicx}
\usepackage{natbib}

\usepackage{amsmath}
\usepackage{amssymb}
\usepackage[makeroom]{cancel}
\usepackage{color}
\usepackage[colorlinks=true,linkcolor=blue,citecolor=blue]{hyperref}

\ifCUPmtlplainloaded \else
  \checkfont{eurm10}
  \iffontfound
    \IfFileExists{upmath.sty}
      {\typeout{^^JFound AMS Euler Roman fonts on the system,
                   using the 'upmath' package.^^J}%
       \usepackage{upmath}}
      {\typeout{^^JFound AMS Euler Roman fonts on the system, but you
                   dont seem to have the}%
       \typeout{'upmath' package installed. JPP.cls can take advantage
                 of these fonts, if you use 'upmath' package.^^J}%
      }
  \else
  \fi
\fi

\ifCUPmtlplainloaded \else
  \checkfont{msam10}
  \iffontfound
    \IfFileExists{amssymb.sty}
      {\typeout{^^JFound AMS Symbol fonts on the system, using the
                'amssymb' package.^^J}%
       \usepackage{amssymb}%
         
       \let\ge=\geqslant  
      }{}
  \fi
\fi

\ifCUPmtlplainloaded \else
  \IfFileExists{amsbsy.sty}
    {\typeout{^^JFound the 'amsbsy' package on the system, using it.^^J}%
     \usepackage{amsbsy}}
    {\providecommand\boldsymbol[1]{\mbox{\boldmath $##1$}}}
\fi

\newcommand{\figref}[1]{figure~\ref{#1}}

\newcommand{\figsand}[2]{figures~\ref{#1}~and~\ref{#2}}
\newcommand{\Figref}[1]{Figure~\ref{#1}}

\newcommand{\Secref}[1]{Section~\ref{#1}}
\newcommand{\secref}[1]{\S\,\ref{#1}}

\newcommand{\secsand}[2]{\S\S\,\ref{#1}~and~\ref{#2}}
\newcommand{\secsdash}[2]{\S\S\,\ref{#1}--\ref{#2}}
\newcommand{\secsref}[1]{\S\S\,\ref{#1}}
\newcommand{\apref}[1]{appendix~\ref{#1}}
\newcommand{\apsand}[2]{appendices~\ref{#1}~and~\ref{#2}}

\newcommand{\apsref}[1]{appendices~\ref{#1}}
\newcommand{\Apref}[1]{Appendix~\ref{#1}}
\newcommand{\Apsand}[2]{Appendices~\ref{#1}~and~\ref{#2}}

\renewcommand{\eqref}[1]{equation~(\ref{#1})}

\newcommand{\eqsdash}[2]{equations~(\ref{#1}--\ref{#2})}
\newcommand{\exref}[1]{(\ref{#1})}
\newcommand{\exsdash}[2]{(\ref{#1}--\ref{#2})}

\newcommand{\partskip}{\\ \hspace*{-5.3mm}}

\newcommand{\bea}{\begin{eqnarray}}
\newcommand{\eea}{\end{eqnarray}}
\newcommand{\beq}{\begin{equation}}
\newcommand{\eeq}{\end{equation}}
\newcommand{\lt}{\left}
\newcommand{\rt}{\right}
\newcommand{\bl}{\bigl}
\newcommand{\br}{\bigr}
\newcommand{\la}{\langle}
\newcommand{\ra}{\rangle}

\newcommand{\mbf}[1]{\boldsymbol{#1}}

\newcommand{\hence}{\quad\Rightarrow\quad}
\newcommand{\rmiff}{\quad\Leftrightarrow\quad}
\newcommand{\rmand}{\quad\mathrm{and}\quad}

\newcommand{\rmor}{\quad\mathrm{or}\quad}

\newcommand{\rmif}{\quad\mathrm{if}\quad}

\newcommand{\rmas}{\quad\mathrm{as}\quad}

\newcommand{\dd}{\partial}
\newcommand{\vdel}{\mbf{\nabla}}
\newcommand{\vdperp}{\vdel_\perp}
\newcommand{\dperp}{\nabla_\perp}
\newcommand{\dpar}{\nabla_\parallel}

\newcommand{\const}{\mathrm{const}}
\newcommand{\rmd}{\mathrm{d}}

\renewcommand{\Re}{\mathrm{Re}}
\newcommand{\tRe}{\widetilde{\Re}}

\newcommand{\eps}{\varepsilon}
\newcommand{\teps}{\tilde\eps}
\newcommand{\epsm}{\eps_\mathrm{m}}
\newcommand{\epsrec}{\epsilon_\mathrm{rec}}

\newcommand{\Rm}{\mathrm{Rm}}
\newcommand{\Rmc}{\Rm_\mathrm{c}}
\newcommand{\Pm}{\mathrm{Pm}}
\newcommand{\Sc}{S_\mathrm{c}}
\newcommand{\tS}{\tilde S}
\newcommand{\tSc}{\tilde S_\mathrm{c}}

\newcommand{\ey}{\hat{\mbf{y}}}
\newcommand{\ez}{\hat{\mbf{z}}}
\newcommand{\vr}{\mbf{r}}
\newcommand{\vrperp}{\vr_\perp}

\newcommand{\rperp}{r_\perp}
\newcommand{\rpar}{r_\parallel}
\newcommand{\vl}{{\mbf{l}}}
\newcommand{\vlam}{{\mbf{\lambda}}}
\newcommand{\lpar}{l_\parallel}
\newcommand{\lparl}{l_{\parallel\lambda}}
\newcommand{\lparlp}{l_{\parallel\lambda'}}
\newcommand{\lparvisc}{l_{\parallel\nu}}
\newcommand{\lperp}{l_\perp}
\newcommand{\lres}{\lambda_\eta}
\newcommand{\lnu}{\ell_\nu}
\newcommand{\lvisc}{\lambda_\nu}
\newcommand{\lviscnew}{\lambda_{\nu,\mathrm{new}}}
\newcommand{\urms}{u_\mathrm{rms}}

\newcommand{\vk}{{\mbf{k}}}
\newcommand{\vp}{{\mbf{p}}}
\newcommand{\vq}{{\mbf{q}}}
\newcommand{\vkperp}{\vk_\perp}
\newcommand{\kperp}{k_\perp}
\newcommand{\vpperp}{\vp_\perp}
\newcommand{\pperp}{p_\perp}
\newcommand{\vqperp}{\vq_\perp}
\newcommand{\qperp}{q_\perp}
\newcommand{\kpar}{k_\parallel}
\newcommand{\dkpar}{\delta\kpar}

\newcommand{\ppar}{p_\parallel}
\newcommand{\qpar}{q_\parallel}
\newcommand{\Lpar}{L_\parallel}
\newcommand{\Lperp}{L_\perp}
\newcommand{\Ipar}{I_\parallel}
\newcommand{\lCB}{\lambda_\mathrm{CB}}
\newcommand{\lD}{\lambda_\mathrm{D}}
\newcommand{\lDsv}{\lambda_\mathrm{D,subvisc}}
\newcommand{\xiD}{\xi_\mathrm{D}}
\newcommand{\xivisc}{\xi_\nu}
\newcommand{\lR}{\lambda_\mathrm{R}}
\newcommand{\lB}{\lambda_B}
\newcommand{\xisub}{\xi'}
\newcommand{\lsub}{\lambda'}
\newcommand{\ellc}{\ell_\mathrm{c}}
\newcommand{\delc}{\delta_\mathrm{c}}
\newcommand{\deleff}{\delta_\mathrm{eff}}
\newcommand{\lrec}{\lambda_\mathrm{rec}}

\newcommand{\vperp}{v_\perp}

\newcommand{\MA}{M_\mathrm{A}}
\newcommand{\vA}{v_\mathrm{A}}
\newcommand{\vAy}{v_{\mathrm{A}y}}
\newcommand{\vAyinf}{v_{\mathrm{A}y,\infty}}

\newcommand{\dby}{\delta b_y}
\newcommand{\dbx}{\delta b_x}
\newcommand{\dB}{\delta B}
\newcommand{\duy}{\delta u_y}
\newcommand{\tout}{t_\mathrm{out}}
\newcommand{\tA}{\tau_\mathrm{A}}
\newcommand{\tAy}{\tau_{\mathrm{A}y}}
\newcommand{\oA}{\omega_\mathrm{A}}
\newcommand{\tnl}{\tau_\mathrm{nl}}
\newcommand{\trec}{\tau_\mathrm{rec}}
\newcommand{\tb}{\tau_b}
\newcommand{\tc}{\tau_\mathrm{c}}
\newcommand{\tres}{\tau_\eta}
\newcommand{\tvisc}{\tau_\nu}
\newcommand{\tviscnew}{\tau_{\nu,\mathrm{new}}}
\newcommand{\dr}{\Delta r}

\newcommand{\vF}{\mbf{F}}
\newcommand{\vGamma}{\mbf{\Gamma}}
\newcommand{\vAA}{\mbf{A}}
\newcommand{\vB}{\mbf{B}}
\newcommand{\vJ}{\mbf{J}}
\newcommand{\vBloc}{\vB_\mathrm{loc}}
\newcommand{\vf}{\mbf{f}}
\newcommand{\vL}{\mbf{L}}
\newcommand{\vu}{\mbf{u}}
\newcommand{\vuperp}{\vu_\perp}
\newcommand{\du}{\delta u}
\newcommand{\vb}{\mbf{b}}

\newcommand{\bperp}{b_\perp}
\newcommand{\vbperp}{\vb_\perp}
\newcommand{\dpsi}{\delta\psi}
\newcommand{\dPsi}{\psi}
\newcommand{\dPhi}{\phi}
\newcommand{\db}{\delta b}

\newcommand{\din}{\delta_\mathrm{in}}

\newcommand{\dM}{\delta M}

\newcommand{\vz}{\mbf{Z}}
\newcommand{\vzperp}{\vz_\perp}
\newcommand{\dvz}{\delta\vz}
\newcommand{\dz}{\delta Z}
\newcommand{\dzmax}{\dz_\mathrm{max}}

\newcommand{\RE}{R_\mathrm{E}}
\newcommand{\RA}{R_\mathrm{A}}
\newcommand{\sigc}{\sigma_\mathrm{c}}
\newcommand{\sigr}{\sigma_\mathrm{r}}

\newcommand{\Ekk}{E_\mathrm{2D}}
\newcommand{\Eres}{E_\mathrm{res}}
\newcommand{\Ereskk}{E_\mathrm{res,2D}}

\newcommand{\cP}{{\cal\hat P}}

\newcommand{\dvu}{\delta\vu}
\newcommand{\dvb}{\delta\vb}

\newcommand{\uperp}{u_\perp}

\title[MHD Turbulence: A Biased Review]{MHD Turbulence: A Biased Review}
\author[A.\ A.\ Schekochihin]%
{Alexander~A.~Schekochihin\thanks{Email: {\tt alex.schekochihin@physics.ox.ac.uk}}}
\affiliation{
Rudolf Peierls Centre for Theoretical Physics, University of Oxford,\\ 
Clarendon Laboratory, Parks Road, Oxford OX1 3PU, UK
\\[\affilskip]
Merton College, Oxford OX1 4JD, UK}

\begin{document}

\maketitle

\begin{abstract}
This (self-contained and aspiring to pedagogy) 
review of scaling theories of MHD turbulence aims to put the developments 
of the last few years in the context of the canonical time line (from 
Kolmogorov to Iroshnikov--Kraichnan to Goldreich--Sridhar to Boldyrev).
It is argued that Beresnyak's (valid) objection that Boldyrev's alignment theory, 
at least in its original form, 
violates the RMHD rescaling symmetry can be reconciled with alignment if the latter 
is understood as an intermittency effect. Boldyrev's scalings, a version of which 
is recovered in this interpretation, and the concept of dynamic alignment (equivalently, 
local 3D anisotropy) are thus an example of 
a qualitative, physical theory of intermittency in a turbulent system. 
The emergence of aligned structures naturally brings into play reconnection 
physics and thus the theory of MHD turbulence becomes intertwined with the 
physics of tearing, current-sheet disruption and plasmoid formation. Recent 
work on these subjects by Loureiro, Mallet et al.~is reviewed and 
it is argued that we may, as a result, finally have a reasonably 
complete picture of the MHD turbulent cascade (forced, balanced, and in the 
presence of a strong mean field)
all the way to the dissipation scale. This picture appears 
to reconcile Beresnyak's advocacy of the Kolmogorov scaling of the dissipation 
cutoff (as $\Re^{3/4}$) with Boldyrev's aligned cascade. 
It turns out also that these ideas open the door to some progress in understanding 
MHD turbulence without a mean field---MHD dynamo---whose 
saturated state is argued to be controlled by reconnection and to 
contain, at small scales, a tearing-mediated cascade very similar to its 
strong-mean-field counterpart (this is a new result). 
On the margins of this core narrative, standard weak-MHD-turbulence theory is argued 
to require some adjustment---and a new scheme for such an adjustment is proposed---to take  
account of the determining part that a spontaneously emergent  
2D condensate plays in mediating the Alfv\'en-wave 
cascade from a weakly-interacting state to a strongly turbulent (critically balanced)  
one. This completes the picture of the MHD cascade at large scales. 
A number of outstanding issues 
are surveyed, most of them concerning variants of MHD turbulence featuring various 
imbalances: between the two Elsasser fields (``cross-helicity'')
or between velocity and magnetic field (``residual energy''); 
subviscous and decaying regimes of MHD turbulence (where there has been dramatic 
progress recently and reconnection 
again turns out to feature prominently) are also reviewed under this heading.  
Some new, if often tentative, ideas about these regimes are proposed along the way 
(a new theory of imbalanced turbulence amongst them). 
Finally, it is argued that the natural direction of research is now away from the fluid MHD 
theory and into kinetic territory---and then, possibly, back again. 
The review lays no claim to objectivity or completeness, focusing on topics and 
views that the author finds most appealing at the present moment. 
\end{abstract}

\newpage 
\tableofcontents
\vskip1cm 

\begin{flushright}
{\small\parbox{8.5cm}{
\dots Oft turning others' leaves, to see if thence would flow\\
Some fresh and fruitful showers upon my sunburn'd brain.\\
But words came halting forth, wanting invention's stay;\\
Invention, Nature's child, fled step-dame Study's blows;\\
And others' feet still seemed but strangers in my way.\\
Thus great with child to speak and helpless in my throes,\\
Biting my truant pen, beating myself for spite,\\ 
``Fool,'' said my Muse to me, ``look in thy heart, and write.''}\\
\vskip2mm
Sir Philip Sidney, {\em Astrophil and Stella}
\vskip5mm
\parbox{8.5cm}{Nothing is more usual and more natural for those, 
who pretend to discover any thing new to the world in philosophy and sciences, 
than to insinuate the praises of their own systems, by decrying all those, 
which have been advanced before them.}
\vskip2mm
David Hume, {\em A Treatise of Human Nature}}
\end{flushright}

\section{Introduction} 
\label{intro}

At times during the last two decades, watching furious debates about the theory of MHD 
turbulence raging over increasingly technical and/or unfalsifiable issues, 
or working hard on minute refinements 
to existing results, one might have been forgiven for gradually losing interest. 
Is MHD turbulence to follow hydrodynamic (isotropic, homogeneous, Kolmogorov) 
turbulence and become a boutique field, ever more disconnected from 
the excitements of ``real'' physics? This perhaps is the fate of any successful 
theory (what more is there to be done?) or indeed of one that stalls for too long 
after initial breakthroughs (all the low-hanging fruit already picked?). 

Most of the reasons for which I have found myself writing this piece 
with a degree of renewed enthusiasm emerged or crystallised in and since 
2017. Enough has happened in these recent years for this text to be entirely different 
than it would have been had it been written before 2017; 
I do not think that the same could have been said during any of the~5, perhaps~10, 
years before that. The last significant conceptual breakthrough predating 2017 
was the dynamic-alignment theory of \citet{boldyrev06} (see \secref{sec:B06}), 
which updated the previous decade's paradigm-changing theory 
of \citet{GS95} (\secref{sec:CB}) and  
was followed by a flurry of numerical activity, sustaining 
the field for nearly 10 years. Some of the furious debates alluded to above had 
to do with the validity of this work---but in the absence of a new 
idea as to what might be going on dynamically, 
the insistence in a series of papers by \citet{beresnyak11,beresnyak12,beresnyak14} 
that Boldyrev's theory failed at small scales (meeting with casual 
dismissal from Beresnyak's opponents and with amused indifference from 
the rest of the community) appeared doomed to be kicked into the long grass, 
waiting for ever bigger computers.\footnote{\citet{beresnyak11} did put forward 
an unassailable, if formal, 
theoretical objection, discussed in \secref{sec:plot}, to Boldyrev's 
original interpretation of dynamic alignment as an angular uncertainty 
associated with field-line wandering. This interpretation is not, 
however, essential for the dynamic alignment itself to remain a feasible 
feature of the turbulent cascade \citep{chandran15,mallet17a}. 
I will put Beresnyak's objection 
to good use in a slightly revised model of the aligned cascade in \secref{sec:revised}.} 

Simultaneously, the community has been showing increasing interest 
and investing increasing resources into studying 
the dissipation mechanisms in MHD turbulence---in 
particular, the role of spontaneously formed current sheets and associated 
local reconnection processes (this was pioneered a long time ago 
by \citealt{matthaeus86} and \citealt{politano89}, 
but has only recently bloomed into an active field:
see references in \secref{sec:disruption}).  
The most intriguing question (which, however, remained mostly unasked---in print---until 2017) 
surely had to be this: if Boldyrev's MHD turbulence consisted of 
structures that were ever more aligned and so ever more sheet-like 
at small scales, was a scale eventually to be reached, given 
a broad enough inertial range, where these sheets 
would become too thin to stay stable and the 
reconnection processes known to disrupt such sheets would kick in? 

Like Boldyrev's theory, 
the full/quantitative realisation that large-aspect-ratio current 
sheets cannot survive also dates back to the first decade of the century, 
if one accepts that the trigger was the paper by \citet{loureiro07} on the plasmoid 
instability (see \apref{app:loureiro}; as always, in retrospect, 
one can identify early precursors, 
notably \citealt{bulanov78,bulanov79}, \citealt{biskamp82,biskamp86} 
and \citealt{tajima97}). This, however, did not translate into a clear 
understanding of the disruption of {\em dynamically forming} sheets 
until the papers by \citet{pucci14} and \citet{uzdensky16} 
(which, in fact, had been around in preprint form 
since 2014, while PRL was undertaking its characteristically thorough 
deliberations on the potential impact of publishing it). 
Once this result was out, it did not take long (even so, it took surprisingly long) 
to apply it to Boldyrev's 
aligning structures---it is this calculation (see \secref{sec:disruption}), 
published in the twin papers by \citet{mallet17b} and \citet{loureiro17}, 
that, in my view, pushed the theory of MHD turbulence forward far enough  
that it is now both closer to a modicum of logical completeness and ripe for a review.  
The outcome appears to be that the Beresnyak vs.\ Boldyrev controversy is 
resolved (both are right, in a sense: see~\secref{sec:diss}), 
Kolmogorov's dissipation scale is back, in a somewhat peculiar way, 
reconnection and turbulence have joined hands,  
and the modellers hunting for current sheets have been vindicated 
and offered further scope for their modelling. 

While emphasising this development 
as conceptually the most exciting amongst the recent ones, 
I will also take the opportunity presented by this review 
to discuss, in~\secref{sec:WT} and \apref{app:WT}, 
my reservations about the standard version of weak Alfv\'en-wave turbulence theory 
and some ideas for how to fix (or interpret) it; 
to summarise, in~\secref{sec:revised}, 
what I view as a set of rather pretty new ideas on the intrinsically intermittent nature 
of aligned turbulence \citep{mallet15,mallet16,chandran15,mallet17a}; 
to explore some old ideas, and propose some new ones, 
on various imbalanced regimes of MHD turbulence 
(with cross-helicity, with residual magnetic energy, subviscous, decaying: 
see \secsdash{sec:imbalanced}{sec:decaying}---in the case of decaying turbulence, 
reconnection has stolen the limelight again and
some very neat new ideas have recently emerged); 
to offer an updated, if tentative, perspective on the saturated state of 
MHD dynamo---i.e., MHD turbulence with no mean field, which 
turns out also to be intertwined with reconnection (\secref{sec:dynamo});  
and to advocate (in \secref{sec:kinetic}) a number of lines of further investigation 
focusing on plasma effects---some of which have started emerging in a 
particularly intriguing way during the last few years. That over a half of this 
review is taken up by these sections, treating of open questions, old confusions, 
new speculative schemes, 
and promising directions for further forays into the unknown, should alert 
my reader to a very heavy caveat attaching to my earlier claim about 
the ``logical completeness'' of our current picture of MHD turbulence: this 
claim in fact only applies to forced, balanced, inertial-range turbulence, with 
collisional dissipation and in the presence of a strong mean field. 
To many disappointed readers, this will seem to be a rarefied version of MHD turbulence 
falling short of anything useful in any global context or indeed 
of appearing anywhere in reality. This is true, but only a few years ago, 
we did not even have a grip on that! 

Because the subject of this review, if not exactly young, is 
still an active one and no one narrative has been settled as definitive, 
my exposition will be chronological, rather than logical, viz., I will discuss 
ideas that have proved to be wrong or incomplete 
before getting to those that as yet have not---not 
least because the latter were strongly influenced by, and 
would not have emerged without, the former. One day, there will be a much shorter 
story told in textbooks, with all intermediate steps forgotten.    
The erudites who already know this history, are uninterested in my prose and 
just want to skim the essential points and check out the new bits can start by reading 
\secsref{sec:revised}, \ref{sec:disruption_scale}--\ref{sec:falsifiable}, 
\ref{sec:imb_new}, \ref{sec:new_res_theory}, \ref{sec:subvisc},
\ref{sec:saffman}, \ref{sec:dynamo_theory}, 
and \apsref{app:bband}, \ref{app:2Dspectra}, \ref{app:multi}, 
\ref{app:chain_spectrum}--\ref{app:rec_driven}, 
and \ref{app:stoch_rec}.
In \secref{sec:halfway}, there is a summary and discussion
of the main takeaways from the material covered up to that point. 

Before proceeding, I would like, by way of a disclaimer, to stress 
the point that is already made in the title of this piece: 
this is a thoroughly biased review. 
Rather than merely recycling the truism that there is no such thing 
as an unbiased review of anything, I am apologising here for this one 
drawing particularly heavily on published papers in which I myself participated.
I hope that I might nonetheless be forgiven on the grounds that the lion's 
share of the credit for those contributions in fact belongs to my co-authors. 
Leaving to more disengaged spectators the task of assigning to these works 
their true measure of (in)significance,
perhaps as minor flecks of colour on the vast canvas of MHD turbulence theory, 
I will instead present this subject as I see it at the moment, 
with those flecks in the foreground.  

\newpage
\part{A Long Road to Kolmogorov}
\label{part:main}
\addcontentsline{toc}{section}{\partskip \sc Part I.\ \nameref{part:main}}

\begin{flushright}
{\small\parbox{8.5cm}{
Omnes autem, quae in rerum natura contingunt, mutationes ita sunt comparatae, 
ut si quid alicui rei accedit, id alteri derogetur. [\dots] 
Quae naturae lex cum sit universalis, ideo etiam ad regulas motus extenditur\dots}
\vskip2mm
M.~V.~Lomonosov, {\em Letter to L.~Euler, 5 July 1748}\,\footnotemark
\footnotetext{``All changes in nature 
occur in such a way that if anything is added anywhere, the same amount 
is subtracted from somewhere else. [\dots] As this is a universal law of nature, 
it extends to the laws of motion\dots''---\citet{lomonosov1748}.}} 
\end{flushright}

\section{K41, IK and GS95} 
\label{sec:prehistory}

The basic starting point for this discussion is to imagine a static, homogeneous 
plasma or, more generally, a conducting continuous medium, threaded by a 
uniform magnetic field. One can think of this situation as describing 
some local patch of a larger system, in which the magnetic field 
and other equilibrium parameters (density, pressure, flow velocity)
are large-scale and structured in some system-dependent way. I am not 
going to be concerned (except much later, in \secsand{sec:decaying}{sec:dynamo})
with the question of what this large-scale structure 
is or how it is brought about---locally, it always looks like our homogeneous 
patch. Within this patch, I shall consider perturbations whose time and 
length scales are short compared to any length scales associated with 
that large-scale structure. Of course, such a local approximation 
is not entirely universal: I am putting aside the cases 
of strong shear, various stratified or rotating systems, etc.---or, to be 
precise, I am excluding from consideration perturbations that are 
sufficiently extended in space and/or time to ``feel'' these background 
gradients. Arguably, in an ideal asymptotic world inhabited by theoretical 
physicists, one can always go to scales small enough for this restriction 
to be justified, 
without hitting dissipation/microphysical scales first (in a real world, 
this is, regrettably, not always true, but let us understand the asymptotically 
idealised reality first). The only large-scale feature that does not thus 
go away at small scales is the magnetic field. This is what makes 
MHD turbulence {\em a priori} different from, for example, rotating or stratified 
turbulence, which, at small enough scales, always reverts to the universal 
Kolmogorov state \citep[see, e.g.,][]{nazarenko11}. 

\subsection{K41}

Let us recall with maximum brevity what this Kolmogorov state is.
Assume that energy is being pumped into the system at large scales and at 
some fixed rate $\eps$. Then, in the inertial range (i.e., at small enough 
scales so the system is locally homogeneous but not small enough for 
viscosity or any other microphysics to matter yet), this same $\eps$ is the 
constant energy flux from scale to scale. Assuming that the cascade 
(i.e., the passing of energy from scale to scale) is local, the energy 
spectrum is, by dimensional analysis,
\beq
E(k)\sim \eps^{2/3} k^{-5/3}, 
\label{eq:E_K41}
\eeq 
the famous Kolmogorov spectrum (\citealt{K41}; henceforth K41), or, in terms 
of typical velocity increments between points separated by a distance $\lambda$: 
\beq
\du_\lambda \sim (\eps\lambda)^{1/3}.
\label{eq:u_K41} 
\eeq  
This is all obvious because the dimensions of the quantities involved are 
\beq
[\eps] = \frac{U^3}{L},\quad
\lt[\int\rmd k E(k)\rt] = [\du_\lambda^2] = U^2,\quad
[k] = [\lambda^{-1}] = L^{-1},  
\eeq
where $U$ is a unit of velocity and $L$ of length. As we will be dealing with 
an incompressible medium (which is always achievable by going to small 
enough scales and so to sufficiently subsonic motions), its density is an irrelevant 
constant.\footnote{In what follows, the considerations leading to scaling laws such 
as \exref{eq:E_K41} or \exref{eq:u_K41}, will require dynamical reasoning 
involving ``cascade times''---thus, \exref{eq:u_K41} is obtained by assuming 
a constant, scale-local energy flux $\du_\lambda^2/\tc\sim\eps$ with the cascade 
time $\tc\sim\lambda/\du_\lambda$ [same as~\exref{eq:u_GS95}]. I am, however,  
starting here with a purely dimensional derivation to emphasise that K41 does not, 
in fact, require us to have any dynamical insight into what is going on in the 
inertial range, and the need for such an insight will arise only once dimensional 
analysis fails to give us a unique answer.}   

\subsection{IK}
\label{sec:IK}

\begin{figure}
\begin{center}
\begin{tabular}{ccc}
\parbox{0.3\textwidth}{
\includegraphics[width=0.3\textwidth]{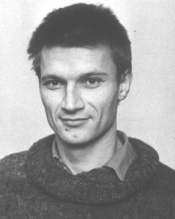}} &
\qquad\qquad&
\parbox{0.3\textwidth}{
\includegraphics[width=0.3\textwidth]{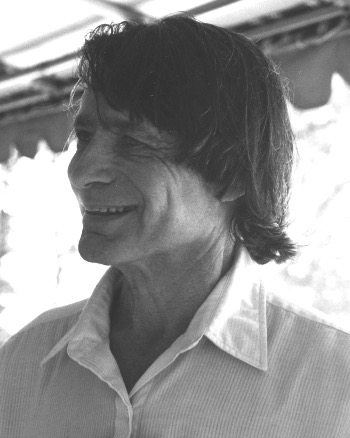}} \\\\
R.~S.~Iroshnikov (1937-1991) & & R.~H.~Kraichnan (1928-2008)\\\\
\parbox{0.3\textwidth}{
\includegraphics[width=0.3\textwidth]{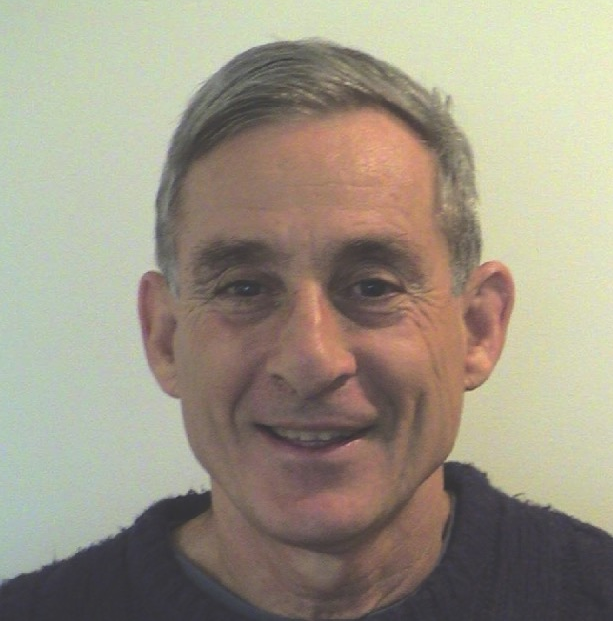}} &
\qquad\qquad&
\parbox{0.3\textwidth}{
\includegraphics[width=0.3\textwidth]{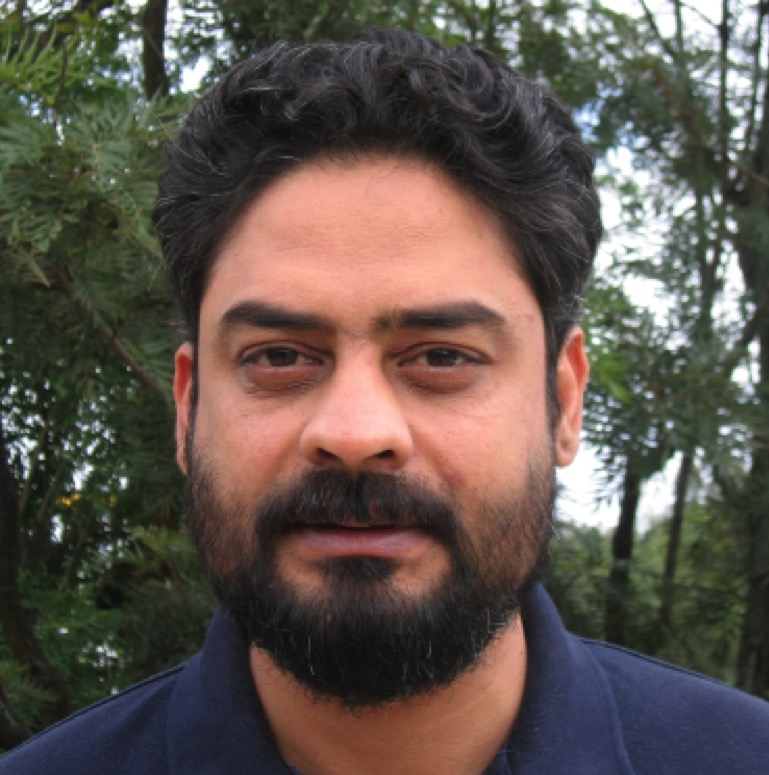}} \\\\
P.~Goldreich & & S. Sridhar
\end{tabular}
\end{center}
\caption{IK and GS (photo of R.~S.~Iroshnikov courtesy of N.~Lipunova and K.~Bychkov, 
Sternberg Astronomical Institute; 
photo of R.~H.~Kraichnan courtesy of the AIP Emilio Segr\`e Visual Archives).} 
\label{fig:IKGS}
\end{figure}

It was \citet{kraichnan65} who appears to have been the first to realise clearly 
the point made above about the irreducibility of the magnetic field. 
He therefore argued that, if the background uniform 
magnetic field $\vB_0$, which in velocity units is called the Alfv\'en speed, 
\beq
\vA = \frac{B_0}{\sqrt{4\pi\rho_0}}
\eeq 
($\rho_0$ is the mass density of the conducting medium), 
was to have a persistent (at small scales) role in the energy transfer from 
scale to scale, then the energy spectrum in the inertial range must be, 
again by dimensional analysis, 
\beq
E(k) \sim (\eps\vA)^{1/2} k^{-3/2}
\quad\Leftrightarrow\quad
\du_\lambda \sim (\eps\vA\lambda)^{1/4}.
\label{eq:IK}
\eeq
This is known as the Iroshnikov--Kraichnan spectrum 
(henceforth IK; \figref{fig:IKGS}).\footnote{\citet{iroshnikov63} got the same result slightly earlier, 
by what one might view as an early weak-turbulence calculation (before weak 
turbulence was properly invented), involving treatment of Alfv\'en waves 
as quasiparticles, opportune closure assumptions and, in the end, dimensional 
analysis. No one seems to have noticed his paper at the time and he 
disappeared into Soviet obscurity. In later years, he worked at the Institute 
of Oceanology and died in 1991, aged~54.} 
The scaling exponent was fixed by the requirement,
put forward with the trademark combination of deep insight and slightly murky
argumentation that one often finds in Kraichnan's papers, that the Alfv\'en time 
$\tA\sim1/k\vA$ was the typical time during which interactions would occur 
(before build-up of correlations was arrested by perturbations propagating away 
from each other), so the energy flux had to be proportional to $\tA$ and, 
therefore, to $1/\vA$---thus requiring them to enter in 
the combination $\eps\vA$.\footnote{Another, much better known and now standard, 
argument for the IK spectrum, 
perhaps less cryptic (but still wrong), is to posit constant flux, 
$\du_\lambda^2/\tc\sim\eps$, where the cascade time is $\tc\sim\tnl^2/\tA$ 
as in \exref{eq:tc_weak}, but, assuming isotropy, $\tnl\sim\lambda/\du_\lambda$, 
$\tA\sim\lambda/\vA$, so $\tc\sim\lambda\vA/\du_\lambda^2$ and \exref{eq:IK} follows. 
Thus, the IK theory follows from the heuristic theory of WT (\secref{sec:oldWT})
for MHD plus the isotropy assumption $\kpar\sim\lambda^{-1}$ (which, however, 
is inconsistent with weak three-wave interactions: see \secref{sec:WTirr}).\label{fn:IK}} 

Kraichnan's prediction was viewed as self-evidently correct for 30 years, 
then wrong for 10 years (\secref{sec:CB}), 
then correct again (in a different sense) for another 10 years (\secref{sec:DA}), 
then had to be revised again, at small enough scales (\secref{sec:disruption}). 
His own interpretation of it (which was also Iroshnikov's, arrived at independently) 
was certainly wrong, as it was based on the assumption---natural for a true Kolmogorovian 
susceptible to the great man's universalist notion 
of ``restoration of symmetries'' at small scales, but, in retrospect, 
illogical in the context of proclaiming the unwaning importance of $\vB_0$ 
at those same small scales---that turbulence sufficiently deep in the inertial range 
would be isotropic, i.e., 
that there is only one $k$ to be used in the dimensional analysis. 
In fact, one both can and should argue that, 
{\em a priori}, there is a $\kpar$ and a $\kperp$, which represent the variation 
of the turbulent fields along and across $\vB_0$ and need not be the same. 
The presence of the dimensionless ratio $\kpar/\kperp$ undermines the dimensional 
inevitability of \exref{eq:IK} and opens up space for much theorising, inspired or otherwise. 

\subsection{GS95}
\label{sec:GS95}

Intuitively, in a strong magnetic field, perturbations with $\kpar \ll \kperp$ 
should be more natural than isotropic ones, as the field is frozen into the motions but 
hard to bend. It turns out that MHD turbulence is indeed anisotropic in this way, 
at all scales, however small. This 
was realised quite early on, when the first, very tentative, experimental and numerical 
evidence started to be looked at \citep{robinson71,montgomery81,shebalin83}, 
but, interestingly, 
it took more than a decade after that for the IK theory to be properly revised. 

\begin{figure}
\centerline{\includegraphics[width=0.9\textwidth]{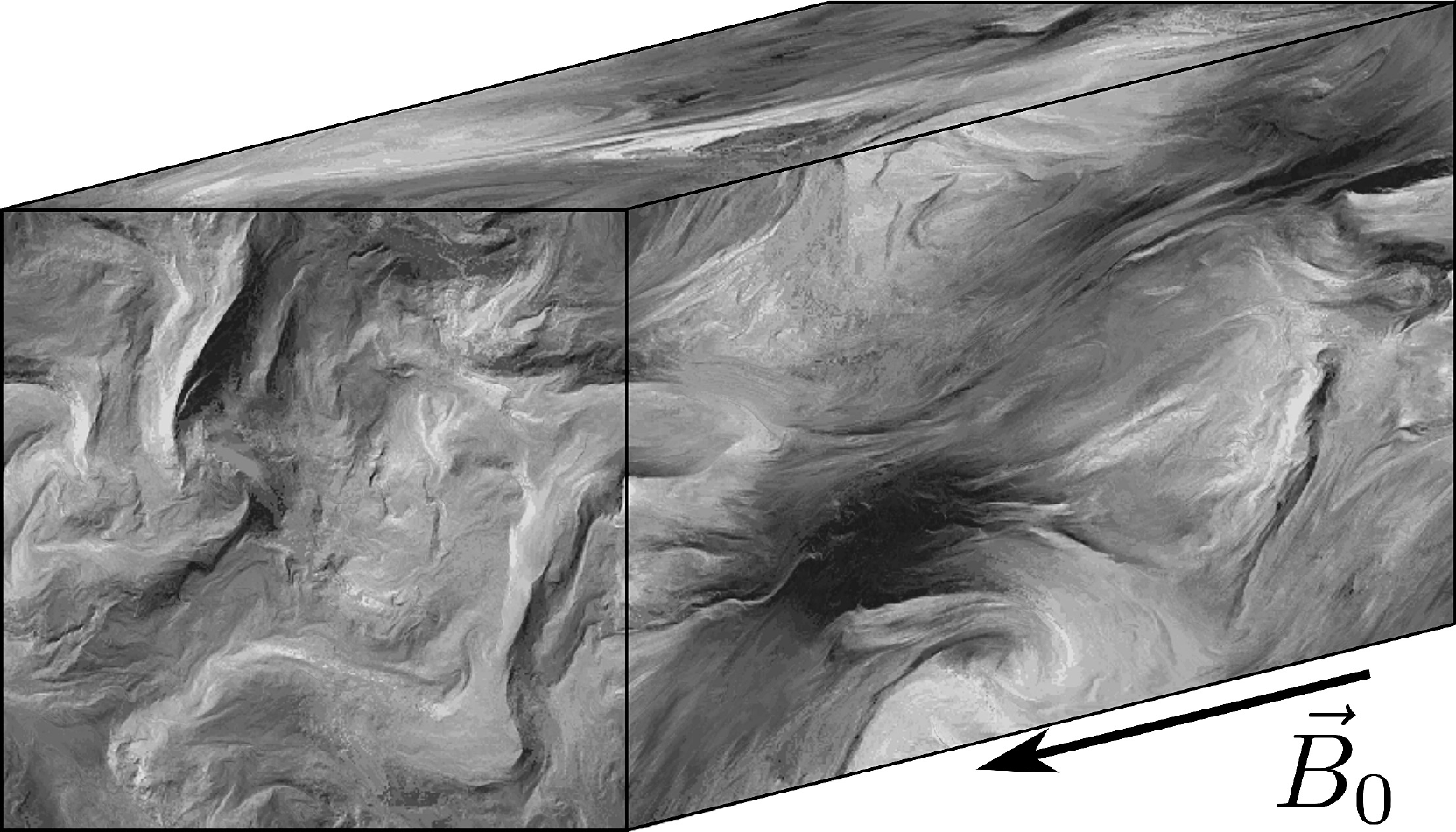}} 
\caption{A visualisation of numerical RMHD turbulence, courtesy of A.~Beresnyak 
(run R5 from \citealt{beresnyak12}, $1536^3$). 
The shades of grey represent the absolute value of $\vzperp^+ = \vuperp + \vbperp$ 
(see \secref{sec:RMHD}).}
\label{fig:beresnyak_cube_rmhd}
\end{figure}
 
Dynamically, 
the parallel variation (on scale $\lpar\sim\kpar^{-1}$) is associated with the propagation 
of \citet{alfven42} waves, the wave period (or ``propagation time'') being 
\beq
\tA \sim \frac{\lpar}{\vA},
\eeq
and the perpendicular variation (on scale $\lambda\sim\kperp^{-1}$) with 
nonlinear interactions, whose characteristic time is na\"ively equal to the 
``eddy-turnover time'': 
\beq
\tnl \sim \frac{\lambda}{\du_\lambda} 
\label{eq:tnl_K41}
\eeq
(we shall see in \secref{sec:DA} why this is na\"ive). 
Here and below, $\du_\lambda$ is used to represent the turbulent field 
on the grounds that, in Alfv\'enic perturbations, $\du_\lambda\sim\db_\lambda$, 
where $\db$ is the magnetic perturbation in velocity units
(see \secref{sec:RMHD} for a discussion with equations). 
Declaring the two times comparable at all scales was an inspired conjecture 
by \citet{GS95,GS97} (henceforth GS95; \figref{fig:IKGS}),\footnote{Anticipated, in fact, 
by \citet{higdon84}, who did not quite connect the dots, but, 
in retrospect, deserves more credit than he is getting.} 
which has come to be known as the 
{\em critical balance} (CB). I shall discuss the physical reasons for it 
properly in \secsand{sec:WT}{sec:CB}, but here let me just postulate it.
Then, naturally, the ``cascade time'' (i.e., the typical time to transfer energy 
from one perpendicular scale $\lambda$ to the next) must be of the same order 
as either of the two other times: 
\beq
\tc \sim \tA \sim \tnl.
\label{eq:CB}
\eeq  
If \exref{eq:tnl_K41} is used for $\tnl$, then \exref{eq:CB} obviates the magnetic field and 
returns us to the K41 scaling \exref{eq:u_K41}, viz., 
\beq
\frac{\du_\lambda^2}{\tc} \sim \eps,\quad
\tc\sim\tnl\sim\frac{\lambda}{\du_\lambda}
\hence
\du_\lambda\sim(\eps\lambda)^{1/3}
\quad\Leftrightarrow\quad
E(\kperp) \sim \eps^{2/3} \kperp^{-5/3}.   
\label{eq:u_GS95}
\eeq
This anisotropic version of K41 is known as the Goldreich--Sridhar (or GS95) spectrum.
Simultaneously, {\em along} the field,\footnote{It turns out that this has to be 
along the exact, perturbed field rather than the mean field \citep{cho00,maron01}---see 
\secref{sec:aniso}.} the velocity increments must satisfy  
\beq
\frac{\du_{\lpar}^2}{\tc} \sim \eps,\quad
\tc\sim\tA\sim\frac{\lpar}{\vA} 
\hence
\du_{\lpar}\sim\lt(\frac{\eps\lpar}{\vA}\rt)^{1/2}.  
\label{eq:upar_GS95}
\eeq

Thus, $\vB_0$'s influence does persist, but its size enters only the parallel 
scaling relations, not the perpendicular ones. Formally speaking, 
\exref{eq:u_GS95} is just the K41 dimensional argument for the perpendicular scale $\lambda$, 
with the CB conjecture used to justify not including $\vA$ and $\lpar$ 
amongst the local governing parameters. The assumption is that the sole role of $\vB_0$ 
is to set the value of $\lpar$ for any given $\lambda$: comparing 
\exref{eq:u_GS95} and \exref{eq:upar_GS95}, we get 
\beq
\label{eq:CB_GS95}
\lpar \sim \vA\eps^{-1/3}\lambda^{2/3}.
\eeq
Physically, this $\lpar$ 
is the distance that an Alfv\'enic pulse travels along the field, at speed $\vA$, 
over the time $\tnl$, given by \exref{eq:tnl_K41}, 
that it takes a turbulent perturbation of size $\lambda$ to break up nonlinearly. 
It is natural to argue, by causality, that this is the maximum distance over 
which any perturbation can remain correlated \citep{boldyrev05,nazarenko11}. 

This narrative arc brings us approximately to the state of affairs in 
mid-1990s, although the GS95 theory did not really become mainstream until the early 
years of this century---and soon had to be revised. Before I move on to discussing 
this revision (\secref{sec:DA}) and the modern state of the subject, I would like to 
put the discussion of what happens dynamically and how CB is 
achieved on a slightly less hand-waving basis than I have done so far.
Indeed, why critical balance? {\em Pace} the causality argument, which sets 
the maximum $\lpar$, why can $\lpar$ not be shorter? 
Is the nonlinear-time estimate \exref{eq:tnl_K41}, crucial for the scaling 
\exref{eq:u_GS95}, justified? What happens dynamically? 

From this point on, my exposition will be more sequential, I will avoid 
jumping ahead to the highlights and adopt a more systematic style, rederiving carefully 
some of the results reviewed in this section (an already 
well educated---or impatient---reader is welcome to skip or skim forward at her own pace). 
  
\section{Reduced MHD}
\label{sec:RMHD}

The theoretical assumption (or numerical/observational evidence) 
that MHD turbulence consists of perturbations that have $\kperp\gg\kpar$ 
but that their Alfv\'enic propagation remains important (so as to allow CB 
should the system want to be in it) leads to the following 
set of equations for these perturbations: 
\beq
\dd_t \vzperp^\pm \mp \vA\dpar\vzperp^\pm + \vzperp^\mp\cdot\vdperp\vzperp^\pm = 
-\vdperp p + \eta \dperp^2\ \vzperp^\pm + \vf^\pm. 
\label{eq:zpm}
\eeq
These are evolution equations for the \citet{elsasser50} fields 
$\vzperp^\pm = \vu_\perp \pm \vbperp$, where $\vuperp$ is the fluid velocity 
perpendicular to the equilibrium field $\vB_0$, and $\vbperp$ is the 
magnetic-field perturbation, also perpendicular to $\vB_0$ and expressed 
in velocity units, i.e., scaled to $\sqrt{4\pi\rho_0}$. The total pressure 
$p$ (which includes the magnetic pressure) is determined by the 
condition that $\vdperp\cdot\vzperp^\pm = 0$, enforcing the solenoidality 
of the magnetic field and the incompressibility of the motions, the latter 
achieved at small enough scales by small enough perturbations. Namely, $p$ 
is the solution of 
\beq
\dperp^2 p = -\vdperp\vdperp:\vzperp^+\vzperp^-, 
\label{eq:ptot}
\eeq
which amounts to multiplying the nonlinear term on the left-hand side of \exref{eq:zpm} 
by a projection operator in Fourier space. 
I have, for simplicity, taken the kinematic viscosity and magnetic 
diffusivity $\eta$ to be the same (but will relax this assumption from 
\secref{sec:revised} onwards). The last term in \eqref{eq:zpm}, 
the body force $\vf^\pm$, stands in for any energy-injection mechanism 
that this small-scale approximation might inherit from the non-universal 
large scales. 

The {\em Reduced MHD equations} \exsdash{eq:zpm}{eq:ptot} 
(RMHD, first proposed by \citealt{strauss76}, but, as often happened in those days, 
also found independently by the Soviets, \citealt{kadomtsev74}),
which also have a compact scalar form (see \apref{app:RMHD_scalar}),
can be derived from the standard compressible MHD equations by ordering all perturbations 
of the equilibrium to be comparable to the Mach number and to $\kpar/\kperp\ll1$ and 
the rate of change of these perturbations to the Alfv\'en 
frequency $\kpar\vA$ (see \citealt{sch07review} or \citealt{schKT}; 
a number of similar, if ever so subtly different, 
schemes exist: see review by \citealt{oughton17} and references therein). 
These equations, apart from the visco-resistive terms, are, in fact, more general than the 
collisional MHD approximation and apply also to 
low-frequency, long-wavelength collisionless perturbations 
near a gyrotropic equilibrium \citep{sch09,kunz15},\footnote{At high $\beta$, the 
amplitudes of these perturbations have to be small enough in order not to run afoul 
of some rather interesting and only recently appreciated spoiler physics 
\citep{squire16,squire17,squire17num,squire19,tenerani17,tenerani18}, 
which I will discuss very briefly in~\secref{sec:PA}.} 
which makes them applicable to the solar wind (notable for being thoroughly measurable) 
and many other, more remote, astrophysical plasmas (only measurable with difficulty, 
but endlessly fascinating to large numbers of curious researchers in gainful employment). 

While, like any nonlinear equations of serious consequence, they are impossible to solve 
except in almost-trivial special cases, 
the RMHD equations possess a number of remarkable properties
that form the basis for all theories of their turbulent solutions. 
\vskip2mm
(i) The perturbations described by them, known as {\em Alfv\'enic}, are nonlinear 
versions of (packets of) Alfv\'en waves: perturbations of velocity and magnetic field 
transverse to $\vB_0$ and propagating at speed $\vA$ along it ($\vzperp^+$ 
in the $\vB_0$ direction, $\vzperp^-$ in the $-\vB_0$ direction). They are 
entirely decoupled from all other perturbations (compressive in the case of 
fluid MHD, kinetic for a collisionless plasma; see \citealt{sch09} and \citealt{kunz15}) 
and can be considered in isolation from them. If evolved via full compressible 
MHD equations, these Alfv\'enic perturbations do not generate motions or fields that 
violate the $\kpar\ll\kperp$ assumption (e.g., higher-frequency fast MHD waves), 
so RMHD appears to be well posed in the sense that it does not break the 
assumptions that it is based on (this was checked numerically by 
\citealt{cho02compr,cho03compr}, who trod in the footsteps of \citealt{matthaeus96}).  
\vskip2mm
(ii) Only counterpropagating fields interact, so the nonlinearity vanishes 
if either $\vzperp^+ = 0$ or $\vzperp^-=0$, giving rise to the so-called 
Elsasser states ($\vuperp = \mp \vbperp$), exact nonlinear solutions that 
are arbitrary-amplitude, arbitrary-shape pulses travelling along $\vB_0$ 
at the velocity $\mp\vA$. 
\vskip2mm
(iii) The energies of the two Elsasser fields
are conserved individually (apart from any injection and dissipation terms), 
viz.,
\beq
\frac{\dd}{\dd t}\frac{\la|\vzperp^\pm|^2\ra}{2} = 
\eps^\pm - \eta\la|\vdperp\vzperp^\pm|^2\ra.
\label{eq:Zpm_conservation}
\eeq
The energy fluxes $\eps^\pm = \la\vzperp^\pm\cdot\vf^\pm\ra$ need 
not be the same and their ratio $\eps^+/\eps^-$ is, in general, a parameter 
of the problem---when it is different from unity, the turbulence is called 
{\em imbalanced} (\secref{sec:imbalanced}).
Another way of framing \exref{eq:Zpm_conservation} is by stating that RMHD 
has two invariants, the {\em total energy} and the {\em cross-helicity}: 
\beq
\frac{\la|\vuperp|^2 + |\vbperp|^2\ra}{2} = 
\frac{\la|\vzperp^+|^2 + |\vzperp^-|^2\ra}{4},\qquad
\la\vuperp\cdot\vbperp\ra = \frac{\la|\vzperp^+|^2 - |\vzperp^-|^2\ra}{4},
\eeq
respectively 
(so imbalanced turbulence is turbulence with non-zero cross-helicity). 
The name of the second invariant has topological origins, alluding, in 
incompressible 3D MHD, to conservation of linkages between flux tubes and vortex 
tubes; in the context of small Alfv\'enic 
perturbations of a strong uniform mean field $\vB_0$, this does not appear 
to be a useful interpretation. 

\vskip2mm
(iv) The amplitudes $\vzperp^\pm$, time and the gradients 
can be arbitrarily but simultaneously rescaled: $\forall \epsilon$ and $a$, 
\beq
\vzperp^\pm \to \epsilon \vzperp^\pm,\quad
\vf^\pm \to \frac{\epsilon^2}{a}\vf^\pm,\quad
\vdperp \to \frac{1}{a}\vdperp,\quad
\dpar \to \frac{\epsilon}{a}\dpar,\quad
t \to \frac{a}{\epsilon} t,\quad
\eta \to \epsilon a \eta. 
\label{eq:RMHD_resc}
\eeq  
This means that $\vzperp^\pm$ and $\dpar$ are, formally speaking, 
infinitesimal compared to $\vA$ and $\dperp$, respectively (perpendicular 
and parallel distances in RMHD are measured ``in different units,'' as are 
the Alfv\'en speed and $\vzperp^\pm$). Any statistical scalings 
or heuristic theories must respect this symmetry \citep{beresnyak11,beresnyak12}---this 
requirement will feature prominently in \secref{sec:revised}. Note that this symmetry 
implies that the parallel-to-perpendicular aspect ratio of the numerical box 
in simulations of RMHD is an arbitrary parameter.
 
\vskip2mm
(v) Defining field increments 
\beq
\dvz_\vlam^\pm = \vzperp^\pm(\vr+\vlam) - \vzperp^\pm(\vr),  
\eeq
where $\vlam$ is a point-separation vector in the perpendicular plane,
assuming statistical isotropy in this plane and considering separations 
$\lambda = |\vlam|$ 
belonging to the inertial range (i.e., smaller than the energy-injection scale but greater 
than the viscous/resistive scale), one finds, in a statistical steady state,\footnote{Write 
an evolution equation for $\dvz_\vlam^\pm$ following directly from \exref{eq:zpm}, 
take its scalar product with~$\dvz_\vlam^\pm$, average to get an evolution equation for 
the second-order structure function $\la|\dvz_\vlam^\pm|^2\ra$, then throw out 
the viscous/resistive terms, assume steady state ($\dd_t=0$), spatial homogeneity 
(correlation functions depend on $\vlam$ but not on $\vr$) and isotropy in the perpendicular 
plane (scalar averaged quantities depend on $\lambda=|\vlam|$ only), 
and, finally, integrate once with respect to $\lambda$.} 
\beq
\la\dz_L^\mp|\dvz_\vlam^\pm|^2\ra = -2\eps^\pm\lambda, 
\label{eq:exact}
\eeq
where $\dz^\mp_L = \dvz^\mp_\vlam\cdot\vlam/\lambda$ is the ``longitudinal'' increment. 
These exact laws are the RMHD version of the exact third-order laws that one 
always gets for turbulent systems with a convective nonlinearity, resembling 
the \citet{K41exact} 4/5 law of hydrodynamic turbulence or (in fact, more closely) 
the \citet{yaglom49} 4/3 law for a passive field (because in RMHD, $\vzperp^+$ advects 
$\vzperp^-$ and vice versa). 
They were derived for incompressible MHD by \citet{politano98a,politano98b} 
assuming spatial isotropy and, 
isotropy having become untenable, adjusted to their RMHD form 
\exref{eq:exact} by \citet{boldyrev09exact}. They provide a useful (although 
not as restrictive as one might have hoped) analytical 
benchmark for any aspiring scaling theory of RMHD turbulence, weak or strong.  
\vskip2mm

Everything in this review concerns turbulence that can be described by RMHD equations, 
with the following exceptions: parts of \secref{sec:decaying}, concerned with various 
types of decaying MHD turbulence, where energy-containing scales are the main object 
of study; 
\secref{sec:dynamo}, which deals with turbulent 
dynamo---a situation in which $\vbperp$ is emphatically not small compared to $\vB_0$ 
(there is no $\vB_0$) and so full MHD equations are needed;
and \secref{sec:kinetic}, where the limitations of the fluid 
description and the importance of kinetic effects are discussed.  

\section{Weak MHD Turbulence}
\label{sec:WT}

Most theory in physics is perturbation theory. In turbulence, the 
available perturbation theory is the ``weak-turbulence'' (WT) approximation 
for wave-carrying systems. Its attraction is that it features a systematic derivation, 
an appealing interpretation of the turbulent 
system as a gas of weakly interacting quasiparticles, or ``quantised'' waves, 
and quantitative predictions for spectra, or occupation numbers, of these waves 
(see textbooks by \citealt{zakharov92}, \citealt{nazarenko11book}, 
\citealt{schKT}, or, for a quick recap, \apref{app:WT_derivation}). 
Putting aside the question of whether the conditions necessary for it to hold 
are commonly (or ever) satisfied by natural turbulent systems, it is still 
interesting---and, arguably, also a matter of due diligence---to inquire
whether such a regime, and such a theory, are relevant for our RMHD system. 
``Such a regime'' 
means small amplitudes---small enough for the nonlinear interactions 
to occur very slowly compared to wave motion. One can certainly imagine, 
at least in principle, driving an RMHD system in a WT way, very gently.  

\subsection{WT is Irrelevant}
\label{sec:WTirr}

\begin{figure}
\centerline{\includegraphics[width=0.75\textwidth]{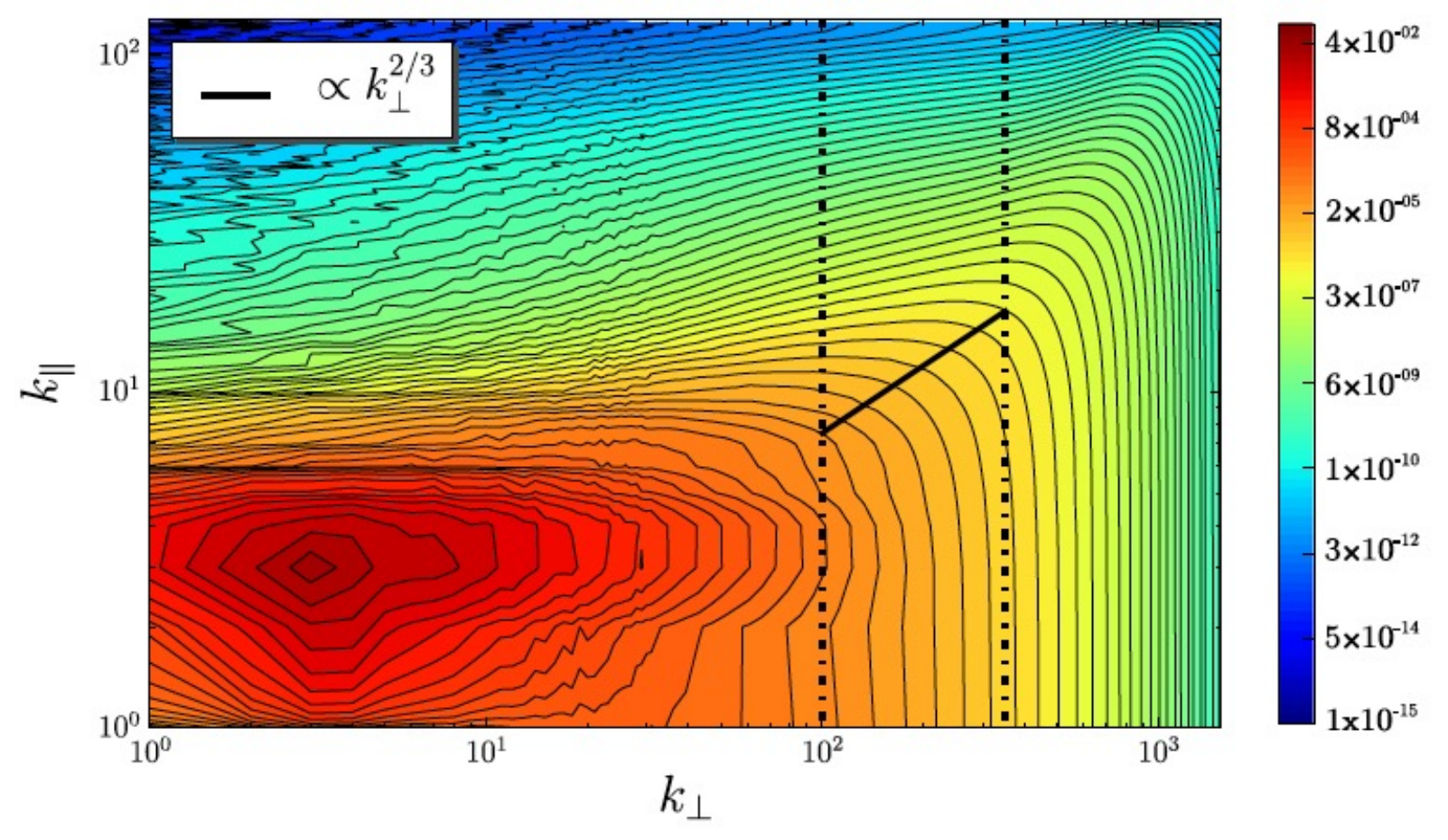}} 
\centerline{\includegraphics[width=0.6\textwidth]{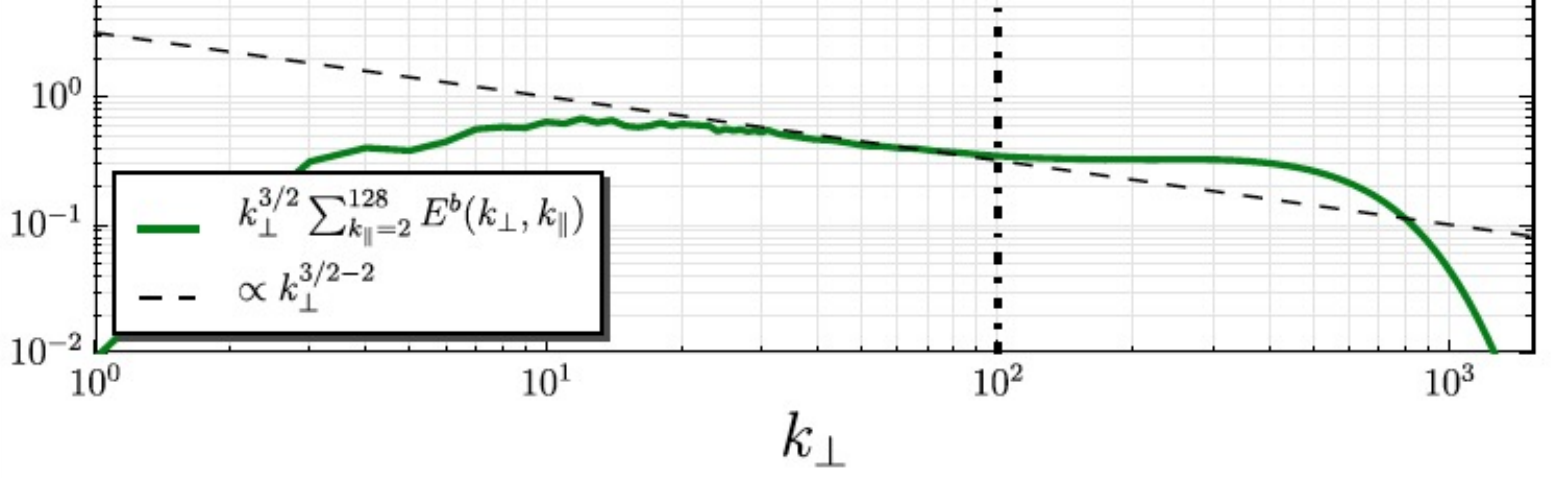}\qquad\quad} 
\caption{(Decaying) MHD simulation of transition from weak to strong turbulence by 
\citet{meyrand16}: the upper panel shows the magnetic spectrum vs.\ $\kpar$ and $\kperp$
(where $\kpar$ is along the global mean field), 
the lower one the same integrated over $\kpar$ and normalised 
by $\kperp^{3/2}$ (see \secref{sec:DA} for why $\kperp^{-3/2}$ rather than $\kperp^{-5/3}$). 
A transition manifestly occurs from a $\kperp^{-2}$ to a $\kperp^{-3/2}$ spectrum and, 
simultaneously, from a state with no $\kpar$ cascade (and a relatively narrow-band 
parallel spectrum) to one consistent with a CB cascade (2D spectra of CB turbulence 
are worked out in~\apref{app:2Dspectra}).
[Reprinted with permission from \citet{meyrand16}, copyright (2016) by the American
Physical Society.]} 
\label{fig:meyrand}
\end{figure}

On a broad-brush qualitative level, one can deal with this possibility as follows. 
Assume that in the energy-injection range, represented by some perpendicular 
scale $\Lperp$ and some parallel scale $\Lpar=2\pi/\kpar$, Alfv\'en waves 
are generated with amplitudes $Z^\pm$ so small that 
\beq
\omega_\vk^\pm = \pm\kpar \vA = \frac{1}{\tA} \gg \frac{1}{\tnl^\pm} \sim \frac{Z^\mp}{\Lperp}. 
\label{eq:WT_cond}
\eeq  
If they are viewed as interacting quasiparticles (``$+$'' can only interact with ``$-$'', 
and vice versa), the momentum and energy conservation in a three-wave interaction 
require
\beq
\lt.
\begin{array}{ll}
&\vp + \vq = \vk,\\
&\omega_\vp^\mp + \omega_\vq^\pm = \omega_\vk^\pm
\hence
-\ppar + \qpar = \kpar
\end{array}
\rt\}\hence
\qpar = \kpar,\quad \ppar = 0. 
\label{eq:kpar_const}
\eeq
Thus, three-wave interaction in fact involves a wave ($\vq$)
scattering off a 2D perturbation ($\ppar=0$, not a wave) and becoming a wave 
($\vk$) with the same frequency (because $\kpar=\qpar$) and a 
different perpendicular wavenumber ($\vkperp = \vpperp + \vqperp$). 
Intuitively, there will be a cascade of the 
waves to higher $\kperp$. If the amplitude of the waves does 
not fall off with $\kperp$ faster than $\kperp^{-1}$, which is equivalent 
to their energy spectrum being less steep than $\kperp^{-3}$, then the nonlinear-interaction 
time will become ever shorter with larger $\kperp$, even as the waves' $\kpar$ and, therefore, 
their frequency stay the same. Eventually, at some perpendicular scale, which I shall 
call $\lCB$, the condition $\tnl\gg \tA$ will 
be broken, so we end up with $\tnl \sim \tA$ and 
can return to considerations of the strong-turbulence regime, critical balance, etc. 
Numerically, this transition was first captured quite recently, 
by \citet{meyrand16}, whose result is shown in \figref{fig:meyrand}. 

The transition scale $\lCB$ is easy to estimate without the need for a specific 
WT theory. In view of \exref{eq:kpar_const}, weak interactions cannot increase 
the characteristic parallel scale of the perturbations, which therefore remains~$\Lpar$. 
Then $\lCB$ is the perpendicular scale corresponding to $\lpar=\Lpar$ 
in~\exref{eq:CB_GS95}, viz., 
\beq
\lCB \sim \eps^{1/2}\lt(\frac{\Lpar}{\vA}\rt)^{3/2}. 
\label{eq:lCB}
\eeq 
In fact, one does not even need to invoke the GS95 CB curve \exref{eq:CB_GS95}, 
because \exref{eq:lCB} is the only dimensionally correct possibility if one 
asks for a scale that depends on $\eps$ and $\tA\sim\Lpar/\vA$ only;  
that $\Lpar$ and $\vA$ must enter in this combination follows from the fact 
that $\dpar$ and $\vA$ only enter multiplying each other in the RMHD 
equations~\exref{eq:zpm}.  

A reader who is both convinced by this argument and regards it as grounds for dismissing 
the WT regime as asymptotically irrelevant, can at this point 
skip to \secref{sec:CB}. The rest of this section is for those restless 
souls who insist on worrying about what happens in weakly forced systems at $\lambda\gg\lCB$. 

\subsection{A Sketch of WT Theory}
\label{sec:oldWT}

A very simple heuristic WT calculation \citep{ng97,GS97}---a useful 
and physically transparent shortcut, and a good starting point for 
discussion---goes as follows. 

Imagine two counterpropagating Alfv\'enic structures of perpendicular size 
$\lambda$ and parallel coherence length $\Lpar$ (which cannot change 
in WT, as per the argument in \secref{sec:WTirr}) passing through 
each other and interacting weakly. Their transit time through each other 
is $\tA\sim\Lpar/\vA$ and the change in their amplitudes during this time is
\beq
\Delta(\dz_\lambda^\pm) 
\sim \dz_\lambda^\pm\frac{\tA}{\tnl^\pm}
\sim \frac{\dz^+_\lambda\dz^-_\lambda}{\lambda}\,\tA,
\label{eq:kick}
\eeq
assuming $\tnl^\pm \sim \lambda/\dz^\mp_\lambda$. By definition of 
the WT regime, $\tnl^\pm \gg \tA$, so the amplitude change in any one interaction 
is small, $\Delta(\dz_\lambda^\pm) \ll \dz_\lambda^\pm$, 
and many such interactions are needed in order to change the amplitude $\dz_\lambda^\pm$ 
by an amount comparable to itself, i.e., to ``cascade'' the energy associated 
with scale $\lambda$ to smaller scales. Suppose that interactions occur all the time 
and that the kicks \exref{eq:kick} accumulate as a random walk. 
Then the cascade time is $\tc^\pm = N\tA$ if after $N$ interactions the amplitude 
change is of order $\dz_\lambda^\pm$: 
\beq
\label{eq:tc_weak}
\Delta(\dz_\lambda^\pm)\sqrt{N}\sim \dz_\lambda^\pm
\hence
\frac{\tA}{\tnl^\pm}\sqrt{\frac{\tc^\pm}{\tA}} \sim 1
\hence
\tc^\pm \sim \frac{(\tnl^\pm)^2}{\tA}. 
\eeq
The standard Kolmogorov constant-flux requirement gives 
\beq
\eps^\pm \sim \frac{(\dz_\lambda^\pm)^2}{\tc^\pm} \sim
\frac{(\dz_\lambda^+)^2(\dz_\lambda^-)^2\tA}{\lambda^2}. 
\label{eq:WT_imb}
\eeq 
Assuming for the moment that $\eps^+\sim\eps^-$ and, therefore, $\dz_\lambda^+\sim\dz_\lambda^-$, 
gets us the classic WT scaling\footnote{For the laterally curious, let me flag here 
one very simple consequence of this result that, however, does not appear to be well known: 
the Lagrangian trajectories in weak Alfv\'enic turbulence diverge diffusively, 
rather than superdiffusively (as they do in strong turbulence)---this is shown 
in~\exref{eq:richardson_WT}.} 
\beq
\dz_\lambda \sim \lt(\frac{\eps}{\tA}\rt)^{1/4}\lambda^{1/2}
\quad\Leftrightarrow\quad
E(\kperp)\sim \lt(\frac{\eps}{\tA}\rt)^{1/2}\kperp^{-2}. 
\label{eq:WT_standard}
\eeq
This scaling is indeed what one finds numerically 
(see \figsand{fig:meyrand}{fig:tarek})---it was first confirmed 
in early, semidirect simulations by \citet{ng97} and \citet{bhatta01}, 
and then definitively by \citet{perez08} and \citet{boldyrev09weak}, 
leading the community to tick off WT as done and dusted.  

As anticipated in \secref{sec:WTirr}, with the scaling \exref{eq:WT_standard}, 
the ratio of the time scales can only stay small above a certain finite scale:
\beq
\frac{\tA}{\tnl} \sim \frac{\tA\dz_\lambda}{\lambda} 
\sim \frac{\tA^{3/4}\eps^{1/4}}{\lambda^{1/2}} \ll 1 
\quad\Leftrightarrow\quad \lambda \gg \eps^{1/2}\tA^{3/2}\sim\lCB,
\label{eq:lCB_WT}
\eeq
where $\lCB$ is transition scale anticipated in \exref{eq:lCB}. 
For $\lambda \lesssim\lCB$, turbulence becomes strong and, presumably, 
critically balanced. Thus, the WT cascade, by transferring energy 
to smaller scales, where nonlinear times are shorter, saws the seeds of its own destruction. 

\subsection{Imbalanced WT}
\label{sec:emb_WTimb}

What if $\eps^+\neq\eps^-$, viz., say, $\eps^+\gg\eps^-$? (If $\eps^+>\eps^-$ but both 
are of the same order, arguably the results obtained for $\eps^+\sim\eps^-$ should still work, 
at least on the ``twiddle'' level.) Alas, \exref{eq:WT_imb} is patently 
incapable of accommodating such a case, an embarrassment first noticed by \citet{dobrowolny80}, 
who were attempting an IK-style, isotropic ($\Lpar\sim\lambda$), imbalanced theory---quite 
wrong, as we now know (\secref{sec:GS95}), 
but they correctly identified the issue with the imbalanced regime.  
They concluded that no imbalanced stationary state was possible except 
a pure Elsasser state. 
This may be true for (certain types of) decaying turbulence 
(see \secref{sec:decay_to_Elsasser}), 
but is certainly not a satisfactory conclusion for a forced case where $\eps^\pm$ 
are externally prescribed. Arguably, the inability to describe the imbalanced 
case casts a shadow of doubt also on the validity of the argument in \secref{sec:oldWT} 
for balanced WT, as introducing the imbalance just lifts a kind of ``degeneracy'' and 
perhaps highlights a problem with the whole story.  

A way out of this difficulty, various versions of which have been explored  
by \citet{galtier00}, \citet{lithwick03}, and \citet{chandran08}, 
is to accept \exref{eq:WT_imb} 
but notice that it allows the two Elsasser fields to have different scaling 
exponents, $\dz^\pm_\lambda \propto\lambda^{\gamma^\pm}$, 
as long as they satisfy $\gamma^+ + \gamma^- = 1$. 
The corresponding 2D spectra of the two fields are 
\beq
\label{eq:WT_spectra}
\Ekk^\pm(\kperp,\kpar) = f^\pm(\kpar) \kperp^{\mu^\pm}, \quad \mu^+ + \mu^- = -4,
\eeq
because $\mu^\pm = -2\gamma^\pm-1$ and, WT permitting no changes in $\kpar$, 
the scaling arguments of \secref{sec:oldWT} apply to each $\kpar$ individually.  
One may then declare that the difference between $\eps^+$ and $\eps^-$ is hidden in the 
prefactors $f^\pm(\kpar)$, which are non-universal, 
inaccessible to ``twiddle'' scaling arguments about local interactions 
in the WT inertial range, and have to be fixed from outside it. 
At the large-scale end, one has to decide whether the outer scales 
for the two Elsasser fields are the same or different \citep{chandran08}
and whether it is the fluxes $\eps^\pm$ or the fields' energies at the outer scale(s) 
that it makes better sense to consider prescribed. 
At the dissipation scale, one has the option of  
``pinning'' the spectra to the same value (an idea due to 
\citealt{grappin83} and revived by \citealt{lithwick03}), 
and it must also be decided whether the two fields are required to start 
feeling viscosity at the same scale or one can do so before the other 
\citep[see discussion in][and, for strong imbalanced turbulence, 
in \secref{sec:pinning}]{beresnyak08}. 
If WT breaks down before the dissipation scale is reached, some other 
set of {\em ad hoc} arrangements is required \citep[see, e.g.,][]{chandran08}. 
Typically, the outcome is that the stronger field has a steeper spectrum than 
the weaker field, but their scalings are non-universal, i.e., they depend on the 
particular set up of the problem, at both macro- and micro-scales. 

Another possibility is that \exref{eq:WT_imb} is wrong. 
Let me observe that the balanced ($\eps^+\sim\eps^-$, $\dz^+_\lambda \sim \dz^-_\lambda$) 
version of this scaling, 
i.e., the statement that the flux $\eps$ is proportional to the fourth power of the 
amplitude, is less likely to be wrong than any particular assignment 
of ``$+$''s and ``$-$''s to these amplitudes: 
all it says is that the flux $\eps$
is what it would have been in the case of strong interactions, 
$\sim\dz_\lambda^2/\tnl$ [cf.~\exref{eq:u_GS95}], times the first 
power of the expansion parameter $\tA/\tnl$, i.e., the lowest order 
that $\eps$ can be in a perturbation expansion in that parameter. 
Thus, one may doubt the validity 
of \exref{eq:WT_imb} for the imbalanced regime without rejecting 
the numerically confirmed $\kperp^{-2}$ scaling of the balanced spectra. 
For example, in the (heuristic) scheme proposed by \citet{sch12}, 
\beq
\eps^\pm \sim \frac{(\dz_\lambda^\pm)^3\dz_\lambda^\mp\tA}{\lambda^2},
\label{eq:WT_imb2}
\eeq 
which changes nothing for balanced WT, but leads to a very different 
situation in the imbalanced case than \exref{eq:WT_imb}, allowing perfectly 
good $\kperp^{-2}$ spectra for both fields.

I do not go through all this in detail because, the WT regime being largely 
irrelevant (\secref{sec:WTirr}), it would also, if it really were non-universal, 
not be very interesting. If it is universal and something like \exref{eq:WT_imb2} holds, 
that {\em is} interesting, but I do not know how to make much progress beyond 
\citet{sch12}, whose theory does not quite match simulations (see \secref{sec:WT_zeromodes}). 
I also do not know how to construct a theory 
of imbalanced WT that would connect smoothly to any believable theory of strong 
imbalanced turbulence (e.g., one presented in \secref{sec:imb_new}). 
An interested reader will find some further, equally unsatisfactory, observations 
in \apref{app:WT_imb}. 

\subsection{2D Condensate}
\label{sec:WT_zeromodes}

\begin{figure}
\centerline{\includegraphics[width=0.8\textwidth]{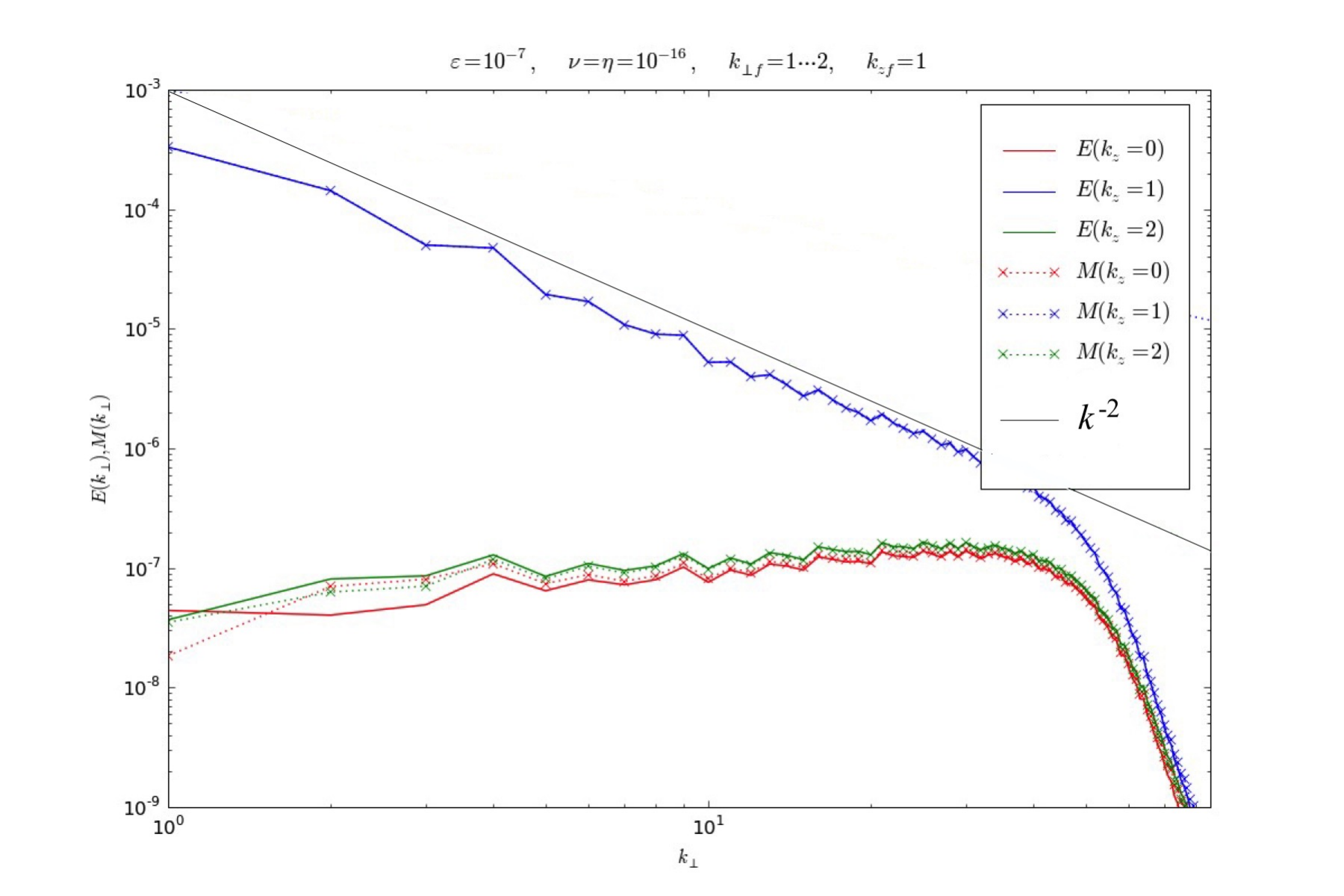}} 
\caption{Kinetic ($E$, solid lines) and magnetic ($M$, dotted lines with crosses) 
energy spectra for $\kpar=0$ (red), $\kpar=2\pi/\Lpar$ (blue) and $\kpar=4\pi/\Lpar$ (green) 
from an unpublished weak RMHD turbulence simulation by \citet{yousef09}. 
The box size was $(\Lperp,\Lpar)$ in the perpendicular and parallel directions, 
respectively, and the forcing was narrow-band, at $\kpar=2\pi/\Lpar$ and 
$\kperp = (1,2)\times2\pi/\Lperp$, deep in the WT regime ($\Lperp\gg\lCB$). 
WT spectra for the case of broad-band forcing can be found 
in \citet{perez08} and \citet{boldyrev09weak} and are discussed 
in \apref{app:bband}.}  
\label{fig:tarek}
\end{figure}

It follows from the discussion in \secref{sec:WTirr} 
that the WT approximation in its standard form cannot, in fact, work for the turbulence 
of Alfv\'en waves, at least not formally, because in every three-wave interaction, 
one of the three waves has $\kpar=0$, so 
is not a wave at all, but a zero-frequency 2D perturbation, 
for which the nonlinear interactions are the dominant influence.  
If such $\kpar=0$ perturbations are forbidden, i.e., if displacements 
vanish at infinity, one must consider four-wave interactions (i.e., go 
to next order in $\tA/\tnl$), which gives rise to an apparently 
legitimate WT state, different from \exref{eq:WT_standard} \citep{sridhar94}. 
There is no particular reason to think, however, that such a restriction 
on displacements is legitimate in a general physical situation \citep{ng96} 
or, even if one starts with no energy at 
$\kpar=0$, that such a state can be maintained, except in a box with 
field lines nailed down at the boundaries---failing such restrictions, 
a 2D ``condensate'' must emerge (and does, in numerical simulations: see 
\citealt{boldyrev09weak}, \citealt{wang11}, \citealt{meyrand15,meyrand16}, 
and \figref{fig:tarek}). 

Mathematically, this becomes quite obvious if we represent the solutions to \exref{eq:zpm}~as 
\beq
\vzperp^\pm(t,\vr) = \sum_{\kpar}\vz^\pm_{\kpar}(t,x,y) e^{i\kpar(z \pm \vA t)} 
\eeq
and separate the $\kpar=0$ modes from the rest: 
\begin{align}
\label{eq:z0}
\dd_t \vz_0^\pm + \cP\vz_0^\mp\cdot\vdperp\vz_0^\pm =
& -\sum_{\kpar\neq 0}\cP\vz_{\kpar}^\mp\cdot\vdperp\vz_{-\kpar}^\pm e^{\mp i 2\kpar\vA t},\\
\dd_t \vz_{\kpar}^\pm + \cP\vz_0^\mp\cdot\vdperp\vz_{\kpar}^\pm = 
& - \sum_{\ppar\neq 0}\cP\vz_{\ppar}^\mp\cdot\vdperp\vz_{\kpar-\ppar}^\pm e^{\mp i 2\ppar\vA t},
\label{eq:zn}
\end{align}
where $\cP$ is the projection operator that takes care of the pressure term 
[see~\exref{eq:ptot}] and has been introduced for brevity; forcing and 
dissipation terms have been dropped. 
The first of these equations, \exref{eq:z0}, describes the condensate---two real fields
$\vz_0^\pm(x,y)$ advecting each other in the 2D plane while subject to an oscillating 
``force'' due to the mutual coupling of the Alfv\'en waves $\vz^\pm_{\kpar}$. These Alfv\'en waves, 
described by \exref{eq:zn}, are advected by the 2D field and also by each other, 
but the latter interaction has an oscillating factor and vanishes in the 
WT approximation. Even if only the Alfv\'en waves are forced and the condensate 
is not, the condensate will nevertheless be built up. 

Returning to three-wave interactions then (where one of the waves is not a wave), 
the traditional approach has been to ignore the inapplicability of the WT approximation 
to the $\kpar=0$ modes by conjecturing that the function  $f^\pm(\kpar)$ 
in \exref{eq:WT_spectra} is flat around $\kpar=0$---the hypothesis 
of ``spectral continuity''.
One can then press on with putting MHD through the WT analytical 
grinder, find an evolution equation for the spectra and show that it has 
steady-state, constant-flux solutions of the form \exref{eq:WT_spectra}. 
This is what was done in the now-classic paper by \citet{galtier00} 
(see \apsand{app:WT_derivation}{app:WT_solution}). 
In balanced turbulence, obviously, $\mu^+=\mu^-=2$, 
and we are back to~\exref{eq:WT_standard}; in imbalanced turbulence, one 
needs further scheming---see references in \secref{sec:emb_WTimb}.  

\citet{nazarenko07} argues that the hypothesis of spectral continuity is certainly  
false if the nonlinear broadening of the waves' frequencies, of order 
$\tnl^{-1}$, is smaller than the linear frequency associated with the 
spacing of the $\kpar$ ``grid'' ($=2\pi/\Lpar$, the inverse 
parallel ``box'' size)---i.e., if the Alfv\'enic perturbations at the 
longest finite parallel scale in the system are already in the WT limit \exref{eq:WT_cond}, 
$\vA/\Lpar\gg\tnl^{-1}$. He is right. \Figref{fig:tarek} is taken from a (sadly, unpublished)
numerical study of weak RMHD turbulence by \citet{yousef09}, who forced Alfv\'en waves 
at $\kpar = 2\pi/\Lpar$, where $\Lpar$ was the box size. It shows that,  
while the $\kperp^{-2}$ scaling of the $\kpar=2\pi/\Lpar$ modes is undeniable, 
the spectra for the unforced modes ($\kpar = 0$ and $\kpar =$~multiples of $2\pi/\Lpar$) 
are dramatically shallower. Similar spectra were reported
by \citet{bigot11} and by \citet{meyrand15}. Qualitatively similar spectra (and a simple 
mechanism for how they might form) were also proposed by \citet{sch12}---but their 
theory fails quantitatively, with the spectra that it predicts for all unforced modes 
at least one power of $\kperp$ steeper than the numerical ones (e.g., their 
$\kpar=0$ condensate has a $\propto \kperp^{-1}$ spectrum, while 
simulations suggest $\propto\kperp^0$).

\citet{nazarenko07} expects that the conventional WT theory 
should survive when $\kpar\vA\gg \tnl^{-1}\gg\vA/\Lpar$. 
This is a situation that should be realisable in a system that is 
weakly and randomly forced in a broad band of frequencies (and, therefore, 
parallel wavenumbers). In \apref{app:bband}, I discuss how, and in what sense, 
one might defend spectral continuity for such a system; I argue that the 2D 
condensate in this case is a strongly turbulent, critically balanced sub-system 
constantly fed by the weakly turbulent waves and developing a falsifiable   
set of scalings, which are, indeed, continuous with the WT scalings. 
While there are some indications (from the simulations by 
\citealt{wang11}; see \apref{app:WT_res}) that these scalings might be right, 
I have not seen spectral continuity corroborated numerically in a definitive fashion, 
as even \citet{perez08} and \citet{boldyrev09weak}, who took great care to force 
in a broad band of~$\kpar$ 
to make sure the conventional WT theory did apply, saw a distinct dip in $f^\pm(\kpar)$ 
at $\kpar = 0$, associated with an emergent condensate (which is magnetically 
dominated; see~\secref{sec:new_res_theory_WT} and \apref{app:WT_res}).  
The same was true in the decaying simulations of 
\citet[][see the upper panel of \figref{fig:meyrand}]{meyrand15,meyrand16},
where an initial small-amplitude (and so WT-compliant) state had the choice 
to evolve towards a continuous parallel spectrum, but refused to do so, 
again developing a $\kpar=0$ condensate with dramatically distinct 
properties, including a high degree of intermittency and a spectrum quite 
similar to~\figref{fig:tarek}.   
  
Thus, the conventional WT theory is at best incomplete and at worst wrong. 
It is discussed further in \apref{app:WT}, where I review the WT's derivation,  
speculate about the structure of the condensate, 
and discuss a number of other WT-related issues. 
Here, having flagged these issues, I want to halt this 
digression into matters that are, arguably, of little impact, 
and move on to the physics-rich core of the MHD-turbulence theory. 

\section{Critical Balance, Parallel Cascade, and Anisotropy}
\label{sec:CB}

\subsection{Critical Balance}
\label{sec:CBCB}

\begin{figure}
\vskip2mm
\centerline{\includegraphics[width=0.85\textwidth]{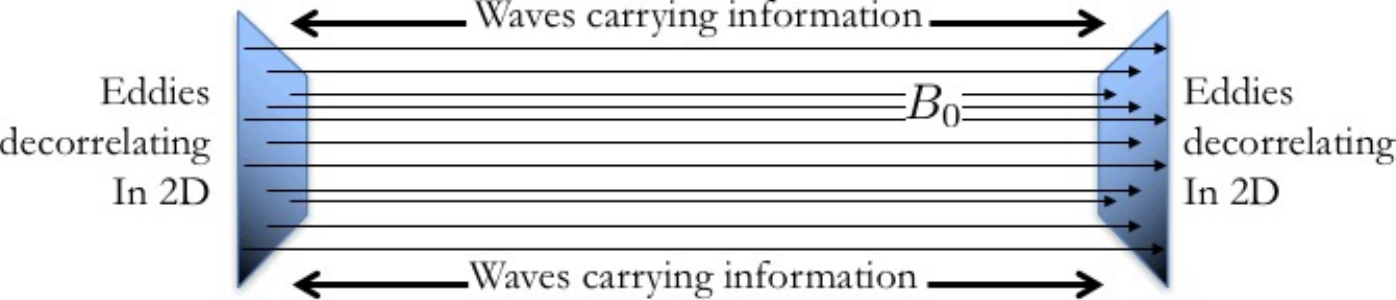}} 
\vskip2mm
\caption{Critical balance in a (2+1)D system supporting both 
nonlinearity and waves (RMHD).}
\label{fig:cb_cartoon}
\end{figure}

\Secref{sec:WT} can be viewed as one long protracted justificatory piece 
in favour of critical balance: even if an ensemble of high-frequency Alfv\'en waves 
is stirred up very gently ($\tnl\gg\tA$), it will, at small enough scales, get itself into 
the strong-turbulence regime ($\tnl\sim\tA$). The opposite limit, a 2D regime with 
$\tnl\ll\tA$, is unsustainable for the very simple reason of causality: as 
information in an RMHD system propagates along $\vB_0$ at speed $\vA$, 
no structure longer than $\lpar\sim\vA\tnl$ can be kept coherent and so 
will break up (see \citealt{boldyrev05}, \citealt{nazarenko11} and \figref{fig:cb_cartoon}). 

\begin{figure}
\centerline{\includegraphics[width=0.55\textwidth]{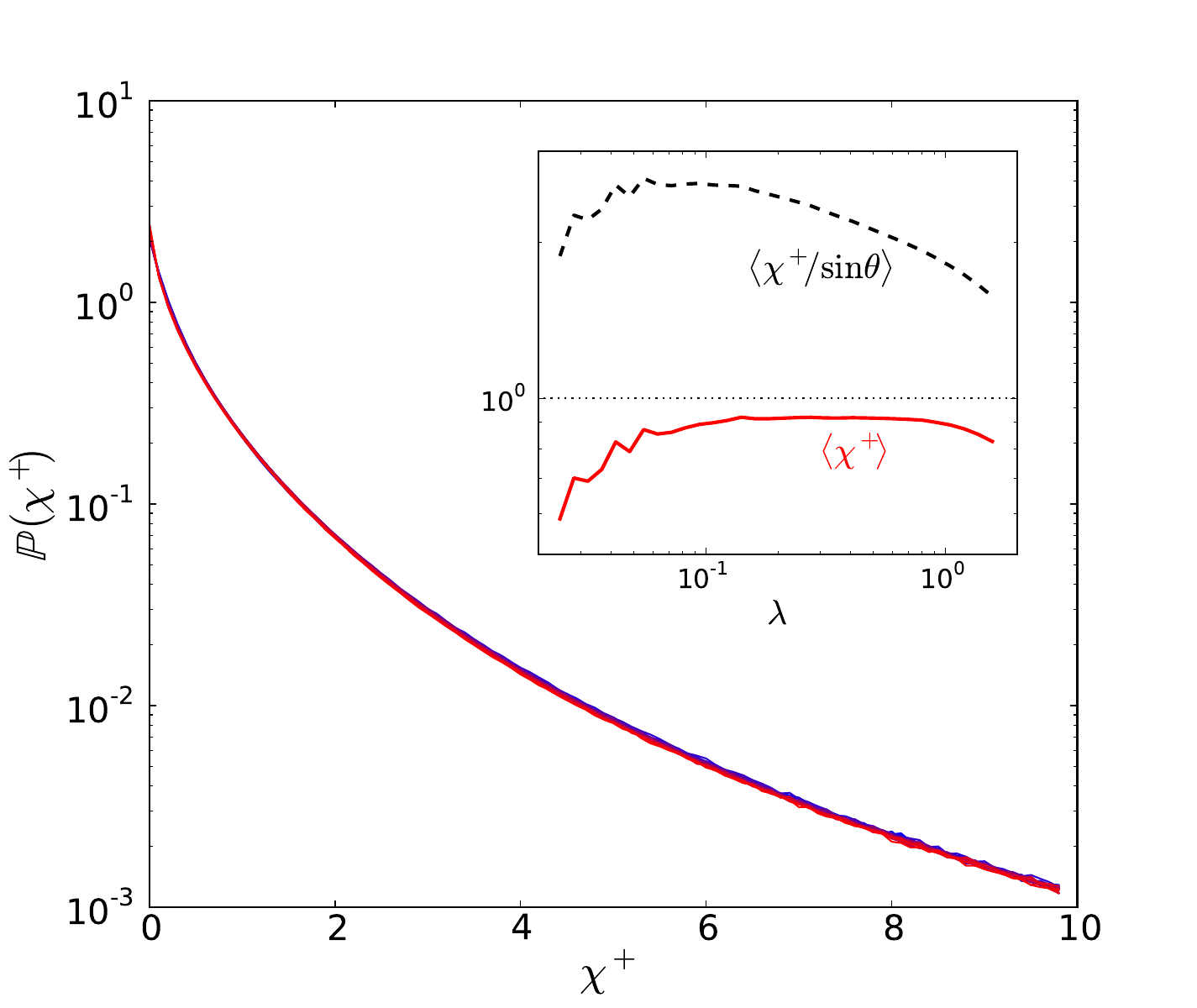}} 
\caption{Refined critical balance: this figure, taken from \citet{mallet15}, 
shows the probability density function (PDF) of the ratio $\chi^+=\tA/\tnl^+$ with 
$\tnl^+$ defined by \exref{eq:tnl_align}. In fact, 17 PDFs are plotted here, 
taken at different scales within an approximately decade-wide inertial range 
(this was a $1024^3$ RMHD simulation)---the corresponding lines are in colour shades 
from blue (smaller scales) to red (larger scales), but this is barely visible because 
the PDFs all collapse on top of each other. The inset shows that the self-similarity 
does not work if $\tnl^+$ is defined without the alignment angle (see \secref{sec:DA}).
[Reprinted from \citet{mallet15} by permission of the Royal Astronomical Society.]}
\label{fig:rcb}
\end{figure}

It is worth mentioning in passing that the CB turns out to be a very robust feature 
of the turbulence in the following interesting sense. 
With a certain appropriate definition of 
$\tnl$ (which will be explained in \secref{sec:align_aniso}),
the ratio $\tA/\tnl$ has been found 
(numerically) by \citet{mallet15} to have a scale-invariant distribution (\figref{fig:rcb}),  
a property that they dubbed {\em refined critical balance (RCB)}. 
It gives a quantitative meaning to the somewhat vague statement $\tA/\tnl\sim1$---and 
becomes important in the (as it turns out, unavoidable) discussion of intermittency 
of MHD turbulence (\secref{sec:MS17}). 

\subsection{Parallel Cascade}
\label{sec:par_cascade}

The most straightforward---and the least controversial---consequence of CB
is the scaling of {\em parallel} increments. I have already derived this result in 
\exref{eq:upar_GS95}, but let me now restate it using Elsasser fields. If it is the case 
that the nonlinear-interaction time and, therefore, the cascade time for $\vzperp^\pm$ 
are approximately the same as their propagation time $\tA\sim\lpar/\vA$, then the parallel 
increments $\dz_{\lpar}^\pm$ satisfy 
\beq
\frac{(\dz_{\lpar}^\pm)^2}{\tA} \sim \eps^\pm
\hence
\dz^\pm_{\lpar} \sim \lt(\frac{\eps^\pm\lpar}{\vA}\rt)^{1/2}
\quad\Leftrightarrow\quad
E^\pm(\kpar) \sim \frac{\eps^\pm}{\vA}\kpar^{-2}.
\label{eq:zpar_GS95}
\eeq 

\citet{beresnyak12,beresnyak15} gives two rather elegant (and related) arguments 
in favour of the scaling \exref{eq:zpar_GS95}, alongside 
robust numerical evidence presented in the latter paper.\footnote{To be precise, the scaling 
he actually observes is closer to $\kpar^{-1.9}$, although he argues that this is 
a finite-resolution effect. Imperfect following of field lines might also conceivably 
be a factor. \citet{meyrand19}, who followed field lines to a higher precision than 
that afforded by linear interpolation at every scale (see \secref{sec:aniso}), 
found a rather good $\kpar^{-2}$ scaling for 
the magnetic-field increments, but a slightly steeper slope for velocities---although 
that too may be a finite-resolution issue.} 
First, he argues that the scaling relation \exref{eq:zpar_GS95} can 
be obtained by dimensional analysis because the RMHD equations \exref{eq:zpm}
stay invariant if $\vA$ and $1/\kpar$ are scaled simultaneously [see \exref{eq:RMHD_resc}]
and so these two quantities must always appear in the combination $\kpar\vA$ 
in scaling relations for any physical quantities---in the case of \exref{eq:zpar_GS95}, 
energy, or field increment. Secondly, \citet{beresnyak15} notes that 
following the structure of the fluctuating field (calculating its increments) along 
the field line (in the positive $\vB_0$ direction) is the MHD equivalent of following 
its evolution forward (for $\vzperp^-$) or backward (for $\vzperp^+$) in time 
and it should, therefore, be possible to infer the parallel spectrum \exref{eq:zpar_GS95}
from the Lagrangian frequency spectrum of the turbulence (as opposed to the Eulerian
one, which is dominated by large-scale sweeping effects: see \citealt{lugones16,lugones19}).
Estimating the energy flux as the rate of change of energy in a fluid element in the Lagrangian 
frame (i.e., excluding sweeping by large eddies), one obtains 
\citep{LL6,corrsin63}
\beq
\eps^\pm \sim (\dz_\tau^\pm)^2\tau^{-1} 
\quad\Leftrightarrow\quad
E^\pm(\omega) \sim \eps^\pm\omega^{-2}, 
\label{eq:Eomega}
\eeq 
where $\dz_\tau^\pm$ is the Lagrangian field increment over time interval $\tau$. 
Then \exref{eq:zpar_GS95} is recovered from \exref{eq:Eomega}
by changing variables $\omega=\kpar\vA$ 
and letting $E^\pm(\omega)\rmd\omega = E^\pm(\kpar)\rmd\kpar$. 

Thus, the parallel cascade and the associated scaling \exref{eq:zpar_GS95} 
appear to be a very simple and solid property of MHD turbulence. 
What happens in the perpendicular direction is a more complicated story. 

\begin{figure}
\begin{center}
\begin{tabular}{cc}
\parbox{0.5\textwidth}{
\includegraphics[width=0.5\textwidth]{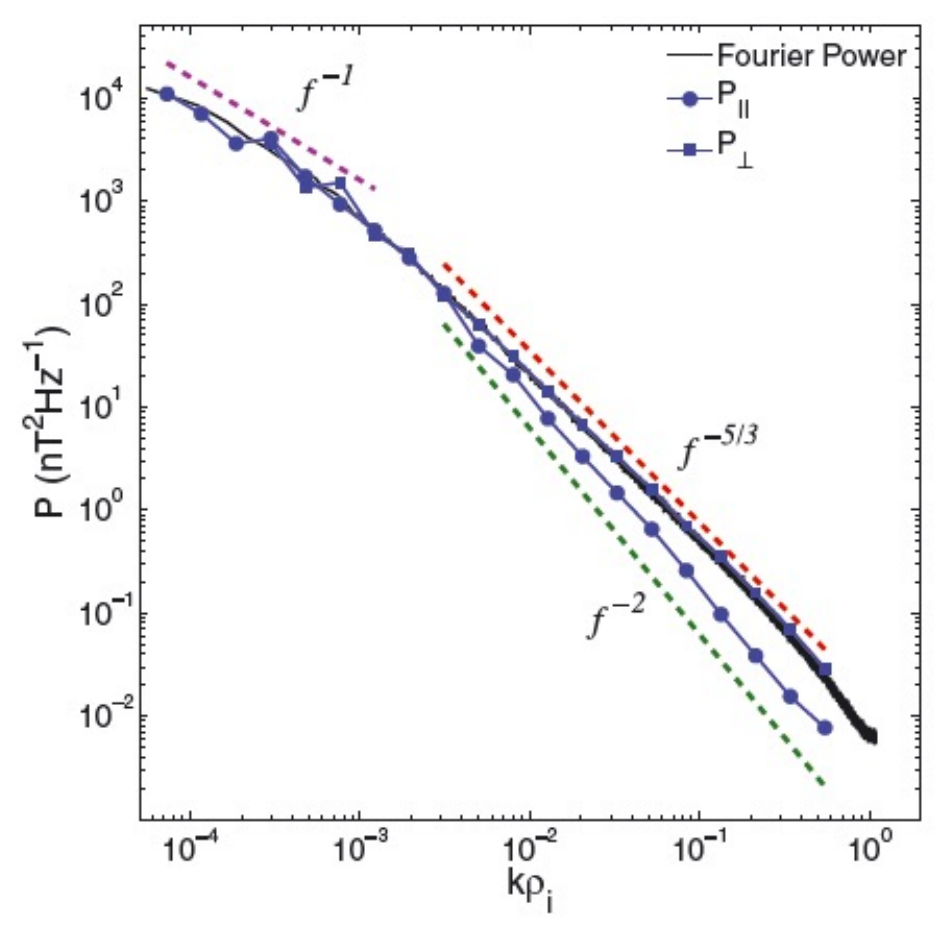}} &
\parbox{0.48\textwidth}{
\includegraphics[width=0.48\textwidth]{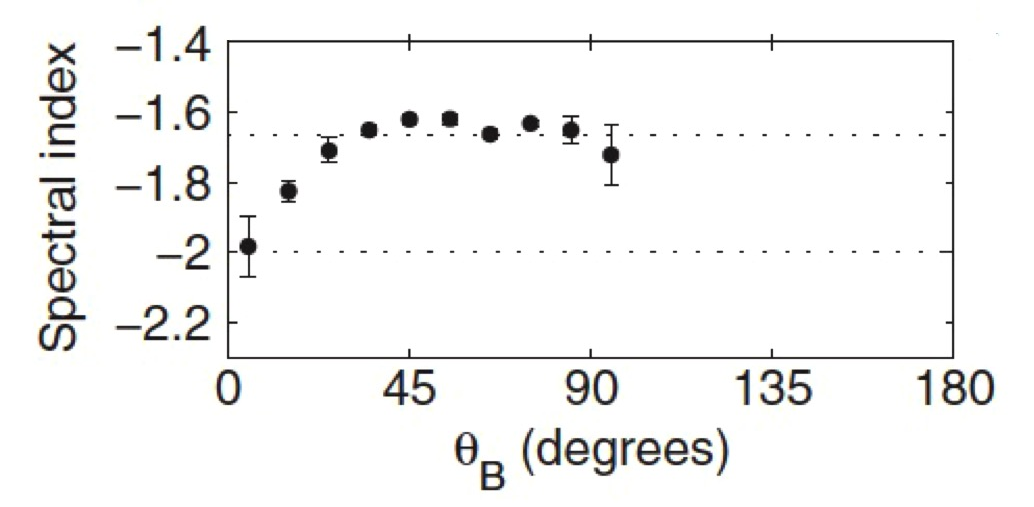}} \\\\
(a) & (b)
\end{tabular}
\end{center}
\caption{(a) Parallel ($P_\parallel$) and perpendicular ($P_\perp$) 
spectra (Fourier and wavelet) of the magnetic fluctuations in the solar wind, measured 
by the Ulysses spacecraft and computed by \citet{wicks10}, with frequencies $f$
converted to wavenumbers $k$ using the Taylor hypothesis
[reprinted from \citet{wicks10} by permission of the Royal Astronomical Society].
(b) An earlier (historic, the first ever) measurement by \citet{horbury08} 
of the spectral index of these spectra as a function of angle to the local mean field
[reprinted with permission from \citet{horbury08}, copyright (2008) by the American
Physical Society].} 
\label{fig:sw_spectra}
\end{figure}

\subsection{Local, Scale-Dependent Anisotropy}
\label{sec:aniso}

Using instead of the parallel increments the perpendicular ones $\dz^\pm_\lambda$ 
and substituting the nonlinear time
\beq
\tnl^\pm \sim \frac{\lambda}{\dz^\mp_\lambda}
\label{eq:tnl_zpm}
\eeq
for the cascade time, we recover \exref{eq:u_GS95}:\footnote{Cf.\ \citet{lithwick07}, 
the imbalanced version of the GS95 scalings (\secref{sec:imb_LGS}). This 
and especially whether the parallel correlations obey \exref{eq:aniso_GS95} is by 
no means uncontroversial. 
I am going to discuss these things in \secref{sec:imbalanced}, 
but here I keep track of $\eps^\pm$ purely for future convenience 
and invite the reader to substitute $\eps^+=\eps^-=\teps^\pm=\eps$ 
whenever thinking of imbalance-related complications becomes too much to bear.}
\begin{align}
\nonumber
\frac{(\dz_\lambda^\pm)^2}{\tnl^\pm} \sim \eps^\pm
&\hence
\frac{\dz_\lambda^+}{\dz_\lambda^-}\sim\frac{\eps^+}{\eps^-},\quad
\dz^\pm_\lambda \sim (\teps^\pm\lambda)^{1/3}, 
\quad \teps^\pm \equiv \frac{(\eps^\pm)^2}{\eps^\mp}\\ 
&\hence
E^\pm(\kperp) \sim (\teps^\pm)^{2/3}\kperp^{-5/3}.
\label{eq:z_GS95}
\end{align}
Treating $\dz^\pm_\lambda$ and $\dz^\pm_{\lpar}$ as increments for the same 
structure, but measured across and along the field, and setting them equal 
to each other, we find a relationship between the parallel and perpendicular 
scales---the scale-dependent anisotropy~\exref{eq:CB_GS95}: 
\beq
\lpar^\pm \sim \vA (\teps^\mp)^{-1/3}\lambda^{2/3}. 
\label{eq:aniso_GS95}
\eeq

The fact of scale-dependent anisotropy of MHD turbulence 
[if, in retrospect, not with the same confidence the scaling \exref{eq:aniso_GS95}] 
was confirmed numerically 
by \citet{cho00} and \citet{maron01} and, in a rare triumph of theory correctly anticipating 
measurement, observed in the solar wind by \citet{horbury08}, 
followed by many others (e.g., \citealt{podesta09aniso,wicks10,luo10,chenmallet11}---a complete 
list is impossible here as this has now become an industry, as successful ideas do; see \citealt{chen16} for a review). \Figref{fig:sw_spectra} shows some of the first of those results.
An important nuance is that, in order to see scale-dependent anisotropy, 
one must measure the parallel correlations along the perturbed, rather than global, mean 
magnetic field.\footnote{This detail was first understood by \citet{cho00} 
and \citet{maron01} (with \citealt{milano01} coming close),
but still needed restating 10 years later \citep{chenmallet11} 
and, it seems, continues (or has until recently continued) to fail 
to be appreciated in some particularly 
die-hard sanctuaries where adherents of the old religion huddle 
for warmth before the dying fire of the isotropic IK paradigm 
(I will refrain from providing citations here---and will, in \secref{sec:DA},
offer some comfort to admirers of Robert Kraichnan, who was, in a certain
sense, less wrong than it appeared in the early 2000s).} The reason for this is as follows. 

\begin{figure}
\vskip2mm
\centerline{\includegraphics[width=0.6\textwidth]{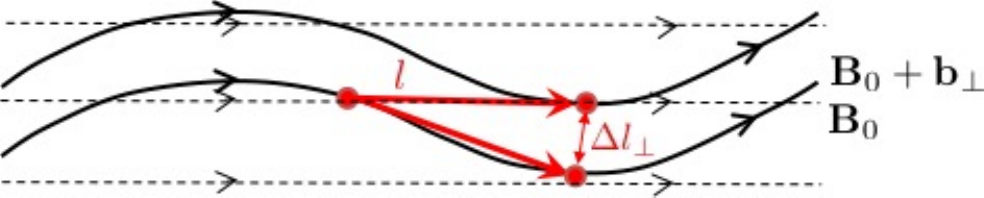}} 
\vskip2mm
\caption{Measuring correlations along local vs.\ global mean field. True 
parallel correlations cannot be captured by a measurement along the global field $\vB_0$ 
if the distance $\Delta l_\perp$ [see \exref{eq:Dlperp}] by which the point-separation 
vector $\vl$ along $\vB_0$ ``slips'' off the exact field line ($\vB_0+\vbperp$) 
is greater than the perpendicular decorrelation length $\lambda$ between 
``neighbouring'' field lines.}
\label{fig:loc_aniso}
\end{figure}

Both the causality argument \citep{boldyrev05,nazarenko11} 
and the Lagrangian-frequency one \citep{beresnyak15} that I invoked 
in \secsref{sec:CBCB} and \ref{sec:par_cascade} 
to justify long parallel coherence lengths of the MHD fluctuations 
rely on the ability of Alfv\'enic perturbations to propagate along the magnetic field. 
Physically, a small such perturbation on any given scale does not know the difference 
between a larger perturbation on, say, a few times its scale, 
and the ``true'' mean field (whatever that is, outside the 
ideal world of periodic simulation boxes). Thus, it will propagate along 
the local field and so it is along the local field that the arguments based 
on this propagation will apply. What if we instead measure correlations along the global 
mean field or, more generally, along some coarse-grained version of the exact field? 
Let that coarse-grained field be the average over all perpendicular 
scales at and below some $\Lperp$ 
(to get the global mean field, make $\Lperp$ the outer scale). 
Define Elsasser-field increments between pairs of points separated by a vector~$\vl$, 
\beq
\dvz_\vl^\pm = \vzperp^\pm(\vr + \vl) - \vzperp^\pm(\vr),
\label{eq:dvz_def}
\eeq
and consider $\vl$ along the exact magnetic field vs.\ $\vl$ along our coarse-grained field. 
The {\em perpendicular} distance by which the latter vector will veer off the field line
(\figref{fig:loc_aniso}) will be dominated by the magnetic perturbation at the largest 
scale that was not included in the coarse-grained field: 
\beq
\Delta\lperp \sim l\, \frac{\db_{\Lperp}}{\vA}. 
\label{eq:Dlperp}
\eeq
If we are trying to capture parallel correlations corresponding to perturbations with 
perpendicular scale $\lambda\ll\Lperp$, then, using CB, $l/\vA\sim\tnl$, and 
\exref{eq:tnl_zpm} with $\dz_\lambda^\pm\sim\db_\lambda$, we conclude that 
\beq
\Delta\lperp \sim \lambda\frac{\db_{\Lperp}}{\db_\lambda} \gg \lambda, 
\eeq
i.e., in such a measurement, the parallel correlations are swamped by perpendicular 
decorrelation, unless, in fact, 
$\lambda \sim \Lperp$ or larger (there is no such problem with 
measuring perpendicular correlations: small changes in a separation vector $\vl$ 
taken perpendicular to the global vs.\ exact field make no difference). 

Consequently, the easiest practical way to extract correlations along the local field 
from either observed or numerically simulated turbulence \citep{chenmallet11}  
is to measure field increments \exref{eq:dvz_def} for many different separation vectors~$\vl$ 
and to calculate for each such increment the angle between $\vl$ 
and the ``local mean field'' $\vBloc$ defined as the arithmetic mean of the magnetic field 
measured at the two points involved:
\beq
\cos\phi = \frac{\vl\cdot\vBloc}{|\vl||\vBloc|},
\quad
\vBloc = \vB_0 + \frac{\vbperp(\vr + \vl) + \vbperp(\vr)}{2}.
\label{eq:Bloc_def}
\eeq
This amounts to coarse-graining the field always at the right scale (just) for the correlations 
that are being probed. 
One can then measure (for example) perpendicular and parallel structure functions 
as conditional averages:
\begin{align}
\la(\dz^\pm_\lambda)^n\ra &= \la|\dvz_\vl^\pm|^n|\phi=90^{\rm o}\ra,\\
\la(\dz^\pm_{\lpar})^n\ra &= \la|\dvz_\vl^\pm|^n|\phi=0\ra,
\end{align}
and similarly for intermediate values of $\phi$ (as explained above, 
the difference between $\vB_0$ and $\vBloc$ matters only for small $\phi$). 
Thus, in general, one measures 
\beq
\la|\dvz_\vl^\pm|^n|\phi\ra \propto l^{\zeta_n(\phi)}.
\eeq  
Alternatively, in simulations, one can simply follow 
field lines to get $\dz^\pm_{\lpar}$ \citep{cho00,maron01} or, as was initially done 
in the solar wind, use local wavelet spectra \citep{horbury08,podesta09aniso,wicks10}. 

It turns out (see references cited above and innumerable others) 
that, quite robustly, $\zeta_2(0)=1$, consistent with \exref{eq:zpar_GS95}, 
whereas $\zeta_2(90^{\rm o})$ is typically between $2/3$ and $1/2$, i.e., between GS95 and IK, 
in the solar wind, and rather closer to $1/2$ in numerical simulations---although this, 
as I will discuss in \secsref{sec:plot} and \ref{sec:revised}, has been hotly disputed 
by \citet{beresnyak11,beresnyak12,beresnyak14,beresnyak19}.  

Thus, while little doubt remains about the reality of scale-dependent anisotropy 
[although not necessarily of the specific scaling \exref{eq:aniso_GS95}] and 
of the $\kpar^{-2}$ spectrum \exref{eq:zpar_GS95}, both arising from the GS95 theory, 
the GS95 prediction for the perpendicular spectrum~\exref{eq:z_GS95} 
has continued to be suspect and controversial. 

\section{Dynamic Alignment, Perpendicular Cascade, and Intermittency}
\label{sec:DA}

Whereas solar-wind turbulence observations 
were, for a period of time, viewed to be consistent with a $-5/3$ 
spectrum,\footnote{\citet{matthaeus82} were possibly the first to say this.
The monumental review by \citet{bruno13}
contains an exhaustive historical bibliography, the shorter
piece by \citet{chen16} reviews the more recent state
of the art: $-3/2$ is back (especially, as we are learning from the Parker
Solar Probe, at lower heliocentric distances; see \citealt{chen20});
solar wind and simulations seem more or less in agreement \citep{boldyrev11}.
Interestingly, the historical $-5/3$ period  
intersected by more than 10 years with the undisputed reign of the 
IK theory, confirming that no amount of adverse evidence can dent a 
dominant theoretical paradigm---or, at any rate, that it takes a long time 
and a hungry new generation entering the field \citep{kuhn}. 
One wonders if, had simulations and observations showing a $-3/2$ spectrum 
been available at the time, IK might have survived forever.} leading ultimately to 
the GS95 revision of the IK paradigm, high-resolution numerical simulations 
of forced, incompressible MHD turbulence, starting with \citet{maron01} 
and \citet{mueller03}, 
have consistently shown scaling exponents closer to $-3/2$ (while strongly 
confirming the local anisotropy; see also \citealt{cho00,cho02}). 
This undermined somewhat the then still young GS95 consensus 
and stimulated hard questioning of the assumptions underlying its treatment 
of nonlinear interactions. The winning idea turned out to be that the nonlinearity 
in MHD turbulence might be depleted in a scale-dependent way by some 
form of alignment between $\vzperp^+$ and $\vzperp^-$ and/or, perhaps,    
between the magnetic and velocity fields. \citet{maron01} commented 
in passing on field alignment in their simulations, and \citet{beresnyak06} 
focused on ``polarisation alignment'' explicitly, putting it on the table 
as a key effect requiring revision of GS95.\footnote{The first inklings of 
correlations naturally arising between $\vuperp$ and $\vbperp$
(the ``Alfv\'enisation'' effect) and affecting the turbulent cascade
in a significant way are, in fact, traceable to \citet{dobrowolny80}, 
\citet{grappin82,grappin83}, \citet{matthaeus83},  
and \citet{pouquet86,pouquet88}, although there was 
perhaps no explicit clarity about any physical distinction between 
alignment as a mechanism for reduction of nonlinearity 
and local imbalance (see \secref{sec:Eimb} and \apref{app:align})---and, 
of course, everybody was chained to the isotropic IK paradigm then.\label{fn:uBcorr}} 
The same possibility was mooted by \citet{boldyrev05}, and a year later  
he came up with a very beautiful (if, as we will see in \secref{sec:plot}, 
somewhat flawed) argument based on the idea of what he referred to as 
``dynamic alignment'' \citep{boldyrev06}, 
which set the direction of the field for the subsequent 10 years and 
which I am now going to discuss. 

\subsection{Alignment and Anisotropy in the Perpendicular Plane}
\label{sec:align_aniso}

Let me, however, 
deviate from Boldyrev's original case for alignment\footnote{In his original case,
\citet{boldyrev06} follows the authors cited in footnote~\ref{fn:uBcorr}
in arguing that, in forced MHD turbulence, strong local fluctuations of cross-helicity
$\vuperp\cdot\vbperp$ would be produced (by the external forcing, on the outer scale), this 
cross-helicity might then cascade to small scales less vigorously than energy
(this is a conjecture), in which case $\vuperp$ and $\vbperp$ would locally align. 
\citet{matthaeus08} did indeed confirm numerically a fast dynamical tendency for
the velocity and magnetic field to align locally, in patches, noting a formal analogy with
the velocity--vorticity alignment in hydrodynamic turbulence
\citep[cf.][]{levich91,levich09,bershadskii19xhel}.
I want to alert the reader that the alignment of $\vuperp$ and $\vbperp$,
on which \citet{matthaeus08},
as well as Boldyrev and his coworkers, focused their numerical investigations,
is not, mathematically, the same thing as the alignment of 
$\vzperp^+$ and $\vzperp^-$ advocated by \citet{chandran15}, \citet{mallet15}, 
and \citet{mallet17a}---the former kind of alignment is, rather, a manifestation
of emergent {\em local imbalance}. In practice, both types of alignment
occur \citep{mallet16}. I shall discuss this topic more carefully
in \apref{app:align}.\label{fn:uBcorr2}} 
and follow instead \citet{chandran15} in positing that, as the two Elsasser fields
advect each other, they will shear each other into an increasingly parallel
configuration. As this occurs, the gradient of the advectee (say, $\vzperp^-$)
in the direction of the advector ($\vzperp^+$) will get smaller than the gradient
of either of them in the direction transverse to both.

\begin{figure}
\centerline{\includegraphics[width=0.5\textwidth]{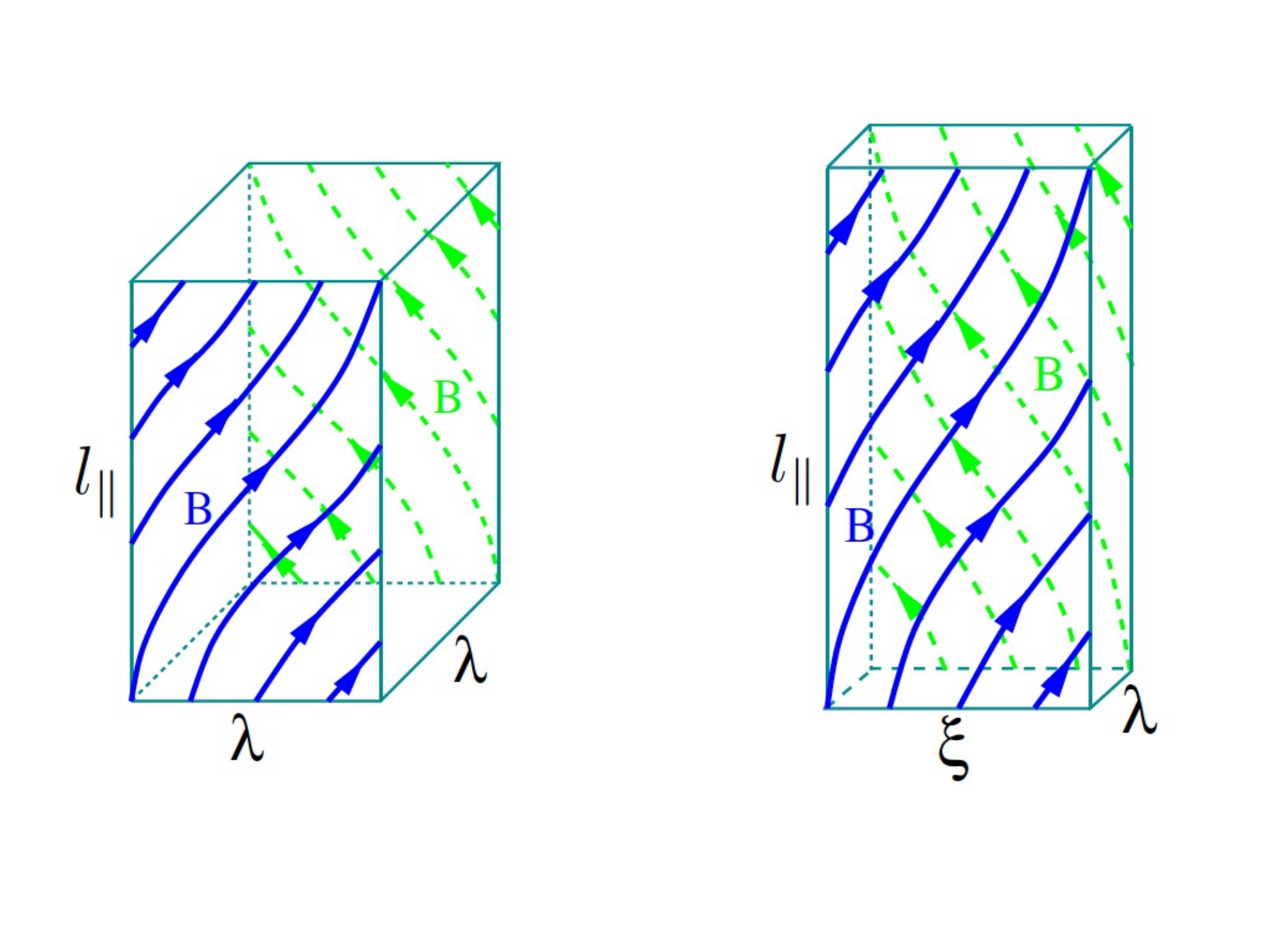}} 
\caption{Cartoon of a GS95 eddy (left) vs.\ a \citet{boldyrev06} aligned eddy (right). 
The latter has three scales: $\lpar\gg\xi\gg\lambda$ 
(along $\vB_0$, along $\vbperp$, and transverse to both). 
This picture is adapted from \citet{boldyrev06}
[reprinted with permission from \citet{boldyrev06}, copyright (2006) by the American
Physical Society]. 
In the context of my discussion, the fluctuation direction should,
in fact, be thought of as along $\vzperp^+$ or $\vzperp^-$ (see \figref{fig:geometry}).}
\label{fig:boldyrev_eddies}
\end{figure}

In general, therefore, we must allow the possibility of a local anisotropy in the 2D plane
perpendicular to the mean magnetic field (\figref{fig:boldyrev_eddies}).
If we do that, our ``twiddle'' estimate of the nonlinear term in~\exref{eq:zpm} becomes
\beq
\vzperp^+\cdot\vdperp\vzperp^- \sim \frac{\dz_\lambda^+\dz_\lambda^-}{\xi^-},
\eeq
where $\xi^-$ is the scale of variation of $\vzperp^-$ (advectee)
{\em in the direction of} $\vzperp^+$ (advector), taken at scale $\lambda$, 
which is the scale of $\vzperp^\pm$'s variation in the direction perpendicular both 
to~$\vzperp^+$ and to~$\vB_0$ (all interactions are still assumed local in scale).
But, by elementary vector calculus, 
\beq
\vzperp^+\cdot\vdperp\vzperp^- - \vzperp^-\cdot\vdperp\vzperp^+
= \vdperp\times\lt(\vzperp^-\times\vzperp^+\rt)
\sim \frac{\dz_\lambda^+\dz_\lambda^-\sin\theta_\lambda}{\lambda},
\label{eq:nlin_id}
\eeq
where $\theta_\lambda$ the angle between the two Elsasser fields taken at scale~$\lambda$. 
Under the scheme whereby the two fields shear each other into alignment in an,
on average, symmetric fashion, we have
$\xi^+ \sim \xi^-$ and 
\beq
\frac{\lambda}{\xi} \sim \sin\theta_\lambda,
\label{eq:theta_lxi}
\eeq
i.e., $\sin\theta_\lambda$ is the aspect ratio of the field structures
in the perpendicular 2D plane.\footnote{Obviously, this is not a rigorous argument.
It is mathematically possible for the two fields to be exactly or approximately
parallel with $\vzperp^+\cdot\vdperp\vzperp^- \approx \vzperp^-\cdot\vdperp\vzperp^+$
but $\lambda/\xi\sim 1$ and the latter ratio unrelated to $\theta_\lambda$, a point
made very forcefully by \citet{bowen22}. Thus, the identification between the alignment
of Elsasser fields and reduction of nonlinearity depends on a plausible dynamical
scenario of how this alignment emerges---hence my insistence on the mutual shearing.
If the reader is wondering whether alignment between $\vuperp$ and $\vbperp$ is a better
conjecture, she will find a discussion of this in \apref{app:align_nlin}.
In short, for balanced turbulence, assuming such an alignment amounts to the same thing,
but for imbalanced one, even locally, Elsasser alignment has more to do with
reducing nonlinearity.} 

In view of all this,
we ought to replace the estimate \exref{eq:tnl_zpm} of the nonlinear time with 
\beq
\tnl^\pm \sim \frac{\xi}{\dz_\lambda^\mp} 
\sim \frac{\lambda}{\dz_\lambda^\mp\sin\theta_\lambda}.
\label{eq:tnl_align}
\eeq
If the last expression is correct, then the more aligned the two fields are
the more the nonlinearity is reduced compared to the ``na\"ive''
GS95 estimate~\exref{eq:tnl_zpm}. The challenge is to work out the scale
dependence of the reduction factor~$\sin\theta_\lambda$, which is
what \citet{boldyrev06} did. 

\subsection{Boldyrev's Alignment Conjecture}
\label{sec:B06}

A version of Boldyrev's argument\footnote{His actual original argument looked somewhat more 
complicated than this, but in the end amounted to the same thing. In later 
papers \citep{perez12,perez14}, 
he does appear to embrace implicitly something closer to the line of 
thinking that I will advocate in~\secref{sec:sym}.} 
is to conjecture that the fields want to be misaligned as little as possible
and that this minimal degree of misalignment 
is set by a kind of uncertainty principle: since the direction of the local magnetic 
field along which these perturbations propagate can itself only be defined within 
a small angle $\sim\db_\lambda/\vA$, the two Elsasser fields 
(or the velocity and the magnetic field) cannot be aligned any more precisely than 
this and so 
\beq
\sin\theta_\lambda \sim \theta_\lambda \sim \frac{\db_\lambda}{\vA}\ll1. 
\label{eq:align_uncert}
\eeq 

Alignment and imbalance (local or global), and, therefore the relative
magnitudes of magnetic, velocity, and Elsasser fields, 
can be linked in a nontrivial way, which is still a matter of some debate,
but I do not wish be distracted and so will postpone the discussion of
that topic to~\secref{sec:imbalanced}. Till then, I shall assume everywhere that
\beq
\eps^+\sim\eps^-\hence 
\dz_\lambda^+\sim\dz_\lambda^-\sim\du_\lambda\sim\db_\lambda.
\eeq 
This allows \exref{eq:align_uncert} to be combined with \exref{eq:tnl_align} 
and yield 
\beq
\tnl \sim \frac{\lambda\vA}{\dz_\lambda^2}. 
\label{eq:tnl_boldyrev}
\eeq
The constancy of flux then implies immediately\footnote{A perceptive reader might 
protest at this point that $\dz_\lambda\propto\lambda^{1/4}$ looks rather 
suspicious in view of the exact law \exref{eq:exact}, 
which seems to imply a $\lambda^{1/3}$ scaling. In fact, there is no 
contradiction: since one of the three Elsasser increments in the exact law \exref{eq:exact}
is the {\em longitudinal} one, the alignment angle successfully insinuates its way in, 
and \exref{eq:exact} should be viewed as saying that $\dz_\lambda^\mp(\dz_\lambda^\pm)^2\sin\theta_\lambda \sim \eps^\pm\lambda$ \citep{boldyrev09exact}. 
This tells us nothing new, other than that the estimate \exref{eq:tnl_align}
for the nonlinear time is reasonable.} 

\beq
\frac{\dz_\lambda^2}{\tnl} \sim \eps
\hence
\dz_\lambda \sim (\eps\vA\lambda)^{1/4}
\quad\Leftrightarrow\quad
E(\kperp) \sim (\eps\vA)^{1/2}\kperp^{-3/2}. 
\label{eq:boldyrev}
\eeq

In dimensional terms, this has brought us back to the IK spectrum \exref{eq:IK}, 
except the wavenumber is now the perpendicular wavenumber
and both anisotropy and CB are retained, although the relationship between 
the parallel and perpendicular scales changes: 
\beq
\tnl \sim \frac{\lpar}{\vA} 
\hence
\lpar \sim \vA^{3/2}\eps^{-1/2}\lambda^{1/2}. 
\label{eq:lpar_boldyrev}
\eeq
Since CB remains in force, 
the parallel cascade stays the same as described in \secref{sec:par_cascade}. 

For imminent use in what follows, let us compute the extent of the inertial range 
that this aligned cascade is supposed to span. Comparing the nonlinear cascade time 
\exref{eq:tnl_boldyrev} with the Ohmic diffusion time (assuming, for convenience 
that the magnetic diffusivity $\eta$ is either the same or larger than the kinematic 
viscosity of our MHD fluid), we find 
\beq
\tnl \sim \lt(\frac{\vA\lambda}{\eps}\rt)^{1/2} 
\ll \tres \sim \frac{\lambda^2}{\eta}
\quad\Leftrightarrow\quad
\lambda \gg \eta^{2/3}\lt(\frac{\vA}{\eps}\rt)^{1/3}\equiv\lres,
\label{eq:lres_boldyrev}
\eeq
where $\lres$ is the cutoff scale---the Kolmogorov scale for this turbulence. 
For comparison, note that the same calculation based on the GS95 
scalings \exref{eq:tnl_zpm} and \exref{eq:z_GS95} gives 
\beq
\tnl^\mathrm{GS95} \sim \eps^{-1/3}\lambda^{2/3} 
\ll \tres \sim \frac{\lambda^2}{\eta}
\quad\Leftrightarrow\quad
\lambda \gg \eta^{3/4}\eps^{-1/4}\equiv\lres^\mathrm{GS95},
\label{eq:lres_GS95}
\eeq
where $\lres^\mathrm{GS95}$ is the classic Kolmogorov scale. 

If one embraces \exref{eq:boldyrev}, one could argue 
that Kraichnan's dimensional argument was actually right, but it should 
have been used with $\kperp$, rather than with $k$, because 
$\kpar$ is not a ``nonlinear'' dimension. This is the style of
reasoning that Kraichnan himself might have found attractive. We are 
about to see, however, that the result \exref{eq:boldyrev} also 
runs into serious trouble and needs revision.

\subsection{Plot Thickens}
\label{sec:plot}

\begin{figure}
\begin{center}
\begin{tabular}{c}
\includegraphics[width=0.65\textwidth]{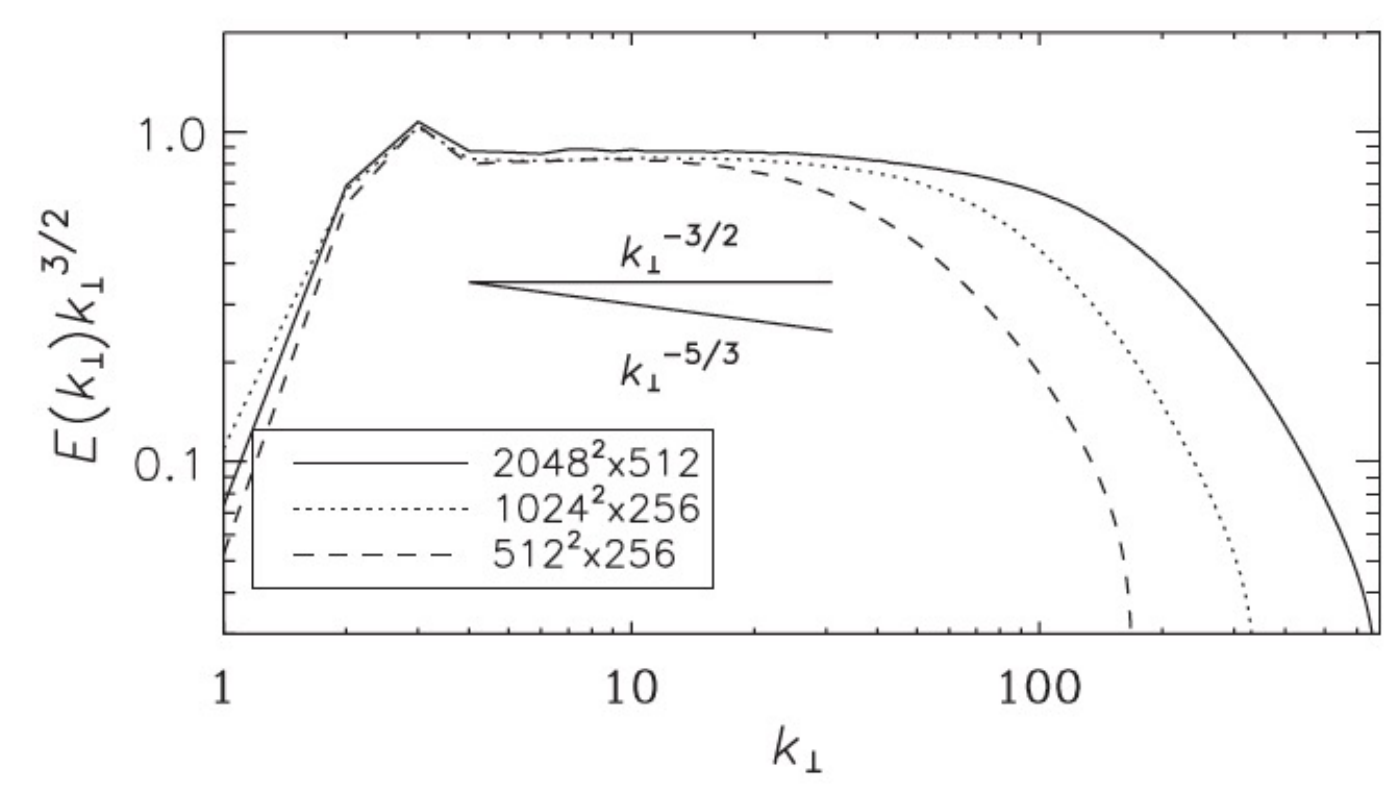}\\
(a) \citet{perez12}\\\\
\includegraphics[width=0.8\textwidth]{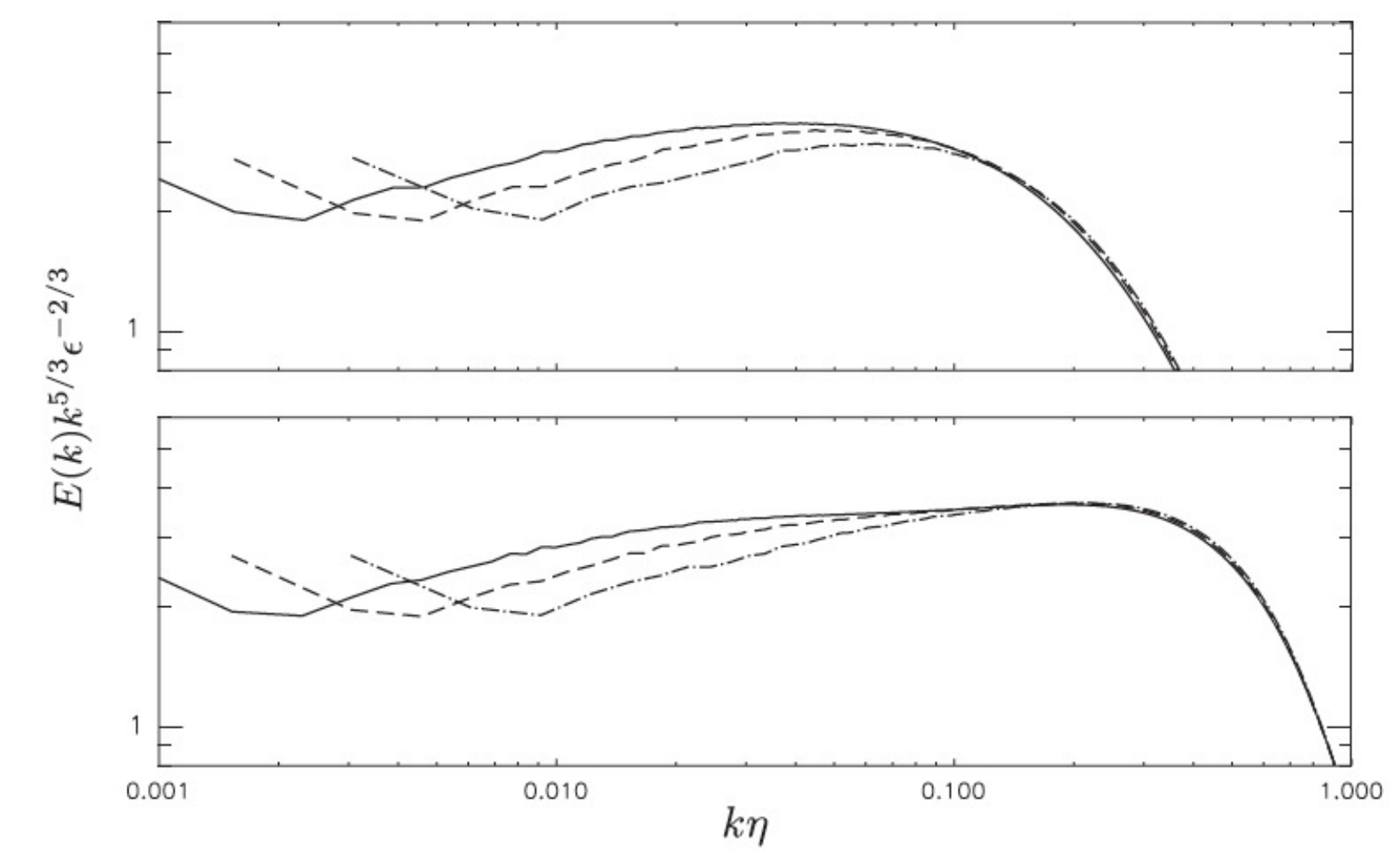}\\
(b) \citet{beresnyak14}
\end{tabular}
\end{center}
\caption{The best-resolved currently available spectra of RMHD turbulence.
(a) From simulations by \citet{perez12} (their figure~1), 
with Laplacian viscosity and resolution up to $2048^2\times512$. 
(b) From simulations by \citet{beresnyak14} (his figure 1, \copyright AAS,
reproduced with permission),
with Laplacian viscosity (top panel) 
and with 4th-order hyperviscosity (bottom panel); the resolution for the 
three spectra is $1024^3$, $2048^3$ and $4096^3$. His spectra are rescaled 
to Kolmogorov scale \exref{eq:lres_GS95} (which he denotes~$\eta$). He finds 
poorer convergence (see his figure 2) when he rescales to Boldyrev's scale~\exref{eq:lres_Rm}. 
\citet{perez12} appear to get a somewhat better outcome (see their figure~8) 
if they plot their spectra vs.\ $\kperp\lres$ where $\lres$ is given by 
\exref{eq:lres_Rm} with $\lCB$ computed in each simulation as the normalisation constant 
in the scaling \exref{eq:MS_normalised} of $\sin\theta_\lambda$ 
(in their analysis, however, this is 
the angle between velocity and magnetic perturbations, not the Elsasser fields).} 
\label{fig:num_spectra}
\end{figure}

This is a very appealing theory, whose main conclusions were rapidly 
confirmed by a programme of numerical simulations undertaken by Boldyrev's group---in 
particular, the angle between velocity and magnetic field, measured in a certain opportune 
way,\footnote{They focused on one particular measure 
of alignment, $\sin\theta_\lambda = \la|\dvu_\vlam\times\dvb_\vlam|\ra/\la|\dvu_\vlam||\dvb_\vlam|\ra$, 
which indeed turns out to scale as $\lambda^{1/4}$ in a certain range of scales. Obviously, 
one can invent other proxies for the alignment angle, involving different fields ($\dvz_\vlam^\pm$) 
and/or different powers of the fields' increments inside the averages. 
This game produces many different 
scalings \citep{beresnyak06,beresnyak09,mallet16} (some of which can be successfully theorised about: 
see \citealt{chandran15}), and it is not {\em a priori} obvious which of these should be most 
representative of the ``typical'' alignment that figures in ``twiddle'' theories.
Perhaps a better handle on the scaling of the alignment 
is obtained when one studies the full distribution of the ``RMHD ensemble'' 
(see \secref{sec:MS17} and \citealt{mallet17a}) and/or the 3D-anisotropic 
statistics (see \secref{sec:3D} and papers by \citealt{chen12}, \citealt{mallet16} 
and \citealt{verdini18,verdini19}).} 
was reported to scale according to $\theta_\lambda\propto \lambda^{1/4}$, 
as implied by \exref{eq:boldyrev} and \exref{eq:align_uncert} 
\citep{mason06,mason08,mason11,mason12,perez12,perez14}. The same papers 
confirmed the earlier numerical claims that the spectrum of MHD turbulence indeed 
scaled as $\kperp^{-3/2}$ (\figref{fig:num_spectra}a). 
However, the legitimacy of this conclusion was 
contested by \citet{beresnyak11,beresnyak12,beresnyak14,beresnyak19}, who
had, in fact, been first to focus on the alignment effect numerically \citep{beresnyak06}, 
but disputed that the numerical spectra exhibiting it were converged and 
argued that systematic convergence tests in fact favoured a trend towards 
a $\kperp^{-5/3}$ spectrum at asymptotically small scales. His point was that 
convergence of spectra with increasing resolution ought to be checked from the dissipative 
end of the inertial interval and that rescaling the spectra in his simulations to 
the Kolmogorov scale~\exref{eq:lres_GS95} gave a better data collapse 
than rescaling them to Boldyrev's cutoff scale~\exref{eq:lres_boldyrev}
(\figref{fig:num_spectra}b; to be precise, in \citealt{beresnyak14}, he
reports that the best numerical convergence obtained by this method is to $\kperp^{-1.7}$). 
Despite the sound and fury of the ensuing debate about the quality of the two  
competing sets of numerics \citep{perez14comment,beresnyak13comment,beresnyak14reply}, 
it would not necessarily be obvious to anyone 
who took a look at their papers that their raw numerical results themselves were 
in fact all that different---certainly not as different as the interpretation of these results
by their authors. Without dwelling on either, however, let me focus instead 
on a conceptual wrinkle in Boldyrev's original argument that \citet{beresnyak11} 
spotted and that cannot be easily dismissed. 

In the RMHD limit (whose applicability to MHD turbulence at sufficiently small 
scales we have no reason to doubt), $\db_\lambda/\vA$ is an arbitrarily small 
quantity, and so must then be, according to \exref{eq:align_uncert}, the 
alignment angle $\sin\theta_\lambda$. Introducing such a large depletion 
of the nonlinearity into \exref{eq:zpm} would abolish it completely in the 
RMHD ordering and render the system linear. The only way to keep the nonlinearity 
while assuming a small angle $\theta_\lambda$ is to take the angle to be small 
but still ordered as unity in the RMHD ordering---in other words, it cannot scale 
with $\epsilon$ under the RMHD rescaling symmetry \exref{eq:RMHD_resc}. 
The same rescaling symmetry implies that any physical scaling that involves 
$\vA$ and $\lpar$ (and no other scales) must involve them in the combination $\lpar/\vA$ 
(see \secref{sec:par_cascade} and \citealt{beresnyak12}), 
which \exref{eq:lpar_boldyrev} manifestly does not. 
All this flies in the face of the fact that a substantial body of 
numerical evidence supporting aligned MHD turbulence was obtained 
by means of RMHD simulations \citep{mason11,mason12,perez12,beresnyak12,mallet15,mallet16}---complemented by explicit evidence that full MHD simulations produce quantitatively the same alignment---so 
the standard recourse to casting a cloud of suspicion on the validity of 
an asymptotic approximation is not available in this case.

In a further blow to the conjecture \exref{eq:align_uncert}, 
it turns out that the alignment angle between the Elsasser fields 
at any given scale is {\em anticorrelated} with their amplitudes \citep{mallet15}, 
supporting the view that the dynamical alignment is indeed {\em dynamical}, being 
brought about by the mutual shearing of the Elsasser fields \citep{chandran15}, 
rather than by the uncertainty principle \exref{eq:align_uncert} (which would 
imply, presumably, a positive correlation between $\theta_\lambda$ and 
$\dz_\lambda$). 

The numerical evidence in favour of alignment appears to be 
real and solid.\footnote{Just to make it all more confusing, the
real {\em observational} evidence for it 
is far from solid: in the solar wind, \citet{podesta09align} and \citet{wicks13align} 
see scale-dependent alignment, but only for fluctuations at large scales---larger 
that what is normally viewed as the outer scale/the start of the inertial range (in the 
solar wind, this shows up as a break between $f^{-1}$ and $f^{-5/3\dots-3/2}$ 
slopes in the frequency spectrum). \citet{osman11} also report alignment, 
on the outer scale, as far as I can tell. 
\citet{chen12} see alignment across the inertial range, but, 
to the best of their measurement, it is not scale-dependent. 
\citet{verdini18,verdini19} managed to extract structure functions 
in three field-dependent directions (see~\secref{sec:3D}) 
that scale in a way that is consistent with scale-dependent alignment, but 
all measures of the alignment angle $\theta_\lambda$ that they tried 
had much shallower (but not flat!) scalings than~$\lambda^{1/4}$. This 
appears to be the first time that scale-dependent alignment at small scales 
has (still quite timidly) shown itself in the solar wind.
Most recently, \citet{bowen22} also measure scale-dependent
alignment, albeit at largish scales, where fluctuations are large
and it is not clear that the inertial 
range has properly started or that the RMHD limit applies. 
Theoreticians must live in hope that, as both instruments and analysis techniques become 
more refined, definite and universal scalings will eventually emerge from this sea 
of uncertainty.\label{fn:alignment_SW}} 
While numerical simulations at currently feasible resolutions 
cannot definitively verify or falsify Beresnyak's expectation that alignment is but 
a transient feature that disappears at asymptotically small scales, they certainly show 
aligned, locally 3D-anisotropic turbulence over a respectable inertial 
subrange at least one order of magnitude wide, and 
probably two. This is approaching the kind of scale 
separations that actually exist in Nature, e.g., in the solar wind (where the
evidence for a $\kperp^{-3/2}$ spectrum has also been firming up: see, e.g., 
\citealt{chen20}), 
and we cannot be casually dismissive of a physical regime, even if transient, 
that occupies most of the phase space that we are able to probe!    

\subsection{Revised Model of Aligned MHD Turbulence}
\label{sec:revised}

\subsubsection{Dimensional and RMHD-Symmetry Constraints}
\label{sec:sym}

Let me make the restrictions implied by Beresnyak's objection more explicit. 
Under the RMHD rescaling symmetry \exref{eq:RMHD_resc}, 
\beq
\dz_\lambda \to \epsilon\dz_\lambda,\quad
\eps \to \frac{\epsilon^3}{a}\eps,\quad
\vA \to \vA,\quad \lambda\to a \lambda 
\label{eq:RMHD_sym}
\eeq
(note that $\eps$ is the energy flux whereas $\epsilon$ is the scaling factor).
Therefore, the scaling relation \exref{eq:boldyrev} becomes  
$\epsilon\dz_\lambda \sim \epsilon^{3/4}(\eps\vA\lambda)^{1/4}$, which 
is obviously a contradiction. Indeed, trialling  
\beq
\dz_\lambda \sim \eps^\mu\vA^\nu\lambda^\gamma
\eeq
and mandating both the symmetry \exref{eq:RMHD_sym} and dimensional consistency, 
one finds that the GS95 solution \exref{eq:z_GS95}, 
$\nu=0$ and $\gamma=\mu=1/3$, is the only possibility, which was 
Beresnyak's point. 

It seems obvious that the only way to rescue alignment is to allow 
another parameter---and the (almost) obvious choice is $\Lpar$, the parallel 
outer scale, which transforms as $\Lpar\to (a/\epsilon)\Lpar$. Then
\beq
\dz_\lambda \sim \eps^\mu\vA^\nu\lambda^\gamma\Lpar^\delta
= \eps^{(1+\delta)/3}\lt(\frac{\Lpar}{\vA}\rt)^\delta\lambda^{(1-2\delta)/3}, 
\label{eq:dz_delta}
\eeq  
where the second expression is the result of imposing on the first 
the RMHD symmetry~\exref{eq:RMHD_sym} and dimensional correctness; 
$\delta=0$ returns us to GS95.\footnote{The weak-turbulence spectrum 
\exref{eq:WT_standard} corresponds to $\delta=-1/4$.} 
The same argument applied to the scaling of $\lpar$ with $\eps$, $\vA$, 
$\lambda$ and $\Lpar$ gives 
\beq
\lpar \sim \eps^{(\sigma-1)/3}\vA^{1-\sigma}\Lpar^\sigma\lambda^{2(1-\sigma)/3},
\label{eq:lpar_sigma}
\eeq 
where $\sigma$ is a free parameter. 
Note that both \exref{eq:dz_delta} and \exref{eq:lpar_sigma} manifestly 
contain the parallel scales and $\vA$ in the solely allowed combinations $\lpar/\vA$ and 
$\Lpar/\vA$. A reassuring consistency check is to ask what perpendicular scale 
$\lambda=\Lperp$ corresponds to $\lpar=\Lpar$: this turns out to be 
\beq
\Lperp \sim \eps^{1/2}\lt(\frac{\Lpar}{\vA}\rt)^{3/2} \sim \lCB,
\label{eq:Lperp}
\eeq
the very same $\lCB$, given by \exref{eq:lCB}, 
at which weak turbulence becomes strong---thus seamlessly 
connecting any strong-turbulence 
theory expressed by \exref{eq:dz_delta} and \exref{eq:lpar_sigma} 
with the WT cascade discussed in \secref{sec:WT}.\footnote{If we had included 
$\Lperp$ with some unknown exponents into \exref{eq:dz_delta} and $\exref{eq:lpar_sigma}$, 
we would have found that $\Lperp$ had to satisfy \exref{eq:Lperp} and so 
could not be treated as an independent quantity. What, might one ask, 
will then happen if I attempt to inject energy at some $\Lperp$ that 
does not satisfy \exref{eq:Lperp}? If this $\Lperp>\lCB$, then the 
cascade set off at the outer scale will be weak and transition to 
the strong-turbulence regime at $\lCB$ as described in \secref{sec:WT}; 
if $\Lperp<\lCB$, then I am 
effectively forcing 2D motions, which should break up by the causality 
argument (\secref{sec:CBCB}) and it is $\Lpar$ that will be determined 
by \exref{eq:Lperp}. Thus, $\lCB$ can be treated without 
loss of generality as the perpendicular outer scale of the CB cascade.}
Notably, if we applied such 
a test to \exref{eq:lpar_boldyrev}, we would find the price of consistency 
to be $\Lperp=\Lpar$, which is allowed but does not have to be the case 
in MHD and certainly cannot be the case in RMHD.  

Finally, since the parallel-cascade scaling \exref{eq:zpar_GS95} remains 
beyond reasonable doubt and, as can be readily checked, respects the rescaling 
symmetry \exref{eq:RMHD_resc} \citep{beresnyak15}, combining it with 
\exref{eq:lpar_sigma} and \exref{eq:dz_delta} fixes 
\beq
\sigma = 2\delta. 
\label{eq:sigma_delta}
\eeq
Alas, CB does not help with determining $\delta$ because, in aligned 
turbulence, the nonlinear time \exref{eq:tnl_align} contains the unknown 
scale $\xi$, or, equivalently, the alignment angle $\theta_\lambda\sim\lambda/\xi$. 
If we did know $\delta$, CB would let us determine this angle: 
\beq
\frac{\lpar}{\vA} \sim \tnl \sim \frac{\lambda}{\dz_\lambda\sin\theta_\lambda} 
\hence
\sin\theta_\lambda \sim \lt(\frac{\lambda}{\lCB}\rt)^{2\delta},  
\eeq  
where $\lCB$ is given by \exref{eq:Lperp}. 
The answer that we want to get---keeping Boldyrev's scalings 
of everything with $\lambda$ but not with $\eps$ or $\vA$---requires 
\beq
\delta = \frac{1}{8}.
\eeq 
Then, instead of \exref{eq:boldyrev}, we end up with  
\beq
\dz_\lambda \sim \eps^{3/8}\lt(\frac{\Lpar}{\vA}\rt)^{1/8}\lambda^{1/4},\quad
\lpar \sim \eps^{-1/4}\vA^{3/4}\Lpar^{1/4}\lambda^{1/2},\quad
\sin\theta_\lambda \sim \eps^{-1/8}\lt(\frac{\vA}{\Lpar}\rt)^{3/8}\lambda^{1/4}, 
\label{eq:MS}
\eeq 
and the dissipation cutoff scale \exref{eq:lres_boldyrev} is 
corrected as follows:
\beq
\tnl \sim \lt(\frac{\Lpar}{\eps\vA}\rt)^{1/4}\lambda^{1/2} 
\ll \tres \sim \frac{\lambda^2}{\eta} 
\quad\Leftrightarrow\quad
\lambda \gg \eta^{2/3}\lt(\frac{\Lpar}{\eps\vA}\rt)^{1/6}\equiv\lres.
\label{eq:lres_MS}
\eeq
Note that, since $\lres\propto\eta^{2/3}$ still, this does not 
address Beresnyak's numerical evidence on the convergence
of the spectra~(\secref{sec:plot})---I shall come back to 
this problem in~\secref{sec:disruption}.

For future convenience, let me recast the scalings \exref{eq:MS} in a somewhat simpler form: 
\beq
\dz_\lambda \sim \lt(\frac{\eps\Lpar}{\vA}\rt)^{1/2}\lt(\frac{\lambda}{\lCB}\rt)^{1/4},
\quad
\frac{\lpar}{\Lpar} \sim \lt(\frac{\lambda}{\lCB}\rt)^{1/2},
\quad
\sin\theta_\lambda \sim \lt(\frac{\lambda}{\lCB}\rt)^{1/4}.
\label{eq:MS_normalised}
\eeq
Defining the magnetic Reynolds number based on the CB scale \exref{eq:Lperp}
and the fluctuation amplitude at this scale, 
\beq
\Rm = \frac{\dz_{\lCB}\lCB}{\eta} \sim \frac{\eps}{\eta}\lt(\frac{\Lpar}{\vA}\rt)^2,
\label{eq:RmCB_def}
\eeq
allows the dissipation scale \exref{eq:lres_MS} to be recast as follows:
\beq
\frac{\lres}{\lCB} \sim \lt(\frac{\Rm}{1+\Pm}\rt)^{-2/3} = \tRe^{-2/3},\quad
\Pm = \frac{\nu}{\eta},\quad
\tRe = \frac{\dz_{\lCB}\lCB}{\nu+\eta}. 
\label{eq:lres_Rm}
\eeq
I have restored the possibility that viscosity $\nu$ might be larger 
than the magnetic diffusivity $\eta$: if that is the case, one must replace 
the latter with the former in the calculation of the dissipative cutoff, 
whereas if $\Pm\lesssim 1$, it does not matter, 
hence the appearance of the magnetic Prandtl number $\Pm$ in 
the combination $(1+\Pm)$. 

Yet another way to write the first of the scaling relations \exref{eq:MS} is 
\beq
\dz_\lambda \sim \eps^{1/3}\lCB^{1/12}\lambda^{1/4}
\quad\Leftrightarrow\quad
E(\kperp)\sim \eps^{2/3}\lCB^{1/6}\kperp^{-3/2}.
\label{eq:Perez}
\eeq 
This is effectively the prediction for the spectrum that 
\citet{perez12,perez14} used in their numerical convergence studies. 
Thus, they and I are on the same page as to what the spectrum of 
aligned turbulence is expected to be, 
although the question remains why it should be that if 
Boldyrev's uncertainty principle \exref{eq:align_uncert} can no longer be used. 

A set of RMHD-compatible scalings \exref{eq:MS}, or \exref{eq:Perez}, 
is also effectively what was deduced by \citet{chandran15} and by \citet{mallet17a} 
from a set of plausible conjectures about the dynamics and statistics 
of RMHD turbulence (they did not 
explicitly discuss the issue of the RMHD rescaling symmetry, but 
used normalisations that enforced it automatically). 
The two papers differed in their strategy for determining the exponent $\delta$; 
my exposition here will be a ``heuristic'' version of \citet{mallet17a}. 

\subsubsection{Intermittency Matters!}
\label{sec:MS17}

The premise of both \citet{chandran15} and \citet{mallet17a} is that in order to 
determine the scalings of everything, including the energy spectrum, 
one must have a working model of intermittency, i.e., 
of the way in which fluctuation amplitudes and their scale lengths 
in all three directions---$\lambda$, $\xi$ and $\lpar$---are distributed
in a turbulent MHD system. It may be disturbing 
to the reader, or viewed by her as an unnecessary complication, 
that we must involve ``rare'' events---as this is what the theory of intermittency 
is about---in the mundane business of the scaling of the energy 
spectra, which are usually viewed as made up from the more ``typical'' fluctuations. 
These doubts might be alleviated by the following observation. 
The appearance of the outer scale $\Lpar$ in \exref{eq:dz_delta} 
suggests that the self-similarity is broken---this is somewhat analogous 
to what happens in hydrodynamic turbulence, where corrections to the 
K41 scaling \exref{eq:u_K41} come in as powers of $\lambda/L$ \citep{K62,frisch95}. 
We may view $\delta$ as just such a correction to the self-similar 
GS95 result, and alignment as the physical mechanism whereby this 
intermittency correction arises. The main difference with the hydrodynamic 
case is that $\delta$ is not all that small (MHD turbulence is 
``more intermittent'' than the hydrodynamic one), the mechanism 
responsible for it has important consequences (\secref{sec:disruption}),  
and so we care.   

\begin{figure}
\begin{center}
\begin{tabular}{cc}
\parbox{0.50\textwidth}{\includegraphics[width=0.50\textwidth]{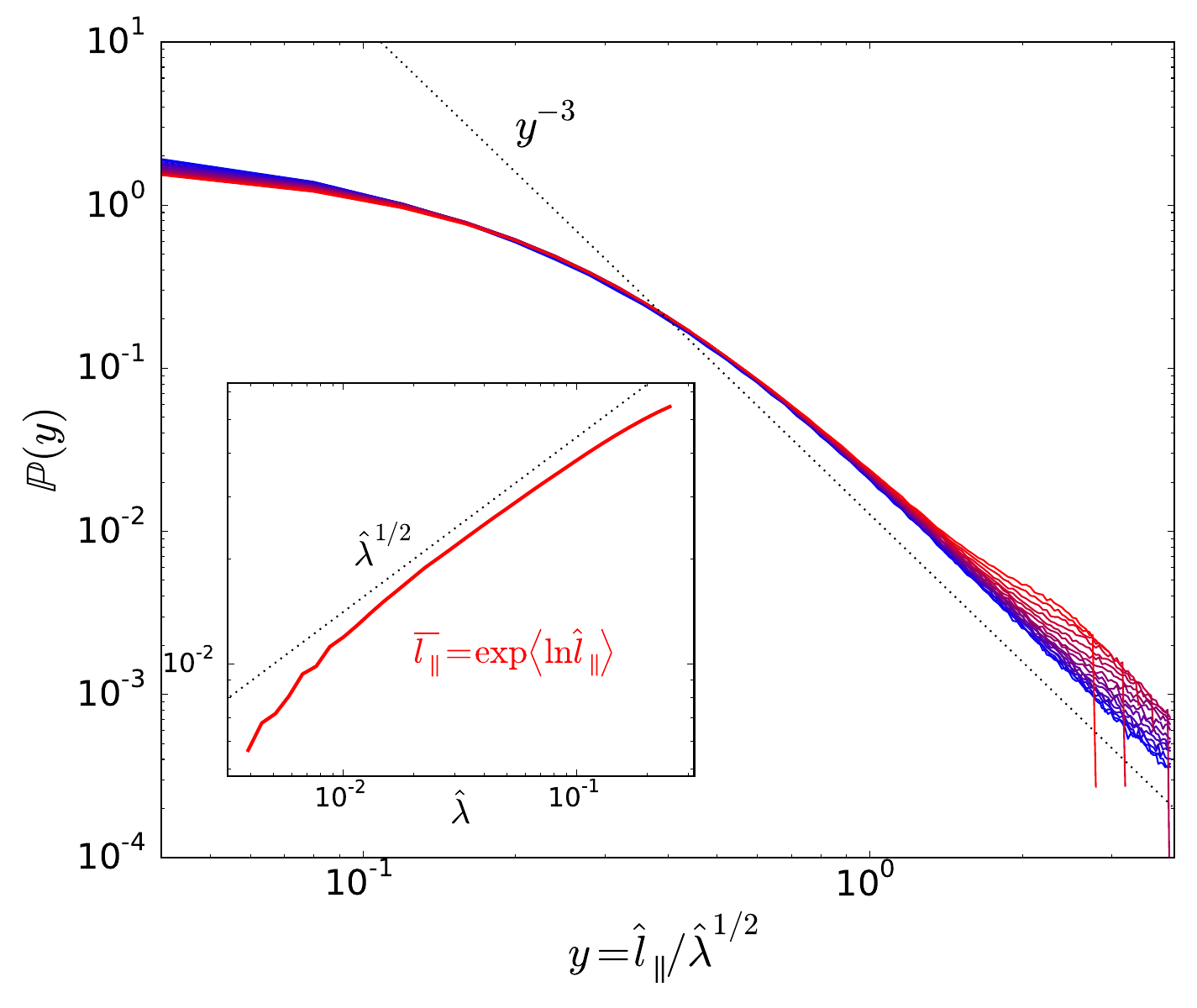}} & 
\parbox{0.49\textwidth}{\includegraphics[width=0.49\textwidth]{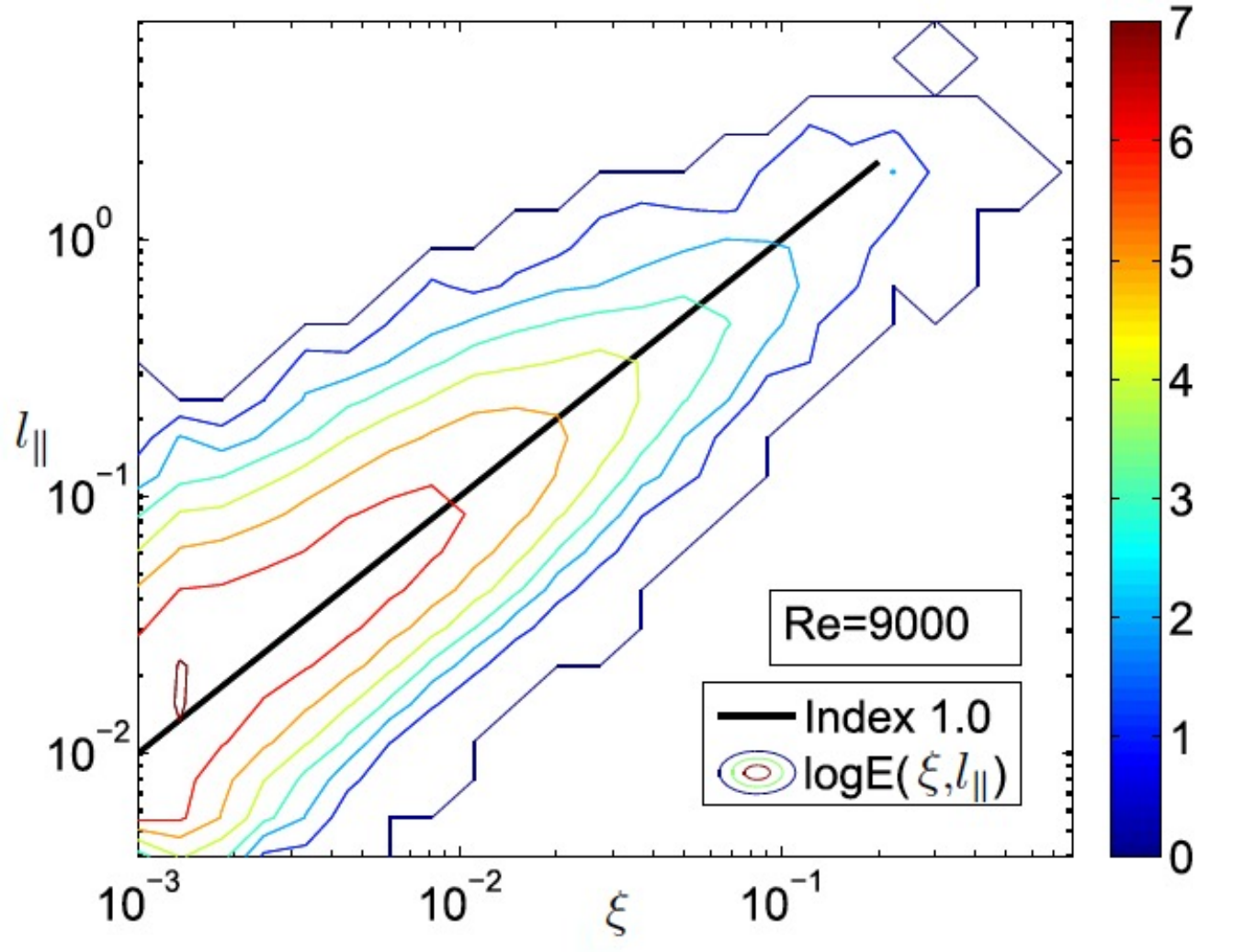}\\}\\
(a) & (b)
\end{tabular}
\end{center}
\caption{(a) Probability distribution of $\lpar/\lambda^{1/2}$ in 
a $1024^3$ RMHD simulation (the shades 
of colour from blue to red correspond to PDFs at increasing scales 
within the inertial range). This plot is taken from \citet{mallet17a} 
(where the reader will find a discussion---somewhat inconclusive---of the 
slope of this PDF) and illustrates how good (or otherwise) is the assumption that 
$\lpar/\lambda^\alpha$ has a scale-invariant distribution 
(the assumption is not as good as RCB, illustrated in \figref{fig:rcb} 
based on data from the same simulation).
(b) Joint probability distribution 
for the length $\lpar$ and width $\xi$ (in my notation) of the most intense 
dissipative structures (adapted from \citealt{zhdankin16}). This shows that 
$\xi\propto\lpar$, in line with~\exref{eq:P_lpar}.
[Reprinted from \citet{zhdankin16} with the permission of AIP Publishing.]
Independent simulations by J.~M.~Stone (private communication, 2018) support this scaling.}
\label{fig:lpar_lambda}
\end{figure}

I shall forgo a detailed discussion of the intermittency model that 
\citet{mallet17a} proposed; 
for my purposes here, a vulgarised version of their argument 
will suffice. They consider the turbulent field as an ensemble of structures, 
or fluctuations, each of which has some amplitude and three scales: 
parallel $\lpar$, perpendicular $\lambda$ and fluctuation-direction $\xi$ 
(they call this the ``RMHD ensemble''). They make certain conjectures 
about the joint probability distribution of these quantities, which then allow 
them to fix scalings. The most crucial (and perhaps also the most arbitrary) of 
these conjectures is, effectively, that for all structures, $\lpar\sim\lambda^\alpha$ 
with the same exponent $\alpha$, i.e., that the quantity $\lpar/\lambda^\alpha$ 
has a scale-invariant distribution (this appears to be confirmed 
by numerical evidence: see \figref{fig:lpar_lambda}a). 
They then determine the exponent $\alpha$ by 
considering ``the most intense structures''\footnote{Often an object of particular 
importance in intermittency theories \citep[e.g.,][]{she94,dubrulle94,she95,grauer94,mueller00,boldyrev02,boldyrev02prl}.}---because it is possible to work 
out what the probability of encountering them is as a function both of $\lambda$ and of~$\lpar$. 

They then conjecture that the most intense structures in the RMHD ensemble are sheets
transverse to the local perpendicular direction. Therefore, if one looks for 
their probability (filling fraction) in any perpendicular plane 
as a function of the perpendicular scale $\lambda$, one expects it to scale as 
\beq
P \propto \lambda. 
\label{eq:P_lambda}
\eeq
If, on the other hand, one is interested in their filling fraction 
in the plane locally tangent to a flux sheet (i.e., defined by the 
local mean field and the direction of the fluctuation vector), it is
\beq
P \propto \xi\lpar.
\label{eq:P_xi}
\eeq
The next conjecture is the ``refined critical balance'' (RCB, already advertised 
in \secref{sec:CBCB}), stating that not only is $\tnl\sim\tA$ in some vague ``typical'' 
sense, but the quantity 
\beq
\chi=\frac{\dz\lpar}{\xi\vA} \sim \frac{\tA}{\tnl} 
\eeq   
has a scale-invariant distribution in the RMHD ensemble---this was discovered 
by \citet{mallet15} to be satisfied with truly remarkable accuracy in numerically 
simulated RMHD turbulence (\figref{fig:rcb}).\footnote{Note that it makes sense then that 
the alignment angle $\sin\theta_\lambda\sim\lambda/\xi$ should be anticorrelated,
and $\xi$, therefore, positively correlated,   
with the fluctuation amplitude $\dz_\lambda$ at any given scale $\lambda$ 
(stronger fluctuations are more aligned---the strongest of them are the sheets 
being discussed here), as I mentioned in \secref{sec:plot} and as \citet{mallet15} 
indeed found.} 
If this is true for all structures, it is true 
for the most intense ones---and a further assumption about those is that 
their amplitude $\dzmax$ is not a function of scale but is simply equal to some typical 
outer-scale value (i.e., the most intense structures are formed by the largest 
perturbations collapsing, or being sheared, into sheets without breaking up 
into smaller perturbations; see \citealt{chandran15}). This, 
together with \exref{eq:P_xi}, implies that for those structures, 
\beq
\xi \sim \lpar \frac{\dzmax}{\vA}
\hence
P \propto \lpar^2 
\label{eq:P_lpar}
\eeq
(\citealt{zhdankin16} confirm that $\xi\propto\lpar$ for the most intense 
dissipative structures: see \figref{fig:lpar_lambda}b).
Comparing \exref{eq:P_lpar} with \exref{eq:P_lambda}, we conclude that $\lpar\propto\lambda^{1/2}$ 
for the most intense structures and, therefore, for everyone else---by the 
conjecture of scale invariance of $\lpar/\lambda^\alpha$, where we now know that $\alpha=1/2$. 
Comparing this with \exref{eq:lpar_sigma}, we see that $\alpha = 2(1-\sigma)/3$, whence 
\beq
\sigma = \frac{1}{4}\hence \delta=\frac{1}{8},
\eeq
the latter by virtue of \exref{eq:sigma_delta}. Q.e.d.: we now have the scalings \exref{eq:MS}.  

I do not know if the reader will find this quasi-intuitive argument more (or less)  
convincing than the formal-looking conjectures and corollaries in \citet{mallet17a}. 
There is no need to repeat their algebra here, but hopefully the above sheds some 
(flickering) light---if not, perhaps a better argument will be invented one day, 
but all I can recommend for now is reading their paper.  
Notably, in their more formal treatment, not just the energy spectrum but the two-point 
structure functions of all orders are predicted---and turn out to be a decent fit 
to numerical data as it currently stands.\footnote{The key tenet of their theory---a log-Poisson 
distribution of field increments, which follows from arguments analogous 
to those advanced in the hydrodynamic-turbulence theory \citep{she94,dubrulle94,she95}---also 
appears to be at least consistent with numerical evidence \citep{zhdankin16logpoisson,mallet17a}.}The same is true about the model proposed in the earlier paper by \citet{chandran15}. 
Their approach is based on a much more enthusiastic engagement with dynamics: 
a careful analysis of how aligned and non-aligned structures might form and interact. 
They get $\delta\approx0.108$, which leads to $\dz_\lambda\propto\lambda^{0.26}$---not 
a great deal of difference with \exref{eq:MS}, considering that all of this is very 
far from being exact science. Their approach does have the distinction, however, 
of emphasising particularly strongly 
the dynamic nature of the dynamic alignment, which arises as Elsasser fields shear 
each other into sheet-like structures.  

\subsection{3D Anisotropy}
\label{sec:3D}

\begin{figure}
\begin{center}
\begin{tabular}{c}
\includegraphics[width=0.95\textwidth]{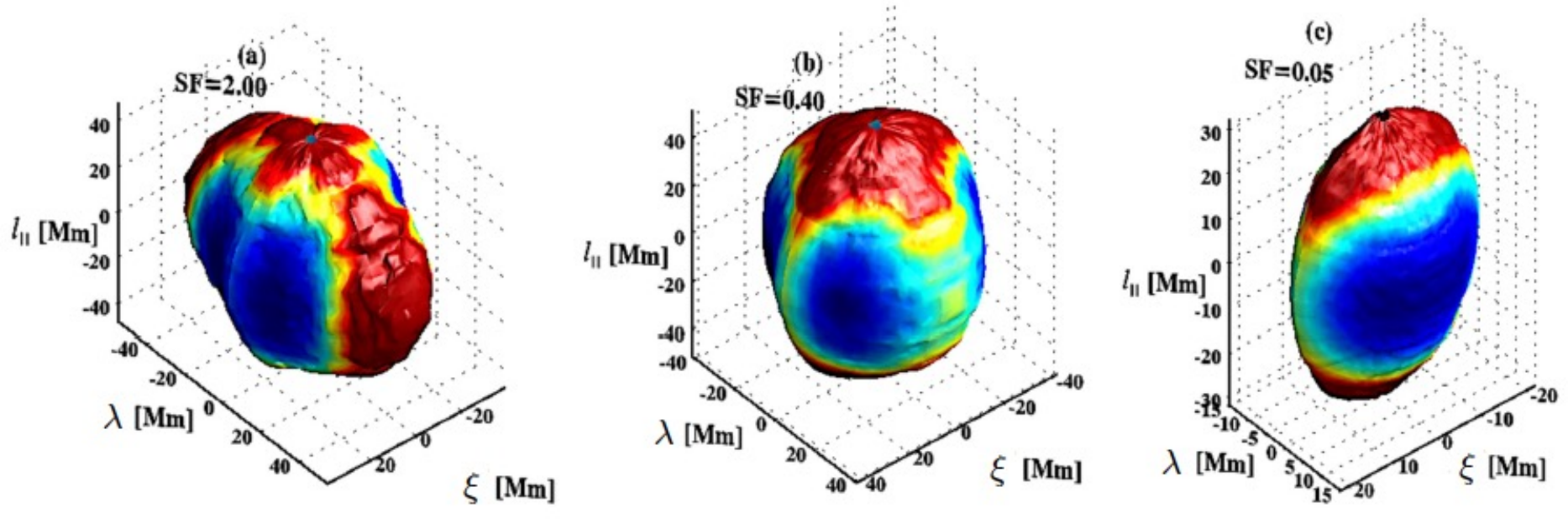}\\
(a) Solar wind (adapted from \citealt{verdini18}, data from {\em Wind} spacecraft)\\\\
\includegraphics[width=0.95\textwidth]{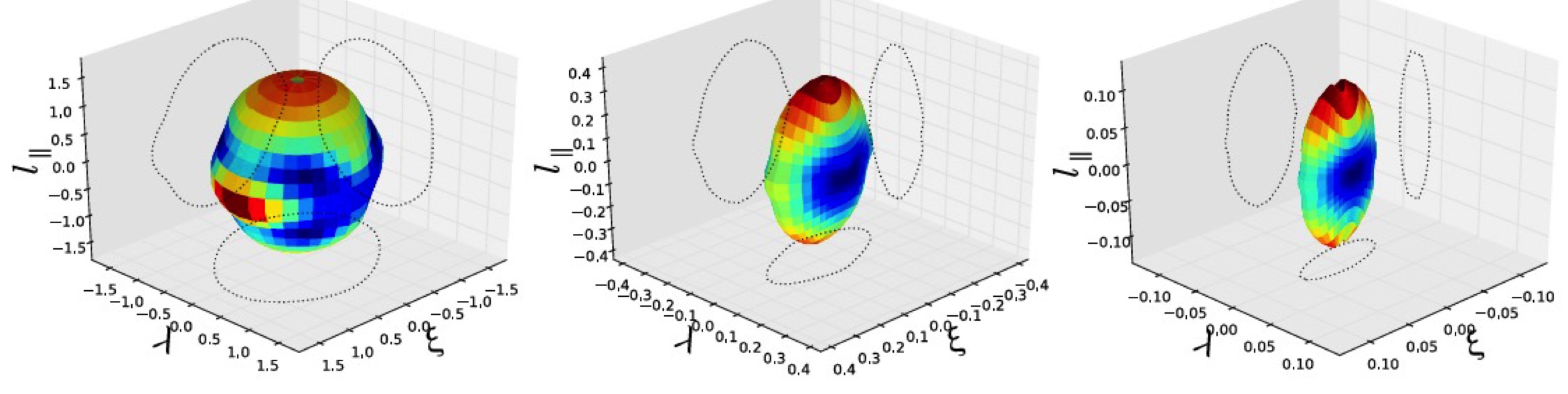}\\
(b) RMHD simulation \citep[from][]{mallet16}
\end{tabular}
\end{center}
\caption{Locally 3D-anisotropic structures in the (a) solar wind
[from \citet{verdini18} \copyright AAS, reproduced with permission]
and (b) numerical 
simulations (here $\lpar$ is normalised to $\Lpar/2\pi$ and $\lambda$ 
and $\xi$ to $\Lperp/2\pi$, hence apparent isotropy at the outer scale).
These are surfaces of constant second-order structure function of the magnetic 
field (a) or one of the Elsasser fields (b). The three images correspond to 
successively smaller fluctuations and so successively smaller scales 
(only the last of the three is firmly in the universal inertial range). 
In both cases, the emergence of statistics with $\lpar\gg\xi\gg\lambda$ 
is manifest. In the solar wind, the route that turbulence takes to this aligned state
appears to depend quite strongly on the solar-wind expansion, which 
distorts magnetic-field component in the radial direction compared 
to the azimuthal ones \citep{verdini15,vech16}. The data shown in panel (a) was 
carefully selected to minimise this effect; without such selection, one sees 
structures most strongly elongated in the $\xi$ direction at the larger scales 
($\xi>\lpar>\lambda$), although they too tend to the universal aligned regime 
at smaller scales (see \citealt{chen12}, where 3D-anisotropy plots like
ones shown here first appeared).} 
\label{fig:3D}
\end{figure}

Before moving on, I would like to re-emphasise the 3D anisotropy of the aligned 
MHD turbulence---and the fact that this anisotropy is local, associated 
at every point with the three directions that themselves depend on the fluctuating 
fields: parallel to the magnetic field ($\lpar$), along the vector direction of the perturbed 
field $\vzperp^\mp$ that advects the field $\vzperp^\pm$ whose correlations we are 
measuring ($\xi$), and the third direction perpendicular to the other two ($\lambda$).
This local 3D anisotropy is measurable\footnote{A sophisticated reader interested in 
how this can be done, might wonder whether the prescription given in \secref{sec:aniso} 
and based on defining the local field $\vBloc$ at each scale 
according to \exref{eq:Bloc_def} is still valid for aligned turbulence: 
indeed, would the distance \exref{eq:Dlperp} by which the point-separation vector $\vl$ 
veered off the exact 
field line not be $\Delta\lperp\gg\lambda$ even when the coarse-graining scale is
$\Lperp\sim\lambda$, because in \exref{eq:Dlperp}, $l/\vA\sim\lambda/\db_\lambda\sin\theta_\lambda$? 
In fact, since $\Delta\lperp$ is clearly in the direction of $\vbperp$, 
the fluctuation direction, 
all we need to do in order to preserve parallel correlations is to ensure $\Delta\lperp\ll\xi$. 
This is indeed marginally satisfied when $\Lperp\sim\lambda$ because, 
in \exref{eq:Dlperp}, $l/\vA \sim \xi/\db_\lambda$. \citet{chen12} and \citet{verdini18,verdini19} 
observationally 
and \citet{mallet16} numerically used this method with apparent success.} 
and has indeed been observed both in the solar wind \citep{chen12,verdini18,verdini19} 
and in numerical simulations \citep{verdini15,mallet16}---both are illustrated 
by \figref{fig:3D}. The main point of discrepancy between the true and virtual reality 
is the scale dependence of the anisotropy: confirmed solidly in simulations 
but only very tentatively in the solar wind (see footnote~\ref{fn:alignment_SW}). 
However, progress never stops, and one can hope for better missions \citep{chen20} 
and even more sophisticated analysis. 

The scaling of the energy spectrum in the parallel direction (\secref{sec:par_cascade}) 
was arguably the most robust and uncontroversial of the results reviewed thus far. 
We then occupied ourselves with the scalings of 
the Elsasser-field increments and of $\lpar$ vs.\ the perpendicular scale $\lambda$, 
culminating in \secref{sec:revised} with a theory that one (hopefully) can believe in. 
The scalings with the third, fluctuation-direction coordinate $\xi$ 
are very easy to obtain because the nonlinear time 
of the aligned cascade \exref{eq:tnl_align} has the same dependence on $\xi$
as it did on $\lambda$ in the unaligned, GS95 theory: see \exref{eq:tnl_zpm}. 
Therefore,  
\beq
\dz_\xi \sim (\eps\xi)^{1/3},\quad
\xi \sim \eps^{1/8}\lt(\frac{\Lpar}{\vA}\rt)^{3/8}\lambda^{3/4} 
\sim \lCB^{1/4}\lambda^{3/4},
\label{eq:xi_scalings}
\eeq 
with the latter formula following from \exref{eq:theta_lxi} and \exref{eq:MS} 
or \exref{eq:MS_normalised}. 
Thus, Elsasser fields' spectra have exponents $-2$ in the $\lpar$ direction, 
$-3/2$ in the $\lambda$ direction and $-5/3$ in the $\xi$ direction
(see the $n=2$ exponents in \figref{fig:intcy}a). Let me note in passing 
that the ``Kolmogorov'' scaling \exref{eq:xi_scalings} will 
play a key part in my discussion, in \apref{app:stoch_rec}, of why 
the \citet{lazarian99} notion of ``stochastic reconnection'' does not 
automatically invalidate the aligned cascade and return us to GS95, 
as an erudite reader might have been worried it would---the idea is that 
the \citet{richardson26} (super)diffusion of Lagrangian trajectories 
in a turbulent field is always determined by the $\xi$-dependent 
scaling~\exref{eq:xi_scalings}, regardless of the nature of the cascade in~$\lambda$. 

A good way of thinking physically of the inevitability of 3D anisotropy 
is to note that, from \exref{eq:tnl_align} and CB, 
\beq
\xi \sim \lpar\frac{\dz_\lambda}{\vA} \sim \lpar \frac{\db_\lambda}{\vA},
\label{eq:xi_displacement}
\eeq 
i.e., $\xi$ is the typical displacement of a fluid element 
and also the typical perpendicular distance a field line wanders within a structure 
coherent on the parallel scale $\lpar$. Fluctuations must therefore 
preserve coherence in their own direction at least on the scale $\xi$. 
They are not constrained in this way in the third direction $\lambda$, and 
the fluctuation direction itself has an angular uncertainty 
of the order of the angle $\theta_\lambda$ between the two fields, so 
it makes sense that the aspect ratio of the structures in the perpendicular 
plane should satisfy \exref{eq:theta_lxi}. 

The dependence of the anisotropy on the local direction of the fluctuating fields 
makes the connection between anisotropy, alignment and intermittency more obvious: 
when we follow perturbed field lines to extract parallel correlations 
or measure one Elsasser field's decorrelation along the direction of another 
Elsasser field, we are clearly not calculating second-order statistics 
in the strict sense---and so, in formal terms, local scale-dependent 
anisotropy always involves correlation functions of (all) higher 
orders.\footnote{It is easy to show that a Gaussian field cannot have 
scale-dependent alignment---although a solenoidal field will naturally 
have some modest scale-independent one \citep{chen12,mallet16}. 
Note also the paper by \citet{matthaeus12}, where the role of higher-order statistics 
in locally parallel correlations is examined with great punctiliousness.} 
Thus, it makes a certain natural sense to speak of the alignment-induced departure 
of MHD-turbulence spectrum from the Kolmogorovian GS95 scaling 
and of the 3D anisotropy of the underlying fluctuation field 
as an intermittency effect, as I have done here. 

\subsection{Higher-Order Statistics}
\label{sec:intcy}

\begin{figure}
\begin{center}
\begin{tabular}{cc}
\includegraphics[width=0.49\textwidth]{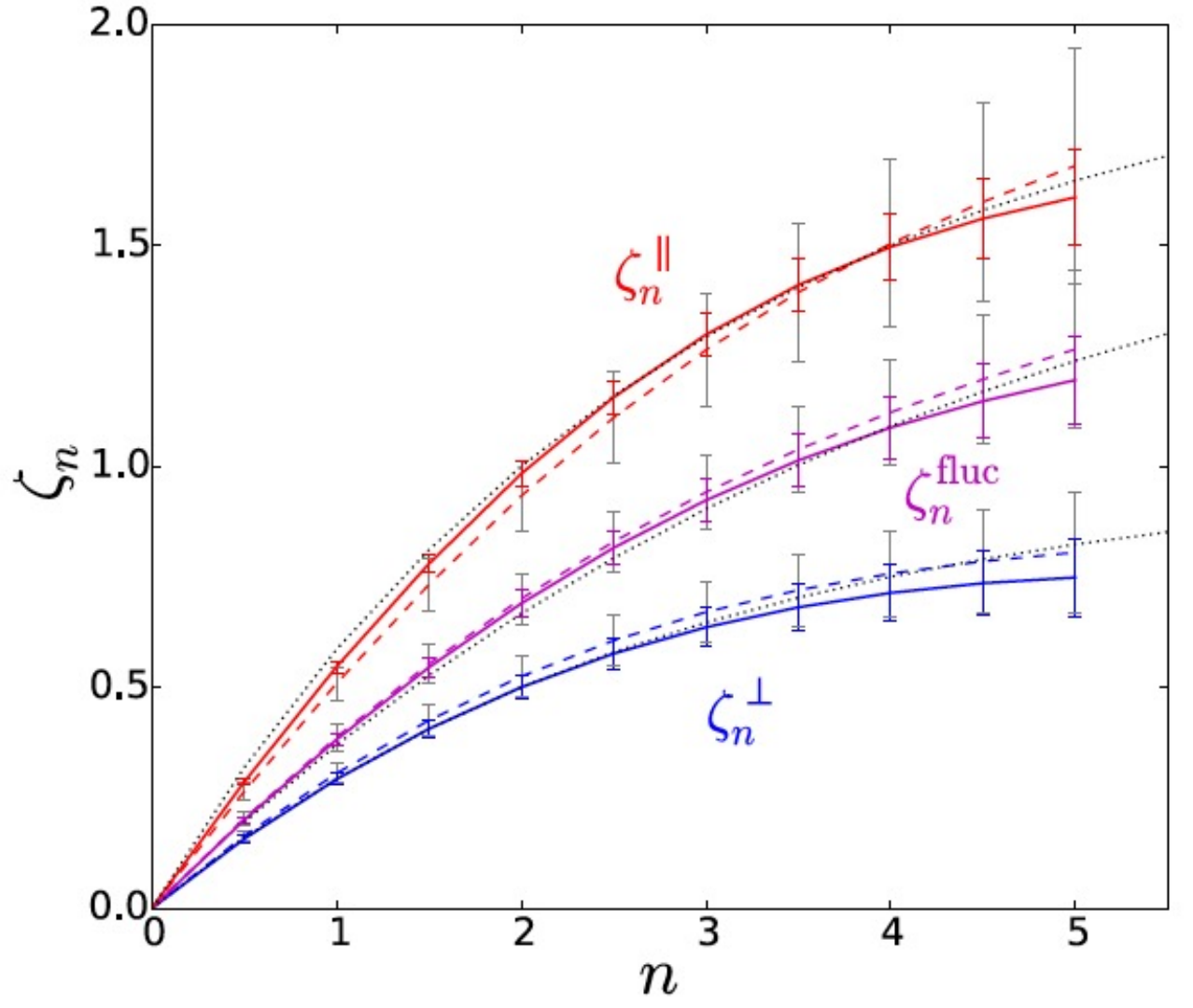} & 
\includegraphics[width=0.49\textwidth]{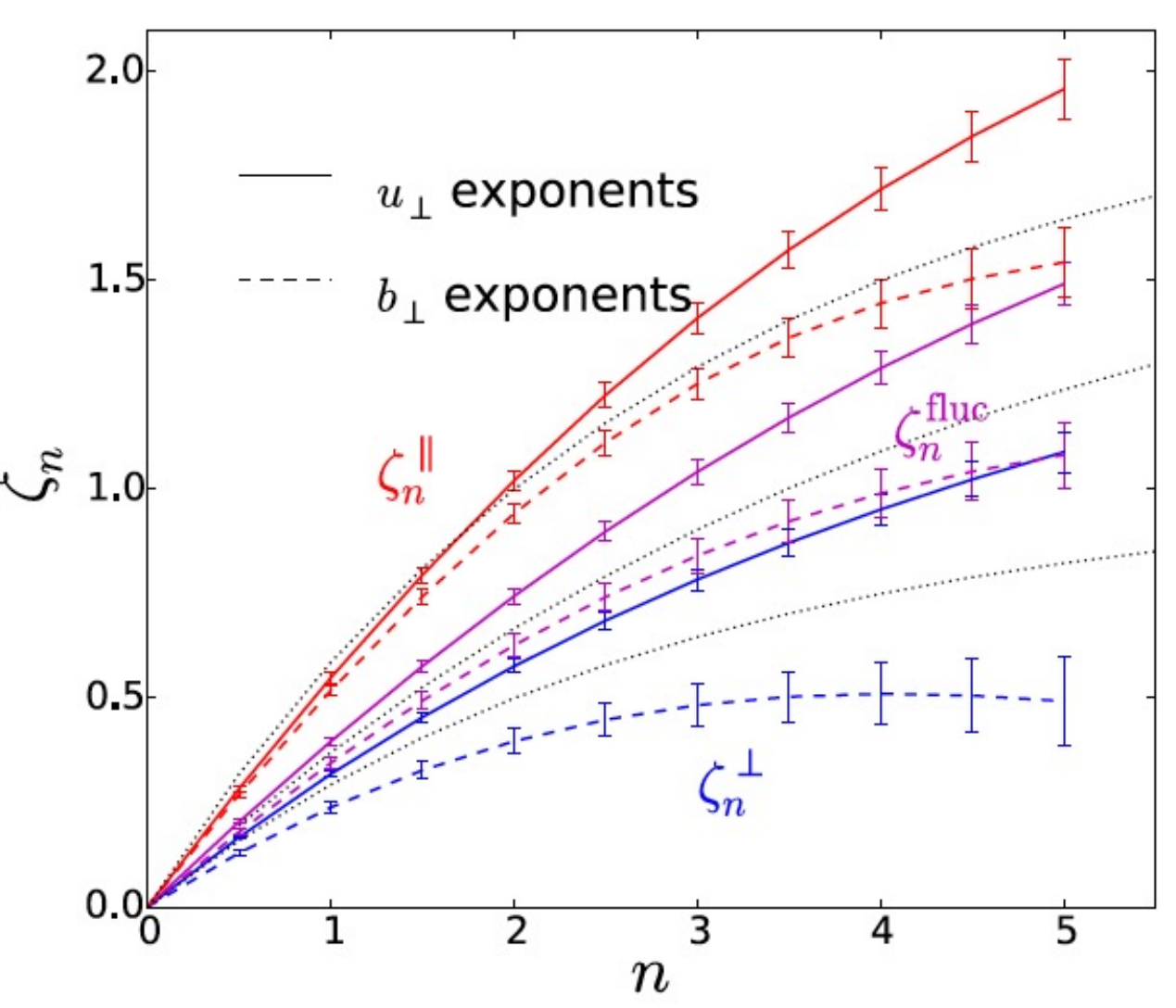}\\
(a) & (b)
\end{tabular}
\end{center}
\caption{Scaling exponents of the structure functions in RMHD turbulence 
simulated by \citet{mallet16} (the plot is reproduced from \citealt{mallet17a}). 
(a) Structure functions of the Elsasser-field increments \exref{eq:dvz_def}: 
by definition, $\la|\dvz_\vl^+|^n\ra \propto l^{\zeta_n}$ and $\zeta_n^\perp$, 
$\zeta_n^\mathrm{fluc}$, $\zeta_n^\parallel$ are exponents for 
$l=\lambda$, $\xi$, $\lpar$, respectively (i.e., all structure functions 
are conditional on the point separation being in one of the three 
directions of local 3D anisotropy; see \secsand{sec:aniso}{sec:3D}). 
The solid lines are for a $1024^3$ simulation (with hyperviscosity), 
the dashed ones are for a $512^3$ simulation, indicating how converged, or otherwise, 
the exponents are, and the dotted lines, in both (a) and~(b), 
are the theoretical model by \citet{mallet17a}.
(b) Similarly defined structure functions of the velocity (solid lines) 
and magnetic-field (dashed lines) increments from the same $1024^3$ simulation. 
The magnetic field is ``more intermittent'' than the Elsasser fields and the 
latter more so than velocity. An early (possibly first) numerical measurement of 
this kind, highlighting the differences between scalings of different fields and 
in different local directions, was done by \citet{cho03}.}
\label{fig:intcy}
\end{figure}

In several places (e.g., in \secsand{sec:aniso}{sec:MS17}), 
I have brushed against the more formal task of the intermittency 
theory---the calculation of the scaling exponents of higher-order structure functions
or, equivalently, of the probability distributions of field increments---and recoiled 
every time, opting for ``twiddle'' algebra and statements about spectra. 
A fair amount of information on these matters is available from simulations 
and from the solar-wind measurements: what intermittency 
looks like in the former is illustrated by \figref{fig:intcy} 
(a survey of previous measurements of structure functions, 
both in simulations and in the solar wind, can be found in \citealt{chandran15}). 
Some of what is known is perhaps understood, but much remains a mystery: 
for example, we do not know why the higher-order scaling exponents are generally 
not the same for velocity, Elsasser and magnetic fields, 
with the latter ``more intermittent'' than the former, as 
is evident in~\figref{fig:intcy}(b) 
(in \secref{sec:new_res_theory}, I will moot a possible connexion between intermittency 
and negative ``residual energy''---an asymmetry between the magnetic and velocity 
spectra seen both in numerical simulations and in the solar wind---but I do not 
know how to translate this into anything resembling~\figref{fig:intcy}b).

Interesting as it is, I will now leave the problem of higher-order statistics alone. 
We know from the (ongoing) history of hydrodynamic-turbulence 
theory that once this becomes {\em the} unsolved problem that everyone is 
working on, the scope for abstract theorising expands to fill all available space 
(and time) while attention paid by the outside world diminishes.\footnote{Let me qualify 
this by mentioning a recent paper by \citet{mallet19} where abstract theory of intermittency 
is converted into insights into particle-heating physics in the solar wind (more of the 
\citealt{chandran10} stochastic heating is found in the more intense, intermittent patches), which 
some might view as a more ``practical'' (in the astrophysical sense) preoccupation.} 
This said, I hasten to dispel any possible impression 
that I do not consider intermittency of MHD turbulence 
an important problem: in fact, as I have argued above, intermittency 
as a physical phenomenon appears to be so inextricably hard-wired into the structure 
of MHD turbulence that any workable theory of the latter has to be a theory of its 
intermittency. 

Finally, let me jump ahead of myself and mention also that we know 
nothing at all of the intermittency in ``tearing-dominated turbulence,'' 
which is about to be introduced (\secref{sec:disruption}), and very little 
of the intermittency in the variants of MHD turbulence 
surveyed in \secsdash{sec:imbalanced}{sec:dynamo}. In particular, 
the relationship between intermittency and Elsasser imbalance, local or global,
appears to me to be a promising object for theoreticians' scrutiny
(see~\secref{sec:Eimb}).

\section{MHD Turbulence Meets Reconnection}
\label{sec:disruption}

\vskip2mm
\begin{flushright}
{\small \parbox{8.5cm}{Finally, we wonder if it is possible that Sweet's mechanism 
might modify somewhat the diffusion and dissipation 
of the magnetic field in hydromagnetic turbulence.}
\vskip2mm
Last sentence of \citet{parker57}} 
\end{flushright}
\vskip5mm

If we accept that MHD turbulence in the inertial range---or, at least, in some subrange 
of the inertial range immediately below the outer scale---has a tendency to organise 
itself into fettuccine-like structures whose aspect ratio in the 2D plane perpendicular 
to the mean magnetic field increases as their size decreases, we are opting for a state 
of affairs that is not sustainable asymptotically, at ever smaller scales. 
These structures are mini-sheets, and sheets in MHD tend to be tearing-unstable. Thus, 
just like WT, strong aligned turbulence too carries in it the seeds of its own 
destruction, making an eventual transition to some new state inevitable at sufficiently 
small scales.\footnote{That this transition can and, generally speaking, will, happen 
{\em within the inertial range} is made obvious by the following rather apt observation  
due to \citet{uzdensky06}. The aspect ratio of an aligned sheet-like structure 
at Boldyrev's cutoff scale \exref{eq:lres_Rm} is $\xi/\lambda \sim \Rm^{1/6}(1+\Pm)^{-1/6}$, 
using \exref{eq:xi_scalings} for $\xi$ and setting $\lambda=\lres$. 
The Lundquist number at this scale is 
$S_\xi = \dz_{\lres}\xi/\eta \sim \Rm^{1/3}(1+\Pm)^{2/3}$. Therefore, 
$\xi/\lambda \sim S_\xi^{1/2}(1+\Pm)^{-1/2}$. Apart from the $\Pm$ dependence, 
this is the aspect ratio of a Sweet--Parker (SP) current sheet, which is $S_\xi^{1/2}(1+\Pm)^{-1/4}$ 
(see \apref{app:SP_rec}). But, provided $S_\xi$ is large enough and $\Pm$ is not too large, 
such a sheet will be violently (i.e., high above threshold) unstable to 
the plasmoid instability, which is a variety of tearing mode and has a growth rate that is 
much larger than the nonlinear rate at which the sheet is formed 
(see \apref{app:loureiro}). Therefore, tearing should muscle its way 
into turbulent dynamics already at some scale that is larger than~$\lres$.} 

The notion that current sheets will spontaneously form in a turbulent MHD fluid 
is not new \citep{matthaeus86,politano89} and the phenomenology of these structures has been 
studied (numerically) quite extensively, e.g.,
by \citet{servidio09,servidio10,servidio11rev,servidio11} in 2D and 
by \citet{zhdankin13,zhdankin14,zhdankin15,zhdankin16} in 3D (see also \citealt{wan14}),
while solar-wind measurements \citep[e.g.,][]{retino07,sundkvist07,osman14,greco16} 
provided motivation and, perhaps, vindication. However, theoretical discussion 
of these results appeared to focus on the association between current 
sheets in MHD turbulence and its intermittent nature, identifying spontaneously 
forming current sheets as the archetypal intermittent events---and effectively segregating this 
topic from the traditional questions about the energy spectrum and the ``typical'' structures
believed to be responsible for it, viz., Alfv\'enic perturbations, 
aligned or otherwise.

In fact, as I argued in \secsand{sec:MS17}{sec:3D}, it is impossible 
to separate the physics of alignment from that of intermittency.
Dynamic alignment produces sheet-like, or ``proto-sheet'', structures that measurably affect
the energy spectrum but are also the intermittent fluctuations that can perhaps 
collapse into proper current sheets. The likelihood that they will do 
so---or, more generally, that aligned structures can survive at all---hinges 
on whether the nonlinear cascade time $\tnl$ at a given scale $\lambda$ 
is longer or shorter than the typical time scale on which a tearing mode can be triggered, 
leading to the break up of the dynamically forming sheets into islands \citep{uzdensky16}. 
Since the growth rate of the tearing mode in resistive MHD is limited by resistivity 
and would be zero in the limit of infinitely small $\eta$, the aligned 
turbulent cascade should be safe from tearing above a certain scale that 
must be proportional to some positive power of~$\eta$. However, this need not 
be the same as Boldyrev's cutoff scale \exref{eq:lres_MS} 
that arises from the competition between the cascade rate 
and vanilla Ohmic (or viscous) diffusion ($\tnl$ vs.\ $\tres$)---so, at the very least, the 
cutoff scale of the aligned cascade may not be what one might have 
thought it was, and what happens below that scale may be more interesting 
than the usual dull exponential petering out of the energy spectrum in the dissipation range.  

This possibility was explored by \citet{mallet17b} 
and \citet{loureiro17} (unaware of each other's converging preoccupations), 
leading to a new scaling for the aligned cascade's cutoff and  
to a new model for the tail end of the MHD turbulence spectrum---mitigating some 
of the unsatisfactory features of the aligned-turbulence paradigm 
and thus providing a kind of glossy finish to the overall picture (despite 
their rather esoteric nature, the two papers appear to have become instant 
classics---so much so as to merit logarithmic corrections: \citealt{comisso18}). 
While the key idea in the two papers was the same, 
their takes on its consequences for the ``tearing-mediated turbulence'' 
were somewhat different---here I will follow \citet{mallet17b}, 
but present their results in a somewhat simpler, 
if less general, form.\footnote{Namely, I will ignore the nuance that, in an 
intermittent ensemble, fluctuations of different strengths that are always present 
even at the same scale will be affected by reconnection to a different degree and so 
more intense structures will be disrupted at larger scales than the less intense ones. 
I will also not present scalings that follow from
the theory of the aligned cascade by \citet{chandran15}, focusing 
for simplicity exclusively on the model by \citet{mallet17a} (which is the same 
as Boldyrev's original theory if the latter is interpreted 
as explained in \secref{sec:revised}). 
In this sense, my exposition in \secref{sec:disruption_scale} is closer 
in style to \citet{loureiro17} than the paper by \citet{mallet17b} was. 
The material difference between the two arises in \secref{sec:recturb} 
and concerns the spectrum of the tearing-mediated turbulence. 
This is now moot, however, as the follow-up 
paper by \citet{boldyrev17} embraced  the \citet{mallet17b} 
spectrum, if not quite the physical model that led to it (see \secref{sec:spectrum_rec}).}  

Before I proceed, let me alert my erudite reader that the profound alteration 
of the MHD cascade by reconnection that I will discuss here has nothing at all to do 
with the concept of ``stochastic reconnection'' in MHD turbulence associated with the
work of Lazarian, Vishniac, Eyink, and their co-workers---this is explained carefully
in~\secref{sec:stoch} and in \apref{app:stoch_rec}. 

\subsection{Disruption by Tearing}
\label{sec:disruption_scale}

The scale at which the aligned structures will be disrupted by tearing can be estimated 
very easily by comparing the nonlinear time \exref{eq:tnl_align}
of the aligned cascade with the growth time 
of the fastest tearing mode that can be triggered in an MHD sheet of a given transverse
scale~$\lambda$. That this growth time is a good estimate for 
the time that reconnection needs to break up 
a sheet forming as a result of ideal-MHD dynamics 
is not quite as obvious as it might appear, but it is true 
and was carefully shown to be so by \citet{uzdensky16}. 
The maximum tearing growth rate~is 
\beq
\gamma \sim \frac{\vAy}{\lambda}\, S_\lambda^{-1/2} (1+\Pm)^{-1/4},
\quad
S_\lambda = \frac{\vAy\lambda}{\eta},
\quad
\Pm = \frac{\nu}{\eta}. 
\label{eq:gmax_TM}
\eeq
How to derive this is reviewed in \apref{app:TM} [see \exref{eq:TM_max}]. 
Here $\vAy$ is the Alfv\'en speed associated with the perturbed magnetic field 
that reverses at scale $\lambda$, $S_\lambda$ is the corresponding Lundquist 
number and $\Pm$ is the magnetic Prandtl number, which only matters if 
the viscosity $\nu$ is larger than the magnetic diffusivity $\eta$. 
In application to our aligned Alfv\'enic structures, we should replace $\vAy\sim\dz_\lambda$. 
Then, using the scalings \exref{eq:MS_normalised} to work out $\tnl$, 
we find that the aligned cascade is faster than tearing as long~as 
\beq
\gamma \tnl \sim \frac{S_\lambda^{-1/2} (1+\Pm)^{-1/4}}{\sin\theta_\lambda} \ll 1
\quad\Leftrightarrow\quad
\lambda \gg 
\Rm^{-4/7}(1+\Pm)^{-2/7}\lCB \equiv \lD, 
\label{eq:lambdaD}
\eeq
where $\Rm \sim S_{\lCB}$, as defined in \exref{eq:RmCB_def}, and 
the scale $\lD$ will henceforth be referred to as the {\em disruption scale}. 
At scales $\lambda\lesssim \lD$, aligned sheet-like structures can no longer retain 
their integrity against the onslaught of tearing.\footnote{This is equivalent to the idea 
of \citet{pucci14} that one can determine the maximum allowed aspect ratio of sheets in MHD 
by asking when the linear-tearing time in the sheet becomes comparable 
to its ideal-MHD dynamical evolution time (see \apref{app:pucci}). 
Careful examination of forgotten ancient texts reveals that nothing is entirely
new under the Moon and the idea that tearing might disrupt 
the MHD cascade in fact appeared first in an early paper by \citet{carbone90}, 
who derived the disruption scale \exref{eq:lambdaD} (without $\Pm$) by comparing 
the tearing growth rate \exref{eq:gmax_TM} with the cascade rate taken from
the IK theory---this gives the same scaling, $\lD\propto\Rm^{-4/7}$, 
because the IK spectrum has the same scaling as Boldyrev's spectrum.
This said, their comparison between the two rates was purely formal: as \citet{boldyrev18}
rightly observe, there is nothing in the IK theory (or in GS95) that makes tearing disruption
inevitable---aligned structures are needed for that.} 

\subsubsection{Some Reservations and Limitations}

Let me observe parenthetically that one can entertain legitimate reservations about 
the validity of \exref{eq:gmax_TM} and other formulae based on 
laminar-tearing theory in a noisy, turbulent environment (see, e.g., the discussion 
and references in footnote~\ref{fn:huang}). Here suffice it to say, that at scales 
where $\gamma\ll\tnl^{-1}$, the laminar formulae are, of course, unjustified, and at scales 
where $\gamma\gg\tnl^{-1}$, if such situations existed, they would be perfectly 
fine, modulo the issue of flows, which I will explain in a moment.\footnote{Another
caveat---there is always another caveat in this business!---is that tearing mode's growth
rate can potentially be modified in at least an order-unity way by small-scale
corrugations superimposed on the unstable magnetic-field profile, even when the size of these
corrugations is relatively small (there is a literature on this: see \citealt{militello09}
and references therein). In the existing solved models, however, these
corrugations are required to have long-time coherence, which is unlikely
in the kind of turbulent environment that we are dealing with here (the context in
the literature is zonal fields generated by drift-wave turbulence in tokamaks,
a different beast).}
The disruption scale~$\lD$ is where $\gamma\sim\tnl^{-1}$, so the 
laminar-tearing formulae are presumably only order-unity wrong, which I acknowledge 
by using ``$\sim$'' instead of~''$=$" everywhere.

Another legitimate reservation concerns application to aligned structures formed by
the MHD cascade of the tearing theory derived for a magnetostatic equilibrium. 
Indeed, these structures are not purely magnetic sheets, 
but rather Elsasser ones, i.e., they have local shear flows 
superimposed on them. These flows are likely somewhat sub-Alfv\'enic 
(the reasons and evidence for that are explained in \secref{sec:new_res_theory}), but, 
technically speaking, it has not been shown that they are sub-Alfv\'enic enough 
to justify application of the no-flow tearing-mode theory or even that they cannot
fatally stabilise tearing (some further discussion 
and references on tearing with flows can be found at the end of \apref{app:loureiro}). 
Since the shear rate in these flows is no larger than $\tnl^{-1}$, 
I think that this difficulty is a technical one, rather than a deal breaker, i.e., 
that what I am doing here is again only order-unity wrong \citep[cf.][]{tolman18}, 
but I cannot prove it rigorously. 
If this admission has not discouraged the reader, let us proceed.

The disruption scale~$\lD$, upon comparison with the putative 
resistive cutoff \exref{eq:lres_Rm} of the aligned cascade turns out to 
supersede it provided $\Pm$ is not too large: 
\beq
\frac{\lD}{\lres} \sim \lt[\frac{\Rm}{(1+\Pm)^{10}}\rt]^{2/21}\gg1.
\label{eq:Pm_max}
\eeq 
In view of the ridiculous exponents involved, this means that in a system 
with even moderately large $\Pm$ and/or not a truly huge $\Rm$, the aligned 
cascade will happily make it to the dissipation cutoff \exref{eq:lres_Rm} 
and no further chapters are necessary in this story.\footnote{This is, in fact, 
not quite true: at $\Pm\gg1$, interesting things can happen between the viscous and 
resistive cutoffs---see \secref{sec:subvisc}. In particular, if the tearing disruption 
fails to occur in the inertial range, it may still occur at subviscous 
scales (\secref{sec:tearing_subvisc}).} 
However, I do want to tell the story in full and so will focus on situations 
in which the condition \exref{eq:Pm_max} is satisfied.

\subsubsection{Debris of Disruption}

I shall turn to the question of what happens at scales below $\lD$ in \secref{sec:recturb}, 
but to do that, it is necessary first to ask 
what becomes of the aligned structures that are disrupted at~$\lD$. 

The tearing instability that 
disrupts them, the so-called Coppi mode, or (the fastest-growing) resistive 
internal kink mode \citep{coppi76}, has the wavenumber [see~\exref{eq:TM_max}]
\beq
k_* \sim \frac{1}{\lambda}\,S_{\lambda}^{-1/4}(1+\Pm)^{1/8}
\sim\frac{1}{\lCB} \Rm^{-1/4}(1+\Pm)^{1/8}\lt(\frac{\lambda}{\lCB}\rt)^{-21/16},
\label{eq:kpeak_TM}
\eeq
where \exref{eq:MS_normalised} has been used to substitute for $\dz_\lambda$ inside $S_\lambda$.
Therefore, at the disruption scale ($\lambda=\lD$), 
\beq
k_*\sim\frac{1}{\lCB} \Rm^{1/2}(1+\Pm)^{1/2}.
\label{eq:kpeakD}
\eeq 
If referred to the length of the sheet $\xiD$,  
which depends on $\lD$ via \exref{eq:xi_scalings}, this wavenumber 
gives us an estimate for the number of islands in the growing perturbation: 
\beq
N \sim k_*\xiD \sim \Rm^{1/14}(1+\Pm)^{2/7}.  
\label{eq:N_islands}
\eeq
As this is always large (or at least $\gtrsim 1$), 
the mode fits comfortably into the sheet 
that it is trying to disrupt.\footnote{Based on \exref{eq:kpeak_TM}, we see that this 
would be the case for tearing perturbations of any inertial-range structure with   
$\lambda \lesssim \Rm^{-4/9}(1+\Pm)^{2/9}\lCB$. 
At larger scales than this, the fastest tearing mode that fits into the sheet is
the FKR mode \citep{furth63} with $\sim$~one growing island of size~$\sim\xi$ 
[see \exref{eq:FKR} and the discussion at the end of \apref{app:TMgammak}]. 
However, both this mode and the secular \citet{rutherford73} evolution that succeeds 
it are always slower than the Coppi mode and, therefore, 
than the nonlinear ideal-MHD evolution of the sheet, so there is no danger 
of disruption at scales greater than~$\lD$.} 

What happens to these islands? When the tearing mode enters 
the nonlinear regime, the island width~is (see \apref{app:TMnlin}) 
\beq
w \sim k_*\lD^2, 
\label{eq:w_nlin}
\eeq
which is smaller than $\lD$ and so, technically speaking, the aligned structures 
need not be destroyed by these islands. 
\citet{uzdensky16} (followed by \citealt{mallet17b} and by \citealt{loureiro17})
argue that, after the tearing mode goes nonlinear, the islands will grow further
while the $X$-points between them collapse into current sheets---all of that 
on the same time scale \exref{eq:gmax_TM} as the mode grew. It seems intuitive that,
in order to break up the aligned structure (ideal-MHD sheet) that spawned them,
the islands would have to get to $w\sim\lD$ \citep{uzdensky16}. If they
did this in a cross-section-area-preserving way, then they would be circular at the
point of disruption: from \exref{eq:w_nlin}, $w k_*^{-1}\sim\lD^2$. Such a set of islands
(flux ropes) would be isotropic in the perpendicular plane.
Another simplistic model is to imagine their width grow due to
further reconnection while their length remains the same, viz.,~$\sim k_*^{-1}$
(simplistic because the lengthening of the inter-island current sheets is ignored). 
The aspect ratio of such a flux rope once it reached the width $\lD$ would be
\beq
\lD k_* \sim N \frac{\lD}{\xiD} \sim N \sin\theta_{\lD} \sim \Rm^{-1/14}(1+\Pm)^{3/14},
\label{eq:align_D}
\eeq
where $N$ is given by \exref{eq:N_islands}. This is a degree of alignment
preserved, but reduced by a (asymptotically large) factor of $N$. 
At the extreme end of the range
of possibilities is the view of \citet{boldyrev17} that, in fact,
islands of size \exref{eq:w_nlin} are already sufficient to break up the
aligned structure, so alignment is not abruptly reduced by tearing disruption. 
Both this idea and some nuances about the $X$-point collapse are discussed
further in \secref{sec:MSC_vs_BL}. The key point for us here is that      
at the disruption scale, the aligned structures 
that cascade down from the inertial range are broken up by reconnection 
into flux ropes that are underaligned and may even be isotropic in the perpendicular plane. 
They are a starting point for a new kind of cascade, which I shall now proceed to consider. 

\subsection{Tearing-Mediated Turbulence}
\label{sec:recturb}

If you accept the argument that the disruption 
by tearing of an aligned structure at the scale~$\lD$ leads to its break-up into 
a number of flux ropes of width~$\sim\lD$, then the natural conclusion is that $\lD$ 
now becomes a kind of ``outer scale'' for a new cascade. There need not be anything 
particularly different about this cascade compared to the standard aligned cascade 
except the alignment angle may now be reset to a greater value. As the disruption-scale 
structures interact with each other and break up into smaller structures, the latter 
should develop the same tendency to align as their inertial-range predecessors did. 
For a while, the structures in this new cascade are safe from tearing as their aspect 
ratio is not large enough, but eventually (i.e., at small enough scales), they too 
will become sufficiently aligned to be broken up by tearing modes. This leads to 
another disruption, another iteration of an aligned ``mini-cascade,'' and so on. 
Thus, if we rename our critical-balance scale $\lCB=\lambda_0$, the 
disruption scale $\lD=\lambda_1$, and call the subsequent disruption scales~$\lambda_n$, 
we can think of the MHD cascade as consisting of a sequence of aligned 
cascades interrupted by disruption episodes. I shall calculate $\lambda_n$ and
the spectral exponents for this new, composite tearing-mediated cascade
in \secref{sec:spectrum_rec},
the final dissipation cutoff in \secref{sec:diss},
the cascade's alignment properties in \secref{sec:align_rec},
and its parallel scalings, again determined by CB, in~\secref{sec:par_rec}.    
  
\subsubsection{Mini-cascades and the Spectrum of Tearing-Mediated Turbulence}
\label{sec:spectrum_rec}

Let us calculate the disruption scales $\lambda_n$, following \citet{mallet17b}. 
Since the ``mini-cascades'' that connect 
them are just the same as the aligned cascade whose disruption we analysed in 
\secref{sec:disruption_scale}, we can use \exref{eq:lambdaD} to deduce a recursion 
relation 
\beq
\lambda_{n+1} \sim S_{\lambda_n}^{-4/7}(1+\Pm)^{-2/7}\lambda_n
\label{eq:lambda_rec}
\eeq
(remembering that $\Rm$ was defined as the Lundquist number at scale $\lCB=\lambda_0$).
To work out the Lundquist number $S_{\lambda_n}$ at scale $\lambda_n$, notice that,
if the alignment angle $\theta_{\lambda}$ just below $\lambda_n$ is increased, 
there must be a downward jump in the amplitude of the turbulent fluctuations 
at~$\lambda_n$: indeed, the nonlinear time \exref{eq:tnl_align} 
shortens compared to what it was in the aligned cascade just above $\lambda_n$, 
and the cascade accelerates. I shall consider the extreme possibility that
the alignment is reset to being order unity, as this will effectively
bracket the range of outcomes for the scalings of this cascade.  
Since it still has to carry the same energy flux, 
we have, for amplitudes just below the disruption scale ($\lambda_n^-$), 
\beq
\frac{\bigl(\dz_{\lambda_n^-}\bigr)^3}{\lambda_n}\sim\eps
\hence
\dz_{\lambda_n^-} \sim (\eps\lambda_n)^{1/3},
\label{eq:z_below}
\eeq
just a Kolmogorov (or GS95) scaling. 
Therefore, the Lundquist number of the $n$-th mini-cascade is
\beq
S_{\lambda_n}\sim \frac{\dz_{\lambda_n^-}\lambda_n}{\eta}
\sim \frac{\eps^{1/3}\lambda_n^{4/3}}{\eta}
\sim \Rm\lt(\frac{\lambda_n}{\lCB}\rt)^{4/3}.
\label{eq:Sn}
\eeq  
In combination with \exref{eq:lambda_rec}, this gives us 
\beq
\frac{\lambda_n}{\lCB} 
\sim \lt[\Rm^{-4/7}(1+\Pm)^{-2/7}\rt]^{\frac{21}{16}\lt[1-\lt(\frac{5}{21}\rt)^n\rt]}
\to \Rm^{-3/4}(1+\Pm)^{-3/8},\quad
n\to\infty. 
\label{eq:lambda_n}
\eeq
I shall return to this obviously suggestive (Kolmogorov!) scaling in~\secref{sec:diss}. 

\begin{figure}
\centerline{\includegraphics[width=0.75\textwidth]{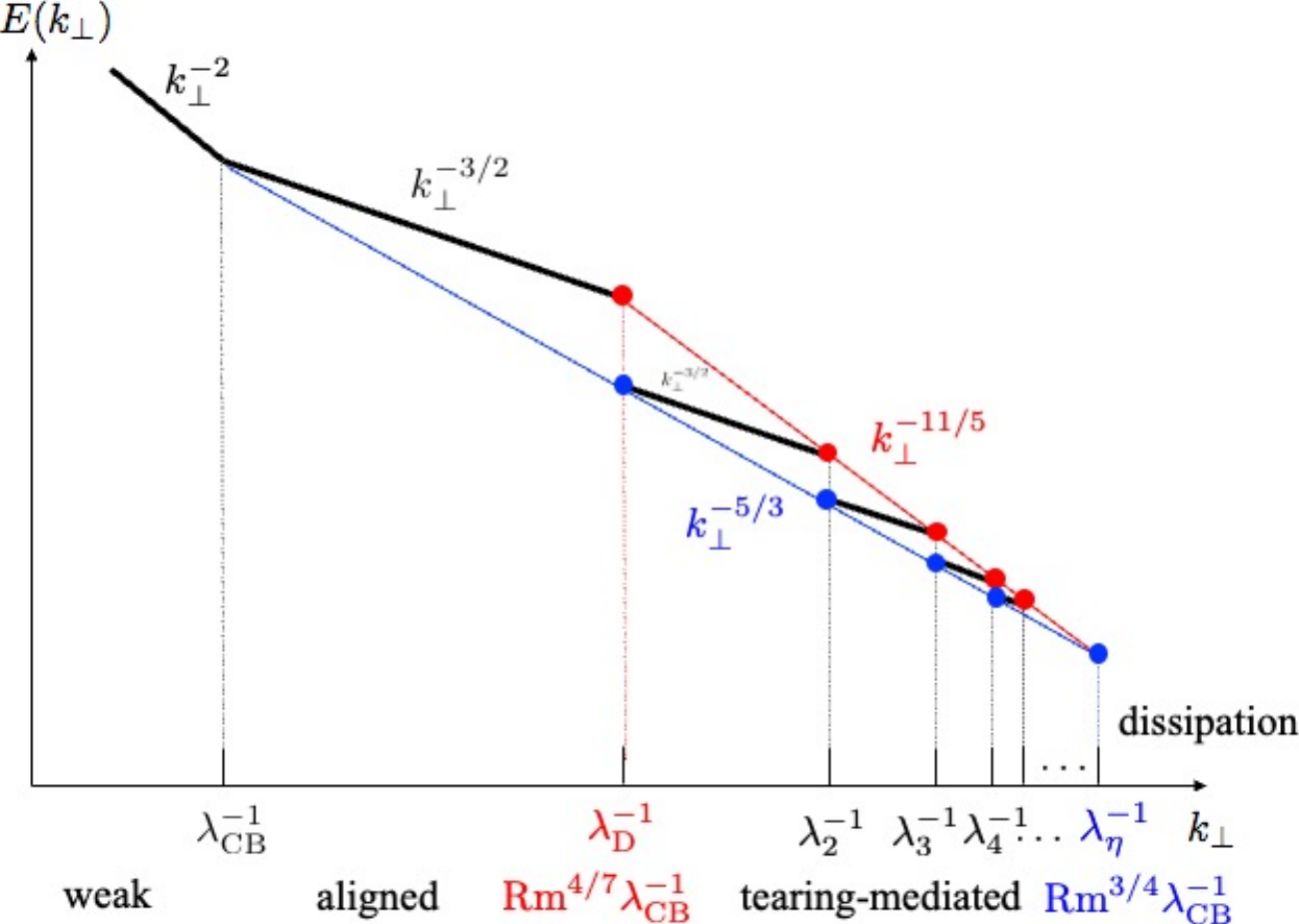}} 
\vskip2mm
\caption{Spectrum of MHD turbulence and transition 
to tearing-mediated cascade [see \exref{eq:Ek_bounds}] (adapted from \citealt{mallet17b}). 
The width of the tearing-mediated range is, of course, exaggerated in this cartoon.
The spectral slopes of the ``mini-cascades'' between $\lambda_n^{-1}$ and $\lambda_{n+1}^{-1}$
are all $\kperp^{-3/2}$, but the overall envelope is $\kperp^{-11/5}$. Note that,
modulo $\Pm$ dependence, 
the disrupted aligned cascade and a putative unaligned GS95 $\kperp^{-5/3}$ spectrum,
starting at $\lCB$, terminate at the same, Kolmogorov, scale~\exref{eq:K_cutoff}.}
\label{fig:spectrum}
\end{figure}

In the picture that I have just painted, the cascade in the tearing-mediated range looks
like a ladder (\figref{fig:spectrum}), 
with amplitude dropping at each successive disruption scale as structures 
become unaligned (or less aligned, in which case the the steps of the ladder are less tall).
In between the disruption scales, there are aligned 
``mini-cascades'' of the same kind as the original one discussed in \secref{sec:revised},
with $\kperp^{-3/2}$ spectra. 
This means that the overall scaling of the turbulent fluctuation amplitudes 
can be constrained between their scaling just below each disruption scale ($\lambda_n^-$),
given by \exref{eq:z_below}, and just above it ($\lambda_n^+$).
The latter scaling, for the 
amplitudes of the structures just before they get disrupted can be inferred 
from the fact that for these structures, the tearing growth 
rate \exref{eq:gmax_TM} must be the same as the nonlinear interaction (cascade) rate: 
letting $\vAy\sim \dz_{\lambda_n^+}$ in \exref{eq:gmax_TM}, we get 
\beq
\tnl^{-1}\sim\gamma 
\sim \bigl(\dz_{\lambda_n^+}\bigr)^{1/2}\lambda_n^{-3/2}\eta^{1/2}(1+\Pm)^{-1/4} 
\label{eq:tnl_n}
\eeq
and, therefore, 
\begin{align}
\nonumber
\frac{\bigl(\dz_{\lambda_n^+}\bigr)^2}{\tnl} \sim \eps
\hence
\dz_{\lambda_n^+} &\sim \eps^{2/5}\eta^{-1/5}(1+\Pm)^{1/10}\lambda_n^{3/5}\\
&\sim \lt(\frac{\eps\Lpar}{\vA}\rt)^{1/2}\lt(\frac{\lD}{\lCB}\rt)^{1/4}
\lt(\frac{\lambda_n}{\lD}\rt)^{3/5}.
\label{eq:tearing_cascade}
\end{align}
The last expression puts this result explicitly in contact with the inertial-range 
scaling~\exref{eq:MS_normalised}. 
Thus, the tearing-mediated-range spectrum is \citep{mallet17b}\footnote{\citet{boldyrev17} have 
a somewhat differently phrased derivation of the $\kperp^{-11/5}$ spectrum,
based on assuming no increase in alignment for the debris of disruption
(see \secref{sec:MSC_vs_BL}). The possibilities associated with fractional
increase, e.g.,~\exref{eq:align_D}, are bracketed by the range~\exref{eq:tearing_cascade}.} 
\beq
\eps^{2/3}\kperp^{-5/3} \lesssim E(\kperp) \lesssim \eps^{4/5}\eta^{-2/5}(1+\Pm)^{1/5}\kperp^{-11/5}. 
\label{eq:Ek_bounds}
\eeq
Since the $-11/5$ upper envelope is steeper than the $-5/3$ lower one, the two 
converge and eventually meet at 
\beq
\lambda_\infty \sim \eta^{3/4}\eps^{-1/4}(1+\Pm)^{-3/8} \sim \lD^{21/16}\lCB^{-5/16},
\label{eq:final_cutoff}
\eeq
which is, of course, the scale \exref{eq:lambda_n} in the limit $n\to\infty$. 
For $\Pm\gg1$, this will, in fact, be superseded by the Kolmogorov cutoff
to be derived in \secref{sec:diss}. 

While in the above construction, the tearing-mediated-range spectrum is pictured 
as a succession of ``steps'' representing the ``mini-cascades'' that connect 
the successive disruption scales (\figref{fig:spectrum}), 
the reality will almost certainly look more like some overall power-law spectrum with 
a slope for which the upper $-11/5$ bound \exref{eq:Ek_bounds} seems to be a good estimate. 
Indeed, the tearing disruptions will be happening within intermittently distributed aligned 
structures of different amplitudes and sizes, on which the disruption scales will depend 
\citep{mallet17b}. Thus, each scale $\lambda_n$ will in fact be smeared over 
some range and, as the successive intervals $(\lambda_n,\lambda_{n+1})$ become 
narrower, this smear can easily exceed their width. Pending a detailed 
theory of intermittency in the tearing-mediated range, perhaps the best way to think 
of the spectrum and other scalings in this range is, therefore, 
in a ``coarse-grained'' sense, focusing on the characteristic dependence of 
all interesting quantities on $\lambda_n$, treated as a continuous variable.  

\subsubsection{Restoration of Kolmogorov Cutoff}
\label{sec:diss}

To calculate the dissipative cutoff for the tearing-mediated cascade, one balances
the larger of the viscous and resistive diffusion rates with the nonlinear
cascade rate~\exref{eq:tnl_n} (this is the longest of the nonlinear time scales
involved): using \exref{eq:tearing_cascade}, one gets 
\beq
\frac{\nu+\eta}{\lres^2} \sim  \gamma
\hence
\lres \sim \eps^{-3/4}\eta^{3/4}(1+\Pm)^{3/2} \sim \lCB\Rm^{-3/4}(1+\Pm)^{3/2}
\equiv\lres^\mathrm{tearing}. 
\label{eq:tearing_cutoff} 
\eeq
For $\Pm\lesssim 1$, this is the same scale as~\exref{eq:final_cutoff}, where
the $-11/5$ and $-5/3$ scalings meet; for $\Pm\gg1$, \exref{eq:tearing_cutoff} is reached
before~\exref{eq:final_cutoff}. 
The condition for the range $[\lD,\lres^\mathrm{K}]$ to be non-empty~is 
\beq
\frac{\lD}{\lres^\mathrm{tearing}}\sim\lt[\frac{\Rm}{(1+\Pm)^{10}}\rt]^{5/28}\gg1.
\label{eq:Pm_max_K}
\eeq   
This is less stringent than \exref{eq:Pm_max},
so will always be satisfied if the disruption occurs in the first place. 

The good news (or, at any rate, the news) is that Kolmogorov's scaling
of the dissipative cutoff is rehabilitated for $\Pm\lesssim 1$. Notably,
it is not quite rehabilitated for $\Pm\gg1$, but that is likely an illusion.
Indeed, while viscosity destroys velocities, aligned magnetic structures
survive unscathed (cf.~\secref{sec:subvisc}), so tearing can keep going in the viscously
dominated regime. If it breaks up the aligned structures at the scale $\lres^\mathrm{tearing}$
into flux ropes of the same width and lesser alignment, the turnover time of these
structures will be shorter than the viscous-diffusion time---in just the same way
as the debris of disruption had shorter turnover times than their mother sheets
in \secref{sec:spectrum_rec}---and the tearing-mediate cascade can continue.
A rough estimate of how small the debris have to get
to be killed completely by diffusion is to set their Lundquist number \exref{eq:Sn}
obtained on the assumption of complete lack of alignment to $S_{\lambda_n}\sim 1+\Pm$
(in other words, the cutoff occurs when 
either $\Rm$ or $\Re$ associated with the $\lambda_n$-scale structures 
is order unity). This gives 
\beq
\lres^\mathrm{K} \sim \lCB \lt(\frac{\Rm}{1+\Pm}\rt)^{-3/4}
= \lCB\tRe^{-3/4} \sim \frac{(\nu+\eta)^{3/4}}{\eps^{1/4}},
\label{eq:K_cutoff}
\eeq 
where $\tRe$, defined in \exref{eq:lres_Rm}, is $\Rm$ when $\Pm\lesssim1$
and $\Re$ when $\Pm\gg1$. This is the proper, classic Kolmogorov scale. 

It is interesting to recall that it is the Kolmogorov scaling at (and of) the dissipation 
scale that was the strongest claim made by \citet{beresnyak11,beresnyak12,beresnyak14,beresnyak19} 
on the basis of a convergence study of his numerical spectra 
(see \secref{sec:plot} and \figref{fig:num_spectra}b). 
While he inferred from that an interpretation of these spectra as showing a $-5/3$ 
scaling in the inertial range, it is their convergence at the dissipative end 
of the resolved range that appeared to be the least negotiable feature of his work. 
He may well have been right. Indeed, his largest simulations (see \figref{fig:num_spectra}) 
fall somewhere in between the condition~\exref{eq:Pm_max_K} 
($\lD/\lres^\mathrm{K} \gtrsim 3$ would require perhaps $\Rm\gtrsim 10^3$ at $\Pm\sim 1$) 
and the more stringent condition \exref{eq:Pm_max} needed to stop Boldyrev's 
cutoff \exref{eq:lres_Rm} from taking over ($\lD/\lres \gtrsim 3$ if $\Rm\gtrsim 10^5$). 
Thus, $\lD$ in Beresnyak's simulations could not have been more than a factor of order unity 
larger than the dissipation scale. An optimist might argue that this could have been
just about enough to pick up the Kolmogorov scaling of the latter. 

\subsubsection{Alignment in the Tearing-Mediated Range}
\label{sec:align_rec}

The structures corresponding to the lower (GS95) envelope \exref{eq:z_below} 
are unaligned, whereas the alignment corresponding to the upper envelope \exref{eq:tearing_cascade} 
is the tightest alignment sustainable in the tearing-mediated range and achieved by each aligned 
``mini-cascade'' just before it is disrupted by tearing at the scale $\lambda_n$. 
This is \citep[cf.][]{boldyrev17}
\beq
\sin\theta_{\lambda_n^+} \sim \frac{\lambda_n/\dz_{\lambda_n^+}}{\tnl} 
\sim \lt(\frac{\lD}{\lCB}\rt)^{1/4}\lt(\frac{\lambda_n}{\lD}\rt)^{-4/5}.
\label{eq:theta_disrupted}
\eeq
Equivalently, the fluctuation-direction coherence scale is 
\beq
\xi_n \sim \frac{\lambda_n}{\sin\theta_{\lambda_n^+}} 
\sim \lCB \lt(\frac{\lD}{\lCB}\rt)^{3/4}\lt(\frac{\lambda_n}{\lD}\rt)^{9/5}.
\eeq
The corresponding spectral exponent is again $-5/3$, which is automatically the case 
given the definitions of $\tnl$, $\theta_\lambda$ and $\xi$ [see \exref{eq:tnl_align}
and~\secref{sec:3D}].

Thus, the smallest possible alignment angle, having reached its minimum at 
$\lD$, gets larger through the tearing-mediated range, according to \exref{eq:theta_disrupted}.
It becomes order unity at the scale~\exref{eq:final_cutoff}, which is the same as
the Kolmogorov cutoff~\exref{eq:K_cutoff} for $\Pm\lesssim 1$ and a bit smaller
than it for $\Pm\gg 1$ (but in fact there is no more alignment below the Kolmogorov
cutoff). 

To the (doubtful) extent that existing numerical evidence can be considered to be probing 
this regime, perhaps we can take heart from 
the numerical papers by both Beresnyak and by Boldyrev's group cited 
in \secref{sec:plot} all reporting that alignment fades away at the 
small-scale end of the inertial range---although this may also be just a banal effect 
of the numerical resolution cutoff. 

\subsubsection{Parallel Cascade in the Tearing-Mediated Range}
\label{sec:par_rec}

As ever, CB should be an enduring 
feature of our turbulence. This means that 
the parallel spectrum \exref{eq:zpar_GS95} will not notice the disruption scale 
and blithely extend all the way through the tearing-mediated range. 
Since the unaligned (or less aligned) flux ropes produced 
in the wake of the disruption of aligned structures have a shorter decorrelation 
time than their aligned progenitors, they should break up in the parallel 
direction (cf.~\citealt{zhou20} and \secref{sec:decay_RMHD}). 
The resulting parallel coherence scale, the same as 
the scale \exref{eq:aniso_GS95} in the GS95 theory, 
is the lower bound on $\lpar$ at each $\lambda_n$. 
The upper bound can be inferred by equating the nonlinear time \exref{eq:tnl_n} 
at $\lambda_n$ to the Alfv\'en time $\lpar/\vA$. The result is
\beq
\vA\eps^{-1/3}\lambda_n^{2/3}\lesssim \lpar\lesssim
\vA\eps^{-1/5}\eta^{-2/5}(1+\Pm)^{1/5}\lambda_n^{6/5}
\sim\Lpar\lt(\frac{\lD}{\lCB}\rt)^{1/2}\lt(\frac{\lambda_n}{\lD}\rt)^{6/5}.
\label{eq:lpar_tearing}
\eeq 
Thus, the upper bound on the parallel anisotropy $\lpar/\lambda$
{\em decreases} with scale in this range (turbulence becomes less anisotropic).

\subsection{Is This a Falsifiable Theory?} 
\label{sec:falsifiable}

\begin{figure}
\centerline{\includegraphics[width=0.85\textwidth]{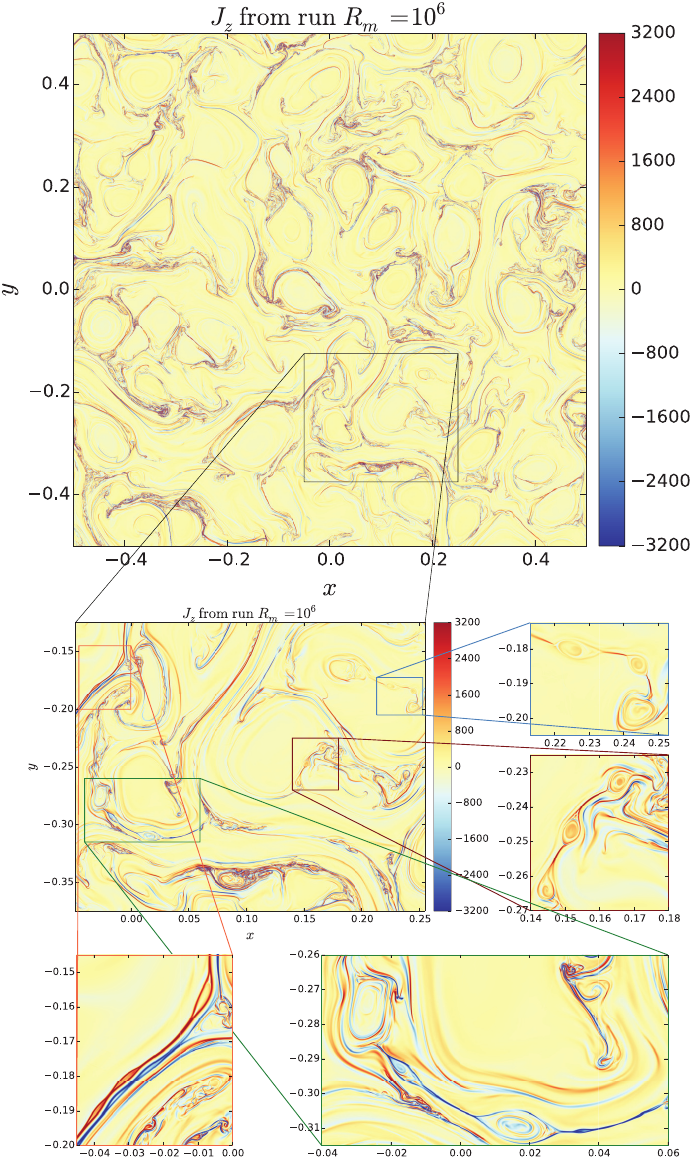}} 
\vskip2mm
\caption{A snapshot of current density ($j_z$) from a 2D, $\Rm=10^6$ (64,000$^2$) MHD simulation 
by \citet{dong18} [reprinted with permission from \citet{dong18}, copyright (2018) by the American
Physical Society; I am grateful to C.~Dong for letting me have the original figure file]. 
Zoomed areas show sheets broken up into plasmoids. The 3D versions of these visualisations 
reported by \citet{dong22} look generally messier, with islands less clearly delineated
(they look a bit like the flux ropes in~\figref{fig:huang3D}), but 
not, at first glance, qualitatively different in a paradigm-changing way.}
\label{fig:dong}
\end{figure}

With considerable difficulty. 

Numerically, anything like a definite confirmation of the tearing disruption (\secref{sec:disruption_scale})
and the existence of a tearing-mediated cascade (\secref{sec:recturb}) requires 
formidably large simulations: the condition \exref{eq:Pm_max} 
demands $\Rm\sim 10^5$ at least (estimated via the frivolous 
but basically sound principle that the smallest large number is~3)---and 
probably quite a bit larger 
if one is to see the scaling of the tearing-mediated-range spectrum \exref{eq:Ek_bounds}. 
However, as was mooted above, an optimist might find cause for optimism  
in the evidence of the MHD turbulence cutoff seen in simulations appearing 
to obey the Kolmogorov scaling \exref{eq:K_cutoff} (see discussion in~\secref{sec:diss})
or in the numerically measured alignment petering out at the small-scale end 
of the inertial range, as required by~\secref{sec:align_rec}. 
While the trouble to which I have gone to 
keep track of the $\Pm$ dependence of the tearing-mediated-range 
quantities did not yield anything qualitatively spectacular, 
there is perhaps an opportunity here for numerical tests: e.g., 
can one obtain Boldyrev's scaling~\exref{eq:lres_Rm} of the dissipation cutoff
in the limit of moderate $\Rm$ and large $\Pm$?---which, in view of \exref{eq:Pm_max}, 
is unlikely to need to be very large to take over and shut down tearing. 

One way to circumvent 
the need for getting into a hyper-asymptotic regime is to simulate directly 
the dynamics of structures that resemble Alfv\'enic sheets deep in the inertial range. 
Such a study by \citet{walker18}, in 2D, of the decay of an Alfv\'enic ``eddy'' 
highly anisotropic in the perpendicular plane, has shown it breaking up promisingly 
into plasmoids and giving rise to a steeper spectrum than exhibited by a larger-$\Rm$ case 
where tearing was too slow. \citet{dong18} went further 
and actually demonstrated a spectral break at the disruption scale and 
a $\kperp^{-11/5}$ spectrum below it, with sheets in a turbulent system very 
vividly breaking up into plasmoids (\figref{fig:dong})---but still in 2D.
So far so good, but the real prize is always for~3D.
The present review has spent so long going through revisions that
3D~results, of which I originally wrote in remote future tense,
are in fact now on the brink of arriving, again due to \citet{dong22},
who have performed the mother of all MHD simulations (with $\Rm$ comfortably over $10^5$)
and are claiming to see a ``sub-inertial range'' with a $-11/5$ spectral slope,
as per~\exref{eq:Ek_bounds}; the spectral break where this scaling starts 
appears to be in the right place, $\lD\propto\Rm^{-4/7}$ as per \exref{eq:lambdaD}.
If I read their results correctly
(based on C.~Dong's presentation at the 2021 APS DPP Meeting), hallelujah. 

Observationally, our best bet for fine measurements of turbulence 
is the solar wind and the terrestrial magnetosphere (e.g., the 
magnetosheath). However, these are collisionless environments, 
so, before any triumphs of observational confirmation can be 
celebrated, all the resistive reconnection physics on which the tearing-mediated-range 
cascade depends needs to be amended for the cornucopia of 
kinetic effects that await at the small-scale end of the cascade 
(see \secref{sec:microphysics}). Once the tearing disruption in MHD was proposed,  
such generalisations were the ripe, low-hanging fruit quite a lot of which 
was immediately picked (\citealt{mallet17c,loureiro17b,boldyrev19}; 
see also \citealt{loureiro18}, where these ideas were ported to pair plasmas). 

Pending all this validation and verification, 
the tearing-mediated cascade remains a beautiful 
fantasy---but one must be grateful that after half a century of scrutiny, 
MHD turbulence still has such gifts to offer.   

\subsection{Tearing Disruption, Plasmoid Chains, Fast Reconnection, 
and Reconnection-Driven Turbulence} 
\label{sec:multilayer}

This section deals with what many readers might feel are fairly esoteric details. 
They are right---and so skipping straight to \secref{sec:halfway} will not subtract 
much from their experience. 

In what until recently was a separate strand of research, much interest 
in the reconnection community 
has focused on stochastic plasmoid chains that arise in current sheets susceptible to 
the plasmoid instability (a sub-species of tearing), where a lively population 
of islands (plasmoids) are born, grow, travel along the sheet with 
Alfv\'enic outflows, occasionally eat each other (coalesce),\footnote{In the process 
of coalescence, they also give rise to transverse secondary current sheets 
and plasmoid chains: see \citet{barta11}.} 
and, as shown by \citet{uzdensky10}, cause reconnection in the sheet that 
they inhabit to be fast,\footnote{The reconnection that is being
referred to here is the true, physical reconnection of the exact magnetic-field lines, not
the effective reconnection of fields coarse-grained at some inertial-range scale~$\lambda$.
In a turbulent environment, the latter, known as ``stochastic reconnection'',
is believed always to be fast (see \citealt{lazarian20} and references therein),
so every cascaded eddy always gets a significant amount of it.
As explained in \secref{sec:stoch} and \apref{app:stoch_rec},
this is plausible, and does not preclude either aligned or tearing-mediated turbulence.} 
meaning independent of $\eta$ as $\eta\to+0$ (a derivation 
of the plasmoid instability, a long list of references on stochastic plasmoid 
chains, and, in \figref{fig:huang}, an example of one, can be found in \apref{app:loureiro};
the \citealt{uzdensky10} argument is reproduced in \apref{app:uls}). 
A stochastic chain can be viewed 
as a kind of ``1D turbulence'', and has some distinctive statistical properties 
(see \apsand{app:plasmoid_pdf}{app:chain_spectrum}). 
Should one imagine the disrupted aligned structures spawning multiple instances of such 
a turbulence, and does the simple theory presented in \secref{sec:recturb} describe 
this situation or does it need to be revised to represent a superposition of 
many fast-reconnecting, plasmoid-infested sheets (as attempted in three different 
ways by \citealt{loureiro17,loureiro20} and \citealt{tenerani20})? 

\subsubsection{Nature of Tearing Disruption}
\label{sec:MSC_vs_BL}

In considering this problem, I first want to return to the question of what the ``disruption''
of the aligned structures actually consists of. There are two lines of thinking on this, 
articulated most explicitly in the papers by \citet{mallet17b} and \citet{boldyrev17}, 
of which I have 
so far stuck with the former. Namely, at the end of \secref{sec:disruption_scale}, 
I followed \citet{uzdensky16}, \citet{mallet17b} and \citet{loureiro17} in invoking the 
collapse of the $X$-points separating the tearing-mode islands as a means of 
consummating the disruption of the aligned 
structure---the theory of the tearing-mediated cascade in~\secref{sec:recturb} 
was then presented as a corollary of this view. 

The collapse of inter-island $X$-points is actually the first step in the formation 
of a stochastic chain. A nuance that I have previously elided is that
comparing the $X$-point collapse rate (which is $\sim$ the tearing growth rate~$\gamma$) 
with the growth rate of a {\em secondary} tearing instability of the same $X$-point 
shows that, at asymptotically large Lundquist numbers, the latter is always greater than the former. 
Therefore, the collapse may itself be disrupted by 
tearing, producing more islands and more $X$-points, followed by the collapse of those, 
also disrupted, and so on. My take on this recursive tearing is presented 
in \apref{app:multi} (alongside a review of prior literature, 
starting with \citealt{shibata01}). I argue there that the smaller-scale islands 
that are produced in this process are not energetically relevant and so we need 
not worry about including them to amend the ``one-level'' scenario of tearing disruption 
presented in \secref{sec:disruption_scale}. 

The recursive tearing proceeds until inter-island current 
sheets are short enough to be stable, at which point the true nonlinear 
plasmoid chain can form, involving not just multiple tearings, but also 
nonlinear plasmoid growth by reconnection, their coalescence, and ejection from the sheet. 
While the multiscale statistics of such a chain may be different from that of a tearing-mediated 
cascade that I described in \secref{sec:recturb}, I assumed there, implicitly, that 
the chain could not survive for a long time, if at all: indeed, the characteristic 
time scale of the process of fully forming the sheet out of an aligned 
structure is the tearing time ($\sim\gamma^{-1}$), and the time to break apart that aligned 
structure by ideal-MHD dynamics is of the same order ($\sim\tnl$). 
So it seems that the ``mother sheet'' (collapsed aligned structure) should break apart 
entirely shortly after (or even before) fully forming and release its  
plasmoids (flux ropes) into the general turbulent wilderness, where they are 
free to interact with each other or with anything else that comes along, and 
are thus no different from turbulent fluctuations of a particular size 
generically splashing around in a large nonlinear system.
This gives rise to the ``mini-cascades'' in \secref{sec:recturb},
with the overall $\kperp^{-11/5}$ spectral envelope~\exref{eq:Ek_bounds}.  

\citet{boldyrev17} also derive the $\kperp^{-11/5}$ spectrum by using \exref{eq:tnl_n} 
as the operational prescription for the cascade time ($\tnl\sim\gamma^{-1}$ at each scale
in the tearing-mediated range). 
However, they have a different narrative about 
what happens dynamically: they do not believe that inter-island 
$X$-points ever collapse, but that, rather, the tearing mode upsets alignment 
by order unity, changing the effective nonlinear cascade rate to the tearing rate.  
This is based on the (correct) observation that the alignment 
angle at the disruption scale \exref{eq:lambdaD} 
\beq
\sin\theta_{\lD}\sim S_{\lD}^{-1/2}(1+\Pm)^{-1/4}
\eeq
is the same (at least for $\Pm\lesssim1$) as the 
angular distortion of the field line caused by the tearing perturbation 
at the onset of the nonlinear regime: indeed, using \exref{eq:w_nlin} and \exref{eq:kpeak_TM} 
at $\lambda=\lD$, 
\beq
\theta_\mathrm{tearing}\sim wk_* \sim (k_*\lD)^2 \sim S_{\lD}^{-1/2}(1+\Pm)^{1/4}.
\eeq  
They think that this is enough to make the aligned structure ``cascade'', in some 
unspecified manner, without much reconnection, production of flux ropes, etc.   
In \citet{loureiro20}, they revise their view a little and 
allow that, since the collapse time, the tearing 
time and, therefore, the cascade time are comparable to each other, {\em some} aligned 
structures might, in fact, collapse into proper reconnecting sheets.\footnote{To 
\citet{loureiro20}, the difference between these reconnection sites and mere tearing 
modes is that the former dissipate a lot of energy. This matters to them because 
they believe that the tearing-mediated cascade can only be a constant-flux 
cascade if it does not involve much reconnection, as reconnection is dissipative---the 
spectrum would have to steepen if reconnection occurred in too many places and thus 
caused a finite energy drain from the cascade. 
Is there really a contradiction between significant reconnection and constant flux? First, 
it is not inevitable (although, in resistive MHD with $\Pm=1$ usually true: see, e.g.,
\citealt{loureiro12}) that reconnection must always involve large dissipation. Secondly, 
and more importantly, if collapse and reconnection of an aligned structure of scale $\lambda$
do lead to significant dissipation, that dissipation does not, in fact, occur at scale $\lambda$, 
but at much smaller scales---the scales of the inter-island sheets and outflows.  
Transfer of energy to those scales could arguably be viewed as part of 
the turbulent cascade.}

It is hard to say whether the two pictures outlined above represent a disagreement 
in substance or merely in the style of presentation. While in the \citet{mallet17b}
interpretation, the collapse of the inter-island $X$-points 
{\em is} the way in which the distortion of alignment caused by tearing leads 
to faster nonlinear break-up of the aligned structures, \citet{loureiro20} think 
this is not necessary but does happen with some finite probability. This might 
not be sufficiently quantifiable a difference to be testable. 

\subsubsection{Onset of Fast Reconnection}
\label{sec:onset}

What may be consequential physically, however, is the onset of fast reconnection in those aligned, 
tearing structures that do manage to collapse into proper sheets. 
\citet{loureiro20} conjecture that when that happens, one 
should start worrying about the resulting reconnecting sheets making a difference 
to the nature of the tearing-mediated (now reconnection-mediated) 
turbulence.\footnote{This is because in the slow-reconnection 
regime (described in \apref{app:SP_rec}), the reconnection time is always longer than 
the tearing time. Indeed, the latter, $\gamma^{-1}$, is given by~\exref{eq:gmax_TM} 
with $\vAy\sim\dz_\lambda$, whereas $\trec$ is given by \exref{eq:fast_onset} but with 
$\epsrec^{-1}\sim \tS_\xi^{1/2}(1+\Pm)^{1/2}$, where $\tS_\xi$ is the Lundquist number based 
on the size of the structure in the fluctuation direction [see~\exref{eq:SP_uy}]; 
therefore, $\gamma\trec \sim (\xi_\lambda/\lambda)^{1/2}\gg 1$.} 
In order for such a transition to be realisable,  
$\Rm$ must be large enough for the characteristic time of fast reconnection 
to be shorter than the cascade time: 
\beq
\trec \sim \epsrec^{-1}\frac{\lambda}{\dz_\lambda} \lesssim 
\tnl \sim \frac{\lambda}{\dz_\lambda} \lt(\sin\theta_\lambda\rt)^{-1}
\rmiff \sin\theta_\lambda \lesssim \epsrec,
\label{eq:fast_onset}
\eeq
where the reconnection time has been estimated 
as the time for all of the flux in a magnetic structure of scale $\lambda$ to be reconnected, 
and $\epsrec\sim 10^{-2}(1+\Pm)^{-1/2}$ is the dimensionless fast-reconnection rate 
in a \citet{uzdensky10} plasmoid chain (see \apref{app:uls}), or some modified 
version of it appropriate for a turbulent environment. The estimate \exref{eq:fast_onset} simply says 
that if the alignment angle (inverse aspect ratio) of a structure manages to become smaller than 
the reconnection rate, such a structure might be capable of becoming a fast-reconnecting sheet. 

Let us recall that the alignment angle decreases with $\lambda$ according 
to~\exref{eq:MS_normalised} in an aligned MHD cascade, reaches its minimum at 
the disruption scale $\lD$ [given by~\exref{eq:lambdaD}], 
and then increases with decreasing $\lambda$ 
according to~\exref{eq:theta_disrupted} in a tearing-mediated cascade. 
The condition \exref{eq:fast_onset} is realisable~if 
\beq
\sin\theta_{\lD} \sim \lt(\frac{\lD}{\lCB}\rt)^{1/4} \lesssim \epsrec 
\rmiff 
\Rm \gtrsim \epsrec^{-7}(1+\Pm)^{-1/2} \sim 10^{14}(1+\Pm)^3. 
\label{eq:10to14}
\eeq  
Obviously, this is never going to be numerically (or indeed experimentally) achievable 
for resistive MHD---unless $\epsrec$ is substantially enhanced in a turbulent setting 
(\citealt{loureiro20} show that this ``nonlinear-reconnecting'' 
regime might also be more easily accessible in certain kinetic settings). 

In the asymptotic world where the condition \exref{eq:fast_onset} can be realised, it will be 
realised for $\lambda\in[\lrec^<,\lrec^>]$, with the two scales lying below and above~$\lD$: 
using \exref{eq:MS_normalised} and~\exref{eq:theta_disrupted} 
for the alignment angle in the aligned and tearing-mediated cascades, respectively, 
one~gets
\beq
\lrec^> \sim \epsrec^4\lCB,\qquad
\lrec^< \sim \epsrec^{-5/4}\lD^{21/16}\lCB^{-5/16}.
\eeq 
One might argue that $\lrec^>$ is irrelevant because at that scale ideal nonlinear 
interactions are faster than tearing, so the instability that assists in the 
formation of a plasmoid chain is too slow to get it going before the energy 
cascades to small scales (if this is not true, then reconnection-mediated turbulence 
sets in at $\lrec^>$, an $\Rm$-independent scale, putting a hard lower bound on 
the allowed alignment angles, $\sin\theta_\lambda\gtrsim\epsrec$, 
and vindicating \citealt{beresnyak19}). 
In contrast, the interval $[\lD,\lrec^<]$ is tearing-dominated, 
so plasmoid chains could be on the cards. 

What might turbulence in such a situation look like? 
To get some idea of that, it is perhaps wise to start with numerical evidence about 
turbulence in individual, ``stand-alone'' plasmoid chains---a subject on which copious 
literature exists, quoted in \apref{app:loureiro}. 

\subsubsection{Reconnection-Driven Turbulence}
\label{sec:turb_sheet}

Much of that literature describes 2D simulations, but there is a handful of papers
dedicated to unstable sheets in~3D. In all of these 3D numerical experiments, 
a large-scale reconnecting configuration---a macroscopic sheet---was set up 
as an initial condition and/or driven by inflows/outflows from/to the boundaries of the domain,  
then went violently unstable, much more so than in 2D, and ended up looking like a strip of 
vigorous turbulence, rather than a quasi-1D chain (see \apref{app:rec_driven} for citations
and further discussion). There does not appear to be any reason for such a configuration to stay 
together without external help, so it is likely that what we are 
witnessing in these numerical simulations is a version of a tearing-mediated 
cascading event prolonged by the bespoke numerical set-up and thus 
guaranteed to go into the fast-reconnecting regime.    

If this is true, then such reconnection-driven MHD 
turbulence and turbulence in a homogeneous box into which energy 
is injected by a body force are different only inasmuch as any two different 
outer-scale, system-specific arrangements for stirring up turbulence are different.  
In the spirit of universality, it is hard to believe 
that small patches of a turbulent sheet would look any different 
in close-up than a generic box of MHD turbulence. One can imagine, however, 
that, due to the macroscopic ``reconnection driving'' of the turbulence in a sheet, 
the turbulent cascade starts off at the outer scale already in a highly aligned, 
tearing-dominated, regime (\citealt{walker18} was an explicit 
attempt to exploit this idea). Indeed, both \citet{barta11} and \citet{huang16} 
see spectra somewhat steeper than~$-2$, perhaps consistent with $-11/5 = -2.2$
(or with the $\kperp^{-2}$ spectrum derived in \apref{app:chain_spectrum}).
In contrast, \citet{beresnyak17} and \citet{kowal17} report small-scale 
statistics very similar to those found in standard MHD turbulence. 
\citet{tenerani20} find the same at a sufficient distance from the neutral line, 
whereas close to it, they see interesting anisotropic scalings dependent on 
the (component of) the field and the direction in which its variation is probed
vis-\`a-vis the orientation of the sheet. 

Moving from turbulence in one sheet to an ensemble of turbulent sheets, 
\citet{tenerani20} speculate about the spectrum 
of a turbulence entirely dominated by reconnecting sheets filling 
a scale-dependent fraction of the volume, and arrive at $\kperp^{-11/5}$ by an 
entirely different route, perhaps a coincidence. \citet{loureiro20}, 
in pursuit of the same idea, amend $\kperp^{-11/5}$ to~$\kperp^{-12/5}$ 
just by assuming the volume-filling fraction $\propto\lambda$ (sheets) 
for energy at every scale $\lambda$ in an otherwise standard tearing-mediated cascade. 

This is the current state of affairs. I do not have a definitive contribution 
to make to the (still wide open) theory of reconnection-driven turbulence.   
In \apref{app:chain_spectrum}, I show, following \citet{barta12} and 
\citet{loureiro16unpub}, how to get a $\kperp^{-2}$ spectrum 
for a stochastic plasmoid chain envisioned by \citet{uzdensky10}. I then 
argue tentatively, in \apref{app:rec_driven}, that if plasmoids 
(in 3D, flux ropes) all go unstable and thus drive small-scale turbulence, 
that turbulence should look like regular (possibly tearing-mediated) 
MHD turbulence, but with a very 
broad driving range featuring a $\kperp^{-1}$ spectrum, in which 
turbulent motions are forced with an alignment angle independent 
of scale and equal to the (dimensionless) fast-reconnection rate~$\epsrec$. 
I have no evidence to back this up.

Finally, let me also flag here the possibility that a good example 
of reconnection-driven turbulence may be the inertial-range turbulence 
in magnetically dominated decaying MHD systems, where decay is controlled 
by reconnection in current sheets that separate outer-scale relaxed structures
(see \secref{sec:decaying}; the 
discussion of spectra is in \secref{sec:decay_spectra}). 
Another (although more uncertain) such example may be the 
saturated state of the turbulent dynamo (\secref{sec:dynamo}) 
if the scheme mooted in~\secref{sec:fast_rec_dynamo} turns out to have merit.  

I will let the subject drop at this point, with the parting message that 
the last word has not been written on the intermittency effects and 
the role of fast plasmoid reconnection in tearing-mediated turbulence.  

\section{Halfway Summary} 
\label{sec:halfway}

\vskip2mm
\begin{flushright}
{\small \parbox{8.5cm}{For the beginning is thought to be more than half of the whole,
and many of the questions we ask are cleared up by~it.}
\vskip2mm
Aristotle, {\em Nicomachean Ethics} (translated by W.~D.~Ross)} 
\end{flushright}
\vskip5mm

\subsection{Is This the End of the Road?}
\label{sec:end}

It never quite is (see \secsdash{sec:onset}{sec:turb_sheet}
and the second part of this review, starting from~\secref{sec:imbalanced}), 
but the basic story looks roughly complete for the first time in years, 
at least as far as forced, balanced RMHD turbulence is concerned.
The principle of critical balance gave us a sound ideology for dealing
with anisotropic turbulence in a system that supports propagation of waves (\secref{sec:CB})
and a firm prescription for the parallel spectrum (\secref{sec:par_cascade}). 
The aligned cascade (\secref{sec:DA}) produced a plausible prediction for the perpendicular
spectrum but used to have an air of unfinished business about it, both in the sense that 
it gave rise to a state that appeared unsustainable at asymptotically small scales 
and in view of the objections, physical and numerical, 
raised by \citet{beresnyak11,beresnyak12,beresnyak14,beresnyak19}.
With the revised interpretation of alignment as an intermittency 
effect (\secref{sec:revised}) and with the tearing-mediated cascade (\secref{sec:disruption})
connecting the inertial-range, aligned cascade 
to the Kolmogorov cutoff~\exref{eq:K_cutoff}, these issues appear to be 
satisfactorily resolved. In what is also an aesthetically pleasing development, 
the tearing-mediated cascade has emerged as an ingenious way in which 
MHD turbulence contrives to thermalise its energy while shedding the excessive 
alignment that ideal-MHD dynamics could not help producing in the inertial range. 
This development joins together in a most definite way 
the physics of turbulence and reconnection---arguably, this
was always inevitable, but it is good that we now appear to have some grip 
on what happens specifically. 

Before I move on to the miscellany of Part~II, let me make a few comments
about the robustness of the general picture presented above and its connection
to other schools of thought on MHD turbulence. 

\subsection{What Can Go Wrong?} 
\label{sec:KHdisaster}

It is only fair to spell out explicitly what is settled and 
what can go catastrophically wrong with this entire picture.

The principle of critical balance and, therefore,
the theory of the parallel cascade (\secref{sec:CB}) are, in my view, quite safe.
They are straightforward physically and have been quite convincingly verified both
observationally and numerically ({\em pace} the ``waves vs.\ structures'' confusion:
see~\secref{sec:waves_vs_structures}). The CB approach also appears to offer
an attractive and credible strategy for dealing with turbulence in wave-carrying systems
other than MHD, in both plasmas and hydrodynamics 
\citep[e.g.,][]{cho04,sch09,sch16,sch19,nazarenko11,barnes11,boldyrev13,chen17,passot17,loureiro18,avsarkisov20,adkins22,skoutnev22},
further bolstering its claim to being a universal physical principle.

In contrast, the aligned perpendicular cascade (\secref{sec:DA}),
its tearing disruption (\secref{sec:disruption_scale}) 
and replacement by a tearing-mediated cascade (\secref{sec:recturb}),
which have formed the bulk of my story so far, are hotly debated concepts. 
Arguably, it is still subject to verification (requiring currently inaccessible
resolutions) that alignment is not a transient, large-scale 
feature, as \citet{beresnyak19} would have it. It seems to me that we do know, 
however, that if we stir up unaligned turbulence, it will get aligned at smaller 
scales (see numerical studies cited in \secref{sec:plot}),
so its possible transient nature can only be due to some secondary instability 
of the aligned structures. The picture presented above relied on this 
being the tearing instability---but it is not entirely impossible that it is, in fact, 
an ideal MHD instability, e.g., some version of Kelvin--Helmholtz (KH) instability. 
The difference is that tearing required resistivity and so the disruption scale 
$\lD\propto\eta^{4/7}$ was asymptotically separated from the outer 
scale $\lCB$ [see \exref{eq:lambdaD}], whereas the KH instability would kick in 
at some $\lambda\sim$~a~finite fraction of $\lCB$. 
The usual expectation is that the KH instability is quenched by the magnetic field 
(and indeed hence perhaps the statistical preponderance of current sheets 
over shear layers; see \secref{sec:new_res_theory}), but this can in principle 
turn out not to be enough. If it does, ideal MHD will take care of limiting 
alignment, without alignment there will be no need for, or dynamical tendency to,
the tearing-mediated remedy to it at small scales, 
and, presumably, we will be back to GS95, in which case I apologise 
to my readers for having wasted their time (this scenario 
is explained in another language at end of \secref{sec:stoch}
and at the beginning of \apref{app:aligned_rec}). 

\subsection{What Is Lost in Translation?}

This section is devoted to misunderstandings and arguments at cross purposes. As any
vibrant, fast-evolving field, MHD turbulence is a subject talked about in many languages,
and differences in vocabulary are sometimes mistaken for fundamental disagreements. 

\subsubsection{Waves vs.\ Structures}
\label{sec:waves_vs_structures}

In the minds of some enthusiasts of current sheets (or, generally, of ``coherent structures'') 
in MHD turbulence, the distinction between the sheets (``structures'') and critically
balanced Alfv\'enic perturbations (``waves'') has become a dichotomy between two allegedly 
incompatible paradigms of how energy is dissipated in MHD turbulence---in strongly dissipative
structures or via wave damping. This is a misunderstanding that
appears to be based on the incorrect perception of CB-based theories as requiring turbulence 
to be an ensemble of random-phased Alfv\'en waves, similar to WT (\secref{sec:WT}). 
No dichotomy, and certainly no mutual exclusivity, between waves and structures,
in fact, exists: while Alfv\'enic perturbations retain 
certain properties associated with the linear-wave response, their turbulence 
is strong (which is the whole point of the CB principle) 
and the tendency to form sheets dynamic 
(mutual shearing of Elsasser fields: see \citealt{chandran15,howes16}). 
This nonlinear, intermittent dynamics perpendicular to~$\vB$ produces
``structures'', while the linear wave-propagation physics gives them coherence along~$\vB$,
via CB as a causality constraint (\secref{sec:CBCB}).

In the recent literature, the most systematic, and sensible, discussion of the ``waves vs.\ structures'' 
issue can be found in \citet{groselj19} (although their focus is on the kinetic, 
rather than MHD, range of scales); the transition---and key differences---between wave
turbulence and CB turbulence are illustrated very vividly in \citet{meyrand16}
(already referred to in~\secref{sec:WT}).

\subsubsection{Cellularisation of Turbulence}
\label{sec:cell}

The interest in dissipative structures has its origin in the
long history of thinking about MHD turbulence (and generally turbulence)
in a language that is, at first glance, very different from the one
in which the preceding sections of this review have been written. This line
of thinking dates back to \citet{montgomery78,montgomery79} and \citet{matthaeus80},
who looked for thermal-equilibrium states in MHD turbulence, subject to conservation
of various secondary invariants---e.g., in 3D, helicity and cross-helicity---and conjectured that
MHD turbulence would tend to such equilibria patch-wise in space.
By a classic variational argument, it is possible show that nonlinear interactions
would be suppressed in the regions---``cells''---of fixed helicity, which would
host force-free magnetic fields \citep{taylor74}. A fixed cross-helicity 
would push velocity and magnetic field into alignment with each other
(which does indeed happen in decaying MHD turbulence: see, e.g., \citealt{matthaeus08}); 
if they were also equal to each other in amplitude, the result would be
a pure Elsasser, non-interacting state.
Global force-free or Elsasser states are usually not achievable
(e.g., because the conserved net helicity and/or cross-helicity of the system are zero),
so the cells would be separated by boundaries where intense nonlinear dynamics 
and hence dissipation would take place (e.g., current sheets). That then would be the structure
of a turbulent state---relaxed cells plus nonlinear, dissipative structures on their boundaries
(non-volume-filling, so intermittent). 

A recent review of this philosophy, from an original source and with copious
references, is \citet{matthaeus15} (a similar thinking, mostly applied to hydrodynamic
turbulence, is reviewed, in a fascinatingly idiosyncratic way, by \citealt{levich09}).
While this approach is most natural in the context of
decaying (freely relaxing) turbulence (and does indeed work there:
see \secref{sec:decaying}; vivid illustrations of this ``cellularisation'' of turbulence
are \figsand{fig:hosking_2D}{fig:hosking_3D}), similar
ideas have been mooted with regard to inertial-range statistics of forced
turbulence: see, e.g., the discussion in~\secref{sec:Eimb} 
of Elsasser-balanced MHD turbulence proving to be a patchwork of locally imbalanced,
heavily cross-helical regions (\figref{fig:perez_patches}).
Dynamic alignment \`a la Boldyrev (\secref{sec:DA}), 
its salient effect being the local reduction of nonlinearity in the MHD inertial range, 
was also originally argued by him \citep{boldyrev06} 
to have to do with conservation and cascading of cross-helicity
(see footnote~\ref{fn:uBcorr2}). 

Thus, perhaps cellularisation and the critically-balanced, aligned dynamics that
I have described above are, in fact, the same thing said in two different languages.
A clear translation between the two is, however, still to be articulated,
especially the kind of translation that would add something genuinely
new to our understanding of the subject. 

\subsubsection{Stochastic Reconnection}
\label{sec:stoch}

I have touted the joining of MHD turbulence and reconnection theories
as a key outcome of the developments described in \secsdash{sec:DA}{sec:disruption},
but of course there exists another school of thought 
for which this very connection has been the defining mantra for many years,
but which has been almost entirely decoupled from those developments.
This school of thought, vaguely anticipated already by \citet{matthaeus85,matthaeus86}
(naturally so, given reconnection's role in the cellularisation picture discussed
in \secref{sec:cell}), was properly launched by \citet{lazarian99}
when they put forward the notion of ``stochastic reconnection''.
\Apref{app:stoch_rec} is my attempt to review this topic (in which task I was
greatly helped by its interpretation by \citealt{eyink11rec})
and its connection to, and lack of contradiction with, the main narrative presented above. 
Here I shall try to summarise it very briefly. 

The notion of turbulent viscosity has been a mainstay of the theory of hydrodynamic
turbulence for nearly as long as this theory has existed
\citep[see, e.g., the textbook by][]{davidson15}. The idea is simply
(but not trivially) that if we ``coarse-grain'' the velocity field at some scale~$\lambda$
in the middle of the inertial range (or even at the outer scale), the effect
of all the motions at scales smaller than $\lambda$ can be modelled, very crudely
but surprisingly adequately, by an effective viscosity~$\sim\du_\lambda\lambda$.
This is because the hydrodynamic cascade is local and (in 3D) direct, so motions
at scale $\lambda$ are broken up at the rate
$\du_\lambda/\lambda = (\du_\lambda\lambda)/\lambda^2$, which can be viewed
as a renormalised diffusion of momentum. The rate at which momentum and
energy are removed from scale $\lambda$ is entirely independent of the true
molecular viscosity. In a similar way, an MHD ``eddy'' at the outer scale
or in the inertial range (but not, according to \secref{sec:disruption},
in the tearing-mediated range) will lose both its momentum and its magnetic flux
on a time scale ($\tnl$, much discussed in the above) that is entirely independent
of the true viscosity and resistivity of the MHD fluid (\citealt{eyink15} argues
this with careful attention to detail). Thus, just like there is a
turbulent viscosity, there is also a turbulent resistivity, and so a turbulent reconnection
with an $\eta$-independent effective rate. This has some interesting and nontrivial 
consequences, both physical and mathematical, 
for a number of problems involving reconnection (see \apref{app:stoch_rec},
the review by \citealt{lazarian20}, and references in both), but it does not mean that
reconnection does not need $\eta$ or small scales---just that those small scales are
reached at an $\eta$-independent rate.

If the turbulent resistivity were isotropic in the plane perpendicular
to the mean magnetic field, no alignment effect would be possible because the aligned
structures would not need tearing to break up, and we would be back
to~\secref{sec:KHdisaster}---no alignment, no tearing-mediated cascade, no need
for this review. That is, essentially, the view of \citet{lazarian20}.
I cannot prove that this is wrong, but I think that it is unlikely (especially
given the numerical evidence announced by \citealt{dong22}; see \secref{sec:falsifiable}).
In \apref{app:aligned_rec}, I show why it is not logically inevitable,
i.e., how stochastic reconnection \`a la Lazarian, 
Vishniac, and Eyink can be reconciled with an aligned cascade. 

\part{Imbalances and Loose Ends}
\label{part:addons}
\addcontentsline{toc}{section}{\partskip \sc Part II.\ \nameref{part:addons}}

\begin{flushright}
{\small\parbox{6.2cm}{As we know,\\
There are known knowns.\\
There are things we know we know.\\
We also know\\
There are known unknowns.\\
That is to say\\
We know there are some things\\
We do not know.\\
But there are also unknown unknowns,\\
The ones we don't know\\
We don't know.}
\vskip2mm
D.~H.~Rumsfeld,\footnotemark\\ {\em U.S.~Department of Defense News Briefing},\\ {\em 12 February 2002}
\footnotetext{Set to verse by \citet{seely03}.}}
\end{flushright}
\vskip5mm

In the remainder of this review, I will survey some of what 
has been done, what remains to be done, and what, in my view, is worth doing 
regarding the regimes of MHD turbulence in which there is 
an imbalance either between the energies of the two Elsasser fields or between 
the kinetic and magnetic energy. Such situations are relevant---and indeed often more 
relevant---in many astrophysical contexts, but remain much less (or even less) 
well understood, than the nice, if somewhat fictional, case in which one 
can just assume $\dz_\lambda^+\sim\dz_\lambda^-\sim\db_\lambda\sim\du_\lambda$. 
Not only the cases of Elsasser~(\secref{sec:imbalanced}) and Alfv\'enic~(\secref{sec:residual}) 
imbalance can be put in this class but also the distinct regimes of MHD 
turbulence that arise below the viscous scale (assuming large $\Pm$;~\secref{sec:subvisc}),
or when the turbulence is allowed to decay freely (\secref{sec:decaying}), 
or when no mean field is imposed (the saturated MHD dynamo;~\secref{sec:dynamo}). 
For the reader's reference, \secsref{sec:imb_new}, 
\ref{sec:new_res_theory}, \ref{sec:subvisc}, \ref{sec:decay_ssim_RMHD}, 
and \ref{sec:dynamo_theory}--\ref{sec:dynamo_scenarios} contain 
some new results and arguments that have not been published elsewhere.

\section{Imbalanced MHD Turbulence}
\label{sec:imbalanced}

\begin{figure}
\begin{center}
\begin{tabular}{cc}
\includegraphics[width=0.49\textwidth]{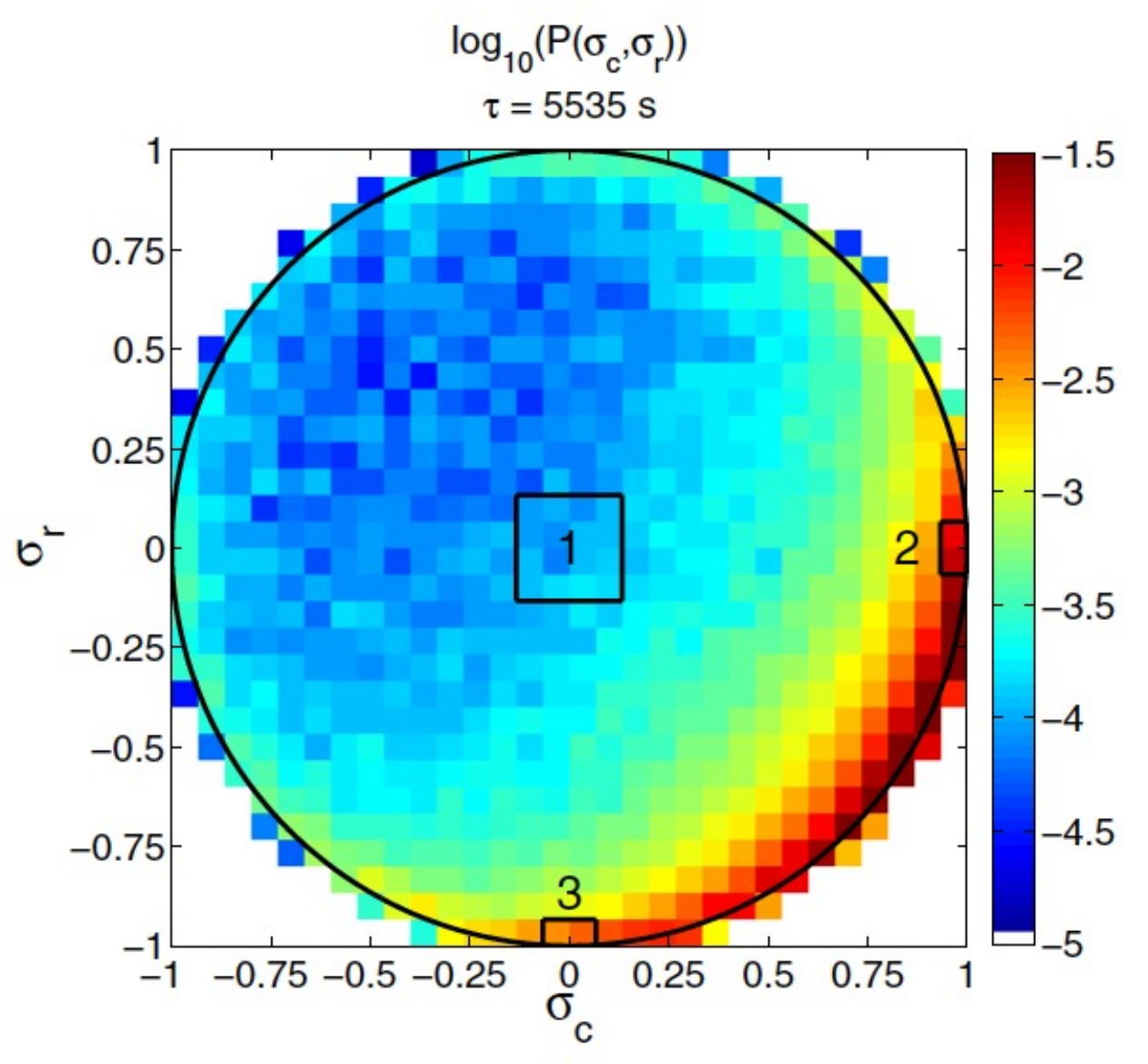} & 
\includegraphics[width=0.49\textwidth]{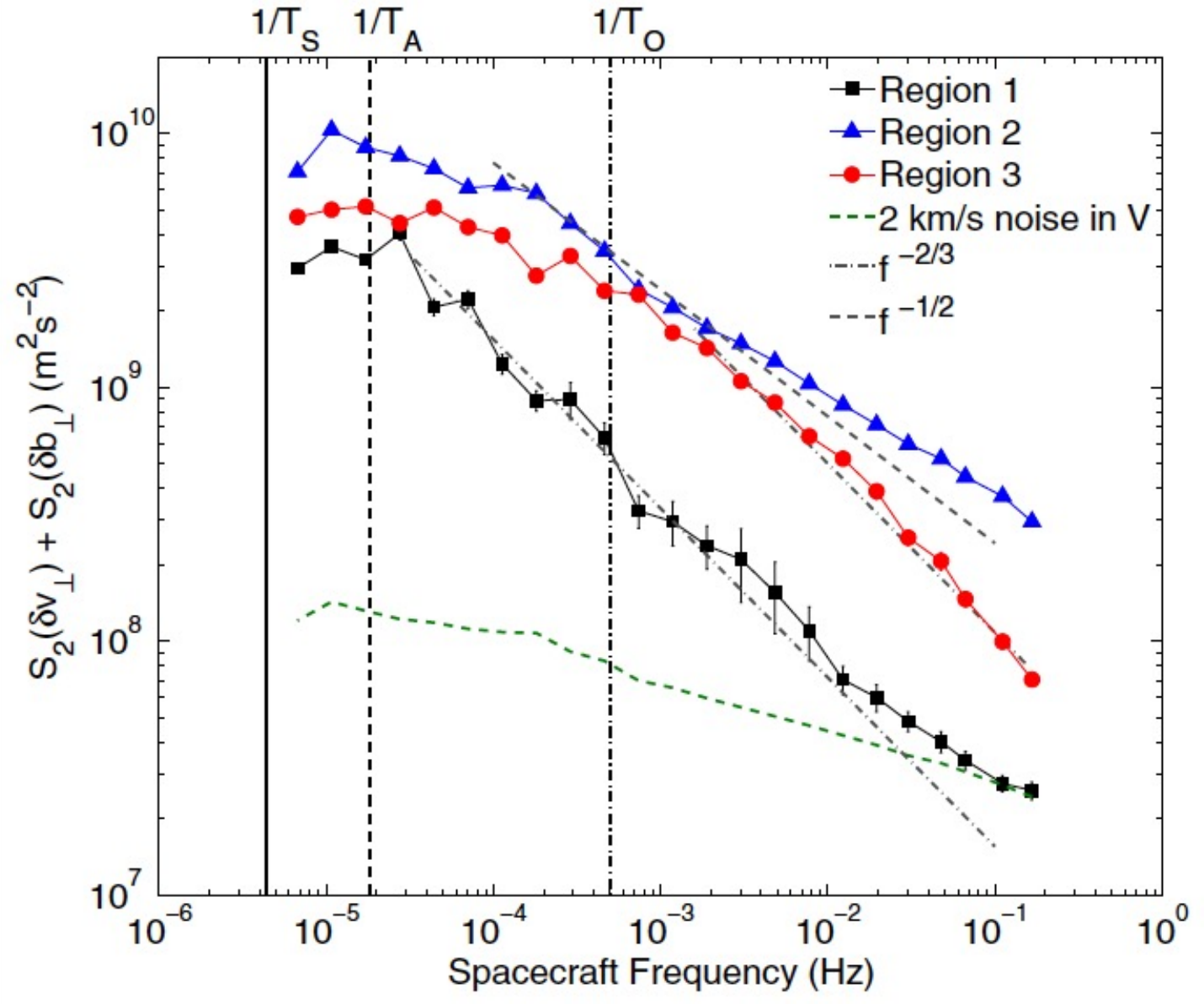}\\
(a) & (b)
\end{tabular}
\end{center}
\caption{(a) Distribution of normalised cross-helicity ($\sigc$) and 
residual energy ($\sigr$) [defined in~\exref{eq:sig_def}] in an interval of fast-solar-wind data 
taken by Wind spacecraft and analysed by \citet{wicks13}, from whose paper both 
plots in this figure are taken (\copyright AAS, reproduced with permission).
(b) Structure functions corresponding to the total 
energy (sum of kinetic and magnetic) conditioned on values of $\sigc$ 
and $\sigr$ and corresponding to Regions 1 (balanced), 2 (Elsasser-imbalanced), 
and~3~(Alfv\'enically imbalanced towards magnetic perturbations), indicated in panel (a). 
The $f^{-2/3}$ slope corresponds to a $\kperp^{-5/3}$ spectrum, 
the $f^{-1/2}$ slope to a $\kperp^{-3/2}$ one.}
\label{fig:wicks}
\end{figure}

\subsection{Imbalance Global and Local}
\label{sec:Eimb}

Since both incompressible MHD and RMHD conserve two invariants---the total energy and 
cross-helicity,---each of the two Elsasser fields $\vzperp^\pm$ has its own conserved energy 
[see \exref{eq:Zpm_conservation}]. The energy fluxes 
$\eps^\pm$ of these fields are, therefore, independent parameters of MHD turbulence. 
Setting them equal to each other makes arguments simpler, but does not, in general, 
correspond to physical reality, for a number of reasons. 

First, everyone's favourite case of directly measurable MHD turbulence is the solar wind, 
where the Alfv\'enic perturbations propagating away from the Sun are launched from 
the Sun \citep{roberts87}, 
while the counterpropagating ones have to be supplied by some mechanism 
that is still under discussion and probably involves Alfv\'en-wave reflection 
as plasma density decreases outwards from the Sun (see \citealt{chandran19} and 
references therein). 
The counterpropagating component is usually energetically smaller, 
especially in the fast wind \citep{bruno13,chen20}. 

Secondly---and, for a theoretical physicist interested in universality, 
more importantly---it is an intrinsic property of MHD turbulence to develop 
regions of {\em local} imbalance. This can be understood dynamically as a desire 
to evolve towards an Elsasser state, $\vzperp^+=0$ or $\vzperp^-=0$, which is an exact solution 
of RMHD equations (confirmed in simulations of decaying RMHD turbulence; see \secref{sec:decay_to_Elsasser}), or statistically as a tendency for 
the local dissipation rates $\eps^\pm$ to fluctuate in space---a mainstay of 
intermittency theories since Landau's famous objection (see \citealt{frisch95}) 
to \citet{K41} and the latter's response in the form of the refined similarity hypothesis, 
accepting a fluctuating~$\eps$ \citep{K62} (the theories of intermittency for 
balanced MHD turbulence proposed by \citealt{chandran15} and \citealt{mallet17a}, 
skimmed through in \secref{sec:MS17}, were based on the same premise). 
In this context, a complete intermittency theory 
for MHD turbulence must incorporate whatever local modification (if any) of the MHD cascade 
is caused by $\eps^+\neq\eps^-$, something that no existing theory has as yet accomplished
or attempted. Another influential school of thought on the root causes of local
imbalance connects it to a tendency for the cross-helicity
to be less vigorously cascaded than energy, leading to local
enhancements of $\vuperp\cdot\vbperp$
(see footnotes~\ref{fn:uBcorr} and~\ref{fn:uBcorr2} and references therein).

That an intimate connection must exist between any verifiable theory of MHD turbulence 
and local imbalance is well illustrated (in \figref{fig:wicks}) 
by the following piece of observational analysis, rather noteworthy, in my (not impartial) view. 
\citet{wicks13} took a series of measurements by Wind spacecraft
of magnetic and velocity perturbations in fast solar wind and sorted them according to the amount 
of imbalance, both Elsasser and Alfv\'enic (\secref{sec:residual}), at each scale. 
They then computed structure functions conditional on these imbalances. 
While the majority of perturbations were imbalanced one way or the other (or both), 
there was a sub-population with $\dz^+_\lambda\sim\dz^-_\lambda\sim\db_\lambda\sim\du_\lambda$.  
Interestingly, the structure function restricted to this sub-population had what 
seemed to be a robust GS95 scaling (corresponding to a $\kperp^{-5/3}$ spectrum), 
even though the structure functions of the imbalanced perturbations 
were consistent with Boldyrev's $\kperp^{-3/2}$ 
aligned-cascade scaling and indeed exhibited some alignment, unlike the GS95 population 
(although not necessarily an alignment with the theoretically desirable scale dependence:
see footnote~\ref{fn:alignment_SW} and \citealt{wicks13align}; \citealt{podesta10xhel} reported analogous results, conditioning 
on the presence of cross-helicity only). 
It is important to recognise that imbalance and alignment of Elsasser fields 
do not automatically imply each other, so balanced fluctuations are not absolutely 
required to be unaligned, or aligned fluctuations to be imbalanced
(see \apref{app:imb_geom}). 
However, as I argued in \secref{sec:MS17}, dynamical alignment 
is an intermittency effect and so there may be a correlation between the emergence 
of imbalanced patches at ever smaller scales and Elsasser 
fields shearing each other into alignment \citep[cf.][]{chandran15}.  

\begin{figure}
\centerline{\includegraphics[width=0.75\textwidth]{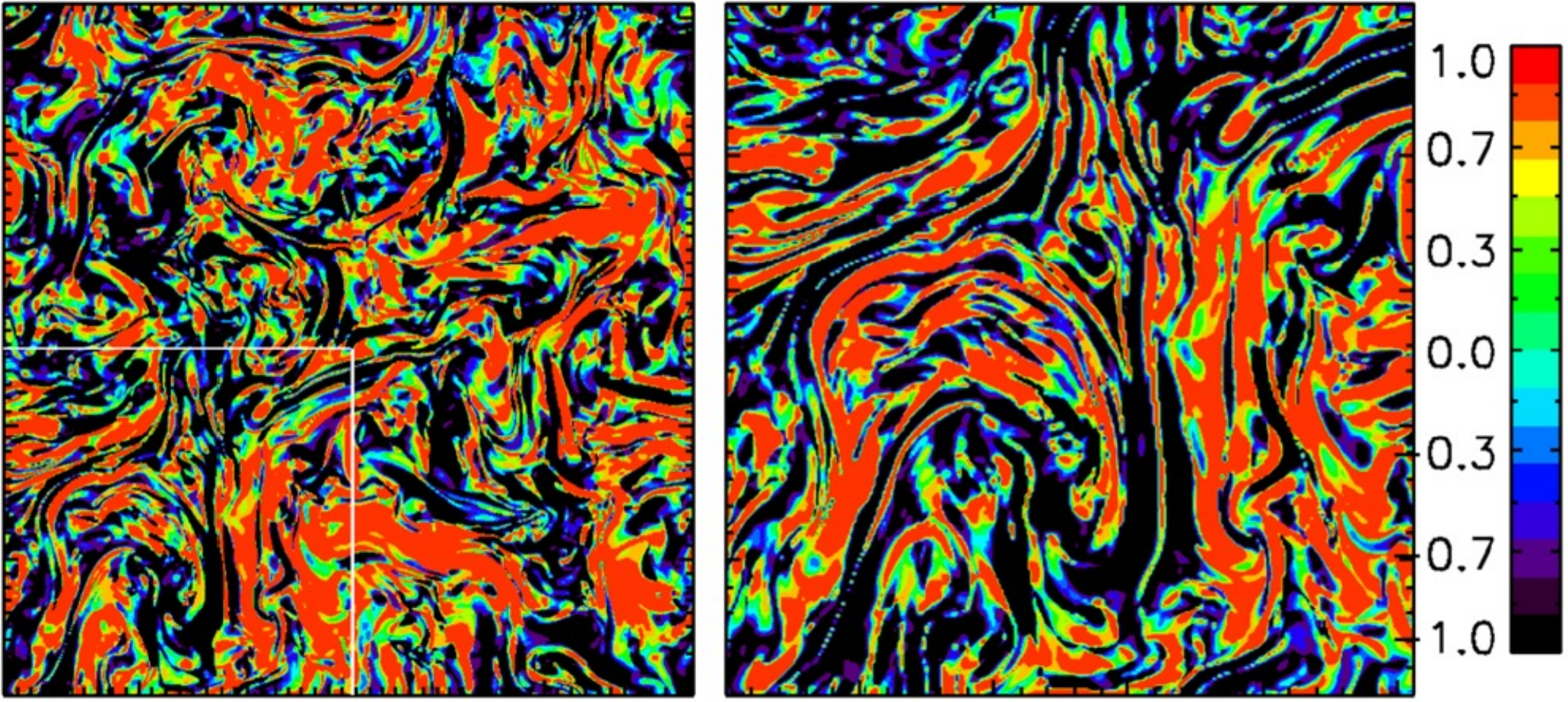}} 
\caption{Cosine of the angle between increments $\dvu_\vlam$ and $\dvb_\vlam$ 
in the $(x,y)$ plane, 
for $\lambda = \Lperp/6$ (left) and $\lambda = \Lperp/12$ (right, corresponding 
to the region demarcated by the white square within the left panel) in a balanced 
RMHD simulation by \citet{perez09} 
[reprinted with permission from \citet{perez09}, copyright (2009) by the American
Physical Society]. 
Since $\dvu_\vlam\cdot\dvb_\vlam =\lt(|\dvz^+_\vlam|^2 - |\dvz^-_\vlam|^2\rt)/4$, 
this is an illustration of patchy local imbalance, as well as of local alignment 
between the velocity and magnetic field.} 
\label{fig:perez_patches}
\end{figure}

Intuitively then, since patches of imbalance are locally ubiquitous even 
in globally balanced turbulence (\citealt{perez09}; see \figref{fig:perez_patches})
and since the theory of balanced turbulence described in \secref{sec:revised}
incorporates intermittency effects in the form of alignment, 
we might expect that this allows for local imbalance---and, therefore, that 
mildly imbalanced turbulence might look largely similar to the balanced one. 
Indeed, how would perturbations in the middle of inertial range ``know'' that 
the local imbalance they ``see'' is local rather than global? Obviously, on average, 
there will not be an imbalance and so the results for $\dz_\lambda$ that one 
derives for balanced turbulence (\secsand{sec:DA}{sec:disruption}) 
are effectively averaged over the statistics 
of the stronger and weaker Elsasser fields---which of $\dz^+_\lambda$ and $\dz^-_\lambda$ 
is which, depends on time and space. 

If we now allow $\eps^+>\eps^-$ on average, it becomes reasonable to expect 
$\dz^+_\lambda > \dz^-_\lambda$ nearly everywhere or, at least, typically---unless 
$\eps^+/\eps^-$ is close enough to unity that fluctuations of local imbalance overwhelm 
the overall global one. In the latter case, presumably the global 
imbalance does not matter---at any 
rate, in the balanced considerations of~\secsand{sec:DA}{sec:disruption}, we only ever 
required $\eps^+\sim\eps^-$, rather than $\eps^+=\eps^-$ exactly. What I am driving at 
here, perhaps with too much faffing about, is the rather obvious point 
that it is only the limit of strong imbalance, 
$\eps^+\gg\eps^-$, that can be expected to be physically distinct, in a qualitative 
manner, from the balanced regime. 

\subsection{Numerical and Observational Evidence}
\label{sec:imb_num}

\begin{figure}
\begin{center}
\begin{tabular}{ccc}
\parbox{0.45\textwidth}{\includegraphics[width=0.45\textwidth]{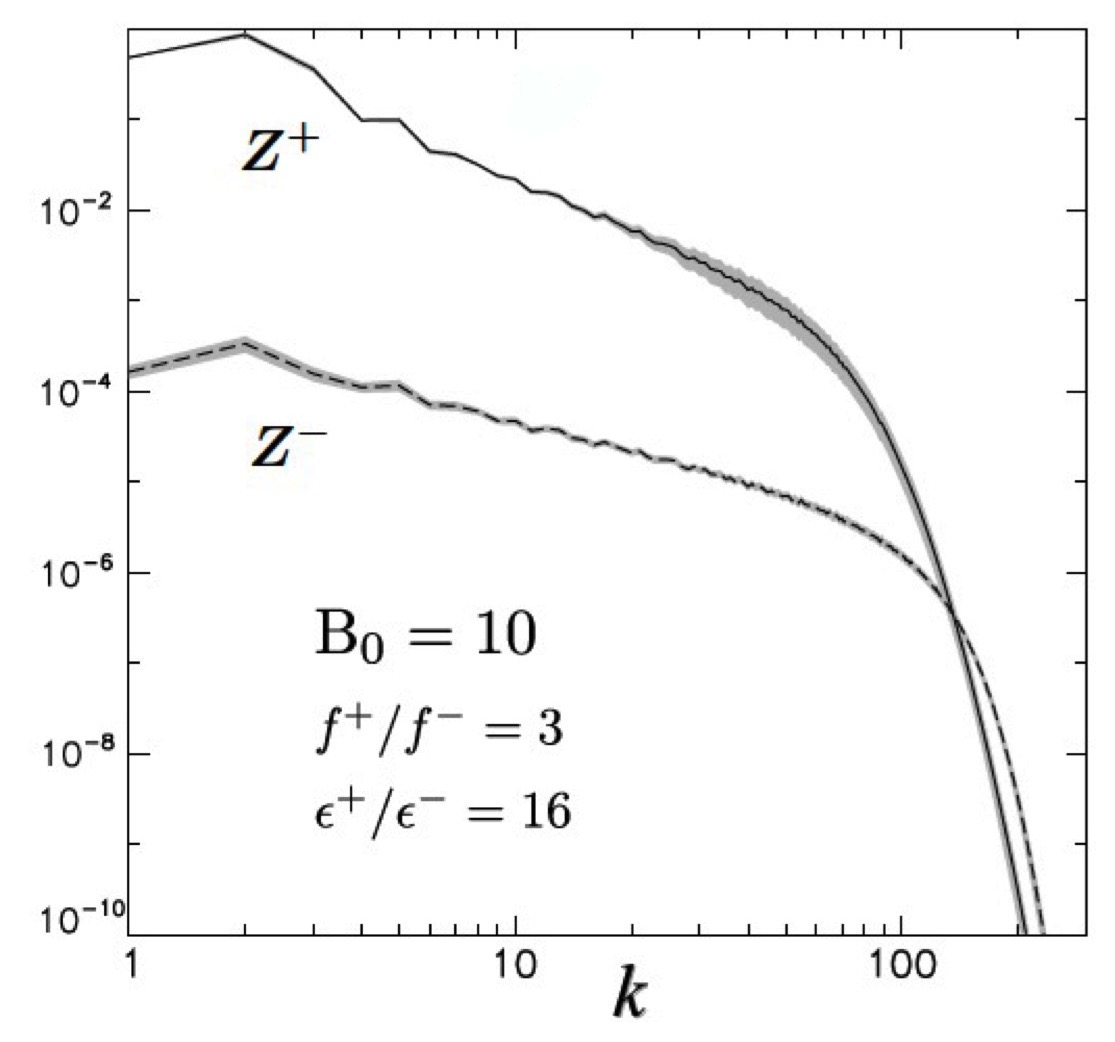}} 
&\qquad\qquad&
\parbox{0.40\textwidth}{\includegraphics[width=0.40\textwidth]{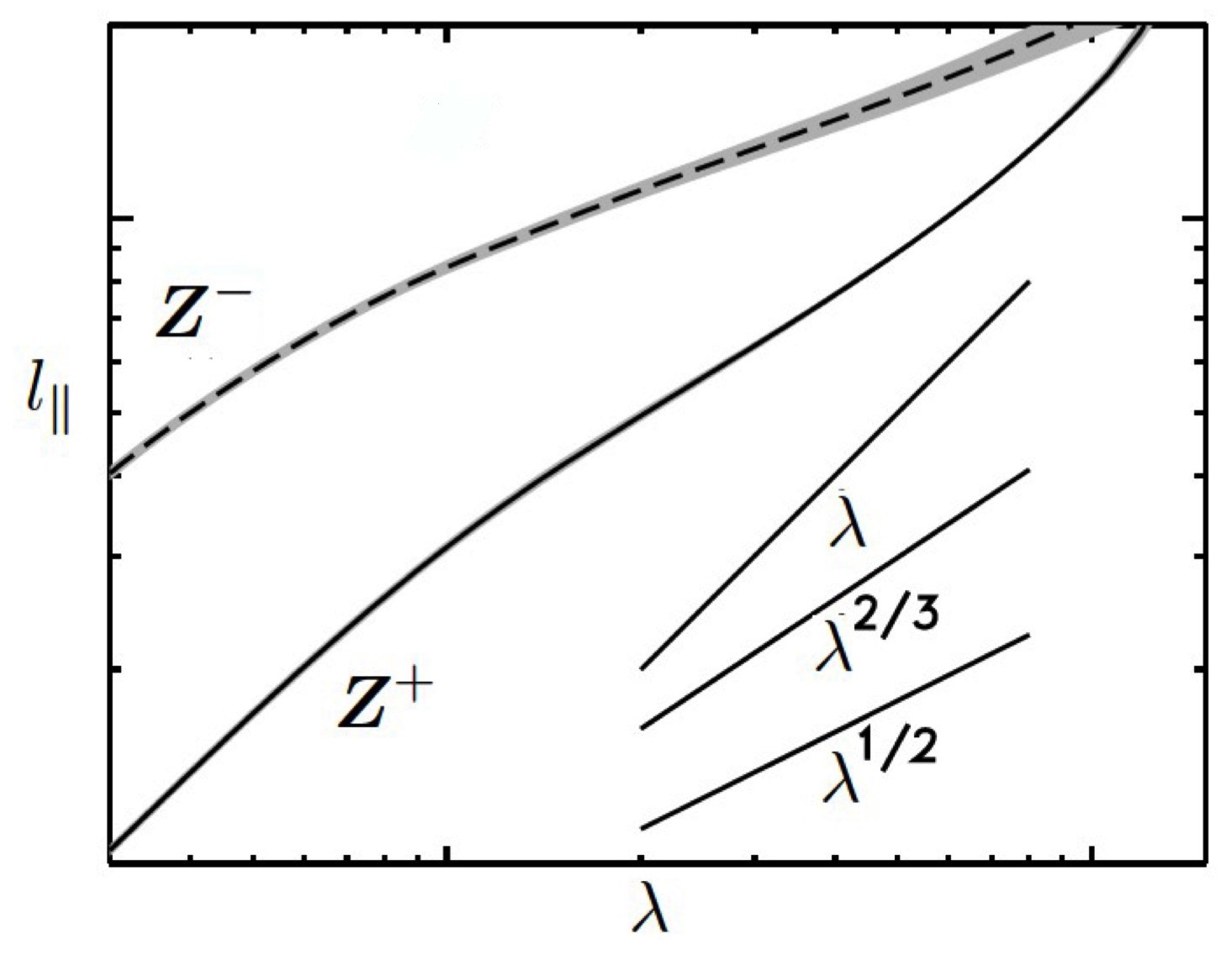}}\\
(a) & & (b)
\end{tabular}
\end{center}
\caption{A typical MHD simulation with large imbalance: (a) spectra, 
(b) anisotropy, $\lpar^\pm$ vs.~$\lambda$. These plots are adapted from 
\citet{beresnyak09} (\copyright AAS, reproduced with permission).
\citet{mallet11} and \citet{meyrand20unpub} 
have qualitatively similar results (although the difference in slopes 
between the weaker and the stronger fields' spectra is much smaller 
in the higher-resolution simulations of \citealt{meyrand20unpub}---see 
an example of that in \figref{fig:meyrand_imb}c, taken from \citealt{meyrand21}).} 
\label{fig:beresnyak_imb}
\end{figure}

As usual, it is this most interesting limit that is also the hardest to resolve numerically 
and so we have little definitive information as to what happens in the strongly imbalanced regime. 
As in the case of the spectra of balanced turbulence, the debate about the 
numerical evidence regarding the imbalanced cascade and its correct theoretical interpretation 
has been dominated by the antagonistic symbiosis of Beresnyak and Boldyrev,
so it is from their papers 
\citep{beresnyak08,beresnyak09,beresnyak10,beresnyak19,perez09,perez10a,perez10b} 
that I derive much of the information reviewed below. 
\citet{perez10a,perez10b} argue that large imbalances are unresolvable 
and refuse to simulate them. \citet{beresnyak09,beresnyak10} do not necessarily 
disagree with this, but believe that useful things can still be learned from 
strongly imbalanced simulations, even if imperfectly resolved.  
Based (mostly) on both groups' simulations, imbalanced MHD turbulence appears to exhibit 
the following distinctive features (which I recount with a degree of confidence as 
they have been reproduced in two sets of independent, unpublished RMHD simulations 
by \citealt{mallet11} and by \citealt{meyrand20unpub}). 
\vskip2mm
(i) The stronger field has a steeper spectrum than the weaker one, with the former steeper 
and the latter shallower than the standard balanced-case spectra 
(\figref{fig:beresnyak_imb}a).
However, it is fairly certain that these spectra are not converged with resolution: 
as resolution is increased, the tendency appears to be for the spectral slopes 
to get closer to each other, both when the imbalance is weak \citep{perez10a}
and when it is strong \citep{mallet11}. This led \citet{perez10a} to argue
that numerical evidence was consistent with the two fields having the same 
spectral slope in the asymptotic limit of infinite Reynolds numbers. 
There is no agreement as to whether the two fields' spectra might be ``pinned'' 
(i.e., equal) 
to each other at the dissipation scale: yes, it seems, in weakly imbalanced simulations of 
\citet{perez10a}, no in the strongly imbalanced ones of \citet{beresnyak09} 
and \citet{meyrand20unpub}.\footnote{Whereas 
the question of pinning may be subject to nontrivial discussion 
\citep{lithwick03,chandran08} in application to MHD turbulence with 
a viscous or resistive cutoff at small scales, it would appear that it is 
more straightforward in a collisionless plasma, e.g., in the solar wind. 
Indeed, there, the decoupling between the two Elsasser fields breaks down 
at the ion Larmor scale, where they are allowed to exchange 
energy \citep{sch09,kunz18}
and, presumably, will not have very different typical amplitudes. Thus, an imbalanced 
turbulence theory with Larmor-scale pinning might be a desirable objective. 
If and when such an outcome proves impossible, this can have interesting implications 
for the very viability of a constant-flux cascade 
(and, at low beta, does, according to \citealt{meyrand21}: see \secref{sec:failed_cascade}).}
In fact, in the latter studies, the dissipation scales of the two fields 
do not appear to be the same: larger for the stronger field, smaller 
for the weaker field (see \figref{fig:beresnyak_imb}a and \figref{fig:meyrand_imb}c). 
\vskip2mm
(ii) The ratio of stronger to weaker field's energies, a crude outer-scale 
quantity that \citet{beresnyak08,beresnyak09,beresnyak10} argue (reasonably, in my view) 
to be more likely to be numerically converged than inertial-range scalings, 
scales very strongly with $\eps^+/\eps^-$: it increases at least as fast as 
\beq
\frac{\la|\vzperp^+|^2\ra}{\la|\vzperp^-|^2\ra} \sim \lt(\frac{\eps^+}{\eps^-}\rt)^2
\label{eq:imb_ratio_sq}
\eeq 
and possibly faster 
(which is inconsistent with the theory of \citealt{perez09}, another 
{\em casus belli} for the two groups; see \secref{sec:imb_PB}). 
\citet[][see \figref{fig:mallet_imb}]{mallet11} and \citet{meyrand20unpub} 
found the same scaling in their simulations for values of 
$\eps^+/\eps^-$ up to 10 (simulations with much higher imbalance are 
numerically suspect). In fact, early numerical evidence for~\exref{eq:imb_ratio_sq} 
appears already in the (decaying) simulations of \citet{verma96} 
(see~\secref{sec:decay_to_Elsasser}). 
\vskip2mm 
(iii) According to \citet{beresnyak08,beresnyak09} and \citet{meyrand20unpub}, 
the stronger field is less anisotropic than the weaker one, in the sense that 
$\lpar^+<\lpar^-$ and $\lpar^+$ drops faster with $\lambda$ than $\lpar^-$ 
(\figref{fig:beresnyak_imb}b). \citet{beresnyak19} notes that this is true 
in his simulations even though he forces the two fields with the same 
parallel scale, i.e., given an opportunity to keep $\lpar^+=\lpar^-$, 
the system refuses to do so.  
\vskip2mm
(iv) \citet[][see \figref{fig:mallet_imb}]{mallet11} 
found that the parallel spectrum of the weaker field 
(measured via its local-field-parallel structure function, as in \secref{sec:aniso}) 
was very robustly $\kpar^{-2}$---to be precise, the exponent varied between $-1.9$ 
and $-2.1$, but in a manner that evinced no systematic dependence on $\eps^+/\eps^-$. 
For the stronger field, they found a gradual 
steepening of the parallel spectrum with higher imbalance.
\vskip2mm
(v) \citet{beresnyak09} found that the alignment angle between 
the Elsasser fields, defined as $\sin\theta$ in \exref{eq:theta_defs}, with numerator 
and denominator averaged separately, decreased with scale roughly as $\lambda^{0.1}$, 
independently of the degree of imbalance. \citet{mallet11} measured the same exponent, 
quite robustly for a wide range of imbalances, but noticed also that 
the scaling exponent depended on the definition of the ``alignment angle'': e.g., 
if root-mean-square numerator and denominator were used, the scaling 
was $\lambda^{0.2\dots0.25}$, closer to the familiar theoretical result~\exref{eq:MS}. 
This is not special to the imbalanced cascades---the same is true in balanced 
turbulence \citep{mallet16}. 
\vskip2mm
(vi) The observational picture is only just emerging. A steeper scaling for the stronger 
field noted in item (i) appears to be consistent with the structure 
functions measured in the fast solar wind by, e.g., \citet{wicks11}, although, besides this, 
they also exhibit low Alfv\'en ratio (see \secref{sec:residual}), which simulations 
do not, and a rather-hard-to-interpret (or, possibly, to trust) scale dependence of the anisotropy.
In contrast, \citet{podesta10xhel} report a scale-independent Elsasser ratio and 
$k^{-3/2}$ spectra for both fields in a number of reasonably imbalanced cases 
of solar-wind turbulence at 1~AU. The same result has been reported by \citet{chen20} 
from the very recent measurements by the Parker Solar Probe made closer to the Sun, 
where the imbalance gets larger ($\la|\vz^+|^2\ra/\la|\vz^-|^2\ra \approx15$)---this may 
be damning for any theory or simulation where the two fields' spectra scale differently, 
at least insomuch as these theories or simulations aspire to apply to the solar wind. 
\vskip2mm
(vii) As the solar wind offers practically the only chance of observational testing
of theory---a chance greatly enhanced by the launch of the Parker Solar Probe---there 
is a growing industry of direct numerical modelling of the 
generation of inward-propagating~($\vz^-$) perturbations by reflection of the 
outward-propagating ones~($\vz^+$), which is what is supposed to happen 
in the expanding solar wind. The latest and most sophisticated study of this 
kind is \citet{chandran19} (who also provide an excellent overview of previous work). 
Their results appear to be quite different from the idealised periodic-box, 
artificially-forced studies discussed above: the stronger field's spectrum is 
actually {\em shallower} than the weaker one's (sometimes as shallow as~$k^{-1}$), 
but both asymptote towards $k^{-3/2}$ with increasing heliocentric distance---good 
news for modelling, in view of what \citet{chen20} have found. \citet{chandran19}
acknowledge, however, that they can break these results by fiddling with 
how their turbulence is forced in the photosphere. Thus, the nature of large-scale  
energy injection appears to matter,\footnote{\citet{chandran19} have a theory 
as to why that is, which will be explained in \secref{sec:imb_LGS}.} 
at least at finite resolutions, perhaps 
reinforcing the doubts expressed above about the convergence of even the 
more idealised simulations. 
\vskip2mm
For a short while still, the field appears set to remain 
open to enterprising theoreticians.

\begin{figure}
\centering
\begin{tabular}{lccccr}
\hline
$\eps^+/\eps^-$ &$\mu^+_\perp$ & $\mu^+_\parallel$ & $\mu^-_\perp$ & $\mu^-_\parallel$ & $\RE$ \\
\hline\hline
1      & -1.6 & -1.9 & -1.6 & -1.9 & 1 \\
2      & -1.6 & -1.9 & -1.5 & -2.0 & 5 \\
5      & -1.8 & -2.0 & -1.5 & -2.0 & 35 \\
8      & -1.8 & -2.1 & -1.5 & -2.0 & 45 \\
10     & -1.9 & -2.2 & -1.4 & -2.1 & 110 \\
100    & -2.3 & -2.3 & -1.4 & -2.0 & 2200 \\
1000   & -2.5 & -2.6 & -1.3 & -2.0 & 13000 \\
\hline
\end{tabular}
\caption{Scalings found in (unpublished) $512^3$ 
RMHD numerical simulations by \citet{mallet11}: 
perpendicular (parallel) spectral indices $\mu_\perp$ ($\mu_\parallel$)
(inferred from structure functions calculated as explained in \secref{sec:aniso}) 
for both fields, denoted by the $\pm$ superscripts. 
In terms of the scaling exponents $\gamma^\pm_{\perp,\parallel}$ of 
the field increments ($\dz^\pm_\lambda\propto\lambda^{\gamma^\pm_\perp}$, 
$\dz^\pm_{\lpar}\propto\lpar^{\gamma^\pm_\parallel}$), these are 
$\mu^\pm_{\perp,\parallel} = -2\gamma^\pm_{\perp,\parallel} - 1$. 
The last column shows the overall Elsasser ratio $\RE=\la|\vzperp^+|^2\ra/\la|\vzperp^-|^2\ra$. 
The parallel scalings of the weaker field were converged with resolution, 
while the perpendicular scalings of the stronger (weaker) field at 
$\eps^+/\eps^- = 10$ became shallower (steeper) as resolution was increased 
from $256^3$ to $512^3$ to $1024^2\times512$; the parallel scaling of the stronger
field also appeared to become shallower. Simulations with 
$\eps^+/\eps^- = 100, 1000$ should be viewed as numerically suspect. 
\label{fig:mallet_imb}}
\end{figure}

\subsection{\citet{lithwick07}} 
\label{sec:imb_LGS}

In what follows, in view of the discussion in \secref{sec:align_aniso} and 
in \apref{app:align_nlin}, I shall 
stick with my use of the Elsasser-field alignment angle $\theta$ in the expression 
\exref{eq:tnl_align} for $\tnl^{\pm}$. This angle is obviously the same for both fields, so
\beq
\tnl^\pm \sim \frac{\lambda}{\dz^\mp_\lambda\sin\theta_\lambda}
\hence
\frac{\tnl^+}{\tnl^-} \sim \frac{\dz^+_\lambda}{\dz^-_\lambda} > 1, 
\label{eq:tnl_ratio}
\eeq
i.e., the cascade of the stronger 
field is slower (because it is advected by the weaker field). 
Assuming nevertheless that both cascades are strong, we infer 
immediately 
\beq
\frac{(\dz^\pm_\lambda)^2}{\tnl^\pm} \sim \eps^\pm
\hence
\frac{\dz^+_\lambda}{\dz^-_\lambda} \sim \frac{\eps^+}{\eps^-}. 
\label{eq:z_ratio}
\eeq
Thus, the two fields' increments have the same scaling with $\lambda$ 
(the same $\kperp$ spectra) and the ratio of their energies is $\sim(\eps^+/\eps^-)^2$, 
in agreement with~\exref{eq:imb_ratio_sq}.  
This is the conclusion at which \citet[][henceforth LGS07]{lithwick07} arrived---they 
considered unaligned GS95-style turbulence ($\sin\theta\sim1$), 
but that does not affect \exref{eq:z_ratio} [note that this result 
already appeared in~\exref{eq:z_GS95}]. 

Things are, however, not as straightforward as they might appear. LGS07 
point out that it is, in fact, counterintuitive that the weaker $\dz_\lambda^-$ perturbation, 
which is distorted by $\dz_\lambda^+$ on a shorter time scale $\tnl^-$, can nevertheless 
coherently distort $\dz_\lambda^+$ for a longer time $\tnl^+$. Their solution to this is to argue 
that, while the weaker field is strongly distorted {\em in space} by the stronger one, 
it remains correlated {\em in time} for as long as the stronger field does
(coherence time of the long-correlated advector inherited by the advectee). 
In other words, during its (long) correlation time $\tnl^+$, 
the stronger field (in its reference frame travelling at $\vA$) 
sees a weak field that has been rendered multiscale by the spatial variation 
of the stronger field, but remains approximately constant 
for a time $\tnl^+$ and so can keep distorting the stronger field 
in a time-coherent~way. 

This argument can only work, it seems, if the long-term coherence of the weaker 
field is not upset by the way in which it is forced, so one must assume 
that it is forced at the outer scale with the same (long!) correlation time as 
the cascade time of the stronger field, in the Alfv\'enic frame of the latter. 
\citet{chandran19} dub this the ``coherence assumption''. 
While hard to justify in general,\footnote{It is, however, worth pointing out
in this context a curious result reported by \citet{lugones19}: they studied
frequency-wavenumber spectra of externally forced (not reflection-driven)
imbalanced MHD turbulence and spotted, in some of their runs (those, it seems
to me, that were more likely to have been in a strong-turbulence regime),
that, counterintuitively, 
the weaker field moved in the same direction as the stronger one, implying perhaps
just the kind of coherence that LGS07 conjectured, without reflection driving. 
\citet{lugones19} interpret this result in terms of reflections off inhomogeneities
of the local mean field, but perhaps that is another way of saying the same thing---that
the weaker field gets locked into long-time coherence with the stronger one.}
it is great for them as, in their model, the weaker field is generated by the
reflection of the stronger one 
as the latter propagates outwards in an expanding solar wind, and so 
one should indeed expect the two fields to be tightly correlated 
at the outer scale. Their endorsement of LGS07---perhaps 
with an amendment that $\theta_\lambda$ should have some scaling 
with $\lambda$ determined by alignment/intermittency (\secref{sec:MS17})---is 
backed up by their numerical results, where both fields' spectra 
approach $\kperp^{-3/2}$ (and their alignment increases) with increasing 
heliocentric distance. 

\subsection{\citet{perez09}} 
\label{sec:imb_PB}

\citet{perez09} disagree with the entire approach leading to \exref{eq:z_ratio}: 
they think that the two Elsasser fields 
should have two different alignment angles $\theta^\pm_\lambda$, both small, 
and posit that those ought to be the angles that they make with the velocity 
field.\footnote{\citet{podesta10imb} base their theory on the same assumption (also unexplained), 
but have a different scheme for generalising Boldyrev's aligned cascade 
to the imbalanced regime. They imagine the geometric configuration 
of the fields to be such that $|\dvu_\vlam|=|\dvb_\vlam|$ and, 
consequently, $\dvz^+_\vlam$ 
and $\dvz^-_\vlam$ are perpendicular to each other. This does not appear to be what actually 
happens, at least in simulations [see \secref{sec:imb_num}, item (v)]. 
\citet{podesta10imb} also inherit from Boldyrev's original construction the incompatibility 
of their scalings with the RMHD symmetry (see \secref{sec:sym}). 
There is an interesting angle in their paper though: they notice, in solar-wind 
observations, that the probabilities with which aligned or anti-aligned (in the sense 
of the sign of $\dvu_\vlam\cdot\dvb_\vlam$) perturbations occur are 
independent of scale throughout the inertial range; they then use the ratio 
of these probabilities as an extra parameter in the theory. This is a step 
in the direction of incorporating patchy imbalance into the game---something that 
seems important and inevitable.} 
Why that should be the case they do not explain, 
but if one takes their word for it, then (as is obvious from the geometry in 
\figref{fig:geometry})
\beq
\dz^+_\lambda\sin\theta^+_\lambda \sim \dz^-_\lambda\sin\theta^-_\lambda 
\hence
\tnl^+\sim\tnl^- \sim \frac{\lambda}{\dz^\pm_\lambda\sin\theta^\pm_\lambda}
\hence 
\frac{\dz^+_\lambda}{\dz^-_\lambda} \sim \sqrt{\frac{\eps^+}{\eps^-}}.
\label{eq:z_ratio_boldyrev}
\eeq
The last result follows from the first relation in \exref{eq:z_ratio} 
with $\tnl^+\sim\tnl^-$. The equality of cascade times also conveniently 
spares them having to deal with the issue, discussed above, of long-time correlatedness, 
or otherwise, of the weaker field (or with $\lpar^+\neq\lpar^-$; see~\secref{sec:imb_lpar}). 

\citet{perez09,perez10a,perez10b} are not forthcoming with any detailed tests of this 
scheme (viz., either of the details of alignment or of the energy-ratio scaling), 
while \citet{beresnyak10} present numerical results that 
contradict very strongly the expectation of the energy ratio scaling 
as $\eps^+/\eps^-$ [as implied by \exref{eq:z_ratio_boldyrev}]
and possibly support $(\eps^+/\eps^-)^2$ [i.e., \exref{eq:z_ratio}].\footnote{\citet{podesta11th} 
collated both groups' data and concluded that the results of \citet{perez10b} were 
entirely compatible with \citet{beresnyak10} and with~\exref{eq:imb_ratio_sq}.}  
\citet{perez10b} reply that \exref{eq:z_ratio_boldyrev} 
should only be expected to hold for local fluctuating values of 
the amplitudes and of $\eps^\pm$ and not for their box averages.
It is not impossible 
that this could make a difference for cases of weak imbalance 
($\eps^+/\eps^-\sim1$), with local fluctuations of energy fluxes  
superseding the overall imbalance, although it seems to me that 
if it does, we are basically dealing with balanced turbulence anyway:  
I do not see any fundamental physical difference between $\eps^+=\eps^-$ 
and $\eps^+\sim\eps^-$ on the level of ``twiddle'' arguments by which everything 
is done in these theories. At strong imbalance, \exref{eq:z_ratio} seems 
to work better \citep{beresnyak09,beresnyak10} for the overall energy ratio, 
but not for spectra, which do not have the same slope (\figref{fig:beresnyak_imb}a). 
\citet{perez10a,perez10b} 
argue that such cases in fact cannot be properly resolved, the limiting factor 
being the weaker field providing too slow a nonlinearity to compete with 
dissipation and produce a healthy inertial range. If so, the 
interesting case is inaccessible and the accessible case is uninteresting, 
we know nothing. 

\subsection{Parallel Scales and Two Flavours of CB} 
\label{sec:imb_lpar}

By the CB conjecture (\secref{sec:CBCB}), 
the parallel coherence lengths of the two fields are, in the ``na\"ive'' 
theory leading to \exref{eq:z_ratio}, 
\beq
\lpar^\pm \sim \vA\tnl^\pm 
\hence
\frac{\lpar^+}{\lpar^-} \sim \frac{\eps^+}{\eps^-} > 1, 
\label{eq:lpar_ratio}
\eeq
whereas in the \citet{perez09} theory \exref{eq:z_ratio_boldyrev}, 
the equality of cascade times implies $\lpar^+\sim\lpar^-$, end of story. 
LGS07 argue that, in fact, also \exref{eq:lpar_ratio} should be replaced by 
\beq
\lpar^+ \sim \lpar^- \sim \vA\tnl^-
\label{eq:lpar_ratio_lithwick}
\eeq
because $\vzperp^+$ perturbations separated by distance $\lpar^-$ in the parallel direction 
are advected by completely spatially decorrelated $\vzperp^-$ perturbations, 
which would then imprint their parallel coherence length on their stronger cousins
(the parallel coherence length of the short-correlated advector imprinted on the advectee). 

Furthermore, if one accepts the LGS07 argument 
that the correlation time of the $\vzperp^-$ field is $\tnl^+$, not $\tnl^-$
(see \secref{sec:imb_LGS}), 
then $\lpar^-\sim\vA\tnl^-$ must be justified not by temporal (causal) decorrelation 
but by the weaker field being spatially distorted beyond recognition on the scale $\lpar^-$, 
even if remaining temporally coherent. This is more or less 
what \citet{beresnyak08} call ``propagation CB'' (the other CB being ``causality CB'').
They note that the typical uncertainty in the parallel 
gradient of any fluctuating field at scale~$\lambda$~is
\beq  
\dkpar\sim\frac{\vbperp\cdot\vdperp}{\vA} \sim \frac{\db_\lambda}{\xi_\lambda\vA}.
\label{eq:kpar_db}
\eeq 
In balanced turbulence, $\dkpar^{-1}\sim\vA\tnl \sim \lpar$ [cf.~\exref{eq:xi_displacement}], 
so this is just a consistency check. In imbalanced turbulence, 
\beq
\db_\lambda\sim\dz^+_\lambda
\hence 
\dkpar^{-1} \sim \frac{\xi_\lambda\vA}{\dz^+_\lambda} 
\sim \vA\tnl^-, 
\eeq
where $\tnl^-$, given by \exref{eq:tnl_ratio}, is 
the spatial-distortion time of $\dz^-_\lambda$, not necessarily its correlation time. 
The parallel scale of any field will be the shorter of $\dkpar^{-1}$ 
and whatever is implied by the causality CB. In the LGS07 theory, the latter is 
$\vA\tnl^+$ for both fields. Since $\tnl^+\gg\tnl^-$, we must set  
$\lpar^+ \sim \lpar^- \sim \dkpar^{-1}$, which is the same as \exref{eq:lpar_ratio_lithwick}. 

Thus, we end up with both Elsasser fields having $\tA \sim \lpar^-/\vA$ 
that is smaller than their correlation time $\tnl^+$ (even though the weaker 
field has a shorter spatial distortion time $\tnl^-\sim\tA$), but their 
cascades are nevertheless strong. 
Whatever you think of the merits of the above arguments, neither 
\exref{eq:lpar_ratio} nor \exref{eq:lpar_ratio_lithwick} appear to be 
consistent with any of the cases reported by 
\citet{beresnyak09}, weakly or strongly imbalanced, 
which all have $\lpar^+<\lpar^-$ (see, e.g., \figref{fig:beresnyak_imb}b).  
No other numerical evidence on the parallel scales 
in imbalanced turbulence is, as far as I know, available in print. 

\subsection{Towards a New Theory of Imbalanced MHD Turbulence}
\label{sec:imb_new}

The \citet{beresnyak08} argument was, in fact, more complicated than 
presented in \secref{sec:imb_lpar}, because they did not agree with 
LGS07 about the long correlation time of the weaker field, 
assumed the stronger field to be weakly, rather than strongly, turbulent,  
and were keen to accommodate $\lpar^+ < \lpar^-$. 
Their key innovation was to allow interactions to be nonlocal. 
I will not review their theory here, because it depends on a number of 
{\em ad hoc} choices that I do not know how to justify, and does not, 
as far as I can tell, lead to a fully satisfactory set of predictions,  
but I would like to seize on their idea of nonlocality of interactions, 
although in a way that is somewhat different from theirs. 
The resulting scheme captures most of the 
properties of imbalanced turbulence observed in numerical simulations 
(\secref{sec:imb_num}) and reduces to the already 
established theory for the balanced case when $\eps^+/\eps^-\sim 1$, 
so perhaps it deserves at least some benefit of the doubt. 

\subsubsection{Two Semi-Local Cascades}
\label{sec:2cascades}

Let me assume {\em a priori} that, as suggested by numerics 
\citep{beresnyak09,beresnyak19}, $\lpar^+\ll\lpar^-$ in the inertial range, 
viz., 
\beq
\frac{\lparl^+}{\lparl^-} \sim \lt(\frac{\lambda}{\Lperp}\rt)^\alpha, 
\label{eq:lpar_ratio_alpha}
\eeq
where $\alpha > 0$ and $\Lperp$ is the perpendicular outer scale 
(so the two Elsasser fields are assumed to have
the same parallel correlation length, $\Lpar$, at the outer scale---e.g., 
by being forced that way, as was done by \citealt{beresnyak19}). 

This implies that, at the same $\lambda$, the stronger field 
$\dz_\lambda^+$ oscillates much faster than the weaker field $\dz_\lambda^-$. 
I shall assume therefore that the interaction between the two fields 
local to the scale $\lambda$ is not efficient: even though 
$\dz_\lambda^-$ is buffeted quite vigorously by the stronger field 
$\dz_\lambda^+$, most of this cancels out. Rather than attempting 
to pick up a contribution arising for the resulting weak interaction, 
let me instead posit that the dominant, strong nonlinear distortion 
of $\dz_\lambda^-$ will be due to the stronger field $\dz_{\lambda'}^+$ 
at a scale $\lambda' > \lambda$ such that 
\beq
\lparlp^+ \sim \lparl^-.
\label{eq:lpar_locality}
\eeq 
In other words, the interaction is nonlocal in $\lambda$ but 
local in $\lpar$.\footnote{\citet{beresnyak08} proposed the same, 
but to describe {\em weak} cascading of $\dz_{\lambda'}^+$ by~$\dz_\lambda^-$. 
Thus, their cascade of the stronger field is weak and nonlocal and that of the 
weaker field is strong and local. In the scheme I am proposing here, both 
cascades are strong and it is the weaker field's one that is nonlocal.} 
The constancy of the flux of the weaker field 
then requires 
\beq
\frac{\lt(\dz_\lambda^-\rt)^2\dz_{\lambda'}^+}{\xi_{\lambda'}} \sim \eps^-,
\label{eq:zm_cascade}
\eeq
where $\xi_{\lambda'}$ has been introduced to account for a possible 
depletion of the nonlinearity due to alignment: 
\beq
\frac{\xi_{\lambda'}}{\Lperp} \sim \lt(\frac{\lambda'}{\Lperp}\rt)^\beta.
\label{eq:xi_beta}
\eeq
In the absence of alignment, $\beta = 1$. For aligned, balanced, 
locally cascading ($\lambda' \sim \lambda$) turbulence, 
$\beta = 3/4$ [see \exref{eq:xi_scalings}].  
By the usual CB argument, the parallel coherence scale of the weaker field~is
\beq
\lparl^- \sim \frac{\vA\xi_{\lambda'}}{\dz_{\lambda'}^+}. 
\label{eq:lparm}
\eeq
Note that, in the terminology of \secref{sec:imb_lpar}, 
this is both the causality CB and the propagation CB, because 
$\dkpar$ for $\dz_\lambda^-$ is determined by the propagation 
of the latter along the ``local mean field'' $\db_{\lambda'}$ 
[see \exref{eq:kpar_db}]. 

Now consider the cascading of the stronger field by the weaker 
one. Since $\lparl^+ \ll \lparl^-$, the $\dz_\lambda^-$ 
fluctuations are, from the point of view of the $\dz_\lambda^+$ ones, 
slow and quasi-2D, and so the weaker field can cascade the strong one 
locally, in the same way as it does in any of the theories described above:  
\beq
\frac{(\dz_\lambda^+)^2\dz_\lambda^-}{\xi_\lambda} \sim \eps^+. 
\label{eq:zp_cascade}
\eeq  
Causality CB would imply $\lparl^+\sim \vA\xi_\lambda/\dz_\lambda^-$, 
but that is long compared to $\dkpar^{-1}$ given by~\exref{eq:kpar_db}, 
so I shall use propagation CB instead, just like LGS07 and \citet{beresnyak08} did:  
\beq
\lparl^+ \sim \frac{\vA\xi_\lambda}{\dz_\lambda^+}. 
\label{eq:lparp}
\eeq
Reassuringly, this choice immediately clicks into consistency with 
the requirement of parallel locality \exref{eq:lpar_locality} if 
$\lparl^-$ is given by \exref{eq:lparm}. 

There are two nuances here. First, in order for the $\dz_\lambda^-$ field 
to be able to distort $\dz_\lambda^+$ according to~\exref{eq:zp_cascade},  
it needs to remain coherent for a time $\sim\xi_\lambda/\dz_\lambda^-$. 
To make it do so, let me invoke the LGS07 argument 
already rehearsed in \secref{sec:imb_LGS}: 
according to~\exref{eq:zm_cascade}, $\dz_\lambda^-$ stays coherent as long 
as $\dz_{\lambda'}^+$ does, which, according to \exref{eq:zp_cascade} 
with $\lambda=\lambda'$, is~$\xi_{\lambda'}/\dz_{\lambda'}^-$---long enough! 

Secondly, in \exref{eq:zp_cascade}, I used the same fluctuation-direction 
scale $\xi_\lambda$ as in \exref{eq:zm_cascade}, except at $\lambda$, rather than 
at $\lambda'$. This may be a somewhat simplistic treatment of alignment in 
local vs.\ nonlocal interactions, but I do not know how to do better, and 
the scalings that I get this way will have all the right properties. 
A reader who finds this unconvincing may assume $\xi_\lambda \sim \lambda$ 
and treat what follows as a GS95-style theory that ignores alignment altogether.  

\begin{figure}
\centerline{\includegraphics[width=0.85\textwidth]{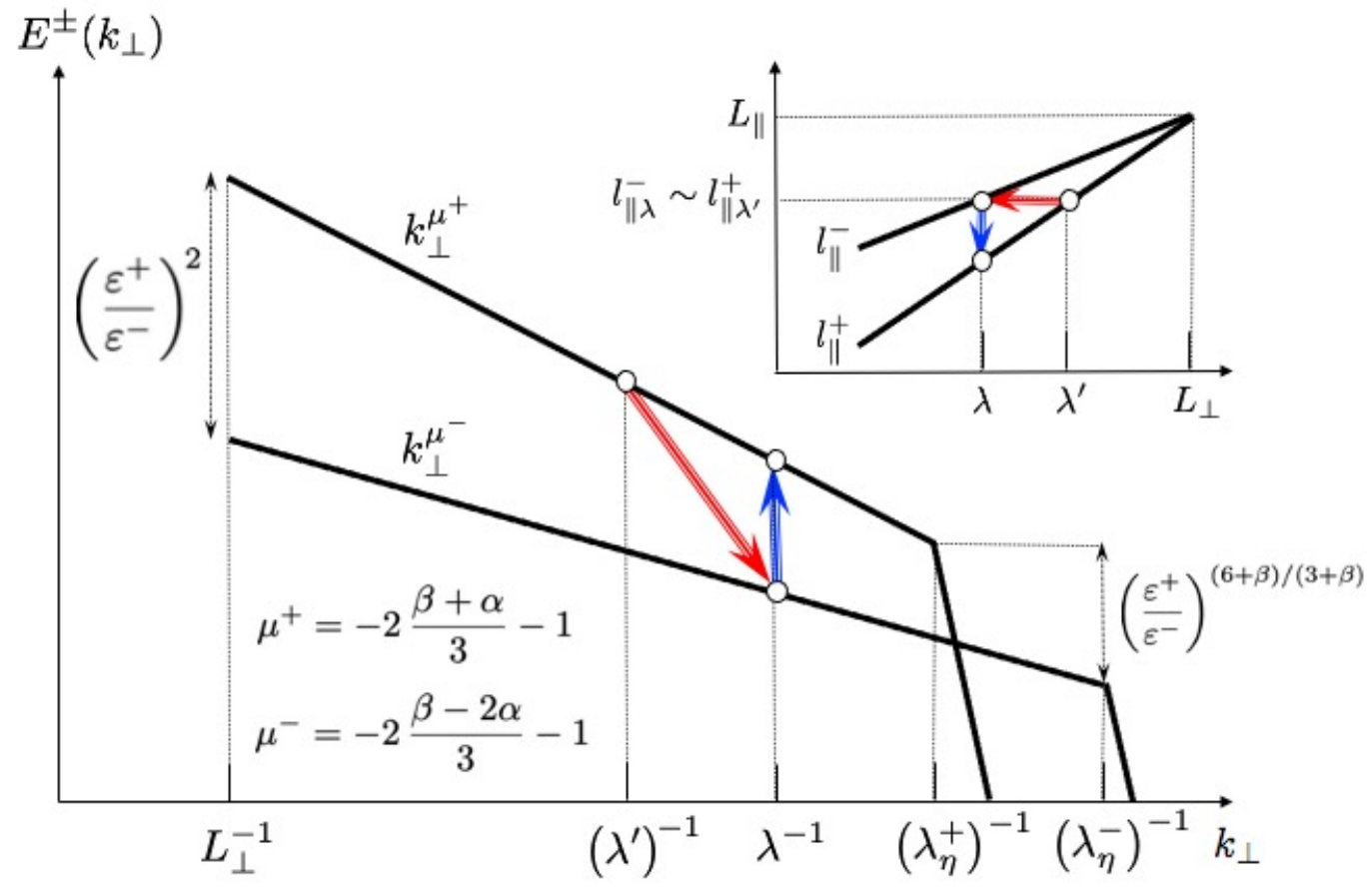}} 
\vskip2mm
\caption{Cartoon of the spectra of imbalanced turbulence. The interactions 
are shown by arrows: red (advection of the weaker field by the stronger field, nonlocal 
in~$\lambda$ but local in~$\lpar$) and blue (advection of the stronger field by the weaker 
field, local in~$\lambda$ but nonlocal in~$\lpar$). The inset shows the parallel 
scales $\lpar^\pm$ vs.\ the perpendicular scale~$\lambda$.}
\label{fig:spectrum_imb}
\end{figure}

To summarise, I am considering here an imbalanced turbulence that consists of 
two ``semi-local'' cascades: that of the stronger field, local in $\lambda$ 
but not in $\lpar$, and that of the weaker one, local in $\lpar$ but not 
in $\lambda$ (\figref{fig:spectrum_imb}).

\subsubsection{Perpendicular Spectra}

In view of \exref{eq:lparm} and \exref{eq:lparp}, 
\exref{eq:zm_cascade} can be rewritten as follows: 
\beq
\frac{\lt(\dz_\lambda^-\rt)^2\dz_\lambda^+}{\xi_\lambda}
\sim \eps^-\frac{\xi_{\lambda'}\dz_\lambda^+}{\xi_\lambda\dz_{\lambda'}^+}
\sim \eps^-\frac{\lparl^-}{\lparl^+} 
\sim \eps^-\lt(\frac{\lambda}{\Lperp}\rt)^{-\alpha}, 
\label{eq:zm_cascade_alpha}
\eeq
the last step being a recapitulation of the assumption \exref{eq:lpar_ratio_alpha}. 
Dividing \exref{eq:zp_cascade} by \exref{eq:zm_cascade_alpha}, one~gets 
\beq
\frac{\dz_\lambda^+}{\dz_\lambda^-} \sim 
\frac{\eps^+}{\eps^-}\lt(\frac{\lambda}{\Lperp}\rt)^\alpha.
\label{eq:zratio_alpha}
\eeq
Thus, the ratio of the energies at the outer scale ($\lambda=\Lperp$) 
is $(\eps^+/\eps^-)^2$, likely the correct scaling [see \secref{sec:imb_num}, 
item (ii) and \secref{sec:imb_PB}], and the spectrum of the stronger field 
is steeper than that of the weaker field, also in agreement with 
numerics [\secref{sec:imb_num}, item (i)]. 

Now, by using \exref{eq:zp_cascade}, \exref{eq:zratio_alpha} and the alignment 
assumption~\exref{eq:xi_beta}, it becomes possible to determine the scalings of both fields: 
\beq
\dz_\lambda^+ \sim \lt[\frac{(\eps^+)^2}{\eps^-}\Lperp\rt]^{1/3}
\lt(\frac{\lambda}{\Lperp}\rt)^{(\beta + \alpha)/3},\qquad
\dz_\lambda^- \sim \lt[\frac{(\eps^-)^2}{\eps^+}\Lperp\rt]^{1/3}
\lt(\frac{\lambda}{\Lperp}\rt)^{(\beta - 2\alpha)/3}.
\label{eq:dz_alpha}
\eeq
Comparing the first of these with \exref{eq:zm_cascade}, one can also 
work out how nonlocal the interactions are: 
\beq
\frac{\lambda}{\lambda'} \sim \lt(\frac{\lambda}{\Lperp}\rt)^{3\alpha/(2\beta-\alpha)}. 
\label{eq:lambdaprime}
\eeq

With $\alpha=0$ and $\beta=1$ in \exref{eq:dz_alpha}, we are back to GS95 
(\secref{sec:aniso}), whereas with $\alpha = 0$ and $\beta=3/4$, 
we recover the aligned theory of~\secref{sec:revised}. 

\subsubsection{Parallel Spectra}

Now, from \exref{eq:lparp}, \exref{eq:xi_beta}, \exref{eq:dz_alpha}, 
and \exref{eq:lpar_ratio_alpha}, the parallel scales are 
\beq
\frac{\lparl^+}{\Lpar} \sim \lt(\frac{\lambda}{\Lperp}\rt)^{(2\beta-\alpha)/3},\qquad
\frac{\lparl^-}{\Lpar} \sim \lt(\frac{\lambda}{\Lperp}\rt)^{2(\beta-2\alpha)/3},
\label{eq:lpar_alpha}
\eeq
where the parallel outer scale is [cf.~\exref{eq:Lperp}] 
\beq
\Lpar = \vA\Lperp^{2/3}\lt[\frac{(\eps^+)^2}{\eps^-}\rt]^{-1/3}.   
\eeq
Combining \exref{eq:lpar_alpha} with \exref{eq:dz_alpha} gives us the 
parallel scalings of the field increments: 
\beq
\dz_{\lpar}^+ \sim \lt[\frac{(\eps^+)^2\Lpar}{\eps^-\vA}\rt]^{1/2}
\lt(\frac{\lpar}{\Lpar}\rt)^{(\beta + \alpha)/(2\beta - \alpha)},\qquad
\dz_{\lpar}^- \sim \lt(\frac{\eps^-\lpar}{\vA}\rt)^{1/2}. 
\eeq
Whereas the stronger field's scaling is (for small $\alpha$, slightly) steeper than $\lpar^{1/2}$, 
the weaker one's is exactly that, corresponding to a $\kpar^{-2}$ spectrum,
as is indeed seen in numerical simulations [\secref{sec:imb_num}, item (iv) 
and \figref{fig:mallet_imb}].  
This makes sense because the weaker field was assumed to have a local parallel cascade 
with the usual CB conjecture, so the standard arguments for its parallel spectrum 
given in \secref{sec:par_cascade} remain valid. 

\subsubsection{Pinning}
\label{sec:pinning}

It turns out that it is possible to determine $\alpha$ by considering 
what happens at the dissipation scale(s). 
The dissipation cutoffs $\lres^\pm$ for the two Elsasser fields 
can be worked out by balancing their fluxes with their dissipation rates: 
\beq
\label{eq:diss_balance}
\eps^\pm \sim \frac{\nu + \eta}{\lt(\lres^\pm\rt)^2}\lt(\dz^\pm_{\lres^\pm}\rt)^2.
\eeq
Using \exref{eq:dz_alpha} to work out the field amplitudes at $\lres^\pm$, one~gets
\beq
\frac{\lres^+}{\Lperp} \sim \lt(\frac{\eps^-}{\eps^+}\tRe\rt)^{-1.5/(3-\beta-\alpha)},
\qquad
\frac{\lres^-}{\Lperp} \sim \tRe^{-1.5/(3-\beta+2\alpha)},
\qquad
\tRe = \frac{\dz^+_{\Lperp}\Lperp}{\nu + \eta}, 
\label{eq:imb_lambda_eta}
\eeq
where, as before, $\tRe$ is the smaller of $\Re$ and $\Rm$. 
There are two possibilities: either $\lres^+ < \lres^-$ or $\lres^+ > \lres^-$. 
The first of these is, in fact, impossible: 
if the weaker field is cut off at $\lres^-$, there is nothing to cascade the stronger 
field at $\lambda < \lres^-$ (locally in $\lambda$, as I assumed in \secref{sec:2cascades}). 
The second possibility is $\lres^+ > \lres^-$. Since the weaker field is cascaded 
by the stronger one nonlocally, a self-consistent situation would be one 
in which $\lres^+$ were the scale $\lambda'$ corresponding to $\lambda=\lres^-$. 
Using~\exref{eq:lambdaprime}, we must therefore ``pin'' the dissipation 
scales together in the following way: 
\beq
\frac{\lres^+}{\Lperp}\sim\lt(\frac{\lres^-}{\Lperp}\rt)^{2(\beta-2\alpha)/(2\beta-\alpha)}
\hence 
1 - \frac{2(\beta-2\alpha)(3-\beta-\alpha)}{(2\beta-\alpha)(3-\beta+2\alpha)} 
= \frac{\ln(\eps^+/\eps^-)}{\ln\tRe}.
\eeq  
Assuming $\alpha\ll 1$, we get 
\beq
\alpha \approx \frac{2(3-\beta)\beta}{3(3+\beta)}\frac{\ln(\eps^+/\eps^-)}{\ln\tRe}.
\label{eq:imb_alpha}
\eeq 
Thus, indeed, $\alpha\to 0$ as $\tRe\to\infty$, but very slowly, with very 
large $\tRe$ needed to achieve a modicum of asymptoticity at larger imbalances.
For the record, this implies, from~\exref{eq:imb_lambda_eta}, 
\beq
\frac{\lres^+}{\Lperp} \sim 
\tRe^{-1.5/(3-\beta)}\lt(\frac{\eps^+}{\eps^-}\rt)^{0.5(9+\beta)/(9-\beta^2)},
\qquad
\frac{\lres^-}{\Lperp} \sim 
\tRe^{-1.5/(3-\beta)}\lt(\frac{\eps^+}{\eps^-}\rt)^{2\beta/(9-\beta^2)}.
\label{eq:imb_lambda_eta_spec}
\eeq
It actually does appear to be the case that $\lres^+>\lres^-$
in \figref{fig:beresnyak_imb}(a) \citep{beresnyak09}
and \figref{fig:meyrand_imb}(c) \citep{meyrand21}---perhaps the strongest 
evidence that we have of the nonlocality of imbalanced cascades.

While what I have proposed above is a kind of ``pinning,'' it is not 
the conventional ``pinning'' that means equating the amplitudes of the two fields 
to each other at the dissipation scale---one of the tenets of the theory 
of weak imbalanced turbulence (\secref{sec:emb_WTimb}). 
Indeed, \exref{eq:diss_balance}~implies that 
the ratio of the Elsasser amplitudes at their respective dissipation scales~is
\beq
\frac{\dz^+_{\lres^+}}{\dz^-_{\lres^-}} \sim 
\sqrt{\frac{\eps^+}{\eps^-}}\frac{\lres^+}{\lres^-}
\sim \lt(\frac{\eps^+}{\eps^-}\rt)^{(3+\beta/2)/(3+\beta)}, 
\eeq
where I have used \exref{eq:imb_lambda_eta_spec}, valid in the limit $\tRe\to\infty$. 
Notably, the amplitude ratio is independent of~$\tRe$ in this limit, but is not 
equal to $(\eps^+/\eps^-)^{1/2}$, as one might have concluded from \exref{eq:diss_balance}
for $\lres^+=\lres^-$ (as \citealt{beresnyak08} did).\\ 

Arguably, this is a rather attractive theory: 
asymptotically, the spectra are parallel, interactions 
are local, etc., but in any finite-width inertial range, there are finite-$\tRe$ 
logarithmic corrections to scalings, locality, etc., accounting for all 
of the distinctive features of imbalanced turbulence seen in non-asymptotic 
simulations (\secref{sec:imb_num}). 

\subsubsection{Alignment, Intermittency, Reconnection}

Like in balanced turbulence, alignment is likely related to intermittency in 
imbalanced turbulence as well. Since, for imbalanced turbulence, we are still 
litigating such basic things as spectra, 
there is not much we know about its intermittency---and I do not propose to engage with 
this topic here any more than I have done already 
with a few throw-away comments in~\secref{sec:Eimb}. 
The argument in \secref{sec:MS17} that led to $\beta=3/4$ depended on assumptions 
about the most intense structures being sheets and on the ``refined critical balance'' 
\citep[][see \figref{fig:rcb}]{mallet15}. 
It seems a worthwhile project to check whether, and in what sense, 
these features survive in imbalanced turbulence.  
 
Since reconnection playing an important role at the small-scale end of the 
inertial range depended on alignment, the equivalent of \secref{sec:disruption} 
for imbalanced turbulence must wait for a better understanding of alignment. 
If tearing disruption does occur at some scale in (strongly) imbalanced turbulence, 
the pinning scheme proposed in \secref{sec:pinning} has to be redesigned. 
Incidentally, it also has to be redesigned (according to \citealt{meyrand21}, 
redesigned quite dramatically) for natural plasmas like the 
solar wind, where the cutoff of the RMHD inertial range is accomplished by kinetic 
effects rather than by Laplacian viscosity---but these matters are outside 
the scope of this review (see \secsand{sec:microphysics}{sec:failed_cascade}). 

\section{Residual Energy in MHD Turbulence}
\label{sec:residual}

\subsection{Observational and Numerical Evidence}
\label{sec:res_evidence}

\begin{figure}
\begin{center}
\begin{tabular}{cc}
\parbox{0.45\textwidth}{\includegraphics[width=0.45\textwidth]{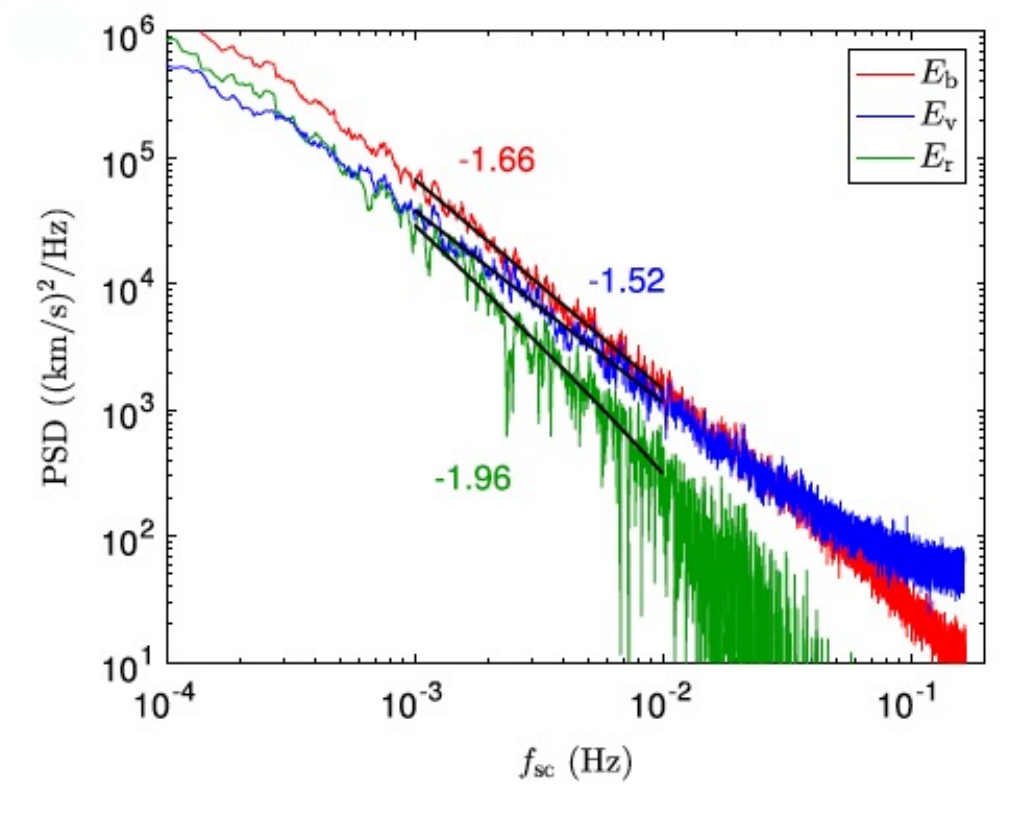}} &
\parbox{0.53\textwidth}{\includegraphics[width=0.53\textwidth]{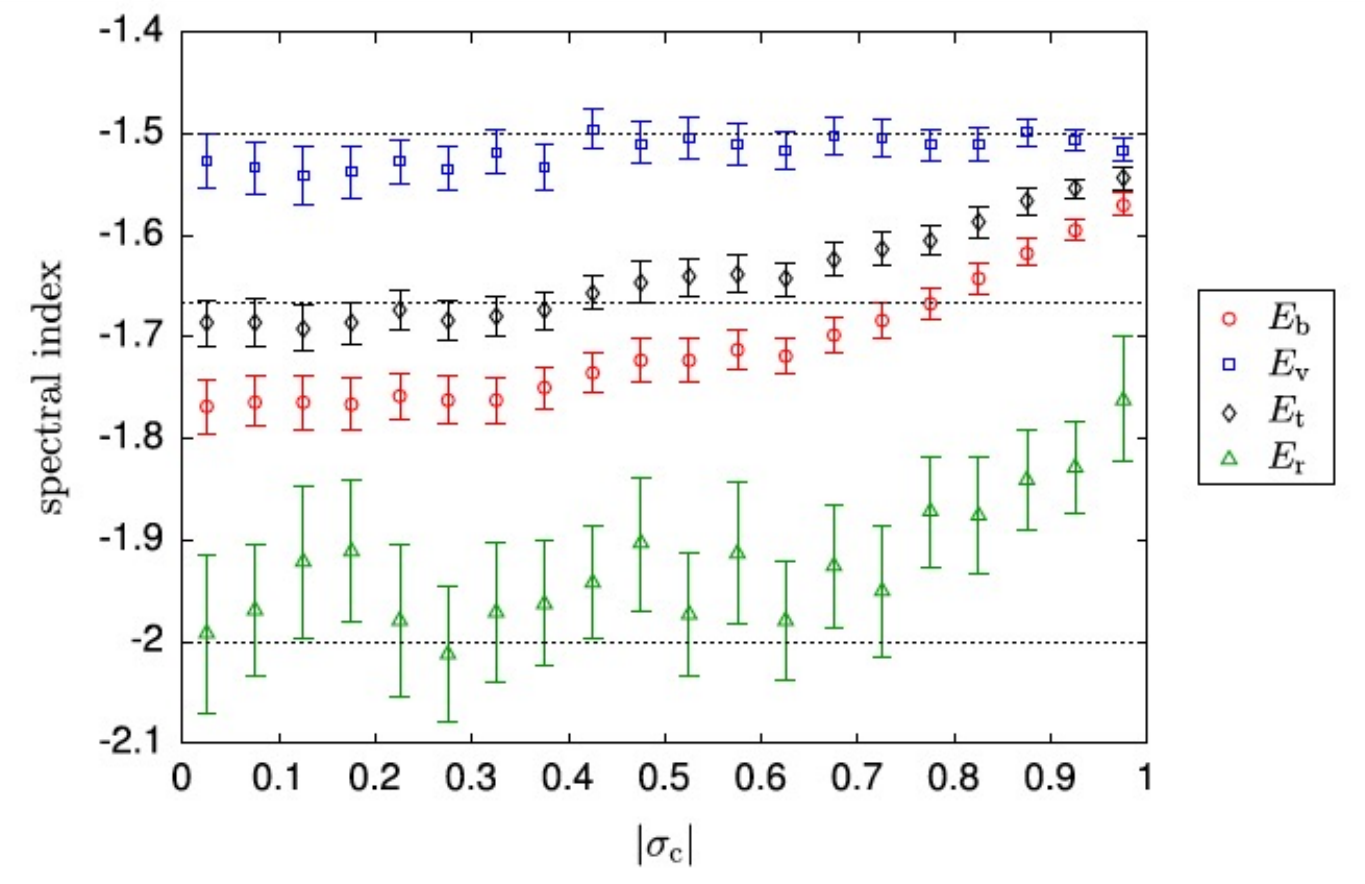}}\\
(a) & (b)
\end{tabular}
\end{center}
\caption{Spectra of magnetic (red), kinetic (blue), total (black) and residual (green) 
energies measured by \citet{chen13res} (figures taken from \citealt{chen16}): (a) typical 
spectra; (b) average spectral indices vs.\ normalised cross-helicity $\sigc$ 
[defined in \exref{eq:sig_def}].}
\label{fig:chen_spectra}
\end{figure}

\begin{figure}
\centerline{\includegraphics[width=0.7\textwidth]{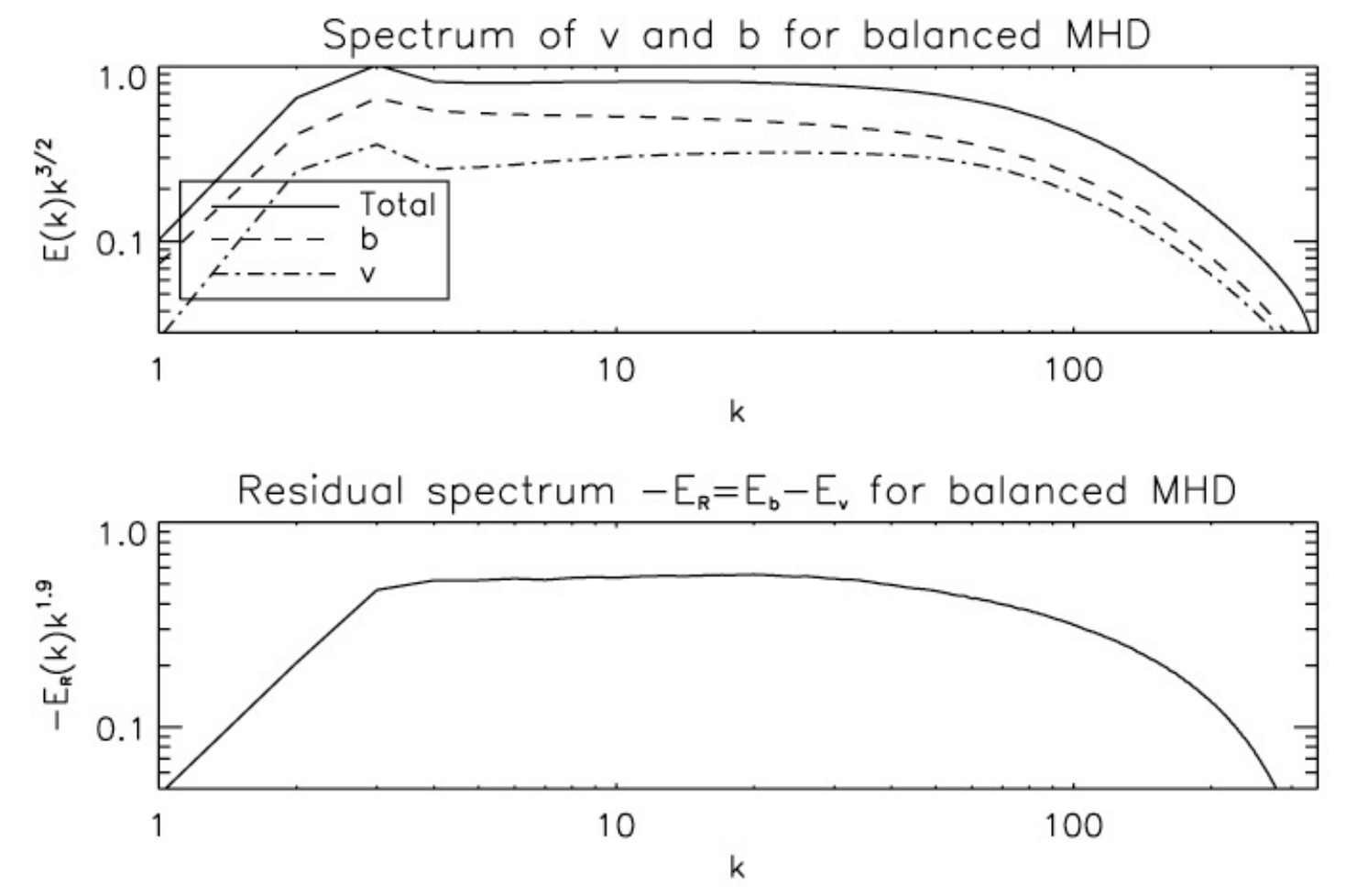}} 
\caption{Spectra of total, magnetic, kinetic (upper panel, solid, dashed and dot-dashed 
lines, respectively, compensated by $\kperp^{3/2}$) and residual energy (lower panel, 
compensated by $\kperp^{1.9}$) in an RMHD simulation by \citet{boldyrev11}
(\copyright AAS, reproduced with permission). \citet{beresnyak14}
does a convergence study by rescaling to the dissipative cutoff and finds instead
a $\kperp^{-1.7}$ scaling to be the best fit to the residual-energy spectrum (see his Figure~3),
so it is possible that what is displayed here is a large-scale transient.} 
\label{fig:boldyrev_res}
\end{figure}

Going back to \figref{fig:wicks}(a), we see that real MHD turbulence observed 
in the solar wind is distributed between cases with a local Elsasser imbalance (cross-helicity) 
and those with an Alfv\'enic one---the latter in favour of the magnetic field. Thus, 
the imbalanced cascades are only half of the story. 
According to the second relation in \exref{eq:imb_limit}, in imbalanced 
turbulence ($|\sigc|\approx1$), it is a geometric inevitability that $|\sigr|\ll1$, 
as illustrated by \figref{fig:wicks}(a) and 
confirmed directly in the statistical study of solar-wind data by \citet{bowen18}. 
In contrast, when the cross-helicity is not large (i.e., when $\sigc$ is not close to~$\pm1$), 
there is flexibility for the perturbations to have finite residual energy: in the event, $\sigr<0$. 
The definitive observational paper on this is \citet{chen13res}, 
confirming negative $\sigr$ over a large data set 
obtained in the solar wind. They also report that residual energy has a spectrum 
consistent with $\kperp^{-2}$ or perhaps a little shallower, but certainly steeper 
than either the kinetic- or magnetic-energy spectra: the scalings 
of all three are reproduced in \figref{fig:chen_spectra}. This seems to be 
in agreement with earlier observational and numerical evidence 
\citep[][\figref{fig:boldyrev_res}]{mueller05,boldyrev11}. 

There are two ways in which a shade of legitimate doubt extends over both
numerical and observational evidence quoted above.  
\vskip2mm
(i) \citet{beresnyak14}, analysing his (largest-ever) simulations, reports that  
the residual energy at the small-scale end of its spectrum scales approximately
as $\kperp^{-1.7}$, same as his kinetic- and magnetic-energy spectra. His conclusion
is that residual energy is merely a scale-independent finite fraction $\approx0.15$
of the total energy.
\vskip2mm
(ii) Solar wind's expansion with heliocentric distance and the resulting
reflection of the outward-propagating Elsasser field leads to an increase in the negative
residual energy, which has nothing to do with nonlinear interactions in the locally
homogeneous turbulence \citep[see, e.g.,][]{perez13}. The steepening of the magnetic field's
spectrum compared to velocity's due to this effect is captured quite clearly in, e.g.,
the expanding-box simulations of \citet{squire20}, to whose paper I also refer the reader
requiring further space-physics references on this subject. In assessing 
solar-wind measurements for evidence of residual energy in MHD turbulence, it may be
nontrivial to separate this expansion effect from the organic local tendency of the nonlinear
interactions to favour negative residual energy. 
\vskip2mm
Alas, nothing is ever clear-cut in this subject, but let me press on. 
Obviously, it cannot be true at asymptotically small scales that, as the solar-wind data suggests, 
the magnetic- and kinetic-energy spectra scale as $\kperp^{-5/3}$ and $\kperp^{-3/2}$, 
respectively, while 
their difference scales as $\kperp^{-2}$---the $\vb$ and $\vu$ spectra must meet somewhere, 
as they indeed do in \figref{fig:chen_spectra}(a). The residual energy appears to peter 
out at the same scale (although that is also where the noise effects kick in)---but 
it would not be asymptotically impossible for it 
to retain the $\kperp^{-2}$ scaling as a subdominant correction to 
approximately equipartitioned $\vb$ and $\vu$ spectra (as suggested 
by \citealt{boldyrev11})---or perhaps to parallel, finitely offset ones (\`a la \citealt{beresnyak14}).
I will discuss a plausible origin for such a correction 
in \secref{sec:new_res_theory}, but first some history. 

\subsection{Old Theories}
\label{sec:old_res_theory}

The first awakening of the MHD turbulence community to the turbulence's 
tendency for residual-energy generation dates back to the dawn of time 
\citep{pouquet76,grappin82,grappin83}, when theories and simulations 
based on isotropic EDQNM\footnote{Eddy-Damped Quasi-Normal Markovian. 
You don't want to know.} closure models of MHD turbulence predicted 
a negative residual energy (i.e., an excess of magnetic energy) scaling 
as a $k^{-2}$ correction to the dominant $k^{-3/2}$ IK spectrum (see \secref{sec:IK}). 
While the isotropic IK theory certainly cannot be relevant 
to MHD turbulence with a strong mean field (see \secref{sec:prehistory}), 
the modern evidence (\secref{sec:res_evidence})
looks very much like those old results, with $k$ replaced by $\kperp$. 
This led \citet{mueller05} to claim a degree of vindication for 
the EDQNM-based theory. This vindication cannot, however, be 
any stronger than the vindication of IK provided by Boldyrev's theory 
(\secref{sec:B06}) and its variants (\secref{sec:revised}): 
same scaling, different physics. 

Below the turgid layers of EDQNM formalism, the basic physical idea 
(best summarised by \citealt{grappin16}) is that  
residual energy is generated from the total energy by nonlinear interactions 
that favour magnetic-field production (the ``dynamo effect'')\footnote{That they 
do favour magnetic-field production and thus promote $\sigr<1$ is confirmed 
quantitatively within the closure theory \citep{grappin82,grappin83,gogoberidze12}. 
Physically, it is possible to argue that simple Alfv\'en-wave interactions will produce 
residual energy \citep{boldyrev12res}---see further discussion in 
\secsand{sec:new_res_theory_WT}{sec:new_res_theory}. 
I am not enthusiastic about dragging the dynamo effect into this.} 
and removed by the ``Alfv\'en effect,'' which tends to equalise 
$\vuperp$ and $\vbperp$ perturbations. A balance of these two effects 
leads to a prediction for the residual-energy spectrum in the form 
\beq
\Eres \sim \frac{\tA}{\tb} E
\sim \lt(\frac{\tA}{\tnl}\rt)^\alpha E,
\label{eq:Eres_dynamo}
\eeq
where $E$ is the total-energy spectrum, 
$\tb$ is the characteristic time scale of the generation of excess 
magnetic energy at a given scale, $\tA$ and $\tnl$ are our 
old friends Alfv\'en and nonlinear times, and the exponent $\alpha$ 
depends on one's theory of how $\tb$ is related to these two basic times. 
For example, in the IK theory, $\tb\sim\tnl^2/\tA$ 
[because IK turbulence is weak; cf.~\exref{eq:tc_weak} and footnote~\ref{fn:IK}], 
so $\alpha=2$. Using the IK scalings \exref{eq:IK} and $\tA/\tnl \sim \du_\lambda/\vA$, 
one then gets from~\exref{eq:Eres_dynamo} 
\beq
\Eres(k) \sim \frac{\eps}{\vA} k^{-2}.
\label{eq:Eres_IK}
\eeq
I know of no unique or obvious way of adjusting this promising 
(but necessarily wrong because IK-based) result to fit a critically balanced cascade: 
indeed, the CB requires $\tA\sim\tnl$, implying $\Eres\sim E$, i.e., 
a scale-independent ratio between the residual and total energy 
(this was also the conclusion of \citealt{gogoberidze12}, 
who undertook the heroic but thankless task of constructing an EDQNM theory of 
anisotropic, critically balanced MHD turbulence). Neither 
solar wind nor MHD simulations appear to agree with this (\secref{sec:res_evidence}). 

Obviously, once we enter the realm of intermittent scalings 
of the kind described in~\secref{sec:sym}, i.e., allow the outer 
scale to matter, there is a whole family of possibilities 
admitted by the RMHD symmetry and dimensional analysis: 
by exactly the same argument as the one that led to \exref{eq:dz_delta}
(and noting that spectrum~$\sim$~amplitude$^2/\kperp$), we must have 
\beq
\Eres(\kperp) \sim 
\eps^{2(1+\delta)/3}\lt(\frac{\Lpar}{\vA}\rt)^{2\delta}\kperp^{-(5-4\delta)/3},
\label{eq:Eres_parametrised}
\eeq
where $\delta$ is some new exponent. In order to determine it, one must input 
some physical or mathematical insight.

\subsection{New Theories: Residual Energy in Weak MHD Turbulence}
\label{sec:new_res_theory_WT}

An interesting step in this direction was made in yet another characteristically 
clever contribution by Boldyrev's group. They showed that 
even weak interactions of AW packets mathematically lead to growth of excess magnetic 
energy and thus of negative residual energy---\citet{boldyrev12res} by analysing 
weak interaction of two model AW packets and \citet{wang11} within 
the framework of traditional WT theory. However, all the action 
in their derivation was in the $\kpar=0$ modes, which hosted the 
excess magnetic energy generated by AW interactions---the 2D magnetic condensate 
whose awkward relationship with WT theory I discussed in \secref{sec:WT_zeromodes}.  

A version of the appropriate derivation is laid out in \apref{app:WT_res}. 
Quantitatively, it cannot be right because the WT approximation does 
not apply to the condensate, which is strongly turbulent (see \apref{app:bband}). 
Qualitatively, the outcome of the WT calculation---growth of excess magnetic 
energy at $\kpar=0$---can be understood as follows. 
Growth of positive (negative) residual energy is the same as growth 
of (anti)correlation between $\vzperp^+$ and $\vzperp^-$: 
\beq
\la\vzperp^+\cdot\vzperp^-\ra = \la|\vuperp|^2\ra - \la|\vbperp|^2\ra.
\eeq  
These correlations are created with particular ease at $\kpar=0$, 
where $\vz_0^+$ and $\vz_0^-$ are forced by the interaction of the 
same pairs of AWs, $\vz_{\kpar}^+$ and $\vz_{\kpar}^-$ 
(which themselves are allowed to be uncorrelated): 
this is obvious from \exref{eq:z0}. The result is that a magnetic condensate 
emerges at $\kpar=0$, giving rise to net negative residual energy---that it should be 
negative is not obvious, but the WT calculation says it is [see~\exref{eq:WT_Eres_scaling}], 
as, perhaps more convincingly, does a qualitative argument that I shall now explain.  

\subsection{New Theories: Residual Energy in Strong MHD Turbulence}
\label{sec:new_res_theory}

In the strong-turbulence regime, no quantitative calculation exists, as usual, but 
a reasonably compelling physical case can be made. 

Emergence of negative residual energy here must be discussed in very different terms than 
in \secref{sec:new_res_theory_WT}. 
As I repeatedly stated in \secref{sec:DA}, my preferred picture of alignment is one 
in which Elsasser fields dynamically shear each other into intermittent structures 
where they are nearly parallel to each other \citep{chandran15}. 
That, of course, means that they become strongly correlated: 
indeed, alignment between Elsasser fields is mathematically impossible 
without non-zero residual energy, as is obvious from the first formula 
in \exref{eq:theta_sig} or from \figref{fig:geometry}. 
That $\db$ should be larger than $\du$ in the 
resulting sheet-like structures is both a selection effect 
and the result of dynamics. 

First, the structures that have 
$\du>\db$---shear layers, rather than current sheets---are prone to 
be destroyed by the Kelvin--Helmholtz instability and to curl up into vortices, 
as they do in hydrodynamic turbulence, whereas in the current sheets, 
the instability is happily stabilised by the magnetic field (at least 
before it all hits the disruption scale and current sheets become 
unstable as well; see \secref{sec:disruption}, 
\apref{app:loureiro} and references therein). 

\begin{figure}
\centerline{\includegraphics[width=0.55\textwidth]{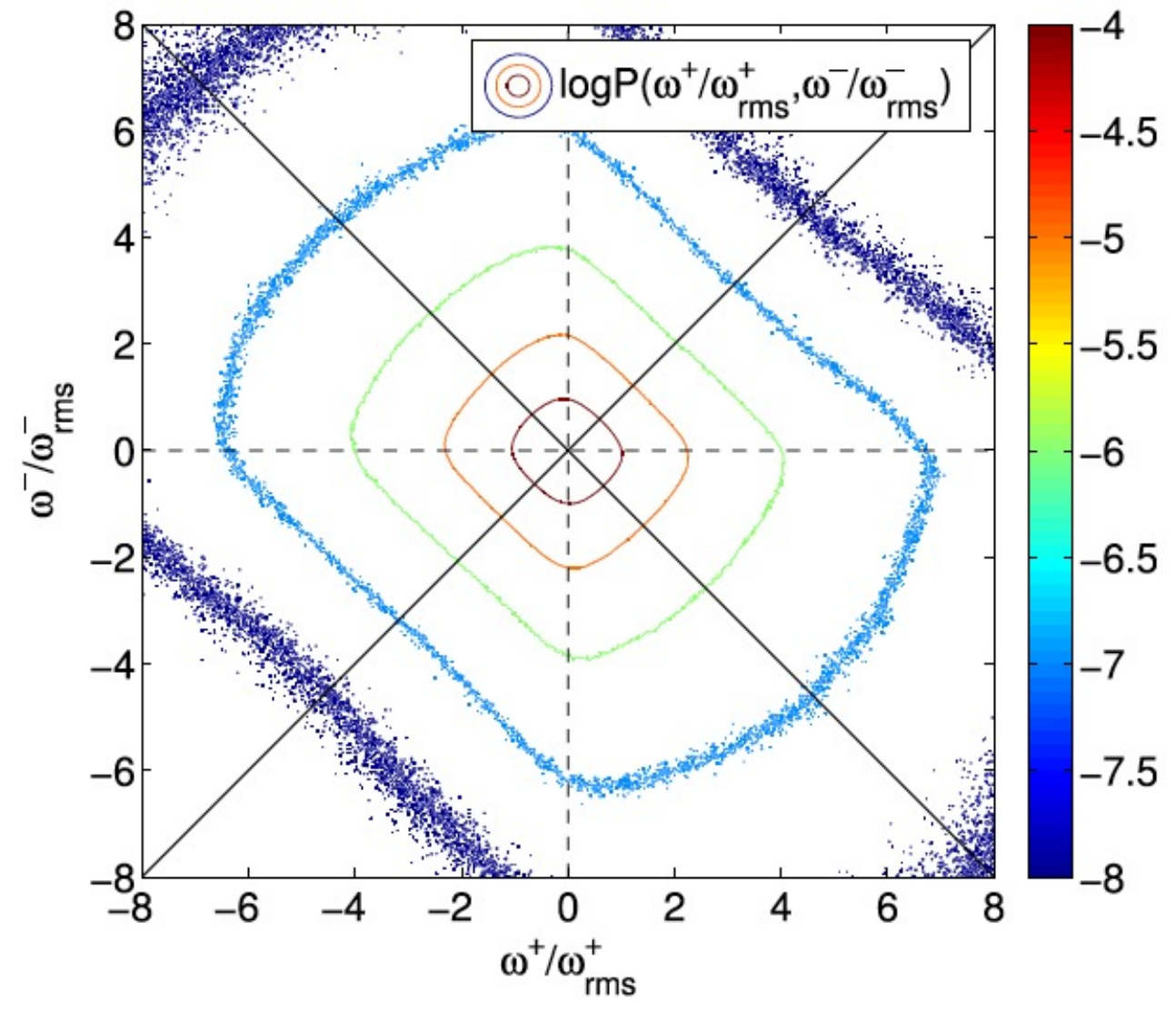}} 
\caption{Joint probability distribution of Elsasser vorticities 
$\omega^\pm = \ez\cdot(\vdperp\times\vzperp^\pm)$ in an RMHD simulation 
by \citet{zhdankin16} [reprinted from \citet{zhdankin16}
with the permission of AIP Publishing]. The contours 
are elongated in the SE-NW direction, indicating $\la\omega^+\omega^-\ra<0$ 
and thus a preponderance of current sheets over shear layers.}
\label{fig:zhdankin_vorticities}
\end{figure}

Secondly, there is a dynamical tendency in RMHD that favours current 
sheets over shear layers: the nonlinearity pushes the ``Elsasser vorticities'' 
$\omega^+ = \ez\cdot(\vdperp\times\vzperp^+)$ and 
$\omega^- = \ez\cdot(\vdperp\times\vzperp^-)$ in opposite directions---this 
is obvious from the evolution equations \exref{eq:zeta} 
for these vorticities, where the nonlinear vortex-stretching terms  
have opposite signs for the two Elsasser fields. 
Thus, $\omega^+$ and $\omega^-$ are ``forced'' equally and in opposite 
directions at every point in space and time. 
The result is a negative correlation between them, 
$\la\omega^+\omega^-\ra<0$, and thus 
a preference for current sheets over shear layers 
(see \figref{fig:zhdankin_vorticities} and \citealt{zhdankin16}). 

Let us now imagine that this effect is strongest in the most intense structures, 
which in \secref{sec:MS17} were all assumed to have the same, scale-independent 
amplitude. If they are current sheets with $\dzmax\sim\db\gg\du$, they would, 
if they were alone in the world, have a spectrum of $\kperp^{-2}$ because they 
are just an ensemble of step functions in $\vbperp$. 
The easiest way to see this is to notice that it is the spectrum of a single 
Heaviside step function. It is also the spectrum of many random steps: 
if the field flips direction randomly, with the number of flips between two points 
separated by a distance $\lambda$ increasing $\propto\lambda$, then 
the field increment will accumulate as a random walk: 
$\la\db_\lambda^2\ra \propto \lambda$, giving a $\kperp^{-2}$ spectrum.\footnote{That 
current sheets naturally forming in a turbulent MHD system do indeed have this spectrum 
was shown by \citet{dallas13ksq,dallas14}, although they only looked at decaying, 
no-mean-field MHD turbulence with a certain class of initial conditions. 
\citet{zhou19} in their decaying 2D RMHD simulations, 
heavily dominated by current sheets, see the same spectrum and explain 
it the same way. Note that if current sheets break up into plasmoid chains, those 
too appear to favour a $\kperp^{-2}$ scaling (see \apref{app:chain_spectrum}).} 
In fact, there are many 
other fluctuations around, whose net spectrum is $\kperp^{-3/2}$ and in which  
$\db\sim\du$. Overall, this shallower scaling would swamp $\kperp^{-2}$. 
However, if the excess magnetic energy is dominated by the most intense 
sheets, one might imagine that the residual-energy spectrum 
would have a $\kperp^{-2}$ scaling. 

Encouragingly, recent analysis of solar-wind data by \citet{bowen18} directly established 
a positive correlation between the most intense, intermittent magnetic structures 
and residual energy. It is perhaps worth observing that 
the fact that a $\kperp^{-2}$ spectrum consistent with the RMHD symmetry 
requires $\delta=-1/4$ in \exref{eq:Eres_parametrised}
and, therefore, the presence of the outer scale $\Lpar$ in the expression 
for $\Eres(\kperp)$, confirms that we must again be dealing with an intermittency effect. 
This said, \figref{fig:intcy}(b) 
appears to be at odds with the notion of a steeper magnetic spectrum, 
showing {\em smaller} scaling exponents for the magnetic field than for velocity---an 
effect of velocity forcing? bad statistical measurement? Sorting this out 
appears to be a worthwhile outstanding task. 
 
Thus, admittedly, all this is less than a theory, but it is something. 

\subsection{Summary}

Perhaps speaking of an ``Alfv\'enically imbalanced regime'' of MHD turbulence 
is misleading. Residual energy is not an RMHD invariant, so this is not something 
that can be viewed as a parameter in the same way as the net Elsasser imbalance can be. 
It is, rather, what appears to be a feature of any (approximately 
balanced) MHD turbulent state (but may be a large-scale transient; see \secref{sec:res_evidence}). 

This feature has so far presented itself in two seemingly distinct manifestations. 
The first one is the tendency for 
sheet-like structures in the inertial range of strong MHD turbulence to be current 
sheets rather than shear layers and thus to have an excess of magnetic energy---it may 
be possible to argue that the most extreme of these structures are responsible
for a subdominant $\kperp^{-2}$ spectrum of residual energy (\secref{sec:new_res_theory}).
The second one is the emergence of a 2D magnetic condensate in weak MHD turbulence 
(\secref{sec:new_res_theory_WT}). 

Are these two different phenomena? Not necessarily: in the WT context, all 
the residual energy is generated amongst $\kpar\approx0$ modes, which are, in fact, 
strongly turbulent (see \apsand{app:bband}{app:WT_res}). Being strongly 
turbulent, this 2D condensate is strongly intermittent and appears to be dominated by 
sheet-like structures \citep{meyrand15}, so the physical mechanism whereby 
an excess of magnetic energy develops in it is likely to be the same as in strong 
MHD turbulence. 

\section{Subviscous MHD Turbulence}
\label{sec:subvisc}

Let me now turn to an interesting, if somewhat boutique, regime of MHD turbulence that 
occurs at scales below the viscous cutoff when $\Pm\gg1$. This was first studied 
by \citet{cho02visc,cho03} and \citet{lazarian04}, and recently picked up again by 
\citet{xulaz16,xulaz17njp}, on the grounds that it is relevant to partially ionised 
interstellar medium, where viscosity is heavily dominated by the neutral 
atoms.\footnote{It has also recently turned out, somewhat unexpectedly, 
that something very similar to this regime might be relevant 
in the context of collisionless gyrokinetic turbulence and ion heating in high-beta 
plasmas \citep{kawazura19}.}
This is a limit in which viscous dissipation 
takes over from inertia in controlling the evolution of the velocity field 
(one might call this ``Stokes,'' or ``Aristotelian'' dynamics; cf.~\citealt{rovelli15}), 
while magnetic field is still happily frozen into this viscous flow and free 
to have interesting 
MHD behaviour all the way down to the resistive scale, which, at $\Pm\gg1$, 
is much smaller than the viscous one. The velocity perturbations below the 
viscous scale will be very small compared to the magnetic ones, so 
this is another MHD turbulent state that features an imbalance between the two fields. 

Below, I am going to present a somewhat updated qualitative theory of the 
subviscous cascade---with tearing disruption and the ubiquitous Kolmogorov cutoff 
yet again making a cameo appearance.   

\subsection{Viscous Cutoff}
\label{sec:visc_cutoff}

When $\Pm\gg1$, there are two possibilities for the nature of turbulence 
at the viscous cutoff. 

The first is that $\Pm$ is large enough 
to break the condition \exref{eq:Pm_max}, viz., $\Pm\gtrsim\Re^{1/9}$, so there is 
no tearing disruption and the (aligned) inertial-range MHD cascade 
encounters viscosity at the Boldyrev cutoff scale \exref{eq:lres_Rm}---for $\Pm\gg1$, 
let me rename~it~$\lvisc$: 
\beq
\lvisc \sim \lCB \Re^{-2/3} \sim \frac{\nu^{3/4}}{\eps^{1/4}}\,\Re^{1/12}
\hence 
\frac{\xivisc}{\lCB} \sim \Re^{-1/2},
\quad
\frac{\lparvisc}{\Lpar} \sim \Re^{-1/3},
\label{eq:lvisc_aligned}
\eeq 
where $\lCB$ is given by \exref{eq:Lperp} and $\Re$ by \exref{eq:RmCB_def}. 
The last two formulae follow via~\exref{eq:xi_scalings} (for the scale 
$\xivisc$ on which the perturbed fields vary along themselves)
and~\exref{eq:MS_normalised} (for the parallel scale~$\lparvisc$).

The second (rather difficult to achieve) possibility is that $1\ll\Pm\ll\Re^{1/9}$, so 
\exref{eq:Pm_max} does hold and we have a tearing-mediated turbulent cascade 
curtailed by the Kolmogorov cutoff~\exref{eq:K_cutoff}---for $\Pm\gg1$, it~is
\beq
\lvisc \sim \lCB\Re^{-3/4} \sim \frac{\nu^{3/4}}{\eps^{1/4}}. 
\label{eq:lvisc_rec}
\eeq

Either way, some finite fraction of $\eps$ is thermalised at $\lvisc$, 
and at $\lambda < \lvisc$ velocity perturbations will have gradients that are 
smaller than the decorrelation rate at $\lvisc$. This decorrelation rate~is
\beq
\tnl^{-1} \sim \frac{\du_{\lvisc}}{\xivisc} 
\sim \tvisc^{-1} \sim \frac{\nu}{\lvisc^2} 
\sim \lt(\frac{\eps}{\nu}\rt)^{1/2}\Re^{-1/6}
\rmor
\lt(\frac{\eps}{\nu}\rt)^{1/2}
\label{eq:tvisc}
\eeq
for \exref{eq:lvisc_aligned} and \exref{eq:lvisc_rec}, respectively. 

\subsection{Magnetic Fields at Subviscous Scales}
\label{sec:mag_subvisc}

In contrast to velocities, magnetic fields are immune to viscosity and 
so can be pushed to scales much smaller than $\lvisc$. However, since 
velocity gradients are suppressed at these scales, these magnetic fields 
will be dominantly interacting with the viscous-scale velocities, 
in a scale-nonlocal fashion. Presumably, since the viscous-scale motions are correlated 
on the parallel scale $\lparvisc$, so will be these magnetic fields, i.e., 
there is no parallel cascade: 
\beq
\lpar \sim \lparvisc = \const.
\label{eq:lpar_subvisc}
\eeq 

\begin{figure}
\begin{center}
\begin{tabular}{cc}
\parbox{0.52\textwidth}{
\includegraphics[width=0.52\textwidth]{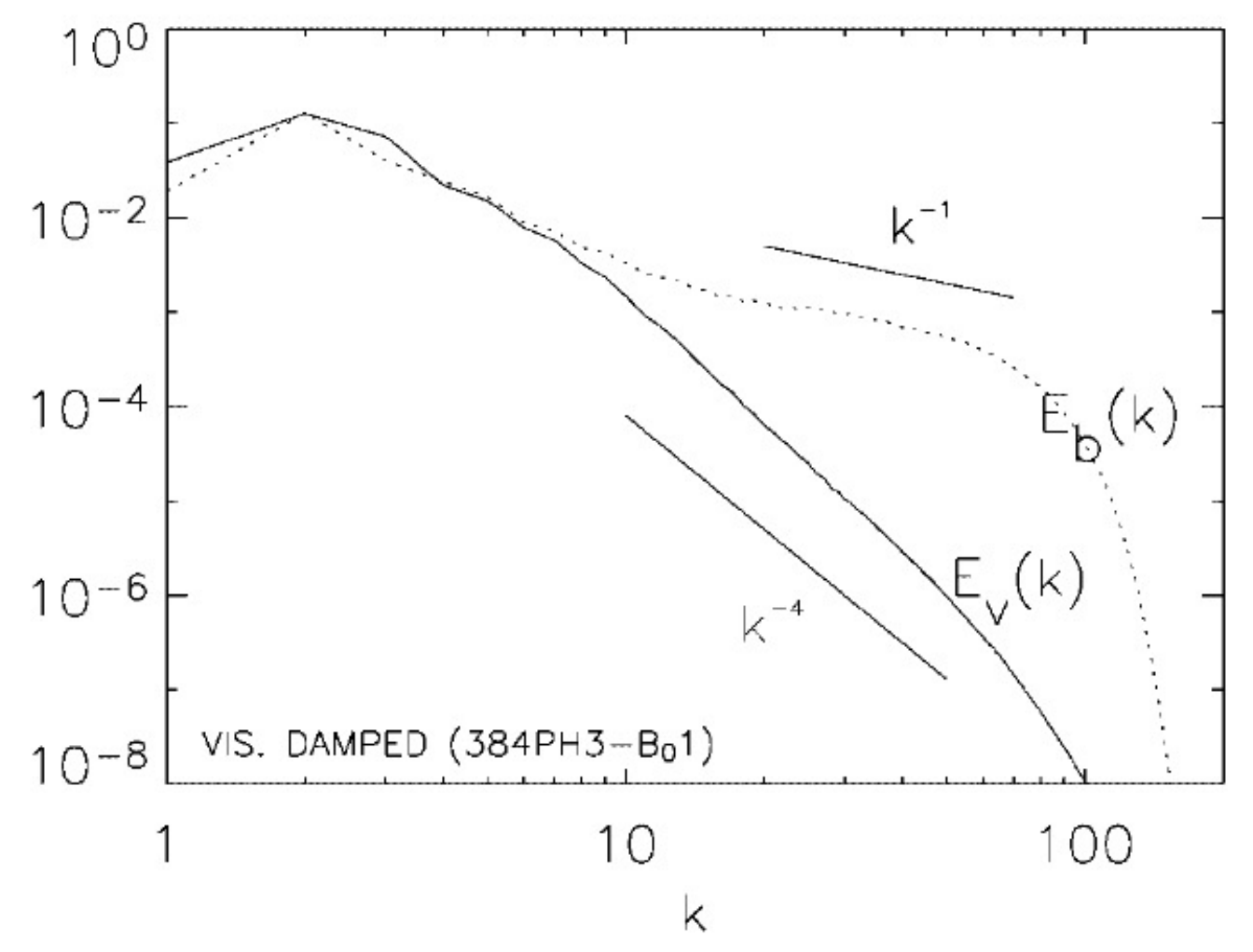}\vskip8mm} & 
\parbox{0.46\textwidth}{
\includegraphics[width=0.46\textwidth]{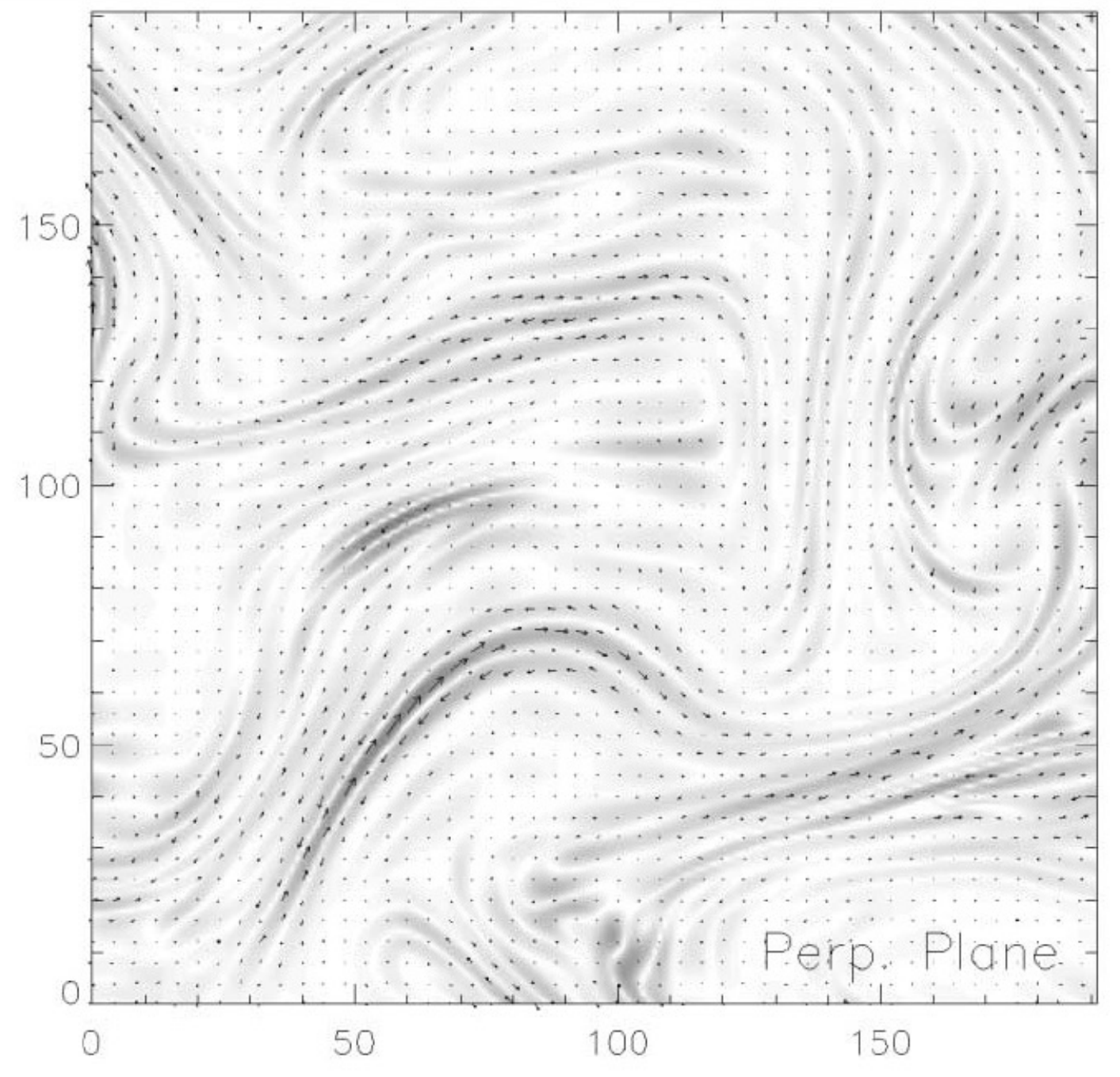}}\\
(a) & (b)
\end{tabular}
\end{center}
\caption{(a) Spectra of magnetic and kinetic energy for subviscous 
turbulence, taken from \citet{cho03}. (b) Magnetic-field strength 
for the filtered $k>20$ part of the field in the same simulation 
(from \citealt{cho02visc}; both figures \copyright AAS, reproduced with permission).
Stripy field structure is manifest.}
\label{fig:cho}
\end{figure}

Numerical simulations \citep{cho02visc,cho03} confirm \exref{eq:lpar_subvisc} 
and show a magnetic spectrum $\propto k^{-1}$. In the mind of any minimally erudite 
turbulence theorist, this cannot fail to trigger a strong 
temptation to consider the whole situation as a variant of \citet{batchelor59} 
advection of a passive field: assuming a cascade of magnetic energy 
with cascade time $\tvisc$ at every scale, one gets (see \citealt{cho02visc} 
and \figref{fig:cho}a):
\beq
\db_\lambda^2 \sim \epsm\tvisc = \const 
\hence E_b(\kperp) \sim \epsm\tvisc \kperp^{-1},
\label{eq:Eb_subvisc}
\eeq   
where $\epsm$ is the part of the turbulent flux that is 
not dissipated at the viscous cutoff (possibly about half of it, 
since velocity and magnetic fields have the same energies at the 
viscous scale and are pushed into viscous dissipation and 
subviscous structure, respectively, at the same rate $\tvisc^{-1}$). 
The spectrum \exref{eq:Eb_subvisc} stretches all the way to the resistive 
scale, where Ohmic dissipation can rival advection: 
\beq
\tvisc^{-1} \sim \frac{\eta}{\lres^2}
\hence 
\lres \sim (\tvisc\eta)^{1/2} \sim \lvisc\Pm^{-1/2}. 
\label{eq:lres_subvisc}
\eeq

The line of reasoning leading to \exref{eq:Eb_subvisc}
should perhaps be viewed with a degree of suspicion.  
In a regime where magnetic fields are nonlocally advected and stretched 
by the viscous-scale velocity field, while the latter experiences 
back reaction from them while constantly being dissipated by viscosity, 
why can one assume that magnetic energy is an independent invariant with 
a constant scale-to-scale flux? While this may be a plausible proposition, 
I do not know how to justify it beyond reasonable doubt---but I do believe 
the scaling \exref{eq:Eb_subvisc} because it is bolstered by the following 
alternative argument of a more dynamical nature.   

The situation at subviscous scales is not entirely dissimilar to a kind of dynamo 
(\secref{sec:dynamo}), 
or rather a 2D version of it in which the perturbed magnetic 
field $\vbperp$ is randomly stretched and sheared by the viscous-scale velocity  
and is excused from the 2D antidynamo theorem \citep{zeldovich56} 
by constant resupply from the inertial range. The the role of $\vB_0$ 
is just to two-dimensionalise the dynamics approximately---maintaining 
all fields at the single parallel correlation scale $\lparvisc$. 
The stretching and shearing of $\vbperp$ leads to a folded magnetic field (\figref{fig:stretch})
forming a stripy pattern, with multiple reversals on small scales 
limited from below only by~$\lres$ (\figref{fig:cho}b).\footnote{
Subviscous-scale fields generated by 
randomly stirred and viscously damped flows in 2D were studied both 
analytically and numerically by \citet{kinney00}, who found them  
to follow a $\kperp^{-1}$ spectrum (which is evident in their Fig.~11, 
even though they do not claim this scaling explicitly).}

Just like in the case of dynamo-generated fields (\secref{sec:kin_dyn}), 
our stripy fields are spatially correlated 
along themselves on scales $\sim\xivisc$ and so can exert coherent Lorentz forces 
back on the viscous-scale velocity field. These forces are tension forces 
consisting of two parts: 
\beq
\vF = \vB_0\cdot\vdel\vbperp + \vbperp\cdot\vdel\vbperp. 
\label{eq:tension_subvisc}
\eeq
Here let us think of $\vbperp$ as just the part of $\vB$ that contains 
subviscous-scale variation and absorb into $\vB_0$ all inertial-range 
fields. The first term in \exref{eq:tension_subvisc} alternates sign 
on the scale $\lres$ (in the direction perpendicular to $\vbperp$) and 
so its effect on the viscous-scale motions should cancel out. 
In contrast, the second term is quadratic in $\vbperp$, 
and its size is $\sim \bperp^2/\xivisc$. In order to be dynamically significant, 
it must be of the same order as the viscous and inertial forces, which 
are similar at the viscous scale: 
\beq
\vbperp\cdot\vdel\vbperp \sim \nu\dperp^2\vuperp \sim \vuperp\cdot\vdel\vuperp
\hence 
\frac{\db_\lambda^2}{\xivisc} \sim \frac{\du_{\lvisc}^2}{\xivisc}
\hence 
\db_\lambda^2 \sim \du_{\lvisc}^2 \sim \eps\tvisc. 
\eeq 
On the face of it, this reproduces \exref{eq:Eb_subvisc} (assuming $\epsm\sim\eps$). 
However, we need not interpret this result as specifically vindicating 
a Batchelor-style cascade. Instead, we could think of the reversal scale 
as always being $\lres$, the size of the reversing field as being 
$\bperp\sim (\eps\tvisc)^{1/2}$,  
and interpret $\db_\lambda$ as the increment 
of a stripy field taken in two points separated by $\lres\ll\lambda\ll\lvisc$. 
The field difference between such two points will always be either 
$\db_\lambda \sim 2\bperp$ or zero, with equal probabilities, 
and so $\la\db_\lambda^2\ra \sim \bperp^2\sim \eps\tvisc$ (this argument is due to 
\citealt{yousef07}, who used it to posit a $k^{-1}$ spectrum for 
dynamo-generated fields at large $\Pm$, which will be visited in~\secref{sec:dynamo_spectra}). 
In other words, cascade or no cascade, $\kperp^{-1}$ can be recovered 
as the spectrum of sharp, repeated stripes.\footnote{To pre-empt a possible 
confusion, let me contrast this with the $\kperp^{-2}$ spectrum that is usually 
associated with a field consisting of sharp discontinuities, e.g., 
the Burgers turbulence of shocks \citep{bec07} or 
an ensemble of current sheets, already discussed in \secref{sec:new_res_theory}. 
I argued there that, in a field of random step-like discontinuities, 
their structure function would accumulate as a random walk: 
$\la\db_\lambda^2\ra \propto \lambda$. 
This is different from the stripy fields posited in this section, 
which are a repeated pattern, giving $\la\db_\lambda^2\ra\sim\const$.} 

\subsection{Velocity Field at Subviscous Scales}
\label{sec:vel_subvisc}

Numerical simulations \citep[][shown in \figref{fig:cho}a]{cho03} reveal that the velocity field 
at subviscous scales is very small and has an approximately $\kperp^{-4}$ 
spectrum. This can be recovered on the basis of the picture that I 
proposed in \secref{sec:mag_subvisc}, in the following way. 
The balance between the viscous and magnetic forces at $\kperp\lvisc\gg1$ 
gives us
\beq
\nu \kperp^2 \vu_{\perp\vk} \sim (\vbperp\cdot\vdel\vbperp)_{\vk}
\hence 
E_u(\kperp) \sim \frac{E_F(\kperp)}{\nu^2\kperp^4} \sim \frac{\const}{\kperp^4},
\label{eq:Eu_subvisc}
\eeq   
where $E_u(\kperp)$ and $E_F(\kperp)$ are the spectra of the velocity and of the tension force, 
respectively. Let me explain why $E_F(\kperp)\sim\const$. 
If $\vbperp$ consists of stripes of field alternating direction 
on the scale $\lres$, then $\vbperp\cdot\vdel\vbperp \sim |\vbperp|^2/\xivisc$ 
consists of a constant field interspersed by sharp downward spikes of 
width $\lres$ across the field and length $\xivisc$ along it. 
At $\kperp\lres\ll1$ and $\kperp\xivisc\gg1$, these are effectively 1D delta functions, 
so $E_F(\kperp)\sim\const$, q.e.d.\footnote{A version of this argument was proposed 
by \citet{sch04dynamo} for dynamo-generated fields. They simulated such fields (in 3D) 
directly and found the spectrum of tension to be flat and the velocity spectrum 
to satisfy \exref{eq:Eu_subvisc} extremely well. \citet{kinney00} argued for, 
and saw, similar behaviour in 2D, although their $E_u$ had a slope closer 
to $\kperp^{-4.5}$. Interestingly, \citet{cho02visc} also reported a steeper spectrum 
like this, although it was perhaps not fully numerically converged 
and so, in \citet{cho03}, they changed their mind in favour of~$\kperp^{-4}$.} 
Note that the contribution of the first term in 
\exref{eq:tension_subvisc} to $E_F$ should scale the same as the spectrum 
of $\vbperp$, viz., $\propto\kperp^{-1}$---or perhaps $\kperp^{-1/2}$ 
from the cross-term, if it does not average to zero---this should produce 
steeper and, therefore, subdominant contributions to $E_u(\kperp)$.\footnote{The mismatch 
of the spectrum obtained this way ($E_u\propto\kperp^{-5}$) 
and the one observed in numerical simulations 
led \citet{lazarian04} to propose an ingenious scheme whereby 
all fields and velocities at subviscous scales had a scale-dependent 
volume-filling fraction, whose scaling was then determined by an additional 
requirement that subviscous velocities had local shears comparable to $\tvisc^{-1}$. 
Although this did give the desired $\kperp^{-4}$ scaling, 
I do not see how such an assumption can be justified.}

\subsection{Disruption by Tearing}
\label{sec:tearing_subvisc}

A reader who still remembers the developments in \secref{sec:disruption} 
might wonder whether these stripy fields are safe against disruption by tearing. 
Setting $\vAy \sim \db_\lambda$ in \exref{eq:gmax_TM}, let us ask whether 
there is a disruption scale $\lDsv$ at which 
the local tearing rate would be larger than the stretching rate by the viscous-scale eddies: 
\beq
\gamma \sim \frac{\db_\lambda^{1/2}}{\lambda^{3/2}}\eta^{1/2}\Pm^{-1/4}
\gtrsim \tvisc^{-1}
\hence 
\lambda \lesssim \epsm^{1/6}\tvisc^{5/6}\eta^{1/2}\nu^{-1/6}\equiv\lDsv, 
\eeq
where \exref{eq:Eb_subvisc} was invoked for $\db_\lambda$. 
Using \exref{eq:lres_subvisc} to estimate the putative resistive cutoff, we~get 
\beq
\frac{\lDsv}{\lres} \sim \epsm^{1/6}\tvisc^{1/3}\nu^{-1/6},
\label{eq:lDsv_subvisc}
\eeq

If we are in the parameter regime where the tearing disruption has already 
occurred in the inertial range ($\Pm\lesssim\Re^{1/9}$) and so \exref{eq:lvisc_rec} holds,
then $\tvisc$ is given by the second expression in~\exref{eq:tvisc}, and 
\exref{eq:lDsv_subvisc} implies 
\beq
\frac{\lDsv}{\lres} \sim \lt(\frac{\epsm}{\eps}\rt)^{1/6} \lesssim 1,
\label{eq:no_disruption}
\eeq
so no new disruption is possible in the subviscous range.\footnote{If in~\secref{sec:visc_cutoff}
I had used the cutoff \exref{eq:tearing_cutoff} instead of \exref{eq:K_cutoff} to calculate~$\tvisc$,
I would have discovered now that $\lDsv \sim (\epsm/\eps)^{1/6}\lvisc\sim\lvisc$, indicating
that the tearing-mediated cascade from \secref{sec:recturb} in fact continued below the cutoff
\exref{eq:tearing_cutoff}---an argument that I already made in~\secref{sec:diss}.} 

In contrast, if the inertial-range cascade was cut off in the aligned 
regime ($\Pm\gtrsim\Re^{1/9}$), so \exref{eq:lvisc_aligned} 
and the first expression in \exref{eq:tvisc} apply, then 
\beq
\frac{\lDsv}{\lres} \sim 
\lt(\frac{\epsm}{\eps}\rt)^{1/6}\Re^{1/18} \gg 1.  
\eeq
Modulo factors of order unity and the ludicrous smallness of the fractional power
of~$\Re$ involved, this means that 
if the tearing disruption did not have the chance to occur in the inertial range, 
it will occur in the subviscous range, and that $\lDsv$ will be the field reversal 
scale, not $\lres$.
In terms of the viscous scale~\exref{eq:lvisc_aligned}, which is $\lvisc\sim\lres\Pm^{1/2}$, 
\beq
\frac{\lD}{\lvisc} \sim 
\lt(\frac{\epsm}{\eps}\rt)^{1/6}\Re^{1/18}\Pm^{-1/2}, 
\eeq 
where I have renamed $\lDsv\to\lD$, since, in this regime,
this is the only disruption scale there is. 

At $\lambda \lesssim \lD$, a local MHD cascade is again ignited, 
just like it was in \secref{sec:recturb}. 
It should not seem strange that inertial motions are again possible: 
viscously dominated tearing of the magnetic sheets will produce $\lD$-sized plasmoids 
whose turnover times are shorter than their viscous-dissipation times
(I already argued this in a similar context in \secref{sec:diss}). 
Indeed, similarly to \exref{eq:z_below}, taking them to be unaligned and 
demanding that they pick up all the available energy flux $\epsm$, 
one gets their amplitude
\beq
\frac{\dz_{\lD}^3}{\lD} \sim \epsm
\hence \dz_{\lD}\sim (\epsm\lD)^{1/3} 
\eeq 
and the associated Reynolds number for the new cascade:
\beq
\Re_{\lD} = \frac{\dz_{\lD}\lD}{\nu} \sim \Re^{5/27}\Pm^{-2/3}\gg 1
\rmif \Re \gg \Pm^{18/5}.  
\eeq 
This cascade is cut off, as usual, at the scale \exref{eq:K_cutoff}, but 
with this new~$\Re$:
\beq
\lviscnew \sim \lD \Re_{\lD}^{-3/4} \sim \frac{\nu^{3/4}}{\epsm^{1/4}},
\label{eq:lvisc_new}
\eeq
the Kolmogorov scale again, obviously. 

Thus, the subviscous cascade turns out to be a complicated transitional 
arrangement for enabling tearing disruption and restoration of the Kolmogorov 
cutoff \exref{eq:lvisc_new}. 
Yet again, below this cutoff, at $\lambda < \lviscnew$, 
we are confronted with a purely magnetic, 
``second subviscous cascade,'' but this time with the (new) viscous-scale 
turnover time given by the formula analogous to the second expression 
in \exref{eq:tvisc}, viz., $\tviscnew \sim (\epsm/\nu)^{1/2}$. All the 
arguments of \secsand{sec:mag_subvisc}{sec:vel_subvisc} apply, but 
with no longer any danger of further disruption [see \exref{eq:no_disruption}].

A reader sceptical of the falsifiability of these arguments 
(given the proliferation of small fractional powers of $\Re$ and the piling up 
of twiddle algebra) might feel this is all a fiction---but it is a logical one!    

\section{Decaying MHD Turbulence}
\label{sec:decaying}

Decaying MHD turbulence belongs to this part of this review because it too tends 
to end up in ``imbalanced'' states dominated either by the magnetic field or by one 
of the Elsasser fields (and because it remains, or has done until recently, 
in certain important respects a ``loose end''). 
On a very crude level, it is perhaps obvious that this should be so, because ideal MHD 
equations have two types of exact solutions for which nonlinear 
interactions vanish: Elsasser states ($\vu = \pm\vB$, or $\vz^\mp = 0$) 
and static force-free magnetic fields ($\vB\times\vJ = 0$, where $\vJ = \vdel\times\vB$). 
If the system finds a way towards either of these solutions, globally or locally, 
concentrated on 
scales large enough to make dissipation small, it may, subject to this small dissipation, 
be able to linger in those states (``may'' because the stability of the force-free states, 
e.g., is not guaranteed: see discussion and references in Appendix A of \citealt{hosking20}). 
We shall see below that both magnetically dominated scenarios and 
convergence to Elsasser states are possible and that recent developments 
point to magnetic reconnection muscling its way into this topic as well. 

The usual theoretical attitude to decaying turbulence, dating back to 
\citet{K41,K41decay}, is to assume that its energy would decay slowly 
compared to the nonlinear interactions at small scales (simply because 
turnover times $\tnl$ are shorter at smaller scales) and hence to expect 
the situation in the inertial range (below the outer scale) to be the same 
as in the forced case: a constant-flux energy cascade, etc. I shall 
discuss the numerical evidence in \secref{sec:decay_spectra}, but for now 
let us accept this philosophy as sound. 
With the small scales thus taken care of, the interesting question 
is the large-scale behaviour: since decaying turbulence is not interfered 
with ``externally'', it has the freedom to decide how fast various types of energy 
(kinetic, magnetic, Elsasser) decay and how the outer scale evolves. 
These are the ``zeroth-order'' questions that any theory of decay must 
be able to answer.\footnote{These are also questions that preoccupy a certain
subcommunity of cosmologists seeking to relate extragalactic magnetic fields
observed in ``cosmic voids'' to theories of primordial genesis of magnetic fields,
of which those observed ones are conjectured to be decayed relics. How this is
done is described in \citet{hosking22cosmo}, where the reader will also find
all the relevant cosmological references. Because of this cosmological ``relevance'',
decaying MHD turbulence has been a popular topic amongst MHD theorists
and simulators for nearly half a century. The resulting literature that I am 
familiar with will be reviewed in what follows. Note that another way in which,
more recently, the decaying MHD turbulence and, especially, the phenomenon of inverse
transfer (\secref{sec:perma}), have been tied to the problem of the origin of cosmic
magnetic fields is via attempts to construct a seed field for the cosmic 
dynamo (\secref{sec:dynamo}) out of decaying magnetic flux tubes generated by
Weibel instability in a primordial plasma \citep{zhou20,zhou22,pucci21}.} 

\subsection{Selective Decay}
\label{sec:sel_decay}

How to answer these questions was, like most other things in turbulence, 
understood already by \citet{K41decay}: one assumes that energy decays 
on some appropriate nonlinear time scale subject to some other invariant(s) 
staying constant (i.e., decaying only due to dissipation, which, at 
$\Re\to\infty$, vanishes at the outer scale); this allows one to 
impose enough constraints on the energy and the outer scale to determine 
the evolution of both. In MHD, this principle is sometimes called 
the ``selective-decay hypothesis'', originating from the idea of 
\citet{taylor74} relaxation---early proponents of this 
view of decaying MHD turbulence were \citet{montgomery78,montgomery79} and \citet{matthaeus80}. 
 
In \secref{sec:saffman}, I will return to what the 
additional invariant was for Kolmogorov, and how that can be generalised to MHD, 
but let me start with the most straightforward 
(and, historically, the earliest and most successful) application of the 
philosophy of selective decay---MHD turbulence in 2D. 

\subsubsection{Decay of 2D MHD Turbulence}
\label{sec:decay_2D}

In 2D MHD, there are two positive-definite invariants: energy and ``anastrophy'' $\la A_z^2\ra$, 
where $A_z$ is the out-of-plane component of the vector potential, which 
in 2D behaves as a passive scalar: 
\beq
\frac{\dd A_z}{\dd t} + \vuperp\cdot\vdperp A_z = \eta\nabla^2 A_z. 
\eeq
Like energy, it is, of course, only truly conserved when $\eta=0$, but 
at finite but small $\eta$, it is ``better conserved'' than energy: assuming 
that the latter is finite as $\eta\to+0$, 
\beq
\frac{\rmd \la A_z^2\ra}{\rmd t} = -2\eta\la B^2\ra \to 0. 
\eeq 
Thus, energy must decay subject to the constraint that 
\beq
 \la A_z^2\ra \sim B^2 L^2 \sim \const,
\label{eq:anastrophy}
\eeq
where $L$ is the outer scale. Assuming (faithfully to \citealt{K41,K41decay}) 
that the (magnetic) energy decays at a rate independent of the dissipation 
coefficients and set instead by the outer-scale nonlinear time $\sim L/U$, 
one gets \citep{hatori84}
\beq
\frac{\rmd B^2}{\rmd t} \sim - \frac{UB^2}{L} \sim -\frac{B^3}{L}
\propto -B^4
\hence 
\la B^2\ra \sim \la U^2\ra \propto t^{-1},\quad 
L \propto t^{1/2},
\label{eq:2D_decay}
\eeq 
where $U$ has been linked to $B$ by assuming ideal Alfv\'enic dynamics ($U\sim B$),
and $L$ to $B$ via anastrophy conservation~\exref{eq:anastrophy}. 
The scalings \exref{eq:2D_decay} were confirmed numerically 
by \citet{biskamp89} in one of the early triumphs of high-resolution MHD simulations
(earlier, lower-resolution studies were by 
\citealt{matthaeus80} and \citealt{ting86}, who explored the 2D MHD decay subject 
to anastrophy conservation as part 
of the early discussions around the ``selective-decay hypothesis''). 

\subsubsection{Decay of Helical MHD Turbulence} 
\label{sec:decay_hel}

There is an immediate, direct generalisation of the above argument to 
3D MHD {\em without a mean field}. Instead of anastrophy, the magnetic 
invariant in 3D MHD is helicity, $\la H\ra = \la\vAA\cdot\vB\ra$, which 
again is ``better conserved'' than energy. Indeed, assuming again that the latter 
decays with time in a manner independent of the dissipation coefficients, 
implies $\eta \la J^2\ra \to \const$ as $\eta\to+0$, whence
\beq
\frac{\rmd \la H\ra}{\rmd t} = -2\eta\la \vB\cdot\vJ\ra \sim O(\eta^{1/2}) \rmas \eta\to+0. 
\label{eq:dHdt}
\eeq
\citet{taylor74} relaxation, which inspired the selective-decay approach to 
decaying MHD turbulence, was precisely the idea that MHD systems would 
decay to states of minimal energy subject to constant helicity. 

If one adopts
\beq
\la H\ra \sim B^2 L \sim \const
\label{eq:Hconst}
\eeq
as the governing constraint, 
the selective-decay calculation exactly analogous to~\exref{eq:2D_decay}~is 
\citep{hatori84,son99}
\beq
\frac{\rmd B^2}{\rmd t} \sim - \frac{UB^2}{L} \sim - \frac{B^3}{L} \propto - B^{5}
\hence \la B^2\ra\sim \la U^2\ra \propto t^{-2/3},\quad
L\propto t^{2/3}, 
\label{eq:hel_decay_ideal}
\eeq
again assuming ideal Alfv\'enic dynamics ($U\sim B$). 

Here, however, what appeared a simple 
and compelling theory crashed against reality, or, at any rate, against what passes for 
reality in the world of MHD turbulence, where figments of theoretical imaginations 
are tested against figments of numerical computations: 
the scalings \exref{eq:hel_decay_ideal} badly 
disagreed with the latter, by \citet{biskamp99,biskamp00}. 
Instead of $U\sim B$, these authors spotted empirically 
in their simulations that $U \propto B^2$ and, by modifying \exref{eq:hel_decay_ideal} 
accordingly, concluded 
\beq
\la B^2\ra \propto t^{-1/2},\qquad
\la U^2\ra \propto t^{-1},\qquad
L \propto t^{1/2}. 
\label{eq:hel_decay_BM}
\eeq 
These scalings did indeed appear to check out numerically, both in their simulations 
and in some later ones \citep{christensson01,banerjee04,frick10,berera14,brandenburg19decdyn}.\footnote{Some of these authors, nevertheless, prefer the scalings \exref{eq:hel_decay_ideal}. 
\citet{banerjee04} hope that \exref{eq:hel_decay_ideal} 
will be recovered at a greater resolution;  
\citet{brandenburg17} deem their turbulence to evolve gradually 
towards, if not quite achieve,~\exref{eq:hel_decay_ideal}; 
\citet{brandenburg19decdyn} think that \exref{eq:hel_decay_BM} 
is a transient regime on the way to~\exref{eq:hel_decay_ideal}. 
In a certain sense, they are right---I will explain this in 
\secsand{sec:decay_hel_rec}{sec:decay_bal}.} 

The status of the scalings \exref{eq:hel_decay_ideal} and \exref{eq:hel_decay_BM} 
will become clear in what follows, but I want to preface what is to come by 
observing that the assumption $U \sim B$ underpinning \exref{eq:hel_decay_ideal} is 
not as intuitively obvious as it might appear to be at first glance. 
Formally, a system relaxing according to J.~B.~Taylor's 
principle will tend to a static state consisting of linear force-free fields. 
Linear force-free fields are one-scale ($\vB\times\vJ = 0$ implies $\nabla^2\vB = -k^2\vB$, 
where $k$ is a single number that depends on the initial $\la H\ra$ 
and boundary conditions), 
so can hardly be thought of as a proper turbulent state, but one could nevertheless 
imagine MHD turbulence decaying through a sequence of magnetically dominated 
states featuring local patches of such fields at large scales, 
with flows constantly re-excited by these local patches crashing into each other 
(and/or going unstable). 
It then makes sense that the kinetic energy should be smaller than the magnetic one, 
and perhaps one could even hope to find magnetically dominated states 
in which $U/B\to 0$ with $t\to\infty$, as in \exref{eq:hel_decay_BM}. 

This argument is not a proof of anything, but it should be enough to motivate one 
to look carefully at the dynamical nature of the processes that underpin 
the decay---and, as I have emphasised repeatedly 
in this review, when one starts to look carefully into the nonlinear dynamics in MHD, 
especially in magnetically dominated configurations, sooner or later reconnection 
makes an entrance. 

\subsection{Reconnection Takes Over}
\label{sec:decay_rec}

\subsubsection{Reconnection-Controlled Decay of 2D MHD Turbulence}
\label{sec:decay_2D_rec}

The argument at the end of \secref{sec:decay_hel} casting doubt on the assumption $U\sim B$ 
in fact applies equally well to the 2D case 
considered in \secref{sec:decay_2D}, with anastrophy substituted for helicity 
(minimum-energy states subject to constant anastrophy are also static, 
linear force-free states). Except in 2D, unlike in 3D, theory \exref{eq:2D_decay} 
was confirmed by numerics, so why worry? 

\begin{figure}
\centerline{\includegraphics[width=0.95\textwidth]{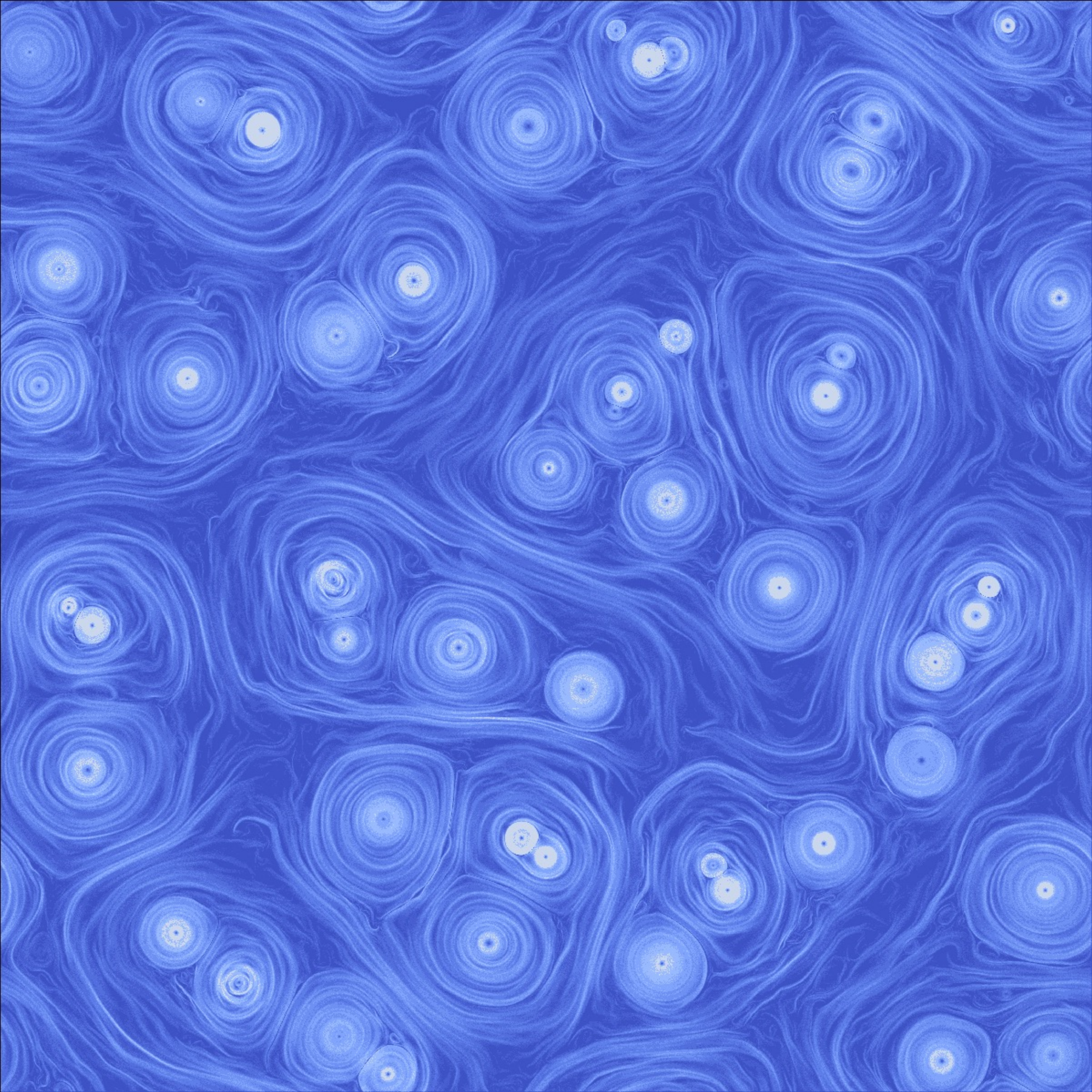}} 
\caption{A snapshot of $|\vB|$ from a decaying 2D MHD turbulence simulation, 
courtesy of D.~Hosking. This run had the resolution of $4608^2$ and 
used $n=4$ equal hyperviscosity and hyperresistivity 
(the run with $\nu_4=\eta_4=2\times10^{-11}$ from \citealt{hosking21}).}
\label{fig:hosking_2D}
\end{figure}

It may well be that the only thing worse for theory than to disagree with numerics 
is to agree with them. It has taken 30 years since \citet{biskamp89} 
for a clear realisation to emerge that the assumption of ideal dynamics ($U\sim B$)
in \secref{eq:2D_decay} was flawed. The self-similar decaying state in which 
2D MHD turbulence ends up is, in fact, quite significantly dominated by magnetic 
fields and, if one examines it visually, consists of magnetic islands separated 
by current sheets (\figref{fig:hosking_2D}), 
via which they reconnect and coalesce into bigger islands---that 
is how $L$ grows. While \citet{biskamp89} did observe that this process was 
occurring in their simulations 
\citep[see also][]{politano89,servidio09,servidio10,servidio11rev,servidio11}, 
it was \citet{zhou19,zhou21} who drew the logical 
conclusion that the characteristic nonlinear time scale for the energy decay 
must then be the reconnection time: 
\beq
\trec \sim \epsrec^{-1}\frac{L}{B},
\qquad \epsrec^{-1} = (1+\Pm)^{1/2}\min\bl\{\tS_L^{1/2},\tSc^{1/2}\br\},
\qquad \tS_L = \frac{BL}{\eta\sqrt{1+\Pm}},
\label{eq:trec_decay}
\eeq
where $\epsrec$ is the dimensionless reconnection rate, 
$\tS_L$ is the Lundquist number (adjusted for a visco-Alfv\'enic 
outflow when $\Pm\gg1$) and $\tSc\sim 10^4$ is its critical 
value above which reconnection switches from the Sweet--Parker (SP) regime (\apref{app:SP_rec}) 
to the fast, plasmoid-dominated regime (\apref{app:uls}).\footnote{If reconnection is stochastic
in the sense advocated by \citet{lazarian20} (see \secref{sec:stoch} and \apref{app:stoch_rec}),
then, presumably, $\epsrec\sim 1$. 
This does not appear to be the case in either 2D or 3D simulations of \citet{zhou19,zhou20}, 
\citet{bhat21}, and \citet{hosking21}, viz., they all see $\epsrec\sim S_L^{-1/2}$, but there is 
no telling what might happen at higher resolutions.} 

This means that the first relation in \exref{eq:2D_decay} 
must, in fact, be rewritten~as 
\beq
\frac{\rmd B^2}{\rmd t} \sim - \frac{B^2}{\trec} \sim - \epsrec\frac{B^3}{L}.
\label{eq:dBdt_rec}
\eeq 
Remarkably, this does not change the decay laws \exref{eq:2D_decay} 
because $\epsrec$ does not change with time: 
in the fast-reconnection regime, because $\tSc$ is a numerical constant, 
and in the SP regime, because $\tS_L \sim \const$ by anastrophy 
conservation~\exref{eq:anastrophy} (which, for pairs of reconnecting/coalescing magnetic 
islands of size~$\sim L$, is just the conservation of their flux~$\sim BL$).
However, by including $\epsrec$ into the rescaling of the time histories 
of $\la B^2\ra$ at different $S_L$, \citet{zhou19} did confirm that it was  
the reconnection time scale \exref{eq:trec_decay},
with its $\eta$ dependence, rather than just the Alfv\'en time $L/B$, 
that controlled the decay in their simulations 
(their $S_L$ was small enough to stay in the SP regime). 
Another way to prove this is to replace the Laplacian 
resistivity with a hyperresistivity (dissipation operator $\eta_n\nabla^n$), 
which, in the ``hyper-SP'' regime, makes $\epsrec$ time-dependent and thus 
produces different decay laws than~\exref{eq:2D_decay}---these indeed turn out to 
be the ones seen in such a numerical experiment 
\citep[][the 2D case is described in their Appendix~A]{hosking21}. 

What about the kinetic energy? The turbulence that \citet{zhou19} studied 
had none at the outset (they started with just a grid 
of magnetic islands), so all flows that emerged with time were due to 
reconnection processes. These are Alfv\'enic outflows 
(ignoring the $\Pm\gg1$ case for now), so $U\sim B$, but their mean 
energy over the system's volume is not same as the magnetic energy because 
they are localised to reconnection sites, i.e., to sheets of length $L$ 
and width $\delta\sim \epsrec L$ \citep{hosking21}. 
Therefore, 
\beq
\la U^2\ra \sim \frac{\delta}{L}\,\la B^2\ra \sim \epsrec\la B^2\ra.
\label{eq:UvsB_rec}
\eeq
The turbulence is magnetically dominated because $\epsrec \ll 1$, 
even though the kinetic and magnetic energies' decay laws are still the same 
because $\epsrec\sim\const$. 

\subsubsection{Reconnection-Controlled Decay of Helical MHD Turbulence}
\label{sec:decay_hel_rec}

The same arguments can be ported immediately to the 3D helical case considered 
in~\secref{sec:decay_hel}. In doing so, I follow the paper by \citet{hosking21}. 

While the 3D dynamics might not be as vividly 
dominated by coalescence of magnetic structures as the 2D one is, let us 
nonetheless assume that the main dynamical process controlling the 
energy decay and transfer to larger scales is reconnection between 
blobs of helicity-conserving magnetic fields. 
To work out the consequences of this assumption, let us again replace 
\exref{eq:hel_decay_ideal} with \exref{eq:dBdt_rec}, where $\epsrec$
is given by \exref{eq:trec_decay}. If reconnection is fast, $\epsrec\sim\const$ 
again, so the decay laws \exref{eq:hel_decay_ideal} survive, with the 
only amendment that kinetic energy, while decaying at the same rate as magnetic, 
is only a small fraction of it given by \exref{eq:UvsB_rec}.

This outcome still disagrees with \exref{eq:hel_decay_BM} and most 
extant numerical simulations, but it is, in fact, highly unlikely that any 
of these simulations are large enough to reach the 
fast-reconnection regime. Let us therefore work out what happens 
when $\epsrec$ is set by the $\tS_L$-dependent SP formula in \exref{eq:trec_decay}.     
As anastrophy is no longer conserved, $\tS_L\neq\const$ and \exref{eq:dBdt_rec} becomes
\beq
\frac{\rmd B^2}{\rmd t} \sim 
- \frac{\eta^{1/2}}{(1+\Pm)^{1/4}}\frac{B^{5/2}}{L^{3/2}} 
\propto -B^{11/2}
\hence \la B^2\ra\propto t^{-4/7},\quad
L\propto t^{4/7}.  
\label{eq:hel_decay_rec}
\eeq
Therefore, $\epsrec\propto \tS_L^{-1/2} \propto (BL)^{-1/2} \propto t^{-1/7}$,  
whence the kinetic energy's decay law is, by~\exref{eq:UvsB_rec},
\beq
\la U^2\ra \sim \epsrec \la B^2\ra \propto t^{-5/7}. 
\label{eq:UvsB_rec_hel}
\eeq 
Note that, since the Lundquist number gets larger with time under this scheme 
(albeit quite slowly), this regime is transient, with reconnection eventually 
becoming fast and the scalings \exref{eq:hel_decay_ideal} and \exref{eq:UvsB_rec} 
with $\epsrec\sim\const$ taking over in the long run. 

Even if we assume that simulations cannot run long enough (or to extend to 
large enough $L$) to reach this state, the transient scalings 
\exref{eq:hel_decay_rec} and \exref{eq:UvsB_rec_hel} are not particularly 
close to the Biskamp--M\"uller scalings~\exref{eq:hel_decay_BM} seen in simulations, 
so it would seem that these new developments bring us no confirmatory joy. In fact, 
things are better then they look, for two reasons. 

First, most numerical simulations use hyperresistivity, and the generalisation 
of the above scaling to the ``hyper-SP'' reconnection does push scaling exponents 
of the magnetic and kinetic energy away from each other and somewhat closer 
to $-1/2$ and $-1$, respectively. At any rate, the important point is that 
the decay laws in hyperresistive simulations that fall short of 
the fast-reconnection regime depend quite sensitively on the order $n$ of the 
hyperresistivity and so the entire idea of reconnection-controlled decay can 
be tested by varying~$n$---\citet{hosking21} did that and found that the decay exponents 
measured in such simulations agreed passably with the theoretical ones obtained 
in the same way as above but for different~$n$, and disagreed fairly decisively with other 
theoretical schemes mooted in the literature (e.g., that of 
\citealt{campanelli04}, discussed in \secref{sec:olesen}). 

Secondly, the reconnection-controlled decay is, in fact, only justifiable if the 
amount of kinetic energy in the initial state is not much larger than 
\exref{eq:UvsB_rec}, i.e., it is a good theory only for magnetically dominated 
MHD turbulence. If one starts with $\la U^2\ra\sim\la B^2\ra$, as 
\citet{biskamp99,biskamp00} did (or $\la U^2\ra \gg \la B^2\ra$ followed by dynamo, 
as in \citealt{brandenburg19decdyn}), a different theory is needed to describe 
the initial stage of the decay (if kinetic energy still ends up decaying faster than 
magnetic, the system will eventually transition to the magnetically dominated regime). 
I shall need to introduce some further new ideas before I discuss such a theory, 
also proposed by \citet{hosking21}, in \secref{sec:decay_bal} and recover  
the precise scalings~\exref{eq:hel_decay_BM}.

\subsubsection{Lack of Rescaling Symmetry for Dissipation Coefficients}
\label{sec:olesen}

Before I do that, let me, as a historical footnote, mention the theory of helical decay by 
\citet{campanelli04}, which can now be falsified in what appears to be 
a definitive way. A reader not interested in history can skip this section, 
as well as \secref{sec:decay_nonhel_old}, and move directly to \secref{sec:saffman}, 
where new things happen.  

The theory in question is formulated in the clever language for discussing self-similar 
decay pioneered by \citet{olesen97}. He observed that MHD (and, indeed, also HD) 
equations have the following rescaling symmetry: $\forall a$ and~$h$,
\beq
\vr \to a\vr,\quad
t \to a^{1-h} t,\quad
\vu \to a^h\vu,\quad
\vB \to a^h\vB,\quad
\nu \to a^{1+h}\nu,\quad
\eta \to a^{1+h}\eta. 
\label{eq:olesen_sym}
\eeq 
He then posited that decaying MHD turbulence would simply go through a sequence 
of these transformations, with the rescaling parameter being a power of time, 
$a=(t/t_0)^{1/(1-h)}$, where $t_0$ is some reference (not necessarily initial) time. 
Then
\beq
\la U^2\ra\propto \la B^2\ra \propto t^{2h/(1-h)},\qquad
L \propto t^{1/(1-h)}.
\label{eq:scalings_h}
\eeq  
The tricky part is, of course, to find~$h$. 
Conservation of helicity, $B^2L\sim\const$, would require $h=-1/2$, 
which gives the scalings \exref{eq:hel_decay_ideal}. 
To get something else, \citet{campanelli04} reasoned that if 
the magnetic field were approximately force-free, it would fall out of the momentum equation, 
and, the induction equation being linear, this force-free magnetic field could, therefore, 
be rescaled  by an arbitrary constant: $\vB\to a^{m}\vB$, 
where $m$ did not need to be the same as~$h$. 
Conservation of helicity combined with the scaling of $L$ in \exref{eq:scalings_h} 
then fixes $m$ in terms of~$h$:  
\beq
B^2 L \sim \const 
\hence  
m = -\frac{1}{2}
\hence 
\la B^2\ra \propto t^{-1/(1-h)}, 
\eeq 
with $h$ still undetermined; $\la U^2\ra$ still satisfies \exref{eq:scalings_h}. 
He then argued that $h=-1$ because 
the dissipation coefficients should stay constant under the rescaling~\exref{eq:olesen_sym}. 
This got him the Biskamp--M\"uller 
scalings~\exref{eq:hel_decay_BM}.\footnote{\citet{christensson05} 
have an argument for $L\propto t^{1/2}$ that is essentially a version of Campanelli's. 
It is based on the self-similar solution \exref{eq:Ek_ssim} for the energy 
spectrum. It is hard-wired into this solution that $L\propto t^{1/(1-h)}$, but
if one now assumes self-similarity all the way down to the dissipation scales, 
the dissipative cutoff must have the same scaling, so $\lres \propto t^{1/(1-h)}$. 
In a nutshell, \citet{christensson05} then set $\lres \sim (\eta t)^{1/2}$ 
by dimensional analysis and hence conclude that $h=-1$. 
It is, of course, far from obvious that $\lres$ depends only on $\eta$ 
and $t$ but not also on $L$, $B$ and/or $U$ (and, therefore, on the initial energy 
and scale of the turbulence), as the Kolmogorov scale does in a turbulence with 
constant energy flux.}
    
Campanelli's argument looks neat, but, on reflection, it is counterintuitive 
that everything should depend on the specific form of dissipation: indeed, if 
one were formally to replace viscosity and resistivity with hyperviscosity 
and hyperresistivity, $\eta\nabla^2 \to \eta_n\nabla^n$, then keeping $\eta_n$ 
unchanged by the scaling \exref{eq:olesen_sym} would require a different value 
of~$h$. Should we then expect different decay laws? This seems unlikely in 
the limit $\eta_n\to+0$. In any event, at finite $\eta_n$, if Campanelli is 
right, his theory, straightforwardly generalised, provides specific predictions for decay laws 
that are different for different~$n$. So does the theory of reconnection-controlled 
decay, in the ``hyper-SP'' regime (\secref{sec:decay_hel_rec})---and its prediction 
is different from Campanelli's. Numerical simulations with hyperdissipation 
are consistent with the former and rule out the latter quite convincingly
\citep{hosking21}. 

The conclusion seems to be that dissipation coefficients need not be invariant 
under the rescaling~\exref{eq:olesen_sym}. This means that any decay theory  
that involves dissipative effects---as is the case for 
reconnection-controlled decay when the reconnection is not fast---need not 
obey Olesen's scalings \exref{eq:scalings_h} for any~$h$ (and indeed 
the helical decay in \secref{sec:decay_hel_rec} does not). 

\subsection{Decay of Non-helical MHD Turbulence: Simulations and Theories}
\label{sec:decay_nonhel_old}

What about 3D MHD turbulence with no mean field and zero helicity?
Unlike anastrophy in 2D, $\la H\ra$ is not a sign-definite quantity and, 
in mirror-symmetric systems, $\la H\ra=0$. This destroys the usefulness of 
the constant-helicity constraint \exref{eq:Hconst} and re-opens 
the problem of the decay laws.\footnote{The constant-helicity constraint 
is also absent when there is an external mean field, 
hence in RMHD, because there is no helicity conservation in this approximation
(the mean field always ``sticks out'' of the volume). I shall 
deal with this regime in \secref{sec:decay_RMHD}.} 
Physically, one might think of the helical MHD turbulence as decaying from 
an initial state produced by dynamo action of a helical flow (see review 
by \citealt{rincon19}); similarly, the starting point for non-helical decay might 
be the saturated state of a non-helical fluctuation dynamo (\secref{sec:dynamo}). 

How to handle the non-helical decay has until recently remained unsettled, 
although quite a lot of numerical evidence about 
what happens in it does exist, alongside some theoretical arguments, 
which I will survey in a moment. I shall get to what I believe to 
be the right way to think about this regime in \secref{sec:saffman}, to which 
an impatient reader is welcome to turn immediately. 

The conversation in the literature about non-helical decay has been heavily 
influenced by the fact that the 2D scalings \exref{eq:2D_decay} 
appear to work quite well also in 3D with zero helicity. 
This was reported with various degrees of certainty in a number 
of numerical experiments \citep{maclow98,biskamp99,biskamp00,christensson01,banerjee04,frick10,berera14,brandenburg15,brandenburg17,reppin17,bhat21}. 
The 2D analogy did not escape their authors: e.g., \citet{brandenburg15} 
speculated about ``near conservation" of a local version of anastrophy
({\em pace} the gauge-non-invariance of~$\la A^2\ra$). 
While it is true that, as observed, e.g., by \citet{bhat21}, 
$\la A^2\ra$, while not a 3D invariant, decays slower with time 
than energy---indeed it does, since  $A\sim BL$, $B$ decays, and $L$ grows 
under any sensible decay law,---it is perhaps a stretch that this is the cause 
rather than a corollary of the latter.   

Another scheme for rationalising $BL\sim\const$ is to argue that dynamically, 
the decay of the magnetic energy and the increase of the scale at which 
it sits are driven by mergers, via reconnection, of long flux tubes, 
as appears to be the case in 3D RMHD 
\citep[][see \secref{sec:decay_RMHD}]{zhou20}---\citet{bhat21} make a long and careful 
empirical case for such a scenario. In fact, in an earlier paper, \citet{reppin17} 
already mention (albeit gingerly and amongst other options) the possibility that
ever larger magnetic structures might be generated via mergers of reconnecting flux ropes, 
an idea that they attribute to \citet{mueller12}, who in turn credit the 2D study 
by \citet{biskamp94}. While it is not at all obvious 
(and, in fact, not true: see \secref{sec:decay_nonhel_new}) that flux tubes in 3D 
MHD without a mean field should reconnect while conserving their ``2D poloidal flux'' 
($\sim BL$), it is true that the 3D non-helical decay 
is controlled by reconnection, just like the 2D and the 3D helical decays turned 
out to be in \secref{sec:decay_rec}---the theory that I will present 
in \secref{sec:decay_nonhel_new} will utilise this.

\citet{campanelli04} also has a theory of non-helical decay. 
This follows from his argument, already rehearsed in \secref{sec:olesen}, 
that the decay exponents are fixed by the requirement that the dissipation 
coefficients in MHD must stay constant under Olesen's rescaling symmetry~\exref{eq:olesen_sym},
implying $h=-1$. In the non-helical case, he does not assume a force-free magnetic field, 
and so does not rescale it separately from \exref{eq:olesen_sym}. The result is 
the scalings \exref{eq:scalings_h}, which, with $h=-1$, are the same as the 
2D scalings~\exref{eq:2D_decay}. 
\citet{olesen15b} commented that under this self-similarity, 
$\la A^2\ra = \const$, so anastrophy would be conserved regardless of the dimensionality 
of the problem (he called this ``dimensional reduction''). 
Just as in the helical case, Campanelli's theory is testable numerically 
via its generalisation to hyperviscous and hyperresistive dissipation, 
and fails the numerical tests in comparison with the reconnection-controlled 
decay described in \secref{sec:decay_nonhel_new} \citep{hosking21}. 

\subsection{Selective Decay Constrained by Saffman Invariants}
\label{sec:saffman}

Let me now outline a new way (or, rather, a new version of a very old way) 
of thinking about the non-helical MHD decay and similar problems, 
proposed by David Hosking \citep{hosking21}.

\begin{figure}
\centerline{\includegraphics[width=0.95\textwidth]{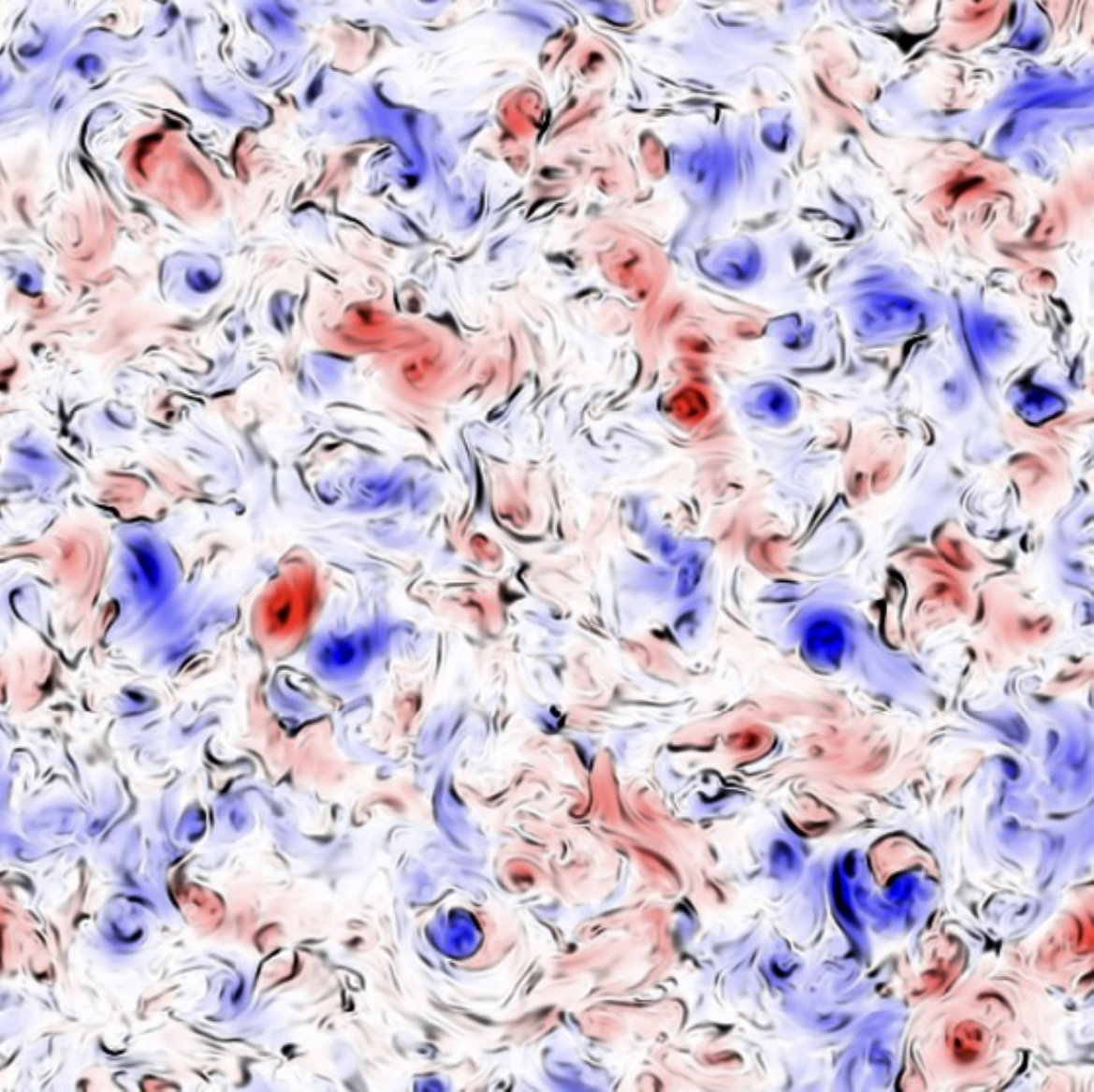}} 
\caption{A cross-section of helicity density $H=\vAA\cdot\vB$ (in Coulomb gauge, $\vdel\cdot\vAA=0$)
from a decaying 3D MHD turbulence simulation ($512^3$) with zero net helicity,
taken from \citet{hosking22cosmo}. Red is $H>0$, blue is $H<0$. The superimposed
grey scale shows the magnitude of the current density $J=|\vdel\times\vB|$, so the black patches
are strong currents between, and on the edges of, local blobs of non-zero helicity---presumably,
these are reconnection sites where decay occurs, at constant Hosking invariant~\exref{eq:hosking},
i.e., conserving the mean square helicity fluctuations.} 
\label{fig:hosking_3D}
\end{figure}

The basic idea is that $\la H\ra=0$ does not mean (except in very artificial set-ups) 
that no part of the turbulence has any local magnetic links or twists 
and thus local helicity (\figref{fig:hosking_3D}). 
While helicity has to sum up to zero over the entire system, 
these local helical features will still impose 
topological constraints on the system's decay. The challenge is to express 
these constraints mathematically. 

We shall draw inspiration from 
the long experience of thinking of such questions that exists in the context 
of (various flavours of) hydrodynamic turbulence.\footnote{A magisterial tutorial 
on this subject, with all the key ideas, nuances, fallacies, historical 
triumphs and setbacks narrated in a friendly and clear style, 
can be found in the books by \citet{davidson13,davidson15}, whose 
own contributions form a significant part of the emerging canon.} 
Since \citet{K41decay}, the decay of hydrodynamic turbulence has been 
treated as selective decay subject to conservation of certain invariants 
related to the conservation of the momentum and angular 
momentum of the motions---but these conserved quantities, while present 
locally in the turbulent ``eddies'', sum up to zero over the 
entire system, so cannot simply be fixed in the same way as $\la A_z^2\ra$ 
or $\la H\ra$ were in \secsand{sec:sel_decay}{sec:decay_rec}. 
Thus, the situation is somewhat analogous to the MHD case with $\la H\ra=0$. 

\subsubsection{Generalised Saffman Invariants}
\label{sec:decay_hydro}

Consider a conserved quantity that has a local density $\psi$ and that is not sign-definite. 
That $\la\psi\ra$ is conserved is expressed 
generically by $\psi$ satisfying a dynamical equation of the form 
\beq
\frac{\dd\psi}{\dd t} + \vdel\cdot \vGamma = \mathrm{diss.~terms},
\label{eq:inv_psi_ev}
\eeq  
where $\vGamma$ is the flux of $\psi$. 
In the context of the upcoming derivations, $\psi$ might be, e.g., (a component of) 
the fluid velocity $\vu$ or the helicity density $H=\vAA\cdot\vB$. Let us now denote 
\beq
\overline{\psi}_V = \int_V\rmd^3\vr\,\psi(\vr), 
\eeq
where $V$ is a finite volume, and assume that 
$\la\psi\ra \equiv \lim_{V\to\infty}\overline{\psi}_V/V = 0$.
The fact that the average of $\psi$ over the infinite space vanishes
does not mean that its integral over a finite volume must do so, but, 
if we take $V\gg L^3$, where $L$ is the 
outer scale (correlation length) of $\psi$, it is reasonable to assume that 
the integral accumulates as a random walk:\footnote{Modulo some subtleties and nuances 
that an interested reader can read about in \citet{davidson13,davidson15} and \citet{hosking21}.} 
\beq
\overline{\psi}_V \sim \lt(\frac{V}{L^3}\rt)^{1/2}\dpsi_L L^3 
\hence 
\lim_{V\to\infty}\frac{\overline{\psi}_V^2}{V} \sim \dpsi_L^2L^3, 
\eeq 
where $\dpsi_L$ is the typical size of the fluctuation of 
$\psi$ at its correlation scale $L$. This suggests that the mean square 
fluctuation of $\psi$ is a finite, $V$-independent quantity. 
It can be expressed in terms of the two-point correlation function of~$\psi$: 
\beq
\lim_{V\to\infty}\frac{\overline{\psi}_V^2}{V} = 
\lim_{V\to\infty}\frac{1}{V}\iint_V\rmd^3\vr\rmd^3\vr'\,\psi(\vr)\psi(\vr') = 
\int\rmd^3\vl\lt\la\psi(\vr)\psi(\vr+\vl)\rt\ra \equiv I_\psi.
\label{eq:Ipsi_def}
\eeq
The integral $I_\psi$ is finite provided correlations decay faster than $1/l^3$ as $l\to\infty$. 
Furthermore, $I_\psi$ is an invariant because, from~\exref{eq:inv_psi_ev}, 
\beq
\frac{\dd}{\dd t} \lt\la\psi(\vr)\psi(\vr+\vl)\rt\ra 
+ \frac{\dd}{\dd\vl}\cdot \lt\la\lt[\vGamma(\vr+\vl)-\vGamma(\vr-\vl)\rt]\psi(\vr)\rt\ra 
= \mathrm{diss.~terms}
\hence
\frac{\rmd I_\psi}{\rmd t} \to 0.
\eeq
The last formula holds as dissipation coefficients~$\to +0$ if  
the dissipation terms can be argued to vanish in this limit 
and if all correlations decay faster than $1/l^3$ (so $I_\psi$ is finite 
and the surface integral left of the flux term vanishes). 

In a system with $\la\psi\ra = 0$ and $\psi\neq 0$ pointwise, 
it is subject to the conservation of $I_\psi$ 
that one ought to assume the selective decay of energy to be happening. 

Following \citet{davidson13,davidson15}, I shall refer to quantities such 
as $I_\psi$ as (generalised) ``Saffman invariants''. The original Saffman
invariant was a measure of conservation of the linear momentum 
of the turbulent eddies, viz., $\psi = \vu$, and was introduced by \citet{saffman67}, 
via a line of reasoning roughly analogous to the above, 
to constrain the decay laws for hydrodynamic turbulence with long-range 
correlations. His calculation, analogous to the selective-decay ones 
in \secref{sec:sel_decay} (but pre-dating them considerably),~is   
\beq
I_{\vu} \sim U^2L^3\sim\const\hence
\frac{\rmd U^2}{\rmd t} \sim -\frac{U^3}{L} \propto -U^{11/3}\hence
\la U^2\ra\propto t^{-6/5},\quad 
L\propto t^{2/5}.
\eeq
\citet{K41decay} used (in fact, pioneered) the same method, except he assumed 
that $I_{\vu}=0$ and conjectured that the decay of turbulence would 
be controlled by the Loitsyansky invariant, which is just $I_\psi$ with 
$\psi = \vr\times\vu \equiv \vL$, expressing the conservation of the angular 
momentum of the eddies \citep[see][]{LL6,davidson13,davidson15}:\footnote{Note that 
$\lt\la\vu(\vr)\cdot\vu(\vr+\vl)\rt\ra$ must decay faster than $1/l^5$ 
as $l\to\infty$ in order for $I_{\vL}$ to be finite, so this situation, 
known as Batchelor turbulence \citep[after][]{batchelor56}, 
describes the decay of turbulent systems 
with shorter-range correlations than in Saffman turbulence. 
It is realised if the initial state is set up to have~$I_{\vu}=0$. 
Note also that the interpretation of the Loitsyansky integral 
in terms of the conservation of angular momentum is somewhat 
shaky, mathematically \citep[see][]{davidson13,davidson15}.} 
\beq
I_{\vL} = \int\rmd^3\vl\, l^2\lt\la\vu(\vr)\cdot\vu(\vr+\vl)\rt\ra 
= \const \hence U^2L^5 \sim \const.
\label{eq:Loits}
\eeq 
With this constraint, one gets the famous \citet{K41decay} laws of decay: 
\beq
\frac{\rmd U^2}{\rmd t} \sim -\frac{U^3}{L} \propto -U^{17/5}\hence
\la U^2\ra\propto t^{-10/7},\quad 
L\propto t^{2/7}.
\label{eq:K41_decay}
\eeq  

\subsubsection{Decay of Magnetically Dominated Non-helical MHD Turbulence}
\label{sec:decay_nonhel_new}

\begin{figure}
\centerline{\includegraphics[width=0.5\textwidth]{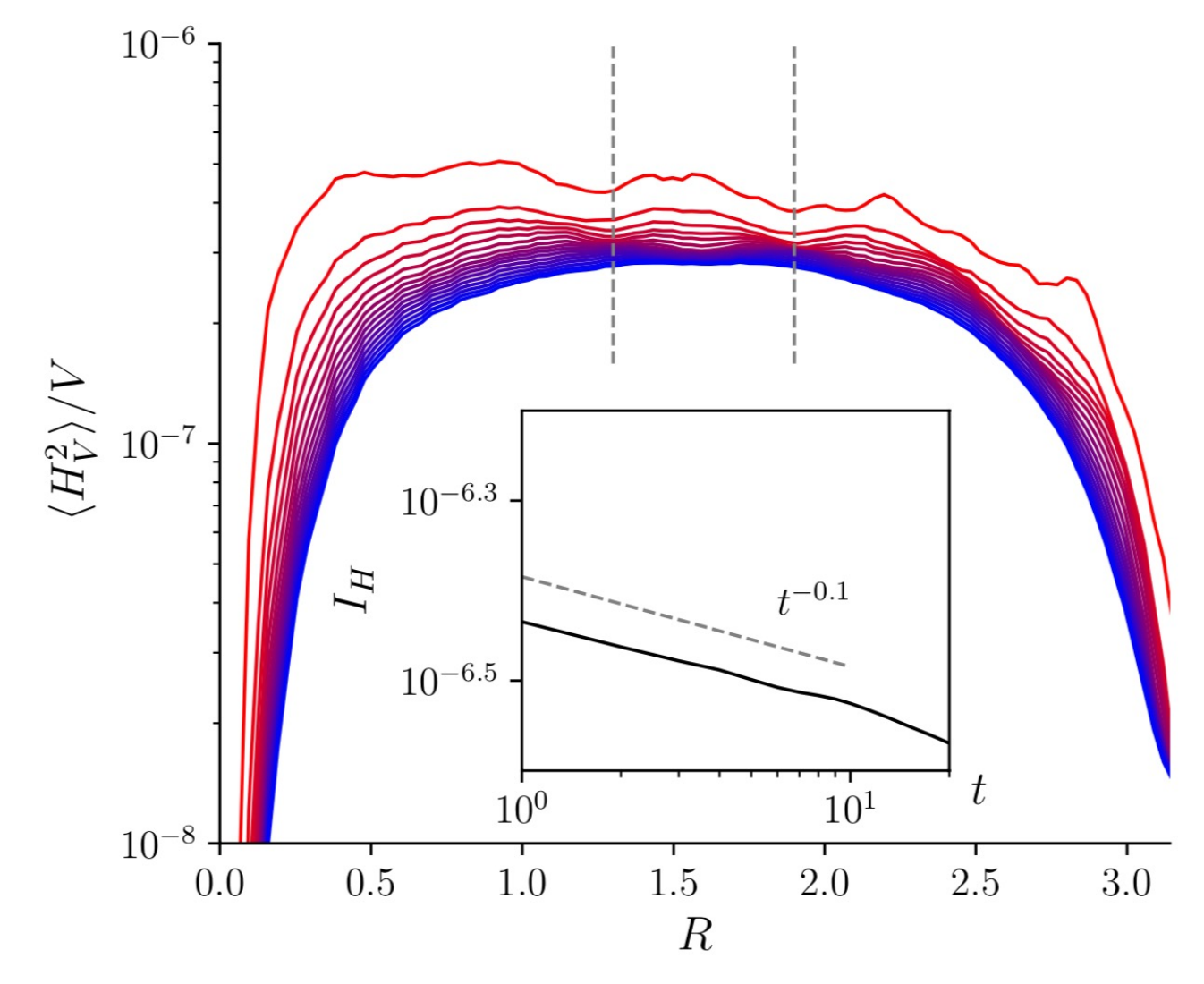}} 
\vskip2mm
\caption{Numerical probe of the conservation of the Hosking invariant~$I_H$, 
defined by~\exref{eq:Ipsi_def} with $\psi = \vAA\cdot\vB$. This plot, 
reproduced from \citet{hosking21}, is 
of the mean square helicity density over a cubic volume $V = (2R)^3$, at different 
times in a decaying, non-helical, 3D MHD turbulence simulation started in 
a magnetically dominated state---red-to-blue curves correspond to 
earlier-to-later times. The simulation box is periodic, which is why $I_H$ is only 
non-zero in a range of $R$ intermediate between the energy-containing scale~$L$ 
and the box size ($2R = 2\pi$). The inset shows the near-constancy of $I_H$ (calculated 
over the interval of $R$ indicated by vertical dashed lines).}
\label{fig:hosking_inv}
\end{figure}

\citet{hosking21} introduced a new invariant, which they called the 
``Saffman helicity invariant'' and to which I will refer as 
the Hosking invariant,~$I_H$ (\figsand{fig:hosking_inv}{fig:hosking_3D}). 
This invariant is $I_\psi$ defined in \exref{eq:Ipsi_def} with 
$\psi = H = \vAA\cdot\vB$, the helicity density (anticipating an erudite 
reader's concern, yes, they did show that it was gauge-invariant).\footnote{It turns out
that a similar idea does exist in the literature with application to pure hydrodynamical
turbulence: the invariant $I_\psi$, with $\psi = \vu\cdot(\vdel\times\vu)$, the
kinetic helicity of the turbulent flow, was introduced by \citet{levich83} and bears their
name; it was used by \citet{frenkel83} to propose an amended turbulence decay theory.
I am grateful to A.~Bershadskii for pointing me to these papers. \label{fn:levich}} 
They then considered the decay of non-helical MHD turbulence subject to 
conservation of $I_H$, 
\beq
I_H \sim (AB)^2L^3 \sim B^4L^5 \sim \const, 
\label{eq:hosking}
\eeq
and with the assumption that the decay occurs 
on the reconnection time scale \exref{eq:trec_decay}. 
When the reconnection is fast ($\epsrec\sim\const$),
\beq
\frac{\rmd B^2}{\rmd t} \sim -\epsrec\frac{B^3}{L} \propto -B^{19/5}
\hence 
\la B^2\ra \propto t^{-10/9},\quad
L \propto t^{4/9},\quad
\la U^2\ra \sim \epsrec\la B^2\ra, 
\label{eq:decay_nonhel_fast}
\eeq
where the last formula is again \exref{eq:UvsB_rec} (in Olesen's language, 
\secref{sec:olesen}, this corresponds to $h=-5/4$). 
When the reconnection is slow, as in \exref{eq:hel_decay_rec}, 
\beq
\frac{\rmd B^2}{\rmd t} \propto - \frac{B^{5/2}}{L^{3/2}} 
\propto -B^{37/10}
\hence \la B^2\ra\propto t^{-20/17},\quad
L\propto t^{8/17},\quad
\la U^2\ra \propto t^{-19/17}.  
\label{eq:decay_nonhel_slow}
\eeq
The decay law for kinetic energy follows from \exref{eq:UvsB_rec}
and $\epsrec\propto \tS_L^{-1/2}\propto (BL)^{-1/2}\propto t^{1/17}$. 
Note that, unlike in the case of helical decay (\secref{sec:decay_hel_rec}), 
the reconnection rate increases (Lundquist number decreases) with time (extremely slowly), 
so the system never gets out of the slow-reconnection regime. In fact, even 
in the case of fast reconnection~\exref{eq:decay_nonhel_fast}, 
the Lundquist number decreases with time, $\tS_L\propto t^{-1/9}$, 
so the asymptotic decay laws are ones for slow reconnection~\exref{eq:decay_nonhel_slow}. 

Whilst neither \exref{eq:decay_nonhel_fast} nor \exref{eq:decay_nonhel_slow} 
are the same as the 2D scalings \exref{eq:2D_decay} that are believed to 
be seen in numerical simulations (\secref{sec:decay_nonhel_old}), they are 
uncannily close---and almost certainly not numerically distinguishable from either 
each other or \exref{eq:2D_decay}. Just as in the case of helical turbulence, 
a way to verify them numerically is to generalise~\exref{eq:decay_nonhel_slow} to 
the hyperresistive case and test the dependence of the scaling exponents 
on the hyperresistivity order $n$---a test that the theory passes reasonably 
well \citep[see][]{hosking21}.

As in the case of helical decay, this theory describes only magnetically dominated 
decay, where the kinetic energy is all due to reconnection outflows. If instead 
the initial state has $\la U^2\ra\sim \la B^2\ra$ (or $\la U^2\ra\gg\la B^2\ra$ 
followed by dynamo), working out its decay laws requires further theoretical arrangements, 
described in \secref{sec:decay_bal}.\footnote{Note that, since $\la U^2\ra$ 
in \exref{eq:decay_nonhel_slow} 
decays a little bit slower than $\la B^2\ra$, it must be the case that, 
if one waits long enough (longer, no doubt, than 
any numerical simulation has ever been able to afford to wait),
the system will get out of the magnetically dominated state and 
into the $\la U^2\ra\sim \la B^2\ra$ territory. 
Amusingly, as we are about to see, that will push it back to a 
magnetically dominated state, so perhaps the system will oscillate between 
\exref{eq:decay_nonhel_slow} and~\exref{eq:hel_decay_xhel}.} 

\subsubsection{Selective Decay Constrained by Saffman Cross-Helicity Invariant} 
\label{sec:decay_bal} 

The method of Saffman invariants (\secref{sec:decay_hydro}) is begging to be applied 
to another non-sign-definite conserved MHD quantity, the cross-helicity, whose 
density is $\vu\cdot\vB \equiv X$. In a balanced turbulence, $\la X\ra = 0$, arguably 
a natural situation in the absence of a mean field.\footnote{This is not to say 
that the case $\la X\ra\neq 0$ has lacked attention: there is a lively literature 
on it, cited in~\secref{sec:decay_to_Elsasser}.} 
There is then the Saffman 
cross-helicity invariant, $I_\psi$ of \exref{eq:Ipsi_def} with $\psi = X$
(which appears to have been first considered, for a different
purpose, by \citealt{bershadskii19xhel}, who was inspired by the paper 
of \citealt{levich83} already mentioned in footnote~\ref{fn:levich}, and
its follow-ups, e.g., \citealt{levich91} and \citealt{levich09}). 
Unlike in the case of $I_H$, it is not obvious (and may not be true) that 
$I_X$ is conserved better than the energy because the fields involved in it
($\vu$ and $\vB$) are the same fields as those whose mean squares make up the energy. 
Let us put this issue aside for further investigation and not allow it to stop us from 
considering what would happen if decay of MHD turbulence were constrained by the 
conservation of $I_X$. 

This is not an entirely frivolous or formalistic exercise 
because there is quite a lot of evidence that MHD turbulence has a tendency to 
break up into patches of non-zero cross-helicity (``imbalance'')---not only in 
the inertial range, as I already noted in~\secref{sec:Eimb}
(see \figref{fig:perez_patches}), 
but also in decaying, zero-mean-field turbulence, on the 
outer scale \citep[e.g.,][]{matthaeus08,servidio08}. This is understandable: nonlinearity 
is likely to be weaker in places where the fields are closer to an Elsasser 
state, so decay in those places would be slower than where $Z^+\sim Z^-$, 
and there would be a natural tendency for the imbalanced patches to survive 
Darwinianly (cf.~\secref{sec:decay_to_Elsasser}). 

If such situations do arise, they must do so when kinetic and magnetic energies 
of the turbulence are (initially) not very different---precisely the case of 
$\la U^2\ra\sim\la B^2\ra$ that is not described by the theories of magnetically 
dominated decay outlined in \secsand{sec:decay_hel_rec}{sec:decay_nonhel_new} 
and that I promised there to deal with later.  

Let me deal with it now, following again \citet{hosking21}. 
Conservation of $I_X$ gives us a constraint, 
\beq
I_X \sim U^2B^2L^3 \sim \const, 
\label{eq:xhel}
\eeq
which in this regime should replace the estimate \exref{eq:UvsB_rec} 
of $\la U^2\ra$ resulting from reconnection outflows. 
Together with the the helicity constraint \exref{eq:Hconst}, $B^2L\sim\const$, 
this gives $U\propto B^2$ which is precisely the relationship 
spotted numerically by \citet{biskamp99,biskamp00}. 
Since kinetic energy is now not small (at least initially), it is perhaps not as 
far-fetched as in the magnetically dominated regime to assume that 
the energy decays on the ideal time scale $\sim L/U$. Doing so 
and making use of our two constraints, we get  
\begin{align}
\frac{\rmd B^2}{\rmd t}\sim - \frac{UB^2}{L} \propto -B^6
\hence 
\la B^2\ra\propto t^{-1/2},\quad
\la U^2\ra\propto t^{-1},\quad
L \propto t^{1/2}. 
\label{eq:hel_decay_xhel}
\end{align}
These are precisely the Biskamp--M\"uller scalings~\exref{eq:hel_decay_BM} 
for helical decay. As I anticipated in \secref{sec:decay_hel_rec}, 
since kinetic energy decays faster than magnetic, these decay laws, if correct, 
must be transient, eventually changing to the magnetically dominated case.  
This is more or less the scenario envisioned by \citet{brandenburg19decdyn} 
based on their simulations that started with $\la B^2\ra \ll \la U^2\ra$, 
exhibited (helical) dynamo action that brought the magnetic field to dynamical 
strength, and then decayed (transiently) according to \exref{eq:hel_decay_xhel}. 

If we instead consider the non-helical case, the helicity constraint \exref{eq:Hconst}
must be replaced by the conservation of the Hosking invariant \exref{eq:hosking}.
Combined with \exref{eq:xhel}, this gives $U\sim B^{1/5}$, i.e., 
magnetic field would decay faster than velocity. But that is not a sustainable 
proposition: by such a decay, magnetic field would be brought down below 
dynamical strength, whereupon it would be re-grown by dynamo action. 
A natural conclusion might be that magnetic field in this regime would not be 
able to constrain decay and instead stay just under dynamical strength and 
follow the velocity field, which would decay according to a purely hydrodynamic 
law, probably~\exref{eq:K41_decay}. There might be some support for this 
scenario in the simulations of \citet{bhat21}. 

Finally, for completeness and without much supporting evidence, but just 
to show the joys of the new toys that are the generalised Saffman invariants for MHD, 
here is what happens in 2D MHD with conserved Saffman cross-helicity---again relevant, 
presumably, for cases in which magnetic energy does not dominate initially. 
In 2D, 
\beq
I_X \sim U^2B^2L^2 \sim \const.
\eeq
But the anastrophy constraint \exref{eq:anastrophy} tells us that $BL\sim\const$, 
so we must have $U\sim\const$, whence  
\beq
\frac{\rmd B^2}{\rmd t} \sim -\frac{UB^2}{L} \propto -B^3
\hence \la B^2\ra \propto t^{-2},\quad L\propto t,\quad 
\la U^2\ra\sim\const. 
\eeq 
Thus, magnetic field will decay quite vigorously and rush to larger scales, 
while kinetic energy will stay constant. Just as in the case of non-helical decay in 3D, 
this regime cannot persist because the magnetic field will soon drop below dynamical  
strength---but there is no dynamo in 2D \citep{zeldovich56}, so it will just decay 
away, exponentially in the kinematic regime, leaving 2D hydrodynamic turbulence 
decaying by its own, slow, 
devices (see \citealt{davidson13,davidson15} for more). Qualitatively, 
this seems to be consistent with what \citet{kinney00} see in their 
``2D dynamo'' simulations. 

\subsection{Permanence of Large Scales vs.\ Inverse Transfer}
\label{sec:perma}

In the introduction to \secref{sec:decaying}, I argued that the inertial-range spectrum
of decaying turbulence is not an interesting new subject (I will return to this 
thought in \secref{sec:decay_spectra}). What is interesting though is its 
large-scale spectrum, at $kL\ll 1$, i.e., the presence and growth, or otherwise, of 
long-range correlations. Consider the 1D spectrum of a statistically isotropic field~$\psi$ 
and Taylor-expand it in small~$k$: 
\beq
E_\psi(k) = 4\pi k^2 \int\frac{\rmd^3\vl}{(2\pi)^3} 
\lt\la\psi(\vr)\psi(\vr+\vl)\rt\ra e^{-i\vk\cdot\vl} 
= \frac{1}{2\pi^2}\lt(I_\psi k^2 + J_\psi k^4 + \dots\rt), 
\eeq   
where $I_\psi$ is the Saffman invariant~\exref{eq:Ipsi_def} and 
$J_\psi = -(1/3)\int\rmd^3\vl\,l^2 \lt\la\psi(\vr)\psi(\vr+\vl)\rt\ra$. 
Obviously, the Taylor expansion can only be extended to $k^2$ and $k^4$
if the coefficients $I_\psi$ and $J_\psi$ are finite, i.e., if correlations 
decay faster than $O(l^{-3})$ and $O(l^{-5})$, respectively. 

If $I_\psi\neq 0$, then the large-scale spectrum is~$\propto k^2$ and, 
moreover, the energy content at low $k$ is frozen 
by the conservation of~$I_\psi$---even though the outer scale $L$ increases during decay, 
there is no energy transfer into larger scales, motions 
at those scales just take longer to decay. This feature is sometimes 
referred to as the ``permanence of large scales'' \citep{davidson13,davidson15}. 

\begin{figure}
\centerline{\includegraphics[width=\textwidth]{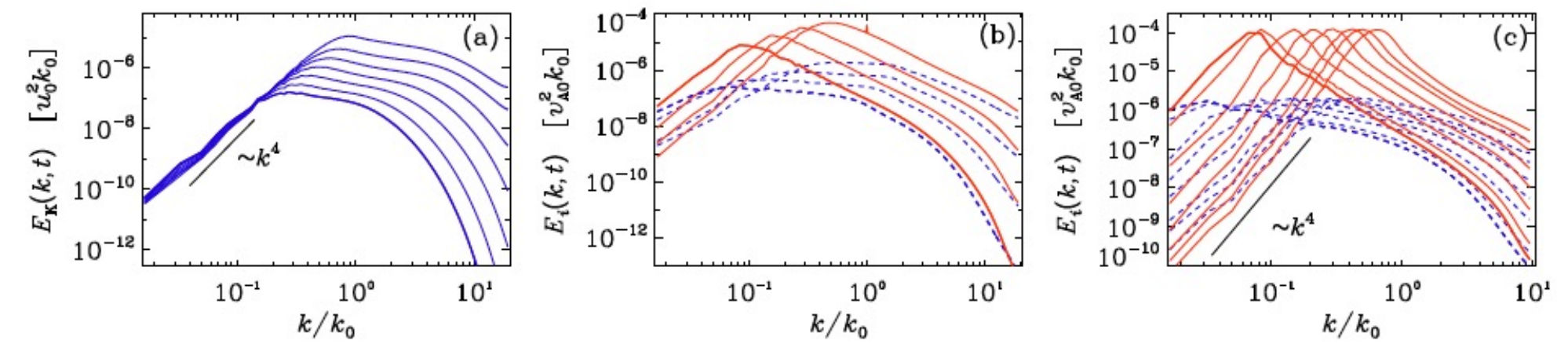}} 
\vskip2mm
\caption{Spectra of kinetic (blue) and magnetic (red) energies in 
decaying turbulence: (a) pure hydrodynamic, (b) MHD with no mean field 
and zero helicity, (c) MHD with no mean field and finite helicity.
The time evolution is from right to left (always towards larger scales).
These plots are from \citet{brandenburg17}.}
\label{fig:brandenburg_decaying}
\end{figure}

What, however, if $I_\psi = 0$, as it would normally be for $\psi = \vu$ 
or $\psi = \vB$? Then the spectrum at large scales is~$\propto k^4$ and, 
if this is K41 turbulence decaying subject to the conservation of the 
Loitsyansky invariant~\exref{eq:Loits}, $I_{\vL}\propto J_{\vu} =\const$, 
there is still permanence of large scales, although with weaker correlations. 
This is illustrated in \figref{fig:brandenburg_decaying}(a).
However, $J_\psi$ need not be an invariant for every $\psi$ 
for which $I_\psi$ is, so, generally speaking, it will change with time.
An example of that is the decay of helical MHD turbulence, 
the evolution of whose magnetic spectrum is shown 
in \figref{fig:brandenburg_decaying}(c): while it still has the $k^4$ long-wavelength
asymptotic, the prefactor $J_{\vB}$ now is manifestly {\em not} 
conserved, but rather grows robustly with time, meaning that magnetic energy is quite 
vigorously transferred to larger scales---an ``inverse cascade'' (non necessarily 
local in $k$) associated with the conservation of magnetic helicity and its transfer 
to large scales, which is a well known phenomenon also in forced turbulence, 
often in the context of helical dynamo action 
\citep{pouquet76,brandenburg01,mueller12,rincon19}.\footnote{In the dynamo case, 
forced \citep{brandenburg01} or decaying \citep{brandenburg19decdyn}, 
a helical velocity field generates a magnetic field from a small seed that initially 
has zero helicity. This field has helicity of one sign at small scales and 
of the opposite sign at large scales (larger than the scale of the velocity), 
keeping overall $H=0$. The small-scale helicity is slowly destroyed by resistivity 
(which possibly makes the whole process very inefficient; see discussion in 
\citealt{rincon19}), while the large-scale helicity is stuck at large scales 
and can, if forcing is switched off or absent from the beginning, 
serve as the starting point for a helical decaying regime---this scenario 
is nicely traced out in \citet{brandenburg19decdyn}.} 
A similar behaviour in non-helical decaying MHD turbulence has recently generated 
a flurry of excitement: an inverse magnetic-energy transfer was discovered there numerically 
by \citet{zrake14} and \citet{brandenburg15} (accompanied by \citealt{berera14} and 
followed by \citealt{reppin17}, \citealt{park17}, 
and \citealt{bhat21})---\figref{fig:brandenburg_decaying}(b) 
is from the non-helical simulation by \citet{brandenburg15}
and shows healthy magnetic-energy growth at low wavenumbers.\footnote{In earlier, 
lower-resolution non-helical simulations by 
\citet{banerjee04}, no inverse transfer was detected.
\citet{reppin17} report that increasing $\Pm$ while holding $\Re$ constant 
also kills the effect. One can imagine that in both of these cases, 
reconnection might not have been able to get going properly.} 

\subsubsection{Scalings for Inverse Transfer of Magnetic Energy}

Since we now do have a theory of the decay of MHD turbulence of any flavour,  
we can easily work out whether and how the large-scale spectrum grows: e.g.,
\beq
J_{\vB} \sim B^2L^5 \sim \lt\{
\begin{array}{ll}
t^{8/3},   & \text{\exref{eq:hel_decay_ideal}, helical, ideal/fast-reconnecting}, B\gg U,\\
t^{16/7},  & \text{\exref{eq:hel_decay_rec}, helical, SP-reconnecting}, B\gg U,\\
t^2,       & \text{\exref{eq:hel_decay_xhel}, helical, ideal}, B\sim U,\\
t^{10/9},  & \text{\exref{eq:decay_nonhel_fast}, non-helical, ideal/fast-reconnecting}, B\gg U,\\
t^{20/17}, & \text{\exref{eq:decay_nonhel_slow}, non-helical, SP-reconnecting}, B\gg U,\\
t^{(5+2m)/(1-h)}, & \text{\secref{sec:olesen}, self-similar}.
\end{array}
\rt.
\label{eq:tk4_regimes}
\eeq 
Thus, in all conceivable regimes, 
there is some inverse transfer of energy to large scales---not 
altogether surprising if the underlying dynamics involves mergers (by reconnection)
of structures, rather than merely slower decay of the larger ones. 

Obviously, the same logic can be applied to spectra of all other fields. 
E.g., \citet{hosking21} observe (and confirm numerically) that, for 
reconnection-dominated decay of non-helical MHD turbulence, 
the spectrum of the quantity $H=\vAA\cdot\vB$ is $\la|H_\vk|^2\ra\propto I_H = \const$ 
at low~$k$, recovering, in a certain sense, the principle of permanence 
of large scales (and providing another way to 
test the conservation of the Hosking invariant~$I_H$).   

\subsubsection{Self-Similar Spectra and Inverse Energy Transfer}
\label{sec:decay_invtrans}

Let me show, for completeness, and for context, in view of the 
discussions that appear in the literature, how to obtain these results in 
the language of self-similar solutions introduced in \secref{sec:olesen}. 
Following \citet{olesen97}, let us work out what 
the symmetry~\exref{eq:olesen_sym} implies for the spectrum of any of the 
fields that have it. For example, for the magnetic field, the spectrum satisfies  
\beq
E(k,t) = 4\pi k^2\int\frac{\rmd^3\vl}{(2\pi)^3}\,e^{-i\vk\cdot\vl}
\la\vB(\vr,t)\cdot\vB(\vr + \vl,t)\ra
= a^{-1-2m}E(a^{-1}k,a^{1-h}t),
\label{eq:olesen_spec}
\eeq
where I used Campanelli's more general rescaling $\vB\to a^m\vB$, where 
$m=-1/2$ with helicity and $m=h$ without. 
A self-similarly evolving solution of \exref{eq:olesen_spec} is 
\beq
E(k,t) = k^{-1-2m}f\bl(kt^{1/(1-h)}\br),
\label{eq:Ek_ssim}
\eeq
where $f(x)$ is some function that needs to be integrable in an appropriate way 
in order for the total energy to be finite: 
\beq
\la B^2\ra = \int_0^\infty\rmd k\, E(k,t) 
= t^{2m/(1-h)}\int_0^\infty\rmd x\,x^{-1-2m} f(x). 
\eeq 
The decay laws \exref{eq:hel_decay_ideal}, \exref{eq:hel_decay_xhel}, 
and \exref{eq:decay_nonhel_fast}
are recovered for $(m,h)=(-1/2,-1/2)$, $(-1/2,-1)$, $(-5/4,-5/4)$, 
respectively [the SP-reconnecting cases 
\exref{eq:hel_decay_rec} and \exref{eq:decay_nonhel_slow} 
are non-self-similar]. 

If $I_{\vB}=0$, the magnetic-energy spectrum must be $\propto k^4$ 
at $kL\ll1$. This requires $f(x)\propto x^{5+2m}$, whence 
\beq
E(k,t) \propto t^{(5+2m)/(1-h)} k^4. 
\label{eq:tk4}
\eeq
This is the same result as the last formula in \exref{eq:tk4_regimes}. 
The solution \exref{eq:Ek_ssim} also implies that the peak of the spectrum, 
at $kL\sim1$, is $E_\mathrm{max} \propto t^{(1+2m)/(1-h)}=\const$ for the 
helical case (manifestly true in \figref{fig:brandenburg_decaying}c) 
and $E_\mathrm{max}\propto t^{-1/2}$ for the non-helical one 
(\figref{fig:brandenburg_decaying}b). \citet{brandenburg17} show that 
rescaling their time-dependent spectra in line with \exref{eq:Ek_ssim}, 
or, to be precise, with the equivalent expression $E(k,t) = L(t)^{1+2m}g\bl(kL(t)\br)$, 
where $L(t)$ is measured directly at every $t$, collapses them all onto a single 
curve, confirming self-similarity.  

Let me observe, finally, that if the prefactor of the low-$k$ 
asymptotic of $E(k)$ changes with time, as it does in \exref{eq:tk4}, 
i.e., if $J_{\vB}$ is not conserved, 
I see no reason to expect that the long-term self-similar evolution 
should be tied to the low-$k$ scaling baked into the initial condition, 
as many authors, starting with \citet{olesen97}, seem to believe. 
The self-similar evolution need not start at $t=0$, 
and it is perfectly possible that it is preceded by some initial 
non-self-similar rearrangement. There appears to be convincing 
numerical evidence that this is indeed what happens 
\citep[e.g.,][]{brandenburg17,brandenburg19decdyn,hosking21}. 

\subsection{Decay of Magnetically Dominated RMHD Turbulence}
\label{sec:decay_RMHD}

Let me now finally return to MHD turbulence with a strong mean field 
(RMHD turbulence), which elsewhere in this review has been my primary preoccupation. 
\citet{zhou20} initialised an RMHD simulation with an array 
of magnetic flux tubes parallel to the mean field 
and found that it behaved rather similarly to the 
magnetically dominated 2D MHD (\secref{sec:decay_2D_rec}), with flux tubes 
reconnecting (coalescing) with each other in the perpendicular plane, 
and the system thus decaying towards a state dominated by ever-larger-scale 
magnetic structures. While the ``perpendicular anastrophy'' $\la A_\parallel^2\ra$
is not formally conserved in RMHD, \citet{zhou20} assumed that RMHD reconnection 
would nevertheless proceed in a quasi-2D fashion, preserving the poloidal ``2D flux'':
\beq
\label{eq:flux_cons}
\bperp\Lperp \sim \const. 
\eeq
If one accepts this, the rest of the argument is exactly the same as 
in 2D MHD (\secsand{sec:decay_2D}{sec:decay_2D_rec}): replacing 
$B\to\bperp$, $U\to\uperp$, and $L\to\Lperp$, one~gets 
\beq
\la\bperp^2\ra \propto t^{-1},
\qquad \la\uperp^2\ra \sim \epsrec\la\bperp^2\ra,
\qquad \Lperp\propto t^{1/2}. 
\label{eq:RMHD_decay}
\eeq 
Since \exref{eq:flux_cons} implies $\tS_{\Lperp}\sim\const$, $\epsrec\sim\const$ 
in both fast and SP reconnection regimes. 

\begin{figure}
\centerline{\includegraphics[width=0.9\textwidth]{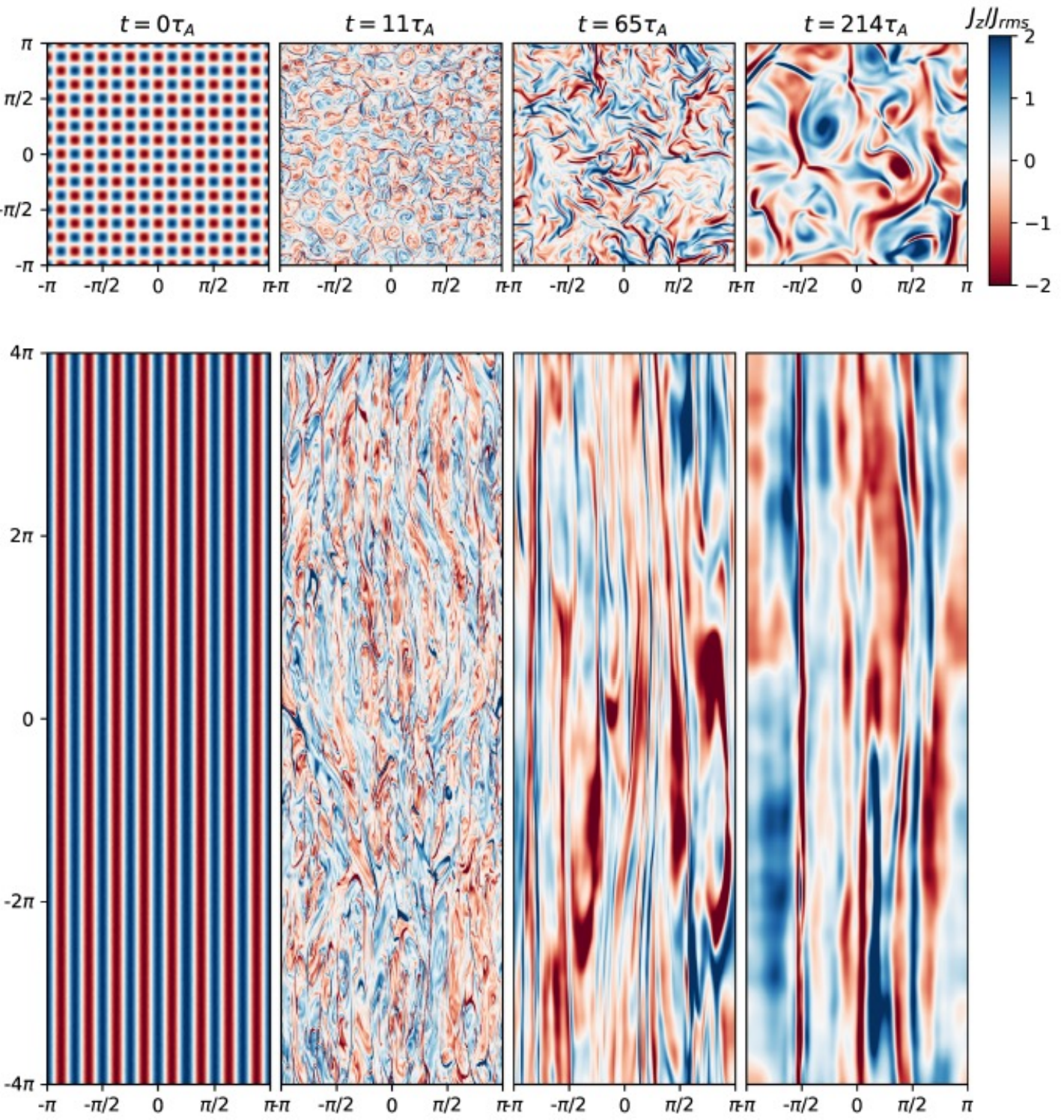}} 
\vskip2mm
\caption{Snapshots of vertical ($z$) current density in the $(x,y,0)$ (top row)
and $(x,0,z)$ (bottom row) planes, taken at a series of subsequent times 
in a decaying RMHD simulation by \citet{zhou20}. This illustrates 
growth of $\Lperp\propto\sqrt{t}$ and $\Lpar\propto t$ (the scalings 
confirmed quantitatively in their paper) and the presence of numerous 
current sheets.} 
\label{fig:muni}
\end{figure}

A new feature is to argue that the length of the tubes along the mean field 
is determined by the CB condition: 
\beq
\tA\sim\frac{\Lpar}{\vA} \sim \trec \propto \frac{\Lperp}{\bperp}
\hence 
\Lpar \propto t. 
\label{eq:decay_CB}
\eeq 
\citet{zhou20} check the scalings \exref{eq:RMHD_decay} and \exref{eq:decay_CB}
in their RMHD simulations (\figref{fig:muni}) and declare success. 
It is interesting to analyse the ingredients of this success in light 
of the long discussion of selective decay in MHD given in the preceding sections. 

\subsubsection{Saffman Cross-Helicity Invariant in RMHD}
\label{sec:RMHD_xhel}

The adoption of \exref{eq:flux_cons} in 3D is, of course, a bit of a sleight of hand. 
This is not a flux-conservation constraint: the 3D poloidal flux through 
the radial cross-section of the flux tube, $\sim \bperp\Lperp\Lpar$,  
does {\em not} stay constant under this scheme;
if it did, critical balance \exref{eq:decay_CB} would then imply $\Lperp\sim\const$, 
which cannot be true. 

There is, however, another way to justify \exref{eq:flux_cons} \citep{hosking21}. 
Since we are dealing with an RMHD system initialised with $\uperp=0$, 
the cross-helicity $\la X\ra=\la\vuperp\cdot\vbperp\ra$
is zero initially and will stay small as long as $\uperp\ll\bperp$. 
The logical application of the principles advocated in \secref{sec:saffman}
is then to posit the conservation of the Saffman cross-helicity invariant: 
\beq
I_X \sim \uperp^2\bperp^2\Lperp^2\Lpar \propto \bperp^3\Lperp^3 \sim \const,
\quad\text{q.e.d.,}
\label{eq:RMHD_xhel}
\eeq  
assuming $\uperp\propto\bperp$ (Alfv\'enic flows) 
and $\Lpar\propto \Lperp/\bperp$ (CB).  

It is an open question whether a decaying RMHD can stay 
in a balanced state, $\la X\ra = 0$, forever, or whether a fluctuation 
of $\la X\ra$ away from zero can eventually tip it over to another 
decaying regime, tending to an Elsasser state (\secref{sec:decay_to_Elsasser}).

\subsubsection{Self-Similar Decay in RMHD}
\label{sec:decay_ssim_RMHD}

The reader who liked Olesen's approach (\secsand{sec:olesen}{sec:decay_invtrans})
and remembers the RMHD symmetry \exref{eq:RMHD_resc} has been straining 
at the leash to apply the former to the latter. It is, of course, the same symmetry 
as \exref{eq:olesen_sym} if one lets $\epsilon = a^h$, except now perpendicular and 
parallel gradients and, therefore, distances transform differently from each 
other:\footnote{Note that 
this scaling of the parallel distances is correct both for distances along the 
global and the local mean field (cf.~\secref{sec:aniso}), 
because $\vbperp\cdot\vdperp \to a^{h-1}\vbperp\cdot\vdperp$.} 
\beq
\vrperp \to a\vrperp,\qquad
\rpar \to a^{1-h}\rpar. 
\eeq
This is just because $\vA$ is now assumed to be an immutable constant, 
so $\rpar$ transforms as time, 
rather than as distance \citep[cf.][and~\secref{sec:par_cascade}]{beresnyak15}.
The resulting scalings, 
\beq
\la\uperp^2\ra \propto \la\bperp^2\ra \propto t^{2h/(1-h)},\qquad 
\Lperp \propto t^{1/(1-h)},\qquad 
\Lpar \propto t,
\eeq 
instantly reproduce \exref{eq:RMHD_decay} and \exref{eq:decay_CB} if $h=-1$. 
A useful observation perhaps is that the scaling of the parallel correlation 
length with time is set just by the requirement of self-similar evolution, without 
the need to conjecture CB explicitly, although of course the RMHD 
symmetry \exref{eq:RMHD_resc} does, in a sense, have CB hard-wired into it. 

Let me make another observation that seems to be of some (possibly marginal) interest. 
In the same vein as \exref{eq:olesen_spec}, one finds, this time for the 2D spectra: 
\beq
\Ekk(\kperp,\kpar,t) = a^{-2-h}\Ekk(a^{-1}\kperp,a^{-1+h}\kpar,a^{1-h}t).
\eeq
The self-similar solution analogous to \exref{eq:Ek_ssim} is, therefore,  
\beq
\Ekk(\kperp,\kpar,t) = \kperp^{-2-h}F\bl(\kperp t^{1/(1-h)},\kpar t\br),
\label{eq:Ek2D_ssim}
\eeq 
with some unknown function $F(x,y)$. The 1D perpendicular spectrum $E(\kperp,t)$ is 
found by integrating \exref{eq:Ek2D_ssim} over all $\kpar$, predictably 
leading to the same result as \exref{eq:Ek_ssim} (with $m=h$); 
the analog of \exref{eq:tk4}~is 
\beq
E(\kperp,t)\propto t^{2(2+h)/(1-h)}\kperp^3,
\eeq 
again manifesting inverse transfer if $h=-1$. Integrating \exref{eq:Ek2D_ssim} 
over $\kperp$ instead, one gets the 1D parallel spectrum: 
\beq
E(\kpar,t) = t^{(1+h)/(1-h)} g(\kpar t), 
\label{eq:Ekpar_ssim}
\eeq
where $g(y)=\int_0^\infty\rmd x x^{-2-h}F(x,y)$. 
This result is interesting for the following reason. $E(\kpar,t)$ is the spectrum 
of a random field reflecting its dependence on a single scalar spatial coordinate, 
the distance along the field. 
The long-wavelength, $\kpar\Lpar \ll1$, asymptotic of this spectrum describes 
the absence of correlations at point separations $\lpar\gg\Lpar$, so 
it is just the spectrum of a 1D white noise (cf.~\apref{app:det_delta}).  
Therefore, $g(y)\to\const$ as $y\to 0$. But \exref{eq:Ekpar_ssim} then implies that 
the energy content of low $\kpar$ is frozen in time if $h=-1$. 
This suggests that RMHD turbulence might have a ``parallel Saffman invariant''~$\Ipar$, 
so $E(\kpar,t) \propto \Ipar\kpar^0 = \const$ at $\kpar\Lpar\ll1$ (cf.~\secref{sec:perma}). 
This invariant should have the form\footnote{Intriguingly, \exref{eq:RMHD_inv} is 
the one-point correlator between the field and its $\kpar=0$ part, evoking the special 
role of the ``2D condensate'' (see \secsand{sec:WT_zeromodes}{sec:new_res_theory_WT}).} 
\beq
\label{eq:RMHD_inv}
\Ipar = \int\rmd\lpar\lt\la\vbperp(\rpar)\cdot\vbperp(\rpar+\lpar)\rt\ra
\sim \bperp^2\Lpar
\eeq 
(instead of $\vbperp$, it may involve some other linear combination of the fields 
$\vbperp$ and $\vuperp$, or~$\vzperp^\pm$). 
If this could be shown to be the relevant conservation law for an RMHD 
selective decay, that would be an alternative way, to \exref{eq:RMHD_xhel}, 
of getting the scalings~\exref{eq:RMHD_decay}.  

\subsection{Decay of Imbalanced MHD Turbulence: Towards Elsasser States}
\label{sec:decay_to_Elsasser}

Finally, let us consider RMHD (or indeed MHD) with $\la X\ra\neq 0$. 
The eventual decay of imbalanced MHD turbulence  
to pure Elsasser states was first mooted by \citet{dobrowolny80},
in response to such states being occasionally observed in the solar wind.  
Since $\vz^+$ and $\vz^-$ advect each other, one can easily imagine that a fluctuation 
of the imbalance at the outer scale in one direction, say in favour of $\vz^+$, 
will lead to $\vz^+$ decaying slower and $\vz^-$ faster, thus increasing the imbalance 
further, until $\vz^-$ disappears and $\vz^+$ is left in splendid isolation. 
The crudest model of this is as follows \citep{maron01}: if $L$ is the energy-containing 
(outer) scale and $Z^\pm$ are the two fields' amplitudes at this scale, then
\beq
\frac{\rmd (Z^\pm)^2}{\rmd t} \sim - \frac{Z^\mp (Z^\pm)^2}{L}
\hence 
Z^+ - Z^- \sim \const,\quad
\frac{\rmd}{\rmd t}\ln\frac{Z^+}{Z^-} \sim \frac{Z^+ - Z^-}{L}.
\label{eq:imb_decay}
\eeq
Thus, an initial imbalance in either direction will cause the (fractional) imbalance 
to get worse with time, until the weaker field has decayed away. 
In other words, cross-helicity $(Z^+)^2 - (Z^-)^2 \sim \const\cdot(Z^+ + Z^-)$ 
decays more slowly than energy, hence the increasing imbalance. The asymptotic 
state is an Elsasser state with 
\beq
Z^+(t\to\infty) \sim (Z^+ - Z^-)(t=0).
\eeq 
Note that this simple model depends on assuming that $L$ 
is the same for both fields and that any alignment effects on the strength 
of the nonlinear interaction can be ignored, which is far from 
obvious and can be hard to sustain \citep[e.g.,][]{hossain95,wan12,bandyopadhyay19}.   

The above scenario did, nevertheless, appear to be confirmed (very slowly in time) in 
the decaying RMHD simulation by \citet{chenmallet11}, initialised 
by first creating a statistically steady, forced, balanced turbulence and then 
switching off the forcing, so the breaking of the symmetry in favour of one 
of the fields arose from an initial fluctuation of the imbalance. 
In full-MHD simulations with a strong mean field, the same result had been found 
in a number of earlier papers \citep{oughton94,maron01,cho02}, 
while in the absence of a mean field, it dates back to even earlier selective-decay 
literature \citep{montgomery78,montgomery79,matthaeus80,ting86,stribling91,hossain95}.  

Let me observe finally that, in the model~\exref{eq:imb_decay}, the energy fluxes  
\beq
\eps^\pm \sim \frac{Z^\mp (Z^\pm)^2}{L} 
\hence 
\frac{\eps^+}{\eps^-} \sim \frac{Z^+}{Z^-}
\eeq  
are in the same relationship with Elsasser energies as they are reported 
to be in forced imbalanced turbulence: see~\exref{eq:imb_ratio_sq}, \exref{eq:z_ratio} 
and \secref{sec:imb_PB} 
(there is also some direct numerical support for this relationship in decaying 
MHD turbulence, going back to~\citealt{verma96}). This is perhaps another 
encouraging sign. 

I do not have any further insights to offer about this regime, so I will stop here. 

\subsection{Inertial-Range Spectra of Decaying MHD Turbulence: Numerical Evidence}
\label{sec:decay_spectra}

The philosophy articulated in the introduction to \secref{sec:decaying} 
with regard to the inertial-range spectra does appear to be vindicated in the RMHD 
simulations of \citet{chenmallet11}, except the perpendicular spectrum was 
steeper than $\kperp^{-3/2}$ (and closer to $\kperp^{-5/3}$) 
and the parallel one steeper than $\kpar^{-2}$---this might actually 
be consistent with what one would expect for a system that moved gradually 
towards greater imbalance (see \secref{sec:imb_new}). 

The evidence from the RMHD simulations by \citet{zhou20} appears to point in the same 
direction: they report $\kperp^{-3/2}$ spectra of both 
magnetic and kinetic energy, presumably of the same origin as those derived in \secref{sec:DA}. 

In the currently available decaying MHD simulations without a mean field, with or 
without helicity, there might not yet be sufficient resolution to tell what the 
asymptotic inertial-range spectra are (see, e.g., \figref{fig:brandenburg_decaying} 
and note particularly that there is no scale-by-scale equipartition between magnetic 
and kinetic energy at these resolutions)---or indeed whether they might be non-universal 
with respect to initial conditions \citep{lee10}, a somewhat disconcerting prospect.
An oft-reported ``non-universal'' spectrum is $k^{-2}$ 
\citep[e.g., by][]{lee10,brandenburg15,brandenburg19decdyn}, which might actually be 
another signature of reconnection (rather than of the WT regime, as some of these authors suggest): 
\citet{dallas13ksq} and \citet[][in 2D]{zhou19} interpret this spectrum geometrically as describing 
an ensemble of current sheets, which are step-like ``discontinuities'' of the magnetic field 
(this is the same argument as I mooted for the residual energy in \secref{sec:new_res_theory}; note that the spectrum of plasmoid chains would also be~$k^{-2}$, as shown in 
\apref{app:chain_spectrum}).  
According to \citet{dallas13sym,dallas14}, however, this scaling gives way to a shallower 
$k^{-5/3}$ or $k^{-3/2}$ slope at sufficiently small scales in simulations with sufficiently  
high resolution, as current sheets curl up and/or break up (see also \citealt{mininni06}), 
so perhaps small-scale universality is safe after all and the current sheets are 
simply the effective energy-containing structures at which the ``true'' inertial range  
starts (cf.~\secref{sec:turb_sheet} and \apref{app:rec_driven}).

\subsection{Summary}

To sum up, there appear to be at least three qualitatively different regimes, 
or, rather, classes of regimes,\footnote{It is formally possible to argue that there 
are many more than three regimes: \citet{wan12} take this to an amusing tongue-in-cheek 
extreme and count $\sim 6,500$ ``conceptually distinct types of possible 
turbulent behaviour'' for 3D MHD with no mean field and $\sim 59,000$ varieties 
with a mean field. No one working on decaying MHD turbulence need fear 
running out of options any time soon!} of decaying MHD turbulence:  
\vskip2mm
(i) RMHD and MHD states with some initial imbalance tend towards 
enduring (i.e., decaying on the viscous/resistive time scale) pure Elsasser 
solutions, due to relatively slower decay of the cross-helicity compared 
to energy (\secref{sec:decay_to_Elsasser}). 
\vskip2mm
(ii) RMHD, 2D MHD, and non-helical, zero-mean-field, 3D MHD turbulence 
starting in magnetically dominated and, 
therefore, balanced configurations, settle into a reconnection-controlled 
decay towards ever-larger-scale magnetic structures accompanied by flows whose 
kinetic energy is a time-independent fraction of the magnetic one 
(\secsref{sec:decay_RMHD}, \ref{sec:decay_2D_rec}, and \ref{sec:decay_nonhel_new}). 
This decay is ``selective'', constrained by the conservation of certain invariants: 
anastrophy in 2D MHD (\secref{sec:decay_2D}), the Hosking invariant 
(the Saffman helicity invariant) in 3D MHD (\secref{sec:decay_nonhel_new}), 
and probably the Saffman cross-helicity invariant in RMHD (\secref{sec:RMHD_xhel}). 
\vskip2mm
(iii) 3D MHD turbulence with no mean field but finite net helicity ends up in 
a decaying state dominated by an approximately force-free magnetic field. 
The decay is reconnection-controlled for magnetically dominated states 
(\secref{sec:decay_hel_rec}). It is constrained by the conservation of helicity 
(\secref{sec:decay_hel}) and, if it starts in a state with comparable magnetic and 
kinetic energies, possibly also by the Saffman cross-helicity invariant  
(\secref{sec:decay_bal}). In the latter case, the kinetic energy decays 
faster than magnetic, eventually pushing the system towards the magnetically 
dominated regime. 
\vskip2mm
The different regimes are distinguished by different scaling laws for energies 
and energy-containing scales vs.~time. In all cases, the energy-containing scale 
grows---and that is the result not just of larger structures decaying slower 
but also of some actual transfer of energy to larger scales, unlike 
in hydrodynamic turbulence (\secref{sec:perma}). In reconnection-controlled 
regimes, this transfer is achieved dynamically by coalescence of magnetic structures. 

At sufficiently small scales, all these different types of decaying turbulence 
probably behave similarly to their forced counterparts, although it remains 
a challenging computational task to confirm this definitively
(\secref{sec:decay_spectra}). Reconnection-controlled decaying MHD turbulence 
below the energy-containing scale is probably a case of (many instances of) 
reconnection-driven turbulence discussed in~\secref{sec:turb_sheet}. 

\section{MHD Dynamo Meets Reconnection}
\label{sec:dynamo}

\vskip2mm
\begin{flushright}
{\small \parbox{8.5cm}{As the field becomes more and more tangled, there will be places
where the field is sharply reversed, and magnetic reconnection may set in, removing
the sharpest kinks.}
\vskip2mm
\citet{kulsrud92}} 
\end{flushright}
\vskip5mm

An interesting and distinct type of MHD turbulence about which I have so far said nothing 
except in the context of its decay (\secref{sec:decaying}) 
is the case of no mean field. Starting with a steady-state, 
forced hydrodynamic turbulence and a dynamically weak, randomly tangled 
magnetic field, one observes exponential 
growth of the latter, a phenomenon known as {\em small-scale dynamo} (or {\em fluctuation dynamo})---expected already by \citet{batchelor50} and \citet{biermann51} 
and since confirmed numerically \citep{meneguzzi81} and experimentally 
\citep{tzeferacos18,bott21,bott22}. 
The system eventually 
saturates with magnetic energy comparable to kinetic, but not, it seems, necessarily equal 
to it scale by scale---what the final state is remains an unsolved problem, both numerically 
(due to lack of resolution) and theoretically (due to lack of theoreticians). 
Furthermore, it matters whether the turbulence possesses net helicity 
(injected into the flow by the forcing) 
and/or has a large-scale shear superimposed on it---if it does, 
small-scale dynamo is accompanied by a {\em mean-field dynamo}, 
leading to growth of a large-scale field (the large scale in question being generally larger 
than the outer scale of the turbulence). 
Saturated states of such dynamos are also poorly understood, for the same reasons. 

Turbulent dynamos deserve 
a separate review---and they have recently received a superb one, by \citet{rincon19}, 
to which I enthusiastically refer all interested public. 
This said, the ideas associated with the role of tearing in RMHD turbulence, 
reviewed in \secref{sec:disruption}, turn out to have some direct bearing on 
the ``purest'' (homogeneous, non-helical, unsheared) small-scale dynamo problem. 
This is, therefore, a natural place for some discussion of~it. 

\subsection{Old Arguments About Saturated Dynamo at Large $\Pm$}
\label{sec:dynamo_old}

\begin{figure}
\centerline{\includegraphics[width=0.75\textwidth]{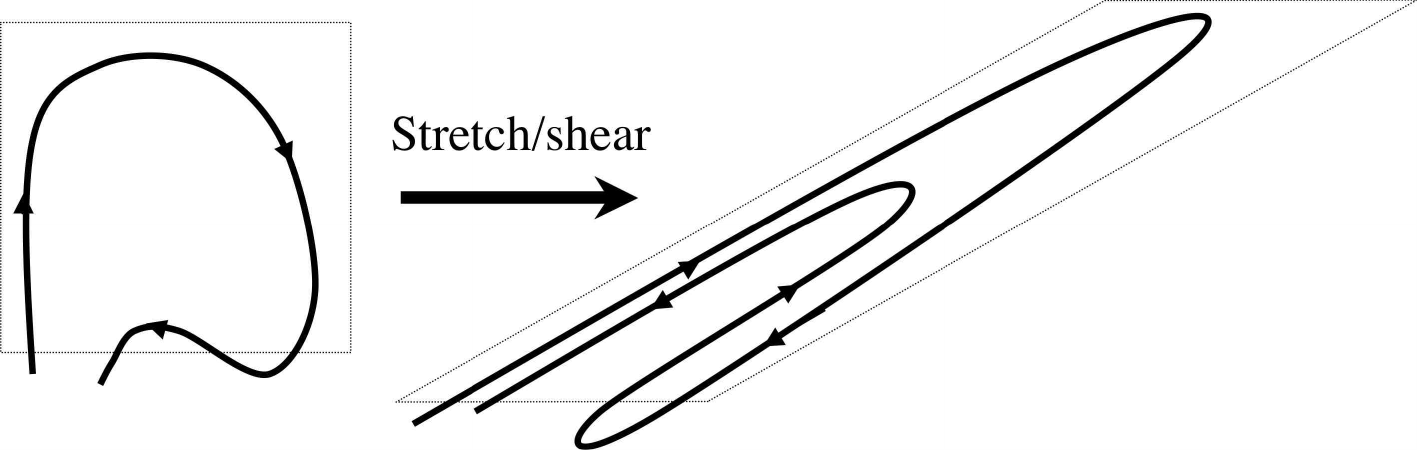}} 
\vskip2mm
\caption{Stretching/shearing a field line produces direction reversals 
(cartoon from \citealt{sch07review}).}
\label{fig:stretch}
\end{figure}

In regimes with $\Pm>1$,\footnote{After I first worked on this problem 
\citep{sch02njp,sch04aniso,sch04dynamo}, 
I grew quite sceptical about the relevance of the $\Pm\gg1$ MHD dynamo to any real-world 
situations: plasmas that formally have high $\Pm$ 
(e.g., the hot interstellar medium or the intergalactic medium in galaxy clusters) 
tend to be very hot and tenuous and, therefore, not very collisional, 
so MHD with Laplacian viscosity cannot possibly apply there \citep[see, e.g.,][and further 
discussion in \secref{sec:PA}]{sch06}. However, recent kinetic simulations 
of dynamo in such plasmas \citep{rincon16,kunz16,stonge18,stonge20} 
appear to be showing many familiar 
large-$\Pm$ features, perhaps because plasma microphysics conspire to produce an effectively 
collisional medium, which might not be entirely dissimilar from a large-$\Pm$ MHD fluid. 
Furthermore, some of the first laboratory plasma dynamos, 
achieved in laser-plasma experiments, 
have turned out to be right in the collisional, $\Pm\gtrsim 1$, MHD regime 
\citep{tzeferacos18,bott21,bott22}. 
Thus, it seems that my scepticism was premature and we ought to tackle the large-$\Pm$ 
dynamo with renewed vigour and sense of relevance. In contrast with $\Pm\gg1$, 
the limit of $\Pm\ll1$ is much 
more straightforwardly relevant: liquid metals and plasmas in convective zones of stars 
{\em are} comfortably collisional MHD fluids, and there are many other examples. This case  
appears, however, to be quite different physically, at least in the kinematic regime, 
and even less well understood, although numerically we do know that 
there is dynamo \citep{iskakov07,sch07dynamo,brandenburg18} and that it has 
some kind of saturated state \citep{brandenburg11,sahoo11}---conclusions to obtain which, 
one still has to push at the resolution limits of currently achievable MHD simulations. 
A massive paper by \citet{sahoo11} contains a wealth of sophisticated statistical 
information but does not answer any of the more basic questions 
(their one distinctive physical conclusion is that the low-$\Pm$ case 
is less intermittent than the high-$\Pm$ one, which is plausible). 
I am not aware of any other systematic numerical 
study of how low-$\Pm$ dynamo saturates---an opportunity for a definitive contribution 
that some enterprising researcher with an MHD code and a large allocation 
of computing time should seize (there is a promise of that in \citealt{mckay19}).} 
small-scale dynamo can be understood as the process of 
a velocity field, restricted to scales above the viscous cutoff, randomly 
stretching and shearing the magnetic field, which is allowed to go to smaller 
scales, limited only by the Ohmic resistivity. Intuitively, 
it is not hard to see that embedding a tangled field line into an ``eddy'' will 
lead to the field line being stretched and folded, resulting in a configuration 
featuring field reversals on ever smaller scales (\figref{fig:stretch}). 
A combination of numerical evidence 
and theoretical arguments (see \citealt{sch07review}, \citealt{rincon19}, 
and references therein)
confirms that this process does indeed lead to net amplification of magnetic energy, 
with that energy residing preferentially in ``folds''---magnetic 
fields that reverse direction across themselves on the resistive scale 
and remain approximately straight along themselves up to the scale of the 
velocity field. When the dynamo saturates, it does so in some poorly understood 
way involving these bundles of alternating fields back-reacting on 
the turbulent flow and arresting further amplification. Whereas in the kinematic-dynamo 
stage (i.e., when the field is dynamically weak), the spectrum of the magnetic energy 
certainly peaks at the resistive scale \citep{sch04dynamo}, 
what exactly happens in the saturated state is a matter 
of some debate. It is tempting to argue, with \citet{biermann51} and
\citet{kraichnan65}, that the system will sort itself out 
into a state where the magnetic energy is at the outer scale, while the smaller 
scales behave in exactly the same way as they would do in the presence of a strong 
mean field. Whether numerical evidence confirms this view is, at the resolutions 
achieved so far, in the eye of the beholder 
(\citealt{kida91,haugen03,haugen04,cho09dynamo,beresnyak09slopes,beresnyak12dynamo,teaca11,eyink13,porter15,grete17,grete21,mckay19,bian19,brandenburg19,seta20}; see \figref{fig:dynamo_spectra}b). 
The alternative possibility 
is that the magnetic energy stays at small scales---not quite as small as in the kinematic 
regime, but still determined by resistivity \citep{sch02njp,sch04dynamo,maron04}. 
The claim is that the folded field structure persists in saturation, with the folds elongating 
to the outer scale ($L$) of the turbulence and direction reversals within folds occurring on 
the scale $\lres\sim L\,\Rm^{-1/2}$, where the stretching rate ($\sim \du_L/L$) 
balances the Ohmic-diffusion rate ($\sim\eta/\lres^2$). 

Despite being associated with the latter point of view, I am not going to defend 
it here in its original form because of certain little known but consequential numerical 
developments, described in \secref{sec:dynamo_num}, that occurred after that debate had 
its heyday. Instead, drawing on the ideas of \secref{sec:disruption}, I will 
propose, in \secsand{sec:dynamo_theory}{sec:dynamo_scenarios}, 
an amended view of the saturated state 
of turbulent dynamo, in which reconnection and MHD turbulence will again meet 
and collaborate.  

\subsection{Numerical Evidence: Reconnection Strikes Again}
\label{sec:dynamo_num}

\begin{figure}
\begin{center}
\begin{tabular}{cc}
\includegraphics[width=0.475\textwidth]{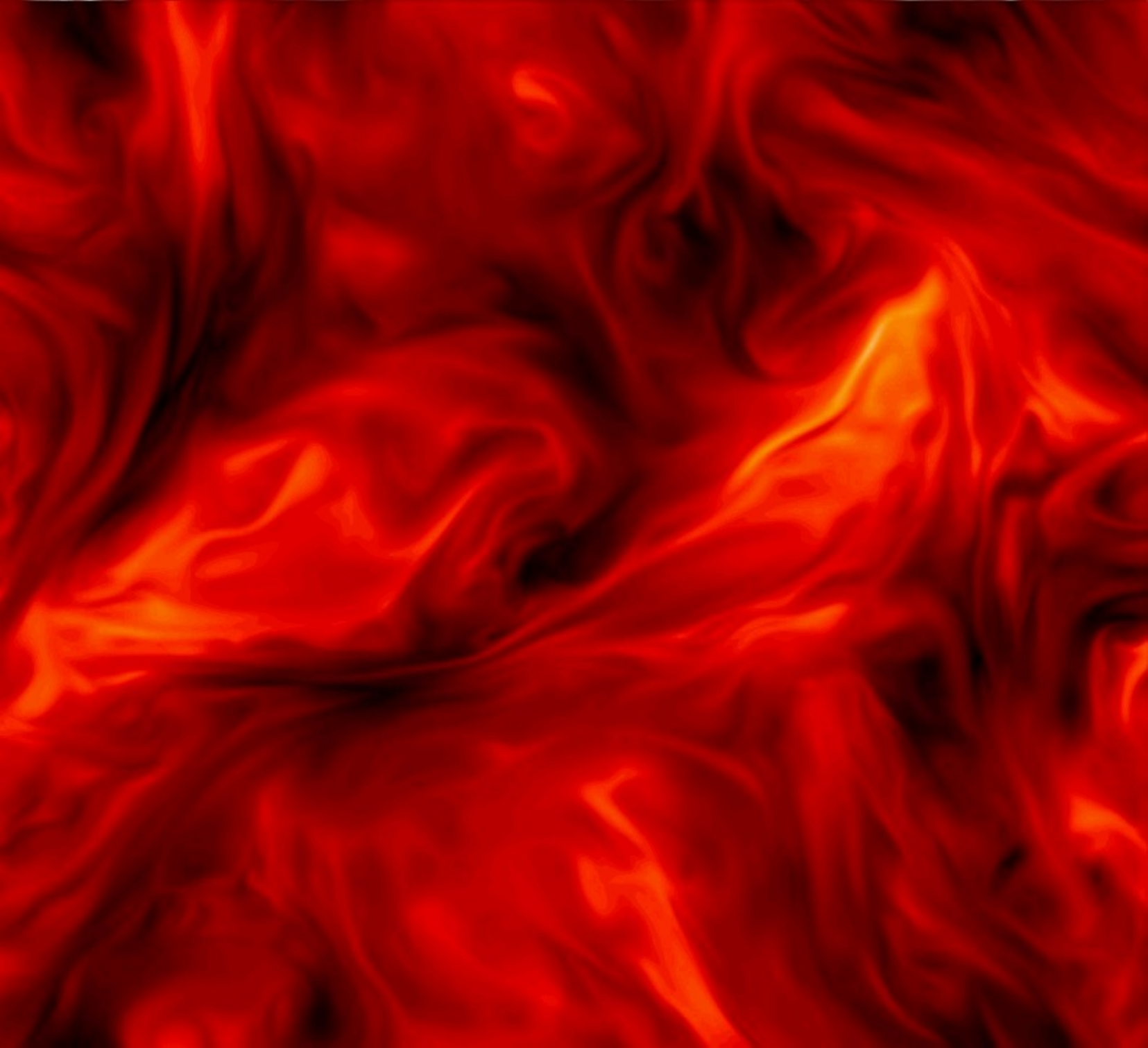} &
\includegraphics[width=0.475\textwidth]{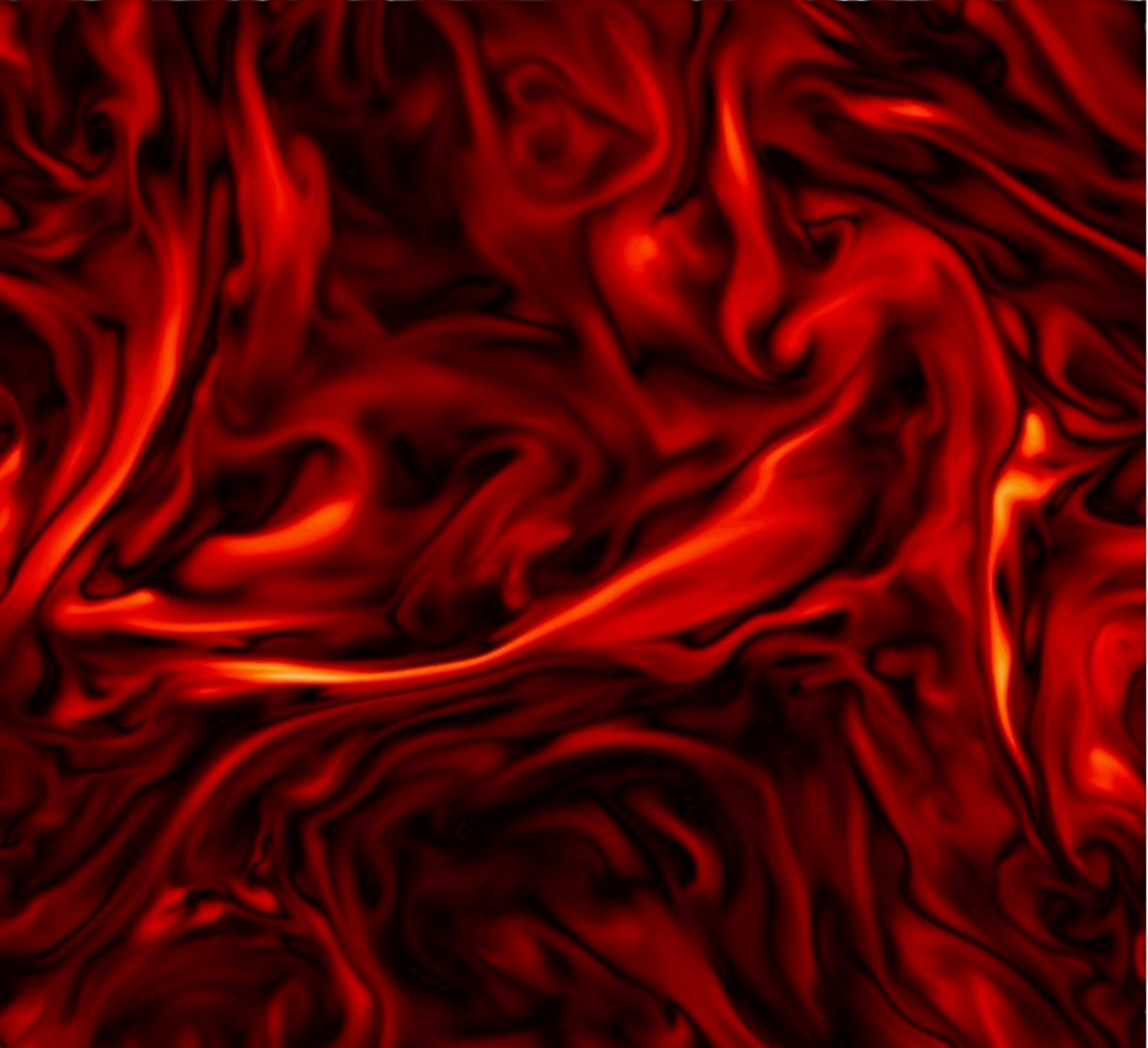} \\ 
(a) $|\vu|$ & (b) $|\vB|$ 
\end{tabular}
\vskip2mm
\includegraphics[width=0.85\textwidth]{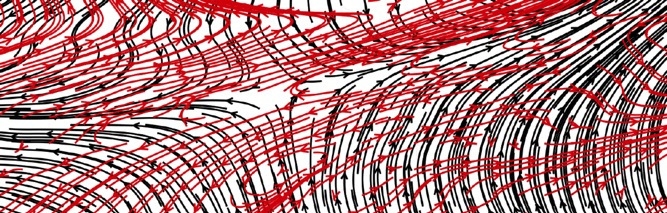}\\
(c) A cut out from above: flows in black, magnetic field lines in red
\end{center}
\caption{From unpublished work by \citet{iskakov08}: a $512^3$ incompressible-MHD 
simulation of saturated fluctuation dynamo, $\Pm=1$, $\Re=1360$ (defined $=\urms/\nu k_0$, 
where $k_0=2\pi$ is the forcing wavenumber, corresponding to the box size; this is the same 
numerical set up as in \citealt{iskakov07} and \citealt{sch07dynamo}). These are 2D cuts from 
instantaneous snapshots of absolute values of (a) velocity, (b) magnetic field. 
Panel (c) is a cut out from these snapshots, zooming in on the horizontal 
fold just down towards the left from the centre of the snapshot. 
Stream lines are in black and field lines are in red.
A reconnecting-sheet structure, with field reversal, inflows and outflows is manifest.
Very pretty 3D visualisations of this kind of reconnecting structure extracted from an 
MHD turbulence simulation can be found in \citet{lalescu15}.}
\label{fig:iskakov}
\end{figure}

\begin{figure}
\begin{center}
\includegraphics[width=0.95\textwidth]{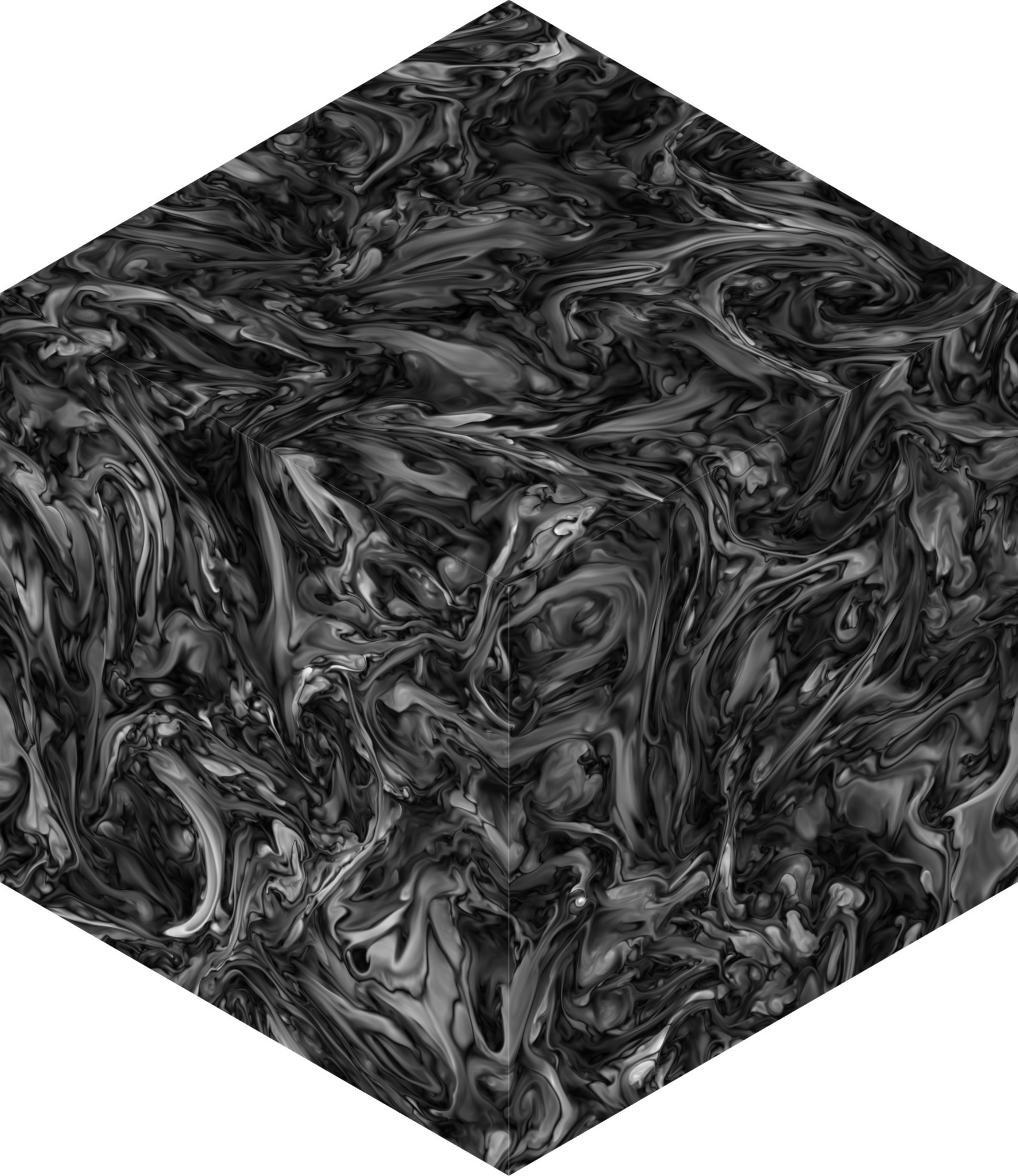}
\end{center}
\caption{From unpublished work by \citet{beresnyak12unpub} (reproduced with the kind 
permission of the author): 
snapshot of the absolute value of magnetic field in a $\Pm=10$ and $\Re\approx500$ 
simulation at $1024^3$ 
(the same numerical set up as in \citealt{beresnyak09slopes} and \citealt{beresnyak12dynamo}; 
note that Beresnyak defines his $\Re$ in terms of the ``true'' integral scale of the flow 
calculated from its spectrum). Plasmoids/fold corrugations galore. 
In his other simulations within this sequence, 
there are even more plasmoid-like-looking features 
at $\Pm=1$ and $\Re\approx6000$, with some sign of them 
breaking up into even smaller structures (cf.~\secref{sec:rec_dyn}). 
In contrast, they start disappearing at $\Pm=10^2$ and $\Re\approx 80$ 
and are gone completely in the ``Stokes'' regime $\Pm=10^4$ and $\Re\approx2$.
Very similar results have been found by \citet{galishnikova22}.}
\label{fig:beresnyak_dynamo}
\end{figure}

The existence of turbulent dynamo was definitively established by \citet{meneguzzi81} in 
what was then a ``hero'' $64^3$ MHD simulation---one of those {\em bona fide} numerical  
discoveries that make computer simulations worthwhile. 20 years later, 
when the debate about the nature of the saturated dynamo state  
focused on interpreting newly accessible improved 
numerical evidence \citep{kinney00,sch04dynamo,maron04,haugen03,haugen04}, 
everyone was staring at not-very-conclusive magnetic spectra with some pronounced 
excess of the magnetic energy over kinetic at small scales, and at visualisations 
of magnetic fields organised in folds (especially at large $\Pm$). 
One could be a believer in universality and think of this as a non-asymptotic state 
that would, at infinite resolution, turn into the usual Kolmogorov-style turbulence spectrum, 
with magnetic energy shifting to the outer scales \citep{haugen03,haugen04,beresnyak09slopes,beresnyak12dynamo}---or one could rely on a different kind of physical 
intuition and argue that there was no obvious physical mechanism for unwrapping 
fields folded at the resistive scale (that was my view). 

In later, sadly unfinished, work, \citet{iskakov08} discovered, however, that, 
in simulations with moderate $\Pm\ge1$ and large $\Re$ (the former being the only affordable 
possibility compatible with the latter), magnetic folds in the nonlinear regime 
became current sheets, with very clear inflow--outflow patterns around 
the field reversals (\figref{fig:iskakov}). 
One might say that this should have been obvious from the start, although 
perhaps less so in the case of $\Pm\gg1$ (see \secref{sec:rec_dyn}).
We also found that the folds became corrugated, and
plasmoid-like structures (probably flux ropes) formed, with (perhaps) approximately circularised 
cross-sections. Larger simulations by \citet{beresnyak12unpub}, 
also unpublished (except for some bits in 
\citealt{beresnyak09slopes} and \citealt{beresnyak12dynamo}), revealed the same 
feature, with the numerical box now teaming with small plasmoid-like structures 
and rippled folds (\figref{fig:beresnyak_dynamo}), a result confirmed 
at even higher resolutions by \citet{galishnikova22}. 
Thus, while the folds could not perhaps be literally unwrapped, they did turn out 
to be prone to breaking up and seeding populations of smaller structures.\footnote{Note that
none of these authors saw any of this happen in the ``Stokes'' regime $\Re\sim1$, $\Pm\gg1$,
which is the only numerically accessible case if one wants very large $\Pm$, and on which much
of the previous physical intuition \citep{sch04dynamo} had been based: there, the saturated state
just consisted of magnetic fields smoothly folded on the resistive scale. I shall argue in \secref{sec:rec_dyn} that this makes sense.} 

There is little definitive analysis of all this available in print. There is, however, 
an intriguing finding by \citet{brandenburg14}, who analysed his own simulations and those 
of \citet{sahoo11} and discovered that the ratio of energy dissipated resistively to that 
dissipated viscously decreased at larger $\Pm$ (\citealt{beresnyak12unpub} 
also had this result, and \citealt{galishnikova22} confirm it;
\citealt{mckay19}, however, raise a degree of doubt as to 
whether it will survive at larger~$\Rm$). 
One might plausibly argue that something like this could happen if kinetic energy, 
first converted into magnetic one as fields were amplified and folded by large-scale 
turbulent flows, were then to be recovered from magnetic energy at smaller scales 
as fluid motions were generated by reconnection and instabilities (presumably, 
tearing instabilities) in the folds. \citet{brandenburg19}, while they do not 
engage with the notion of reconnecting folds, do confirm explicitly that, 
in larger-$\Pm$ simulations, there is increasing net transfer of magnetic to kinetic 
energy at small scales, with kinetic energy's viscous thermalisation 
increasingly dominating the overall dissipation rate (this was checked, and checked out,
in the simulations by \citealt{galishnikova22}).   

\begin{figure}
\begin{center}
\begin{tabular}{cc}
\parbox{0.49\textwidth}{
\includegraphics[width=0.49\textwidth]{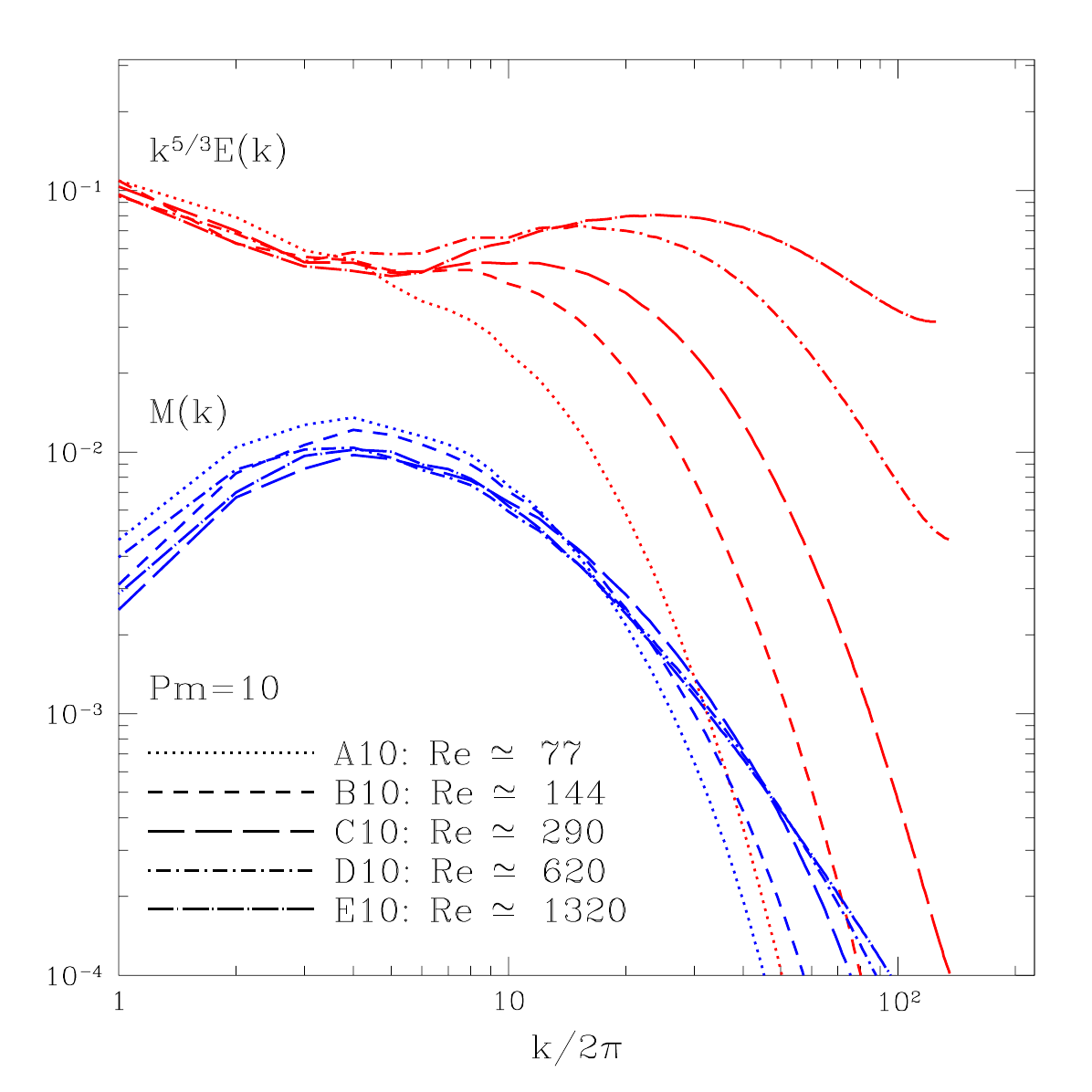}} & 
\parbox{0.49\textwidth}{
\includegraphics[width=0.49\textwidth]{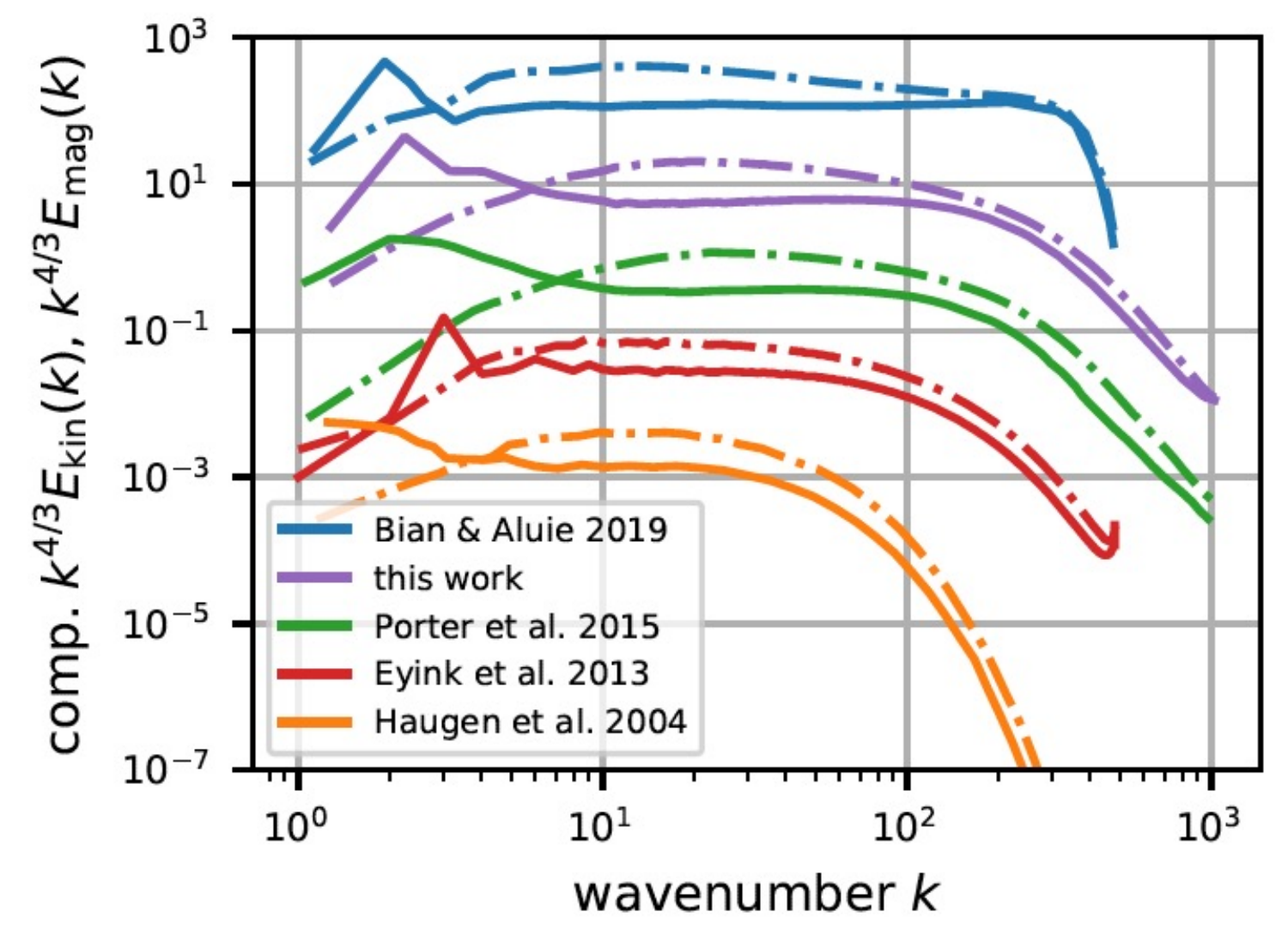}\\\\\\\\}\\
(a) \citet{iskakov08} & (b) \citet{grete21}
\end{tabular}
\end{center}
\caption{(a) Spectra of saturated MHD dynamo: kinetic-energy (red, compensated by $k^{5/3}$) 
and magnetic-energy (blue) spectra from a series of incompressible-MHD simulations 
with $\Pm=10$ and increasing $\Re$ by \citet{iskakov08}.  
The numerical set up is the same as in \citet{sch07dynamo};  
the resolution is $512^3$ (so the highest-$\Re$ run may be numerically suspect). 
(b) A summary plot from \citet{grete21} (\copyright AAS, reproduced with permission) 
of their and several other numerical studies, viz., 
from top to bottom, \citet{bian19,grete21,porter15,eyink13,haugen04}. The kinetic-energy (solid lines) 
and magnetic-energy (dot-dashed lines) spectra are all compensated by $k^{4/3}$
to highlight the shallow kinetic-energy spectrum at small scales. 
\citet{galishnikova22} report very similar spectra and find also 
the magnetic spectrum steepening at even smaller scales,
perhaps vindicating the prediction of \secref{sec:rec_dyn}.}
\label{fig:dynamo_spectra}
\end{figure}

A signature of this behaviour is discernible if one examines the magnetic- and, especially, 
kinetic-energy spectra in saturated dynamo simulations at relatively high resolutions 
($512^3$ and up), without attempting to see what one might want to see, e.g., 
scale-by-scale equipartition or $k^{-5/3}$. \Figref{fig:dynamo_spectra}(a), taken from 
the unpublished simulations by \citet{iskakov08}, shows that the kinetic-energy spectrum 
steepens at large scales compared to the hydrodynamic case 
(the empirical slope is $k^{-7/3}$; see \citealt{sch04dynamo} and \citealt{stonge20}), 
but picks up around the same wavenumber where the magnetic-energy spectrum has its peak 
and becomes shallower than Kolmogorov---\citet{grete21} find $k^{-4/3}$ to be a good fit, 
both in their simulations and, in retrospect, in many previous ones (\figref{fig:dynamo_spectra}b). 
By analysing energy transfers, they attribute this to significant transfer of energy 
from magnetic to kinetic, by means of the tension force---this is analogous to 
the result of \citet{brandenburg19} and again entirely consistent with 
(although does not amount to a definitive confirmation of) the idea that reconnection 
in the folds drives small-scale motions, which come to dominate the kinetic-energy spectrum
at those scales. This is perhaps reinforced by their \citep{grete21} observation 
(in disagreement with \citealt{bian19}) that the energy cascade at small scales is almost 
entirely controlled by magnetic forces, rather than by hydrodynamic advection. 
The paper by \citet{grete21} appears to be the first one for well over a decade that, 
having measured different kinetic- and magnetic-energy spectra, 
dares to consider the possibility that this might mean something physical, 
rather than merely insufficient resolution for expected asymptotic recovery of $k^{-5/3}$ or~$k^{-3/2}$.

Thus, reconnection appears to have caught up with dynamo, just as it did with 
Alfv\'enic turbulence in \secref{sec:disruption}, the general principle at work 
in both cases being that while 
large-scale motions push magnetic fields into small-scale, 
direction-reversing configurations, 
resistive effects invariably manage to break those up (provided $\Re\gg1$; see below).   

\subsection{Towards a New Theory of Reconnection-Limited Dynamo} 
\label{sec:dynamo_theory}

\subsubsection{Kinematic Dynamo and Onset of Tearing}
\label{sec:kin_dyn}

Consider first a weak, tangled magnetic field being stretched by fluid motions whose scale 
is $\ell$ (why I call it $\ell$ rather than $\lambda$ is about to become obvious).
Let us imagine that these fluid motions are part of vanilla Kolmogorov turbulence, 
described, inevitably, by~\exref{eq:u_K41}: 
\beq
\du_\ell\sim(\eps\ell)^{1/3}.
\label{eq:dynamo_K41}
\eeq 
Balancing the associated stretching rate with the Ohmic-diffusion rate gives one the resistive 
scale: 
\beq
\tnl^{-1} \sim \frac{\du_\ell}{\ell} \sim \frac{\eps^{1/3}}{\ell^{2/3}}
\sim\tres^{-1}\sim\frac{\eta}{\lres^2}
\hence
\lres(\ell)\sim(\eta\tnl)^{1/2} \sim \ell\, \Rm_\ell^{-1/2},\quad
\Rm_\ell = \frac{\du_\ell \ell}{\eta}.
\label{eq:lres_dynamo}
\eeq 
The scale $\lres(\ell)$ is the reversal scale of the magnetic field generated by the
dynamo of the eddies of size~$\ell$; this field's 
typical coherence scale {\em along itself} will be~$\ell$. 

Imagine now a general configuration in which magnetic field $B_\lambda$ 
(as usual, in velocity units) 
reverses direction on some scale $\lambda$, not necessarily equal to $\lres$.  
It will be subject to tearing at 
the rate \exref{eq:gmax_TM}, but with $\vAy$ replaced by~$B_\lambda$: 
\beq
\label{eq:tearing_dynamo}
\gamma \sim \frac{B_\lambda}{\lambda}\,S_\lambda^{-1/2}(1+\Pm)^{-1/4}
\sim \frac{B_\lambda^{1/2}}{\lambda^{3/2}}\,\eta^{1/2}(1+\Pm)^{-1/4}. 
\eeq
When $B_\lambda$ is infinitesimally small, as it would be in the kinematic stage 
of the dynamo, the tearing rate is small, $\gamma\ll\tres^{-1}$. 
It will become comparable to the resistive-diffusion rate 
at $\lambda=\lres$ when the fields reversing at this scale grow to be at least   
\beq
\label{eq:nl_threshold}
B_{\lres} \sim \frac{\eta}{\lres}(1+\Pm)^{1/2} \sim \du_\ell\tRe_\ell^{-1/2},
\quad
\tRe_\ell = \frac{\du_\ell \ell}{\nu+\eta} = \frac{\Rm_\ell}{1+\Pm}. 
\eeq
Here $\tRe_\ell$ is equal to the usual Reynolds number~$\Re_\ell$ when 
$\Pm\gtrsim1$ and to the magnetic Reynolds number~$\Rm_\ell$ 
when $\Pm\ll1$ [cf.~\exref{eq:lres_Rm}]. 
In the former case, since the stretching rate $\tnl^{-1}$ at the viscous scale 
$\ell\sim \lnu = \eps^{-1/4}\nu^{3/4}$ is the largest, 
it is the viscous-scale eddies that will play the dominant role 
in amplifying an infinitesimally small magnetic field, 
but the dynamo will go nonlinear as soon as the field's energy becomes comparable 
to the energy of the viscous-scale motions, $B_{\lres}\sim\du_{\lnu}$. 
Since, by definition of $\lnu$, $\Re_{\lnu}\sim1$, 
the estimate \exref{eq:nl_threshold} also turns into $B_{\lres}\sim\du_{\lnu}$, i.e.,
tearing in the folds will start outpacing 
Ohmic diffusion at exactly the same moment as the nonlinearity kicks in
(this is perhaps obvious because tearing needs Lorentz force: see \apref{app:TM}). 
Thus, a nonlinear dynamo is also a reconnecting dynamo, which I shall 
call ``tearing-limited''. 

In the limit of $\Pm\ll1$, the fastest eddies capable of field amplification  
are at the resistive scale, $\ell\sim\lres$ \citep[e.g.,][]{boldyrev04}. 
Since $\tRe_{\lres}\sim\Rm_{\lres}\sim1$, 
the estimate \exref{eq:nl_threshold} becomes $B_{\lres}\sim\du_{\lres}$, so it
again tells us that the nonlinearity and tearing become important at the same time. 
Admittedly, there is no longer a scale separation between $\ell$ and $\lres$ in this 
situation, so the magnetic field is not, strictly speaking, ``folded'' 
(this is quite obvious from the snapshots of growing fields in \citealt{sch07dynamo}), 
although one 
might still speculate that tearing is possible across generic $X$-point configurations. 
I shall keep my discussion general, but it might be easier for a doubtful reader  
just to think of large $\Pm$ in all cases. 

\subsubsection{Self-Similar Dynamo}
\label{sec:ssim_dyn}

It has been argued by \citet{sch02njp,sch04dynamo} and \citet{maron04} 
(with later variants by \citealt{beresnyak12dynamo} and \citealt{xulaz16}) 
and numerically confirmed in a conclusive fashion by \citet{cho09weakB} and \citet{beresnyak12dynamo} 
that, once the dynamo goes nonlinear, the field continues to be amplified, but 
by ever larger-scale motions that are, at a given time, just as energetic as the 
field is at that time.\footnote{There is a nice direct demonstration of that in the paper by 
\citet{brandenburg19}, who measure the energy transfer from the velocity to the 
magnetic field and show that the sign of this transfer reverses at a scale that increases with time: 
the eddies above that scale act as a dynamo, while below that scale, the dynamo-generated 
fields drive some secondary flows, which then dissipate viscously. 
Analogous conclusions, by analogous means, were reached by \citet{bian19}, 
\citet{stonge20}, \citet{grete21}, and \citet{galishnikova22}. The experimental
dynamo observed in laser plasmas may be in this regime \citep{bott21,bott22}.} 
That is, the scale $\ell(t)$ of the motions amplifying the field at time~$t$ is set by the condition 
\beq
\du_{\ell(t)} \sim B_\lambda(t).
\eeq 
This leads, neatly, to a self-similar amplification regime: 
\beq
\frac{\rmd B_\lambda^2}{\rmd t} \sim \frac{\du_\ell}{\ell}B_\lambda^2 
\sim \frac{\du_\ell^3}{\ell} \sim \eps
\hence
B_\lambda(t)\sim (\eps t)^{1/2}, \quad \ell(t)\sim \eps^{1/2}t^{3/2}. 
\label{eq:dynamo_ssim}
\eeq
After a few outer-scale eddy-turnover times, $t\sim L/\du_L$, the field's energy becomes 
comparable to that of the flow, $B_\lambda\sim\du_L$, and the dynamo saturates.
At any time during the self-similar growth, the cascade below $\ell$ presumably looks 
just like the cascade in the saturated state, whereas above $\ell$, the turbulence is 
still hydrodynamic.  

\subsubsection{Tearing-Limited Dynamo: Universality Regained}
\label{sec:rec_dyn}

\citet{sch02njp,sch04dynamo} calculated the field-reversal scale $\lambda$ 
in the self-similar and saturated dynamo regimes by balancing $\du_\ell/\ell$ 
with the Ohmic-dissipation rate $\eta/\lambda^2$. 
I now know, thanks to the argument in \secref{sec:kin_dyn} (obvious in retrospect!), 
that the folds generated by this process will in fact tear faster than they diffuse. 
So let me therefore balance the tearing rate \exref{eq:tearing_dynamo} 
with $\du_\ell/\ell$ and obtain a scale familiar from the ``ideal-tearing'' 
condition \exref{eq:pucci} \citep{pucci14,tenerani15visc}:
\beq
\label{eq:lambda_rev}
\lambda(\ell) \sim \ell\, \Rm_\ell^{-1/3}(1+\Pm)^{-1/6}
\sim \eps^{-1/9}\ell^{5/9}\eta^{1/3}(1+\Pm)^{-1/6}.
\eeq
In order for tearing to supersede Ohmic diffusion, we must have\footnote{This condition 
means that the ``Stokes'' ($\Re\lesssim 1$) 
simulations of \citet{kinney00} and \citet{sch04dynamo} could not have captured this effect.
In the simulations of \citet{iskakov08} and \citet{beresnyak12unpub}, one can see very clearly 
that when $\Pm$ is increased, which, at fixed finite resolution, has to happen at the expense 
of $\Re$, the magnetic folds become ever smoother and plasmoids/fold corrugations 
ever fewer, until they disappear entirely. Interestingly, at a given $\Re$, larger 
values of $\Pm$ appear to promote the break up of the folds---perhaps 
because their aspect ratio $\ell/\lambda$ is, according to \exref{eq:lambda_rev}, 
larger when $\Pm$ is larger, and so is the number of islands \exref{eq:N_dynamo} 
produced by the fastest-growing tearing mode.}  
\beq
\lambda(\ell) \gg \lres(\ell) \quad\Leftrightarrow\quad\tRe_\ell^{1/6}\gg1,
\eeq
where $\lres(\ell)$ was taken from \exref{eq:lres_dynamo}. 
Note that $\lambda\ll\ell$ always, except, for low $\Pm$, at the start 
of the self-similar regime, when $\Rm_\ell\sim1$ (this seems to suggest 
that even a low-$\Pm$ dynamo may form reconnecting folds in the nonlinear regime). 

Let us imagine for now that that the self-similar evolution \exref{eq:dynamo_ssim} has run 
its course and the dynamo has saturated in a state where the only motions that are 
responsible for the (re)generation of the folds are on the outer scale, viz., $\ell\sim L$, 
while the motions below this scale no longer affect the magnetic field (I will experiment 
with relaxing this assumption in \secref{sec:multifolds}). The reversal scale 
of the folds is then set by \exref{eq:lambda_rev} with $\ell\sim L$. 
I shall call this scale 
\beq
\label{eq:lamL}
\lR = \lambda(L) \sim L\, \Rm_L^{-1/3}(1+\Pm)^{-1/6}.
\eeq  

Consider a fold of length $L$ and width $\lR$. 
Its tearing will produce islands whose number can be inferred from \exref{eq:kpeak_TM}: 
\beq
N \sim k_*L \sim \frac{L}{\lR} S_{\lR}^{-1/4}(1+\Pm)^{1/8}
\sim \Rm_L^{1/6}(1+\Pm)^{1/3}.  
\label{eq:N_dynamo}
\eeq 
Just as I did at the end of \secref{sec:disruption_scale}, 
I can argue here that these islands will grow, perhaps circularise, and turn into plasmoids 
(flux ropes) of transverse size~$\lR$, while their mother fold is destroyed.  
Similarly to \secref{sec:recturb}, I can entertain the possibility that they are the outer-scale 
structures of a new turbulent cascade, seeded by the tearing of the fold at the scale~$\lR$. 
At scales below $\lR$, this new cascade is of the usual RMHD
kind considered in \secsdash{sec:CB}{sec:disruption}---the mean field now is $B_{\lR}$, 
assuming that fields that make up 
the folds are unlikely to be exactly anti-parallel and so there is some component of the 
folded field, generally of the same order as its reversing component, pointing in the direction 
perpendicular both to the latter and to the direction of reversal.\footnote{There is perhaps 
a whiff of evidence for this in \citet{sch04dynamo}, who found that $\la|\vB\cdot\vJ|^2\ra$ 
in the nonlinear regime of the dynamo had the same $\Rm$ scaling as $\la|\vB\times\vJ|^2\ra$,
where $\vJ=\vdel\times\vB$. Precisely anti-parallel fields would have had $\vB\cdot\vJ=0$.}
 
Let the flux rope have a circulation velocity $\du_{\lR}$ and a perturbed 
field $\db_{\lR}\sim\du_{\lR}$. One can estimate these quantities by the same logic 
as led to \exref{eq:z_below}: if this new cascade is to carry (a finite fraction of) 
the same energy flux as produced the fold,\footnote{I am assuming here that tearing, while 
destroying the folds, does not dissipate a significant amount of energy directly: the role of 
resistivity in the process of tearing is to break magnetic field lines, not to remove magnetic 
energy. This is not necessarily obvious, but is perhaps backed up by the following unsurprising 
estimate of the fraction of energy dissipated by resistivity in magnetic structures 
of width $\lR$: 
$\eps_{\lR}/\eps \sim \eta B_{\lR}^2/\lR^2\eps \sim (L/\lR)^2\Rm_L^{-1} \sim\tRe_L^{-1/3}\ll1$.} 
then 
\beq
\frac{\du_{\lR}^3}{\lR}\sim\eps\hence
\du_{\lR}\sim(\eps\lR)^{1/3}\sim\du_L\lt(\frac{\lR}{L}\rt)^{1/3}
\sim \du_L\Rm_L^{-1/9}(1+\Pm)^{-1/18}\sim\db_{\lR}. 
\label{eq:u_dynamo}
\eeq
Finally, the length of the flux rope (its ``parallel'' scale) is set by critical balance: 
the scale over which coherence can be maintained by propagating information at the Alfv\'en 
speed $\sim B_{\lR}\sim\du_L$~is 
\beq
\lpar\sim\frac{B_{\lR}\lR}{\du_{\lR}}\sim L^{1/3}\lR^{2/3}
\sim L\, \Rm_L^{-2/9}(1+\Pm)^{-1/9}. 
\label{eq:lpar_dynamo}
\eeq 

\begin{figure}
\centerline{\includegraphics[width=0.75\textwidth]{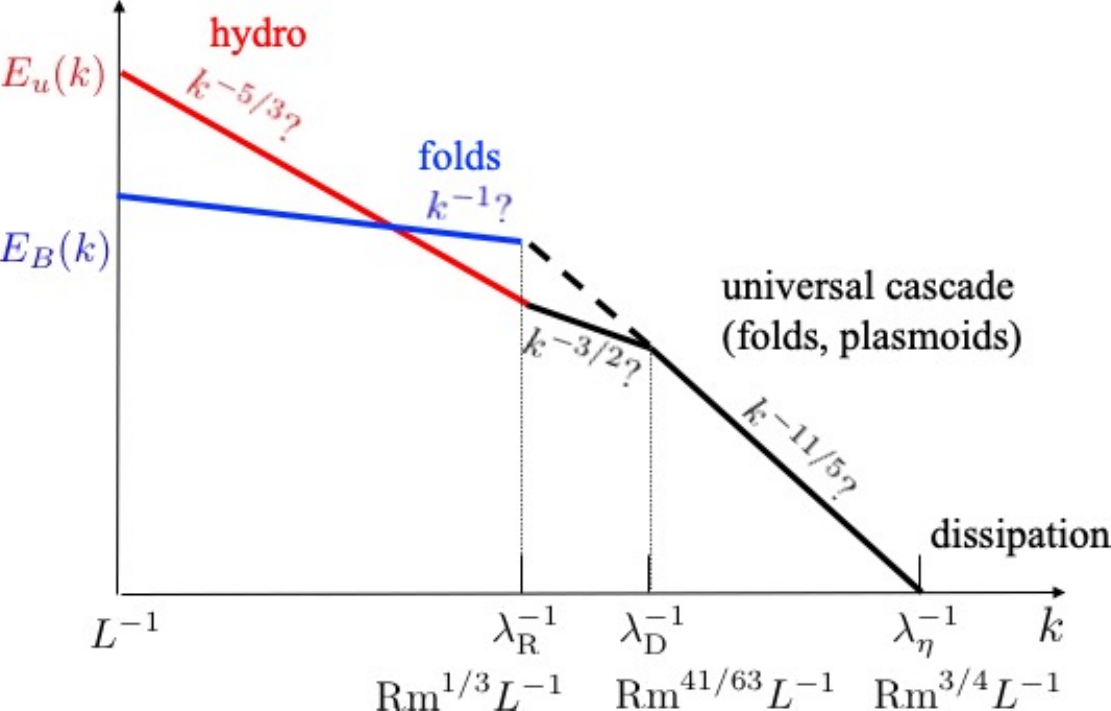}} 
\vskip2mm
\caption{Spectrum of isotropic MHD turbulence, which is the saturated state 
of small-scale dynamo. The universal cascade below the fold-reversal scale $\lR$ 
[see~\exref{eq:lamL}]
is described in~\secref{sec:rec_dyn}. Various options for the spectrum  
at $k\lR < 1$ are discussed in~\secref{sec:dynamo_scenarios}; in particular,
the $k^{-1}$ magnetic spectrum is~\exref{eq:EB_yousef}.
The velocity spectrum may well be steeper than $k^{-5/3}$ at the largest
scales and then shallower at smaller ones (see \secref{sec:dynamo_num}),
before steepening again at $k\lR\gtrsim 1$.
The disruption ($\lD$) and dissipation ($\lres$) scales are given by 
\exref{eq:dynamo_lD} and \exref{eq:dynamo_lres}, respectively. 
Whether the $k^{-11/5}$ spectrum starts at $\lR$ (\secref{sec:multifolds}) 
or at $\lD$ (\secref{sec:rec_dyn}) is not obvious because how the spectra 
at scales below and above~$\lR$ are connected remains an open question.}
\label{fig:dynamo_spectrum}
\end{figure}

Thus, we have got ourselves a critically balanced RMHD-type cascade, 
with $\du_L\sim B_{\lR}$ being the Alfv\'en speed, 
$\lR$ given by \exref{eq:lamL} playing the role of $\lCB$, 
$\lpar$ given by \exref{eq:lpar_dynamo} in the role of the parallel outer scale $\Lpar$, 
and the outer-scale amplitude $\du_{\lR}$ given by~\exref{eq:u_dynamo}. 
The RMHD ordering parameter for this cascade is, therefore, 
\beq
\epsilon \sim \frac{\du_{\lR}}{\du_L}\sim\frac{\lR}{\lpar} 
\sim \Rm_L^{-1/9}(1+\Pm)^{-1/18}\ll 1,
\eeq 
not terribly small in any real-world situation, but perfectly legitimate in principle. 
The Reynolds number of this cascade will generally large: 
\beq
\tRe_{\lR} \sim \frac{\du_{\lR}\lR}{\nu+\eta}\sim\Rm_L^{5/9}(1+\Pm)^{-11/9}\gg1
\quad\Leftrightarrow\quad
\tRe_L \gg (1+\Pm)^{6/5}. 
\eeq
The latter inequality might not always be satisfied when $\Pm\gg1$, but again is a perfectly 
legitimate limit. When it is {\em not} satisfied, the motions produced by the 
tearing of the folds will be quickly dissipated by viscosity and thus cannot seed a proper 
cascade; one option then is to invoke \secref{sec:multifolds} to deal with 
the sub-$\lR$ structure. 

The RMHD cascade seeded by the production of flux ropes in tearing folds, 
as just described, will do what an RMHD cascade does: push energy to smaller scales, 
become aligned and be eventually disrupted by tearing. The arguments of \secref{sec:recturb} 
apply: a succession of mini-cascades will be seeded, etc., as per \figref{fig:spectrum}. 
One expects a $k^{-3/2}$ spectrum \exref{eq:Perez} down to the disruption scale set 
by \exref{eq:lambdaD}, viz., 
\beq
\label{eq:dynamo_lD}
\lD \sim \lR\Rm_{\lR}^{-4/7}(1+\Pm)^{-2/7} \sim L\,\Rm_L^{-41/63}(1+\Pm)^{-41/126}. 
\eeq
Below this scale, the mini-cascades will have a $k^{-11/5}$ spectral envelope described
by~\exref{eq:Ek_bounds} (\figref{fig:dynamo_spectrum}). 
The final resistive cut-off is then determined by \exref{eq:K_cutoff}: 
\beq
\lres \sim \lR \tRe_{\lR}^{-3/4} \sim \lR\Rm_L^{-5/12}(1+\Pm)^{11/12} 
\sim L\,\tRe_L^{-3/4}, 
\label{eq:dynamo_lres}
\eeq
the Kolmogorov scale again---reassuringly, some things in the world never change. 

Thus, turbulence in the saturated state of small-scale dynamo is,
at scales below $\lR$, likely to be similar 
to the tearing-mediated turbulence of \secref{sec:recturb}, the only difference being 
that the direction of the ``local mean field'' will be fluctuating strongly and all 
the statistics will be isotropic overall (although certainly not isotropic with  
respect to this fluctuating local mean field, as indeed spotted
by~\citealt{beresnyak09slopes} and, most recently, studied by \citealt{stonge20};
in a decaying, no-mean-field set up, this was shown already by \citealt{milano01}). 
An aesthetically pleasing conclusion from all this is that 
universality is regained at small scales: even without the crutch of the mean field, 
MHD turbulence manages to turn itself into RMHD (Alfv\'enic) turbulence, 
at least in local patches where it is seeded by the tearing of the folds. 

\subsubsection{Outstanding Question} 

Just as I did in \secref{sec:falsifiable}, let me ask first whether this is a falsifiable 
theory. The numerical evidence that I described in \secref{sec:dynamo_num} was what encouraged 
me to bring reconnection into the dynamo game. However, it is clear that 
the state-of-the-art numerical spectra shown in \figref{fig:dynamo_spectra} 
describe a still non-asymptotic situation, at best only just starting to capture 
reconnection in the folds. Is there hope to do better any time soon? 
Surprisingly, the answer is yes. Just like in the case of the tearing-mediated
cascade with a mean field (\secref{sec:falsifiable}), numerical verification
that, in the earlier version of this review, I referred to as a distant hope,
is becoming reality before the final version goes into print. The just-published 
dynamo simulations by 
\citet{galishnikova22}, which aimed specially to assess the fold-tearing scenario 
proposed above and pushed the numerical resolution upwards another notch (to $2240^3$), 
appear to offer some tentative evidence for the disruption of folds by tearing, 
the $\lR\propto \Rm^{1/3}$ scaling of the field-reversal scale,\footnote{Or, as they would have it,
$\lR\propto \Rm^{3/10}\Pm^{1/5}$. Their quantitative comparisons were done for tearing
of sinusoidal, rather than Harris-sheet, magnetic-field profile across the folded
structure: $n=2$ in \apref{app:dprime_n}, rather than $n=1$ adopted in the main text of this review.}
and a steeper-than-Kolmogorov spectrum below that scale ($\propto k^{-2}$ or thereabouts).
Thus, I present this scenario now with 
greater confidence than I originally thought I could. The theory appears to be 
on the brink of being testable by state-of-the-art numerics. 

The key open question, however, both theoretically and numerically, 
is still what has been the central question of small-scale-dynamo 
theory since the founding papers of \citet{batchelor50} and \citet{biermann51}
and preoccupied the authors of the long string of papers cited in \secref{sec:dynamo_old}: 
will magnetic energy get stuck at small scales [in my current scheme, 
at the reversal scale $\lR$ given by \exref{eq:lamL}] or will it make its way to the outer 
scale~$L$ via some further step in the nonlinear evolution? 
The latter would be everyone's preferred outcome: astrophysicists 
could then have their cosmic magnetic fields at the scales 
where they are observed to be,\footnote{I am referring to extragalactic astrophysicists 
(the galactic ones 
get their large-scale field from mean-field dynamo; see \citealt{rincon19})---see 
\citet{vacca18}, \citet{vazza21}, and references therein for magnetic fields 
in clusters of galaxies and beyond. Interestingly, laboratory turbulent dynamo, 
recently achieved in a laser plasma, also appears to have its magnetic 
energy at the outer scale, in apparent contradiction with MHD simulations of the same 
experiment \citep{tzeferacos18,bott21}.} 
while theoretical physicists could stop worrying about no-mean-field vs.\ strong-mean-field MHD 
upsetting their universalist dream by exhibiting different kinds of turbulence---the 
fluctuating field at $L$ would just be the effective mean field in the inertial range, 
as \citet{kraichnan65} prophesied. 

As I already said in \secsand{sec:dynamo_old}{sec:dynamo_num}, 
a definitive demonstration of an inverse cascade in forced, isotropic MHD turbulence 
has remained elusive, leaving space for disagreement over how to interpret 
insufficiently asymptotic simulations; the highest-resolution-to-date study by 
\citet{galishnikova22} does not nail the issue completely but would be much 
easier to interpret if a compelling argument in favour of an $\Rm$-independent 
magnetic-energy-containing scale were available after all. 
I do not currently have such an argument---but I am able to offer, in \secref{sec:dynamo_scenarios}, 
a small catalogue of speculations, mine and other people's, 
about some relevant physical processes---note that they are not 
necessarily all mutually exclusive and may coexist in a saturated dynamo state, 
perhaps in different spatial patches or at different times. 

\subsection{Saturation Scenarios} 
\label{sec:dynamo_scenarios}

\vskip2mm
\begin{flushright}
{\small \parbox{8.5cm}{Wenn es aber Wirklichkeitssinn gibt, und niemand wird 
bezweifeln, da{\ss} er seine Daseinsberechtigung hat, 
dann mu{\ss} es auch etwas geben, das man M\"oglichkeitssinn 
nennen kann.
Wer ihn besitzt, sagt beispielweise nicht: Hier ist dies oder 
das geschehet, wird geschehen, mu{\ss} geschehen; 
sondern er erfindet: Hier k\"onnte, sollte oder m\"u{\ss}te 
geschehen; und wenn man ihm von irgend etwas erkl\"art, 
da{\ss} es so sei, wie es sei, dann denkt er: Nun, es k\"onnte 
wahrscheinlich auch anders sein. So lie{\ss}e sich 
der M\"oglichkeitssinn geradezu als die F\"ahigkeit definieren, 
alles, was ebensogut sein k\"onnte, zu denken und das, was ist, 
nicht wichtiger zu nehmen als das, was nicht ist. 
Man sieht, da{\ss} die Folgen solcher sch\"opferischen Anlage 
bemerkenswert sein k\"onnen\dots}
\vskip2mm
Robert Musil, {\em Der Mann ohne Eigenshaften}\footnote{``But if there is 
a sense of reality,---and no one 
will doubt that it has every right to exist,---then there must 
also be something that one could call a sense of possibility.
A person who possesses it does not, for example, say: here this 
or that has happened, will happen, must happen; no, he rather 
starts inventing: here might, should, or could happen 
something or other; and if he is explained about something 
that it is so and so, and how it is, then he thinks: well, 
it could have probably also been different. Thus, one may 
define the sense of possibility as the ability to perceive 
everything that can be, and not to attach more importance 
to what is than to what is not. It is evident that the 
consequences of such creative faculty can be quite 
remarkable\dots''---Robert Musil, {\em The Man Without Qualities}.}} 
\end{flushright}
\vskip5mm

To start with, let me imagine for the time being that the state with field reversals 
at the scale $\lR$ conjectured in \secref{sec:rec_dyn} {\em is} the saturated 
state and explore what happens between $\lR$ and the outer scale~$L$. I shall 
discuss this in \secsdash{sec:multifolds}{sec:magnetoelastic} before moving 
on to schemes for bringing magnetic energy to the outer scale 
in \secsdash{sec:fast_rec_dynamo}{sec:sheardynamo}. A reader wary of 
speculations (who has, nevertheless, inexplicably, got this far) may wish 
just to read \secsand{sec:dynamo_spectra}{sec:fast_rec_dynamo} and move on to~\secref{sec:kinetic}.

\subsubsection{Multiscale Folds?}
\label{sec:multifolds}

In \secref{sec:rec_dyn}, I assumed that in the saturated state, the only motions 
capable of stretching magnetic fields into folds were the outer-scale motions and
that, consequently, all folds had length $L$ and reversal scale $\lR=\lambda(L)$
[see \exref{eq:lambda_rev} and \exref{eq:lamL}]. Let me now relax this assumption and 
inquire what would happen if motions across some scale range $\ell<L$ 
produced different, independent folds. As I argued in \secref{sec:ssim_dyn}, 
the fields produced by any given motion cannot be stronger than this motion, 
so let us take a bold leap and guess that, for $\lambda < \lR$,   
\beq
B_\lambda \sim \du_{\ell(\lambda)} \sim \lt[\eps\ell(\lambda)\rt]^{1/3}
\sim \eps^{2/5}\eta^{-1/5}(1+\Pm)^{1/10}\lambda^{3/5}, 
\label{eq:fold_cascade}
\eeq 
where $\ell(\lambda)$ is obtained by inverting the $\lambda(\ell)$ 
dependence \exref{eq:lambda_rev}. 
In a triumph of dimensional inevitability, this is just the same as the scaling 
\exref{eq:tearing_cascade}, leading to the familiar $k^{-11/5}$ spectrum \exref{eq:Ek_bounds}
(cf.~the relationship between the $\xi$ and $\lambda$ scalings in~\secref{sec:align_rec}).
One can now again fantasise about these folds breaking up into flux ropes as described
by~\secref{sec:rec_dyn}, seeding mini-cascades similar to those produced by 
the successive disruptions of the RMHD cascade---those too have a $k^{-11/5}$ upper 
envelope, so perhaps this scaling emerges as an inevitable outcome at small scales 
of pretty much any scenario that involves resistivity. 

Note that in the scheme leading to \exref{eq:fold_cascade}, 
interactions between velocities and magnetic 
fields are nonlocal in scale: velocities at scale $\ell$ interact with magnetic 
fields at scale $\lambda(\ell)\ll\ell$, and vice versa.\footnote{It is easy to see 
how a large-scale flow directly produces small-scale fields (\figref{fig:stretch}). 
Since the fields' {\em parallel} scale is $\ell$, these formally smaller-scale fields can 
in turn exert $\ell$-scale Lorentz forces: these are quadratic 
in the field, $\sim\vB\cdot\vdel\vB$, so they do not know about direction reversals but 
do depend on the parallel scale \citep{sch04dynamo}---consequently, they are
able to fight back coherently against the $\ell$-scale flow. 
This is, of course, only a heuristic argument and one could legitimately 
wonder if it might be simplistic and misleading. It might be, but not, it seems 
to me, for any of the reasons that have so far been aired in the literature. 
The most categorical statement of locality in MHD turbulence can be found in 
\citet{aluie10}. Their proof depends on the assumption that both velocity and magnetic 
spectra have scaling exponents in the range $(-3,-1)$---equivalently, 
that $\du_\lambda\propto\lambda^{\gamma^u}$ and $B_\lambda\propto\lambda^{\gamma^B}$ 
with $\gamma^u,\gamma^B\in(0,1)$ 
(it is probably also true, conversely, that if interactions are local, the 
scaling exponents should be in this range). This makes sense 
because, in very simple terms, the contribution from field increments at a larger 
scale ($\Lambda$) to those at a smaller scale ($\lambda$) is 
$\du_{\Lambda\to\lambda}\sim\lambda\du_\Lambda/\Lambda
\sim \du_\lambda(\lambda/\Lambda)^{1-\gamma}\ll\du_\lambda$
(provided $\gamma<1$)
and the contribution from the smaller-scale increments to the larger-scale ones is 
$\du_{\lambda\to\Lambda}\sim\du_\lambda
\sim \du_\Lambda(\lambda/\Lambda)^\gamma\ll\du_\Lambda$
(provided $\gamma>0$). 
In RMHD turbulence, all this holds and interactions are indeed likely to be local
(as I always assumed them to be in \secsdash{sec:WT}{sec:disruption}).
In the case of saturated dynamo, however, the unresolved 
issue is precisely whether velocity and magnetic field have 
scaling exponents $\in(0,1)$ across the same range of scales---and also 
whether velocities at every scale are of the kind, dynamically, that can 
stretch magnetic fields at the same scale.} 
This nonlocality is more extreme than, e.g., \citet{beresnyak12dynamo} would have it, 
because $\lambda/\ell$ in \exref{eq:lambda_rev} scales with $\Rm_\ell$ and $\Pm$, 
which are asymptotically large numbers 
(although it scales with quite modest fractional powers of them). 
Below~$\lR$, however, this nonlocality should get swamped by the local 
RMHD cascade proposed in~\secref{sec:rec_dyn}. 

\subsubsection{Spectra Above the Reversal Scale}
\label{sec:dynamo_spectra}

In the absence, as yet, of any mechanism for magnetic structures 
at scales longer than~$\lR$ to emerge dynamically, 
the magnetic spectrum at scales above $\lR$ 
should reflect the volume-filling properties of folds and the 
distribution of their reversal scales (all of this is certain to be highly 
intermittent). The simplest, perhaps too simple, guess is that it will be 
\citep{yousef07},
\beq
E_B(k) \propto k^{-1},
\label{eq:EB_yousef}
\eeq
which follows if one assumes that magnetic increments across any point separation $>\lR$ will
tend to have, in an averaged sense, the same value, roughly equal to the rms field $B_{\lR}$, 
i.e., that there is the same magnetic energy at every scale $\in[\lR,L]$ 
(cf.~\secref{sec:mag_subvisc}). The spectrum \exref{eq:EB_yousef} has the unique
property that, while its peak would sit at some $k_\mathrm{peak}\sim L^{-1}$, its
energy is weakly (logarithmically) dominated by its small-scale cutoff, viz.,
$k\sim\lR^{-1}$. Remarkably, this appears to be consistent, or at least
not inconsistent, with what was found by \citet{galishnikova22}, who see
the spectral peak at a wavenumber (approximately) independent of $\Rm$,
but a fairly flat profile of $k E_B(k)$ up to $k\sim \lR^{-1}$. 

There is still the question of what the velocity field does in the interval $[\lR,L]$. 
In~\exref{eq:fold_cascade}, I blithely assumed that it continued 
to obey the Kolmogorov scaling \exref{eq:dynamo_K41}. 
If this were true, that would connect nicely onto 
the flux-rope amplitude \exref{eq:u_dynamo} (and hence onward to the universal 
tearing-mediated cascade). Admittedly, however, 
the justification for a Kolmogorov scaling in that case is difficult 
as \exref{eq:fold_cascade} implies energy exchanges with the folds at smaller 
scales and thus undermines the assumption of a constant flux through the scale range 
between $L$ and $\lR$. If the energy flux $\eps$ were depleted in favour of the folds 
at each scale $\ell$ on the same typical time scale $\ell/\du_\ell$ as the 
cascade of $\du_\ell$ proceeded, then the velocity field would have
a steeper-than-Kolmogorov spectrum. The most elementary way to see this is to model
the evolution of the spectrum $E_k$ as \citep[cf.][]{batchelor53}
\beq
\frac{\dd E_k}{\dd t} = -\frac{\dd \eps_k}{\dd k} - \gamma_k E_k,
\qquad \eps_k = \frac{kE_k}{\tau_k},\qquad
\tau_k^{-1} = \const\times k\sqrt{kE_k},
\label{eq:Ek_model}
\eeq
where $\eps_k$ is the energy flux through wavenumber $k$, $\tau_k$ is the cascade time ($\sim\tnl$),
and $\gamma_k$ is the rate of transfer of the kinetic energy into the folds.
Letting $\gamma_k\tau_k = \sigma = \const$ and seeking a steady-state solution gives
\beq
\frac{\dd\eps_k}{\dd k} = -\sigma\frac{\eps_k}{k}
\hence \eps_k \propto k^{-\sigma}
\hence E_k \propto k^{-(5+2\sigma)/3}.
\eeq
Note that, via a calculation analogous to \exref{eq:fold_cascade}, this would lead to 
a steeper-than-$k^{-11/5}$ spectrum of folds, meaning that folds with reversal 
scales smaller than $\lR$ would get swamped by the tearing-mediated 
cascade originating from the longest, $\lR$-scale folds, and we would be back to
the scenario described in \secref{sec:rec_dyn}.

\citet{cho09weakB} and \citet{beresnyak12dynamo} report
that the fraction of the energy flux transferred into magnetic fields 
during the self-similar regime described in \secref{sec:ssim_dyn} is numerically quite 
small---between 0.04 and 0.07. 
This suggests that any steepening of the velocity spectrum 
compared to the Kolmogorov scaling should, in theory, be very slight. 
The numerical evidence on velocity spectra was reviewed at the 
end of~\secref{sec:dynamo_num} (see \figref{fig:dynamo_spectra})---there is some steepening 
at low $k$ and shallowing at higher $k$, as magnetic fields' back reaction kicks in,
effectively making $\gamma_k<0$ in \exref{eq:Ek_model}, 
but none of the extant simulations is likely to have reached asymptotically large $\Re$ or~$\Rm$. 
   
\subsubsection{Magnetoelastic Turbulence?}
\label{sec:magnetoelastic}

Let us now explore what happens in the scale interval $[\lR,L]$
if we abandon \exref{eq:fold_cascade} and return to the scenario in which 
the velocity field at the outer scale $L$ constantly passes a certain 
fraction of the injected power $\eps$ to the folds with reversals at $\lR$ 
and hence into the tearing-mediated cascade, while  the rest of the injected power 
goes into some motions on scales $[\lR,L]$ that do not exchange 
energy with that cascade, i.e., do not stretch the field or cause it to develop 
sub-$\lR$ structure. What kind of motions can these be? 

In search of the answer to this question, I wish to revisit the old idea 
\citep{moffatt86,gruzinov96,chandran97phd,sch02njp,maron04} 
that a tangled mess of small-scale magnetic fields provides an elastic background 
through which larger-scale Alfv\'en waves can propagate isotropically. 
The relevant calculation is straightforward. Consider the equations of 
incompressible MHD without a mean field: 
\begin{align}
\label{eq:NSEq}
&\dd_t u_i + u_j\dd_j u_i = - \dd_i p + \dd_j M_{ij},\\
&\dd_t B_i + u_n\dd_n B_i = B_n\dd_n u_i,
\label{eq:ind}
\end{align}
where the equation for pressure $p$ is $\dd_i u_i = 0$ and 
$M_{ij}=B_i B_j$ is the Maxwell stress tensor (the magnetic field is in velocity units). 
We can recast the induction equation~\exref{eq:ind} in terms of $M_{ij}$ and forget about $B_i$: 
\beq
\label{eq:Mstress}
\dd_t M_{ij} + u_n\dd_n M_{ij} = M_{nj}\dd_n u_i + M_{in}\dd_n u_j. 
\eeq
The information about magnetic fields' reversals is now hidden away 
and only their ability to exert Lorentz force, quadratic in $B_i$, is retained. 
Let us expand the flows and the Maxwell stresses around a time- and space-averaged state: 
\beq
\la u_i\ra = 0,\quad
\la M_{ij}\ra = \vA^2\delta_{ij}, 
\quad \vA^2 = \frac{1}{3}\la B^2\ra,
\quad M_{ij} = \la M_{ij}\ra + \dM_{ij}. 
\eeq
Linearising \exref{eq:NSEq} and \exref{eq:Mstress} around this ``equilibrium'' filled 
with tangled fields, we get isotropically propagating Alfv\'en waves whose 
dispersion relation and eigenvector are\footnote{\citet{sch02njp} argued that if 
the small-scale magnetic fields were organised in long-scale folds, these Alfv\'en waves 
would propagate as a kind of ripple along these folds, thus making them locally anisotropic. 
Mathematically, this led to the disappearance of the factor of $1/3$ in $\vA^2$, 
because the tensor of magnetic-field directions $B_iB_j/B^2$ was a long-scale object. 
Since, however, I now propose that the folds will break up into 
flux ropes, etc., it seems more logical to think of the resulting magnetic tangle 
as an isotropic mess, at least from the point of view of long-scale perturbations.} 
\beq
\omega^2 = k^2\vA^2,\quad
\dM_{ij} = \vA^2 (\dd_i\xi_j + \dd_j\xi_i),
\label{eq:omega_magel}
\eeq
where $\xi_i$ is the displacement ($\dd_t\xi_i = u_i$). These can be dubbed 
{\em magnetoelastic waves} to highlight the formal mathematical \citep{ogilvie03}
and obvious physical analogy between a magnetised plasma and certain types 
of polymeric fluids. Admittedly, this analogy between magnetic field lines 
and polymer strands moving with the fluid and elastically back-reacting on it 
becomes precarious if one looks beyond the ideal description: there is no 
such thing as ``antiparallel'' polymers strands, and so there is no reconnection. 
It is not obvious whether fast reconnection of field lines can foil 
their ability to make plasma an elastic medium: do tangled fields spring back when 
pushed at or just reconnect quickly to accommodate the push? 
Here, I shall imagine that they do spring back and explore the consequences. 

One of the consequences appears to be a surprising return of the IK turbulence 
(\secref{sec:IK}), 
which I have so far thoroughly dismissed---perhaps an indication that a clever idea, however 
wrong, never goes to waste. The reason that the IK scheme was wrong in the presence 
of a strong mean magnetic field was that Alfv\'en waves could not be legitimately expected 
to run around isotropically at small scales. Well, according to~\exref{eq:omega_magel}, 
the magnetoelastic waves do run 
around isotropically, and so the IK theory is back in business. While Kraichnan's 
dimensional argument leading to \exref{eq:IK} may or may not be compelling, 
the version of the IK theory outlined in footnote~\ref{fn:IK} is perhaps sensible. 
Indeed, whereas at the outer scale $L$, the nonlinear time $\tnl\sim L/\du_L$ 
and the Alfv\'en time $\tA\sim L/\vA$ are certainly comparable 
(because $\la B^2\ra\sim\du_L^2$ for saturated dynamo), the former will 
shorten less quickly than the latter at smaller scales ($\tA\propto\ell$, 
while $\tnl\propto$~a fractional power of $\ell$). Thus, at scales $\ell\ll L$, 
the magnetoelastic turbulence might be expected to be weak. The cascade time is then 
worked out from the random-walk argument \exref{eq:tc_weak}, and the spectrum follows from 
the constancy of flux: 
\beq
\tc\sim\frac{\tnl^2}{\tA}\sim\frac{\ell\vA}{\du_\ell^2},\quad
\frac{\du_\ell^2}{\tc} \sim \eps
\hence
\du_\ell \sim (\eps\vA\ell)^{1/4}
 \quad\Leftrightarrow\quad
E(k) \sim (\eps\vA)^{1/2} k^{-3/2}. 
\label{eq:scaling_magel}
\eeq
Presumably, this cascade 
terminates when it hits $k\sim\lR^{-1}$, where the scale separation between 
the magnetoelastic waves and the magnetic fields associated with the tearing-mediated 
cascade of \secref{sec:rec_dyn} breaks down.

There is some numerical evidence in favour of an isotropic $k^{-3/2}$ spectrum 
of perturbations with a sound-like isotropic dispersion relation 
$\omega\propto k$---the MHD fast (magnetoacoustic) waves: 
see \citet{cho02compr,cho03compr}, 
who were inspired by the same scaling derived for weak turbulence of sound waves 
by \citet{zakharov70}; a later study by \citet{kowal10} 
appears to be less certain about the scaling exponent. 

One might have thought that some evidence as to how much of a fiction, or otherwise, 
the spectrum \exref{eq:scaling_magel} were in an elastic medium, could 
be found in simulations of polymer-laden turbulence. Surprisingly, the state 
of the art in this area features much smaller resolutions than in MHD. 
The most recent relevant numerical papers appear to be \citet{valente16}
and \citet{fathali19} (see references therein for the paper trail). 
They report significant energy transfer in the inertial range from the motions of the 
solvent fluid to the elastic polymer admixture and back (quite a lot of it nonlocal in~$k$, 
from large-scale flows to small-scale polymer structure, perhaps analogous to dynamo); 
they also see spectral exponents in the $[-5/3,-3/2]$ range, at modest resolutions. 
They do not appear to be aware of, or interested in, the possibility 
of elastic waves.\footnote{In contrast, \citet{balkovsky01} and \citet{fouxon03} 
are fully aware of it, as well as of the MHD analogy with Alfv\'en waves. They have 
a theory of turbulence of these waves at scales where elasticity is important, 
below the so-called \citet{lumley69} scale (this is set by the balance between the turbulent 
rate of strain and the polymer relaxation time, a quantity without a clear 
MHD analogue because magnetic field lines have no interest in curling up the way 
polymers do, entropically; 
in our problem, the corresponding scale should be the outer scale~$L$).  
They think that in this scale range, the waves will be nonlocally advected 
by the Lumley-scale motions, resulting in spectra steeper than $k^{-3}$ 
because otherwise the nonlocality assumption fails. I do not see why 
such an assumption should hold, either for polymer-laden turbulence or in MHD.\label{fn:fouxon}} 

Since polymer-laden turbulence has the advantage of being (relatively) easy to 
set up and measure in the laboratory, numerical simulations are not the only evidence 
available---there is a lively history of real experiments, amongst which some recent 
ones appear to have been important breakthroughs. Thus, \citet{varshney19} have, 
for the first time, it seems, managed to excite and measure elastic waves experimentally.
\citet{zhang21} report detailed measurements of scalings in the inertial range that 
imply an energy flux, increasing with $k$, from the fluid motions into the elastic 
energy of the polymers, and kinetic-energy spectra that have a well-developed 
power law, $k^{-2.38}$, which is steeper than what numerics show but shallower 
than what theoreticians (cited in footnote~\ref{fn:fouxon}) predict. An open field for further 
theorising then, with not much more known definitively than in MHD.     

In MHD turbulence, even whether the magnetoelastic cascade \exref{eq:scaling_magel}, 
or indeed the magnetoelastic waves, exist at all remains an open 
question. \citet{hosking20} have shown numerically that magnetoelastic waves 
do exist in certain tangled, force-free magnetic configurations, and are 
well described by \exref{eq:omega_magel} (modulo some further nuance that 
can mean that $\vA$ is somewhat reduced for tangled fields that are spatially 
intermittent). What is still unknown is whether they can propagate 
against the background of a saturated dynamo state or are 
quickly damped by small-scale motions and thus rendered irrelevant. 

\subsubsection{Fast-Reconnection-Limited Dynamo?}
\label{sec:fast_rec_dynamo}

The narrative arc that in \secref{sec:onset} led me to examine, with 
\citet{loureiro20}, the possibility of a competition between tearing and 
fast plasmoid reconnection (reviewed in \apref{app:uls}), 
naturally invites doing the same for dynamo. 
For magnetic folds with reversals at scale $\lambda$ driven by motions at scale $\ell$
and energetically comparable to those motions, 
let us compare the fast-reconnection time given by a formula analogous to~\exref{eq:fast_onset}
with the characteristic time of the flow:  
\beq
\trec \sim \epsrec^{-1}\frac{\lambda}{B_\lambda} \sim \epsrec^{-1}\frac{\lambda}{\du_\ell}
\lesssim \frac{\ell}{\du_\ell} 
\rmiff
\lambda \lesssim \epsrec\ell \equiv \lrec, 
\label{eq:lrec_dynamo}
\eeq
where $\epsrec\sim\tSc^{-1/2}(1+\Pm)^{-1/2}$ and $\tSc\sim 10^4$ (see~\apref{app:loureiro}). 
In order for any possible effect of the onset of fast reconnection to matter, 
$\lrec$ must be larger than the reversal scale \exref{eq:lambda_rev} set by tearing, 
which is achieved provided
\beq
\Rm_\ell \gtrsim \epsrec^{-3}(1+\Pm)^{-1/2}\sim 10^6(1+\Pm), 
\label{eq:fast_rec_dynamo}
\eeq   
a tough ask, but not as bad as~\exref{eq:10to14}. Note that this condition, if satisfied, 
amply guarantees that the fast-reconnection regime actually is reached, i.e., 
that the slow-reconnection rate (see \apref{app:SP_rec}) is slower than fast one,~or 
\beq
\tSc \lesssim \tS_\ell \sim \Rm_\ell(1+\Pm)^{-1/2} 
\rmiff
\Rm_\ell \gtrsim 10^4 (1+\Pm)^{1/2}. 
\eeq

What would happen if \exref{eq:fast_rec_dynamo} were achieved? 
In principle, this means that a fast-reconnecting plasmoid chain could be 
formed out of such a fold, seeding a reconnection-driven cascade 
(\secref{sec:turb_sheet}). In order for this to happen, the folds, 
rather than being destroyed by tearing and/or swept away by the flow, 
would have to stick around long enough for
the plasmoids (flux ropes) in them to gobble up more flux and grow 
to bigger sizes than the reversal scale set by tearing ($\lR$ for $\ell = L$), 
eventually to~$\lrec$. If this were to be the fate of most folds, 
$\lrec$ would be the effective reversal scale, independent of~$\Rm$, 
as ideally desired. 

I shall not go as far as claiming that this outcome accords 
with simulations---both because they do not usually have $\Rm$ anywhere 
close to \exref{eq:fast_rec_dynamo} and because \exref{eq:lrec_dynamo} 
with $\ell = L$ would imply a magnetic-energy-containing scale $\sim 10^2$ 
shorter than the outer scale, which is rather too short. 
However, if $\epsrec$ were closer to $10^{-1}$ 
than to $10^{-2}$ in a turbulent environment characteristic of the saturated 
dynamo, both the condition~\exref{eq:fast_rec_dynamo} and the 
prediction~\exref{eq:lrec_dynamo} would start looking much more realistic
(alternatively, another method of increasing the field's scale is needed; 
see, e.g., \secref{sec:sheardynamo}). 
Thus, a promising scenario, but still very much to be confirmed---and it is
far from clear that it can be confirmed at the numerical resolutions likely
to be available any time soon \citep[cf.][]{galishnikova22}. 

It cannot have escaped the reader that an even more attractive possibility 
would be $\epsrec \sim 1$, which would absolve us from any constraints 
on $\Rm$ and set $\lrec\sim L$ in the saturated state. That possibility 
is effectively the one associated with the other type of fast 
reconnection---stochastic reconnection (\apref{app:stoch_rec})---and is 
the currently trending dynamo-evolution scenario examined in~\secref{sec:xulaz}. 

\subsubsection{\citet{xulaz16}}
\label{sec:xulaz}

\citet{xulaz16} propose that, in a $\Pm\gg1$ system, 
once the magnetic energy has grown to be comparable to the 
energy of the viscous-scale eddies ($B_{\lres}\sim\du_{\lnu}$; cf.~\secref{sec:kin_dyn}), 
its spectrum will embark on a rearrangement exercise in which its spectral peak moves 
from the resistive to the viscous scale while the overall magnetic energy stays constant 
(I will discuss this proposition in a moment). 
Once it reaches the viscous scale, a self-similar secular regime follows, 
of the kind described in \secref{sec:ssim_dyn},\footnote{\citet{xulaz16} believe that 
they can derive very precisely the fraction of the energy flux going into magnetic 
fields quoted at the end of \secref{sec:dynamo_spectra} (it is~$=3/38$, they say) 
from a semi-quantitative theory that contains adjustable constants of order unity and 
is a variant of the dynamo-with-reconnection model by \citet{kulsrud92} 
(who also derived that number)---see further discussion at the end of~\secref{sec:subra}.} 
except the scale of $B_\lambda$ 
is now the same as the scale of the motions that are performing the dynamo action, 
$\lambda\sim\ell(t)$, whereas below that scale, a GS95-type turbulent 
cascade forms, with $B_{\ell(t)}$ playing the role of the mean field (this is 
also the view of \citealt{beresnyak12dynamo}). 
As time advances, $\ell(t)\to L$, and the dynamo saturates with scale-by-scale 
equipartitioned $k^{-5/3}$ magnetic and velocity spectra, just like everyone 
since \citet{biermann51} has always wanted it to do. 

In the narrative of \citet{xulaz16}, this pleasing outcome  
depends on the assumption, unproven, but not in principle impossible, that 
fast stochastic reconnection (reviewed in \apref{app:stoch_rec}) 
will always provide just enough turbulent magnetic 
diffusivity to prevent the dynamo-generated field from organising into folds 
with reversals at scales much below $\ell(t)$. I cannot rule this out definitively 
without a clear dynamical picture of the turbulence in the presence 
of dynamically significant dynamo-generated fields.\footnote{It may, 
however, be worth observing that, according to the numerical 
results reported by \citet{busse07}, Lagrangian particles in 
MHD turbulence without a mean field tend to separate along the 
local field direction, rather than across it. An enthusiast of 
field-line folding might interpret this as an indication that 
stochastic reconnection might find it difficult to prevent fold creation. 
\citet{eyink11dynamo}, a strong advocate of stochastic reconnection, 
notes his puzzlement at this result.} 

If this assumption proves true, the \citet{xulaz16} scenario for the $\Pm\gg 1$ case 
still needs the earlier transitional stage to move the magnetic energy 
(i.e., the field-reversal scale) from the resistive to the viscous scale.
They justify this by arguing that, since the 
spectrum of the magnetic field is $\propto k^{3/2}$ at $k\ll\lres^{-1}$ 
\citep{kazantsev68,kulsrud92}, the magnetic modes with $k\ll\lres^{-1}$ can continue being 
amplified by the viscous-scale motions after those with $k\sim\lres^{-1}$ 
have reached energetic equipartition with those motions---if the overall magnetic 
energy is assumed to stay constant, this then leads to a gradual ``overturning'' 
of the spectrum and shifts its peak towards the viscous scale. 

I do not think this argument is entirely satisfactory, for two reasons. 
First, the $k^{3/2}$ spectrum is a Fourier-space representation 
of the growing, folding field---it seems dubious to me to disaggregate 
it into individual Fourier modes and view each of them 
as an independent entity that back-reacts on the velocity field or 
is amplified by the latter separately from all others (amplification 
of the field is always accompanied by a change in its scale and the 
$k^{3/2}$ spectrum is the resulting mean distribution of the magnetic 
energy amongst wavenumbers during its growth; see, e.g., 
lecture notes by \citealt{schKT}).  
Secondly, I do not see why the energy should stay constant, putting  
the self-similar stage on hold until magnetic fields and motions are 
at the same scale, rather than proceeding to grow in the way described 
in \secref{sec:ssim_dyn}. 
It seems to me that if the \citet{xulaz16} scheme were correct, 
we should see their spectral rearrangement at constant energy 
already in numerical simulations with $\Pm\gg 1$ 
and $\Re\sim 1$, which is the only glimpse 
of truly scale-separated large-$\Pm$ dynamics that we currently have. 
In the event, we do see the spectrum in the nonlinear stage of such simulations 
become shallower than $k^{3/2}$ and shift a little towards 
larger scales---but not all the way to the flow's scale, with the peak still, it seems, 
at the resistive scale (this behaviour appears 
to be accounted for adequately by assuming that the magnetic folds locally anisotropise 
the viscous-scale flow and thus stymie its ability to amplify them: 
see \citealt{sch04aniso,sch04dynamo} and \citealt{stonge20}). 

Thus, in my view, the \citet{xulaz16} scenario, while attractive 
if true, remains at least as much of a speculation 
as anything that my exasperated reader will find elsewhere in this section.
 
\subsubsection{\citet{subramanian99}}
\label{sec:subra}

While I am not proposing to review here the entire history of dynamo saturation schemes, 
it is useful to describe a model proposed a long time ago by \citet{subramanian99}, 
which, in a certain general sense, anticipated the \citet{xulaz16} scenario and many 
other similar schemes, including~\secref{sec:fast_rec_dynamo}. 

\citet{subramanian99} conjectured that the effect of nonlinear back-reaction  
of magnetic fields on the flow would be to increase the effective magnetic diffusivity  
of the turbulent medium: 
\beq
\eta_\mathrm{eff} = \eta + \tau\la B^2\ra,
\label{eq:eta_eff}
\eeq 
where $\tau$ is some adjustable constant with dimensions of time. 
This would go on until the effective magnetic Reynolds number reached 
the critical value at which the small-scale dynamo was at its threshold---known 
to be $\Rmc\sim10^{1\dots2}$ \citep[see, e.g.,][]{sch07dynamo}. He then proposed that 
the saturated state of the dynamo would simply be the marginal state of a kinematic 
dynamo with $\Rm_\mathrm{eff}=\Rmc$, and with magnetic energy therefore sitting at the scale 
\beq
\lB \sim L\Rmc^{-1/2}. 
\eeq
This is not far from where, quantitatively, numerical simulations have been 
placing the peak of the magnetic-energy spectrum in finite-resolution boxes for the last two decades 
(see references in \secsand{sec:dynamo_old}{sec:dynamo_num}). 
Indeed, Subramanian's prescription was operationalised 
by \citet{schober15} to produce a fully-fledged modelling tool for unfailingly successful 
comparisons with simulation outputs for specific values of $\Re$ and~$\Rm$, 
as well as of the Mach number.  

The key conceptual point is that under this scheme, there is no 
dependence of $\lB$ on~$\Rm$. A slight wrinkle is that the magnetic energy 
in the saturated state is 
\beq
\la B^2\ra \sim \frac{\du_L L}{\tau \Rmc} \sim \frac{\du_L^2}{\Rmc}
\eeq
if one makes the most obvious choice $\tau\sim L/\du_L$, i.e., $\la B^2\ra$ is uncomfortably 
smaller than the kinetic energy (although not outrageously inconsistent with numerical evidence).
Since $\Rmc$ is, however, merely a constant, 
this can be fudged by judicious modelling choices or, following \citet{subramanian99}, 
by arguing that if the magnetic energy is concentrated in flux ropes of radius 
$\lB$ and length $L$, then $\la B^2\ra \sim B^2(\lB/L)^2 \sim B^2\Rmc^{-1}$, so 
the actual magnetic fields are, in fact, locally as strong as the flow.   

This approach does not address any rearrangements that might be caused by 
the Alfv\'enic dynamics, nor does it produce a specific mechanism for the enhancement 
of magnetic diffusivity, but it does capture, conceptually, a broad general class 
of physically plausible models, to which both \secref{sec:fast_rec_dynamo} and 
\citet{xulaz16} also belong 
(stochastic reconnection is, of course, turbulent magnetic diffusivity by another name). 
To explain the enhancement of $\eta$, \citet{brandenburg05} introduce an effective 
``nonlinear drift'' $\propto\vJ\times\vB$, justified by analogy with ambipolar effects
(a version of that was also explored by \citealt{subramanian03}, where he complicated 
his model somewhat to produce turbulent magnetic hyperdiffusion). Interestingly, 
an even earlier paper by \citet{kulsrud92} already contained calculations 
of nonlinear dynamo both for a model with ambipolar damping, featuring an effective 
magnetic diffusivity of the form \exref{eq:eta_eff}, and with 
magnetic reconnection (see their~\S4.2). In the latter case, instead of an effective 
magnetic diffusivity, they have a damping rate of the form $-\epsrec\la B^2\ra^{1/2} k$ 
modelling the removal of magnetic structure by fast reconnection with dimensionless 
rate $\epsrec$ (this is just $\trec^{-1}$ of \secref{sec:fast_rec_dynamo}), 
but the outcome is similar. 
Their calculation of the self-similar dynamo regime with reconnection 
appears to be functionally equivalent to that of \citet{xulaz16}, and produces 
the same result. 

To put all this in context, the other, competing general class of models includes 
schemes in which dynamically strong magnetic fields locally 
change the nature of the flow to stop it from amplifying them further 
\citep[e.g.,][]{cattaneo96,zienicke98,kim99,sch04aniso,sch04dynamo,cattaneo09,baggaley10,rempel13,seta20,stonge20}. It is if one wants to make a theory of this type work 
that one is required to produce a dynamical mechanism for transferring magnetic energy from small 
scales to large. 

\subsubsection{Inverse Magnetic-Energy Transfer via Sporadic Decay?}
\label{sec:inverse}

In pursuit of such mechanisms, an interesting recent development 
came from simulations of {\em decaying} MHD turbulence without a mean field: 
as I already mentioned in~\secref{sec:perma}, \citet{zrake14} and 
\citet{brandenburg15} discovered numerically that such a turbulence, 
even without net helicity, could support a certain amount of inverse transfer 
of magnetic energy from small to large scales (as expected 
theoretically: see \secref{sec:perma}).\footnote{The existence of such an inverse transfer
in the case of non-zero net helicity is well known and well simulated
(see references in \secref{sec:perma}), 
but is not relevant here because it is just a nonlinear counterpart of the helical 
mean-field dynamo, a topic reviewing which I leave to \citet{rincon19}.} 
In \secref{sec:decay_nonhel_new}, I argued, following \citet{zhou20}, \citet{bhat21}, 
and \citet{hosking21}, 
that the dynamical mechanism by which large-scale magnetic fields are generated 
in decaying, non-helical MHD turbulence starting from a magnetically dominated state 
is the merger of reconnecting flux ropes. 
Let me explore what would happen if 
the same mechanism were to apply locally to the magnetic structures 
at the reversal scale $\lR$, which {\em are} flux ropes (plasmoids), 
released from disintegrating folds.  

Imagine that, instead of being continuously 
forced everywhere, our saturated dynamo were to be left alone for a period 
of time (and/or in a region of space)---this could be due to the natural 
spatiotemporal intermittency of the system or to a method of forcing 
leading to sporadic energy-injection events with quiescent periods of 
decaying turbulence in between \citep[e.g., in galaxy clusters:][]{roh19}. 
With the arrival of each quiescent period, mergers between the flux ropes 
should push magnetic energy to larger scales. 
 
The salient bit of theory that is needed to assess this effect 
is that the magnetic-energy-containing scale will grow with time 
as a power law during the decay of the turbulence: thus, if the field starts 
at scale $\lR$, its scale after a period of decay will~be  
\beq
\lB \sim \lR\lt(\frac{t}{\trec}\rt)^\alpha,
\qquad
\trec \sim \epsrec^{-1}\frac{\lR}{B_{\lR}},
\eeq
where $0<\alpha<1$ in 
all conceivable circumstances ($\alpha=4/9$ for fast-reconnection-controlled decay; 
see \secref{sec:decay_nonhel_new}), 
$\trec$ is the characteristic reconnection time \exref{eq:trec_decay}
at the beginning of the decay, with the initial energy-containing 
scale $\lR$ given by \exref{eq:lamL} and $B_{\lR}\sim\du_L$ 
(same as the outer-scale velocity field). 
Suppose the decay is allowed to proceed for about one outer-scale turnover time $L/\du_L$. 
The magnetic field's scale after that will be 
\beq
\lB \sim \epsrec^{\alpha}\lR^{1-\alpha} L^\alpha. 
\label{eq:lB_decay}
\eeq
If reconnection is fast (\apref{app:uls}), or stochastic (\apref{app:stoch_rec}), 
$\lB$ will have a weaker $\Rm_L$ 
scaling than $\lR$, but it is not a triumph of inverse transfer. 
Another way to reach the same tepid conclusion is  
by asking how long it would take to get the 
magnetic field to the outer scale, $\lB\sim L$. The answer is 
\beq
t \sim \trec\lt(\frac{L}{\lR}\rt)^{1/\alpha} 
\sim \frac{L}{\du_L}\,\epsrec^{-1}\lt(\frac{L}{\lR}\rt)^{(1-\alpha)/\alpha}
\gg \frac{L}{\du_L}, 
\label{eq:time_to_L}
\eeq
quite a long time, as expected, i.e., the forcing would have to be very sporadic 
to achieve this. 

Obviously, all this does not amount to much more than an initial ``back-of-the-envelope'' 
assessment, and a more sophisticated treatment might still yield a more pleasing outcome. 

\subsubsection{Local Shear Dynamo?}
\label{sec:sheardynamo}

Let me complete my catalogue of speculations regarding the structure of
the saturated dynamo state by invoking another piece of dynamo physics 
that, despite being of potentially fundamental and ubiquitous nature, 
emerged relatively late in the game. 
A combination of small-scale turbulence and a large-scale shear 
generically leads to the emergence of large-scale magnetic field, 
even when the turbulence has no net helicity---an effect known as the ``shear dynamo''. 
This was mooted theoretically in several early mean-field-dynamo 
schemes and then confirmed numerically by \citet{yousef08prl,yousef08an} 
(see references therein for the precursor theories, numerics and counter-arguments). 
This result turned out to be due to a form of ``stochastic $\alpha$ effect'' 
\citep{heinemann11,jingade18}, depending, therefore, on fluctuating helicity in the flow. 
Interestingly, the shear dynamo turned out to work also when the small-scale 
turbulence was magnetic, i.e., by the combination of a large-scale shear and 
the saturated state of small-scale dynamo \citep{yousef08an}. 
\citet{squire15,squire16dynamo} made sense of that by discovering 
semi-analytically the ``magnetic shear-current effect'' and showing that 
small-scale magnetic fields were actively helpful in enabling the shear dynamo. 

The outcome of \secref{sec:rec_dyn} was a situation in which the
outer-scale ($L$) field-stretching motions (plus possibly some sort of kinetic-energy 
cascade to smaller scales) coexisted with MHD turbulence produced by 
the break up of the folds, with an effective outer scale $\lR\ll L$ 
(in \secsref{sec:fast_rec_dynamo}, \ref{sec:subra}, and \secref{sec:inverse},
this scale was increased, but remained smaller than~$L$).
It seems to be an attractive speculation that 
the combination of this turbulence with the local shears associated with the 
``hydrodynamic'' scales $>\lR$ might act as a local shear dynamo and create 
``local mean fields'' on scales $>\lR$. It would be interesting to investigate 
whether such a mechanism exists and, if it does, whether it 
can push the magnetic-energy-containing scale closer to~$L$.\\

To conclude, there are plenty of potential theories---far too many, so no 
convincing one theory yet. Hero numerics reaching for asymptoticity, and intelligently analysed, 
might help pare down this field and finally give our understanding of the saturated MHD dynamo 
a modicum of completeness to match what has been achieved for MHD turbulence 
with a mean field. 

\section{The Frontier: Kinetic Turbulence} 
\label{sec:kinetic}

\vskip2mm
\begin{flushright}
{\small \parbox{8.5cm}{We can measure the globula of matter and the distances 
between them, but Space plasm itself is incomputable.}
\vskip2mm
Vladimir Nabokov, {\em Ada, or Ardor}} 
\end{flushright}
\vskip5mm

\subsection{Sundry Microphysics at Low Collisionality}
\label{sec:microphysics}

I ended the first part of this review with a proclamation in \secref{sec:end} that 
the story of MHD turbulence looked reasonably complete (before spending five 
chapters on the loose ends!). Since the main 
reason for this triumphalism was that MHD cascade finally made sense at 
the dissipation scales---and the key role in making it make sense belonged to reconnection, 
a dissipative phenomenon---it is an inevitable complication that 
microphysics of dissipation may matter. The visco-resistive MHD 
description adopted here 
does apply to some natural plasmas, e.g., stellar convective zones or colder 
parts of accretion discs. These are mostly low-$\Pm$ environments. 
Whereas I have made an effort 
to keep all results general and applicable to the high-$\Pm$ limit, 
it is, in fact, quite hard to find naturally occurring high-$\Pm$ plasmas 
for which the standard visco-resistive MHD equations are a good model: this would 
require the particles' collision rate to be larger than their Larmor frequency, which 
rarely happens at high temperatures and low densities needed to achieve high 
$\Pm$ (one exception, quite popular these days, is plasmas created in laser 
experiments: see, e.g., \citealt{bott21,bott22}). 
In fact, most of the interesting (and observed) plasmas in this hot, rarefied 
category are either ``dilute'' (an apt term coined by \citealt{balbus04} to describe 
plasmas where turbulence is on scales larger than the mean free path, but the Larmor 
motion is on smaller scales that it---a good example is galaxy clusters; see, 
e.g., \citealt{melville16} and references therein) or downright collisionless 
(i.e., everything happens on scales smaller than the mean free path; the most 
obvious example is the solar wind: see the mega-review by \citealt{bruno13} 
or a human-sized one by \citealt{chen16}). 
In either case, between the 
``ideal-MHD scales'' and the resistive scale, there is a number of other scales 
at which the physics changes. These changes are of two distinct kinds. 

The first is the appearance of dispersion in the wave physics: 
Alfv\'en waves become kinetic Alfv\'en waves (KAWs),  
with a different linear response and, therefore, a different variety  
of critically balanced cascade \citep{cho04,sch09,sch19,boldyrev12,boldyrev13,chen17,passot17,milanese20}. 
The culprits here are the ion inertial scale (at which the Hall effect comes in), 
the ion sound scale (at which the electron-pressure-gradient force 
becomes important in Ohm's law), 
and the ion Larmor scale (at which the finite size of ion Larmor orbits starts playing a role). 
Which of these matters most depends on plasma beta and on the ratio of the ion and 
electron temperatures, but they all are essentially ion-electron decoupling effects 
and lead to more or less similar 
kinds of turbulence, at least in what concerns the KAW cascade. 
Note that the subviscous regime (\secref{sec:subvisc}) is, of course, irrelevant 
for such plasmas---except possibly, in a somewhat exotic way, at high beta \citep{kawazura19}. 

The second important modification of MHD is that reconnection in a collisionless plasma 
need not be done by resistivity, but can be due to other 
physics that breaks flux conservation, viz., electron inertia, electron finite 
Larmor radius (FLR) and, more generally, other kinetic features of the 
electron pressure tensor. Tearing modes are different in such plasmas, 
with a double ion-electron layer structure and a variety of scalings 
in a variety of parameter regimes.\footnote{Appendix B.3 of \citet{zocco11} 
has a review of standard results for collisionless and semicollisional 
tearing modes at low beta (using a convenient minimalist set of dynamical equations as a vehicle), 
as well as all the relevant references of which we were aware at the time.
There is a huge literature on semicollisional and collisionless reconnection 
and, short of dedicating this review to name-checking 
it all (which would be a noble ambition, but a doomed one, as the literature is multiplying 
faster than one can keep track), I cannot give proper credit to everyone who deserves it. 
A useful recent treatment of electron-only tearing done with applications to 
space turbulence in mind is \citet{mallet20}.} 
Since tearing is important for rounding off the MHD cascade, all these 
effects must be considered and appropriate modifications 
worked out for the theory of tearing-mediated turbulence described 
in \secref{sec:disruption}---some of this has been done by 
\citet{mallet17c} and by \citet{loureiro17b}. 
It is going to be interesting to find out whether, where, and when 
any of this matters or if perhaps the aligned MHD cascade just segues directly 
into the KAW cascade (see, however, a discussion in a moment as to what that means). 
Since there are some mysteries still outstanding with regard to the scale at which the spectrum 
of solar-wind turbulence has a spectral break between the 
inertial range and the ``kinetic'' (KAW) range \citep{chen14break,boldyrev15}, 
perhaps something interesting can be done here (e.g., is the break set by the onset 
of reconnection, rather than by the Larmor scale?---see \citealt{vech18}). 

Furthermore, KAW turbulence in the kinetic range and {\em its} relationship with 
reconnection is a topic that is rapidly becoming very popular with both numerical 
modellers \citep[e.g.,][]{tenbarge13a,banon16,cerri17a,franci17,franci18} 
and observational space physicists \citep[e.g.,][]{greco16}. 
There is a promise of interesting physics---interesting both conceptually and because 
it is eminently measurable in space. 
In the context of the prominent role that was given in \secref{sec:disruption} 
to the break up of MHD sheets in setting up the tail end of the 
MHD cascade, I want to highlight an intriguing suggestion (implicitly) contained 
in the paper by \citet{cerri17a} and further fleshed out by \citet{franci17}. 
They look (numerically) at the formation 
of current sheets in kinetic turbulence and the disruption of these sheets 
by tearing (plasmoid) instabilities---and discover that it is precisely these 
processes that appear to seed the sub-Larmor-scale cascade with a steep 
(steeper than in the inertial range) energy spectrum usually associated  
with KAW turbulence. One might wonder then if such a KAW cascade is 
an entirely distinct phenomenon from a collisionless version of tearing-mediated turbulence.
If we allow ourselves to get excited about 
this question, we might speculate that it rhymes nicely with the idea on which 
\citet{boldyrev12} relied to advocate a steeper ($-8/3$) slope of KAW 
turbulence than the $-7/3$ implied by the standard CB-based theory 
\citep{cho04,sch09}. They argued 
that the energetically dominant perturbations at each scale were concentrated in 
2D structures, thus making turbulence non-volume-filling (and perhaps 
monofractal; cf.\ \citealt{kiyani09} and \citealt{chen14}). While \citet{boldyrev12} did 
not appear to think of these 2D structures as reconnecting sheets, 
an interpretation of them as such does not seem {\em a priori} unreasonable. 
So perhaps this is what happens in collisionless turbulence: 
sheet-like structures form in the usual (MHD) way, get disrupted 
by collisionless tearing and/or related instabilities and seed sub-Larmor 
turbulence,\footnote{See \citet{mallet17c} for a discussion of what else they seed.} 
which stays mostly concentrated in those sheets or their remnants, 
with an effectively 2D filling fraction. Another possibility---or a version of 
this scheme---is to abandon the old KAW cascade altogether and declare 
sub-Larmor turbulence to be entirely controlled by 
(collisionless) tearing in a similar way to the tearing-mediated cascade 
of \secref{sec:recturb}, with spectral slopes between $-3$ and $-8/3$, 
still consistent with observations and simulations \citep{loureiro17b,boldyrev19}.

What I have said about kinetic physics so far might not sound like 
a true conceptual leap: basically, at small scales, we have different linear physics and 
a zoo of possibilities, depending on parameter regimes; 
one could work productively on porting some of the basic ideas developed 
in the preceding sections to these situations.\footnote{Another class of kinetic
situations for which such porting appears to be a successful strategy is relativistic
plasma turbulence. As full-Vlasov kinetic simulations have gradually become more computationally
affordable over the last decade, the first plasmas for which they became affordable
were pair plasmas moving at relativistic speeds---computations made easier
on account of the mass ratio being unity and
of the frequencies associated with the turbulent motion being closer to the highest plasma-wave
frequencies needing to be resolved in a Vlasov simulation anyway.
A lively industry has developed and, in certain respects,
kinetic, relativistic plasma turbulence is now better documented than its slower-moving,
harder-to-compute counterpart.
I am not aware of any qualitatively dramatic way in which the small-scale structure
of the former has proven to be much different from that of the latter, or of the
vanilla-MHD prototype of either---but students of this topic
have not been preoccupied primarily with the structure of the turbulence,
but rather followed the (observationally well-motivated) astrophysical
obsession with nonthermal particle acceleration. Of this, relativistic turbulence has indeed
been found guilty, provided it was stirred up with $\dB\sim B_0$
\citep{zhdankin17,comisso18relat}. 
The most recent spate of papers on this topic, which represent the state of the art
and from which the historical paper trail can be followed,
are \citet{comisso21},
\citet{vega22a,vega22b},
\citet{hankla22} (imbalanced turbulence),
\citet{nattila22} (small $\dB/B_0$, no nonthermal acceleration),
\citet{zhdankin21} (non-unity mass ratio, ion vs.\ electron heating), and
\citet{chernoglazov21} (alignment and current sheets in fluid but relativistic MHD turbulence).
It is interesting that in this context again, reconnecting structures spontaneously
generated by turbulence appear to play a prominent role---this time as sites of
particle acceleration.\label{fn:relativistic}} 
There are, however, ways in which kinetic physics does bring in something altogether new. 
Four examples of that, chosen in a very biased way, are given in what follows. 

\subsection{Cascade Fragility}
\label{sec:failed_cascade}

\begin{figure}
\centerline{\includegraphics[width=0.65\textwidth]{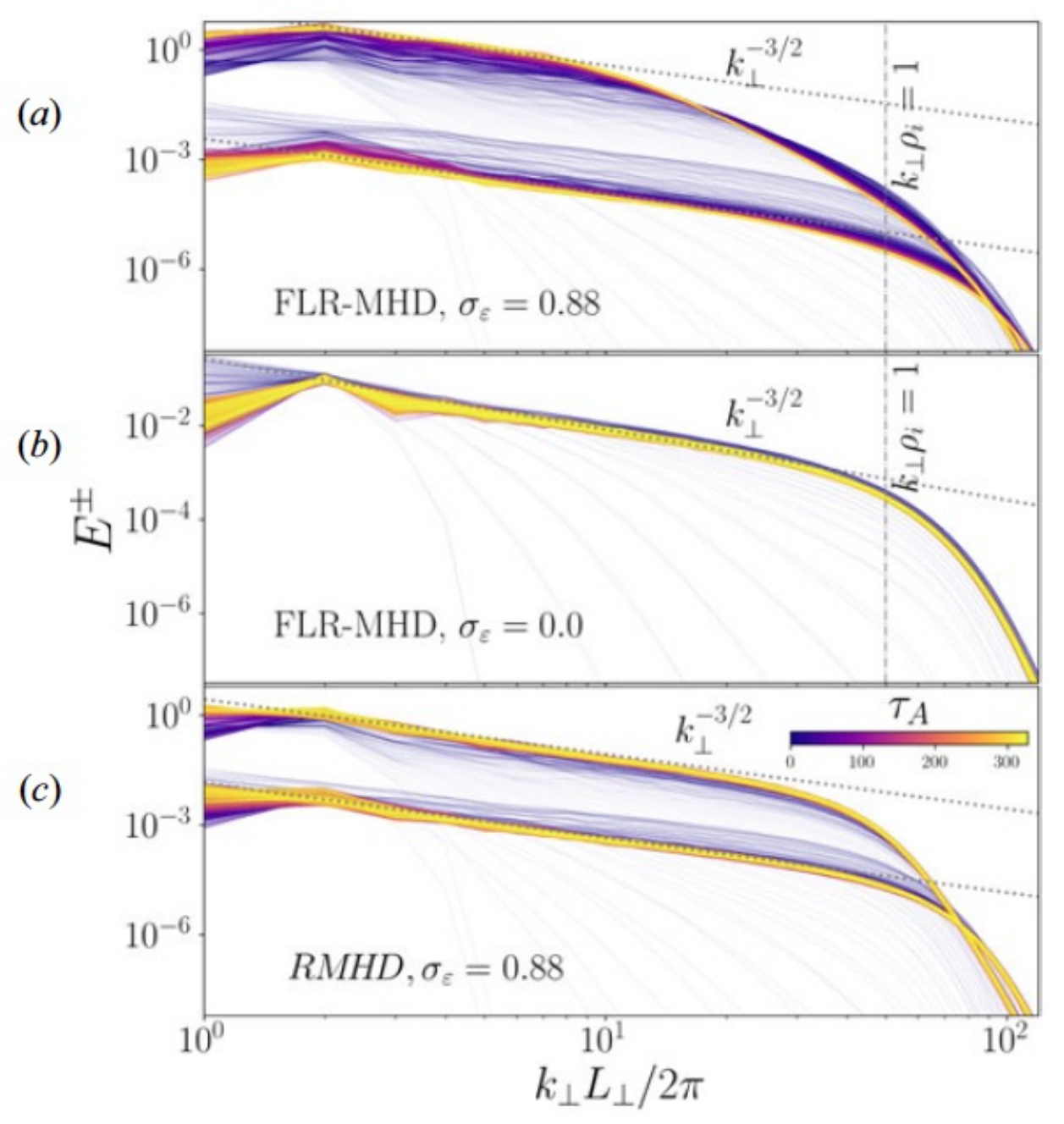}} 
\vskip2mm
\caption{Spectra of Elsasser field for 
(c) heavily imbalanced RMHD turbulence ($\eps^+/\eps^-\approx 16$), 
(a) similarly imbalanced 
turbulence in an RMHD-like model of a low-beta collisionless plasma with 
FLR effects, and (b) balanced turbulence in the same model. Purple-to-yellow 
colour scale shows time evolution of the spectra. The spectrum of the stronger 
Elsasser field in (a) does not reach a steady state, with spectral 
break moving to larger scales. Measurement of energy fluxes in these 
simulations shows that (a) is not a constant-flux solution, whereas (b)~is. 
These results are from \citet{meyrand21}.}
\label{fig:meyrand_imb}
\end{figure}

FLR effects do not just change how linear waves propagate at sub-Larmor scales. They also 
change the nature of the second conserved quantity (the first being energy) possessed 
by the plasma: (R)MHD cross-helicity (imbalance) becomes magnetic helicity in the 
transition from the inertial to the sub-Larmor scale range. The trouble is that 
the KAW helicity is a quantity that naturally wants to cascade inversely, from 
small scales to large \citep{sch09,cho11,kim15,cho16,miloshevich21}. 
In low-beta plasmas, there is no dissipation at the ion Larmor scale \citep{sch19}, 
so an imbalanced cascade arriving from the inertial range would get 
thoroughly ``confused'' by the sudden need to reverse the direction of the 
helicity cascade. The result, it turns out, is a mighty blow back from the small 
scales to large and a failure to achieve a constant-flux steady state 
(see \figref{fig:meyrand_imb}, taken from \citealt{meyrand21}; some evidence of strange 
behaviour of energy fluxes in imbalanced solar-wind turbulence does appear 
to exist: see, e.g., \citealt{smith09}). The imbalanced cascades are fragile.

What this means for the real plasmas that these models aspire to describe
(low beta, high Elsasser imbalance---typical solar-wind conditions close to the Sun,
currently sampled with gusto by the Parker Solar Probe; see, e.g., \citealt{chen20})
is that the fluid approximation is broken---not just in the sense of requiring FLR bolt-ons,
but in the sense that the system cannot accommodate a steady-state turbulent cascade 
while still respecting the $\kpar\ll\kperp$ and low-frequency 
(viz., $\omega\ll$~the ion cyclotron frequency) ordering associated 
with the gyrokinetic approximation and its various fluid-like reductions 
\citep{howes06,sch09,sch19}. A fully kinetic system deals with this problem
by building up energy in the stronger Elsasser field and, therefore, shortening
the nonlinear cascade times~$\tnl$, until critical balance
pushes $\kpar$ high enough to break the gyrokinetic approximation
and open up a new dissipation channel via the ion-cyclotron
resonance---a happy reconciliation, it seems, between ample observational
evidence both for ion-cyclotron heating and for low-frequency cascades in
the solar wind \citep{squire22}. 

This is a relatively rare example of plasma microphysics seriously upsetting 
system-scale macrophysics (another possible set of such examples is flagged
in \secref{sec:PA}). Sadly, this renders much of \secref{sec:imbalanced} almost  
completely irrelevant for such plasmas---unless an effective way can be found 
of modifying the dissipation-scale pinning scheme (\secref{sec:pinning}) 
to account for the kinetic dissipation mechanism described above. 

\subsection{Phase-Space Turbulence}
\label{sec:phspace_turb}

What is turbulence? Some energy is injected into some 
part of the phase space of a nonlinear system (in fluid systems, that simply means 
position or wavenumber space), which is, generally speaking, not the part of the phase 
space where it can be efficiently thermalised. So turbulence is a process 
whereby this energy finds its way from where it is injected to where it can be 
dissipated, and its means of doing this is nonlinear coupling, usually from 
large to small scales (I am now putting to one side the upsetting example 
of the failure of this process discussed in \secref{sec:failed_cascade}). 
What kind of coupling is possible and at what rate 
the energy can be transferred from scale to scale 
then determines such things as energy spectra in 
a stationary state with a constant flux of energy. 

The same principle applies to kinetic turbulence, but now the phase space 
is 6D rather than 3D: the particle distribution depends on positions and velocities, 
and energy transfer can be from large to small scales (or vice versa) in all 
six coordinates. The transfer of (free) energy to small scales in velocity space, 
leading ultimately to activation of collisions, however small the collision rate, 
is known as ``phase mixing''. It is not always a nonlinear phenomenon: 
the simplest (although not necessarily very simple) phase-mixing process is 
the linear \citet{landau46} damping. In a magnetised plasma, this is the {\em parallel} 
(to $\vB_0$) phase mixing, whereas the {\em perpendicular} phase mixing is nonlinear and 
has to do with particles on Larmor orbits experiencing different electromagnetic 
fields depending on the radius of the orbit (the Larmor radius is a kinetic 
variable, being proportional to $\vperp$). The latter phenomenon 
leads to an interesting phase-space ``entropy cascade'' 
(\citealt{sch08,sch09,tatsuno09,plunk10,cerri18,eyink18,kawazura19}; 
cf.\ \citealt{pezzi18}), 
which is one of the more exotic phenomena that await a curious researcher at sub-Larmor scales. 
Its importance in the grand scheme of things is that it channels turbulent 
energy into ion heat, while the KAW cascade heats electrons---the question 
of which dissipation channel is the more important one, and when, being both 
fundamental and ``applied'' (in the astrophysical sense of the word---e.g., to accretion 
flows: see \citealt{quataert99}, \citealt{EHT19}). Understanding how energy is transferred 
between scales in phase space requires thinking somewhat outside the standard 
turbulence paradigm and so perhaps counts as conceptual novelty. 
Not much of it has been done so far and it is worth doing more. 

Returning to parallel phase mixing, this too turns out to be interesting 
in a nonlinear setting, even though it is a linear phenomenon itself. 
First theoretical \citep{sch16,adkins18} and numerical \citep{parker16,meyrand19} analyses 
suggest that, in a turbulent system, parallel phase mixing is 
effectively suppressed by the stochastic plasma echo, perhaps rendering 
kinetic systems that are notionally subject to Landau damping effectively 
fluid, at least in terms of their energy-flow budgets. In the context of 
inertial-range MHD turbulence, this is relevant to the compressive 
(``slow-mode'') perturbations, which, in a collisionless plasma, are 
energetically decoupled from, and nonlinearly slaved to, 
the Alfv\'enic ones, while the latter are 
still governed by RMHD \citep{sch09,kunz15}. Linearly, 
these compressive perturbations must be damped---but nonlinearly they are 
not \citep{meyrand19}, 
thus accounting for them exhibiting a healthy power-law 
spectrum and other fluid features in the solar wind 
\citep{chen16,verscharen17}. 
In this vein, one might also ask whether the Landau damping of KAWs 
at sub-Larmor scales is always efficient or even present at all---and 
if it is, as \citet{tenbarge13a}, \citet{banon16}, \citet{kobayashi17}, 
and \citet{chen19} all say, then what is different at these scales. 
Given that \citet{loureiro13colless} see a characteristic signature of phase mixing 
in collisionless reconnection, reconnection might yet again turn out to be the key player, 
as indeed it has been conjectured to be at these scales (see \secref{sec:microphysics}). 

The broader question is whether there is generally Landau damping in turbulent systems
and whether, therefore, to put it crudely, ``all turbulence is fluid.'' 
While it might be a little disappointing if it is, the way and the sense in which this 
seems to be achieved are surprising and pleasingly nontrivial---and possibly 
soon to be amenable to direct measurement if the first MMS results on velocity-space 
(Hermite) spectra in the Earth's magnetosheath \citep{servidio17} 
are a good indication of the possibilities that are opening up. 

\subsection{Statistical Thermodynamics of Collisionless Plasma}

\begin{flushright}
{\small\parbox{6.2cm}{
At quite uncertain times and places,\\
\null\quad The atoms left their heavenly path,\\
And by fortuitous embraces,\\
\null\quad Engendered all that being hath.\\
And though they seem to cling together,\\
\null\quad And form "associations" here,\\
Yet, soon or late, they burst their tether,\\
\null\quad And through the depths of space career.}
\vskip2mm
J.~C.~Maxwell, {\em Molecular Evolution}}
\end{flushright}
\vskip5mm

Phase-space turbulence is interesting not just in its own right, but also
as a means to answering an even broader, and trickier, question, or class of questions, 
which concerns the equilibria underlying 
the fluid-like (or kinetic) dynamics of plasma turbulence. The meaning of the word 
``thermalised,'' which I used to describe the fate of turbulent energy at the beginning 
of~\secref{sec:phspace_turb}, is really only fully clear if the underlying equilibrium 
is Maxwellian---i.e., if significant departures from thermal equilibrium occur, 
and, therefore, a kinetic description is required, only for the turbulent fluctuations. 
This is certainly not the case for many natural plasmas, where the overall distribution 
functions both of ions and of electrons, can be order-unity non-Maxwellian---the 
prime and most accessible example of that being the solar 
wind \citep[e.g.,][]{marsch06,marsch18,martinovic20}. 
A number of questions then arise, all of them both unsolved and 
extremely fundamental on a level that should appeal to any physicist: 
\vskip2mm
(i) Do universal (i.e., not sensitive in detail to initial conditions) 
collisionless equilibria exist? 
\vskip2mm
(ii) If they do, is it possible to find them by constructing a statistical 
thermodynamics of collisionless 
plasma based on such staples as the maximum-entropy principle? And if so, 
what is the correct definition of entropy for a collisionless plasma
\citep[cf.][]{fowler68,eyink18,matthaeus20,zhdankin22a,chavanis22,ewart22}?  
\vskip2mm
(iii) Is it then possible to have a theory of plasma relaxation to these 
equilibria in terms of some ``collisionless collision integrals"? What 
is the relaxation rate (i.e., the ``effective collision rate'')? 
It is at this step that the structure of phase-space 
turbulence, which was flagged as interesting in its own right in \secref{sec:phspace_turb}, 
comes in as an essential ingredient: in the same way as the ``fluid'' theory of 
turbulent transport in, e.g., a gyrokinetic plasma, requires knowledge 
of the statistical properties of turbulence, which determine
turbulent fluxes of conserved ``fluid'' quantities \citep[e.g.,][]{abel13}, 
the kinetic theory of collisionless relaxation requires one to know 
the phase-space correlation function of the perturbed distribution 
function, which determines the turbulent flux of phase density
\citep{kadomtsev70,chavanis22,ewart22}.    
\vskip2mm
(iv) And if all, or some, of the above is accomplished, how can one then describe 
the passage of energy from turbulent fluctuations into the equilibrium 
distribution (``thermalisation'')\footnote{Of course, collisionless equilibrium distributions
do not have to be ``thermal'', i.e., they can have extended tails at high
energies---such tails are indeed observed by astronomers 
and so nonthermal particle acceleration by turbulence is an object of intense interest
to astrophysicists: see references in footnote~\ref{fn:relativistic}.
Mathematically, distribution functions with nonthermal, power-law
tails can be derived as solutions of kinetic equations containing
phase-space advection and diffusion with velocity-dependent coefficients
\citep[see, e.g.,][and references therein]{wong20,vega22a,uzdensky22}
or as maximisers of a judiciously chosen entropy \citep[e.g.,][]{zhdankin22b,ewart22}.
In a complete theory, they should appear as fixed points of the ``collisionless
collision integrals''.}
in terms of some generalised free-energy 
cascade (one example of how one might approach this question 
can be gleaned from a comparison between 
\citealt{sch09}, where a Maxwellian equilibrium is assumed, and  
\citealt{kunz15,kunz18}, where the equilibrium is allowed to be 
pressure-anisotropic). 
\vskip2mm
These questions are not new---there was some very interesting activity 
around them half a century ago \citep[e.g.,][]{fowler68,kadomtsev70,dupree72}, 
which produced a flurry of follow-ups, 
but the subject has since gone into abeyance, due, in large part, 
to the impossibility at the time 
of numerical or observational testing of any of the theories.
This is now changing: computers are big enough to handle the vastness of the 6D phase 
space, and the heliosphere is teeming with spacecraft full of eager and 
discerning instruments.
This is a exciting research frontier if there has ever been one. Here is not 
the place to review this topic in any further detail---my own take on it, 
as pedagogical as I can manage, alongside a more complete set of 
references, can be found in \citet{schKT}.

\subsection{Macro- and Microphysical Consequences of Pressure Anisotropy}
\label{sec:PA}

Another line of inquiry pregnant with conceptual novelty concerns
the effect of self-generated pressure anisotropy 
on MHD dynamics. Pressure anisotropies are generated in response 
to any motion in a magnetised collisionless or weakly collisional plasma 
as long as this motion 
leads to a change in the strength of the magnetic field. The conservation 
of the magnetic moment ($\propto \vperp^2/B\propto$
the angular momentum of Larmor-gyrating particles) 
then causes positive (if the field grows) or negative (if it decreases) 
pressure anisotropy to arise \citep[see, e.g.,][]{sch10}. 
This is usually quite small, but it becomes relevant at high beta, when even small
anisotropies (of order $1/\beta$) can have a dramatic effect, in two ways.
Dynamically, pressure anisotropy supplies additional stress, which, 
when the anisotropy is negative ($\pperp<\ppar$), can cancel Maxwell's 
stress and thus remove magnetic tension---the simplest way 
to think of this is in terms of the Alfv\'en speed 
being modified so: 
\beq
\vA\to\sqrt{\vA^2 + \frac{\pperp-\ppar}{\rho}}.
\label{eq:vA_aniso}
\eeq
Kinetically, pressure anisotropy is a source of free energy and 
will trigger fast, small-scale instabilities, most notably mirror 
and firehose (see \citealt{kunz14} and references therein). 
The firehose corresponds to the Alfv\'en speed \exref{eq:vA_aniso} 
turning imaginary, i.e., it is an instability caused by negative tension; 
the mirror is not quite as simple to explain, but is fundamentally 
a result of effective magnetic pressure going negative by means 
of some subtle resonant-particle dynamics (see \citealt{southwood93}, \citealt{kunz15}
and references therein). These instabilities in turn can regulate 
the anisotropy by scattering particles or by subtler, more devious means 
(see \citealt{melville16} and references therein).
   
Investigating of the dynamics of a simple finite-amplitude 
Alfv\'en wave in a collisionless, high-beta plasma, \citet{squire16,squire17,squire17num} 
showed that both of these effects did occur and altered the wave's behaviour drastically: 
it first slows down to a near halt due to the removal of magnetic tension, transferring much 
of its kinetic energy into heat and then, 
having spawned a colony of particle-scattering Larmor-scale perturbations, 
dissipates as if it were propagating in a plasma with a large \citet{braginskii65} 
parallel viscosity. Sound waves in a collisionless plasma 
get similarly infested by firehoses and mirrors, except 
the resulting effective collisionality helps them propagate 
in a medium that they thus render more fluid and, therefore, 
incapable of Landau damping (a different mechanism than discussed 
in \secref{sec:phspace_turb}, but a similar outcome; see \citealt{kunz20}).

These effects occur provided the amplitude of the waves is above a certain 
limit that scales with plasma beta: this is because 
pressure anisotropy must be large enough to compete with tension in \exref{eq:vA_aniso} 
and the amount of anisotropy that can be generated is of the order of the 
field-strength perturbation. For an Alfv\'en wave, the latter is quadratic in 
the wave's amplitude:
\beq
\lt(\frac{\db}{\vA}\rt)^2\sim \frac{\pperp-\ppar}{p}
\gtrsim \frac{\vA^2}{p/\rho}\sim \frac{1}{\beta}.
\label{eq:PA_threshold}
\eeq 
In formal terms, this means that in high-beta collisionless plasmas, 
the small-amplitude and high-beta limits do not commute. The conventional picture 
of Alfv\'enic turbulence simply obeying RMHD equations, even in a collisionless 
plasma \citep{sch09,kunz15}, may then have to be seriously revised for such
environments as galaxy clusters, for example, where $\beta \sim 10^2$ \citep{sch06}
(a first step in this direction has been taken by 
\citealt{squire19}, who found that MHD turbulence with Braginskii viscosity, 
while looking in many respects similar to the usual Alfv\'enic turbulence, 
nevertheless manages to minimise changes in the magnetic-field 
strength to a much greater extent---a property they dubbed ``magneto-immutability''; 
cf.~\citealt{tenerani20constB}, who do not like the term, but explore useful 
dynamical scenarios for achieving just such a state).\footnote{Let me mention 
parenthetically that at the small-scale end of the turbulent cascade, 
electron pressure anisotropies lie in wait to mess with the way in which 
reconnection occurs. I will not go into this here, referring the reader 
to a review by \citet{egedal13}. This is another microphysical effect 
that may need to be inserted into the sub-Larmor dynamics (see \secref{sec:microphysics}).}  

In situations where the inequality \exref{eq:PA_threshold} is not satisfied, but
the pressure anisotropy is a feature of the equilibrium, 
it seems we can live with the current theory as long as $\beta$ is not too large. 
This is the case for most instances of the solar wind, where negative pressure anisotropy
is driven by expansion away from the Sun (so $B$ decreases with heliocentric distance) and 
$\beta$ ranges from $\ll1$ closer to the Sun to $\gtrsim 1$ in our own
neighbourhood. A recent investigation by \citet{bott21fh} of Alfv\'enic turbulence in an expanding
box (mimicking the solar wind) at $\beta \approx 2-4$ found that a kinetic
version of the firehose instability, which, at these $\beta$ (unlike at $\beta\gg1$),
kicks in at a (negative) value of $\pperp-\ppar$ order-unity short of
zeroing out the magnetic tension, generated a sea of microscale magnetic perturbations
that, by scattering particles, adjusted the effective collisionality
of the plasma to keep it exactly marginal to that kinetic firehose;
the critically balanced, Alfv\'enic cascade survived unscathed, albeit with
the Alfv\'en speed modified according to~\exref{eq:vA_aniso}. The (potentially)
more interesting things that happen at $\beta\gg1$ are costlier to tackle
numerically, but, it seems, soon will be. 

Existing understanding of another basic high-beta MHD process, 
the small-scale dynamo, which I discussed at length in \secref{sec:dynamo}, is 
also potentially endangered by ubiquitous pressure anisotropies---but has  
survived the first contact with direct numerical experimentation, which 
required extra-large, ``hero'' kinetic simulations 
\citep{rincon16,kunz16,stonge18}.\footnote{``Hero'' they might have been, but 
they were only kinetic as far as ions were concerned, whereas electrons were fluid, 
with the usual resistive unfreezing of flux. There is as yet no demonstration of a dynamo 
in a plasma with kinetic electrons, and the only paper that tried to get such a dynamo, 
failed, because magnetic fields got Landau-damped away \citep{pusztai20}, but in
a system with fairly low scale separations. Two other recent studies \citep{pucci21,zhou22}
have got a little farther, showing that electron pressure anisotropy generated by externally
driven flows in an unmagnetised plasma 
will lead to generation of microscale magnetic fields via Weibel
instability (another pressure-anisotropy-driven instability, even smaller-scale
than firehose and mirror, and, unlike them, requiring particles {\em not} to be magnetised);
their plan is for these fields to serve as seed for a fully two-species-kinetic dynamo.}
So far it appears that in this problem as well, 
changing magnetic fields render plasma more collisional 
in some effective sense and so large-$\Pm$ dynamo remains a relevant paradigm. 
The same conclusion was reached by \citet{stonge20}, who simulated dynamo 
action and saturation in MHD with Braginskii stress (the collisional limit 
of pressure-anisotropic dynamics). 
  
This line of investigation may be particularly rich in surprises because 
pressure-anisotropy stress undermines much of our basic intuition for 
ideal-MHD dynamics, not just modifies microscale plasma physics.  
This said, it is not entirely inconceivable that, at the end of the day 
(or of the decade), 
in some grossly coarse-grained sense, turbulent plasmas will just turn out 
to supply their own effective collisionality even where Coulomb collisions 
are rare---and so astrophysicists, with their focus on large-scale 
motions, need not be too worried about the validity of fluid models beyond requiring
a few easily implementable tweaks. I hope life is not quite so boring, 
although, as a theoretical physicist and, therefore, a believer in universality, 
I should perhaps expect to be pleased by such an outcome.  

\section{Conclusion}
\label{sec:conc}

Let us stop here. The story of MHD turbulence is a fascinating one---both the story 
of what happens physically and the story of how it has been understood. 
It is remarkable how long it takes to figure out simple things, obvious in retrospect. 
It is even more remarkable (and reassuring) that we get there 
after all, in finite time. This story now looks reasonably complete, 
at least in broad-brush outline (\secref{sec:end}) and modulo numerous caveats
and many interesting loose ends (\secsdash{sec:imbalanced}{sec:dynamo}). 
Is this an illusion? Is it all wrong again? We shall know soon enough, but 
in the meanwhile, the siren call of kinetic physics is too strong to resist and the 
unexplored terrain seems vast and fertile (\secref{sec:kinetic}). 
Is everything different there? Or will it all, 
in the end, turn out to be the same, with Nature proving itself a universalist bore and 
contriving to supply effective collisions where nominally there are few? 
Is turbulence always basically fluid or do subtle delights await us in phase space? 
Even if we are in danger of being disappointed by the answers to these questions, 
getting there is proving to be a journey of amusing twists and turns. 

\begin{acknowledgments} 
For a topic as broad as this, it is difficult to list all the people from 
whom I have learned what I know (or think I know) of this subject. 
The most important such influence has been Steve~Cowley. 
The views expressed in the first part of this paper (\secsdash{sec:CB}{sec:disruption}) 
were informed largely by my collaboration with Alfred~Mallet and Ben~Chandran 
and by conversations with Andrey Beresnyak (even if he is likely to disagree
with my conclusions), Nuno~Loureiro and Dmitri Uzdensky.
I have learned most of what I know of reconnection from 
Nuno and Dmitri and of the solar wind from Chris~Chen, Tim~Horbury, and Rob~Wicks. 
I owe the first epigraph of this paper to the erudition of Richard McCabe 
and the second (as well as the epigraph of \secref{sec:disruption}) to that of Matt Kunz. 
The sections on weak turbulence (\secref{sec:WT}, \apref{app:WT} and especially 
\apref{app:bband}) would have been different had Thomas Foster not 
convinced me to make a modicum of peace with the traditional theory. 
Ben Chandran, Nuno Loureiro, and Andrey Beresnyak have also helped me think coherently of 
imbalanced turbulence---without necessarily endorsing the outcome (\secref{sec:imbalanced}). 
David Hosking demolished comprehensively my first version of \secref{sec:decaying}  
(on decaying MHD turbulence) and forced a complete rewrite---in the process, 
he wrote what I consider to be one of the more beautiful papers that 
I have had the good fortune of being involved with \citep{hosking21}. 
The contents of \secref{sec:dynamo} were inspired by discussions 
with Andrey~Beresnyak, Alisa Galishnikova, Fran\c{c}ois~Rincon, and Matt Kunz, 
re-examining my views on small-scale dynamo that had been formed in the early 2000s. 
The views expressed in \secref{sec:kinetic} are in large measure a result of 
my recent and current excursions to the kinetic frontier in the company   
of Toby Adkins, Lev Arzamasskiy, 
Michael Barnes, Archie Bott, Andrew Brown, Bill Dorland, Robbie Ewart,
Yohei Kawazura, Matt Kunz, Romain Meyrand, Michael Nastac, Eliot Quataert, and Jono Squire. 
Besides the colleagues and friends mentioned above, conversations with 
Allen Boozer, Axel Brandenburg, Mike Brown, Peter Davidson, Daniele Del Sarto, Greg Eyink,   
Henrik Latter, Alex Lazarian, Sergey Nazarenko, Maurizio Ottaviani, 
Felix Parra, Marco Velli, Mahendra Verma, and Muni Zhou have helped 
me work out what to say, and how to say it, in various bits of this review. 
I would also like to thank the authors whose figures appear in 
the text for giving me permission to reproduce their art. 
I hasten to add that none of those mentioned here bear any responsibility 
for my many enduring confusions and biases. 
If, nevertheless, this paper occasionally manages to make sense to its readers, 
its six referees deserve some significant share of credit for that---they all provided 
reviews both spirited and helpful, ranging from spotting multitudinous typos 
to calling out preposterous stylistic bloopers to identifying 
important logical pitfalls, and even, in one instance, offering detailed 
suggestions as to the most appropriate wine selection to go with each section.

Visits to a number of pleasant places have helped bring this work to completion. 
I~am delighted to acknowledge the hospitality of the Wolfgang Pauli Institute, 
Vienna, where, in meetings held annually for 12 years (2007-19), 
many key interactions took place and ideas were hatched.  
This paper started as an ``opinion piece'' written for the 1st JPP Frontiers 
of Plasma Physics Conference at the Abbazia di Spineto in 2017, when the 
news of tearing-mediated turbulence was very fresh. 
My extended stay in 2018 at the Niels Bohr International Academy, Copenhagen, 
where some nontrivial {\em i}'s were dotted and {\em t}'s crossed, 
was supported by the Simons Foundation (via Martin Pessah, to whom I am grateful 
for offering me NBIA's hospitality). 
Another place whose hospitality, in 2019, proved germane to making progress  
was the Kavli Institute of Theoretical Physics, Santa Barbara, 
during its programme on ``Multiscale Phenomena in Plasma Astrophysics'' 
led by Anatoly Spitkovsky.
In the UK, my work was supported in part by grants from STFC (ST/N000919/1 and ST/W000903/1) 
and EPSRC (EP/M022331/1 and EP/R034737/1). 
The manuscript was finally finished during the first Covid-19 lockdown 
(in 2020) and revised during the second (2021), but I offer no thanks to the virus. 

{\em Declaration of Interests.} The author reports no conflict of interest.
\end{acknowledgments} 

\appendix

\part*{Appendices}
\label{part:apps}
\addcontentsline{toc}{section}{\partskip \sc \nameref{part:apps}}

\section{Splendours and Miseries of WT Theory}
\label{app:WT}

\subsection{RMHD in Scalar Form}
\label{app:RMHD_scalar}

It is convenient to rewrite the RMHD equations \exref{eq:zpm} in terms of two scalar fields, 
so-called Elsasser potentials $\zeta^\pm$, 
which are the stream functions for the 2D-solenoidal fields $\vzperp^\pm$ \citep{sch09}, 
viz., 
\beq
\vzperp^\pm = \ez\times\vdperp\zeta^\pm, 
\label{eq:zeta_def}
\eeq
where $\ez = \vB_0/B_0$. Then $\zeta^\pm$ satisfy, as shown by taking 
the curl of \exref{eq:zpm} and using \exref{eq:zeta_def}, 
\beq
\frac{\dd\omega^\pm}{\dd t} \mp \vA\dpar\omega^\pm 
= - \lt\{\zeta^\mp,\omega^\pm\rt\} + \lt\{\dd_j\zeta^\pm,\dd_j\zeta^\mp\rt\},
\label{eq:zeta}
\eeq
where $\omega^\pm = \ez\cdot(\vdperp\times\vzperp^\pm) = \dperp^2\zeta^\pm$ are 
Elsasser vorticities, all dissipative terms have been dropped, and 
\beq
\lt\{\zeta^\mp,\omega^\pm\rt\} = \frac{\dd\zeta^\mp}{\dd x}\frac{\dd\omega^\pm}{\dd y} 
- \frac{\dd\zeta^\mp}{\dd y}\frac{\dd\omega^\pm}{\dd x}
= \vzperp^\mp\cdot\vdperp\omega^\pm. 
\eeq 
Note that I have written 
\exref{eq:zeta} in a slightly different (but equivalent) form than in \citet{sch09}. 
The present version emphasises that the two physical influences of the 
nonlinearity on the Elsasser vorticities are advection 
by the other Elsasser field $\vzperp^\mp$ (the first term on the right-hand side) 
and ``vortex stretching'' (the second term) \citep[cf.][]{zhdankin16}---this is 
a useful way to write these equations for, e.g., 
the argument in \secref{sec:new_res_theory} about the build-up of negative 
correlations between $\omega^+$ and~$\omega^-$ (residual energy).  

In Fourier space, \exref{eq:zeta} has a nicely generic form  
\beq
\label{eq:zetak}
\dd_t\zeta^\pm_\vk \mp i\kpar\vA\zeta^\pm_\vk = 
\sum_{\vp\vq} M_{\vk\vp\vq}\delta_{\vk,\vp+\vq}\zeta^\mp_\vp\zeta^\pm_\vq,
\eeq
with the coupling coefficients 
\beq
M_{\vk\vp\vq} = \ez\cdot\lt(\vkperp\times\vqperp\rt)\frac{\vkperp\cdot\vqperp}{\kperp^2}
= \qperp^2\sin\phi\cos\phi,
\eeq
where $\phi$ is the angle between $\vkperp$ and $\vqperp$. 

\subsection{Classic WT Calculation}
\label{app:WT_derivation}

Our objective is to derive an evolution equation for 
the spectra $C^\pm_\vk = \lt\la|\zeta^\pm_\vk|^2\rt\ra$. 
Multiplying \exref{eq:zetak} by $\zeta^{\pm*}_\vk$ and adding to the resulting 
equation its complex conjugate, we~get 
\beq
\dd_t C^\pm_\vk = 2\Re \sum_{\vp\vq} 
M_{\vk\vp\vq}\delta_{\vk,\vp+\vq}\lt\la\zeta^\mp_\vp\zeta^\pm_\vq\zeta^{\pm*}_\vk\rt\ra. 
\label{eq:WT2}
\eeq
Similarly, the evolution equation for the triple correlator appearing in the 
right-hand side of~\exref{eq:WT2}~is 
\begin{align}
\nonumber
\dd_t \lt\la\zeta^\mp_\vp\zeta^\pm_\vq\zeta^{\pm*}_\vk\rt\ra 
\mp i2\ppar\vA \lt\la\zeta^\mp_\vp\zeta^\pm_\vq\zeta^{\pm*}_\vk\rt\ra = & 
\sum_{\vk'\vk''} \lt[ M_{\vp\vk'\vk''}\delta_{\vp,\vk'+\vk''}
\lt\la\zeta^\pm_{\vk'}\zeta^\mp_{\vk''}\zeta^\pm_\vq\zeta^{\pm*}_\vk\rt\ra \rt.\\
\nonumber
&\quad\lt. +\, M_{\vq\vk'\vk''}\delta_{\vq,\vk'+\vk''}\rt.
\lt\la\zeta^\mp_\vp\zeta^\mp_{\vk'}\zeta^\pm_{\vk''}\zeta^{\pm*}_\vk\rt\ra\\
&\quad\lt. +\, M_{\vk\vk'\vk''}\delta_{\vk,\vk'+\vk''}
\lt\la\zeta^\mp_\vp\zeta^\pm_\vq\zeta^{\mp*}_{\vk'}\zeta^{\pm*}_{\vk''}\rt\ra\rt]
\equiv A_{\vk\vp\vq}, 
\label{eq:WT3}
\end{align}
where, in working out the linear term, it has been opportune 
to take account of $\kpar = \ppar + \qpar$. 
To lowest order in the WT expansion, with $A_{\vk\vp\vq}$ approximated as constant in time, 
the solution to this equation~is
\beq
\lt\la\zeta^\mp_\vp\zeta^\pm_\vq\zeta^{\pm*}_\vk\rt\ra = 
\frac{1 - e^{\mp i2\ppar\vA t}}{\pm i2\ppar\vA}\,A_{\vk\vp\vq} 
\to \frac{\pi\delta(\ppar)}{2\vA}\, A_{\vk\vp\vq}
\rmas t\to\infty.
\label{eq:WT3_delta}
\eeq 
This is the moment when it turns out that every interaction must involve 
the $\ppar=0$ mode, for which the WT approximation is, in fact, broken.  

Pressing on regardless, let us adopt the random-phase approximation, 
as always in WT \citep{zakharov92,nazarenko11book}.  
Namely, to lowest order in the WT expansion, 
any wave field is only correlated with itself at the same $\vk$, all odd 
correlators vanish [which is why I had to iterate from \exref{eq:WT2} to \exref{eq:WT3}], 
and all even correlators are split into products of quadratic ones, viz., 
\begin{align}
&\lt\la \zeta^\pm_\vk\zeta^\pm_{\vk'}\rt\ra = C^\pm_\vk\delta_{\vk,-\vk'},\\
\label{eq:RE_zero}
&\lt\la \zeta^+_\vk\zeta^-_{\vk'}\rt\ra = 0,\\
\label{eq:triple1}
&\lt\la\zeta^\pm_{\vk'}\zeta^\mp_{\vk''}\zeta^\pm_\vq\zeta^{\pm*}_\vk\rt\ra = 0,\\
&\lt\la\zeta^\mp_\vp\zeta^\mp_{\vk'}\zeta^\pm_{\vk''}\zeta^{\pm*}_\vk\rt\ra = 
C^\mp_\vp\delta_{\vp,-\vk'} C^\pm_\vk\delta_{\vk'',\vk},\\
&\lt\la\zeta^\mp_\vp\zeta^\pm_\vq\zeta^{\mp*}_{\vk'}\zeta^{\pm*}_{\vk''}\rt\ra = 
C^\mp_\vp\delta_{\vp,\vk'} C^\pm_\vq\delta_{\vq,\vk''}.
\label{eq:triple3}
\end{align}
Therefore, noticing that $M_{\vq,-\vp,\vk}=-M_{\vk\vp\vq}\kperp^2/\qperp^2$, we get 
\beq
\label{eq:Akpq}
A_{\vk\vp\vq} = M_{\vk\vp\vq}\delta_{\vk,\vp+\vq} C^\mp_\vp\lt(C^\pm_\vq - 
\frac{\kperp^2}{\qperp^2}\,C^\pm_\vk\rt).  
\eeq
Combining \exref{eq:Akpq} with \exref{eq:WT3_delta} and putting the latter back 
into~\exref{eq:WT2}, we arrive at the classic WT equation derived by \citet{galtier00}: 
\beq
\dd_t N^\pm_\vk = \frac{\pi}{\vA}\sum_{\vp\vq} 
\frac{\kperp^2 M^2_{\vk\vp\vq}}{\pperp^2\qperp^2}\,\delta_{\vk,\vp+\vq}\,
\delta(\ppar)\, N^\mp_\vp\lt(N^\pm_\vq - N^\pm_\vk\rt), 
\label{eq:WT_Galtier}
\eeq
where $N^\pm_\vk = \kperp^2 C^\pm_\vk = \lt\la|\vz^\pm_{\perp\vk}|^2\rt\ra$.

\subsection{Solution of WT Equation}
\label{app:WT_solution}

The wavenumber sum in \exref{eq:WT_Galtier} is turned into an integral 
in the usual fashion: taking account of the restriction $\vk = \vp+\vq$ 
and of the fact that the integrand is even in~$\phi$,   
\beq
\sum_{\vp\vq}\lt(\dots\rt) = 2\,\frac{V}{(2\pi)^3}\int_{-\infty}^{+\infty}\rmd\qpar
\int_0^\infty\rmd\qperp\qperp\int_0^\pi\rmd\phi \lt(\dots\rt), 
\eeq
where $V = \Lperp^2\Lpar$ is the volume of the box. 
The angle integral can be recast as an integral with respect to~$\pperp$:
\beq
\pperp^2 = \kperp^2 + \qperp^2 - 2\kperp\qperp\cos\phi
\hence
\int_0^\pi\rmd\phi\sin\phi\lt(\dots\rt) = 
\int_{|\kperp-\qperp|}^{\kperp+\qperp}\frac{\rmd\pperp\pperp}{\kperp\qperp}\lt(\dots\rt).
\eeq
Finally, defining the 2D spectra $\Ekk^\pm(\kperp,\kpar) = \kperp N^\pm_\vk V/(2\pi)^2$, 
we~get
\begin{align}
\nonumber
\dd_t \Ekk^\pm(\kperp,\kpar) = &\ \frac{1}{\vA}\int_0^\infty\rmd\qperp
\int_{|\kperp-\qperp|}^{\kperp+\qperp}\rmd\pperp
\frac{\kperp^2\qperp^2}{\pperp}\sin\phi\cos^2\phi\\
&\ \times \frac{\Ekk^\mp(\pperp,0)}{\pperp}
\lt[\frac{\Ekk^\pm(\qperp,\kpar)}{\qperp} - \frac{\Ekk^\pm(\kperp,\kpar)}{\kperp}\rt],
\label{eq:WT_Epm}
\end{align}
where $\cos\phi = (\kperp^2 + \qperp^2 - \pperp^2)/2\kperp\qperp$ 
and $\sin\phi = (1 - \cos^2\phi)^{1/2}$. 

Let us now, as anticipated in \exref{eq:WT_spectra}, assume 
\beq
\Ekk^\pm(\kperp,\kpar) = f^\pm(\kpar)\kperp^{\mu^\pm},\qquad
\Ekk^\mp(\kperp,0) = f^\mp(0)\kperp^{\mu_0^\mp},
\label{eq:E_powers}
\eeq
substitute these into the right-hand side of \exref{eq:WT_Epm} 
and non-dimensionalise the integral by changing the integration variables 
to $x = \qperp/\kperp$ and $y = \pperp/\kperp$: 
\begin{align}
\label{eq:WT_flux}
\dd_t E^\pm(\kperp,\kpar) &= \frac{f^\mp(0)f^\pm(\kpar)}{\vA}\,I(\mu^\pm,\mu_0^\mp)\,
\kperp^{\mu^\pm + \mu_0^\mp + 3} \equiv -\frac{\dd\Pi^\pm(\kperp,\kpar)}{\dd\kperp},\\
I(\mu,\mu_0) &= \int_0^\infty\rmd x\int_{|1-x|}^{1+x}\rmd y\, 
y^{-2 + \mu_0} x^2 \lt(x^{\mu - 1} - 1\rt) \sin\phi\cos^2\phi, 
\label{eq:WT_I}
\end{align}
where $\cos\phi = (1+x^2-y^2)/2x$. 
The energy flux formally introduced in \exref{eq:WT_flux}~is
\beq
\Pi^\pm(\kperp,\kpar) = -\frac{f^\mp(0)f^\pm(\kpar)}{\vA}
\frac{I(\mu^\pm,\mu_0^\mp)}{\mu^\pm + \mu_0^\mp + 4}\,
\kperp^{\mu^\pm + \mu_0^\mp + 4}. 
\label{eq:WT_Pi}
\eeq
It is assumed here that the flux in $(\kperp,\kpar)$ space is in the $\kperp$ 
direction only (no parallel cascade in WT). 
In order for \exref{eq:WT_Pi} to be independent of $\kperp$, 
it must be the case that\footnote{Or $\mu^\pm=1$, in which case $I=0$, so $\Pi^\pm = 0$. 
This is a (UV-divergent) thermal equilibrium spectrum, which irrelevant for 
a forced problem below the forcing scale, but will, in a certain sense, be resurrected 
in \apref{app:det_beta}, at large scales.} 
\beq
\mu^\pm + \mu_0^\mp = - 4,
\label{eq:mumu4}
\eeq
but then, in order for the expression in \exref{eq:WT_Pi} to have 
a finite value, it must also be the case that 
$I(\mu,\mu_0) \to 0$ when $\mu + \mu_0 + 4 \to 0$.     
That this is indeed the case is shown by changing the integration variables 
to $\xi = 1/x$, $\eta = y/x$, a change that leaves the domain of integration 
in \exref{eq:WT_I} the same (a {\em Zakharov transformation}; see \citealt{zakharov92}). 
In these new variables, 
\beq
I(\mu,\mu_0) = -\int_0^\infty\rmd\xi\int_{|1-\xi|}^{1+\xi}\rmd\eta\, 
\eta^{-2 + \mu_0} \xi^{-\mu-\mu_0-2} \lt(\xi^{\mu - 1} - 1\rt)
\sin\phi\cos^2\phi,
\eeq
where $\cos\phi = (1+\xi^2-\eta^2)/2\xi$. When $\mu + \mu_0 = -4$, this is 
exactly the same integral as~\exref{eq:WT_I}, except with a minus sign, so 
$I = - I = 0$, q.e.d. 

The problem with this otherwise respectable-looking calculation is that 
$\Ekk^\mp(\pperp,0)$, which plays a key role in \exref{eq:WT_Epm}, 
is the spectrum of zero-frequency, $\ppar=0$ modes, for which 
the WT approximation cannot be used, so $\mu_0^\mp$ is certainly not 
determinable within WT, the random-phase approximation should not 
have been used for these modes (and has been explicitly shown not to 
hold for them by \citealt{meyrand15}), and so 
it is at the very least doubtful that \exref{eq:WT_Epm} can be 
used for the determination of $\mu^\pm$, the scaling exponents 
for the waves, either. For the moment, let me put aside the latter 
doubt and act on the assumption that if I can figure out $\mu_0^\mp$ 
in some way, $\mu^\pm$ will follow by~\exref{eq:mumu4}. 

\subsection{Case of Broad-Band Forcing: Spectral Continuity} 
\label{app:bband}

The argument that is about to be presented here is heuristic and routed in the 
ideas about the treatment of strong turbulence described in 
\secsref{sec:GS95} and \ref{sec:CB}---it turns out that,  
to understand weak turbulence, one must understand strong 
turbulence first. I will, therefore, not attempt to deal with 
imbalanced WT---because, even though I did, in~\secref{sec:imb_new}, 
attempt to construct a coherent picture of strong imbalanced turbulence, 
it is too tentative and too fiddly to be inserted into what follows, which will 
be tentative and fiddly in its own right. 
Thus, the ``$\pm$'' tags are now dropped everywhere. 

\begin{figure}
\centerline{\includegraphics[width=0.55\textwidth]{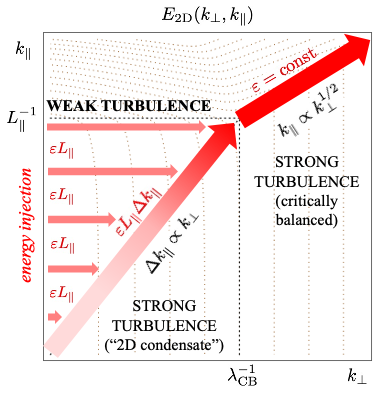}} 
\vskip2mm
\caption{Cartoon of the 2D spectrum of broad-band-forced weak turbulence. 
Schematic contour lines of $\Ekk(\kperp,\kpar)$ are the brown dotted lines.
Red arrows are energy fluxes: $\Pi(\kperp,\kpar)\sim\eps\Lpar$ arriving from 
the forcing wavenumbers to the ``2D condensate'' at each $\kpar$, 
$\eps\Lpar\Delta\kpar$ flowing through the condensate [see \exref{eq:cond_flux}], 
and $\eps = \const$ after the transition to critically balanced cascade 
(cf.~\figref{fig:2Dspectra}a).}
\label{fig:wt_spectrum}
\end{figure}

In reality, the delta function $\delta(\ppar)$ in \exref{eq:WT3_delta} has a width 
equal to the characteristic broadening of the frequency resonance 
due to nonlinear interactions, $\Delta\kpar \sim \tnl^{-1}/\vA$, to~wit,
\beq
\label{eq:delta_ppar}
\delta(\ppar) = \frac{\Delta\kpar}{\pi}\frac{1}{\ppar^2 + \Delta\kpar^2}
\eeq  
(in the WT approximation, $\Delta\kpar \to 0$). 
At $\ppar \lesssim\Delta\kpar$, the ``$\ppar=0$" condensate resides, 
whose turbulence is strong (\figref{fig:wt_spectrum}). 
Let us work out the structure of this turbulence. 

Let us assume that our WT is forced in a broad band of parallel wavenumbers 
$\kpar \in (0,2\pi/\Lpar)$ (obviously, the parallel size of ``the box'' must 
be $\gg\Lpar$). This can happen, e.g., if the forcing is completely
random with parallel coherence length $\Lpar$, in which case its $\kpar$ spectrum 
at $\kpar < 2\pi/\Lpar$ is flat (a white noise). Thus, the same amount of energy 
is injected into each $\kpar$, this energy is cascaded weakly in $\kperp$ 
(by the still-to-be-worked-out condensate) 
without change in $\kpar$ until it arrives at the CB scale 
associated with this $\kpar$, i.e., at the $\kperp$ for which 
$\Delta\kpar(\kperp) \sim \kpar$ (equivalently, $\tnl^{-1}\sim\kpar\vA$),   
where it joins the condensate. Therefore, the flux of energy into, and via, 
the condensate is not scale-independent: at any given $\kperp$, it is 
\beq
\int_0^{\Delta\kpar(\kperp)}\rmd\kpar \Pi(\kperp,\kpar) \sim \eps\Lpar\Delta\kpar(\kperp),
\qquad
\Delta\kpar(\kperp) \sim \frac{\tnl^{-1}}{\vA},
\label{eq:cond_flux}
\eeq
where I have assumed that $\Pi(\kperp,\kpar)\sim\eps\Lpar$ is a constant 
in both of its arguments (constant in $\kperp$ because the WT cascade is 
a constant-flux one and constant in $\kpar$ because the amount of energy injection 
is the same at every $\kpar$). Then, for the condensate at scale $\lambda \sim \kperp^{-1}$, 
\beq
\frac{\dz_\lambda^2}{\tnl} \sim \eps\Lpar\Delta\kpar(\kperp) \sim \frac{\eps\Lpar}{\tnl\vA}
\hence
\dz_\lambda \sim \lt(\frac{\eps\Lpar}{\vA}\rt)^{1/2}
\hence
E_0(\kperp) \sim \frac{\eps\Lpar}{\vA}\,\kperp^{-1},
\label{eq:E0_1D}
\eeq 
where $E_0(\kperp)$ is the condensate's 1D spectrum. 

This 1D spectrum is the 2D spectrum $E_0(\kperp,\kpar)$ integrated over all 
parallel wavenumbers belonging to the condensate, viz., 
\beq
E_0(\kperp) \sim \int_0^{\Delta\kpar(\kperp)}\rmd\kpar E_0(\kperp,\kpar) 
\sim E_0(\kperp,\kpar) \Delta\kpar(\kperp).
\label{eq:E0_1D2D}
\eeq
The last step is valid on the assumption that $E_0(\kperp,\kpar)$ is, in fact, 
independent of $\kpar$, because by the usual CB assumption, there cannot 
be any correlations at parallel scales $\kpar^{-1}>\vA\tnl\sim\Delta\kpar^{-1}(\kperp)$ 
and so the corresponding $\kpar$ spectrum is that of a white noise 
(cf.~\apref{app:det_delta}). 

The cascade time for the condensate (which advects itself) is 
\beq
\tnl^{-1} \sim \frac{\dz_\lambda}{\lambda} \sim \lt(\frac{\eps\Lpar}{\vA}\rt)^{1/2}\lambda^{-1}
\hence
\Delta\kpar(\kperp) \sim \frac{\lt(\eps\Lpar\rt)^{1/2}}{\vA^{3/2}}\,\kperp. 
\label{eq:tnl_cond}
\eeq
I am assuming that there is no dynamic alignment (\secref{sec:DA}) for the condensate because 
the condensate is effectively forced at every scale by the WT cascade---this is not a proof, 
but a conjecture, adopted for its simplicity and plausibility. 
Finally, \exref{eq:tnl_cond}, via \exref{eq:E0_1D2D} and \exref{eq:E0_1D}, leads to
\beq
E_{0,\mathrm{2D}}(\kperp,\kpar) \sim \lt(\eps\Lpar\vA\rt)^{1/2}\kperp^{-2},
\qquad
\kpar \lesssim  \Delta\kpar(\kperp).
\label{eq:E0_2D}
\eeq
Thus, $\mu_0= -2$ for reasons that have little to do with weak interactions, 
and, therefore, by \exref{eq:mumu4}, $\mu = -2$ as well (in \apref{app:WT_2D}, 
the same results are rederived in a slightly different way, which may or may 
not shed more light). 

Thus, there is, in fact, no difference between the WT spectrum 
at $\kpar>\Delta\kpar(\kperp)$ and the condensate's spectrum at 
$\kpar < \Delta\kpar(\kperp)$, 
even though the nature of turbulence in these two regions is quite different. 
The above construction can thus be viewed as a physical argument in support of 
spectral continuity. It does not make the derivation of the WT equation 
in \apref{app:WT_derivation} {\em formally} correct but it does perhaps 
lend it some credibility. 

\subsection{Residual Energy in WT} 
\label{app:WT_res}

If one takes this appearance of WT credibility seriously, there is another result 
that can be ``derived'' within it. The random-phase 
approximation for Alfv\'en waves implied the absence of correlations between the 
counterpropagating Elsasser fields, \exref{eq:RE_zero}. 
What if we relax this assumption---and 
{\em only} this assumption!---while still splitting fourth-order correlators into 
second-order ones? Namely, let us set 
\beq
\kperp^2\lt\la\zeta^\pm_\vk\zeta^\mp_{\vk'}\rt\ra = R^\pm_\vk\delta_{\vk,-\vk'},
\label{eq:Rk_def}
\eeq 
where $R_\vk^{-*} = R_\vk^+\equiv R_\vk$, and work out the WT evolution 
equation for $R_\vk$. This is interesting, {\em inter alia}, because 
$\Re R_\vk$ is the 3D residual-energy spectrum and so the derivation 
I am about to present (which is a version of what \citealt{wang11} did)
has a claim to providing theoretical backing 
to the presence of negative residual energy both in observed and 
in numerically simulated MHD turbulence (see \secref{sec:residual}).  

From the field equation \exref{eq:zetak}, straightforwardly,  
\beq
\dd_t R_\vk - 2i\kpar\vA R_\vk = \kperp^2\sum_{\vp\vq} M_{\vk\vp\vq}\delta_{\vk,\vp+\vq}
\lt(\lt\la\zeta^-_\vp\zeta^+_\vq\zeta^{-*}_\vk\rt\ra 
+ \lt\la\zeta^{+*}_\vp\zeta^{-*}_\vq\zeta^+_\vk\rt\ra\rt). 
\label{eq:Rk}
\eeq
Following the same protocol as in \apref{app:WT_derivation}, let us 
write the evolution equation for the third-order correlators in \exref{eq:Rk} 
in terms of fourth-order correlators and then split the latter into second-order ones, 
but now allowing non-zero 
correlations between different Elsasser fields according to~\exref{eq:Rk_def}: 
\begin{align}
\nonumber 
\dd_t\lt\la\zeta^\mp_\vp\zeta^\pm_\vq\zeta^{\mp*}_\vk\rt\ra 
& \mp i2\qpar\vA \lt\la\zeta^\mp_\vp\zeta^\pm_\vq\zeta^{\mp*}_\vk\rt\ra 
= \delta_{\vk,\vp+\vq}\frac{\ez\cdot\lt(\vkperp\times\vqperp\rt)}{\kperp^2\pperp^2\qperp^2}
\lt[\vkperp\cdot\vpperp N^\pm_\vq\lt(N^\mp_\vk - N^\mp_\vp\rt)\rt.\\
&\quad+ \lt.\vkperp\cdot\vqperp\lt(R^\mp_\vp R^\pm_\vq - R^\pm_\vk N^\mp_\vp\rt)
+ \vpperp\cdot\vqperp\lt(R^\pm_\vk R^\pm_\vq - R^\mp_\vp N^\mp_\vk\rt)\rt].
\label{eq:WT3_Rk}
\end{align}
The presence of the first term is proof that $R_\vk=0$ is, generally speaking, 
not a sustainable solution. However, since growth of correlations between 
counterpropagating Elsasser fields contradicts the random-phase approximation 
and thus undermines WT, perhaps we could hope (falsely, as I will show shortly) 
that $R_\vk$ might be small and so the terms containing $R_\vk$ in \exref{eq:WT3_Rk} 
could be neglected for the time being. Then the solution of \exref{eq:WT3_Rk}~is
\beq
\lt\la\zeta^\mp_\vp\zeta^\pm_\vq\zeta^{\mp*}_\vk\rt\ra = 
\frac{1 - e^{\pm i2\qpar\vA t}}{\mp i2\qpar\vA}\,
\delta_{\vk,\vp+\vq}\frac{\ez\cdot\lt(\vkperp\times\vqperp\rt) \vkperp\cdot\vpperp}{\kperp^2\pperp^2\qperp^2}
N^\pm_\vq\lt(N^\mp_\vk - N^\mp_\vp\rt). 
\eeq 
Substituting this into \exref{eq:Rk}, solving that in turn, and denoting 
\beq
B_{\vk\vp\vq} = \frac{\lt|\vkperp\times\vqperp\rt|^2 \vkperp\cdot\vpperp}{\kperp^2\pperp^2\qperp^2}
\lt[N^+_\vq\lt(N^-_\vk - N^-_\vp\rt) + N^-_\vq\lt(N^+_\vk - N^+_\vp\rt)\rt],
\eeq 
we find 
\begin{align}
\nonumber 
\Re R_\vk &= \Re\sum_{\vp\vq}\delta_{\vk,\vp+\vq}
\frac{1}{i2\qpar\vA}\lt(\frac{1-e^{i2\kpar\vA t}}{i2\kpar\vA} 
- e^{i2\qpar\vA t}\frac{1-e^{i2\ppar\vA t}}{i2\ppar\vA}\rt) B_{\vk\vp\vq},\\
&\to \frac{\pi^2}{4\vA^2}\sum_{\vp\vq}\delta_{\vk,\vp+\vq}
\delta(\ppar)\delta(\qpar) B_{\vk\vp\vq}
\rmas t\to \infty.
\end{align}
The 2D spectrum of residual energy is, therefore, 
\begin{align}
\nonumber
& \Ereskk(\kperp,\kpar) = \frac{V\kperp\Re R_\vk}{(2\pi)^2} = 
- \frac{\pi \delta(\kpar)}{4\vA^2}\int_0^\infty\rmd\qperp
\int_{|\kperp-\qperp|}^{\kperp+\qperp}\rmd\pperp\,
\frac{\kperp^2\qperp^2}{\pperp}\,\sin\phi\cos^2\phi\\
&\times\lt\{\frac{E^+(\qperp,0)}{\qperp}\lt[\frac{E^-(\kperp,0)}{\kperp} - 
\frac{E^-(\pperp,0)}{\pperp}\rt]
+ \frac{E^-(\qperp,0)}{\qperp}\lt[\frac{E^+(\kperp,0)}{\kperp} - 
\frac{E^+(\pperp,0)}{\pperp}\rt]\rt\},
\end{align}
where the wavenumber integrals have been manipulated in exactly the same 
way as they were in \apref{app:WT_derivation}, in the lead-up to \exref{eq:WT_Epm}. 
Again assuming the power-law solutions~\exref{eq:E_powers}, we get 
\beq
\Ereskk(\kperp,\kpar) = - \const\,\frac{f^+(0)f^-(0)}{\vA^2}\,
\kperp^{\mu_0^+ + \mu_0^- + 3}\delta(\kpar).
\label{eq:WT_Eres_scaling}
\eeq
One establishes that the prefactor is negative (i.e., $\const > 0$) by computing the 
wavenumber integral directly \citep{wang11}. 

What does this result tell us? Primarily, it tells us that the WT calculation that has
led to it is formally invalid and can, at best, be interpreted as a qualitative indication 
of what is going on. All the action has turned out to be concentrated in 
the $\kpar=0$ condensate, while for Alfv\'en waves with $\kpar\neq 0$, there is 
no residual energy. That we were going to end up with $\delta(\kpar)$ was, in fact, 
already obvious from the presence of the oscillatory term in \exref{eq:Rk}. 
Nevertheless, without a claim to mathematical rigour, one can, 
as I did in \apref{app:bband}, interpret the delta function in 
\exref{eq:WT_Eres_scaling} as having a width $\Delta\kpar \sim \tnl^{-1}/\vA$, 
where $\tnl\propto\kperp^{-1}$ is the cascade time for the condensate,
worked out in~\exref{eq:tnl_cond}. Taking $\mu_0^+ + \mu_0^- = -4$ and 
$\delta(\kpar)\sim \Delta\kpar^{-1} \propto \kperp^{-1}$ in \exref{eq:WT_Eres_scaling}
gets us 
\beq
\Ereskk(\kperp,\kpar) \propto - \kperp^{-2},
\label{eq:Eres_smeared}
\eeq
whereas the 1D spectrum can be calculated either by integrating out the delta 
function in \exref{eq:WT_Eres_scaling} or by integrating its broadened version 
in \exref{eq:Eres_smeared} over its width $\Delta\kpar~\propto~\kperp$: 
\beq
\Eres(\kperp) = \int\rmd\kpar \Ereskk(\kperp,\kpar)
= - \const\,\frac{f^+(0)f^-(0)}{\vA^2}\,\kperp^{-1}.
\label{eq:Eres_1D}
\eeq
This is the result of \citet{wang11}, who, however, go to slightly greater 
lengths in setting up a quasi-quantitative calculation in which they introduce by hand 
a nonlinear relaxation rate $\tnl^{-1}\propto\kperp$ into \exref{eq:Rk} and thus 
get their $\delta(\kpar)$ to acquire the Lorentzian 
shape \exref{eq:delta_ppar}. They attribute 
this relaxation to the $R_\vk$-dependent terms in \exref{eq:WT3_Rk}, which is 
qualitatively correct, but quantitatively just as invalid as is generally the 
application of the WT approximation (i.e., correlator splitting) to the 
strongly turbulent condensate. 

\begin{figure}
\centerline{\includegraphics[width=0.75\textwidth]{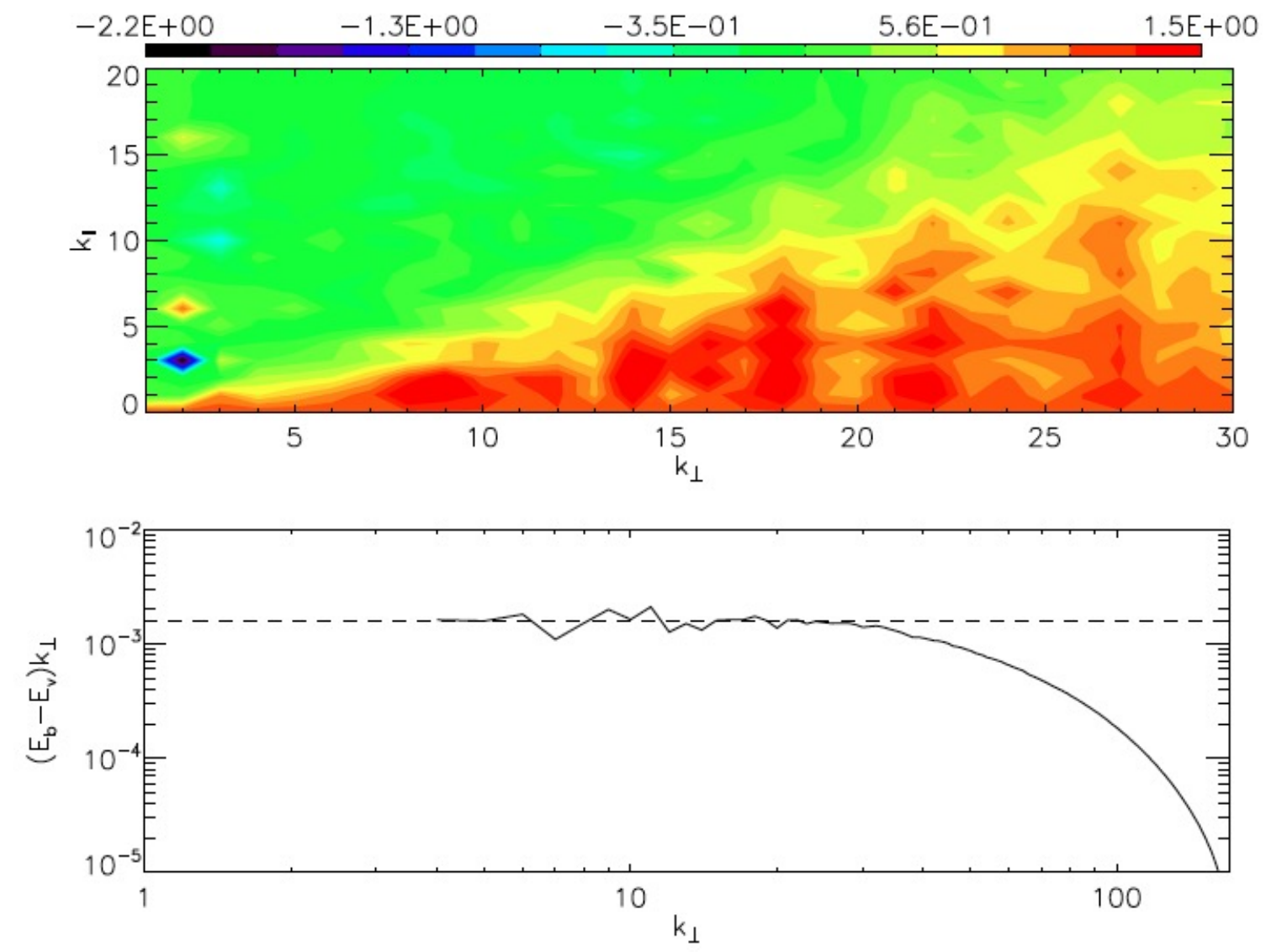}} 
\caption{The 2D spectrum, $\Ereskk(\kperp,\kpar)/\Ereskk(\kperp,0)$ (upper panel) 
and the 1D, $\kperp^{-1}$-compensated spectrum (lower panel) 
of residual energy from WT simulations by \citet{wang11} ($512^3$, 
broad-band forced at $\kpar = 1,\dots,16$ and $\kperp=1,2$;
\copyright AAS, reproduced with permission).} 
\label{fig:wang}
\end{figure}

Note that \exref{eq:Eres_1D} is, in fact, the same result as \exref{eq:E0_1D}---by comparing 
\exref{eq:E_powers} with~\exref{eq:E0_2D}, or just by dimensional analysis, it is 
easy to confirm that $f^\pm(0)\sim (\eps\Lpar\vA)^{1/2}$, 
so the dimensional prefactors match. 
Thus, all we have learned from the above calculation is that the 
condensate has residual energy and that the amount of the latter is 
comparable, at every scale, to the amount of energy in the condensate. 
One might argue that the added value of the WT calculation was in confirming 
that this residual energy was negative---although the negativity of the 
prefactor in \exref{eq:WT_Eres_scaling} is a quantitative result, not a 
qualitative one (one just has to calculate the appropriate integral 
and discover it to be negative, as \citealt{wang11} did), and so, in principle,
cannot be guaranteed to hold for the true, strongly turbulent condensate. 
I find the qualitative argument 
for the development of negative vorticity correlation $\la\omega^+\omega^-\ra<0$ 
explained in \secref{sec:new_res_theory} more compelling. 
The WT calculation above basically just confirms that growth of residual energy 
is a strong-turbulence effect. 

The qualitative considerations presented above are given some credence 
by the numerical simulations of WT reported by \citet{wang11}: their residual 
energy does indeed have a $\kperp^{-1}$ spectrum and concentrates in a wedge 
of wavenumber space $\kpar \lesssim \Delta\kpar\propto \kperp$, quite convincingly 
(\figref{fig:wang}).  

\subsection{Imbalanced WT}
\label{app:WT_imb}

As I acknowledged in \secref{sec:emb_WTimb}, I do not know how to construct 
a good theory of imbalanced WT. 
If imbalanced WT, like the balanced one, spawns a 2D condensate that is predominantly 
magnetic, that may be a helpful insight, as the presence of significant residual 
energy would impose geometric constraints (\apref{app:imb_geom}) on 
the ``$+$" and ``$-$" components of the condensate. \citet{boldyrev09weak} 
do find a magnetic condensate in an imbalanced simulation, but they only have 
results for order-unity imbalance. They also point out that if the cross-correlations 
\exref{eq:Rk_def} are retained in the derivation of the 
WT equation \exref{eq:WT_Galtier} for $N_\vk$, 
this makes the evolution equation \exref{eq:WT_Epm} for $\Ekk^\pm(\kperp,0)$
acquire terms under the integral containing 
$\Ekk^\pm(\kperp,0)\Ereskk(\pperp,0) + \Ekk^\pm(\pperp,0)\Ereskk(\kperp,0)$. 
Steady-state solutions then turn out to be possible only if 
\beq
\Ereskk(\kperp,0) \propto \kperp^{-2},\qquad
\Ekk^\pm(\kperp,0) \propto \kperp^{-2},
\eeq 
i.e., the degeneracy of the $\mu_0^+ + \mu_0^- = -4$ solution is lifted 
and all scalings are fixed. Perhaps this points us in the right direction, 
despite the fact that the WT equation for $\Ekk^\pm(\kperp,0)$, whose derivation 
requires correlator splitting etc., is not, in fact, quantitatively valid 
for the condensate. 

In their mildly imbalanced WT simulation, \citet{boldyrev09weak} 
find that $E^+(\kperp)$ and $E^-(\kperp)$ have, respectively, a steeper and 
a shallower slope than $\kperp^{-2}$, but the spectra appear to be pinned (equal) 
at the dissipation scale and thus get closer to each other with increased resolution.
Thus, if one wants a theory that describes finite-resolution simulations, 
some scheme like the one I proposed in \secref{sec:imb_new} would need to be 
invented for the WT regime, generalising \apref{app:bband} to the imbalanced case.

\section{Alignment, Imbalance, and Reduction of Nonlinearity}
\label{app:align}

These topics have cropped up repeatedly (e.g., in \secsand{sec:align_aniso}{sec:Eimb}).
This appendix is an attempt to treat them carefully.

\subsection{Geometry and Types of Alignment} 
\label{app:imb_geom}

\begin{figure}
\centerline{\includegraphics[width=0.75\textwidth]{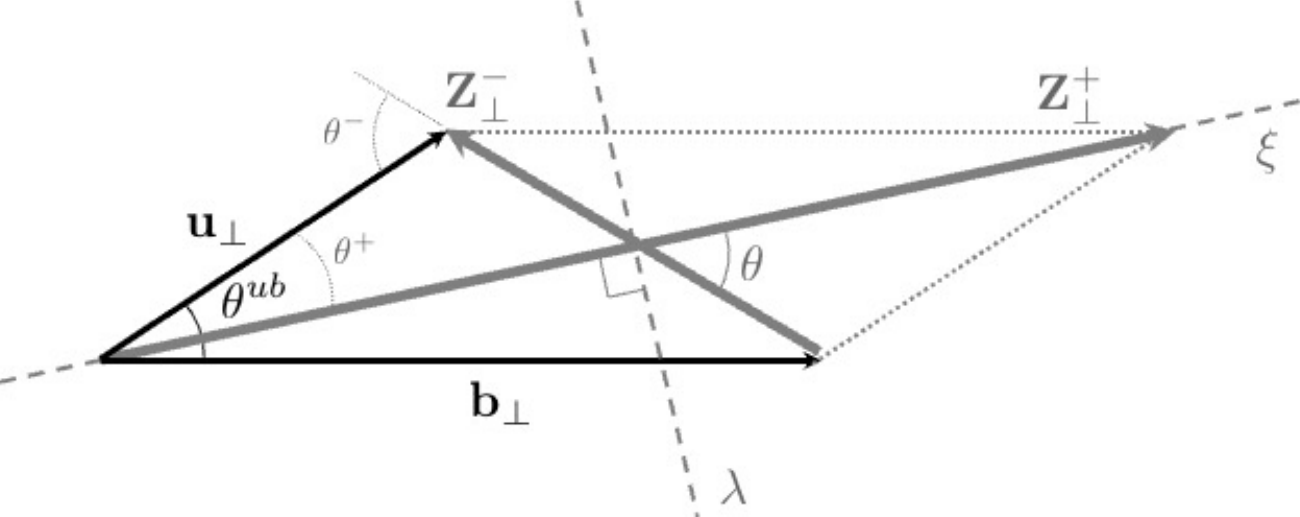}} 
\caption{Geometry of velocity, magnetic and Elsasser fields ($\vB_0$ is perpendicular 
to the page). All four fields are aligned: 
the angles $\theta$, $\theta^{ub}$, $\theta^\pm$ are all small (although they do not have to be). 
Also shown are the axes along which the $\lambda$ and $\xi$ scales in \exref{eq:tnl_align} 
are meant to be calculated (along and across $\vzperp^+$, respectively). 
The angle between these axes is $\phi = \pi/2 - \theta$ and so $\cos\phi = \sin\theta$.} 
\label{fig:geometry}
\end{figure}

Let us consider the formal relationship between imbalance, residual energy, 
and (the two kinds of) alignment. The first salient fact is purely 
geometric (\figref{fig:geometry}):
the two alignment angles (defined for a particular pair of field increments)
\beq
\label{eq:theta_defs}
\sin\theta = \frac{|\dvz^+_\vlam\times\dvz^-_\vlam|}{|\dvz^+_\vlam| |\dvz^-_\vlam|},
\quad 
\sin\theta^{ub} = \frac{|\dvu_\vlam\times\dvb_\vlam|}{|\dvu_\vlam| |\dvb_\vlam|},
\eeq 
and the Elsasser and Alfv\'en ratios 
\beq
\RE = \frac{|\dvz^+_\vlam|^2}{|\dvz^-_\vlam|^2},
\quad
\RA = \frac{|\dvu_\vlam|^2}{|\dvb_\vlam|^2}
\eeq
are related by the following equations      
\beq
\label{eq:sin_th}
\sin^2\theta = \frac{\sin^2\theta^{ub}}{\sin^2\theta^{ub} + (1-\RA)^2/4\RA},
\quad
\sin^2\theta^{ub} = \frac{\sin^2\theta}{\sin^2\theta + (1-\RE)^2/4\RE},
\eeq
so only two of these four quantities are independent. 
Equivalently, in terms of the normalised local cross-helicity and residual energy,
defined by 
\beq
\label{eq:sig_def}
\sigc = \frac{|\dvz^+_\vlam|^2 - |\dvz^-_\vlam|^2}{|\dvz^+_\vlam|^2 + |\dvz^-_\vlam|^2} 
= \frac{\RE - 1}{\RE + 1}
\rmand
\sigr = \frac{|\dvu_\vlam|^2 - |\dvb_\vlam|^2}{|\dvu_\vlam|^2 + |\dvb_\vlam|^2}
= \frac{\RA - 1}{\RA + 1},
\eeq
respectively, the alignment angles are \citep{wicks13align}\footnote{Note the
connection implied by the first of these formulae 
between non-zero residual energy and Elsasser alignment (cf.~\secref{sec:new_res_theory}).} 
\beq
\cos\theta = \frac{\sigr}{\sqrt{1-\sigc^2}},
\quad
\cos\theta^{ub} = \frac{\sigc}{\sqrt{1-\sigr^2}}.
\label{eq:theta_sig}
\eeq
This means that alignment between the velocity and magnetic field 
is not formally the same thing as alignment between the Elsasser fields,
and it is a nontrivial decision which of these one believes to matter
for the determination of $\tnl^\pm$. 

\begin{figure}
\centerline{\includegraphics[width=0.85\textwidth]{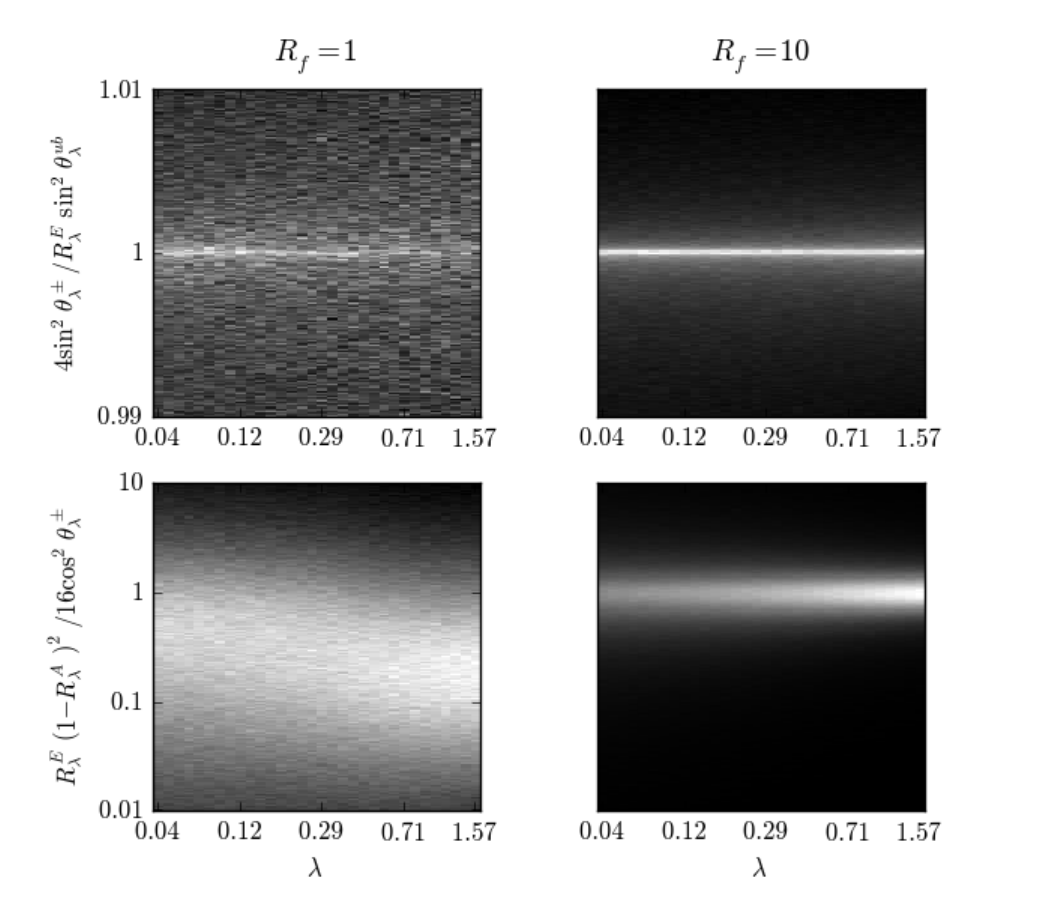}} 
\caption{A direct test of the relations \exref{eq:imb_limit}
in globally balanced ($\eps^+/\eps^-=1$, left column)
and imbalanced ($\eps^+/\eps^-=10$, right column) RMHD simulations
by \citet{mallet11} (these are the same unpublished simulations as tabulated
in \figref{fig:mallet_imb}). The plots show histograms of
$4\sin^2\theta/\RE\sin^2\theta^{ub}$ (upper row)
and $(1-\RA)^2\RE/16\cos^2\theta$ (lower row) vs.~perpendicular point separation~$\lambda$. 
Note that the first relation in \exref{eq:imb_limit} is reasonably well
satisfied even in the globally balanced simulation.}  
\label{fig:mallet_geom}
\end{figure}

Before taking a side on this question in \apref{app:align_nlin}, let me consider
a strongly imbalanced situation, where 
$\RE\gg1$, i.e., the cross-helicity is large ($\sigc\approx1$)---in view
of the discussion in~\secref{sec:Eimb},
this is probably an asymptotic case best related to the generic situation,
at least locally. In this limit, \exref{eq:sin_th} gives us
\beq
\sin^2\theta^{ub} \approx \frac{4\sin^2\theta}{\RE}\ll1,
\quad
(1-\RA)^2 \approx \frac{16\cos^2\theta}{\RE}\ll1. 
\label{eq:imb_limit}
\eeq
\citet{mallet11} found, unsurprisingly, that these relations were extremely
well satisfied in their RMHD simulations
(the first relation even in balanced ones---see \figref{fig:mallet_geom}). 
Thus, (local) imbalance implies that $\vuperp$ and $\vbperp$ are both closely aligned and 
have nearly the same amplitude (this is geometrically obvious from \figref{fig:geometry}), 
but whether or not the Elsasser fields are aligned is up to the turbulence to decide. 
It does seem to decide to align them [see \secref{sec:imb_num}, item (v)], 
hence the way in which I drew the field increments in \figref{fig:geometry}. 

If $\RE$ (equivalently, $\sigc$) is independent of scale in the inertial range, 
as reported for the solar wind by \citet{podesta10xhel} and \citet{chen20}, 
then the first relation in \exref{eq:imb_limit} implies that $\theta^{ub}$ 
and $\theta$ should have the same scaling.
In numerical simulations, they appear to do so, approximately, in balanced 
turbulence \citep{mallet16}, which, of course, is patch-wise imbalanced~(\secref{sec:Eimb}),  
but not in the strongly imbalanced cases studied by \citet{beresnyak09} 
and \citet{mallet11} (see~\secref{sec:imb_num}). 
Accordingly, \citet{mallet11} found the dependence of $\theta^{ub}$ on $\lambda$ getting 
shallower with increased imbalance, as $\RE$ vs.\ $\lambda$ got steeper
(see \figref{fig:mallet_imb}) and $\theta$ stayed approximately the same.
Alas, those simulations are in all probability not in the 
asymptotic regime, so a follow-up study at higher resolutions would be very welcome. 

\subsection{Alignment and Reduction of Nonlinearity} 
\label{app:align_nlin}

The nonlinear term in RMHD~\exref{eq:zpm} can be expressed so:
\begin{align}
\nonumber
\vzperp^+\cdot\vdperp\vzperp^- &= \vzperp^-\cdot\vdperp\vzperp^+
+ \vdperp\times\lt(\vzperp^-\times\vzperp^+\rt)\\
&= \vdperp\cdot\lt(\vuperp\vuperp - \vbperp\vbperp\rt)
+ \vdperp\times\lt(\vuperp\times\vbperp\rt).
\label{eq:nlin_id_ub}
\end{align}
The first of these identities already appeared as~\exref{eq:nlin_id}.
As I already explained in \secref{sec:align_aniso}, it says, schematically, 
\beq
\frac{1}{\xi^-} \sim \frac{1}{\xi^+} + \frac{\sin\theta}{\lambda},
\label{eq:xi_theta}
\eeq
which makes the connection between Elsasser-field alignment ($\sin\theta\ll1$)
and reduction of nonlinearity ($\xi^\pm\gg \lambda$) a consistent choice
(although not a mathematically inevitable one). This conclusion
is valid regardless of the degree of imbalance.

Consider now the second identity in~\exref{eq:nlin_id_ub}.
Without imbalance, i.e., if $\dz_\lambda^+\sim\dz_\lambda^-\sim\du_\lambda\sim\db_\lambda$,
it implies, analogously to~\exref{eq:xi_theta},  
\beq
\frac{1}{\xi^-} \sim \frac{1}{\xi^u} + \frac{1}{\xi^b} + \frac{\sin\theta^{ub}}{\lambda},
\eeq
where $\xi^u$ and $\xi^b$ are the characteristic scales of variation
of $\vuperp$ and $\vbperp$, respectively, along themselves.
Again it is a consistent choice to posit that
$\xi^u\sim\xi^b\sim\xi^\pm$ and $\lambda/\xi\sim\sin\theta^{ub}\ll 1$.
In contrast, if (locally or globally)
$\sqrt{\RE}=\dz_\lambda^+/\dz_\lambda^- \gg1$, we have
\beq
\frac{\lt(\dz_\lambda^+\rt)^2}{\xi^-\sqrt{\RE}} \sim
\frac{\du_\lambda^2(1-\RA)}{\xi^{u,b}} + 
\frac{\du_\lambda\db_\lambda\sin\theta^{ub}}{\lambda}
\sim \lt(\frac{\cos\theta}{\xi^{u,b}} + \frac{\sin\theta}{\lambda}\rt)
\frac{\lt(\dz_\lambda^+\rt)^2}{\sqrt{\RE}},
\eeq
where at the last step, I used both of the geometric relations~\exref{eq:imb_limit}
and $\du_\lambda\sim\db_\lambda\sim\dz_\lambda^+$. Again, the connection
between reduced nonlinearity and alignment is consistent, but it is
the alignment of Elsasser fields that matters (if $\sin\theta\sim 1$,
$\xi^-\sim\lambda$, so no reduction). The alignment between
$\vuperp$ and $\vbperp$ is geometrically inevitable, but does not
by itself imply anything about a reduction of nonlinearity. 

This is all perhaps too obvious to have needed spelling out. To summarise,
in balanced turbulence (or rather in balanced regions within turbulence), which 
type of alignment one prefers to think about, or measure, does not appear to
make a qualitative difference, whereas in the presence of imbalance, Elsasser
alignment is a natural choice. Since both globally balanced and imbalanced 
turbulence is likely to be imbalanced locally (see \secref{sec:Eimb}),
and indeed since it is in locally imbalanced patches that most of the ``action''
is likely to be, I have made Elsasser alignment
synonymous with ``alignment'' everywhere in this review.

As I stated repeatedly in the main text, the supporting physical (dynamical) argument
is to think of alignment as a result of mutual shearing of Elsasser fields, 
following \citet{chandran15}. Such a shearing will
lead to simultaneous reduction of $\lambda/\xi$ and $\sin\theta$. 
This approach is circumstantially supported by the ``refined critical balance'' 
discovered by \citet{mallet15}---the remarkable self-similarity shown by 
the ratio $\tA/\tnl^\pm$, with $\tnl^\pm$ defined by \exref{eq:tnl_align}, 
using the angle between the Elsasser fields (see \figref{fig:rcb}). 
Arguably, this says that it is this 
$\tnl^\pm$ that $\tA$ (and, therefore, $\lpar^\pm$) ``knows'' about, so it is this 
$\tnl^\pm$ that should be viewed as the decorrelation (cascade) time of
the turbulent structures. 
The same paper spotted a pronounced anticorrelation, at a given~$\lambda$, 
between the Elsasser alignment angle~$\theta$ and the magnitudes of the
Elsasser-field increments---as noted in \secref{sec:plot}, this is
again consistent with the shearing origin of alignment.

Another useful way to look at the relationship between alignment
and nonlinearity is afforded by casting RMHD in the form~\exref{eq:zeta}, in terms
of Elsasser potentials and vorticities. Alignment between $\vzperp^+$
and $\vzperp^-$ is equivalent to alignment between $\vdperp\zeta^+$ and
$\vdperp\zeta^-$, so small $\sin\theta$ will imply smallness of the
second term on the right-hand side of~\exref{eq:zeta}. The first term,
$\lt\{\zeta^\mp,\omega^\pm\rt\}$, is small if $\vdperp\zeta^\mp$ and
$\vdperp\omega^\pm$ are approximately aligned, i.e., if contours of constant
$\zeta^\mp$ and $\omega^\pm$ approximately coincide. The latter condition
is indeed satisfied for perturbations that resemble tubes or sheets---and thus
for perturbations that are likely to emerge from mutual shearing of Elsasser
fields (this observation is due to \citealt{bowen22}). 

Finally, let me reiterate that Elsasser alignment is not formally obliged to be
co-located with locally imbalanced regions, even if observations
by \citet{wicks13} discussed in~\secref{sec:Eimb} suggest that it might be,
and even though the reduction of nonlinearity in the aligned MHD cascade
has originally been argued to be connected to local enhancements of cross-helicity
(see footnote~\ref{fn:uBcorr2} and references therein). 
In numerical simulations, it remains to be checked whether such an intrinsic
connection does exist.

\section{2D Spectra of RMHD Turbulence}
\label{app:2Dspectra}

As we trade in $\kperp$ (or $\lambda$) and $\kpar$ (or $\lpar$) scalings, it is only 
natural that we might wish to have 2D spectra of RMHD turbulence, $\Ekk(\kperp,\kpar)$. 
It is quite easy to work them out, given the information we already have about the $\lambda$ 
and $\lpar$ scalings of the Elsasser increments. 

Since, as I explained in \secref{sec:aniso}, the physically meaningful parallel correlations 
are along the local mean field, we should think of our Elsasser fields $\vzperp^\pm$ as being 
mapped on a grid of values of $(\vrperp,\rpar)$, where $\rpar$ is the distance measured 
along the exact field line (what matters here is not that the parallel distances are 
slightly longer  
than their projection on the $z$ axis---the difference is small in the RMHD ordering---but 
that we probe correlations along the exact field line rather than slipping off it; 
see \figref{fig:loc_aniso}). The Fourier transform 
of $\vzperp^\pm(\vrperp,\rpar)$ is a function of $\kperp$ and $\kpar$, $\vzperp^\pm(\vkperp,\kpar)$, 
and the 2D spectrum is defined to~be 
\beq
\Ekk(\kperp,\kpar) = 2\pi\kperp \la|\vzperp^\pm(\vkperp,\kpar)|^2\ra.
\label{eq:Ekk_def}
\eeq

Let us start with the premise that $\Ekk(\kperp,\kpar)$ will be a product of 
power laws in both of its arguments and that the scaling exponents of these power laws 
will be different depending on where we are in the $(\kperp,\kpar)$ 
space vis-\`a-vis the line of critical balance, which is also 
a power-law relation, between $\kperp$ and $\kpar$: 
\beq
\tnl \sim \tA \quad\Leftrightarrow\quad 
\kpar \sim \kperp^\sigma.
\eeq 
We shall treat the wavenumbers as dimensionless, $\kpar\Lpar\to\kpar$, $\kperp\lCB\to\kperp$. 
According to \exref{eq:MS_normalised}, 
\beq
\sigma = \frac{1}{2}. 
\eeq
Thus, we shall look for the 2D spectrum in the form   
\beq
\label{eq:Ekk_gen}
\Ekk(\kperp,\kpar) \sim \lt\{
\begin{array}{ll}
\kpar^{-\alpha}\kperp^\beta, & \kpar \gtrsim \kperp^\sigma,\\\\
\kpar^{\delta}\kperp^{-\gamma}, & \kpar \lesssim \kperp^\sigma.
\end{array}
\rt.
\eeq
The four exponents $\alpha$, $\beta$, $\gamma$, and $\delta$
can be determined by the following arguments, broadly analogous to those proposed by 
\citet{sch16} for drift-kinetic turbulence except for the calculation of~$\beta$
(and hence of~$\alpha$), 
which will be significantly modified here in light of some new theoretical 
developments (\apref{app:det_beta}). 

\subsection{Determining $\delta$: Long Parallel Wavelengths}
\label{app:det_delta}

At long parallel wavelengths, $\kpar\ll\kperp^\sigma$, the $\kpar$ spectrum measures  
correlation between points along the field line that are separated by longer distances 
than an Alfv\'en wave can travel in one nonlinear time ($\tA\gg\tnl$) and, consequently, 
are causally disconnected (\secref{sec:CBCB}). Therefore, their parallel correlation 
function is that of a 1D white noise and the corresponding spectrum is flat:
\beq
\delta = 0. 
\eeq

It may be worth belabouring this point: the flat $\kpar$ spectrum at 
$\kpar\lesssim\kperp^\sigma$ (\figref{fig:2Dspectra}b) is the Fourier-space signature 
of CB turbulence, not an indication of the presence of quasi-2D motions or of 
failure of local-in-scale interactions (as, e.g., \citealt{meyrand16} appear to imply). 
This highlights the fact that the wavenumbers where energy is present 
are not quite the same thing as the correlation scales of the turbulent field,
and so one should not expect that CB requires a spectrum peaked 
at $\kpar\sim\kperp^\sigma$ (a fallacy that has made it into a number 
of published texts, rigorous peer review notwithstanding). 
The same argument applies to frequency spectra, should one want to plot them: 
there must be a flat spectrum at $\omega\lesssim\tnl^{-1}$ because instances 
separated by times longer than $\tnl$ are uncorrelated and will, therefore, 
have white-noise statistics. 

\subsection{Determining $\gamma$: Short Perpendicular Wavelengths}
\label{app:det_gamma}

Let us calculate the 1D $\kperp$ spectrum: if we assume (and promise to check later) that 
$\alpha>1$, then the $\kpar$ integral over $\Ekk(\kperp,\kpar)$ is dominated by the 
region $\kpar\lesssim\kperp^\sigma$ and the 1D spectrum is mostly determined by the 
CB scales $\kpar\sim\kperp^\sigma$ (as is indeed argued in the GS95 theory and its descendants 
reviewed in the main text): 
\beq
E(\kperp) \sim \int_0^{\kperp^\sigma}\rmd\kpar \Ekk(\kperp,\kpar) \sim 
\kperp^{-\gamma+\sigma}. 
\label{eq:Eperp1D}
\eeq
Then the amplitude of an Elsasser field at scale $\lambda=\kperp^{-1}$~is
\beq
\dz_\lambda^2 \sim \int_{\kperp}^\infty\rmd\kperp' E(\kperp') 
\sim \kperp E(\kperp) \sim \kperp^{-\gamma+\sigma+1},
\label{eq:dz_vs_kperp}
\eeq 
assuming $\gamma-\sigma > 1$. On the other hand, 
the usual Kolmogorov constant-flux condition coupled with the CB conjecture gives~us
\beq
\frac{\dz_\lambda^2}{\tnl} \sim \const,\quad
\tnl^{-1} \sim\tA^{-1} \propto \kpar \sim \kperp^{\sigma}
\hence
\dz_\lambda^2 \sim \kperp^{-\sigma}.
\eeq
Comparing this with \exref{eq:dz_vs_kperp}, we get  
\beq
\gamma = 2\sigma + 1 = 2. 
\eeq 
The 1D spectral exponent in \exref{eq:Eperp1D} 
is then $-\gamma+\sigma=-3/2$, as it should be [see \exref{eq:MS_normalised}]. 

\subsection{Determining $\beta$: Long Perpendicular Wavelengths}
\label{app:det_beta}

This turns out to be a somewhat subtle issue, connected to an interesting recent 
development in turbulence theory.    

\subsubsection{Kinematics}
\label{app:beta_kin}

Consider first the following na\"ive, purely kinematic calculation. 
Let us write the desired spectrum \exref{eq:Ekk_def}~as
\begin{align}
\nonumber
\la|\vzperp^\pm(\vkperp,\kpar)|^2\ra 
&= \int\frac{\rmd^2\vrperp}{(2\pi)^2}\,e^{-i\vkperp\cdot\vrperp} 
\la\vzperp^\pm(\vrperp,\kpar)\cdot\vzperp^{\pm *}(0,\kpar)\ra\\
&= \frac{1}{2\pi}\int_0^\infty\rmd\rperp\rperp J_0(\kperp\rperp) C^\pm(\rperp,\kpar), 
\label{eq:int_with_J0}
\end{align}
where $C^\pm(\rperp,\kpar)$ is the two-point correlation 
function of $\vzperp^\pm(\vrperp,\kpar)$. It is only a function of 
the point separation $\rperp$ 
because of statistical homogeneity and isotropy in the perpendicular plane. 
For any given~$\kpar$, it might seem reasonable, by the CB conjecture, 
to estimate the correlation length of the field to be $\lambda \sim \kpar^{-1/\sigma}$
and assume therefore that 
the integral in \exref{eq:int_with_J0} is effectively restricted 
by $C^\pm(\rperp,\kpar)$ to $\rperp\lesssim\lambda$. If we now let 
$\kperp\lambda\ll1$ (equivalently, $\kperp^\sigma\ll\kpar$), 
then the Bessel function can be expanded in small 
argument: $J_0(\kperp\rperp) = 1 - \kperp^2\rperp^2/4 + \dots$. 
The spectrum \exref{eq:Ekk_def} is then
\begin{align}
\label{eq:Ekk_exp}
&\Ekk(\kperp,\kpar) = \frac{\kperp}{2\pi}\lt(C_0 + C_2 \kperp^2 + \dots\rt),\\
&C_0 = 2\pi\int_0^\infty\rmd\rperp\rperp C^\pm(\rperp,\kpar),\quad 
C_2 = - \frac{\pi}{2}\int_0^\infty\rmd\rperp\rperp^3 C^\pm(\rperp,\kpar).
\end{align}
The first of these coefficients, 
$C_0 = \int\rmd^2\vrperp\la\vzperp^\pm(\vrperp,\kpar)\cdot\vzperp^{\pm*}(0,\kpar)\ra$,
vanishes if $\int\rmd^2\vrperp\vzperp^\pm(\vrperp,\kpar)=0$, which 
seems to be a reasonable assumption for a solenoidal field [see \exref{eq:zeta_def}] 
in a box.
This leaves us with the series \exref{eq:Ekk_exp} for $\Ekk$ starting at the 
second term and so $\Ekk\propto \kperp^3$ to lowest order. Thus, 
\beq
\beta = 3?
\label{eq:beta_kin}
\eeq

I have added a question mark to this statement because it is, in fact, not as 
straightforwardly obvious as the kinematic argument suggests.  

\subsubsection{Thermodynamics}
\label{app:beta_thermo}

What if one applies to the long perpendicular scales the same logic as I did to 
long parallel scales in \apref{app:det_delta}? If the fields are decorrelated 
at scales longer than the CB scale, i.e., at $\kperp\ll \kpar^{1/\sigma}$, assuming 
a 2D white-noise spectrum (at fixed $\kpar$) would imply that 
\beq
\Ekk(\kperp,\kpar) = f(\kpar) \kperp \hence \beta = 1. 
\label{eq:beta_thermo}
\eeq
This is just a thermal equilibrium spectrum, with energy equipartitioned 
amongst all available $\vkperp$'s. The tendency for a thermal spectrum to emerge at large 
scales has indeed been recently noticed in 3D forced hydrodynamic turbulence 
(where the corresponding spectrum is $\propto k^2$; see \citealt{alexakis19} 
and references therein).

\subsubsection{Dynamics}
\label{app:beta_dyn}

How then does one reconcile the thermodynamical result~\exref{eq:beta_thermo}
with the kinematic one~\exref{eq:beta_kin}? Let me give here the RMHD analog of 
the argument proposed by \citet{hosking22hydro} for hydrodynamic turbulence. 
The effect of the energy-containing scales (in our case, of the CB scales 
$\kpar\sim\kperp^\sigma$) on the longer perpendicular scales is two-fold: 
modes with larger $\vkperp$'s 
couple (``beat'') to feed those with smaller $\vkperp$'s, thus creating long-scale 
perturbations; and those long-scale perturbations are mixed to even longer scales 
by the turbulent diffusion arising from the energy-containing-scale fields. 
It turns out that the balance between these two effects produces the thermal 
spectrum~\exref{eq:beta_thermo}, while pushing the kinematic asymptotic~\exref{eq:beta_kin}   
to ever smaller $\kperp$'s as time goes on.

Thanks to \apref{app:WT}, I already have the analytical tools to demonstrate this 
(semi)quantitatively. Since we are dealing with modes for which 
$\tnl^{-1}\ll\kpar\vA$, it is not unreasonable to apply the WT approximation 
to their description (with the usual disclaimer about its formal breakdown 
wherever it involves CB perturbations). 
Let us therefore consider \exref{eq:WT_Epm}, but now at wavenumbers 
$\kperp$ that are much smaller than the wavenumbers at 
which $\Ekk^\pm(\qperp,\kpar)$ or $\Ekk^\pm(\qperp,0)$ contain most of their energy. 
Then, in \exref{eq:WT_Epm}, the integral over $\qperp$ can be assumed to be 
dominated by $\qperp\gg \kperp$, and the integral over $\pperp=\qperp + \kappa$ 
is, therefore, 
\begin{align}
\nonumber
\int_{|\kperp-\qperp|}^{\kperp+\qperp}\rmd\pperp
\frac{\kperp^2\qperp^2}{\pperp^2}\sin\phi\cos^2\phi\,\Ekk^\mp(\pperp,0)
&\approx 
\Ekk^\mp(\qperp,0)
\int_{-\kperp}^{+\kperp}\rmd\kappa\,\kappa^2\sqrt{1-\frac{\kappa^2}{\kperp^2}}\\
& = \frac{\pi}{8}\kperp^3 \Ekk^\mp(\qperp,0). 
\end{align}
This turns the WT equation~\exref{eq:WT_Epm} into 
\beq
\frac{\dd\Ekk^\pm}{\dd t} + D^\pm\kperp^2\Ekk^\pm = \kperp^3 F^\pm(\kpar).
\label{eq:Epm_thermo}
\eeq
The left-hand side of \exref{eq:Epm_thermo} features the usual turbulent 
diffusion with diffusivity
\beq
D^\pm = \frac{\pi}{8\vA}\int_0^\infty\rmd\qperp\Ekk^\mp(\qperp,0),
\eeq
due entirely to the $\qpar=0$ modes---but these, in the context, are just 
the CB modes with $\qpar\sim\qperp^\sigma$ (cf.~\apref{app:bband}).  
The right-hand side of \exref{eq:Epm_thermo} contains the effect 
of energy-containing-scale modes coupling to feed the long-scale 
spectrum: 
\beq
F^\pm(\kpar) = \frac{\pi}{8\vA}\int_0^\infty\frac{\rmd\qperp}{\qperp}
\Ekk^\mp(\qperp,0)\Ekk^\pm(\qperp,\kpar). 
\eeq 
The fact that this enters in~\exref{eq:Epm_thermo} with a prefactor of $\kperp^3$ 
is a reflection of the fields being solenoidal and thus requiring the 
kinematic asymptotic \exref{eq:beta_kin} at $\kperp\to 0$. 

The solution of \exref{eq:Epm_thermo} (starting from zero initial condition)~is 
\beq
\Ekk^\pm(\kperp,\kpar) = \kperp\lt(1 - e^{-D^\pm\kperp^2 t}\rt)
\frac{F^\pm(\kpar)}{D^\pm}
\to\lt\{\begin{array}{ll}
\kperp^3 tF^\pm(\kpar), & \kperp \ll (D^\pm t)^{-1/2},\\\\
\kperp F^\pm(\kpar)/D^\pm, & \kperp \gg (D^\pm t)^{-1/2},\\
\end{array}\rt.
\eeq
the latter asymptotic being the steady-state solution. 
Thus, the kinematic asymptotic \exref{eq:beta_kin} does exist, 
but is constantly pushed to larger scales as time goes on. 
Formally this means that the correlation scale $\lambda$ above which 
the correlation function $C^\pm(\rperp,\rpar)$ in~\exref{eq:int_with_J0} 
decays sufficiently fast with $\rperp$ 
to enable the expansion~\exref{eq:Ekk_exp}, is not the CB scale 
$\lambda \sim \kpar^{-1/\sigma}$ as na\"ively assumed in \apref{app:beta_kin}, 
but rather $\lambda\sim (D^\pm t)^{1/2}$. Since I am interested in asymptotically 
long times, I will adopt the thermal spectrum~\exref{eq:beta_thermo} 
in what follows. 

\subsection{Determining $\alpha$: Short Parallel Wavelengths}
\label{app:det_alpha}

Finally, $\alpha$ is determined simply by the requirement that the 2D spectra 
match along the CB line: substituting $\kpar\sim\kperp^\sigma$ into \exref{eq:Ekk_gen} 
and equating powers of $\kperp$, we~get 
\beq
\alpha = \frac{\beta+\gamma}{\sigma} - \delta = 6. 
\eeq
This somewhat ridiculous exponent\footnote{Amazingly, this scaling has just been
confirmed by \citet[][this is mentioned in passing in their \S A.2]{squire22}, applying a new
field-line-following method to the 2D spectra in the MHD inertial range of their kinetic turbulence
simulation.} suggests that there is very little energy indeed in wave-like perturbations with $\tA\ll\tnl$. 

\begin{figure}
\begin{center}
\includegraphics[width=0.6\textwidth]{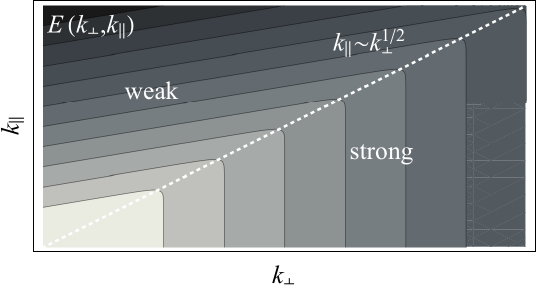}\\
(a)\\
\vskip5mm
\begin{tabular}{cc}
\includegraphics[width=0.45\textwidth]{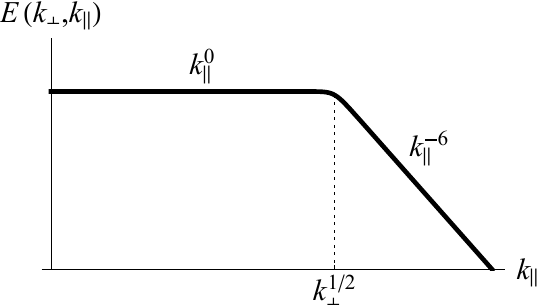} & 
\includegraphics[width=0.45\textwidth]{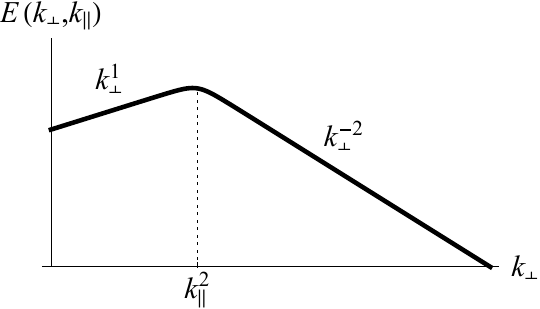}\\
(b) & (c)
\end{tabular}
\end{center}
\caption{Sketch of the 2D spectra \exref{eq:Ekk} of RMHD turbulence: 
(a) in the 2D wave-number plane; 
(b) at constant $\kperp$; 
(c) at constant $\kpar$. Note that $\kpar$ here is 
measured along the perturbed field, not the $z$ axis (see discussion in \secref{sec:aniso}).}
\label{fig:2Dspectra}
\end{figure}

Note that the consistency of what I have done above can be checked by calculating the 
1D $\kpar$ spectrum: 
\beq
E(\kpar) = \int\rmd\kperp\Ekk(\kperp,\kpar) 
\sim \int_0^{\kpar^{1/\sigma}}\rmd\kperp \kpar^{-\alpha}\kperp^\beta 
+ \int_{\kpar^{1/\sigma}}^\infty\rmd\kperp \kpar^\delta\kperp^{-\gamma} 
\sim \kpar^{-\zeta},
\eeq
where 
\beq
\zeta = \alpha - \frac{\beta+1}{\sigma} = \frac{\gamma-1}{\sigma}-\delta = 2, 
\label{eq:det_zeta}
\eeq
as it should be (see \secref{sec:par_cascade}).\\

To summarise, the 2D spectrum \exref{eq:Ekk_gen} of critically balanced Alfv\'enic turbulence~is
\beq
\label{eq:Ekk}
\Ekk(\kperp,\kpar) \sim \lt\{
\begin{array}{ll}
\kpar^{-6}\kperp^1, & \kpar \gtrsim \kperp^{1/2},\\\\
\kpar^0\kperp^{-2}, & \kpar \lesssim \kperp^{1/2},
\end{array}
\rt.
\eeq
leading to 1D spectra $E(\kperp)\sim \kperp^{-3/2}$ and $E(\kpar)\sim\kpar^{-2}$. 
The spectra \exref{eq:Ekk} are sketched in \figref{fig:2Dspectra}. 

I leave it as an exercise for the reader to show that if the same scheme is applied 
to the tearing-mediated turbulence described in \secref{sec:recturb} [starting with 
\exref{eq:lpar_tearing} for~$\sigma$], the exponents in \exref{eq:Ekk_gen}~are 
\beq
\sigma = \frac{6}{5}, \quad
\delta = 0,\quad
\gamma = \frac{17}{5},\quad
\beta = 1,\quad
\alpha = \frac{11}{3},\quad 
\eeq
and $\zeta=2$, unchanged from \exref{eq:det_zeta} (as it should be, 
according to~\secref{sec:par_rec}).

\subsection{2D Spectrum of WT}
\label{app:WT_2D}

The 2D spectrum of broad-band-forced WT determined in \apref{app:bband} 
can easily be obtained by arguments analogous to the above: 
\beq
\delta = 0 
\eeq
for the same reason as in \apref{app:det_delta}, 
\beq
\gamma = \sigma + 1 
\eeq
similarly to \apref{app:det_gamma}, but now 
employing the non-constant-flux argument \exref{eq:E0_1D}, 
\beq
\beta = -4 + \gamma = \sigma - 3
\eeq
by the WT condition \exref{eq:mumu4} with 
$\mu_0 = -\gamma$ and $\mu = \beta$ instead of the 
calculation of \apref{app:det_beta} (which would only apply above 
the perpendicular scale of the forcing), 
and, finally, matching the spectra as in \apref{app:det_alpha}, we~get 
\beq
\beta + \gamma = (\alpha + \delta)\sigma \hence  
2(\sigma - 1) = \alpha\sigma.
\label{eq:WT_matching}
\eeq 
Here we can either set $\sigma = 1$ by assuming a nonaligned 
cascade of the condensate, as in \exref{eq:tnl_cond}, and hence get $\alpha = 0$, 
or set $\alpha = 0$ by assuming no parallel cascade and 
equal forcing at all $\kpar$ in the WT regime, in which case 
the matching condition \exref{eq:WT_matching} requires $\sigma=1$ 
(and so I would have had to contend with discontinuous spectra 
if, in \apref{app:bband}, I had allowed the condensate to have alignment).
Thus, 
\beq
\sigma = 1,\qquad \delta = 0,\qquad \gamma = 2,\qquad \beta = -2,\qquad \alpha = 0.
\eeq
Note that the 2D spectrum in the strongly nonlinear region $\kpar \lesssim \kperp^\sigma$ 
does not actually change at the transition from the WT to the CB turbulence (although 
the CB boundary does)---spectral continuity vindicated.

\section{A Reconnection Primer}
\label{app:reconnection}

Since it is now clear that reconnection phenomena play an essential role in MHD turbulence, 
it is useful to provide a series of shortcuts to the key results. 
I will not do any precise calculations of the kind that make the theory 
of resistive MHD instabilities such a mathematically accomplished subject
(what better example on which to teach an undergraduate class to solve ODEs with 
boundary layers than the many incarnations of the tearing mode!), 
but will instead go for ``quick and dirty'' ways of getting 
at the right scalings. Readers yearning for more exactitude will find it, e.g., 
in a recent treatment by \citet{boldyrev18}; those who prefer to get their 
instruction from the original source should start with the foundational papers 
by \citet{furth63} and \citet{coppi76}. 
     
When dealing with resistive MHD instabilities, it is convenient to write the RMHD equations 
in their original form \citep{strauss76}, in terms of the stream (flux) functions for 
the velocity and magnetic fields: 
\beq
\vuperp = \ez\times\vdperp\Phi,\quad
\vbperp = \ez\times\vdperp\Psi.
\label{eq:PhiPsi_def}
\eeq
Since the Elsasser potentials introduced in~\exref{eq:zeta_def} are just 
$\zeta^\pm = \Phi \pm \Psi$, one can recover the equations for $\Phi$ and~$\Psi$ 
from \exref{eq:zeta}, 
or, indeed, use \exref{eq:PhiPsi_def} and derive them directly from the momentum and 
induction equations of MHD (see \citealt{sch09}, \citealt{oughton17}, and references therein): 
\begin{align}
\label{eq:Phi}
\frac{\dd}{\dd t}\dperp^2\Phi + \lt\{\Phi,\dperp^2\Phi\rt\} 
&= \vA\dpar\dperp^2\Psi + \lt\{\Psi,\dperp^2\Psi\rt\} + \nu\dperp^4\Phi,\\
\frac{\dd}{\dd t}\Psi + \lt\{\Phi,\Psi\rt\} 
&= \vA\dpar\Phi + \eta\dperp^2\Psi, 
\label{eq:Psi}
\end{align}
where the difference between the Ohmic diffusivity $\eta$ 
and viscosity $\nu$ has been restored. 

\subsection{Tearing Instability}
\label{app:TM}

Let us ignore the parallel derivatives in \exsdash{eq:Phi}{eq:Psi} and consider small 
perturbations of a simple static equilibrium in which the in-plane magnetic field 
points in the $y$ direction and reverses direction at $x=0$: 
\beq
\Phi =\dPhi(x,y) e^{\gamma t},\quad 
\Psi = \Psi_0(x) + \dPsi(x,y) e^{\gamma t}\hence
\vbperp = \ey b_0(x) + \ez\times\vdperp\dPsi e^{\gamma t},
\eeq
where $b_0(x)=\Psi_0'(x)$ is an odd function describing the reversing field profile 
and $\gamma$ is the rate at which perturbations of this profile 
will grow (if they are interesting). Now linearise the RMHD \eqsdash{eq:Phi}{eq:Psi} 
and Fourier-transform them in the $y$ direction: 
\begin{align}
\label{eq:dPhi}
\lt[\gamma - \nu(\dd_x^2-k_y^2)\rt](\dd_x^2-k_y^2)\dPhi &= 
ik_y\lt[b_0(x)(\dd_x^2-k_y^2) - b_0''(x)\rt]\dPsi,\\
\lt[\gamma - \eta(\dd_x^2-k_y^2)\rt]\dPsi & = ik_yb_0(x)\dPhi. 
\label{eq:dPsi}
\end{align}
When $\eta$ is small, this system has a boundary layer around $x=0$, 
of width $\din$, outside which the solution is an ideal-MHD one and inside 
which resistivity is important and reconnection occurs. 

\subsubsection{Outer Solution}
\label{app:outer}

If we assume that the outer-region solution has scale $\lambda$ and
\beq
\tres^{-1} \equiv \frac{\eta}{\lambda^2}
\sim \tvisc^{-1}\equiv \frac{\nu}{\lambda^2}
\ll \gamma \ll \tAy^{-1}\equiv\frac{\vAy}{\lambda}, 
\eeq
where $\vAy \equiv \lambda b_0'(0)$, then the outer solution satisfies 
\beq
\dd_x^2\dPsi = \lt[k_y^2 + \frac{b_0''(x)}{b_0(x)}\rt]\dPsi,
\quad
\dPhi = -\frac{i\gamma}{k_y b_0(x)}\,\dPsi.
\label{eq:outer}
\eeq
The first of these equations is the (perturbed) 
magnetic-force balance (inertial terms are neglected), 
the second describes the ideal-MHD advection of the equilibrium magnetic field by 
the perturbed flow. 

\begin{figure}
\centerline{\includegraphics[width=0.5\textwidth]{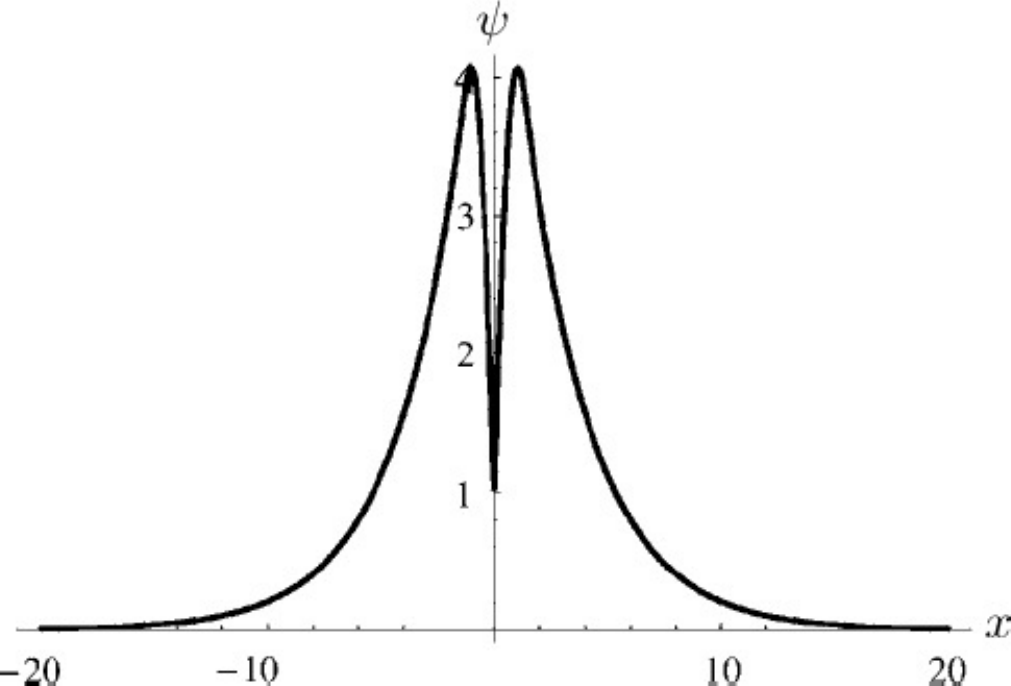}} 
\caption{The outer solution for a tearing mode in a large-aspect-ratio sheet 
(adapted from \citealt{loureiro07}). $\Delta'$ measures the discontinuity 
of $\dd_x\psi$ at $x=0$ [see \exref{eq:Dprime}].}
\label{fig:tearing}
\end{figure}

Since $\dPsi$ is even and the magnetic field $b_y = \dd_x\dPsi$ must reverse direction 
at $x=0$, $\dPsi$~has a discontinuous derivative (\figref{fig:tearing}). 
This corresponds to a singular 
current that is developed by the ideal-MHD solution as it approaches the boundary 
layer---with the singularity resolved inside the layer by resistivity. 
The solutions outside and inside the layer are matched to each other by 
equating the discontinuity in the former to the total change in $\dd_x\dPsi$ 
calculated from the latter: 
\beq
\Delta' \equiv \frac{\lt[\dd_x\dPsi_\mathrm{out}\rt]_{-0}^{+0}}{\dPsi_\mathrm{out}(0)}
= \frac{2}{\din}\int_0^\infty\rmd X\,
\frac{\dd_X^2\dPsi_\mathrm{in}(X)}{\dPsi_\mathrm{in}(0)},
\label{eq:Dprime}
\eeq 
where $\dPsi_\mathrm{out}(x)=\dPsi(x)$ is the outer solution, 
$\dPsi_\mathrm{in}(X)=\dPsi(X\din)$ is the inner one, and 
$X = x/\din$ is the ``inner'' variable, rescaled to the current layer's width~$\din$.  

To find $\Delta'$ from the outer solution, one must solve \exref{eq:outer} for 
some particular form of~$b_0(x)$. For our purposes, all that is needed is the asymptotic 
behaviour of $\Delta'$ in the limit of $k_y\lambda\ll1$, where $\lambda$ is the 
characteristic scale of $b_0(x)$. While in general this asymptotic depends 
on the functional form of $b_0(x)$, it~is 
\beq
\Delta' \sim \frac{1}{k_y\lambda^2}
\label{eq:Dprime_outer}
\eeq
if one can assume that $b_0(x)$ varies faster at $|x|\lesssim\lambda$, in the region where 
it reverses direction, than at $|x|\gg\lambda$, where it might be approximately 
flat (see \apref{app:dprime}). 
An example of such a situation is the exactly solvable and ubiquitously 
useful \citet{harris62} sheet $b_0(x) = \vAy \tanh(x/\lambda)$. 
This situation might be particularly relevant because in ideal MHD, 
field-reversing configurations of the kind that we need to support a tearing mode 
tend to be collapsing sheets, with $\lambda$ shrinking dynamically compared to 
the characteristic scales in the $y$ direction or indeed in the $x$ direction 
away from the field-reversal region (see further discussion 
in \apref{app:CS}).

\subsubsection{Scaling of $\Delta'$}
\label{app:dprime}

A reader who is happy to accept \exref{eq:Dprime_outer} can now skip to \apref{app:inner}. 
For those who would like to see a more detailed derivation leading 
to \exref{eq:Dprime_outer}, let me put forward the following argument, 
which is adapted from \citet{loureiro07,loureiro13kh}. 

Consider first $|x|\sim \lambda$. Since $b_0''/b_0 \sim 1/\lambda^2 \gg k_y^2$, 
we may neglect the $k_y^2$ term in \exref{eq:outer} and seek a solution in the form 
$\dPsi = b_0(x)\chi(x)$. This allows us to integrate the equation directly, with 
the result 
\beq
\dPsi = b_0(x)\lt[C_1^\pm + C_2^\pm\int_{\pm x_0}^x\frac{\rmd x'}{b_0^2(x')}\rt],
\label{eq:dPsi_outin}
\eeq
where $\pm$ refer to solutions at positive and negative $x$, respectively, 
$C^\pm_{1,2}$ are integration constants and $x_0\sim\lambda$ is some 
integration limit, whose precise value does not matter (any difference that
it makes can be absorbed into $C_1^\pm$). Since $b_0(x)$ is an odd function, 
\beq
b_0(x) \approx \frac{x}{\lambda}\,\vAy
\quad\text{at}\quad
|x|\ll\lambda.
\label{eq:b0_in}
\eeq
Taking $x\to0$ in \exref{eq:dPsi_outin}, we can, therefore, fix the constant $C_2^\pm$ via
\beq
\dPsi(0) = -C_2^\pm\frac{\lambda}{\vAy}.  
\label{eq:C2}
\eeq
Now formally taking $x\to\pm\infty$ and assuming that $b_0(x)\to\pm\vAyinf=\const$ 
in these limits, we find that the solution \exref{eq:dPsi_outin} asymptotes to  
\beq
\dPsi \approx \pm C_1^\pm \vAyinf \mp \dPsi(0)\frac{\vAy}{\vAyinf}\frac{x}{\lambda}. 
\label{eq:dPsi_outin_inf}
\eeq

Let us return to \exref{eq:outer} and consider $|x|\gg\lambda$. 
At such distances, $b_0''/b_0\to 0$ by assumption, so we must now solve 
\exref{eq:outer} neglecting the $b_0''/b_0$ terms while retaining $k_y^2$, 
and then match the resulting ``outer-outer'' solution to the large-$|x|$ 
asymptotic \exref{eq:dPsi_outin_inf} of the ``outer-inner'' solution~\exref{eq:dPsi_outin}. 
The solution that vanishes at infinity~is 
\beq
\dPsi = C_3^\pm e^{\mp k_y x}
\eeq 
and its $k_y x\ll1$ asymptotic~is
\beq
\dPsi \approx C_3^\pm \mp C_3^\pm k_y x. 
\eeq
Demanding that this match \exref{eq:dPsi_outin_inf}, we get 
\beq
C_3^\pm = \frac{\vAy}{\vAyinf}\frac{\dPsi(0)}{k_y\lambda},
\quad
C_1^\pm = \pm \frac{C_3^\pm}{\vAyinf}. 
\label{eq:C1C3}
\eeq 

Finally, returning to \exref{eq:dPsi_outin}, which, with 
\exref{eq:C2} and~\exref{eq:C1C3}, has become
\beq
\dPsi = b_0(x)\frac{\vAy\psi(0)}{\lambda}\lt[\pm \frac{1}{k_y\vAyinf^2} 
- \int_{\pm x_0}^x\frac{\rmd x'}{b_0^2(x')}\rt],
\eeq
and using \exref{eq:b0_in}, we obtain, for $k_y\lambda\ll1$, 
\beq
\Delta' = \frac{\dPsi'(+0) - \dPsi'(-0)}{\dPsi(0)} \approx 
2\,\biggl(\frac{\vAy}{\vAyinf}\biggr)^2\frac{1}{k_y\lambda^2}
\sim \frac{1}{k_y\lambda^2},\quad\text{q.e.d.}
\eeq
Pending detailed insight into the functional form of the aligned fluctuations 
in MHD turbulence, I am going to treat this scaling of $\Delta'$ 
with $k_y$ and $\lambda$ as generic. A more general scaling 
\beq
\Delta'\lambda \sim \frac{1}{(k_y\lambda)^n} 
\label{eq:Dprime_n}
\eeq 
corresponds, for $n>1$, to $b_0(x)$ decaying to zero at large $x$ on the same 
scale as it reverses direction around $x=0$: e.g., one gets 
$n=2$ for $b_0(x)=\vAy\tanh(x/\lambda)/\cosh^2(x/\lambda)$ \citep{porcelli02} 
or for a simple sinusoidal profile \citep{ottaviani93,boldyrev18}. 
There is some space for discussion 
as to whether $n=1$ or $n=2$ is the best model for what happens 
in a typical MHD-turbulent structure \citep[cf.][]{walker18}.
Generalising all the scalings derived throughout this review  
to arbitrary $n$ is a tedious but straightforward exercise 
\citep{pucci18,singh19,boldyrev20}, 
which I have opted to forgo, to avoid bulky $n$-dependent exponents 
everywhere. A meticulous reader who wishes to do this exercise will find 
the tearing-mode scalings for arbitrary $n$ in \apref{app:dprime_n} 
and take it from there. For the tearing-mediated RMHD cascade (\secref{sec:disruption}),
the resulting turbulence scalings were worked out by \citet{boldyrev17,boldyrev20};
for the tearing-mediated dynamo (\secref{sec:dynamo_theory}), by
\citet{galishnikova22}.

\subsubsection{Inner Solution}
\label{app:inner}

In the inner region, whose width is $\din$,  
we can approximate the equilibrium magnetic field's profile~by \exref{eq:b0_in}. 
Since $k_y\ll \dd_x\sim\din^{-1}$, the equations \exref{eq:dPhi} and \exref{eq:dPsi} 
for the tearing perturbation become
\begin{align}
(\gamma - \nu\dd_x^2)\dd_x^2\dPhi &= ik_y\frac{x}{\lambda}\vAy\dd_x^2\dPsi,\\
(\gamma - \eta\dd_x^2)\dPsi &=ik_y\frac{x}{\lambda}\vAy\dPhi.
\end{align}
The first of these is the balance of inertia, viscous and Lorentz forces, the 
second is Ohm's law in resistive MHD. Combining them, we get 
\beq
\dd_x^2\dPsi = -\lt(\frac{\gamma\lambda}{k_y\vAy}\rt)^2
\frac{1}{x}\lt(1 - \frac{\nu}{\gamma}\,\dd_x^2\rt)
\dd_x^2\,\frac{1}{x}\lt(1 - \frac{\eta}{\gamma}\,\dd_x^2\rt)\dPsi.
\label{eq:dPsi_inner}
\eeq
This immediately tells us what the width of the boundary layer is: 
\begin{align}
\label{eq:din_res}
\frac{\nu}{\gamma\din^2}\ll1 
\quad&\Rightarrow\quad
\lt(\frac{\gamma\lambda}{k_y\vAy}\rt)^2\frac{\eta}{\gamma\din^4}\sim 1
\hence
\frac{\din}{\lambda}\sim\lt(\frac{\gamma\tAy^2}{\tres}\rt)^{1/4}\frac{1}{(k_y\lambda)^{1/2}},\\
\frac{\nu}{\gamma\din^2}\gg1 
\quad&\Rightarrow\quad
\lt(\frac{\gamma\lambda}{k_y\vAy}\rt)^2\frac{\eta\nu}{\gamma^2\din^6}\sim 1
\hence
\frac{\din}{\lambda}\sim\lt(\frac{\tAy^2}{\tres\tvisc}\rt)^{1/6}\frac{1}{(k_y\lambda)^{1/3}}.
\label{eq:din_visc}
\end{align}
The latter regime, in which viscosity is large, is a slightly less popular version 
of the tearing mode, but it can be treated together with the classic limit \exref{eq:din_res} 
at little extra cost. 

Let us now rescale $x=X\din$ in \exref{eq:dPsi_inner}. 
Then $\dPsi_\mathrm{in}(X) \equiv \dPsi(X\din)$ satisfies 
\begin{align}\label{eq:inner_res}
\frac{\nu}{\gamma\din^2}\ll1 
\quad&\Rightarrow\quad
\dd_X^2\dPsi_\mathrm{in} = -\frac{1}{X}\,\dd_X^2\frac{1}{X}
\lt(\Lambda - \dd_X^2\rt)\dPsi_\mathrm{in}, \quad
\Lambda = \lt(\frac{\gamma\lambda}{k_y\vAy}\rt)^2\frac{1}{\din^2},\\
\frac{\nu}{\gamma\din^2}\gg1 
\quad&\Rightarrow\quad
\dd_X^2\dPsi_\mathrm{in} = \frac{1}{X}\,\dd_X^4\frac{1}{X}
\lt(\Lambda - \dd_X^2\rt)\dPsi_\mathrm{in},\quad
\Lambda = \lt(\frac{\gamma\lambda}{k_y\vAy}\rt)^2\frac{\nu}{\gamma\din^4}.
\label{eq:inner_visc}
\end{align} 
In both cases, the inner solution depends on a single dimensionless parameter $\Lambda$ 
(the eigenvalue). 
In view of \exsdash{eq:din_res}{eq:din_visc}, this parameter is, in both cases, just the 
ratio of the growth rate of the mode to the rate of resistive diffusion across a layer 
of width $\din$, with the appropriate scaling of~$\din$:
\beq
\Lambda \sim \frac{\gamma\din^2}{\eta} \sim \lt\{
\begin{array}{ll}
\displaystyle\frac{\gamma^{3/2}\tres^{1/2}\tAy}{k_y\lambda}
\sim\frac{(\gamma\tAy)^{3/2}}{k_y\lambda}\,S_\lambda^{1/2},& 
\displaystyle\frac{\nu}{\gamma\din^2}\sim\frac{\Pm}{\Lambda}\ll1,\\\\
\displaystyle\frac{\gamma\tres^{2/3}\tvisc^{-1/3}\tAy^{2/3}}{(k_y\lambda)^{2/3}}
\sim \lt[\frac{(\gamma\tAy)^{3/2}}{k_y\lambda}\,(S_\lambda\Pm)^{1/2}\rt]^{2/3},
& \displaystyle\frac{\nu}{\gamma\din^2}\sim\frac{\Pm}{\Lambda}\gg1,
\end{array}
\rt.
\label{eq:Lambda_def}
\eeq 
where the Lundquist number (associated with scale $\lambda$)
and the magnetic Prandtl number are defined as follows: 
\beq
S_\lambda = \frac{\tres}{\tAy} = \frac{\vAy\lambda}{\eta}, 
\quad
\Pm = \frac{\tres}{\tvisc} = \frac{\nu}{\eta}.
\eeq

\subsubsection{Peak Growth Rate and Wavenumber}
\label{app:TMgammak}

Whatever the specific form of the solution of \exref{eq:inner_res} \citep{coppi76,boldyrev18} 
or \exref{eq:inner_visc}, 
$\Delta'$ calculated from it according to \exref{eq:Dprime} (and non-dimensionalised) 
must be a function only of~$\Lambda$: 
\beq
\Delta'\din = f(\Lambda). 
\label{eq:tearing_dr}
\eeq
Equating this to the value \exref{eq:Dprime_outer} calculated from the outer solution, 
we arrive at an equation for~$\Lambda$: 
\beq
f(\Lambda) \sim \frac{\din}{k_y\lambda^2} \sim \lt\{
\begin{array}{ll}
\displaystyle\frac{\gamma^{1/4}\tAy^{1/2}\tres^{-1/4}}{(k_y\lambda)^{3/2}}
\sim \Lambda^{1/6}\lt(k_y\lambda\, S_\lambda^{1/4}\rt)^{-4/3},&
\Pm\ll\Lambda,\\\\
\displaystyle \frac{\tAy^{1/3}(\tres\tvisc)^{-1/6}}{(k_y\lambda)^{4/3}}
\sim\lt(k_y\lambda\, S_\lambda^{1/4}\Pm^{-1/8}\rt)^{-4/3}, 
& \Pm\gg\Lambda.
\end{array}
\rt.
\label{eq:Lambda_eq}
\eeq
Since the function $f(\Lambda)$ does not depend on any parameters apart from 
$\Lambda$, one might intuit that the maximum growth of the tearing mode should 
occur at $\Lambda\sim1$, when $f(\Lambda)\sim1$ (I shall confirm this momentarily). 
Using these estimates in \exref{eq:Lambda_eq} and \exref{eq:Lambda_def}, we find
\beq
k_y\lambda \sim S_\lambda^{-1/4}(1+\Pm)^{1/8}\equiv k_*\lambda
\hence
\gamma\tAy \sim S_\lambda^{-1/2}(1+\Pm)^{-1/4},
\label{eq:TM_max}
\eeq 
where $\Pm$ only matters if it is large. 
Note that if $S_\lambda\gg (1+\Pm)^{1/2}$, the assumption $k_y\lambda\ll1$ is confirmed.
These are the maximum growth rate and the corresponding wavenumber of the 
tearing mode.\footnote{I picked up the general idea of this argument from J.~B.~Taylor 
(2010, private communication); it is a slight generalisation of his treatment 
of the tearing mode in \citet{taylor15}.} Note that, for this solution, 
since $f(\Lambda)\sim1$, \exref{eq:Lambda_eq} gives~us  
\beq
\frac{\din}{\lambda} \sim k_*\lambda. 
\label{eq:din_TMmax}
\eeq

If setting $\Lambda\sim1$, $f(\Lambda)\sim1$ did  
not feel inevitable to the reader, perhaps the following considerations will help. 
Let us consider two physically meaningful limits that do not satisfy these assumptions. 

First, let us ask what happens if $\Lambda\ll1$. This means that 
the mode grows slowly compared to the Ohmic diffusion rate
in the current layer, $\gamma\ll \eta/\din^2$, 
a situation that corresponds, in a sense that is to be quantified in a moment, 
to small $\Delta'$. In this limit, $f(\Lambda)\sim\Lambda$ to lowest order in the Taylor expansion
(I have not shown this rigorously, but hopefully it is plausible to the reader; for 
a derivation, see, e.g., \citealt{taylor15} or \citealt{boldyrev18}). 
Putting this into \exref{eq:Lambda_eq} and using \exref{eq:Lambda_def} to unpack $\Lambda$, 
we find 
\beq
\gamma\tAy \sim 
\lt\{
\begin{array}{ll}
\displaystyle S_\lambda^{-3/5}(k_y\lambda)^{-2/5}, &
k_y\lambda \ll S_\lambda^{-1/4}\Pm^{-5/8},\\\\
\displaystyle S_\lambda^{-2/3}\Pm^{-1/6}(k_y\lambda)^{-2/3}, & 
k_y\lambda \gg S_\lambda^{-1/4}\Pm^{-5/8}.
\end{array}
\rt.
\label{eq:FKR}
\eeq
This is the famous FKR solution (\citealt{furth63}; see also \citealt{porcelli87} 
for the large-$\Pm$ case). Since, to get it, $\Lambda\ll1$ was assumed, 
substituting \exref{eq:FKR} into \exref{eq:Lambda_def} tells us that the approximation 
is valid at wavenumbers exceeding the wavenumber \exref{eq:TM_max} of peak growth, 
$k_y\gg k_*$. Note that this imposes an upper bound on $\Delta'$: 
\beq
\Delta'\lambda \sim \frac{1}{k_y\lambda} \ll \frac{1}{k_*\lambda}. 
\label{eq:small_Dprime}
\eeq
This is sometimes (perhaps misleadingly) called the ``small-$\Delta'$'' (or weakly driven) limit. 

\begin{figure}
\centerline{\includegraphics[width=0.5\textwidth]{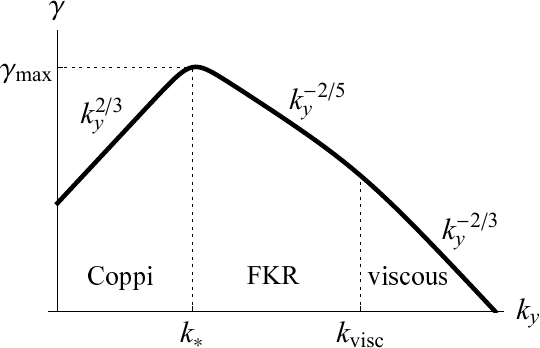}} 
\caption{Tearing growth rate $\gamma$ vs.~$k_y$: 
the \citet{coppi76} solution~\exref{eq:Coppi}
for $k_y\ll k_*$, where $k_*$ is given in~\exref{eq:TM_max}, 
and the FKR solution~\exref{eq:FKR} at $k_y\gg k_*$. 
The viscous version of the latter takes over at $k_y\gg k_\mathrm{visc}$, 
where $k_\mathrm{visc}\lambda\sim S_\lambda^{-1/4}\Pm^{-5/8}$.
This cartoon is for $\Pm\ll 1$;  
if $\Pm\gg 1$, the viscous-FKR scaling starts at~$k_*$.}
\label{fig:tearing_rate}
\end{figure}

Let us now consider the limit opposite to \exref{eq:small_Dprime}, i.e., 
when $\Delta'$ is very large and $k_y \ll k_*$. In \exref{eq:Lambda_eq}, 
this corresponds to $f(\Lambda)\to\infty$ and we argue that this limit must be reached 
for some value $\Lambda\sim1$ (it is not physically reasonable to expect that $\Lambda\gg1$, 
i.e., that the growth rate of the mode can be much larger than the Ohmic diffusion rate 
in the current layer; this reasoning is confirmed by the exact solution---see \citealt{coppi76}). 
This implies, with the aid of~\exref{eq:Lambda_def}, 
\beq
\gamma\tAy \sim S_\lambda^{-1/3}(1+\Pm)^{-1/3}(k_y\lambda)^{2/3}.
\label{eq:Coppi}
\eeq
This long-wavelength 
(``infinite-$\Delta'$,'' or strongly driven) 
limit of the tearing mode was first derived by \citet{coppi76} 
(and by \citealt{porcelli87} for the large-$\Pm$ case).  

We see that the small-$k_y$ asymptotic \exref{eq:Coppi} is an ascending and 
the large-$k_y$ one \exref{eq:FKR} a descending function of $k_y$ (\figref{fig:tearing_rate}). 
The wavenumber $k_*$ of peak growth lies in between, where these two asymptotics 
meet, which is quite obviously the solution~\exref{eq:TM_max}. 

The applicability of this solution is subject to an important caveat. 
The Harris-like equilibrium that was used to obtain it is a 1D configuration, 
implicitly assumed to extend as far in the $y$ direction as the mode requires to develop. 
In reality, any sheet-like configuration  
forming as a result of (ideal) MHD dynamics will have a length, as well as width: 
$\xi\gg\lambda$, but still finite. The finiteness of $\xi$ will limit the 
wavenumbers of the tearing perturbations that can be accommodated. The fastest-growing mode 
\exref{eq:TM_max} will only fit into the sheet~if 
\beq
k_*\xi \gtrsim 1 
\quad\Leftrightarrow\quad
\frac{\xi}{\lambda} \gtrsim S_\lambda^{1/4}(1+\Pm)^{-1/8}. 
\label{eq:ky_to_fit}
\eeq
If this condition fails to be satisfied, i.e., if the aspect ratio of the sheet 
is too small, the fastest-growing mode will be the FKR mode \exref{eq:FKR} with 
the smallest possible allowed wavenumber $k_y\xi\sim1$. Thus, low-aspect-ratio 
sheets will develop tearing perturbations comprising just one or two islands, 
whereas the high-aspect-ratio ones will spawn whole chains of them, consisting of 
$N \sim k_*\xi$ islands. 

\subsubsection{Case of Arbitrary Scaling of $\Delta'$}
\label{app:dprime_n}

As promised at the end of \apref{app:dprime}, here is the generalisation 
of the main tearing-mode scalings to the case of $\Delta'$ scaling 
according to \exref{eq:Dprime_n}. For $\Lambda\sim1$, \exref{eq:Lambda_def}, which 
is independent of $n$, implies 
\beq
\gamma\tAy \sim (k_*\lambda)^{2/3} S_\lambda^{-1/3}(1+\Pm)^{-1/3},
\quad
\frac{\din}{\lambda}\sim(k_*\lambda)^{-1/3}S_\lambda^{-1/3}(1+\Pm)^{1/6}.
\label{eq:TMk_n}
\eeq
Using \exref{eq:Dprime_n} in \exref{eq:tearing_dr} and setting $f(\Lambda)\sim1$ 
gets~us, instead of \exref{eq:din_TMmax},
\beq
\frac{\din}{\lambda} \sim (k_*\lambda)^n.
\label{eq:din_TMmax_n}
\eeq
Combining this with \exref{eq:TMk_n} leads to the generalised version 
of~\exref{eq:TM_max}:
\beq
k_*\lambda \sim S_\lambda^{-1/(3n+1)}(1+\Pm)^{1/2(3n+1)},
\quad
\gamma\tAy \sim S_\lambda^{-(n+1)/(3n+1)}(1+\Pm)^{-n/(3n+1)}. 
\label{eq:TM_max_n}
\eeq 
Using these scalings instead of \exref{eq:TM_max} in 
\exref{eq:kpeak_TM} and \exref{eq:gmax_TM} introduces $n$ dependence everywhere 
in \secsref{sec:disruption}, \ref{sec:tearing_subvisc}, and~\ref{sec:dynamo_theory},
but does not appear to change anything qualitatively 
\citep{boldyrev17,boldyrev20,galishnikova22}. 

\subsection{Onset of Nonlinearity}
\label{app:TMnlin}

The tearing mode normally enters 
a nonlinear regime when the width $w$ of its islands becomes comparable to $\din$. 
The islands then grow secularly \citep{rutherford73} until $w\Delta'\sim1$. 
As we saw in \apref{app:TMgammak}, for the fastest-growing Coppi mode,  
$\Delta'\sim\din^{-1}$, so the secular-growth stage is skipped. 
The width of the islands at the onset of the nonlinear regime is, therefore,  
\beq
\frac{w}{\lambda} \sim \frac{\din}{\lambda}
\sim\frac{1}{\Delta'\lambda} \sim (k_*\lambda)^n. 
\label{eq:w_onset}
\eeq
There is little overhead here for keeping $n$ general, as in~\exref{eq:din_TMmax_n}, 
so I will. 

The amplitudes $\dbx$ and $\dby$ of the tearing perturbation at the onset of nonlinearity 
can be worked out by observing that the typical angular distortion of a field line 
due to the perturbation is
\beq
wk_* \sim \frac{\dbx}{b_0(x)|_{x\sim w}} 
\label{eq:w_bx}
\eeq 
and that, by solenoidality, 
\beq
k_* \dby\sim\frac{\dbx}{w}.
\label{eq:w_by}
\eeq
Since, from \exref{eq:b0_in}, $b_0(x)|_{x\sim w}\sim (w/\lambda)\vAy$, we have
\beq
\frac{\dbx}{\vAy} \sim \frac{w^2 k_*}{\lambda} \sim (k_*\lambda)^{2n+1},\quad
\frac{\dby}{\vAy} \sim \frac{w}{\lambda} \sim (k_*\lambda)^n. 
\label{eq:dby}
\eeq
Note that \exref{eq:w_bx} and \exref{eq:w_by} together 
imply that $\dby\sim b_0(x)|_{x\sim w}$, i.e., 
the perturbed field is as large as the equilibrium field is 
locally at $x\sim w$ (but not as the upstream field $\vAy$ at $x\sim\lambda$). 
 
Let us confirm that \exref{eq:w_onset} was a good estimate for the onset of nonlinearity, 
i.e., that, once it is achieved, the characteristic rate of the nonlinear evolution 
of the tearing perturbation becomes comparable to its linear growth rate \exref{eq:TM_max_n}. 
The nonlinear evolution rate can be estimated as $k_* \duy$, where $\duy$ is 
the outflow velocity from the tearing region. When $\Pm\lesssim1$, this is obviously 
Alfv\'enic, $\duy\sim\dby$. When $\Pm\gg1$, the situation is more subtle as 
the viscous relaxation of the flows is in fact faster than their Alfv\'enic evolution 
(as we are about to see). Then the outflow velocity must be determined from the 
force balance between viscous and magnetic stresses: using \exref{eq:dby}, 
\beq
\frac{\nu}{w^2} \duy \sim k_*\dby^2 
\hence
\frac{\duy}{\dby} \sim \frac{k_* w^2 \dby}{\nu}
\sim \frac{k_* w^3 \vAy}{\lambda \nu} 
\sim (k_*\lambda)^{3n+1}\frac{S_\lambda}{\Pm} 
\sim \frac{1}{\sqrt{\Pm}}. 
\label{eq:visc_bal_TM}
\eeq
Combining the small- and large-$\Pm$ cases, we get 
\beq
\label{eq:uout}
\duy \sim \frac{\dby}{\sqrt{1+\Pm}}
\hence
k_*\duy \sim \frac{(k_*\lambda)^{n+1}\vAy/\lambda}{\sqrt{1+\Pm}}\sim\gamma.
\eeq
In the last expression, \exref{eq:dby} was used to estimate $\dby$ 
and then \exref{eq:TM_max_n} to ascertain that the nonlinear and linear rates 
are indeed the same. 

\subsection{What Happens Next?}
\label{app:what_next}

Once nonlinear effects come in, the tearing perturbation 
becomes subject to ideal-MHD evolution (for $\Pm\gg1$, also to viscous forces). 
This leads to collapse of the $X$-points separating the islands 
of the tearing perturbation into current sheets 
\citep{waelbroeck93,jemella03,jemella04}. The time scale for this 
process is the same as that for the Coppi mode's growth \citep{loureiro05}
(which, as we have just seen, is the same as the ideal-MHD time scale for a 
perturbation that is gone nonlinear). 

In order to explain what happens as a result of this, I must cover some essential
material regarding current sheets and reconnection in them, but let me preview
the main ingredients of the overall story first.
\vskip2mm 
(i) The standard, resistively limited, reconnecting current sheet towards which
the $X$-point collapse leads is described in \apref{app:SP_rec}.
\vskip2mm
(ii) At asymptotically large Lundquist numbers, such a sheet is, in fact, violently
unstable (\apref{app:loureiro}).
\vskip2mm
(iii) Therefore, it cannot, in fact, exist and will be disrupted in mid-formation,
spawning a population of multiscale islands, or plasmoids (\apref{app:CS}).
\vskip2mm
(iv) If it does not fall apart as a result---e.g., if it is held together by some
dynamics external to it---it turns into a stochastic plasmoid chain, which is
a site of fast MHD reconnection (\apref{app:uls}).
\vskip2mm
All of these things will be described, somewhat painstakingly, in what follows---they
are the background to the blithe assumptions made at the end of \secref{sec:disruption_scale}
about the nature of the debris left in the wake of the tearing disruption of the aligned
turbulent structures---assumptions that I needed to construct the model of the 
tearing-mediated turbulence presented in~\secref{sec:recturb}. 

\subsection{Sweet--Parker Sheet}
\label{app:SP} 

\begin{figure}
\begin{center}
\begin{tabular}{cc}
\includegraphics[width=0.35\textwidth]{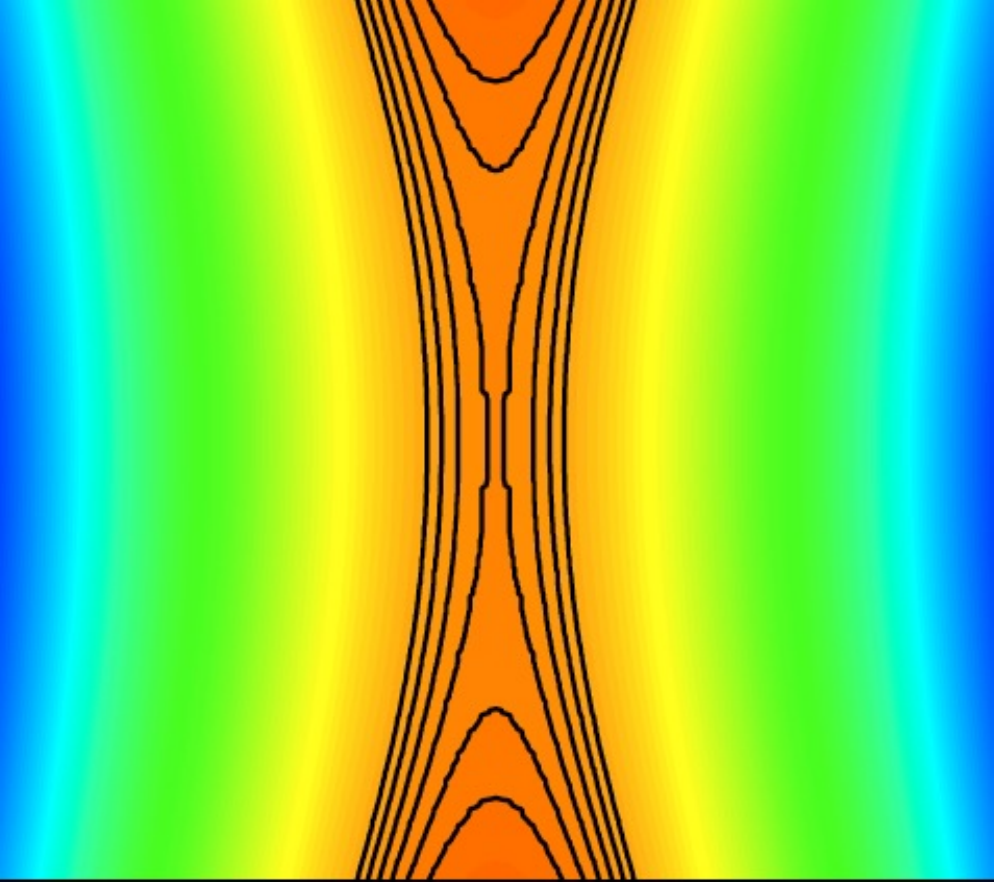} & 
\includegraphics[width=0.35\textwidth]{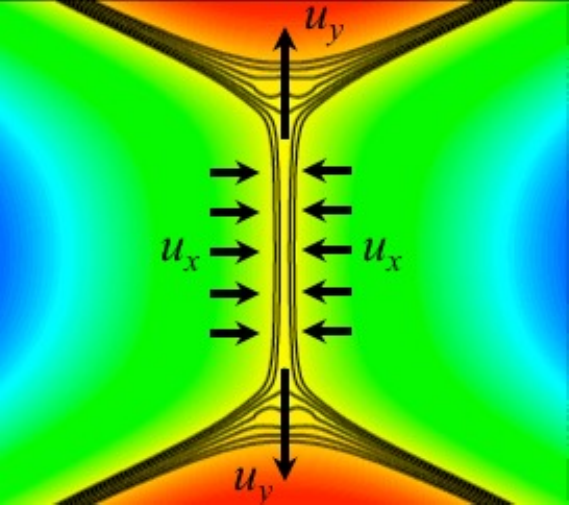}\\
(a) & (b)
\end{tabular}
\end{center}
\caption{(a) An $X$-point, shown during the nonlinear stage of tearing mode,  
(b) SP current sheet, formed later on, upon collapse of that 
$X$-point (adapted from a 2D RMHD numerical simulation by \citealt{loureiro05}). 
The black lines are magnetic-field lines (constant-flux contours). 
The in-plane field reverses direction along the middle of the domain that is shown.
In the notation of \apref{app:SP_rec}, the length of the sheet is $\ell$ 
and its width is~$\delta$.}
\label{fig:SP}
\end{figure}

Let me flesh out what was meant by the $X$-point collapse in \apref{app:what_next}. 
The idea is that, once the nonlinearity takes hold and Alfv\'enic (or visco-Alfv\'enic)
outflows from the reconnection region develop, the reconnecting site will suck 
plasma in, carrying the magnetic field with it, thus leading to formation of an extended 
sheet, which is a singularity from the ideal-MHD viewpoint, resolved, of course, by 
resistivity and acting as a funnel both for magnetic flux 
and plasma (\figref{fig:SP}).\footnote{To avoid a misunderstanding, let me anticipate 
here the discussion in \apref{app:CS} 
and say that the $X$-point collapse at the end of the evolution of a tearing 
mode is not the unique scenario that can lead to the formation of a sheet, which 
is in fact a fairly generic dynamical feature of the evolution of $X$-points 
in {\em ideal}~MHD. The developments in this sub-section are not in any way restricted 
to sheets formed from tearing modes.} 
After the collapse has occurred and a sheet has been formed, the magnetic field just outside 
the resistive layer (the ``upstream field'') is now the full equilibrium field, 
brought in by the incoming flow $u_x$ of plasma 
(in terms of the discussion in \apref{app:TMnlin}, this means that now $\dby\sim\vAy$). 

\subsubsection{Sweet--Parker Reconnection}
\label{app:SP_rec}

The flux brought in by the inflow $u_x$ must be destroyed by resistivity (reconnected and 
turned into $b_x$). This translates into what formally 
is just a statement of balance between the advective and resistive terms 
in the induction equation: 
\beq
u_x\vAy \sim \eta j_z \sim \eta\frac{\vAy}{\delta}
\hence 
\delta \sim \frac{\eta}{u_x} \sim \frac{\ell}{S_\ell}\frac{\vAy}{u_x},
\quad
S_\ell = \frac{\vAy\ell}{\eta},
\label{eq:flux_bal}
\eeq
where $\delta$ is the resistive layer's width and $u_x$ the inflow velocity. 
I have, in line with 
the prevailing convention (and physics) of the reconnection theory, introduced a 
Lundquist number based on the sheet length $\ell$ (in the context of a sheet formed 
between two islands of a tearing perturbation, this length is $\ell\sim k_*^{-1}$).  

Since the sheet has to process matter as well as flux and since matter must be 
conserved, we may balance its inflow ($u_x$) and outflow ($u_y$): 
\beq
u_x \ell \sim u_y \delta
\hence 
u_x \sim \frac{\delta}{\ell}u_y
\hence 
\delta \sim \frac{\ell}{\sqrt{S_\ell}}\lt(\frac{\vAy}{u_y}\rt)^{1/2},  
\label{eq:flow_bal}
\eeq
where the third equation is the result of combining the second with \exref{eq:flux_bal}. 

Finally, the outflow velocity is inevitably Alfv\'enic in the absence of viscosity: 
this follows by balancing Reynolds and Maxwell stresses (inertia and tension) 
in the momentum equation (in either $y$ or $x$ direction; note that $b_x\sim \vAy\delta/\ell$). 
Physically, this is just saying that the tension in the ``parabolic''-shaped 
freshly reconnected magnetic field line (manifest in \figref{fig:SP}a)
will accelerate plasma and propel it 
out of the sheet. In the presence of viscosity, i.e., when $\Pm\gg1$, we must instead balance 
the magnetic stress with the viscous one, exactly like I did in \exref{eq:visc_bal_TM}, 
but with a narrower channel and a greater upstream field:  
\beq
\frac{\nu}{\delta^2}u_y \sim \frac{\vAy^2}{\ell}
\hence
\frac{u_y}{\vAy}\sim \frac{\vAy\delta^2}{\ell\nu} \sim \frac{1}{\sqrt{\Pm}}. 
\eeq
To get the last expression, $\delta$ had to be substituted from \exref{eq:flow_bal}. 
Just as I have done everywhere else, I will combine the low- and high-$\Pm$ 
cases:\footnote{Note that replacing in this argument $\ell\to k_*^{-1}$, 
$u_y\to\duy$, $\vAy\to\dby\sim\vAy w/\lambda \sim \vAy k_*\lambda$ 
gives us back the scalings associated 
with the tearing mode at the onset of nonlinearity (\apref{app:TMnlin}), 
with $\delta\sim\din$. This is, of course, inevitable as both theories 
are based on the same balances in the reconnection region, except 
the tearing before $X$-point collapse had a smaller upstream field~$\dby$.\label{fn:onset}}  
\beq
u_y \sim \frac{\vAy}{\sqrt{1+\Pm}}
\hence
\frac{\delta}{\ell} \sim \frac{(1+\Pm)^{1/4}}{\sqrt{S_\ell}} \equiv\frac{1}{\sqrt{\tS_\ell}}, 
\quad\tS_\ell = \frac{u_y \ell}{\eta},
\label{eq:SP_uy}
\eeq 
where $\tS_\ell$, the Lundquist number based on the outflow velocity, is an obviously 
useful shorthand. Other relevant quantities can now be calculated, e.g., 
the rate at which flux is reconnected: 
\beq
\frac{\dd\Psi}{\dd t}\sim 
u_x\vAy \sim \frac{u_y\vAy}{\sqrt{\tS_\ell}} \sim \frac{\vAy^2}{(1+\Pm)^{1/4}\sqrt{S_\ell}}. 
\label{eq:rec_rate}
\eeq

The argument that I have just presented is one of the enduring classics of the genre 
and is due to \citet{sweet58} and \citet{parker57} (hereafter SP; the large-$\Pm$ extension 
was done by \citealt{park84}). While the argument is qualitative, it does work, 
in the sense both that one can construct unique solutions of the SP kind, 
in a manner pleasing to rigorous theoreticians \citep{uzdensky96,uzdensky00}, 
and that SP reconnection has been measured and confirmed experimentally 
\citep{ji98,ji99} (\figref{fig:mrx} shows an SP sheet measured in their MRX 
experiment at Princeton). 

\subsubsection{Plasmoid Instability}
\label{app:loureiro}

\begin{figure}
\centerline{\includegraphics[width=\textwidth]{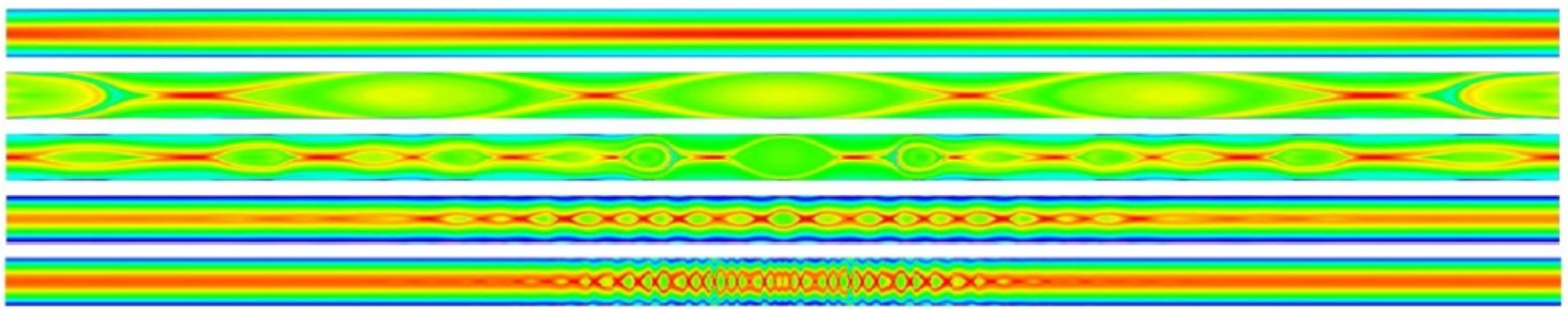}} 
\caption{Plasmoid instability in current sheets with, from top to bottom, 
$S_\xi=10^4, 10^5, 10^6, 10^7, 10^8$. The domain shown is $0.12$ of the 
full length of the sheet. This plot is adapted from \citet{samtaney09}, 
who confirmed the scalings \exref{eq:loureiro} numerically.
[Reprinted with permission from \citet{samtaney09}, copyright (2009) by the American
Physical Society.]} 
\label{fig:samtaney}
\end{figure}

However, like for any sheet, one can 
work out a tearing instability for an SP sheet (this is not the same tearing instability as 
might have given rise to the sheet as suggested at the beginning of \apref{app:SP}---the SP 
sheet is now the underlying equilibrium). The results of \apref{app:TMgammak} can be 
ported directly to this situation (with some caveats that I will discuss in the next 
paragraph), by identifying $\delta=\lambda$ and $\ell = \xi$. This gives instantly 
\beq
\gamma \sim \frac{u_y}{\xi}\tS_\xi^{1/4},\quad
k_*\xi \sim \tS_\xi^{3/8},\quad
\frac{\din}{\delta}\sim \tS_\xi^{-1/8}. 
\label{eq:loureiro}
\eeq
This is the so-called {\em plasmoid instability} 
(\citealt{tajima97,loureiro07,loureiro13kh,bhatta09,comisso16}; see \figref{fig:samtaney}). 
The realisation that SP sheets must be unstable can be traced back 
to \citet{bulanov78,bulanov79}, with the first numerical demonstration 
achieved by \citet{biskamp86} (see also \citealt{biskamp82}, 
\citealt{steinolfson84}, \citealt{matthaeus85}, \citealt{dahlburg86}, \citealt{lee86}, 
and \citealt{malara92}). 
However, this knowledge did not seem to have impacted 
the field as much as it should have done\footnote{A reader interested in history 
will find a useful review of secondary-tearing literature in Appendix~B of \citet{delsarto17}.} 
until the appearance of the analytical 
paper by \citet{loureiro07} and the rise of the plasmoid-chain simulation 
industry in 2D \citep{lapenta08,daughton09,daughton09pop,cassak09b,samtaney09,huang10,huang12,huang13,huang17,barta11,loureiro12,shen13,tenerani15,tenerani20}, followed, more recently, 
by its more turbulent counterpart in 3D 
\citep{oishi15,huang16,beresnyak17,kowal17,stanier19,yang20,daldorff22}\footnote{In 3D, plasmoids become 
flux ropes, which are prone to going kink-unstable and breaking up. 
Their coherence length along the mean field should then be set by 
a CB-style argument---a balance between the Alfv\'enic propagation time along the 
field and some typical perpendicular circulation time [see, e.g., 
\exref{eq:chain_CB}]. This has not, to my 
knowledge, been carefully checked (except, in a different set up, by \citealt{zhou20}).
Note that most of the 3D simulations cited here were outside the RMHD regime
of strong guide field---see footnote~\ref{fn:3Dsheet_sims}.} 
and even some experimental undertakings \citep{jaraalmonte16,hare17,hare17pop,hare18,peterson19}. 
Perhaps this was because plasmoids had to wait for their 
moment in the sun until they could be properly resolved numerically---and that 
required fairly large simulations. 
Indeed, for an SP sheet to start spawning plasmoids, a sizeable Lundquist number 
is needed: asking for 
$\din/\delta$ to be a reasonably small number, say, at least $1/3$, 
\exref{eq:loureiro} gives us 
\beq
\tS_\xi \gtrsim \tS_{\xi,\mathrm{c}}^\mathrm{(plasmoid)}\sim10^4,
\label{eq:Sc}
\eeq  
the critical Lundquist number for the plasmoid instability 
\citep{biskamp86,loureiro05,loureiro07,samtaney09,ni10,shi18}. 

Let me pause briefly to discuss how (in)valid the direct application of \apref{app:TM} 
to an SP sheet in fact was. The good news is that, for such an equilibrium configuration, 
it is certainly true that $n=1$ in \exref{eq:Dprime_n}, because the 
scale of the equilibrium field's reversal within the sheet 
is much smaller than that of its variation outside it, and so the derivation 
in \apref{app:dprime} applies---indeed it was originally invented by 
\citet{loureiro07,loureiro13kh} for this exact problem. The bad news is that an SP sheet 
is, in fact, not a static equilibrium considered in \apref{app:TM}, but a dynamic 
one, featuring an Alfv\'enic (or, at $\Pm\gg1$, sub-Alfv\'enic) outflow~$u_y$ 
[see~\exref{eq:SP_uy}], 
which is strongly sheared transversely to the sheet and has a positive outward gradient along it. 
Thus, technically, one must re-derive the tearing mode for this new equilibrium
before jumping to conclusions. 
Doing so leads to scalings different from \exref{eq:FKR} in the FKR regime, but 
does not affect the fastest-growing mode \exref{eq:TM_max} (a nice semi-qualitative derivation 
of this result is offered by \citealt{boldyrev18}; the previous, quite sophisticated, 
if not in all cases penetrable, paper trail on tearing with flows is \citealt{bulanov78,bulanov79}, 
\citealt{paris83}, \citealt{hofman75}, \citealt{dobrowolny83}, 
\citealt{einaudi86,einaudi89}, \citealt{chen90}, \citealt{loureiro13kh}, \citealt{shi18},
\citealt{tolman18}). This gives me licence not to worry about this complication here. 

Admittedly, in an extended SP sheet, the presence of flows can also lead to a very 
fast Kelvin--Helmholtz instability that appears at the periphery of the sheet 
and is actually even faster than the plasmoid instability \citep{loureiro13kh}. 
This, however, does not change the most important conclusion from \exref{eq:loureiro}, 
and arguably the only relevant one, 
which is that the instability of an SP sheet is massively supercritical: 
at large enough $\tS_\xi$, it is nowhere near marginal stability.  
The question therefore really is whether we should expect SP sheets ever to be 
formed in natural circumstances. This brings us to our next topic. 

\subsection{Formation and Disruption of Sheets}
\label{app:CS} 

\begin{figure}
\centerline{\includegraphics[width=0.6\textwidth]{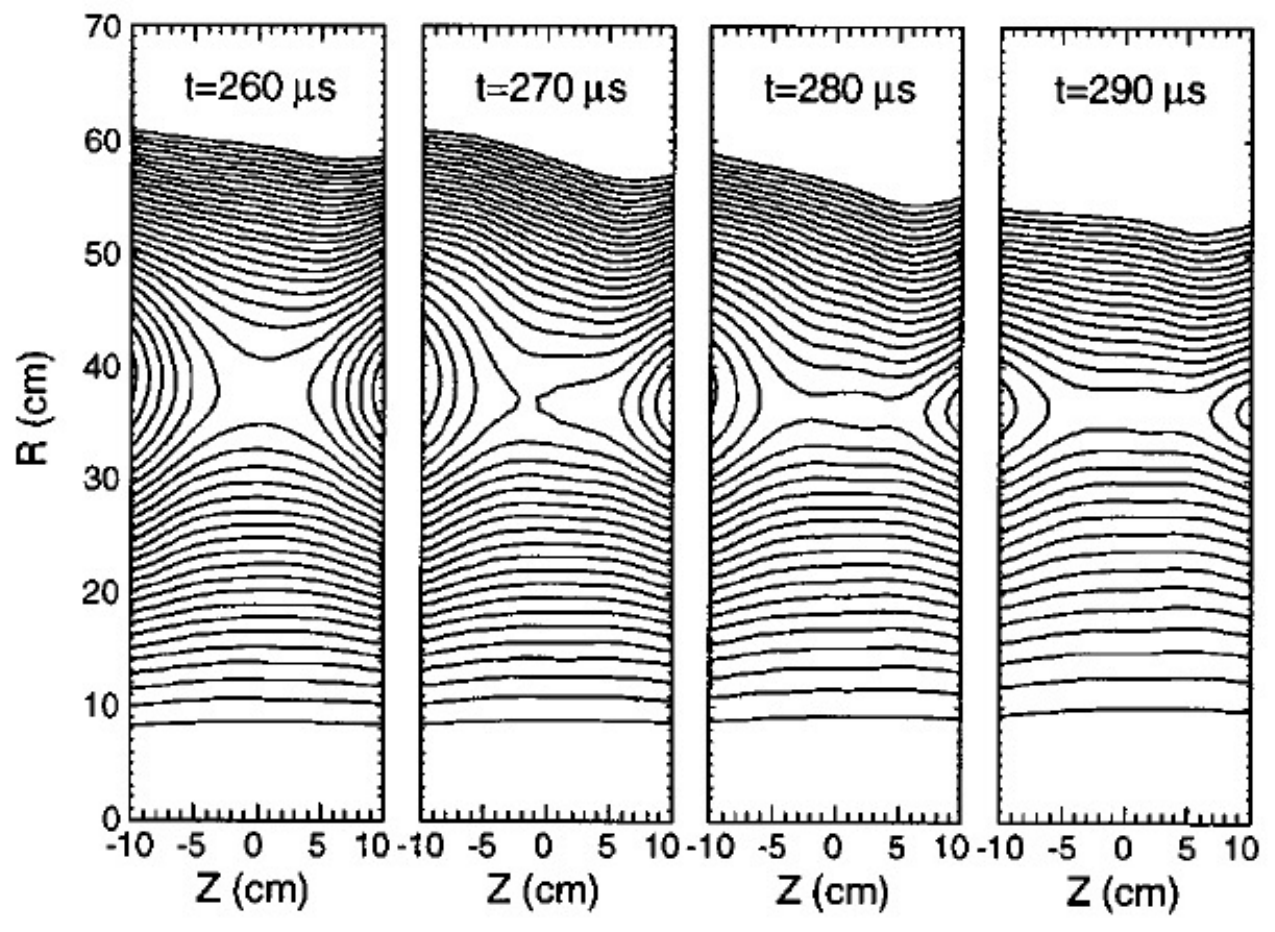}} 
\caption{Formation of a sheet from an $X$-point in the MRX experiment 
at Princeton [reprinted from \citet{yamada97} with the permission of AIP Publishing].}
\label{fig:mrx}
\end{figure}

Let us put SP sheets aside and talk more generally about MHD sheets 
of the kind envisioned in \apref{app:TM} as the background equilibrium for tearing.  
The naturally occurring tearing-unstable ideal-MHD solutions 
are, of course, not static equilibria: they arise, 
basically, because of the dynamical tendency in MHD for $X$-points to collapse 
into sheets (which I invoked in \apsref{app:what_next}), illustrated in \figref{fig:mrx}.  
An elementary example is the classic \citet{chapman63} collapsing solution 
of MHD equations: 
\beq
\Phi_0 = \Gamma(t) xy,\quad
\Psi_0 = \frac{\vAy}{2}\lt[\frac{x^2}{\lambda(t)} - \frac{y^2}{\xi(t)}\rt].
\label{eq:CKsln}
\eeq
Here $\Gamma(t)$ can be specified arbitrarily and then 
$\lambda(t)$ and $\xi(t)$ follow upon direct substitution of \exref{eq:CKsln} 
into the RMHD equations \exsdash{eq:Phi}{eq:Psi} (with $\eta=0$). 
The original \citet{chapman63} version of this was the exponential collapse:
\beq
\Gamma(t) = \Gamma_0 = \const,\quad
\lambda(t) = \lambda_0 e^{-2\Gamma_0 t},\quad
\xi(t) = \xi_0 e^{2\Gamma_0 t}.
\label{eq:CKsheet}
\eeq 
A later, perhaps more physically relevant example, due to \citet{uzdensky16}, 
is obtained by fixing the outflow velocity at the end of the sheet to be 
a constant parameter: if $u_y = \dd\Phi_0/\dd x = u_0 y/\xi$, then
\beq
\Gamma(t) = \frac{u_0}{\xi(t)},\quad 
\lambda(t) = \frac{\lambda_0\xi_0}{\xi_0 + 2u_0 t},\quad
\xi(t) = \xi_0 + 2u_0 t. 
\label{eq:ULsheet}
\eeq
In this, or any other conceivable model of sheet formation, the aspect ratio 
increases with time as the sheet's width $\lambda$ decreases and its length 
$\xi$ increases. 

The traditional thinking about sheets in MHD held that an ideal collapsing 
solution such as \exref{eq:CKsln} (or an explosively collapsing 
one obtained by \citealt{syrovatskii71} for compressible MHD) 
would culminate in a steady-state current sheet, which, from the ideal-MHD point 
of view, would be a singularity, resolved in resistive MHD by Ohmic 
diffusion, leading to an SP sheet. 
One could then discuss magnetic reconnection in such a sheet (\apref{app:SP_rec}). 
However, as we saw in \apref{app:loureiro}, this object is massively unstable
to tearing perturbations and will break up into a multitude of islands (``plasmoids''). 
\citet{uzdensky16} and \citet{pucci14} argued that, in fact, it 
would never form, because tearing perturbations growing against the background 
of a collapsing ideal-MHD solution will disrupt it before it reaches 
its steady-state, resistive SP limit. 

The detailed demonstration of this result involves realising that not only 
does the instantaneous aspect ratio of a forming sheet decide what types of 
tearing perturbations are allowed (single-island FKR modes or multi-island 
fastest-growing, ``Coppi'' modes; see \apref{app:TMgammak}), 
but that, in principle, this can change as the sheet evolves, 
that many different modes can coexist and that these perturbations will grow on different 
time scales not only linearly but also nonlinearly (the FKR modes having to go through the 
secular \citealt{rutherford73} regime, the Coppi ones not). A careful analysis of all 
this can be found in the paper by \citet{uzdensky16} (the follow-up by 
\citealt{tolman18} deals with the effect of the sheet-forming flows on the tearing mode). 
The summary that will suffice 
for my purposes here is that if the fastest-growing linear mode \exref{eq:TM_max} 
fits into the sheet, it will also be the one that first reaches the nonlinear 
regime and disrupts the formation of the sheet. Note that 
at the onset of the nonlinear regime of the tearing mode, the width of the islands 
is given by \exref{eq:w_onset}. Since $w\ll\lambda$, 
islands of this size are, in fact, short of what is needed to disrupt the sheet. 
\citet{uzdensky16} argue that the collapse of the inter-island $X$-points,  
already mooted in \apref{app:what_next}, will eventually---on the same, or faster, 
time scale as that of the growth of the mode---produce islands of size $w\sim\lambda$,
which is what they have to get to in order to be properly disruptive. 
This is a key ingredient for the picture of ``tearing-mediated turbulence'' 
advocated in \secref{sec:recturb} (but see caveats in \secref{sec:MSC_vs_BL}). 

\begin{figure}
\centerline{\includegraphics[width=\textwidth]{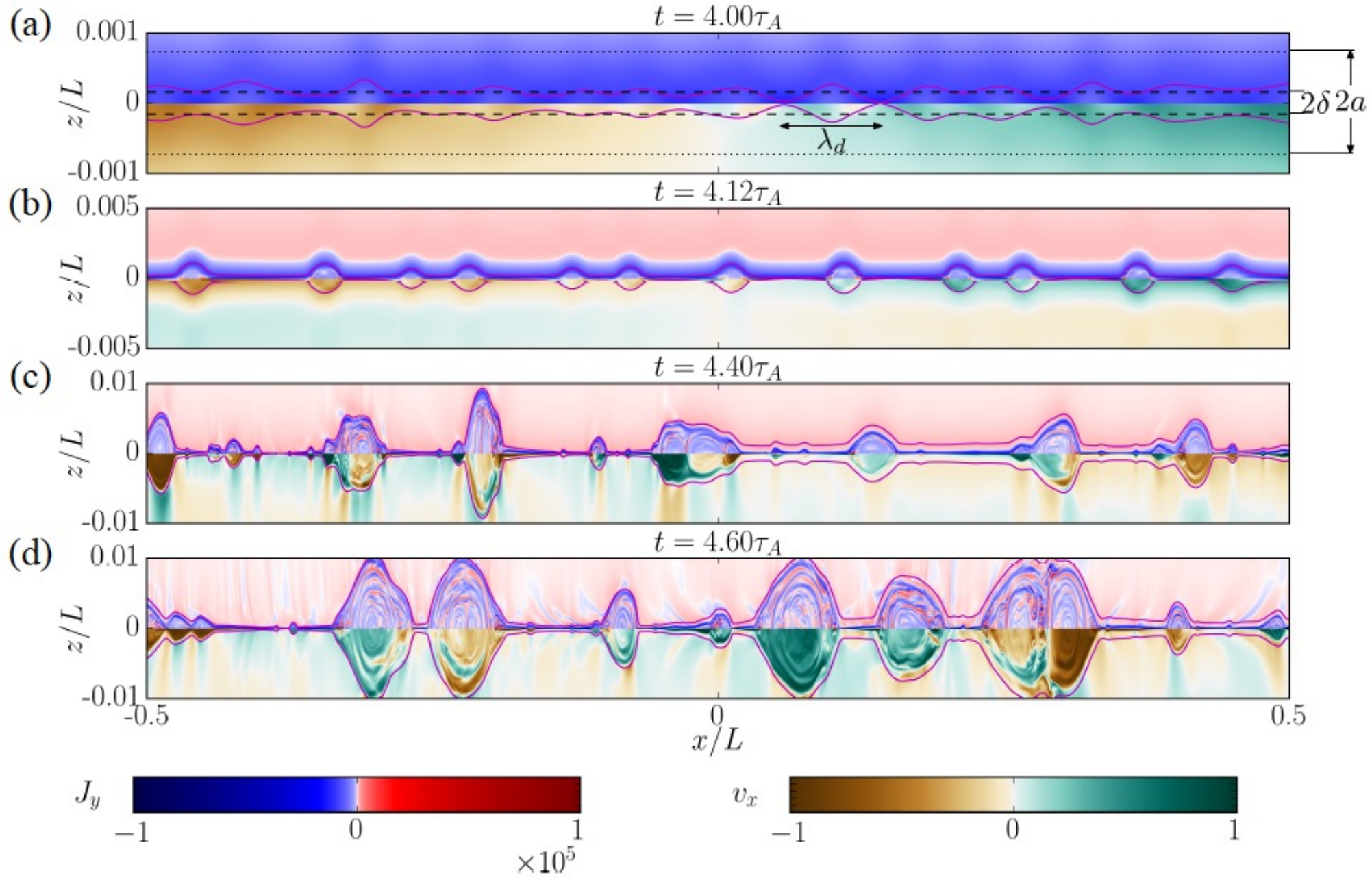}} 
\caption{This is a plot from \citet{huang17} (\copyright AAS, reproduced with permission)
illustrating the evolution of tearing 
perturbations of an evolving sheet in a 2D MHD simulation with $S_\xi\sim 10^6$ and $\Pm\ll1$. 
Their $(x,y,z)$ are my $(y,z,x)$, their $L$ is my $\xi$ (sheet length), 
their $a$ is my $\lambda$ (sheet width), their $\tau_\mathrm{A}$ 
is my $\Gamma^{-1}\sim \xi/\vA$ (characteristic time of the sheet evolution), 
their $\delta$ is my $\din$ (width of the tearing inner layer). 
The colour in the upper halves of their panels shows out-of-page current 
(colour bar ``$J_y$'') and in the lower halves the outflow velocity along 
the sheet (colour bar ``$v_x$''). The solid magenta lines 
are separatrices demarcating two ``global'' coalescing islands that they set up 
to form the sheet. The four snapshots are
(a) at the moment when the tearing mode goes 
nonlinear ($w\sim\din$; see \apref{app:TMnlin}), 
(b) a little later, showing formation of secondary sheets 
(and so collapse of inter-island $X$-points),
(c) later on, with a secondary instability 
of these sheets manifesting itself as more plasmoids appear 
(cf.\ \apref{app:multi}), 
and (d) in saturation, which for them is the period of stochastic but statistically 
steady and fast (with a rate independent of $S_\xi$) reconnection and which 
obviously also corresponds to islands reaching the width of the sheet 
and starting to form a stochastic chain, moving and coalescing 
(see \apref{app:uls}). Note that all of this evolution happens 
within one Alfv\'en time, although the initial-growth stage does need a few 
Alfv\'en times to get going.} 
\label{fig:huang}
\end{figure}

\subsubsection{``Ideal Tearing''} 
\label{app:pucci}

So what kind of sheets can form before disruption occurs? 
Namely, what aspect ratio can a sheet reach before 
the growth rate of the tearing mode triggered in the sheet becomes larger than 
the rate at which the sheet is collapsing via its ideal-MHD evolution? 
The former rate is given by \exref{eq:TM_max} and the latter is 
$\Gamma\sim\vAy/\xi$, as is illustrated by the Uzdensky--Loureiro 
solution \exref{eq:ULsheet}.\footnote{Assuming an Alfv\'enic outflow. 
This is fine even when $\Pm\gg1$ as long 
as the sheet is macroscopic, i.e., viscosity is unimportant at the scale $\lambda$. 
If instead one is considering a microscopic ``equilibrium,'' like the secondary 
$X$-points between the islands of a tearing perturbation (\apref{app:TMnlin}), 
one should use $\Gamma \sim u_y/\xi$, where $u_y$ is the visco-Alfv\'enic 
outflow: see \exref{eq:SP_uy}. The condition \exref{eq:pucci} then 
becomes $\xi/\lambda \gtrsim\tS_\lambda^{1/2} = S_\lambda^{1/2}(1+\Pm)^{-1/4}$.} 
Then
\beq
\gamma \gtrsim \Gamma 
\quad\Leftrightarrow\quad
\frac{\xi}{\lambda} \gtrsim S_\lambda^{1/2}(1+\Pm)^{1/4} 
\quad\Leftrightarrow\quad
\frac{\xi}{\lambda} \gtrsim S_\xi^{1/3}(1+\Pm)^{1/6}. 
\label{eq:pucci}
\eeq
The last expression contains the Lundquist number referred to the length $\xi$ rather 
than to the width $\lambda$ of the sheet, as customary in magnetic-reconnection 
theory (cf.~\apref{app:SP}).
Note that the assumption that it is the fastest-growing Coppi mode~\exref{eq:TM_max} that 
should be used in this estimate is confirmed {\em a posteriori} by checking that the mode 
does fit into the sheet [cf.~\exref{eq:ky_to_fit}]: for $\xi$ satisfying 
the equality in~\exref{eq:pucci},
\beq
k_*\xi \sim S_\lambda^{1/4}(1+\Pm)^{3/8} \sim S_\xi^{1/6}(1+\Pm)^{1/3}\gg1. 
\label{eq:pucci_k}
\eeq 

The scaling \exref{eq:pucci} of the aspect ratio of the sheet with $S_\xi$ was 
proposed by \citet{pucci14} to be the maximum possible attainable one before the sheet 
is destroyed by what they termed ``ideal tearing,'' i.e., by tearing modes that grow 
on the same time scale as the ideal-MHD sheet evolves (this result was 
checked numerically by \citealt{landi15} and \citealt{delzanna16}, extended
to $\Pm\gg1$ by \citealt{tenerani15visc}, and generalised by \citealt{pucci18} to the case 
of arbitrary scaling of $\Delta'$ with $k_y$ that I explained in \apref{app:dprime_n}). 
The conclusion that the sheet is 
indeed destroyed depends on the $X$-point-collapse argument described above, 
because the tearing modes by themselves do not produce islands
as wide as the sheet (see \apref{app:TMnlin}).  

The argument in \secref{sec:disruption_scale} about the disruption of MHD turbulence 
by tearing is essentially the application of the criterion \exref{eq:pucci} to 
the aligned structures of which Boldyrev's MHD turbulent cascade consists. 

Since the aspect ratio of the sheet described by \exref{eq:pucci} is smaller than that 
of the SP sheet~\exref{eq:SP_uy} ($S_\xi^{1/3}$, rather than $S_\xi^{1/2}$), 
\citet{pucci14} argued that global SP sheets could never form. 
An extensive numerical study by \citet{huang17} of the instability of 
forming current sheets 
has indeed confirmed explicitly that the plasmoid-instability scalings \exref{eq:loureiro} 
derived for an SP sheet only survive up to a certain 
critical value 
\beq
S_{\xi,\mathrm{c}}^\mathrm{(ideal)}\sim 10^5-10^6
\eeq 
[which obviously had to be bigger than the critical 
Lundquist number \exref{eq:Sc} for the plasmoid instability itself], with 
the ``ideal-tearing'' scalings \exref{eq:pucci} and \exref{eq:pucci_k} taking 
over at $S_\xi \gtrsim S_{\xi,\mathrm{c}}^\mathrm{(ideal)}$.\footnote{They also find that 
$S_{\xi,\mathrm{c}}^\mathrm{(ideal)}$ gets smaller when larger initial background noise 
is present in the system and that the onset of the tearing instability 
(and, therefore, of fast reconnection) is generally 
facilitated by such noise (the same is true for the plasmoid instability of SP 
sheets: see \citealt{loureiro09} and \citealt{sun22}; note also an earlier
paper on the same subject by \citealt{fan04}).
Their paper is written in a way that might give one the impression that 
they disagree profoundly with both \citet{uzdensky16} and \citet{pucci14}: the main point 
of disagreement is their observation that the disruption of the sheet happens when 
$\gamma$ is equal a few times $\Gamma$, rather than $\gamma/\Gamma \approx 1$ 
[see \exref{eq:pucci}], and that exactly how many times $\Gamma$ it must be 
depends on the initial noise level. In the context of the turbulence-disruption 
arguments advanced in \secref{sec:disruption}, this may be a useful practical 
caveat pointing to the value of $\lD$ [see \exref{eq:lambdaD}] 
possibly being an overestimate by a factor of order unity. However, 
all theory in this review is order-unity-inaccurate ``twiddle'' theory, so I am not 
as bothered by this complication as someone attempting a quantitative 
numerical study might be. In any event, the fact that the disruption of the 
sheet is helped by more noise is surely a good thing for the validity of $\gamma/\Gamma \sim 1$ 
as the disruption criterion in a turbulent environment, where there is noise aplenty. 
Another (related) complication that matters quantitatively but probably not qualitatively 
is the possible presence of logarithmic corrections and other subtleties 
in the tearing-instability scalings for 
time-dependent sheets \citep{comisso16b,comisso17,comisso18,tolman18,huang19}.\label{fn:huang}} 
\Figref{fig:huang}, taken from their paper, is an excellent illustration of the evolution 
of tearing perturbations and plasmoid chains. 

\subsubsection{Recursive Tearing} 
\label{app:multi}

It is not a difficult leap to realise that if a collapsing 
``global'' MHD sheet-like configuration (which, the way it was introduced 
at the beginning of \apref{app:CS}, was manifestly an $X$-point configuration) 
is unstable to tearing, the secondary $X$-points generated by this tearing 
can also be unstable to (secondary) tearing and thus might not ``complete'' 
the collapse into ``proper'' SP sheets that was posited for them 
above. This can happen if the secondary tearing has 
a shorter growth time than the primary one, which, as we are about to see, 
is always the case. This conjures up an image of recursive tearings 
piling up on top of each other {\em ad infinitum}  
or, rather, until the inter-island $X$-points have small enough Lundquist numbers to be stable. 
At that point, they can all collapse properly into mini-SP-sheets 
and we are left with a multiscale population of fully nonlinear islands, 
which can now break up the ``mother sheet'' and/or start interacting with each other
and form a stochastic chain described in~\apref{app:uls}. 
For the purposes of the discussion in the main text (\secref{sec:recturb}), 
the issue is whether we should be concerned that the outcome 
is not just a number of flux ropes of one size $\sim\lambda$, 
but a whole multiscale distribution of them. 

\begin{figure}
\centerline{\includegraphics[width=0.55\textwidth]{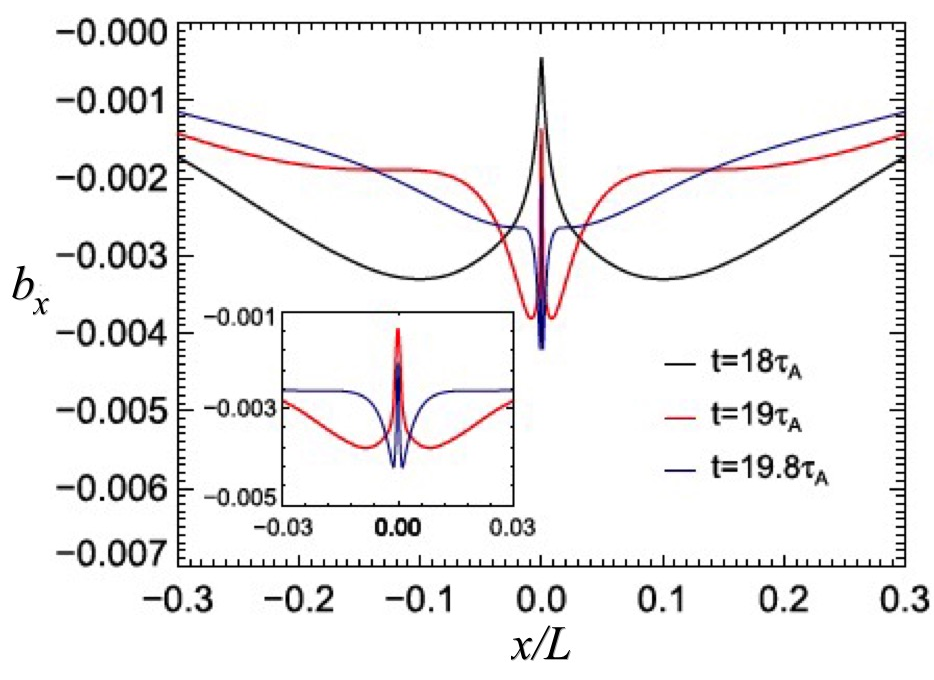}} 
\caption{A plot, adapted from \citet{tenerani15}
(\copyright AAS, reproduced with permission),
of the $b_x=-ik_y\psi(x)$ profiles (cf.\ \figref{fig:tearing}) 
for nested tearing modes: primary (black), secondary (red) and tertiary (blue).
They extracted these from a direct numerical simulation of a recursively tearing sheet.
This is a remarkably clean example of the similarity of tearing at ever smaller scales.} 
\label{fig:tenerani}
\end{figure}

The first model of recursive tearing was proposed in the seminal
paper by \citet{shibata01}, and was more recently numerically tested and amended
by \citet{tenerani15,tenerani16}
(both models are usefully reviewed by \citealt{singh19}). I shall explain
their argument first, and then propose a certain modification of it that, in my view,
makes it more plausible (and does lead to one important consequence).

Let us work on the assumption that the secondary tearing of an inter-island $X$-point
works in the same way as the primary tearing described in \apref{app:TMgammak}
(see \figref{fig:tenerani}). Therefore, we can assign our old equilibrium parameters
to the $i$-th level of tearing:
\beq
v_i \equiv \vAy,\quad
\lambda_i \equiv \lambda,\quad
\xi_i \equiv \xi.
\label{eq:eqm_i}
\eeq
The same quantities at the $(i+1)$-st level are then determined by the resulting
perturbation in a way that depends on a number of assumptions. 

First, as announced above, let us assume that the ``equilibrium''~\exref{eq:eqm_i} 
will produce a tearing perturbation with growth rate and wavenumber 
given by~\exref{eq:TM_max}: 
\beq
\gamma_i \sim \frac{u_i}{\lambda_i}\,\tS_i^{-1/2},\quad
k_{*,i}^{-1} \sim \lambda_i \tS_i^{1/4},
\label{eq:tearing_i}
\eeq
where [cf.~\exref{eq:uout}]
\beq
u_i = \frac{v_i}{\sqrt{1+\Pm}},\quad
\tS_i = \frac{S_i}{\sqrt{1+\Pm}},\quad 
S_i = \frac{v_i\lambda_i}{\eta}.
\eeq
This is an assumption, not a certainty, because 
the local ``equilibrium'' set up by the $i$-th tearing perturbation, 
which features flows as well as fields, is not {\em a priori} obliged to
be tearing unstable in exactly the same way as a very simple static equilibrium used
in \apref{app:TM}. Flows are expected to have a seriously stabilising effect~if
\beq
\frac{u_i}{\xi_i}\sim\gamma_i
\rmiff
\frac{\xi_i}{\lambda_i} \sim \tS_i^{1/2}
\label{eq:gi_marginal}
\eeq
\citep{bulanov78,bulanov79,biskamp86,shi18,tolman18}.
This is also the ``ideal-tearing'' threshold \exref{eq:pucci}, at which
the tearing only just outperforms the $X$-point collapse 
(except, for $\Pm\gg1$, the Alfv\'enic outflow here is tempered by viscosity, because
secondary-sheet dynamics, as well as tearing, happen at scales where viscosity matters). 
Our second assumption, therefore, 
following both \citet{shibata01} and \citet{tenerani15,tenerani16},
will be that \exref{eq:gi_marginal} holds at every level, i.e., tearing at every level
is ideal and marginal. It makes sense that it should not be much faster than
that, otherwise the $i$-th-level ``equilibrium'' would have gone unstable
earlier in its evolution. 

The next steps are less obvious. We need to determine three quantities at the $(i+1)$-st
level of tearing: $v_{i+1}$, $\lambda_{i+1}$ and $\xi_{i+1}$. Imposing \exref{eq:gi_marginal}
at every level reduces this to two, so we need two further assumptions.
\citet{shibata01} propose, first, that the length of the $(i+1)$-st sheet
is the wavelength of the $i$-th tearing mode~\exref{eq:tearing_i}, viz.,
\beq
\xi_{i+1} \sim k_{*,i}^{-1} \sim \lambda_i\tS_i^{1/4},
\label{eq:xi_is_k} 
\eeq
and, secondly, that the $X$-point collapse proceeds far enough for the reconnecting field
to be the same at all levels of tearing:
\beq
v_i \sim v_0. 
\label{eq:vi_same}
\eeq
Both of these assumptions can be doubted and revised, but, before doing this, 
let us see what they imply.

Using \exref{eq:vi_same}, \exref{eq:gi_marginal}, and \exref{eq:xi_is_k}, in that order, 
we get
\beq
\frac{\tS_{i+1}}{\tS_i} = \frac{v_{i+1}\lambda_{i+1}}{v_i\lambda_i}
\sim \frac{\lambda_{i+1}}{\lambda_i}
\sim \frac{\xi_{i+1}\tS_{i+1}^{-1/2}}{\lambda_i}
\sim \tS_i^{1/4}\tS_{i+1}^{-1/2}
\hence
\tS_{i+1}\sim \tS_i^{5/6}.
\label{eq:S_i_ST}
\eeq
This implies, {\em inter alia}, that the $(i+1)$-st tearing starts only when
islands grow a bit larger than the width $w\sim\din$ 
that they achieved when the $i$-th tearing went nonlinear:
using \exref{eq:w_onset} with $n=1$, we can estimate 
\beq
\frac{\lambda_{i+1}}{\delta_{\mathrm{in},i}} \sim \frac{\lambda_{i+1}}{\lambda_i\tS_i^{-1/4}}
\sim\tS_i^{1/12}. 
\label{eq:price_ST}
\eeq
It is not obvious that it should be so, but it is reassuring that the secondary
instability happens {\em after}
the onset of the nonlinear stage, and not before (the conviction that, in fact,
the secondary tearing must start right at the onset of nonlinearity led
\citealt{tenerani15,tenerani16} to their version of recursive tearing, which I will
discuss at the and of this section).

The assumption~\exref{eq:vi_same} is perhaps questionable. 
I do not see why the local $X$-point ``equilibria'' produced in the nonlinear stage 
of the primary tearing should stay stable until $X$-point collapse makes 
$v_i\sim v_0$ (of course, knowing definitely whether they do so requires
a careful quantitative theory of the secondary tearing, currently unavailable).
It seems more natural to assume that the effective ``equilibrium'' upstream field for
the secondary tearing ($v_{i+1}$) is reduced from that for the primary one ($v_i$)
by the fact that the islands generated by the latter are not as wide ($\lambda_{i+1}$)
as the primary sheet ($\lambda_i$). Then 
\beq
v_{i+1}\sim v_i\,\frac{\lambda_{i+1}}{\lambda_i}.
\label{eq:embedding}
\eeq
This is derived in the same way as $\db_y/\vAy\sim w/\lambda$ was from
\exref{eq:w_bx} and \exref{eq:w_by}, but now $\db_y\to v_{i+1}$, $\vAy\to v_i$,
$w\to \lambda_{i+1}$ and $\lambda\to \lambda_i$. 
This idea, sometimes called the ``embedding effect'', goes back to 
\citet{cassak09a}, whose simulations appeared to support the notion that 
secondary tearing would get going in these circumstances
(a more recent paper by \citealt{delsarto17} took the same view). 

Adopting \exref{eq:embedding} instead of \exref{eq:vi_same} in the \citet{shibata01}
scheme for recursive tearing amounts to replacing~\exref{eq:S_i_ST}~with   
\beq
\frac{\tS_{i+1}}{\tS_i} \sim \frac{v_{i+1}\lambda_{i+1}}{v_i\lambda_i} 
\sim \lt(\frac{\lambda_{i+1}}{\lambda_i}\rt)^2
\sim \tS_i^{1/2}\tS_{i+1}^{-1}
\hence
\tS_{i+1}\sim \tS_i^{3/4}.
\label{eq:S_i_STplus}
\eeq
This new scheme does not lose any of the properties of the old one that made
the latter plausible:
\vskip2mm 
(i) since \exref{eq:price_ST} becomes 
\beq
\frac{\lambda_{i+1}}{\delta_{\mathrm{in},i}} \sim \frac{\lambda_{i+1}}{\lambda_i\tS_i^{-1/4}}
\sim\tS_i^{1/8}, 
\label{eq:price_STplus}
\eeq
the $(i+1)$-st tearing still sets on during the nonlinear stage of the $i$-th one;
\vskip2mm
(ii) the Lundquist number gets smaller at every level:
\beq
\tS_i \sim \tS_0^{(3/4)^i} \to 1 \rmas i\to\infty;
\label{eq:S_i}
\eeq
\vskip2mm
(iii) the tearing growth rate nevertheless increases:
\beq
\gamma_i \sim \frac{u_0}{\lambda_0}\tS_0^{-(3/4)^i/2} \to \frac{u_0}{\lambda_0}
\rmas i\to\infty,
\eeq
so the perturbations always grow faster than the underlying ``equilibria'' evolve
($\gamma_{i+1}\gg\gamma_i$). 
Thus, the entire hierarchy of islands is created very quickly, on the time scale 
of (a few times) $\gamma_0^{-1}\sim \xi_0/u_0$. 
If there is some critical Lundquist number $\tSc$ required for 
these tearing modes to be unstable, \exref{eq:S_i} allows us to work out 
the maximum number of times that the recursive tearing will be 
iterated before $X$-points can collapse unimpeded into proper, stable,  
reconnecting current sheets: 
\beq
i_\mathrm{max}\sim \ln \frac{\ln\tS_0}{\ln\tSc}.
\eeq
It is obvious that in practice this will not be a large number at all.

The embedding assumption~\exref{eq:embedding} enables me to argue that,
while islands at all scales below $\lambda_0$ are produced, they do not contain much energy. 
Indeed, the effective energy density in the $i$-th-level islands~is
\beq
v_{i,\mathrm{eff}}^2 \sim v_i^2\frac{\lambda_i}{\lambda_0} 
\sim v_0^2\lt(\frac{\lambda_i}{\lambda_0}\rt)^3,
\label{eq:vi_eff}
\eeq 
where the extra factor of $\lambda/\lambda_0$ represents the fraction of the 
volume that these islands fill, assuming that they are arranged neatly in a row. 
When translated into a spectral slope, this gives $\kperp^{-4}$,
which is easily dominated by the $\kperp^{-11/5}$ spectrum of the tearing-mediated 
turbulence derived in~\secref{sec:spectrum_rec}. 
If this is true, we should be allowed to dismiss recursive tearing as a side show 
in the context of the tearing-mediated cascade. 

Let me conclude this section with some further nuances and caveats. After the entire
tearing hierarchy has formed, the $X$-points fully collapse
and the sheet is perhaps broken up by its spawn of islands, of which only the largest
ones are energetically of any consequence [see \exref{eq:vi_eff}]. 
Were it instead to persist for some time
(which is not not impossible: see \secsand{sec:MSC_vs_BL}{sec:onset}), everything would change 
in the course of the subsequent dynamics of its plasmoid (island) population:  
plasmoid shapes, their number (they travel along the sheet, coalesce, and eventually get 
ejected from the sheet), field amplitudes in them (reconnection continues via elementary 
inter-plasmoid current sheets that are short enough to be stable).
Such stochastic plasmoid chains have been studied numerically by many people 
(see references in \apref{app:loureiro}).
Their statistical steady state is, I believe, correctly described by 
the theoretical model of \citet{uzdensky10}, reviewed in \apref{app:uls}.
In this context, an important caveat to the theory of recursive tearing discussed above
is that imagining that all those nonlinear processes happen {\em after} recursive tearing
has run its course was surely a gross idealisation. Namely, 
I assumed implicitly that secondary tearing would be 
the first instability to kick in once the primary tearing mode became nonlinear---and thus 
ignored, e.g., the possibility, raised some time ago by \citet{malara92}, that the islands 
produced by tearing might start coalescing before secondary tearing destabilised 
the inter-island $X$-points. \citet{tenerani15,tenerani16} claim to see this in
their simulations. 

The alternative version of recursive tearing proposed by \citet{tenerani15,tenerani16} 
was, it seems, inspired by the notion that the nonlinear rearrangements of the island population
that occurred in a plasmoid chain gave one licence to 
discard the \citet{shibata01} assumption~\exref{eq:xi_is_k} that the number of islands
at each level was decided by the wavenumber of the fastest-growing tearing mode at that level. 
Instead they propose (and claim confirmed in their simulations) 
that the width of the $(i+1)$-st sheet is, in fact, 
the island width $w\sim\din$ of the $i$-th tearing mode right 
at the onset of nonlinearity, viz., using \exref{eq:w_onset} with $n=1$, 
\beq
\lambda_{i+1} \sim \delta_{\mathrm{in},i} \sim \lambda_i\tS_i^{-1/4}.
\label{eq:lambda_is_delta} 
\eeq 
Then, instead of~\exref{eq:S_i_ST},
\beq
\frac{\tS_{i+1}}{\tS_i} \sim \frac{\lambda_{i+1}}{\lambda_i} 
\sim \tS_i^{-1/4} 
\hence
\tS_{i+1}\sim \tS_i^{3/4},
\label{eq:S_i_TV}
\eeq
where the assumption \exref{eq:vi_same} that the upstream field is the equilibrium
field has been retained. The scaling of the Lundquist number is the same as \exref{eq:S_i},
but for a different reason. If \exref{eq:vi_same} were to be replaced by the embedding
assumption~\exref{eq:embedding}, one gets instead 
\beq
\frac{\tS_{i+1}}{\tS_i} \sim \frac{v_{i+1}\lambda_{i+1}}{v_i\lambda_i} 
\sim \lt(\frac{\lambda_{i+1}}{\lambda_i}\rt)^2
\sim \tS_i^{-1/2} 
\hence
\tS_{i+1}\sim \tS_i^{1/2}.
\label{eq:S_i_TVplus}
\eeq
Under this scheme, one does not need to know $\xi_i$, but if one wants to know it, 
it is determined from the ideal-tearing condition~\exref{eq:gi_marginal}
rather than from the wavelength of the tearing mode~\exref{eq:xi_is_k}
(otherwise one gets a tearing growth rate $\gamma_i$ that far exceeds the marginal
level $\sim u_i/\xi_i$). One somewhat awkward implication  
is that the $i$-th tearing is expected to find some way of producing more numerous, 
smaller islands than allowed by the wavenumber $k_*$ of its fastest-growing mode:
from \exref{eq:gi_marginal}, \exref{eq:tearing_i}, and \exref{eq:S_i_TV}, 
\beq
\xi_{i+1}k_{*,i}\sim \tS_{i+1}^{1/2}\tS_i^{-1/4}\frac{\lambda_{i+1}}{\lambda_i}
\sim \tS_i^{-1/8} \ll 1, 
\label{eq:price_TV}
\eeq
or $\xi_{i+1}k_{*,i}\sim \tS_i^{-1/4}$ if \exref{eq:S_i_TVplus} is used instead of \exref{eq:S_i_TV}.

Thus, four recursive-tearing scenarios are available: \exref{eq:S_i_ST},
\exref{eq:S_i_STplus}, \exref{eq:S_i_TV}, and \exref{eq:S_i_TVplus}. Which of these
you believe depends on which of the plausible assumptions discussed above you find
most plausible---or best verified numerically (a hard task). 
Do the differences between them really matter? 
Certainly not for the qualitative picture of recursive 
tearing quickly leading to the formation of a fully nonlinear 
plasmoid chain. Once this has happened, i.e., once all $X$-points have fully 
collapsed, \exref{eq:vi_same}~will certainly be true at all levels 
\citep{uzdensky10,loureiro12}. 
However, if it were also true and, consequently, \exref{eq:embedding} untrue, 
during the initial recursive tearing, as  
\citet{shibata01} and \citet{tenerani15,tenerani16} would have it, 
then I would not be able to wave away the 
role of the secondary islands in the disruption process, as I did in 
\secref{sec:MSC_vs_BL}. Indeed, modifying \exref{eq:vi_eff} to have $v_i\sim v_0$,
one~gets 
\beq
v_{i,\mathrm{eff}}^2 \sim v_0^2\frac{\lambda_i}{\lambda_0}, 
\label{eq:vi_eff_v0}
\eeq
which corresponds to a spectrum of $\kperp^{-2}$ (cf.~\apref{app:chain_spectrum}). 
This could swamp the $\kperp^{-11/5}$ spectrum of the tearing-mediated turbulence 
(\secref{sec:spectrum_rec}) unless mitigated by some volume-filling effects
\citep[as, e.g., in][]{tenerani20}. 
In any case, there would then be a legitimate question of how  
all these islands might modify, or even completely determine, 
the tearing-mediated-range statistics. These issues are discussed in~\secref{sec:turb_sheet} 
and \apref{app:rec_driven}.

\subsection{Fast MHD Reconnection}
\label{app:uls}

I have referred several times already to a fully nonlinear plasmoid chain 
being a possible end result of recursive tearing (\apref{app:multi}) 
and making reconnection fast (\secref{sec:multilayer}). Let me reproduce 
here, in broad brush, the theory of this regime by \citet{uzdensky10}. 

Once all the $X$-points at all levels of recursive tearing have collapsed, 
the current sheet becomes a chain of plasmoids of different sizes connected 
by the longest SP sheets that can remain stable, i.e., ones whose ``critical'' 
length and width~are
\beq
\ellc \sim \frac{\tSc\eta}{u_y},\qquad
\delc \sim \ellc\,\tSc^{-1/2},
\label{eq:ellc}
\eeq
where $u_y\sim\vAy/\sqrt{1+\Pm}$ is the outflow velocity [see~\exref{eq:SP_uy}] 
and $\tSc$ is the critical Lundquist number \exref{eq:Sc} for the plasmoid instability. 
The inter-plasmoid sheets cannot be any longer than $\ellc$ because the moment they get 
stretched longer they go unstable and break up into more plasmoids. 
Thus, the number of plasmoids typically 
found in a sheet of length $\ell$, in steady state, is just $N\sim\ell/\ellc \sim \tS_\ell/\tSc$. 
These plasmoids are all of different sizes, having been generated at various levels of 
recursive tearing or as a result of coalescence of earlier-generation plasmoids. 
One can think of them as belonging to many hierarchical levels, with plasmoids 
of the $n$-th level living in ``local'' sheets bounded by pairs of $(n-1)$-st-level 
plasmoids. At every level, they are all moving along their local sheet with 
a mean (visco-)Alfv\'enic outflow $u_y$, the same at every level, 
eventually getting ejected into (coalesced with) the previous-level plasmoids. 

\begin{figure}
\centerline{\includegraphics[width=0.55\textwidth]{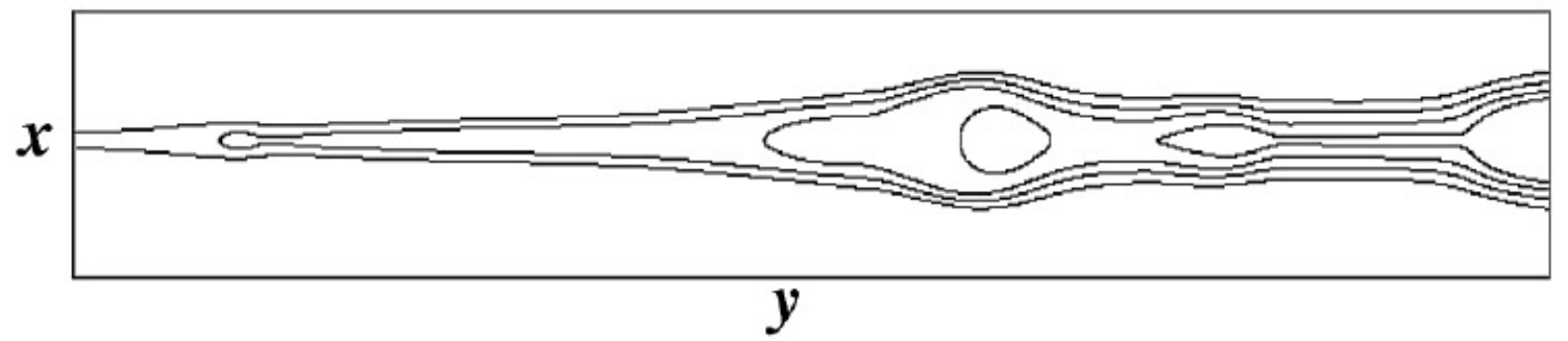}} 
\vskip2mm
\caption{Contour plot of the magnetic flux function illustrating the open flux. 
This is taken from a section of a 2D MHD simulation of a plasmoid chain; 
the centre of the sheet is somewhere far away on the left.
[Reprinted with permission from \citet{uzdensky10}, copyright (2010) by the American
Physical Society.]}
\label{fig:uls}
\end{figure}

It is surprisingly easy to argue that reconnection in such 
a system (illustrated by the lowest panel of \figref{fig:huang}) is fast.    
First notice that the plasmoids travelling along the sheet and eventually 
ejected from it would carry no reconnected flux (no $\db_x$) if they only contained  
closed field lines. However, since the upstream (reconnecting) field 
$\vAy$ decreases gently from the centre of the sheet ($y=0$) outwards along $y$, 
the reconnection on the smaller-$|y|$ side of each plasmoid is slightly 
faster than on the larger-$|y|$ side of~it. Therefore, each plasmoid carries some 
open flux (\figref{fig:uls}) and it is the ejection of this open flux 
that contributes to the overall reconnection rate.      
At every level $n$ in the plasmoid hierarchy, reconnection in a sheet containing 
$n$-th-and-higher-level plasmoids and bounded by two $(n-1)$-st-level ones 
adds to the open flux enveloping the $(n-1)$-st-level plasmoid on the larger-$|y|$ 
side and subtracts from the one on the smaller-$|y|$ side. 
The overall reconnected flux is the sum over these contributions, all of which 
cancel each other except for the one from the centre of the sheet. 
Thus, the overall reconnection rate is just the reconnection rate in the central 
elementary sheet, given by \exref{eq:rec_rate} with the critical 
Lundquist number~$\tSc$: 
\beq
\frac{\dd\Psi}{\dd t} \sim \tSc^{-1/2}u_y\vAy \sim 10^{-2}(1+\Pm)^{-1/2}\vAy^2,
\label{eq:rec_rate_fast}
\eeq
independent of the actual Lundquist number $\tS_\ell$ and the same at every level 
in the hierarchy. 

This result can be rederived (or re-interpreted) 
as a modification of the SP reconnection, proposed by \citet{shibata01}, 
in which the effective width $\deleff^{(n-1)}$ of the sheet (whose length is $\ell^{(n-1)}$)
connecting the $(n-1)$-st-level 
plasmoids, for the purposes of mass (and with it, flux) ejection, is the width of the 
largest plasmoids in that sheet, which are the $n$-th-level plasmoids. 
Then the reconnection rate in such a sheet, i.e., the rate of growth of the flux 
$\Psi^{(n-1)}$ in the $(n-1)$-st-level plasmoids,~is  
\beq
\frac{\dd\Psi^{(n-1)}}{\dd t} \sim u_x^{(n-1)}\vAy 
\sim\frac{\deleff^{(n-1)}}{\ell^{(n-1)}}\,u_y\vAy,\qquad
\deleff^{(n-1)} \sim w^{(n)},
\label{eq:rec_rate_eff}
\eeq
where the inflow velocity $u_x^{(n-1)}$ has been calculated from mass conservation, 
as in \exref{eq:flow_bal}, and the (visco-)Alfv\'enic outflow $u_y$ is 
the same at every level of the hierarchy, because the inter-plasmoid $X$-points 
are all fully collapsed, so the upstream field is $\vAy$ at every level. 
The $n$-th-level plasmoids' width is then found by letting the 
perturbed field inside them be comparable to that upstream field: 
\beq
\db_y^{(n)} \sim \frac{\Psi^{(n)}}{w^{(n)}} \sim \vAy
\hence 
w^{(n)} \sim \frac{\Psi^{(n)}}{\vAy}.
\label{eq:width_est}
\eeq
Finally, the flux typically contained inside the $n$-th-level plasmoid can be estimated 
as the reconnection rate at that level times the time that an $n$-th-level 
plasmoid will take to travel out of the $(n-1)$-st-level sheet:
\beq
\Psi^{(n)}\sim \frac{\ell^{(n-1)}}{u_y}\frac{\dd\Psi^{(n)}}{\dd t}. 
\label{eq:flux_est}
\eeq  
Combining \exsdash{eq:rec_rate_eff}{eq:flux_est}, we get 
\beq
\frac{\dd\Psi^{(n-1)}}{\dd t} \sim \frac{\dd\Psi^{(n)}}{\dd t},
\label{eq:rec_rate_inv}
\eeq 
so the reconnection rate is the same at every level and thus equal 
to the reconnection rate \exref{eq:rec_rate_fast} at $n\to\infty$, i.e., 
in the most elementary sheet, q.e.d. 

Thus, the basic reason for reconnection becoming fast in this way is that 
plasmoids make the SP sheet effectively fatter, relieving the severe constraint 
that pumping mass and flux through a narrow funnel would otherwise impose. 
The only remaining constraint is the need to get the SP sheet to be at least as long 
as $\ellc$ in order for it to be able to break up into plasmoids. 

\begin{figure}
\begin{center}
\begin{tabular}{cc}
\parbox{0.47\textwidth}{
\includegraphics[width=0.47\textwidth]{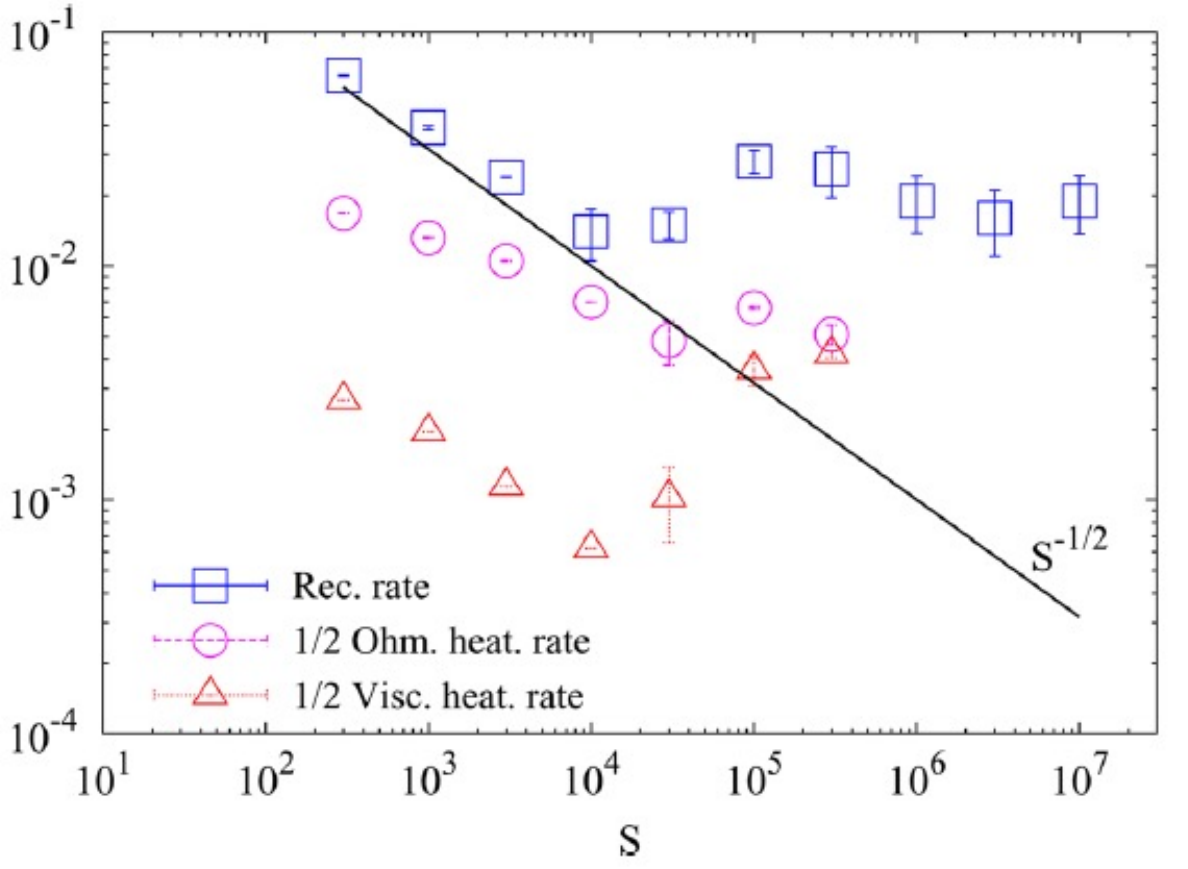}} &
\parbox{0.50\textwidth}{
\includegraphics[width=0.50\textwidth]{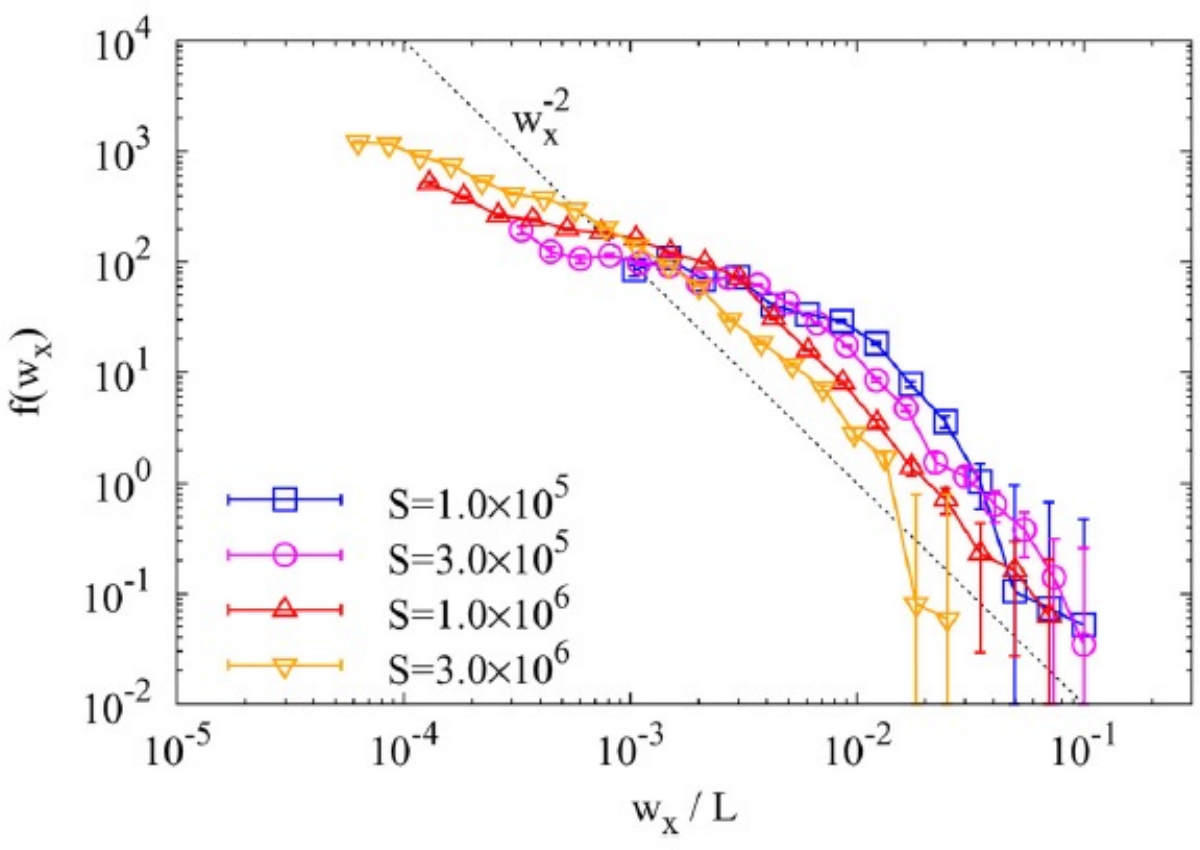}} \\\\
(a) & (b)
\end{tabular}
\end{center}
\caption{(a) Reconnection rate (blue squares), 
normalised, in my notation, to $u_y\vAy$, in 2D MHD $\Pm=1$ simulations by \citet{loureiro12}. 
Transition at $S_\ell\sim10^4$ from the SP scaling to the fast-reconnection regime 
\exref{eq:rec_rate_fast} is manifest. (b) Plasmoid-width distribution function 
in the same simulations, from the same paper, confirming the scaling 
predicted by \citet{uzdensky10} (see \apref{app:plasmoid_pdf}).
[Reprinted from \citet{loureiro12} with the permission of AIP Publishing.]} 
\label{fig:loureiro_fast}
\end{figure}

The fact that SP reconnection transitions to a fast, plasmoid-dominated 
regime at $S_\ell\gtrsim\Sc\sim 10^4$, with the reconnection rate set by $\Sc$,  
was confirmed numerically by 
\citet{bhatta09}, \citet[][see \figref{fig:loureiro_fast}a]{loureiro12} 
and in numerous subsequent simulations of bespoke reconnecting configurations 
(many of them cited in \apref{app:loureiro}). Whether this kind of fast 
reconnection and statistically steady plasmoid chains occur as local features 
of MHD turbulence is a tricky question: see \secsand{sec:MSC_vs_BL}{sec:onset}. 
Note that, in \apref{app:stoch_rec}, 
I will discuss another, quite different, sense in which reconnection 
is sometimes described as fast when it occurs in a turbulent environment.

\subsubsection{Plasmoid Flux and Width Distribution}
\label{app:plasmoid_pdf}

The relation \exref{eq:flux_est}, combined with the constant 
reconnection rate~\exref{eq:rec_rate_fast}, allows one to determine the 
distribution function of the plasmoid fluxes---not necessary for the above 
argument, but a nice, testable result, which will prove useful in what follows. 
The number of plasmoids with $\Psi>\Psi^{(n)}$, or, equivalently, $w > w^{(n)}$, 
in the sheet of overall length $\ell$~is 
\beq
N^{(n)}\sim \frac{\ell}{\ell^{(n-1)}} 
\sim \frac{\tSc^{-1/2}\vAy\ell}{\Psi^{(n)}}
\sim \frac{\tSc^{-1/2}\ell}{w^{(n)}}, 
\label{eq:plasmoid_pdf}
\eeq
the last expression following from~\exref{eq:width_est}. Therefore, the 
plasmoid-flux distribution function is $f(\Psi)\propto\Psi^{-2}$ and 
the plasmoid-width distribution function is $f(w)\propto w^{-2}$. These scalings 
are indeed corroborated numerically \citep[][see \figref{fig:loureiro_fast}b]{loureiro12}.

\subsubsection{Spectrum of Plasmoid Chain}
\label{app:chain_spectrum}

It is instructive to work out the spectrum of the plasmoid chain 
imagined by \citet{uzdensky10}. This chain is a multi-scale structure, but 
not one that is naturally described as a local constant-flux cascade: rather, 
energy is brought into it by the incoming (reconnecting) field $\vAy$ and 
deposited by the reconnection processes (which include coalescence) 
into plasmoids at all levels. The field inside these plasmoids is always~$\vAy$,
independently of their size [see~\exref{eq:width_est}]. 
However, the plasmoids are arranged 
in a row along the chain, rather than randomly in space, so the smaller 
ones fill less space than the larger ones and thus the effective energy 
density associated with them is smaller, viz., for $n$-th-level plasmoids, 
\beq
\bl[\db_\mathrm{eff}^{(n)}\br]^2 \sim \vAy^2\frac{w^{(n)}}{w^{(1)}},
\label{eq:db_eff_chain}
\eeq
where $w^{(1)}$ is the width of the largest plasmoids and, therefore, 
of the chain.\footnote{This is exactly the same argument as I used in 
obtaining~\exref{eq:vi_eff} and~\exref{eq:vi_eff_v0}.} 
Note that the size of the plasmoids in the $y$ direction (along the chain)
does not matter here because the smaller plasmoids are more numerous than the 
larger ones by the exact same factor by which they are shorter in length 
($\sim\ell/\ell^{(n)}$). Now let $\db_\mathrm{eff}^{(n)}=\db_{\mathrm{eff},\lambda}$, 
$\lambda = w^{(n)}$. By \exref{eq:width_est}, \exref{eq:flux_est}, 
\exref{eq:rec_rate_inv} and \exref{eq:rec_rate_fast}, 
\beq
w^{(n)} \sim \tSc^{-1/2}\ell^{(n-1)}, 
\eeq 
so we have $w^{(1)}\sim\tSc^{-1/2}\ell$. From \exref{eq:db_eff_chain}, therefore, 
\beq
\db_{\mathrm{eff},\lambda} \sim \vAy\tSc^{1/4}\lt(\frac{\lambda}{\ell}\rt)^{1/2}
\hence
E(k_x) \sim \vAy^2\tSc^{1/2}\ell^{-1}k_x^{-2}. 
\label{eq:chain_spectrum}
\eeq
I have used $k_x$, rather than $\kperp$, because the direction 
of maximum variation here is very obviously $x$, transverse to the sheet;
it is, of course, true that~$\kperp\sim k_x$.  

The total energy flux into the chain, per unit volume, is 
\beq
\eps \sim \frac{\vAy^2 u_x\ell}{\ell w^{(1)}} 
\sim \frac{\vAy^2 u_y}{\ell}. 
\label{eq:chain_eps}
\eeq 
It is, therefore, possible to recover \exref{eq:chain_spectrum} 
formally from a constant-flux argument in which the nonlinear time 
is the life time (ejection time) of an $n$-th-level plasmoid \citep{loureiro16unpub}: 
\beq
\frac{[\db_\mathrm{eff}^{(n)}\br]^2}{\tnl^{(n)}} \sim \eps,\quad 
\tnl^{(n)}\sim \frac{\ell^{(n-1)}}{u_y}\sim\frac{\tSc^{1/2} w^{(n)}}{u_y}
\hence 
\db_{\mathrm{eff},\lambda} \sim 
\lt(\frac{\eps}{u_y}\rt)^{1/2}\tSc^{1/4}\lambda^{1/2},
\label{eq:chain_spectrum_loureiro}
\eeq
which is the same as \exref{eq:chain_spectrum}, by way of~\exref{eq:chain_eps}.
Yet another way to derive the same result is via the plasmoid-width distribution 
function~\exref{eq:plasmoid_pdf} \citep{barta12}: 
the energy in the plasmoids of width $\lambda\sim k_x^{-1}$~is, per unit volume, 
\beq
E(k_x)\rmd k_x 
\sim \frac{\vAy^2 \tSc^{1/2}\lambda^2\rmd N(\lambda)}{\tSc^{-1/2}\ell^2} 
\sim \vAy^2\tSc^{1/2}\ell^{-1}\rmd\lambda
\sim \vAy^2\tSc^{1/2}\ell^{-1}k_x^{-2}\rmd k_x,
\label{eq:chain_spectrum_barta}
\eeq
whence follows the spectrum~\exref{eq:chain_spectrum}.

All of this works on the assumption that parallel (to the mean field) 
dynamics do not upset things (plasmoid-width distribution etc.) in a major way. 
By the usual CB argument, plasmoids, 
which are flux ropes in 3D, cannot extend much farther along the mean 
field than an Alfv\'en wave can travel in some characteristic nonlinear 
time associated with the plasmoid. The most obvious estimate of this 
time is one in~\exref{eq:chain_spectrum_loureiro}, whence~\citep{loureiro16unpub} 
\beq
\frac{\lpar^{(n)}}{\vA}\sim\tnl^{(n)} \sim \frac{\tSc^{1/2}w^{(n)}}{u_y}
\hence
\lpar \sim \frac{\vA}{u_y}\,\tSc^{1/2}\lambda
\hence 
E(\kpar) \sim \frac{\vAy^2 u_y}{\vA\ell}\,\kpar^{-2}.  
\label{eq:chain_CB}
\eeq 
Thus, the chain's parallel spectrum is the same as the perpendicular one---this appears 
to be what \citet{huang16} report for their 3D turbulent plasmoid chain. 
The ``fluctuation-direction'' spectrum (cf.~\secref{sec:3D})
is the same again, because, clearly,  
\beq
\xi\sim\tSc^{1/2}\lambda
\hence
E(k_y) \sim \vAy^2\ell^{-1} k_y^{-2}. 
\eeq 
The ``turbulence'' of plasmoids in a plasmoid chain has a fixed, scale-independent 
alignment angle equal to the reconnection rate,~$\sim\tSc^{-1/2}$. 

A sceptical reader might observe that a $k_x^{-2}$ spectrum for a plasmoid chain 
is in fact no big revelation because plasmoids are connected by elementary 
sheets \exref{eq:ellc}, which, being step-like ``discontinuities'' of width~$\delc$ 
in~$x$, already on their own should give rise to a $k_x^{-2}$ spectrum for 
all $k_x\lesssim\delc^{-1}$ 
(the same argument already appeared in \secsand{sec:new_res_theory}{sec:decay_spectra}). 
This is true, and so the point of the above calculation is that plasmoids' 
energy distribution does not swamp the $k_x^{-2}$ scaling of the spectrum; 
also, the same scaling in $k_y$ and $\kpar$ is a property of the plasmoids, 
not of the inter-plasmoid sheets. 

\subsubsection{Reconnection-Driven Turbulence}
\label{app:rec_driven}

\begin{figure}
\centerline{\includegraphics[width=0.99\textwidth]{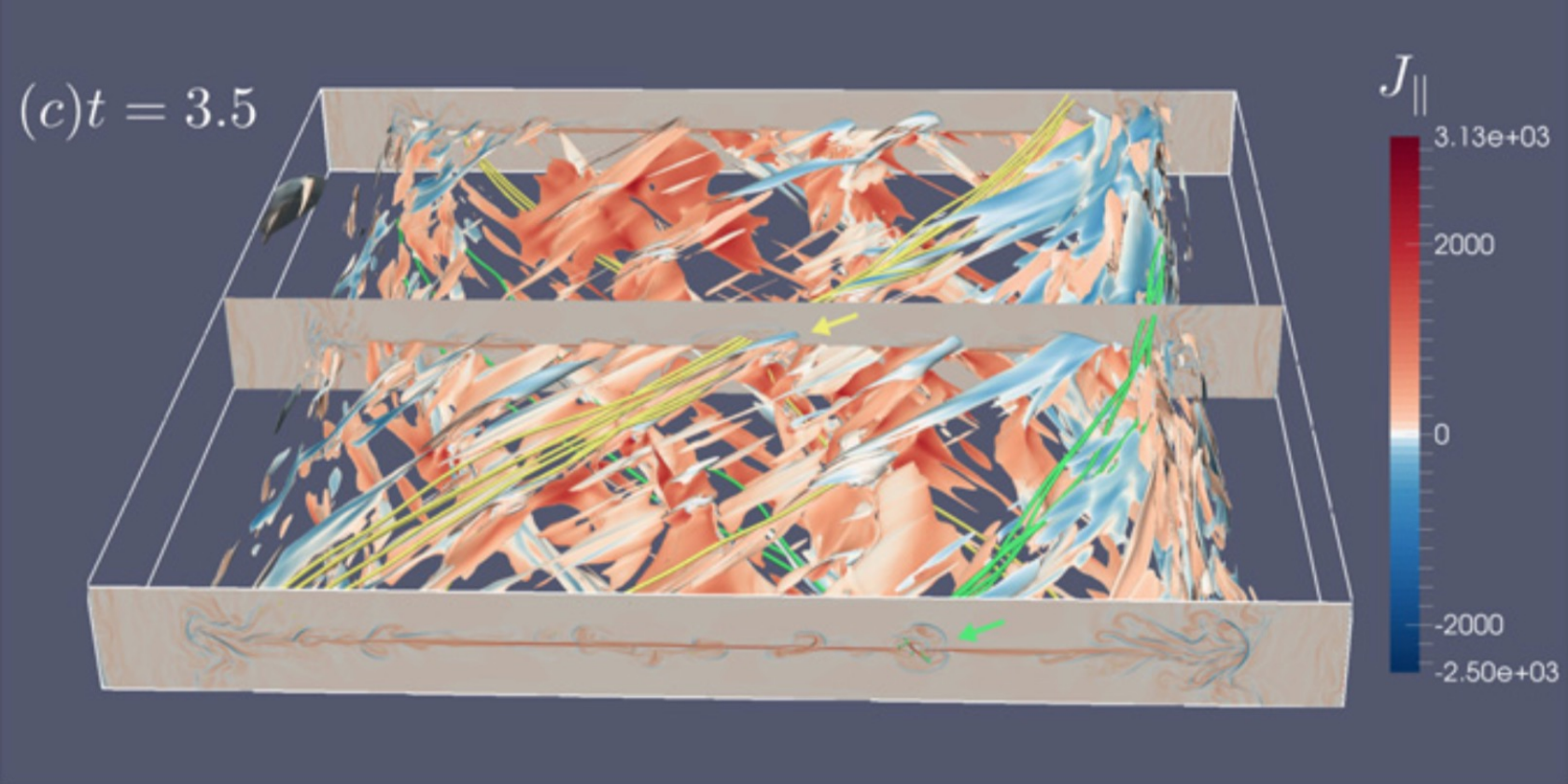}} 
\vskip2mm
\caption{A 3D, turbulent plasmoid (flux-rope) chain obtained in the simulations of 
\citet{huang16} (\copyright AAS, reproduced with permission).}
\label{fig:huang3D}
\end{figure}

In fact, the reconnecting sheets seen in 3D simulations do not resemble all 
too closely the relatively orderly procession of multi-scale plasmoids 
envisioned by \citet{uzdensky10} and seen quite clearly in 2D simulations 
(see \figref{fig:huang} and papers cited in \apref{app:loureiro}). 
In 3D, the chain looks like a strip of vigorous turbulence (\figref{fig:huang3D}), 
even if flux ropes (3D plasmoids) may be identifiable there, 
at least when a mean field is imposed.\footnote{According to \citet{kowal20}, 
this turbulence is driven primarily by Kelvin--Helmholtz instability, not tearing---at 
least at large scales, long times, and in the outflow regions 
(for a theory of KH instability in reconnecting sheets, see \citealt{loureiro13kh}).
However, the simulations by \citet{kowal17} from which that conclusion was drawn 
had an anti-parallel, reconnecting field 10 times larger than the guide field. 
This is the opposite of the RMHD regime that one expects to find locally in the kind 
of MHD turbulence that I have discussed so far, 
where the in-plane field is always small, $\bperp\ll B_0$. 
Of the rest of the 3D papers cited in \apref{app:loureiro},
\citet{oishi15} and \citet{yang20}
had $B_0=0$, and the others $\bperp\sim B_0$, with the exception of \citet{daldorff22},
who probed down to $\bperp\sim0.2 B_0$ and compared 3D and 2D runs. 
Their conclusion is that 3D is generally
messier due to flux ropes getting interlinked with each other. This
somewhat impedes their ability to flow out along the sheet and be
ejected from it in the way plasmoids do in 2D simulations and
in the fast-reconnection model of \citet{uzdensky10}---leading,
as a result, to somewhat slower reconnection. \citet{daldorff22} observe, however,
that, unsurprisingly, cases with lower $\bperp/B_0$ look more like the 2D case.
There still does not appear to exist a systematic 3D study of a reconnecting sheet
in RMHD---besides settling the question of convergence of reconnection rates
between the 3D and 2D cases, it would be more directly relevant 
to aligned structures that arise in the inertial
range of MHD turbulence. \label{fn:3Dsheet_sims}}  
One can think of this situation as a result of the flux ropes constantly 
going unstable and thus seeding turbulent cascades on multiple scales. 
A na\"ive but perhaps instructive model of such ``reconnection driving'' 
of turbulence (promised in \secref{sec:turb_sheet}) 
can be constructed by assuming that the energy density 
\exref{eq:db_eff_chain} in the plasmoids at each level~$n$ describes the 
``outer scale'' of a mini-cascade, which then proceeds in some regular 
MHD manner. For this argument, it does not matter what exactly this 
cascade is like, just that it has some power-law scaling below its mother plasmoid's 
scale~$w^{(n)}$: 
\beq
\dz^{(n)}_\lambda \sim 
\db_\mathrm{eff}^{(n)}\lt(\frac{\lambda}{w^{(n)}}\rt)^\gamma
\sim \vAy\tSc^{1/4}\frac{\lt[w^{(n)}\rt]^{(1 - 2\gamma)/2}\lambda^\gamma}{\ell^{1/2}},
\qquad\lambda < w^{(n)}, 
\eeq 
where, e.g., $\gamma=1/4$ for an aligned MHD cascade (\secref{sec:revised}) 
and $\gamma=3/5$ for a tearing-mediated one (\secref{sec:spectrum_rec}).
The cumulative energy density at scale $\lambda$ from all the mini-cascades is then
\beq
\dz_\lambda^2 \sim \sum_{w^{(n)}>\lambda} \bl[\dz^{(n)}_\lambda\br]^2 
\sim \frac{\vAy^2\lambda^{2\gamma}}{\ell\tSc^{-1/2}}
\int_{\max\{\lambda,\,\delc\}}^{w^{(1)}}\!\!\!\!\!\!\!\!\!\!\!\!
\rmd N(w)\,w^{1-2\gamma}
\sim \vAy^2\min\lt\{1,\lt(\frac{\lambda}{\delc}\rt)^{2\gamma}\rt\},
\eeq
where $N(w)$ is the plasmoid-width distribution~\exref{eq:plasmoid_pdf}. 
The upper bound of the integral is the width $w^{(1)}\sim\ell \tSc^{-1/2}$ 
of the largest plasmoids, 
and the lower bound is the larger of the scale $\lambda$ and 
the width $\delc\sim\ellc\tSc^{-1/2}$ 
of the smallest plasmoids, i.e., ones associated with the 
critical SP sheet~\exref{eq:ellc}. This result means that there 
is a $\kperp^{-1}$ spectrum in the reconnection-driving range~$\kperp\delc\lesssim 1$ 
followed by an MHD turbulence 
spectrum $\propto \kperp^{-2\gamma-1}$ at~$\kperp\delc \gtrsim 1$. 

What this spectrum is depends on whether the cascade is ideal/aligned or 
tearing-dominated. In order to work that out with confidence, 
one has to know rather more than we (or, at least, I) 
currently do about the turbulent dynamics of flux ropes. 
Recall however, that the structures in the driving range  
are already aligned with angle~$\sim\tSc^{-1/2}$. 
It is, therefore, possible, and indeed plausible, 
that the MHD cascade seeded by unstable flux ropes might be  
tearing-mediated, as mooted in~\secref{sec:turb_sheet}. 

To conclude, under the simplistic scheme explored above, 
reconnection-driven turbulence at small enough scales 
appears to be some form of regular MHD turbulence, 
but with a broad scale range into which energy 
is injected directly by reconnection processes---broad asymptotically 
but certainly not captured in full asymptotic glory 
by any realisable numerical experiment. 
This might explain a degree of discord in the scalings reported 
in the papers cited in~\secref{sec:turb_sheet}. 

\subsection{Stochastic Reconnection and MHD Turbulence}
\label{app:stoch_rec}

As promised in \secref{sec:stoch}, here is a (biased) review of
``stochastic reconnection'', the notion primarily associated today with
\citet[][henceforth~LV99]{lazarian99}---a widely cited paper, which, however, 
has acquired the reputation of being rather hard to understand. 
\citet{eyink11rec} seem to me to have succeeded in explaining it with a degree of clarity 
by adopting somewhat different terms, based on a sizeable body of precursor work 
by \citet{eyink09,eyink11dynamo}. There are many self-reviews of this school of thought, 
of which the most recent and comprehensive one is \citet{lazarian20}, but my attempt below 
may be the first by an external observer. Their main idea is roughly as follows. 

First, let us note that, instead of \exref{eq:flux_bal}, we may 
follow \citet{kulsrud05book} and start our consideration 
of an SP sheet by stating that the width of the sheet must be equal 
to the typical distance that the field lines would diffuse resistively 
in the direction ($x$) transverse to the upstream field (which points in $y$)
over the time that it takes the plasma to transit through the sheet and be ejected out 
of it, viz., 
\beq
\delta \sim (\eta \tout)^{1/2},\quad \tout \sim \frac{\ell}{u_y}
\hence 
\delta \sim \lt(\frac{\eta\ell}{u_y}\rt)^{1/2} 
\sim \frac{\ell}{\sqrt{S_\ell}}\lt(\frac{\vAy}{u_y}\rt)^{1/2},
\label{eq:SP_via_Kulsrud}
\eeq 
which is the same expression as~\exref{eq:flow_bal}. 

LV99, as interpreted by \citet{eyink11rec},\footnote{For the connoisseurs, 
there is, in fact, not complete equivalence between the argument of LV99 
and its interpretation by \citet{eyink11rec}. The former paper, together with
many of its successors and citers, 
believe that their stochastic-reconnection mechanism can only work in~3D,
because magnetic-field 
lines are too topologically constrained in 2D. For \citet{eyink11rec}, there is no 
problem in 2D as Lagrangian trajectories in 2D MHD turbulence still separate quickly. 
\citet{loureiro09} did report fast reconnection in an SP sheet buffeted by 2D turbulence; 
\citet{kulpadybel10} disagreed. A recent study by \citet{sun22} sides 
with \citet{loureiro09} but reports a scaling of the reconnection rate
with the injected turbulent power that is less strong than the $\eps^{1/2}$
originating from \exref{eq:eyink} and explained in detail in what follows. Note that
a key (if possibly not sole) role in the acceleration of reconnection by turbulence
in both \citet{loureiro09} and \citet{sun22} appeared to be played by the
formation of plasmoids, encouraged by turbulence and thus setting in at a lower
critical Lundquist number than in a laminar SP sheet [see~\exref{eq:Sc}]. Thus, 
it is possible that what they see is, in fact, the kind of fast reconnection
described in~\apref{app:uls}.} 
argue that if the sheet is embedded in a turbulent environment,  $\delta$ should instead be 
calculated as the distance by which two magnetic field lines 
initially starting arbitrarily close-by, separate after time $\tout$, 
and that this distance is the same as the distance by which two Lagrangian 
fluid particles separate.
It is this identification between stochastic 
particle trajectories and field lines that requires all the work contained in 
\citet{eyink09,eyink11dynamo}. 
In formal terms, he is able to prove that, in the presence 
of resistivity, the magnetic field at any point in space and time is 
an average over those realisations of a certain random field that end up at that point 
after evolving as ``virtual'' magnetic fields  ``frozen'' into a stochastic flow 
that is the superposition of the Lagrangian turbulent velocity field and a white 
noise with the diffusion constant $\eta$. However small is $\eta$, such fields 
diverge in the same way as Lagrangian trajectories do. 
\citet{eyink13} successfully tested this proposition in a large 
numerical simulation of MHD turbulence.\footnote{\citet{eyink15} takes another look
at this topic, this time explaining in some mathematical detail
how inertial-range motions and fields can be ``coarse-grained'' and shown to be
subject to a kind of renormalised rate of reconnection controlled only by
ideal MHD dynamics. The main idea is, I believe, as I summarised it
in~\secref{sec:stoch}---admittedly, in a simplistic, ``coarse-grained'' way.} 

In fluid dynamics, the stochastic separation of Lagrangian trajectories 
is known as Richardson diffusion: one argues, with \citet{richardson26}, 
that the rate of change of the typical square distance $\dr^2$ between them is 
the turbulent diffusivity associated with velocities at the scale~$\dr$:  
\beq
\frac{\rmd \dr^2}{\rmd t} \sim D(\dr) \sim \du_{\dr}^2\tc
\sim \frac{\du_{\dr}^4}{\eps} \sim \eps^{1/3}\dr^{4/3}. 
\label{eq:richardson}
\eeq 
The last two steps follow from $\du_{\dr}^2/\tc \sim \eps$ 
and $\du_{\dr}\sim(\eps\dr)^{1/3}$. The latter relation might appear 
to be valid only for standard K41 or GS95 turbulence \exref{eq:u_GS95}. 
In fact, it is also valid for the aligned turbulence because 
the velocities in \exref{eq:richardson} must be in the direction 
of the particle separation $\dr$, so we must use the scaling 
of $\du$ with $\xi$, not with $\lambda$---and that is always the 
Kolmogorov scaling, including for the aligned cascade [see \exref{eq:xi_scalings}] 
and even for the tearing-mediated one (see \secref{sec:align_rec}). 
Integrating \exref{eq:richardson} gets us ``superdiffusion'':
\beq
\dr(t) \sim \eps^{1/2}t^{3/2},
\label{eq:superdiff}
\eeq
as long as $\dr(t)<\lCB$ (the outer scale of the strong turbulence), 
or, equivalently, as long as $t$ is shorter than the nonlinear time~$\tnl$ 
at scale $\lCB$. A salient feature here is that $\dr(t)$ is independent 
of the initial separation, which can be arbitrarily small.   
Therefore, the width of the reconnecting sheet and the 
inflow speed are
\beq
\delta \sim \dr(\tout) \sim \eps^{1/2}\lt(\frac{\ell}{u_y}\rt)^{3/2}
\hence
u_x \sim \frac{\delta}{\ell}\,u_y\sim\lt(\frac{\eps\ell}{u_y}\rt)^{1/2}. 
\label{eq:eyink}
\eeq 

This result is all I need to work out what LV99 means 
for the theories of turbulent cascade presented in the main text. I will do this 
in \apsand{app:GS95_rec}{app:aligned_rec}, to which a reader only interested 
in the effect of reconnection on turbulence can safely skip. 
\Apsand{app:LV99strong}{app:LV99weak} are 
for those who are also interested in the effect of turbulence
on reconnection,\footnote{Let me 
mention very briefly another case for fast reconnection, made very vigorously 
in a number of recent papers by \citet[][see references therein to 
precursor papers]{boozer21}. The main idea is similar 
to LV99 and \citet{eyink11dynamo}: fast separation of trajectories and, therefore, 
of field lines, leads to accelerated reconnection. The difference is that 
Boozer is thinking of large-scale, system-specific motions and argues 
that those {\em laminar} flows, which nevertheless will generically have 
chaotic Lagrangian trajectories, are already very efficient mixers of 
any frozen-in, advected fields---temperature, generic passive scalar, 
magnetic field---and thus will do the job. The separation between trajectories 
is exponential in this case, at the rate $\gamma \sim u/\ell$, viz., 
$\dr(t) \sim \dr_0 e^{\gamma t}$. Since it depends on the initial separation, 
which has to be chosen to be the resistive scale, say, 
$\dr_0\sim (\eta/\gamma)^{1/2}$, such a scheme would not help reconnection 
in a sheet: choosing $\tout\sim 1/\gamma$, we would just recover 
\exref{eq:SP_via_Kulsrud}. However, Boozer wants us to abandon our fixation 
on reconnection in sheets, which he argues are a red herring, 
and instead think of generic 3D field configurations. He then estimates 
the reconnection timescale as the time when $\dr(\trec)\sim\ell$, viz., 
$\trec \sim \gamma^{-1}\ln(\ell/\dr_0) \sim (\ell/u)\ln\Rm$. 
This is all fairly plausible---if he is actually wrong, that would be 
for some very subtle reason, e.g., 
if the things that a dynamically strong magnetic field had to do to 
a laminar chaotic flow to stop being amplified by it 
(see papers cited at the end of~\secref{sec:subra})
turned out also to impede the exponential separation of field lines. To be fair, 
\citet{boozer21example} did produce a specific example of 
a laminar flow that does what Boozer wants it to do, so his scheme appears  
to be realisable. However, in turbulent systems 
of the type that interests me here, I am not considering mean-flow 
effects on large scales, while on small scales, presumed aligned structures 
in RMHD limit are likely to have a tendency to collapse (or try to collapse) 
into sheets. Thus, the turbulent separation of trajectories, and field lines, 
according to \exref{eq:superdiff} is a more relevant situation for me 
(and the validity of which Boozer does not reject).} 
and in the status of the numerical evidence on the subject.

\subsubsection{Stochastic Reconnection Mediated by Strong Turbulence}
\label{app:LV99strong} 

LV99 formulated their prediction for the reconnection rate not in terms of 
its scaling with $\eps$, as in \exref{eq:eyink}, but with the Alfv\'enic 
Mach number, which, for the purposes of this exposition, I will define as 
\beq
\MA \equiv \frac{\du_{\Lperp}}{\vAy}, 
\eeq
where $\vAy$ is the in-plane Alfv\'en speed associated with the upstream 
(reconnecting) magnetic field and $\du_{\Lperp}$ is the turbulent velocity field 
at the outer scale $\Lperp$. How to express the prediction \exref{eq:eyink} 
in terms of $\MA$ depends on how the turbulence is driven. 

In LV99, it is driven weakly, so, using the standard WT scaling~\exref{eq:WT_standard}, 
they have
\beq
\du_{\Lperp} \sim \lt(\frac{\eps\vA}{\Lpar}\rt)^{1/4}\Lperp^{1/2}
\hence
\eps \sim \frac{\du_{\Lperp}^4\Lpar}{\vA\Lperp^2}
= \MA^4 \frac{\vAy^4\Lpar}{\vA\Lperp^2}. 
\label{eq:eps_MA}
\eeq
In their model, in fact, the driving is isotropic and the mean field 
is of the same order as the in-plane field: $\Lpar \sim \Lperp \equiv L$ 
and $\vAy\sim\vA$ (but $\MA\ll1$, so the turbulence is indeed weak). 
This is outside the RMHD limit, but it is probably 
fine to extrapolate ``twiddle'' scalings to this regime: $\eps \sim \MA^4\vA^3/L$.
In what follows, I shall keep the anisotropic scaling~\exref{eq:eps_MA} 
but show how the main results simplify in the isotropic case and reduce to LV99.
 
There are, obviously, two distinct possibilities: when $\Lperp$ is much larger 
and much smaller than the sheet width~$\delta$. I shall deal with the latter 
in \apref{app:LV99weak}. In the former regime, the formula \exref{eq:eyink} applies 
if $\delta$ is small enough that turbulence is already in the strong 
regime at that scale, viz.,~if 
\beq
\Lperp \gg \lCB=\eps^{1/2}\lt(\frac{\Lpar}{\vA}\rt)^{3/2} \gtrsim \delta, 
\label{eq:Lperp_lCB_delta}
\eeq 
where the CB scale $\lCB$ is given by \exref{eq:lCB}. 
Using \exref{eq:eyink}, we find that the second inequality 
in \exref{eq:Lperp_lCB_delta} is equivalent to 
\beq
\frac{\delta}{\lCB} \sim \lt(\frac{\ell\vA}{\Lpar u_y}\rt)^{{3/2}} \lesssim 1. 
\label{eq:ST_cond}
\eeq
If this is true, then, even though turbulence is driven weakly at scale~$\Lperp$, 
stochastic reconnection is mediated by strong turbulence at scale~$\delta$.   
Assuming \exref{eq:ST_cond} to be true and using \exref{eq:eyink} again, 
one gets the reconnection rate 
\beq
\frac{u_x}{\vAy} \sim \MA^2\frac{\vAy\Lpar}{\vA\Lperp}
\lt(\frac{\ell\vA}{\Lpar u_y}\rt)^{1/2}
\sim \MA^2\lt(\frac{\ell}{L}\rt)^{1/2},
\label{eq:LVstrong}
\eeq
the last expression having been obtained for the case of isotropic driving 
and taking also the outflow to be Alfv\'enic, $u_y\sim\vAy$. 

This is the well-known LV99 prediction $u_x\propto\MA^2$, which, to numericists, 
has all the attraction of a testable result. 
In a recent study by \citet{yang20}, they tried to test it 
and found instead $u_x\propto\MA$, causing some concern about the validity 
of LV99. In fact, while their result disagrees with \exref{eq:LVstrong}, it 
seems to me to be perfectly consistent with \exref{eq:eyink} if one can assume 
that motions comparable in size to the box-wide rms velocity exist on the scale 
of the sheet width---either because the sheet generates its own turbulence 
(as it does in such simulations when they are not externally driven: 
see \secref{sec:turb_sheet} and references therein), or because it is able 
locally to chop down the driven energy-containing motions to its own scale
(\citealt{yang20} did force their turbulence externally). 
Namely, let us assume that, effectively, for the sheet,  
\beq
\Lperp\sim\lCB\sim\delta \hence \frac{\ell}{u_y}\sim\frac{\Lpar}{\vA}. 
\label{eq:lCB_delta}
\eeq
The latter relationship is a kind of CB: the time scale for the incoming matter and 
flux to be taken out of the sheet is the same as for an Alfv\'en wave to travel 
the correlation length along the guide field. Using $\Lperp\sim\lCB$
in the strong-turbulence scaling~\exref{eq:MS_normalised}, we~get 
\beq
\du_{\Lperp}\sim\lt(\frac{\eps\Lpar}{\vA}\rt)^{1/2}\hence
\eps \sim \frac{\du_{\Lperp}^2\vA}{\Lpar} \sim \MA^2\frac{\vA\vAy^2}{\Lpar}. 
\label{eq:eps_MA_CB}
\eeq
The formula \exref{eq:eyink} now gives 
\beq
\frac{u_x}{\vAy} \sim \MA\lt(\frac{\ell\vA}{\Lpar u_y}\rt)^{1/2} \sim \MA,
\eeq
the last expression following from \exref{eq:lCB_delta}. This result is almost 
trivial: $u_x\sim\du_{\Lperp}$, so the sheet sucks in flux 
at the typical velocity of turbulent motions inside~it. 

\subsubsection{Stochastic Reconnection Mediated by Weak Turbulence}
\label{app:LV99weak} 

\begin{figure}
\begin{center}
\begin{tabular}{cc}
\parbox{0.4\textwidth}{
\includegraphics[width=0.4\textwidth]{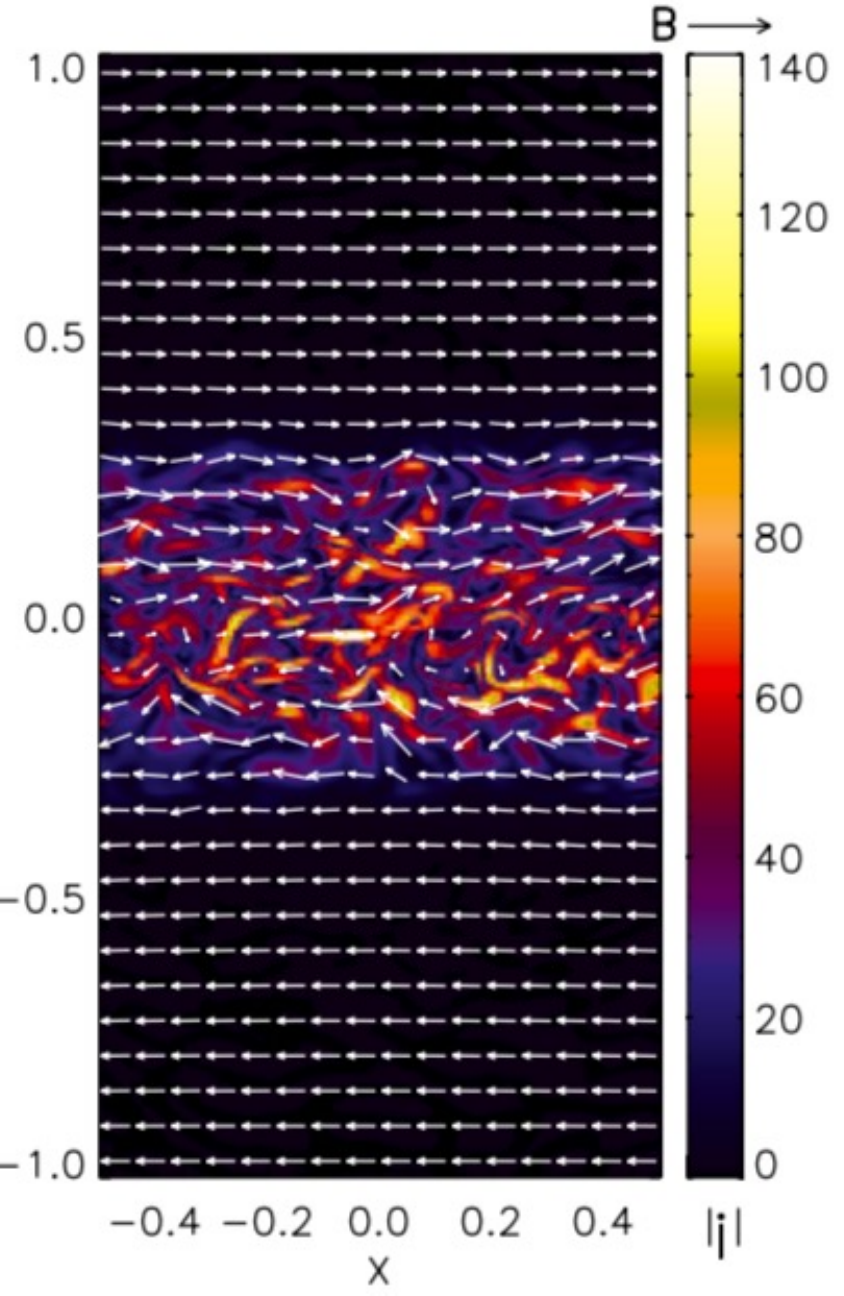}} &
\parbox{0.55\textwidth}{
\includegraphics[width=0.55\textwidth]{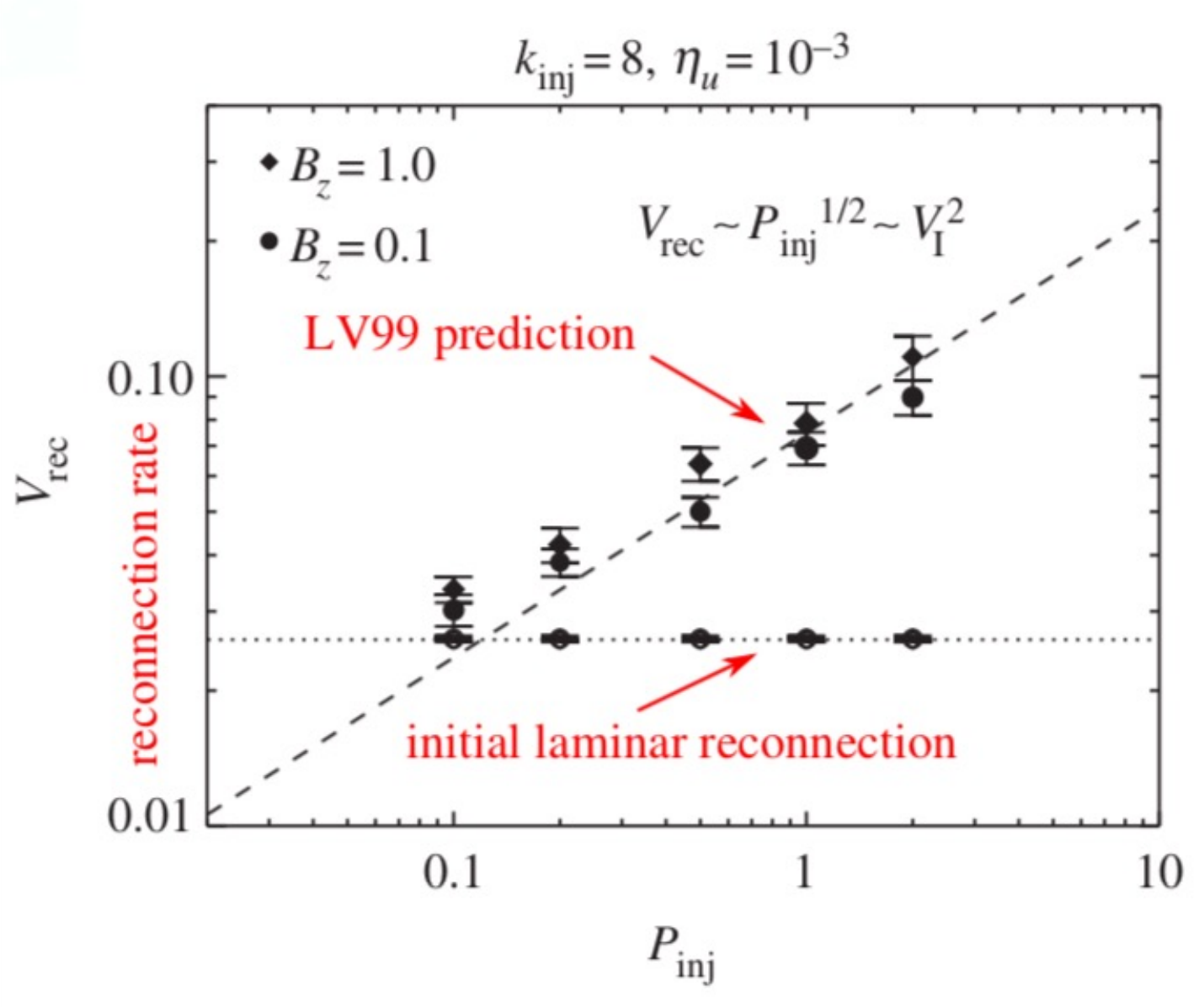}} \\\\
(a) & (b)
\end{tabular}
\end{center}
\caption{Simulations of stochastic reconnection by \citet{kowal09}: 
(a) arrows are magnetic fields, colour shows (turbulent) currents; 
(b) reconnection rate $V_\mathrm{rec}$ vs.\ injected power $P_\mathrm{inj}$, 
which, in my notation, are $u_x$ and $\eps$, respectively---this plot, 
taken from \citet{lazarian15}, shows $u_x\propto\eps^{1/2}$, 
in accordance with~\exref{eq:eyink_weak}.} 
\label{fig:kowal}
\end{figure}

Historically the first suite of confirmatory numerics for the LV99 theory was 
reported by \citet{kowal09,kowal12} (usefully summarised by \citealt{lazarian15}), 
who, in their simulations of a macroscopic magnetic field reconnecting in 
the presence of externally injected turbulence, did indeed see the scaling 
$u_x\propto\eps^{1/2}$, independent of~$\eta$ (\figref{fig:kowal}). 
Most of their simulations, however, seem to be in the second regime 
considered by LV99, one in which $\Lperp \lesssim \delta$. 

Since the turbulence that they are injecting into the sheet is still weak,  
and its scale is smaller than the width of the sheet,  
the expression for the turbulent diffusion coefficient used in 
\exref{eq:richardson} no longer applies (unless the sheet manages to generate 
its own turbulence, which, in their simulations, it does not appear to do---perhaps 
because it is quite short?). Instead, we have 
\beq
\frac{\rmd\dr^2}{\rmd t} \sim D(\Lperp) \sim \frac{\du_{\Lperp}^2}{\tc\oA^2} 
\sim \frac{\eps}{\oA^2},
\label{eq:richardson_WT}
\eeq
where $\oA\sim\vA/\Lpar$ is the Alfv\'en frequency and 
$\tc$ is the correlation time of WT [it is given by \exref{eq:tc_weak}, 
but we do not need to know this to get to the last expression]. 
An easy way to understand why these two characteristic time scales 
appear in \exref{eq:richardson_WT} in the way they do is as follows. 
The turbulent diffusion coefficient is just the time integral of the 
two-time correlation function of the velocity field, and we can calculate 
it by assuming a simple functional form for this correlation function 
characteristic of a slowly decorrelating but fast-oscillating wave field: 
\beq
D \sim \int_0^\infty\rmd \tau\la\vuperp(t)\cdot\vuperp(t-\tau)\ra
\sim \du_{\Lperp}^2 \int_0^\infty\rmd \tau\,e^{-\tau/\tc}\cos(\oA\tau)
= \frac{\du_{\Lperp}^2}{\tc(\oA^2 + 1/\tc^2)}. 
\eeq    
In the absence of waves, or in CB ($\oA\sim\tc^{-1}$), one recovers 
the usual formula $D\sim\du_{\Lperp}^2\tc$, analogous to \exref{eq:richardson}, 
whereas for $\oA\gg\tc^{-1}$, one gets~\exref{eq:richardson_WT}. 

Thus, adopting~\exref{eq:richardson_WT}, we get, 
in the same way as we did in~\exref{eq:eyink},   
\beq
\dr(t) \sim \frac{(\eps t)^{1/2}}{\oA}
\hence 
\delta \sim \lt(\frac{\eps\ell}{u_y}\rt)^{1/2}\frac{\Lpar}{\vA},
\qquad
u_x \sim \biggl(\frac{\eps u_y}{\ell}\biggr)^{1/2}\frac{\Lpar}{\vA}. 
\label{eq:eyink_weak}
\eeq
With the aid of \exref{eq:eps_MA}, the reconnection rate in terms of $\MA$~becomes
\beq
\frac{u_x}{\vAy}\sim \MA^2\frac{\vAy\Lpar}{\vA\Lperp}
\lt(\frac{\Lpar u_y}{\ell\vA}\rt)^{1/2}
\sim \MA^2\lt(\frac{L}{\ell}\rt)^{1/2},
\label{eq:LVweak}
\eeq
the last expression being for an isotropic forcing and Alfv\'enic outflows, 
as in \exref{eq:LVstrong}. This prediction too appears in LV99, although 
derived by a slightly different route. Yet again, $u_x\propto \eps^{1/2} \propto \MA^2$, 
which is what \citet{kowal09} confirmed (\figref{fig:kowal}). 

Note that in \exref{eq:richardson_WT}, the $\Lperp$ dependence disappeared 
from the expression for turbulent diffusivity. 
This means that actually the above calculation remains valid also for 
$\Lperp \gg \delta$---as long as $\delta\gg\lCB$, i.e., as long as the turbulence 
is weak at the scale $\delta$. 
In this case, one must replace $\Lperp$ with $\dr(t)$ in \exref{eq:richardson_WT}, 
but the result is the same because, as we have just seen, $D(\dr)$ is independent of $\dr$ 
for WT---the divergence of Lagrangian trajectories in WT is always just diffusive, 
Richardson superdiffusion \exref{eq:superdiff} being a distinctive property 
of strong turbulence. Therefore, \exref{eq:LVweak} only breaks down and transitions 
into~\exref{eq:LVstrong} when $\lCB$ becomes comparable to $\delta$, 
i.e., when \exref{eq:ST_cond} becomes true. In summary,  
\beq
\frac{u_x}{\vAy}\sim \MA^2\frac{\vAy\Lpar}{\vA\Lperp}
\min\lt\{\lt(\frac{\ell\vA}{\Lpar u_y}\rt)^{1/2},
\lt(\frac{\Lpar u_y}{\ell\vA}\rt)^{1/2}\rt\},
\eeq
the LV99 magic formula for the rate of stochastic reconnection, generalised to 
the case of anisotropic driving. 

\subsubsection{Stochastic Reconnection in GS95 Turbulence}
\label{app:GS95_rec}

Let us now apply \exref{eq:eyink} to a typical turbulent structure in which 
$\ell=\xi$ and $u_y\sim\du_\xi\sim(\eps\xi)^{1/3}$. This instantly implies 
\beq
\delta \sim \xi,\quad u_x \sim \du_\xi. 
\label{eq:GS95_rec}
\eeq
For GS95 turbulence ($\xi\sim\lambda$), this means that reconnection 
of field lines within each ``eddy'' occurs on the same time scale as the 
``eddy'' turns over---this is, I think, what \citet{lazarian15} mean 
when they say that stochastic reconnection makes GS95 turbulence 
``self-consistent''.

\subsubsection{Stochastic Reconnection in Aligned Turbulence}
\label{app:aligned_rec}

What if the turbulence is aligned? According to my argument above, 
in view of \exref{eq:xi_scalings}, it might seem that 
\exref{eq:GS95_rec} remains valid. This is worrisome: indeed 
this tells us that the width of the layer over which stochastic reconnection 
would be happening is larger than the width of the aligned structure: 
$\delta\sim\xi\gg\lambda$! If this were true, writing this review would 
have been a waste of time: 
aligned structures would be quickly broken up by stochastic reconnection, 
so there would be no aligned cascade. This would invalidate all of \secref{sec:DA}  
and, consequently, obviate any consideration of tearing-mediated turbulence 
in \secref{sec:disruption}---the cascade would just be GS95 all the way to 
the dissipation scale. \citet{lazarian15} (and their previous papers referenced 
therein) might then be excused for (politely) ignoring 
all the newfangled turbulence theory postdating GS95, 
and \citet{beresnyak11,beresnyak12,beresnyak14,beresnyak19} would be vindicated 
much more completely than I could offer to do in~\secref{sec:diss}. 
I cannot prove formally that this does not or cannot happen (cf.~\secref{sec:KHdisaster}), 
but I can show that aligned turbulence is, in fact, not ruled out by stochastic reconnection. 

Let us imagine that an aligned structure has emerged with transverse (to fluctuating fields) 
scale $\lambda$ and longitudinal (fluctuation-direction) scale $\xi\gg\lambda$. 
In order for stochastic reconnection to destroy it quickly, 
there must be turbulent structures within the layer of width $\lambda$ 
whose longitudinal scales are as large as $\sim\lambda$.  
But within this layer, there is an intense shear $\sim\du_\lambda/\lambda$, 
which should suppress any turbulent motions whose typical nonlinear time 
scales are longer than the inverse of this shear. 
For small enough structures, the nonlinear times will be short and eventually 
overcome the shear. Let us find the longitudinal scale $\xisub$ of the largest possible 
such motions: their nonlinear decorrelation rate is
\beq
\frac{\du_{\xisub}}{\xisub} \sim \frac{\du_\lambda}{\lambda}
\hence 
\frac{\xisub}{\lambda} \sim 
\lt(\frac{\lambda}{\xi}\rt)^{1/2}
\sim \lt(\frac{\lambda}{\lCB}\rt)^{1/8} \ll 1,
\label{eq:xisub}
\eeq 
where I have used \exref{eq:xi_scalings} for $\du_{\xisub}\sim(\eps\xisub)^{1/3}$ 
and $\du_\lambda=\du_\xi\sim(\eps\xi)^{1/3}$.\footnote{NB: $\xi$ is a function of $\lambda$; 
by $\du_\xi$, I mean $\du_\lambda$ expressed in terms of $\xi$, 
not $\du_\lambda$ taken at $\lambda=\xi$.} 
If these motions are aligned in the usual way, with transverse 
scale $\lsub$, then, using \exref{eq:xi_scalings} again, 
$\lsub/\lambda\sim(\lambda/\lCB)^{1/2}$. 

Going back to \exref{eq:richardson}, one must now integrate 
this equation up to time $\tout\sim\xi/\du_\xi\sim\eps^{-1/3}\xi^{2/3}$, 
which is longer than the nonlinear time $\xisub/\du_{\xisub} \sim \eps^{-1/3}\xisub^{2/3}$
of the largest turbulent structures within our layer. This gives 
conventional turbulent diffusion: 
\beq
\delta \sim \dr(\tout) \sim \lt(\eps^{1/3}\xisub^{4/3}\tout\rt)^{1/2}
\sim \xisub^{2/3}\xi^{1/3}\sim\lambda, 
\label{eq:delta_aligned}
\eeq
where the last step follows from \exref{eq:xisub}. 
Just like in \exref{eq:GS95_rec}, the width of the stochastically   
reconnecting layer is the same as the width of the (now aligned) structure, 
so the magnetic fields in the aligned cascade reconnect just as fast as 
the turbulent structures decorrelate. Thus, the aligned cascade is consistent 
with stochastic reconnection. 

The same is going to be true of the 
tearing-mediated cascade of \secref{sec:recturb} because, 
in the argument leading to \exref{eq:delta_aligned}, 
all I needed was the Kolmogorov scaling 
of the turbulent velocities in the fluctuation direction, which 
is always preserved (\secref{sec:align_rec}).  
The competition between the nonlinear decorrelation rate 
and the tearing rate that leads to disruption~(\secref{sec:disruption_scale}) 
is unaffected by all this because disruption happens precisely 
at the scale where tearing becomes ``ideal''. Any smaller-scale turbulence, 
ambient or created by the tearing, can presumably be viewed as providing seed 
perturbations for the instability. 

The overall conclusion appears to be that stochastic reconnection, while 
a useful notion in the treatment of large-scale magnetic-field configurations 
with externally imposed turbulence, 
does not undermine (or modify) the existing theory of the aligned or tearing-mediated 
turbulence, but rather plays an interpretative role: it provides 
a further insight into the behaviour of tangled magnetic fields in a turbulent 
environment and reassures us that, whatever topological rearrangements 
are necessary for the cascade to proceed, they can always occur on the time 
scales of the cascade. 

\bibliography{../JPP/bib_JPP}{}
\bibliographystyle{jpp}

\end{document}